%% file: panda_tdr_mvd.tex
\documentclass[pdftex]{sty/panda_book}
\markpanda{Strong interaction studies with antiprotons}{Technical Design Report - MVD}
\usepackage[sectionbib]{bibunits}
\usepackage[toc,page]{appendix}
\usepackage{wrapfig}
\usepackage{comment}

\usepackage{caption}
\usepackage{subfig}

\usepackage{hyperref}  
\hypersetup{
			pdfborder=0 0 0,
		    pdftitle={PANDA Technical Design Report - MVD},
			pdfauthor={The PANDA Collaboration}
}

\newcounter{dummy}

\begin{document}

%
% Main Sections
%
%\pagenumbering{roman}
\input{main/titlepages}

\cleardoublepage
\pagenumbering{arabic}
\setcounter{page}{1}
\bibliographyunit[\chapter]
\defaultbibliographystyle{unsrt}
%\defaultbibliographystyle{plain}

% Extra command definitions to become uniform in selected style questions
\input{sty/mvd_commands}

%\cleardoublepage
\input{introduction/MainIntro/intro}

\cleardoublepage \input{introduction/introduction}

\cleardoublepage \input{pixelpart/pixelpart}

\cleardoublepage \input{strippart/strippart}

\cleardoublepage \input{integration/offdetector}

\cleardoublepage \input{simulations/simulations}

\cleardoublepage \input{projectmanagement/projectmanagement}

\appendix
\appendixpage
%\addappheadtotoc

\cleardoublepage \input{appendix/bonn-tracking-station}
\cleardoublepage \input{appendix/juelich-readout-system}
\cleardoublepage \input{appendix/sim-vertexing}
\cleardoublepage \input{appendix/aliTech}

\cleardoublepage
\refstepcounter{dummy}
\input{main/acronyms}
\cleardoublepage
\refstepcounter{dummy}
\addcontentsline{toc}{chapter}{List of Figures}
\listoffigures

\cleardoublepage
\refstepcounter{dummy}
\addcontentsline{toc}{chapter}{List of Tables}
\listoftables

\end{document}

%% file: main/titlepages.tex
%\pagenumbering{roman}
\thispagestyle{empty}
\onecolumn
%
% PAGE I - Title and Abstract and Figure
%
%\vspace*{0.5cm}
\begin{center}
{\bfseries \sffamily \huge Technical Design Report for the:\\ \ \\ \Panda{} \\  Micro Vertex Detector \\
{\sffamily \small (Anti\underline{P}roton \underline{An}nihilations at \underline{Da}rmstadt)}\\
\ \\ Strong Interaction Studies with Antiprotons}
\vskip 1cm
{\large \sffamily \Panda{} Collaboration}
%
%removed for final version (date)
%\vskip 0.5cm
%\fbox{\today}
%
\vskip 2cm
\end{center}
\vskip 1cm

\begin{center}
% put a typical picture here
\includegraphics[width=1.7\swidth]{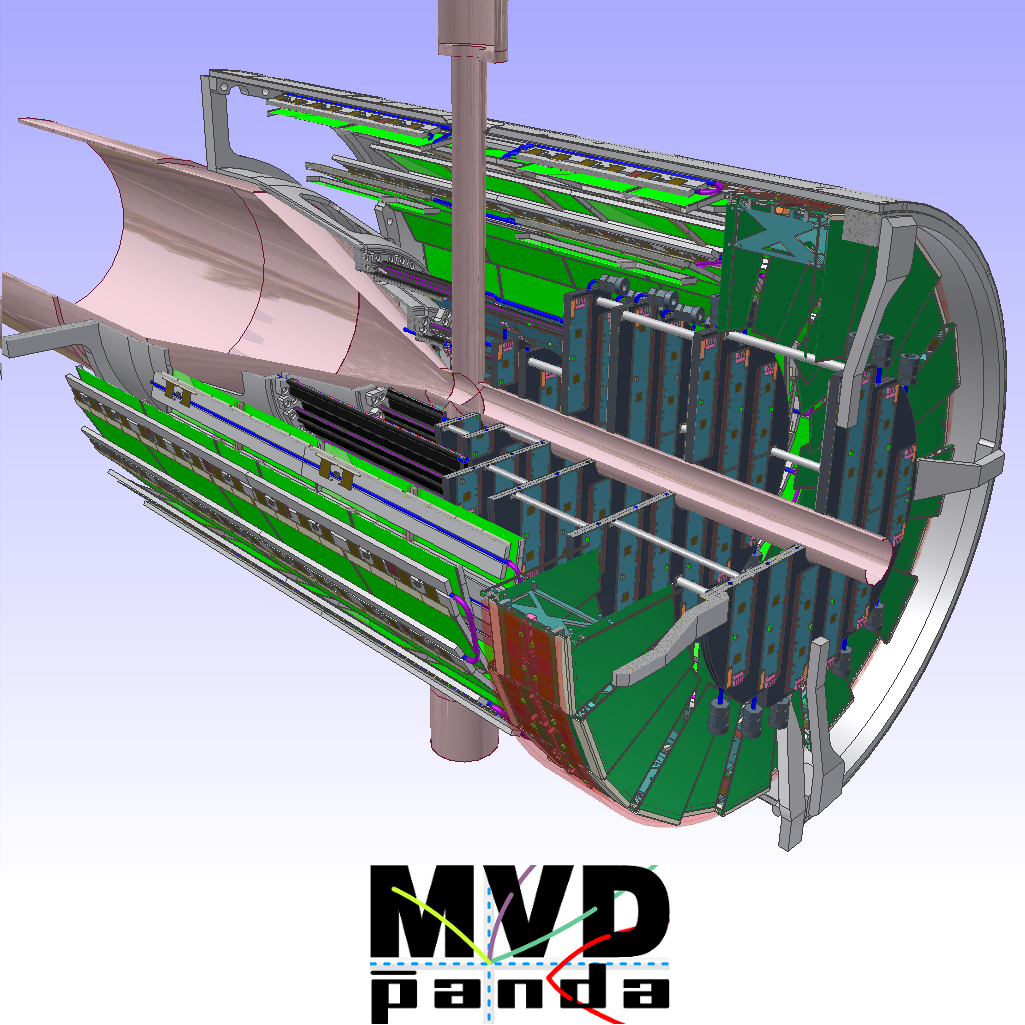}
\end{center}
\vfill
\clearpage{\pagestyle{empty}\cleardoublepage}
%\cleardoublepage
%
% new second page with date
%
\pagenumbering{roman}
\thispagestyle{empty}
\onecolumn
\vspace*{1.5cm}
\begin{center}
{\bfseries \sffamily \huge Technical Design Report for the:\\}
\vskip 2.5cm
{\bfseries \sffamily \huge \Panda{} \\  Micro Vertex Detector \\}
{\bfseries \sffamily \huge {\Large (Anti\underline{P}roton \underline{An}nihilations at \underline{Da}rmstadt)}\\}
\vskip 2.5cm
{\bfseries \sffamily \huge Strong Interaction Studies with Antiprotons}
\vskip 4cm
{\huge \sffamily \Panda{} Collaboration}
\vskip 2cm
\end{center}
\vskip 2cm

\begin{center}
% put a typical picture here
{\LARGE \sffamily 30$\mathsf{^{th}}$ November 2011}
\end{center}
\vfill
%
% PAGE II-IV - Collaboration
%
\clearpage{\pagestyle{empty}\cleardoublepage}

%\cleardoublepage
\begin{center}
\vspace*{3mm }
{\LARGE \bfseries \sffamily The \Panda{} Collaboration}
\vskip 7mm
\input{./main/authors}
\end{center}
%
% Spokespersons
%
\vfill
\hrulefill\\
\begin{tabbing}
Editors:  \hspace{3cm} \= K.-Th. Brinkmann   \hspace{1cm}  \= Email: \verb$Brinkmann@hiskp.uni-bonn.de$ \\
                       \> D. Calvo \>  Email: \verb$calvo@to.infn.it$ \\
                       \> T. Stockmanns  \>  Email: \verb$t.stockmanns@fz-juelich.de$ \\
\\
Technical Coordinator: \> Lars Schmitt  \> Email: \verb$l.schmitt@gsi.de$\\
Deputy:                       \> Bernd Lewandowski \> Email:  \verb$b.lewandowski@gsi.de$\\
\\
Spokesperson:  \> Ulrich Wiedner  \> Email:  \verb$ulrich.wiedner@ruhr-uni-bochum.de$\\
Deputy: \> Paola Gianotti \> Email:  \verb$paola.gianotti@lnf.infn.it$
\end{tabbing}
\hrulefill\\

%\clearpage

%\hrulefill\\
MVD Group: 
\begin{tabbing}
\hspace{0.1\textwidth}\=\hspace{0.3\textwidth}\= \hspace{0.3\textwidth} \=\hspace{0.3\textwidth}\\

\>L.~Ackermann$^2$ \>             H.~Kleines$^6$ \>             M.~Ramm$^6$ \\
\>M.~Becker$^1$\>             R.~Kliemt$^1$\>             O.~Reinecke$^2$\\
\>S.~Bianco$^1$\>             K.~Koop$^1$\>             J.~Ritman$^4$ \\
\>K.-Th.~Brinkmann$^1$\>        M.~Kosmata$^2$\>           A.~Rivetti$^8$  \\       
\>L.~Busso$^9$\>             F.~Kr\"uger$^2$\>             R.~Schnell$^1$\\
\>D.~Calvo$^8$ \>             T.~Kugathasan$^9$ \>     H.~Sohlbach$^3$\\        
\>S.~Coli$^8$  \>             S.~Marcello$^9$\>            T.~Stockmanns$^4$\\ 
\>F.~De~Mori$^9$ \>             G.~Mazza$^8$  \>             J.~Tummo$^1$\\
\>P.~De~Remigis$^8$\>            M.~Mertens$^4$\>             P.~Vlasov$^1$\\ 
\>D.~Deermann$^4$\>             M.~Mignone$^8$  \>         R.~Wheadon$^8$ \\    
\>S.~Esch$^4$ \>             O.~Morra$^7$\>             Th.~W\"urschig$^1$\\
\>A.~Filippi$^8$  \>             J.~Pfennings$^5$\>          P.~W\"ustner$^6$ \\  
\>G.~Giraudo$^8$  \>             R.~Pietzsch$^2$\>      H.-G.~Zaunick$^1$\\
\>D.~Grunwald$^5$\>             A.~Pitka$^1$\>               L.~Zotti$^9$ \\
\>R.~J\"akel$^2$\>             D.-L.~Pohl$^4$ \\
\>V.~Jha$^4$ \>             T.~Quagli$^9$\\

\end{tabbing} % centering

$^1$ HISKP, Universit\"at {\bf Bonn}, Germany\\
$^2$ IKTP, TU {\bf Dresden}, Germany\\
$^3$ FH S\"udwestfalen, Fachbereich Informatik und Naturwissenschaften, {\bf Iserlohn}, Germany\\
$^4$ Forschungszentrum J\"ulich, Institut f\"ur Kernphysik, {\bf J\"ulich}, Germany\\
$^5$ ZAT, {\bf J\"ulich}, Germany\\
$^6$ ZEL, {\bf J\"ulich}, Germany\\
$^7$ INAF-IFSI and INFN, Sezione di~Torino, {\bf Torino}, Italy\\
$^8$ INFN, Sezione di~Torino, {\bf Torino}, Italy\\
$^9$ Universit\`{a} di Torino and INFN, Sezione di~Torino, {\bf Torino}, Italy\\

\hrulefill\\

\vfill
%
% preamble
%
\cleardoublepage
\input{./main/preamble}
%
% Table of contents
%
\cleardoublepage
\tableofcontents
\newpage
\thispagestyle{empty}
\mbox{}
\clearpage{\pagestyle{empty}\cleardoublepage}
%
% EOF
%

%% file: main/authors.tex
%
%list of institutions
%
\institem{Universit\"at {\bf Basel}, Switzerland}
\authitem{W.~Erni},
\authitem{I.~Keshelashvili},
\authitem{B.~Krusche},
\authitem{M.~Steinacher}\lastitem
\institem{Institute of High Energy Physics, Chinese Academy of Sciences, {\bf Beijing}, China}
\authitem{Y.~Heng},
\authitem{Z.~Liu},
\authitem{H.~Liu},
\authitem{X.~Shen},
\authitem{Q.~Wang},
\authitem{H.~Xu}\lastitem
\institem{Universit\"at {\bf Bochum}, I. Institut f\"ur Experimentalphysik, Germany}
\authitem{M.~Albrecht},
\authitem{J.~Becker},
\authitem{K.~Eickel},
\authitem{F.~Feldbauer},
\authitem{M.~Fink},
\authitem{P.~Friedel},
\authitem{F.H.~Heinsius},
\authitem{T.~Held},
\authitem{H.~Koch},
\authitem{B.~Kopf},
\authitem{M.~Leyhe},
\authitem{C.~Motzko},
\authitem{M.~Peliz\"aus},
\authitem{J.~Pychy},
\authitem{B.~Roth},
\authitem{T.~Schr\"oder},
\authitem{J.~Schulze},
\authitem{M.~Steinke},
\authitem{T.~Trifterer},
\authitem{U.~Wiedner},
\authitem{J.~Zhong}\lastitem
% ...
%
% EOF
%

%% file: main/preamble.tex
% preamble.tex
%
\begin{center}
\vspace*{2cm}
{\Large \bfseries \sffamily Preface}\addcontentsline{toc}{chapter}{Preface}
\vskip 2cm
\begin{minipage}[t]{8cm}
\sloppy\large
This document illustrates the technical layout and the expected performance
of the Micro Vertex Detector (MVD) of the \Panda experiment. The MVD will
detect charged particles as close as possible to the interaction zone. 
Design criteria and the optimisation process as well as the technical 
solutions chosen are discussed and the results of this process are
subjected to extensive Monte Carlo physics studies. The route towards
realisation of the detector is outlined.\\
% \COM{to be continued}

\end{minipage}
\end{center}
\vspace*{2cm}
\centerline{
% add a detector picture here
%\includegraphics[width=0.8\dwidth]{./main/STT.pdf}
}
\vfill

\clearpage
\vspace*{18cm}
\hrulefill\\
\vspace*{2cm}\\
\begin{minipage}[t]{10cm}
\sloppy
The use of registered names, trademarks, \etc in this publication does not
imply, even in the absence of specific statement, that such names are exempt
from the relevant laws and regulations and therefore free for general use.
\end{minipage}
\vfill
% EOF
%

%% file: sty/mvd_commands.tex
% Some definitions (esp. to avoid mixed-up naming):
\newcommand{\ra}{$\rightarrow$}
\newcommand{\rt}{ROOT }
\newcommand{\rts}{ROOTs }
\newcommand{\frt}{FairRoot }
\newcommand{\pndrt}{PandaRoot }
\newcommand{\pnd}{\Panda}

% alert command
\newcommand{\alert}[1]{\textcolor{red}{#1}} 
\newcommand{\authalert}[1]{} % empty to hide
\newcommand{\MARK}[1]{\textcolor{red}{#1}} 
% Bold note
\newcommand{\Note}[1]{\textcolor{red}{\footnotesize{\bf{#1}}}}
% Italic note: \MARK already defined in panda_commands

% Referencing figures with full Word or abbreviation? Set here.
\newcommand{\Figref}[1]{Figure~\ref{#1}}
\newcommand{\figref}[1]{figure~\ref{#1}}
\newcommand{\Figsref}[1]{Figures~\ref{#1}}
\newcommand{\figsref}[1]{figures~\ref{#1}}
\newcommand{\Tabref}[1]{Table~\ref{#1}}
\newcommand{\tabref}[1]{table~\ref{#1}}

% lazy definitions to worry with pictures later
\newcommand{\mypicture}[5]{\begin{figure}[]\centering\includegraphics[width=#2\textwidth]{#1} \caption[#3]{#4}\label{#5}\end{figure}}
\newcommand{\mypicturebig}[5]{\begin{figure*}[]\centering\includegraphics[width=#2\textwidth]{#1} \caption[#3]{#4}\label{#5}\end{figure*}}
\newcommand{\pict}[1]{\mypicture{simulations/pictures/Pandaroot-Logo1.png}{0.2}{}{\alert{put picture #1}}{}}
\newcommand{\prepic}[1]{\mypicture{#1}{0.4}{}{\alert{a picture}}{}}
\newcommand{\prepicbig}[1]{\mypicturebig{#1}{0.8}{}{\alert{a picture}}{}}

% code formats
%\small %\footnotesize %\scriptsize
\newcommand{\code}[1]{\texttt{\small{#1}}}
\newcommand{\codeline}[1]{\begin{quote}\texttt{\small{#1}}\end{quote}}
\newcommand{\codetable}[1]{\noindent\texttt{\small{\begin{tabular}{|p{\textwidth}|}\hline#1\hline\end{tabular}}}}

% EOF

%% file: introduction/MainIntro/intro.tex
\chapter{The \Panda Experiment and its Tracking Concept}
%\addcontentsline{toc}{chapter}{Overview of the \Panda experiment and the overall tracking concept}

%Author: T. Würschig , contact: t.wuerschig-at-hiskp.uni-bonn.de

The following sections contain a general introduction to the \Panda experiment
and, in particular, a short description of the implemented overall tracking concept.
They belong to a common introductory part for the volumes 
of all individual tracking systems.

\section{The \Panda Experiment}
%\section*{I.\hspace{6mm} The \Panda experiment}
%\addcontentsline{toc}{section}{I.\hspace{6mm} The \Panda experiment}

\begin{figure}[!b]
\begin{center}
\vspace{-4mm}
 \includegraphics[width=7.5 cm]{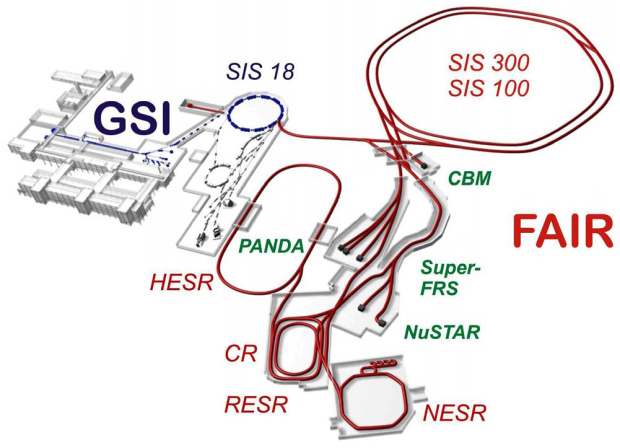}
% Pic3-01_MVD-BTS-Geometry.png: 1239x931 pixel, 125dpi, 25.18x18.92 cm, bb=0 0 714 536 
\caption[Overview of the future \Fair facility]
{Overview of the future \Fair facility. 
The upgraded accelerators of the existing GSI facility will act as injectors. 
New accelerator and storage rings are highlighted in red, 
experimental sites are indicated with green letters.
}
\label{Pic-FAIR}
\end{center}
\end{figure}

The \Panda (Anti\textbf{P}roton \textbf{AN}nihilation 
at \textbf{DA}rmstadt) experiment~\cite{PANDA-LOI}
is one of the key projects at the future 
\textbf{F}acility for \textbf{A}ntiproton and \textbf{I}on \textbf{R}esearch (\Fair)~\cite{FAIR1}\,\cite{Greenpaper}, 
which is currently under construction at GSI, Darmstadt. 
For this new facility the present GSI accelerators 
will be upgraded and further used as injectors. 
The completed accelerator facility will feature 
a complex structure of new accelerators and storage rings. 
An overview of the \Fair facility is given in \figref{Pic-FAIR}. 
Further details of the accelerator complex are described in~\cite{FAIR-Accelerator}. 
The \Fair accelerators will deliver primary proton and ion beams 
as well as secondary beams of antiprotons 
or radioactive ions, all with high energy, high intensity and high quality. 
Experiments to be installed at the facility will address 
a wide range of physics topics in the fields of 
nuclear and hadron physics as well as in atomic and plasma physics. 
An executive summary of the main \Fair projects can be found in~\cite{FAIR1} and~\cite{FAIR2}.

\begin{figure*}[tp]
\begin{center}
\includegraphics[width=\dwidth]{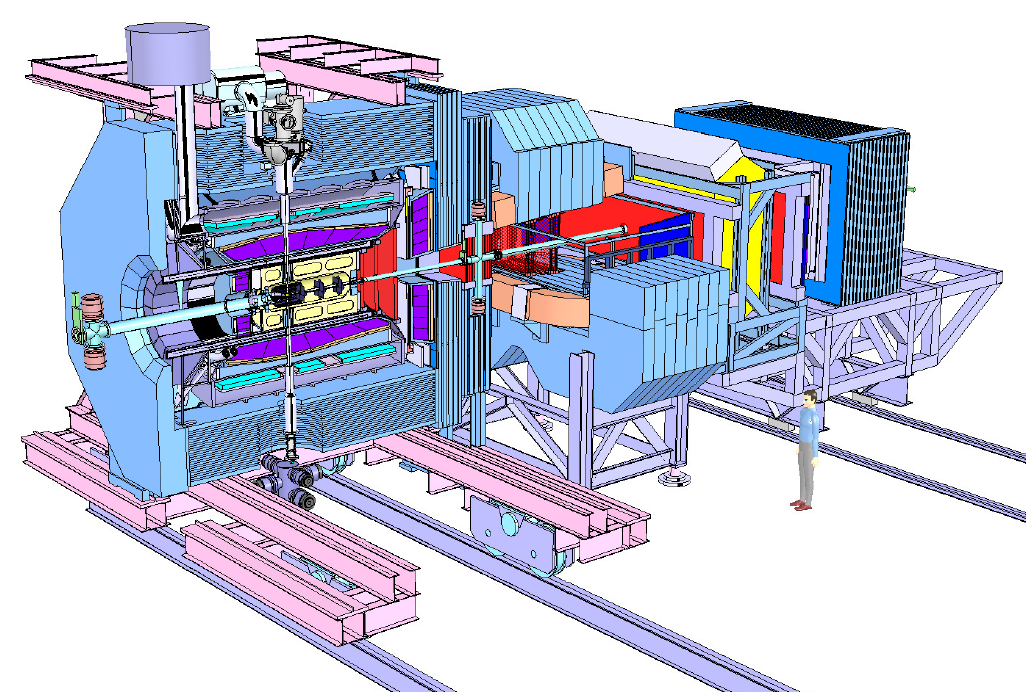}
\caption[Layout of the \Panda detector]
{
Layout of the \Panda detector consisting of a Target Spectrometer, 
surrounding the interaction region, and a 
Forward Spectrometer to detect particles emitted in the forward region. 
The \HESR antiproton beam enters the apparatus from the left side.
}
\label{fig:int:det}
\end{center}
\end{figure*}   

The \Panda experiment will perform precise studies of antiproton-proton annihilations 
and reactions of antiprotons with nucleons of heavier nuclear targets. 
It will benefit from antiproton beams with unprecedented intensity and quality. 
The covered centre-of-mass energy between 1~\gev and 5~\gev
allows for very accurate measurements, especially in the charm quark sector. 
Based on a broad physics program, studying the non-pertubative regime,
it will be possible to explore the nature of the strong interaction
and to obtain a significant progress in our understanding 
of the QCD spectrum and hadron structure.

Nowadays these studies are carried out mainly at electron machines that offer 
the advantage of kinematically clean reactions 
but at the price of a reduced set of final states and reduced cross-sections.
Also the future experiments currently planned as upgrade 
at existing high-energy physics facilities will not deliver 
high-precision data over the full charm spectrum.
In this context, the \Panda experiment will be a unique tool 
to improve both statistics and precision of existing data 
and to further explore the physics in the charm quark sector.
Moreover, the \Panda collaboration is in the ideal situation 
to be able to benefit from the expertise gained during
the construction of the LHC detectors and of the B-factory experiments,
which have determined a significant progress in the detector technology
due to the performed optimisation or the introduction of
completely new concepts and instruments.

In the first section of this chapter 
the scientific program of \Panda will be summarised. 
It ranges from charmonium spectroscopy 
to the search for exotic hadrons and the study of nucleon structure, 
from the study of in-medium modifications of hadron masses to the physics of 
hypernuclei.
Therefore, antiproton beams in the momentum range from 
1.5~\gevc to 15~\gevc
will be provided by the high-energy storage ring (\HESR) to the experiment.
An overview of this accelerator and storage ring will be given in the second section.
To explore the broad physics program, the \Panda collaboration wants 
to build a state-of-the-art general purpose detector 
studying annihilation reactions of antiprotons with protons (\pbarp) 
and in nuclear matter (\pbarA). 
The different target systems will be discussed in section~\ref{intro:target}.
The \Panda apparatus consists of a set of systems surrounding 
an internal target placed in one of the two straight sections 
of the \HESR.
\Figref{fig:int:det} shows the layout of the \Panda detector. 
It consists of a 4~m long and 2~T strong superconducting solenoid 
instrumented to detect both charged and neutral particles 
emitted at large and backward angles (Target Spectrometer, TS) 
and of a 2~Tm resistive dipole magnetic spectrometer 
to detect charged and neutral particles emitted at angles between 
zero and twenty degrees (Forward Spectrometer, FS) with respect to the beam axis.
A complex detector arrangement is necessary in order to reconstruct 
the complete set of final states, relevant to achieve the proposed physics goals.
With the installed setup, a good particle identification with an
almost complete solid angle will be combined 
with excellent mass, momentum and spatial resolution.  
More details of the \Panda detector will be described in section~\ref{s:over:panda}.

\input{introduction/MainIntro/physicsintro}
\input{introduction/MainIntro/hesr}

\input{introduction/MainIntro/targets}

\input{introduction/MainIntro/luminosity}
\input{introduction/MainIntro/detector}

\begin{figure*}[tp]
\begin{center}
\includegraphics[width=\dwidth]{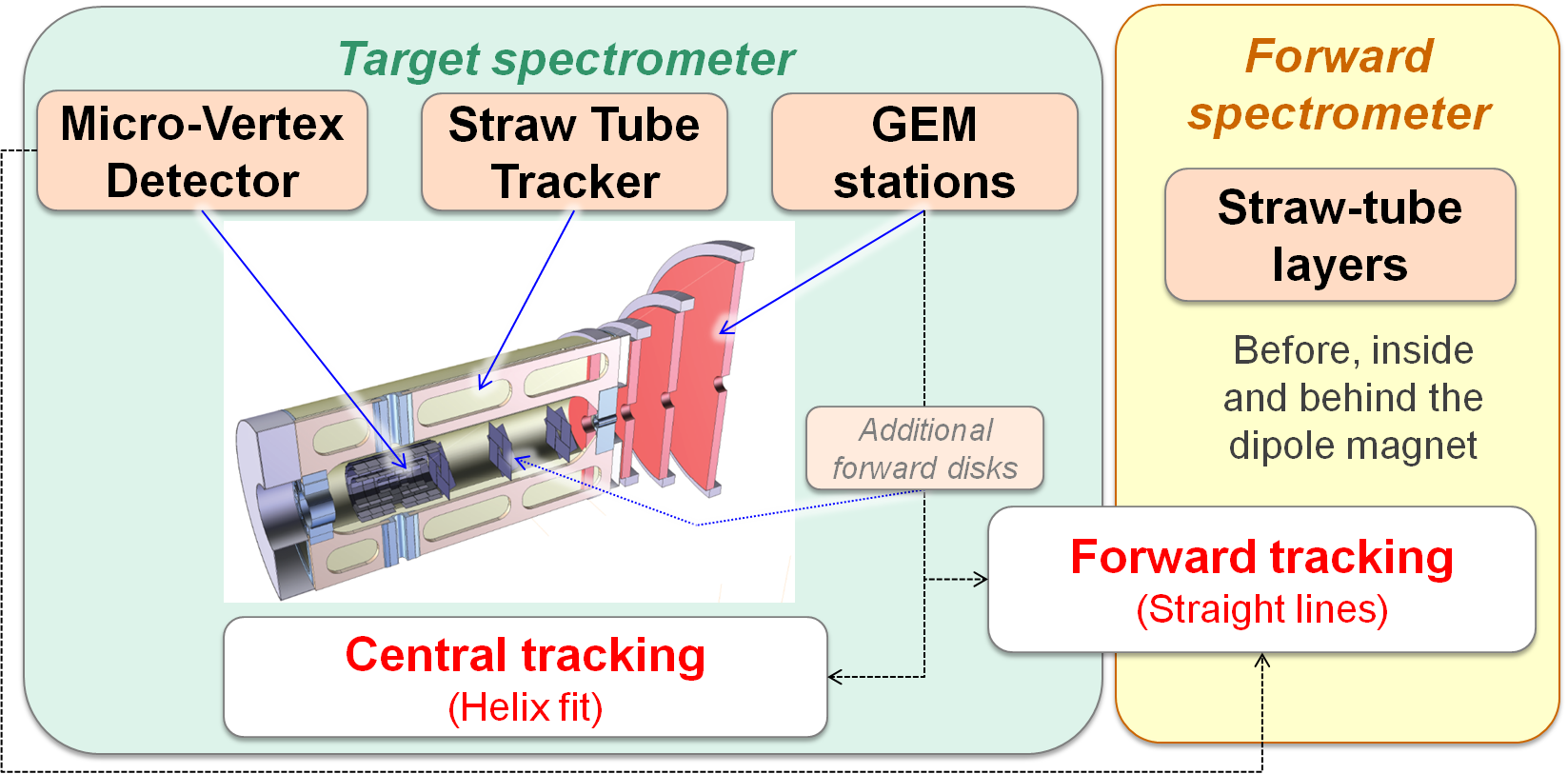}
\caption[Overview of the \PANDA tracking system.]
{
Overview of the \PANDA tracking system, including the option of the additional forward disks.
}
\label{fig:int:tracking}
\end{center}
\end{figure*}   

\section{The Charged Particle Tracking System}
%\section*{II.\hspace{4mm} The charged particle tracking system}
%\addcontentsline{toc}{section}{II.\hspace{4mm} The charged particle tracking system}

There are different tracking systems for charged particles at \PANDA,
positioned inside the target spectrometer 
and in the forward region around the dipole magnet. 
Main tasks of the global tracking system are 
the accurate determination of the particle momenta, 
a high spatial resolution of the primary interaction vertex 
and the detection of displaced secondary vertices. 
Therefore, measurements of different subdetectors 
have to be merged in order to access the full tracking information.

\subsection{Basic Approach}

The magnetic solenoid field in the target spectrometer results in 
a circular transverse motion of charged particles with non-zero transverse momentum. 
The particle momentum then can be extracted via the determination of the bending radius. 
However, tracks with a small polar angle will exit the solenoid field too soon to be measured properly.
For this case, the particle deflection induced by the subsequent dipole magnet 
is used to measure the particle momentum. 
Basically it can be deduced from a combined straight line fit before and after the dipole.

Due to the different analysing magnets, 
different track fitting algorithms 
have to be applied for central and forward tracks.  
Central tracks are reconstructed by combining hit points in the MVD layers 
with the hit information of the STT or the GEM stations. 
For the reconstruction of small angle tracks the straw tube layers 
in the forward spectrometer have to be used.
In overlap regions the MVD, the additional forward disks or the GEM stations 
can contribute to the forward tracking
because the delivery of an additional track point closer to the IP 
significantly improves the precision of the fitting results. 
After the global identification of individual tracks 
an event mapping have to be performed to match  
different tracks of the same event to a common vertex 
which either corresponds to the primary interaction vertex 
or a delayed decay of short-lived particles.

The luminosity monitor at the downstream end of the experiment 
is a tracking device of its own right.
It was introduced to measure the time integrated luminosity, 
which is essential for the determination of cross sections for different physics processes. 
Therefore, elastically scattered antiprotons are measured 
under small angles corresponding to small momentum transfers. 
The associated differential cross sections are well known 
and thus provide an ideal reference channel. 
Additional information from the MVD will eventually improve 
the measurement by taking advantage of the reconstructed  
slow recoil proton at polar angles of around 90$^{\circ}$, 
which is correlated with the highly energetic antiproton 
detected in the luminosity monitor.

\subsection{Optimisation Criteria}
\input{introduction/MainIntro/req}

\putbib[lit_intMAIN]

%% file: introduction/MainIntro/physicsintro.tex
% FILE: physicsintro.tex
%
\subsection{The Scientific Program}
%\subsection*{I.1\hspace{4mm} The scientific program}
%\addcontentsline{toc}{subsection}{I.1\hspace{4mm} The scientific program}
%\COM{Author(s): D. Bettoni}
\label{sec:sciprog}
One of the most challenging and fascinating 
goals of modern physics is the achievement of a fully quantitative
understanding of the strong interaction, which is the subject of hadron physics.
Significant progress has been achieved over the past few years thanks to
considerable advances in experiment and theory. New experimental results
have stimulated a very intense theoretical activity and a refinement of
the theoretical tools. 

Still there are many fundamental questions which remain basically
unanswered.
Phenomena such as the confinement of quarks, the existence of glueballs 
and hybrids, the origin of the masses of hadrons in the context of the
breaking of chiral symmetry are long-standing puzzles and represent
the intellectual challenge in our attempt to understand the nature of the
strong interaction and of hadronic matter.

Experimentally, studies of hadron structure can be performed with different
probes such as electrons, pions, kaons, protons or antiprotons.
In antiproton-proton annihilation, particles with gluonic degrees of freedom
as well as particle-antiparticle pairs are copiously produced,
allowing spectroscopic studies with very high statistics and precision.
Therefore, antiprotons are an excellent tool to address the open problems.

The \PANDA experiment is being designed to fully exploit the extraordinary 
physics potential arising from the availability of high-intensity, cooled
antiproton beams.
Main experiments of the rich and diversified hadron physics program
are briefly itemised in the following.
More details can be found in the \panda physics booklet~\cite{PANDA:PhysBooklet}.
\begin{itemize}
\item \textbf{Charmonium Spectroscopy}\\
A precise measurement of all states below and above the
open charm threshold is of fundamental importance for a better understanding of QCD.
All charmonium states can be formed directly in \pbarp annihilation.
At full luminosity \PANDA will be able 
to collect several thousand \ccbar states per day.
By means of fine scans it will be possible to measure masses with accuracies 
of the order of 100~\kev and widths to 10\% or better.
The entire energy region below and above the open charm threshold will be explored.

\item \textbf{Search for Gluonic Excitations}\\
One of the main challenges of hadron physics
is the search for gluonic excitations, 
i.e.~hadrons in which the gluons can act as principal components.
These gluonic hadrons fall into two main categories: glueballs, 
i.e.~states of pure glue, and hybrids, which consist of a \qqbar pair and excited glue.
The additional degrees of freedom carried by gluons allow these hybrids and glueballs
to have \JPC exotic quantum numbers: in this case mixing effects with nearby
\qqbar states are excluded and this makes their experimental identification easier.
The properties of glueballs and hybrids are determined by the long-distance features
of QCD and their study will yield fundamental insight into the structure of the QCD
vacuum.
Antiproton-proton annihilations provide 
a very favourable environment in which to look for gluonic hadrons. 

\item \textbf{Study of Hadrons in Nuclear Matter}\\
The study of medium modifications of hadrons embedded in hadronic matter
is aiming at understanding the origin of hadron masses in the context
of spontaneous chiral symmetry breaking in QCD and its partial restoration
in a hadronic environment. 
So far experiments have been focussed on the light quark sector.
The high-intensity \pbar beam of up to 15~\gevc will allow an extension of
this program to the charm sector both for hadrons with hidden and open charm.
The in-medium masses of these states are expected to be affected primarily
by the gluon condensate.

Another study which can be carried out in \PANDA is the measurement of \jpsi
and \D meson production cross sections in \pbar annihilation
on a series of nuclear targets. The
comparison of the resonant \jpsi yield obtained from \pbar annihilation
on protons and different nuclear targets allows to deduce the
\jpsi-nucleus dissociation cross section, a fundamental parameter to 
understand \jpsi suppression in relativistic heavy ion collisions interpreted
as a signal for quark-gluon plasma formation. 

\item \textbf{Open Charm Spectroscopy}\\ 
The \HESR, running at full luminosity and at \pbar
momenta larger than 6.4~\gevc, would produce a large number
of \D meson pairs.
The high yield and the well defined production kinematics of \D meson pairs 
would allow to carry out a significant charmed meson spectroscopy program 
which would include, for example, the rich \D and \Ds meson spectra.

\item \textbf{Hypernuclear Physics}\\
Hypernuclei are systems in which 
neutrons or protons are replaced by hyperons. 
In this way a new quantum number, strangeness, 
is introduced into the nucleus.
Although single and double {\Lambdaplain}-hypernuclei were discovered many decades ago, only 6 double 
{\Lambdaplain}-hypernuclei are presently known. 
The availability of \pbar beams at \FAIR will allow efficient production
of hypernuclei with more than one strange hadron, making \PANDA competitive
with planned dedicated facilities. This will open new perspectives for
nuclear structure spectroscopy and for studying the forces between hyperons
and nucleons. 

\item \textbf{Electromagnetic Processes}\\
In addition to the spectroscopic studies described above, \PANDA will be able to
investigate the structure of the nucleon using electromagnetic processes, such as
Deeply Virtual Compton Scattering (DVCS) and the process \pbarp~$\to$~\ee, which 
will allow the determination of the electromagnetic form factors of the proton 
in the timelike region over an extended $q^2$ region.
Furthermore, measuring the Drell Yan production of muons 
would give access to the transverse nucelon structure.
\end{itemize}

%% file: introduction/MainIntro/hesr.tex
\begin{figure*}[thb]
\begin{center}
\vspace{-2mm}
\includegraphics[width=14.5cm]{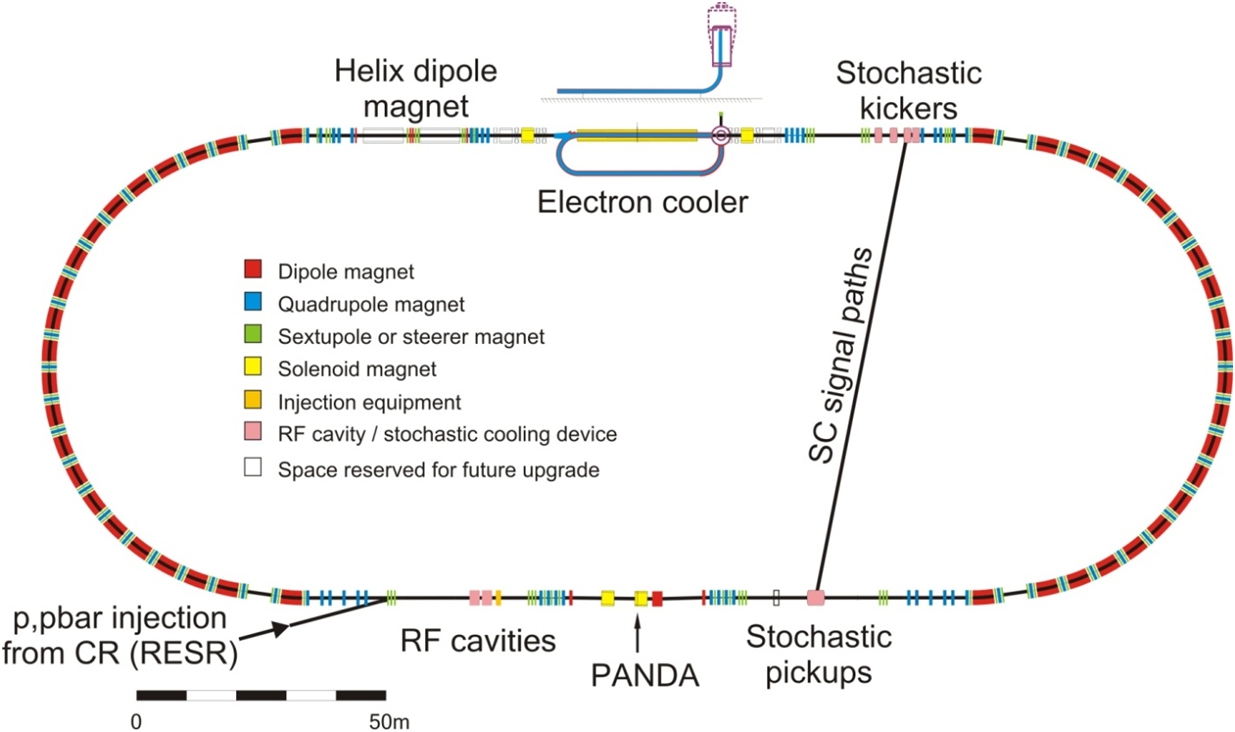}
\vspace{-1mm}
\caption[Layout of the High Energy Storage Ring \HESR]
{Layout of the High Energy Storage Ring \HESR. 
The beam is injected from the left into the lower straight section. %Stochastic cooling and electron cooling is foreseen.  
The location of the \PANDA target is indicated with an arrow.}
\label{f:over:hesr}
\end{center}
\end{figure*}

\subsection{High Energy Storage Ring -- \HESR}
%
%\subsection*{I.2 \hspace{4mm} High Energy Storage Ring -- \HESR}
%\addcontentsline{toc}{subsection}{I.2 \hspace{4mm} High Energy Storage Ring -- \HESR}
\label{s:over:hesr}

\begin{table*}
\begin{center}
\scalebox{0.85}{
\begin{tabular}{l l} 
\hline
  \multicolumn{2}{c}{\textbf{Experimental Requirements}}\\ 
\hline
Ion species & Antiprotons \\
$\bar{p}$ production rate &
$2\cdot 10^7 ${/s} ($1.2\cdot 10^{10}$ per 10~min) \\
Momentum / Kinetic energy range &
1.5 to 15~\gevc / 0.83 to 14.1~\gev \\
Number of particles &
$10^{10}$ to $10^{11}$ \\
%Target thickness &
%$4\times 10^{15} ${atoms/cm$^2$} (H$_2$ pellets) \\
%Transverse emittance &
%$<$ 1 {mm$\times$mrad} \\
%Beam size (radius) at IP & $\sim$ 1\,mm (RMS) \\
Betatron amplitude at IP &
1~m to 15~m \\
Betatron amplitude E-Cooler &
25~m to 200~m \\
\vspace{-2mm} & \\ 
%\hline
\hline
  \multicolumn{2}{c}{\textbf{Operation Modes}}\\ 
\hline
High resolution (HR) & Peak Luminosity of 2$\cdot
10^{31}${cm$^{-2}$s$^{-1}$} for $10^{10}\
\bar{p}$ \\
 & assuming $\rho_{target}$ = $4\cdot 10^{15}$ atoms/cm$^2$\\
 & RMS momentum spread $\sigma_p / p \leq 4\cdot 10^{-5}$, \\
 & 1.5 to 8.9 \gevc \\
High luminosity (HL) & Peak Luminosity up to 2$\cdot
10^{32}${cm$^{-2}$s$^{-1}$} for $10^{11}\
\bar{p}$ \\
 & assuming $\rho_{target}$ = $4\cdot 10^{15}$ atoms/cm$^2$\\
 & RMS momentum spread  $\sigma_p / p \sim 10^{-4}$,\\
 & 1.5 to {15} \gevc  \\
 \hline
\end{tabular}
}%scalebox
\caption[Experimental requirements and operation modes of \HESR]{Experimental requirements and operation modes of \HESR for the full \Fair version.}
\label{t:hesr1}
\end{center}
\end{table*}

The \HESR is dedicated to supply \PANDA with 
high intensity and high quality antiproton
beams over a broad momentum range from 1.5~\gevc to 15~\gevc~\cite{HESR:FAIR:2008}.
\Tabref{t:hesr1} summarises the experimental requirements 
and main parameters of the two operation modes for the full \Fair version.
The High Luminosity (HL) and the High Resolution (HR) mode 
are established to fulfil all challenging specifications
for the experimental program of \PANDA~\cite{hesr:lehrach:2009}.
The HR mode is defined in the momentum range from 1.5~\gevc to 9~\gevc. 
To reach a relative momentum spread down to the order of
$10^{-5}$, only $10^{10}$ circulating particles in the ring are
anticipated. 
The HL mode requires an order of magnitude higher beam
intensity with reduced momentum resolution to reach a peak luminosity
of 2$\cdot 10^{32}${cm$^{-2}$s$^{-1}$} 
in the full momentum range up to 15~\gevc.
To reach these beam parameters a very powerful phase-space cooling is needed.
Therefore, high-energy electron cooling~\cite{Ecool08} 
and high-bandwidth stochastic cooling~\cite{HESR-CoolingScenario} will be utilised.

The \HESR lattice is designed as a racetrack shaped ring with a
maximum beam rigidity of 50~Tm (see \figref{f:over:hesr}). 
It consists of two 180$^{\circ}$ arcs and two 155~m long straight sections 
with a total circumference of 575~m~\cite{HESR-Lattice}. 
The arc quadrupole magnets will allow for a flexible adjustment of transition energy, 
horizontal and vertical betatron tune as well as horizontal dispersion.
In the straight section opposite to the injection point, an electron cooler will be installed. 
The \panda detector with the internal target is placed at the other side. 
Further components in the straight \panda section are 
beam injection kickers, septa and multi-harmonic RF cavities. 
The latter allow for a compensation of energy losses due to the beam-target interaction, 
a bunch rotation and the decelerating 
or accelerating of the beam. 
Stochastic cooling is implemented via several kickers and opposing high-sensitivity pick-ups 
on either side of the straight sections. 

\begin{figure}[!b]
\vspace{-2mm}
\begin{center}
\includegraphics[width=\swidth]{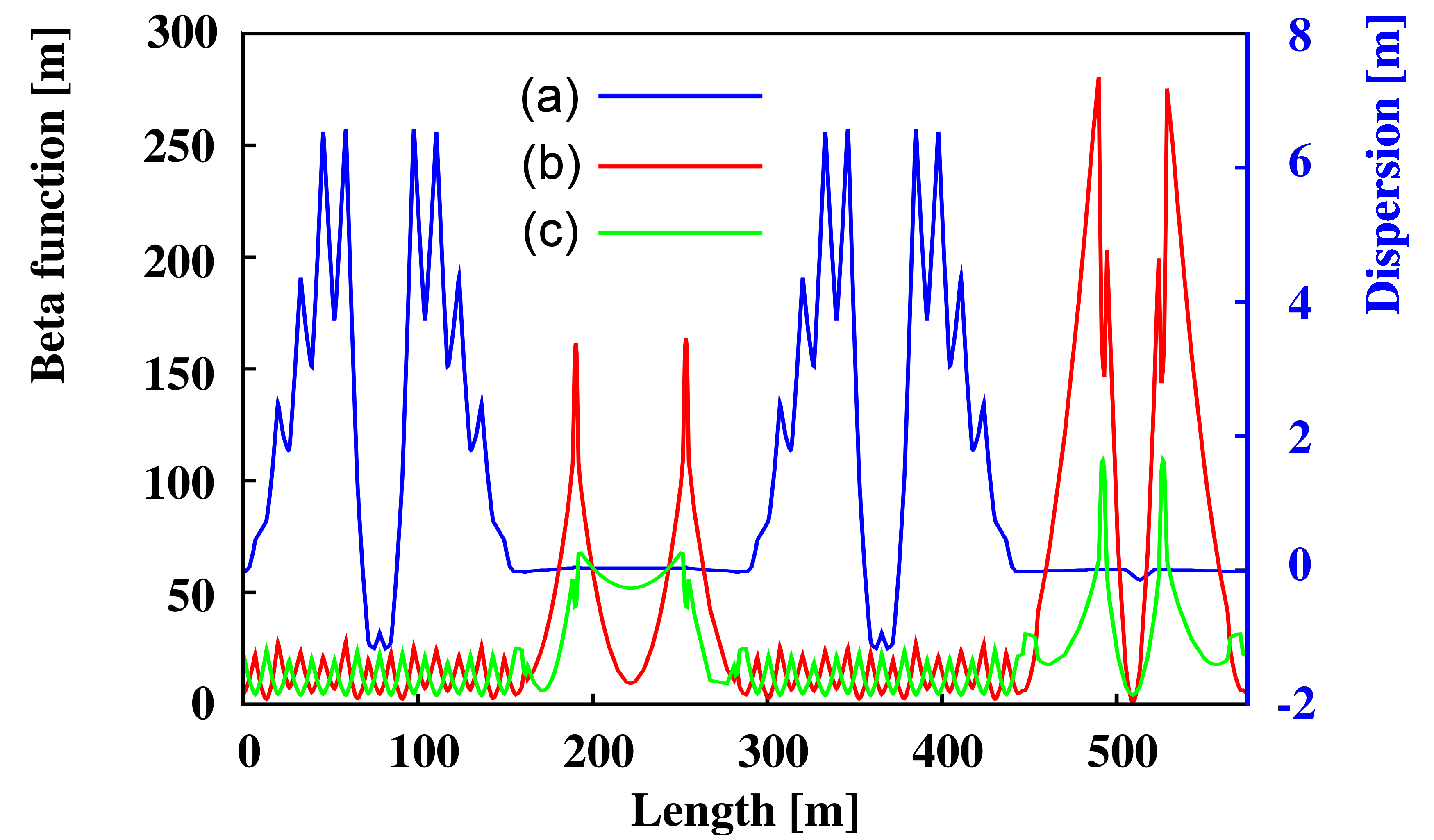}
\vspace{-2mm}
\caption[Optical functions of \HESR lattice for $\gamma_{tr}$ = 6.2]
{Optical functions of the $\gamma_{tr}$ = 6.2 lattice:
Horizontal dispersion (a), horizontal (b) and vertical (c) betatron function. 
Electron cooler and target are located at
a length of 222~m and 509~m, respectively.
}
\label{f:hesr:optics}
\end{center}
\end{figure}

Special requirements for the lattice are low dispersion in the
straight sections and small betatron amplitudes in the range between 
1~m and 15~m at the internal interaction point (IP) of the \PANDA detector. 
In addition, the betatron amplitude at the electron cooler 
must be adjustable within a large range between 25~m and 200~m.
Examples of the optical functions for one of the defined optical settings are shown in \figref{f:hesr:optics}. 
The deflection of the %large aperture 
spectrometer dipole magnet of the \Panda detector 
will be compensated by two dipole magnets that create a beam chicane.
These will be placed 4.6~m upstream and 13~m downstream the \Panda IP
thus defining a boundary condition for the quadrupole elements closest to the experiment.
For symmetry reasons, they have to be placed at $\pm$14~m with respect to the IP. 
The asymmetric placement of the chicane dipoles will result 
in the experiment axis occurring at a small angle with respect
to the axis of the straight section.
The \PANDA solenoid will be compensated by one solenoid magnet. 
%Another four 
Additional correction dipoles 
have to be included around the electron cooler 
%to correct the beam deflection 
due to the toroids  that will be used to overlap the electron beam with the antiproton beam.
Phase-space coupling induced by the electron cooler solenoid will be 
compensated by two additional solenoid magnets.

Closed orbit correction and local orbit bumps at dedicated locations in the ring 
are crucial to meet requirements for the beam-target interaction 
in terms of maximised ring acceptance and optimum beam-target overlap \cite{Wel08}.
The envisaged scheme aims on a reduction of
maximum closed orbit deviations to below 5~mm 
while not exceeding 1~mrad of corrector strength.
Therefore, 64 beam position monitors and 
48 orbit correction dipoles are intended to be used.
Because a few orbit bumps will have to be used %at a few positions 
in the straight parts of the \HESR,
all correction dipoles therein %for orbit corrections in these sections 
are designed to provide an additional deflection strength of 1~mrad. 

Transverse and longitudinal cooling will be used to compensate a transverse beam blow up 
and to achieve a low momentum spread, respectively. %~\cite{HESR-CoolingScenario}.
While stochastic cooling will be applicable in the whole momentum range, 
electron cooling is foreseen in a range from 1.5~\gevc to 8.9~\gevc with a possible upgrade to 15~\gevc.
The relative momentum spread can be further improved by combining both cooling systems.
Beam losses are dominated by hadronic interactions between antiprotons and target protons, 
single large-angle Coulomb scattering in the target and energy straggling induced 
by Coulomb interactions of the antiprotons with target electrons. 
%The relative momentum acceptance of the \HESR ring is restricted to about 1$\cdot$10$^{-3}$. 
Mean beam lifetimes for the \HESR range between 1540~s and 7100~s. 
The given numbers correspond to the time, after which the initial beam intensity is reduced by a factor of $1/e$.
A detailed discussion of the beam dynamics and beam equliibria for the \HESR can be found 
in~\cite{hesr:lehrach:2009,lehrach:2006,hinterberger:2006,BoineFrankenheim:2006ci}.
Advanced simulations have been performed for both cooling scenarios.
In case of electron cooled beams the RMS relative momentum spread obtained for the HR mode 
ranges from $7.9\cdot 10^{-6}$ (1.5~\gevc) to $2.7\cdot 10^{-5}$ (8.9~\gevc), 
and $1.2\cdot 10^{-4}$ (15~\gevc) \cite{Rei07}.
With stochastic cooling in a bandwidth of 2~GHz to 6~GHz, 
%are based on a Fokker-Planck approach.
%In this case, 
the RMS relative momentum spread for the HR mode results in 
$5.1\cdot 10^{-5}$ (3.8~\gevc), $5.4\cdot 10^{-5}$ (8.9~\gevc)
and $3.9\cdot 10^{-5}$ (15~\gevc)~\cite{Sto08}.
In the HL mode a RMS relative momentum spread of roughly $10^{-4}$ can be expected.
Transverse stochastic cooling can be adjusted independently to
ensure sufficient beam-target overlap.

%% file: introduction/MainIntro/targets.tex
\begin{figure*}[thb]
\begin{center}
\includegraphics[width=15 cm]{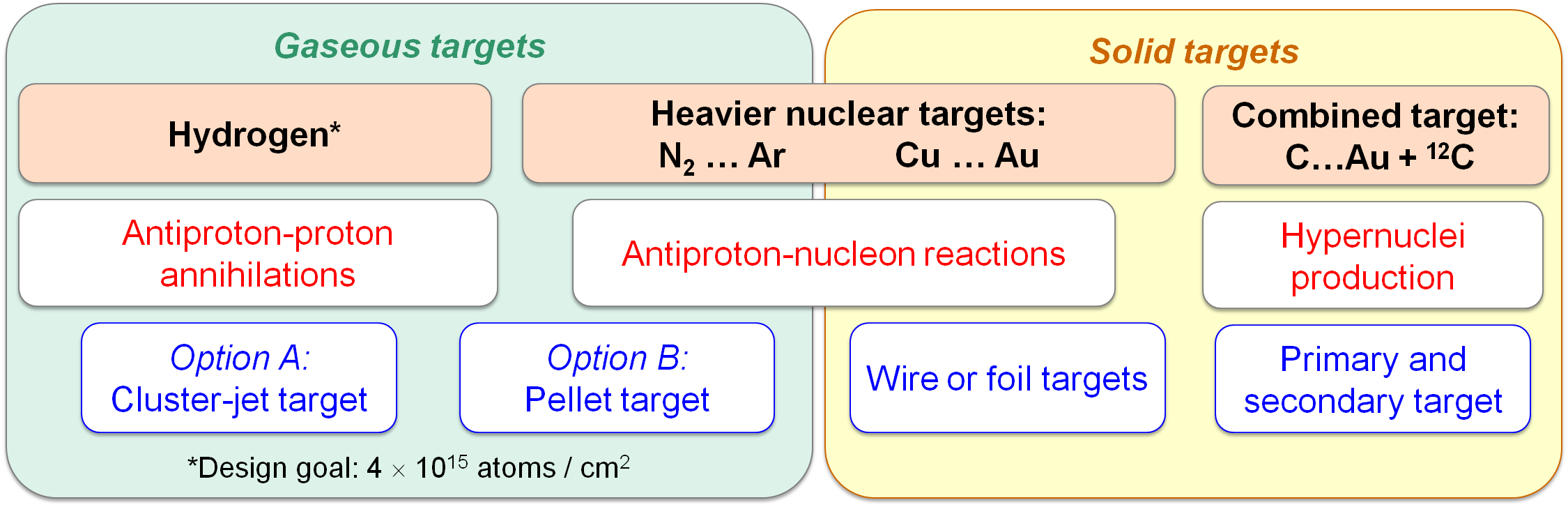}
\caption
[Summary of the different target options foreseen at \panda]
{Summary of the different target options foreseen at \panda.}
\label{pic-target:options}
\end{center}
\end{figure*}

\subsection{Targets}
\label{intro:target}
%\subsection*{I.3 \hspace{4mm} Targets}
%\addcontentsline{toc}{subsection}{I.3 \hspace{4mm} Target systems}
The design of the solenoid magnet allows for an implementation of different target systems. 
\panda will use both gaseous and non-gaseous targets. 
A very precise positioning of the target is crucial for the exact definition of 
the primary interaction vertex. 
In this context, big challenges for either system result from the long distance of 
roughly 2~m between the target injection point and the dumping system. 
Hydrogen target systems will be used for the study of antiproton-proton reactions. 
A high effective target density of about $4\cdot10^{15}$ hydrogen atoms per square centimetre 
must be achieved to fulfill the design goals of the high luminosity mode. 
Besides the application of hydrogen as target material, an extension to heavier gases 
such as deuterium, nitrogen or argon is planned 
for complementary studies with nuclear targets. 

At present, two different solutions are under development: 
a cluster-jet and a pellet target. 
Both will potentially provide sufficient target thickness but exhibit different properties 
concerning their effect on the beam quality and the definition of the IP. 
Solid targets are foreseen for hyper-nuclear studies and the study of antiproton-nucleus 
interaction using heavier nuclear targets. 
The different target options are shortly described in the following.
\Figref{pic-target:options} gives an overview to all target option foreseen at \panda.

\subsubsection*{Cluster Jet Target}
%\addcontentsline{toc}{subsection}{\hspace{4mm}...  Cluster Jet Target}

Cluster jet targets provide a homogeneous and adjustable target density without any time structure. 
Optimum beam conditions can be applied in order to achieve highest luminosity. 
The uncertainty of the IP in a plane perpendicular to the beam axis is defined by 
the optimised focus of the beam only. 
An inherent disadvantage of cluster-jet targets is the lateral spread of the cluster jet leading 
to an uncertainty in the definition of the IP along the beam axis of several millimetres. 

For the target production a pressurised cooled gas is injected into vacuum through a nozzle. 
The ejected gas immediately condensates and forms a narrow supersonic jet of molecule clusters.  
The cluster beam typically exposes a broad mass distribution which strongly depends on the gas 
input pressure and temperature.
In case of hydrogen, the average number of molecules per cluster varies from 10$^3$ to 10$^6$. 
The cluster-jets represent a highly diluted target and offer a very homogenous density profile. 
Therefore, they may be seen as a localised and homogeneous monolayer of hydrogen atoms
being passed by the antiprotons once per revolution, i.e.~the antiproton beam can be focused at highest phase space density.
The interaction point is thus defined transversely 
but has to be reconstructed longitudinally in beam direction. 
%The possibility of adjusting the target density along
%with the gradual consumption of antiprotons for running at constant
%luminosity will be an important feature.
At a dedicated prototype cluster target station an effective target density 
of $1.5\cdot 10^{15}$ hydrogen atoms per square centimetre 
has been achieved using the exact \panda geometry \cite{taeschner:2011}.
This value is close to the maximum number required by \panda. 
Even higher target densities seem to be feasible and are topic of ongoing R\&D work.

\subsubsection*{Hydrogen Pellet Target}
%\addcontentsline{toc}{subsection}{\hspace{4mm}...  Hydrogen Pellet Target}

Pellet targets provide a stream of frozen molecule droplets, called pellets, 
which drip with a fixed frequency off from a fine nozzle into vacuum.
The use of pellet targets gives access to high effective target densities. 
The spatial resolution of the interaction zone can be reduced by skimmers to a few millimetres. 
A further improvement of this resolution can be achieved  by tracking the individual pellets. 
However, pellet targets suffer from a non-uniform time distribution, 
which results in larger variations of the instantaneous luminosity as compared to a cluster-jet target. 
The maximum achievable average luminosity is very sensitive to deviations 
of individual pellets from the target axis. 
The beam must be widened in order to warrant a beam crossing of all pellets. 
Therefore, an optimisation between the maximum pellet-beam crossing time on the one hand 
and the beam focusing on the other is necessary.   

The design of the planned pellet target is based on the one currently used at the WASA-at-COSY 
experiment~\cite{WASA-Pellet}.
The specified design goals for the pellet size and the mean lateral spread of the pellet train 
are given by a radius of 25~$\tcmu$m to 40~$\tcmu$m and 
a lateral RMS deviation in the pellet train
of approximately 1~mm, respectively. 
At present, typical variations of the interspacing of individual pellets range 
between 0.5~mm and 5~mm. 
A new test setup with an improved performance has been constructed~\cite{PANDA-PelletTarget}. 
First results have demonstrated the mono-disperse and satellite-free droplet production 
for cryogenic liquids of H$_2$, N$_2$ and Ar~\cite{PANDA-PelletTarget2009}. 
However, the prototype does not fully include the \panda geometry. 
The handling of the pellet train over a long distance still has to be investigated in detail. 
The final resolution on the interaction point is envisaged to be in the order of 50~$\tcmu$m. 
Therefore, an additional pellet tracking system is planned.

\subsubsection*{Other Target Options}
%\addcontentsline{toc}{subsection}{\hspace{4mm}...  Other Target Options}

In case of solid target materials the use of wire targets is planned. 
The hyper-nuclear program requires a separate target station in upstream position. 
It will comprise a primary and secondary target. 
The latter must be instrumented with appropriate detectors. 
Therefore, a re-design of the innermost part of the \panda spectrometer becomes necessary.
This also includes the replacement of the MVD. 
%However, measurements with this modified setup are only foreseen at 
%a later stage of the experiment.

%% file: introduction/MainIntro/luminosity.tex
\subsection{Luminosity Considerations}
%\subsection*{I.4 \hspace{4mm} Luminosity Considerations}
%\addcontentsline{toc}{subsection}{I.4 \hspace{4mm} Luminosity considerations}
\label{lumi-considerations}

%All luminosity considerations in this section are discussed for the full \FAIR version. 
The luminosity $L$ describes the flux of beam particles convolved with the target opacity. 
Hence, an intense beam, a highly effective target thickness 
and an optimised beam-target overlap are essential to yield 
a high luminosity in the experiment.
The product of $L$ and the total hadronic cross section $\sigma_H$ 
delivers the interaction rate $R$, 
i.e.~the number of antiproton-proton interactions in a specified time interval,
which determines the achievable number of events for 
all physics channels and allows the extraction 
of occupancies in different detector regions. 
These are needed as input for the associated hardware development. 

Obviously, the achievable luminosity is directly linked 
with the number of antiprotons in the \HESR.
The particles are injected at discrete time intervals. 
The maximum luminosity thus depends on the antiproton production rate $R_{\bar p}$~=~d$N_{\bar p}$/d$t$.
Moreover, a beam preparation must be performed before the target can be switched on.
It includes pre-cooling to equilibrium, 
the ramping to the desired beam momentum and a fine-tuned focusing 
in the target region as well as in the section for the electron cooler.
Therefore, the operation cycle of the \HESR can be separated into two sequences
related to the beam preparation time  $t_{\textsf{\tiny{prep}}}$ (target off) 
and the time for data taking $t_{\textsf{\tiny{exp}}}$ (target on), respectively.
The  beam preparation time  $t_{\textsf{\tiny{prep}}}$ also contains 
the period between the target switch-off and the injection,
at which the residual antiprotons are either dumped 
or transferred back to the injection momentum.

\subsubsection*{Macroscopic Luminosity Profile}

\begin{figure*}[hbtp]
  \centering
    \resizebox{\dwidth}{!}{\includegraphics{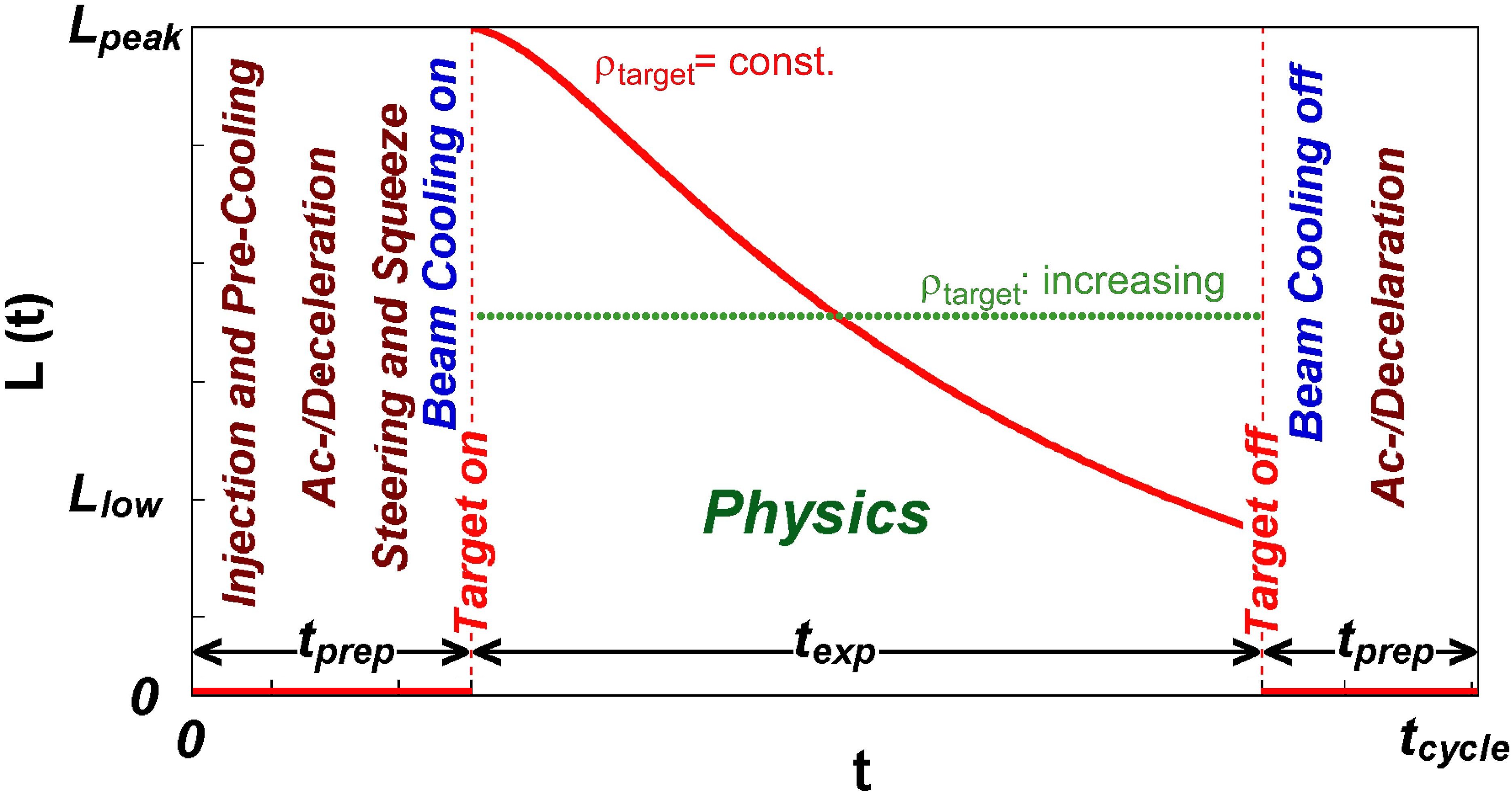}}
  \caption[Time dependent macroscopic luminosity profile~$L(t)$ in one operation cycle]
{
Time dependent macroscopic luminosity profile~$L(t)$ in one operation cycle 
for constant (solid red) and increasing (green dotted) target density $\rho_{\textsf{\tiny{target}}}$.
Different measures for beam preparation are indicated.
Pre-cooling is performed at 3.8 \gevc. 
A maximum ramp of 25~mT/s is specified for beam ac-/deceleration.} 
%To calculate the cycle average luminosity, machine cycles and beam
%preparation times have to be specified. After injection, the beam
%is pre-cooled to equilibrium (with target off) at 3.8 \gevc.
%The beam is then ac-/decelerated to the desired beam momentum. A
%maximum ramp rate of 25~mT/s is specified. After reaching the
%final momentum beam steering and focusing in the target and beam
%cooler region takes place. Total beam preparation time $t_{prep}$
%ranges from 120~s for 1.5 \gevc to 290~s for 15 \gevc. 
  \label{fig:lumcycle}
\end{figure*}

A schematic illustration of the luminosity profile during one operation cycle
is  given in \figref{fig:lumcycle}.
The maximum luminosity is obtained directly after the target is switched on.  
During data taking the luminosity decreases due to hadronic interactions,
single Coulomb scattering and energy straggling of the circulating beam in the target.
Compared to beam-target interaction, minor contributions are related to single intra-beam scattering (Touschek effect).
Beam losses caused by residual gas scattering can be neglected, if the vacuum is better than 10$^{-9}$~mbar. 
A detailed analysis of all beam loss processes can be found in~\cite{lehrach:2006,hinterberger:2006}.
The relative beam loss rate $R_{\textsf{\tiny{loss}}}$ for the total cross section $\sigma_{\textsf{\tiny{tot}}}$ 
is given by the expression
\begin{equation}
R_{\textsf{\tiny{loss}}} = \tau^{-1} = f_0 \cdot n_t \cdot \sigma_{\textsf{\tiny{tot}}}
\end{equation}
where $\tau$ corresponds to the mean (1/$e$) beam lifetime, 
$f_0$ is the revolution frequency of the antiprotons in the ring
and $n_t$ is the effective target thickness 
defined as an area density given in atoms per square centimetre. 
For beam-target interactions, the beam lifetime is independent of the beam intensity.
The Touschek effect depends on the beam equilibria and beam intensity. 
At low momenta the beam cooling scenario and the ring acceptance have large impact 
on the achievable beam lifetime.
%Beam lifetimes are ranging from 1540~s to 7100~s. 

\begin{table*}
\centering {\small
\begin{tabular}{lccc}
 & 1.5 \gevc & 9 \gevc & 15 \gevc \\ \hline
\vspace{-3mm}&&&\\
Total hadronic cross section/ mbarn &
$100$&
$57$&
$51$\\ 
\hline
\multicolumn{4}{c}{\parbox[0pt][4mm][c]{10cm}{\textbf{Cluster jet target}}}\\ 
\hline
\vspace{-3mm}&&&\\
Target density: /cm$^{-2}$ &
$8 \cdot 10^{14 \hspace{1mm}*}$ &
$8 \cdot 10^{14 \hspace{1mm}*}$ &
$8 \cdot 10^{14 \hspace{1mm}*}$ \\ 
Antiproton production rate: /s$^{-1}$ &
$2 \cdot  10^{7}$ &
$2 \cdot  10^{7}$ &
$2 \cdot  10^{7}$ \\
Beam preparation time: /s &
$120$ &
$140$ &
$290$ \\
Optimum cycle duration: /s &
$1280$ &
$2980$ &
$4750$ \\ 
Mean beam lifetime: /s 
&
$\sim 5920$ &
$\sim 29560$ &
$\sim 35550$ \\ 
Max Cycle Averaged Luminosity: /cm$^{-2}$s$^{-1}$ &
$0.29 \cdot  10^{32}$ &
$0.38 \cdot  10^{32}$ &
$0.37 \cdot  10^{32}$ \\
%\end{tabular}
%}
%\end{table*}

%\begin{table*}
%\caption{Max. cycle averaged luminosity for a H$_2$ pellet target.}
%\label{tab:lifetimes} \centering {\small
%\begin{tabular}{lccc}
 % & \multicolumn{3}{c}{$(\tauloss^{-1})$ / s$^{-1}$} \\
\hline
\multicolumn{4}{c}{\parbox[0pt][4mm][c]{10cm}{\textbf{Pellet target}}}\\ 
\hline
\vspace{-3mm}&&&\\
Target density: / cm$^{-2}$&
$4 \cdot 10^{15}$ &
$4 \cdot 10^{15}$&
$4 \cdot 10^{15}$ \\
Antiproton production rate: /s$^{-1}$&
$2 \cdot  10^{7}$ &
$2 \cdot  10^{7}$ &
$2 \cdot  10^{7}$ \\
Beam preparation time: /s &
$120$ &
$140$ &
$290$ \\
Optimum cycle duration: /s &
$4820$ &
$1400$ &
$2230$ \\ 
Mean beam lifetime: /s 
&
$\sim 1540$ &
$\sim 6000$ &
$\sim 7100$ \\ 
Max Cycle Averaged Luminosity: /cm$^{-2}$s$^{-1}$ &
$0.53 \cdot  10^{32}$ &
$1.69 \cdot  10^{32}$ &
$1.59 \cdot  10^{32}$ \\ \hline
\end{tabular}
}
\caption[Maximum achievable cycle averaged luminosity for different H$_2$ target setups]
{
Calculation of the maximum achievable cycle averaged luminosity
for three different beam momenta: 
Input parameters and final results for different H$_2$ target setups.
(* Lower limit, cf. chapter~\ref{intro:target}). 
}
\label{table_HydroAvLumi}
\end{table*}

\subsubsection*{Cycle Average Luminosity}

\begin{table*}[t]
\begin{center}
\small
\begin{tabular}{|c|c|c|c|}
\hline
\vspace{-3mm}
  & \hspace{1.5cm}
    & \hspace{1.5cm}
      & \hspace{1.5cm} \\
\vspace{-3mm}&&&\\
Target material 
  & $\bar L$ ($p_{\textsf{\tiny{beam}}}$=1.5~\gevc)
    & $\bar L$ ($p_{\textsf{\tiny{beam}}}$=15~\gevc) 
      & $n_t$\\
 & [cm$^{-2}$s$^{-1}$] 
    & [cm$^{-2}$s$^{-1}$] 
      & [atoms/cm$^2$]
\parbox[0pt][6mm][t]{0cm}{}\\
\hline
\vspace{-3mm}&&&\\
deuterium 
  & $5\cdot 10^{31}$ 
    & $1.9\cdot 10^{32}$
      & $3.6\cdot 10^{15}$\\
\vspace{-3mm}&&&\\
argon 
  & $4\cdot 10^{29}$
    & $2.4\cdot 10^{31}$
      & $4.6\cdot 10^{14}$\\
\vspace{-3mm}&&&\\
gold 
  & $4\cdot 10^{28}$
    & $2.2\cdot 10^{30}$
      & $4.1\cdot 10^{13}$
\parbox[0pt][6mm][t]{0cm}{}\\
\hline
\end{tabular}
\caption[Expected luminosities for heavier nuclear targets at \Panda]
{Expected maximum average luminosities, $\bar L$, 
and required effective target thickness, $n_t$, 
for heavier nuclear targets at \panda
at minimum and maximum beam momentum $p_{\textsf{\tiny{beam}}}$.
Given numbers refer to an assumed number of $10^{11}$ antiprotons in the \HESR.}
%\vspace{-6mm}
\label{table_NuclearAvLumi}
\end{center}
\end{table*} 

In physics terms, the time-averaged cycle luminosity is most relevant.
The maxi\-mum average luminosity depends on the ratio of the antiproton production rate 
to the loss rate and is thus inversely proportional to the total cross section.
It can be increased if the residual antiprotons after each cycle 
are transferred back to the injection momentum and then merged with the newly injected particles.
Therefore, a bucket scheme utilising broad-band cavities is foreseen 
for beam injection and the refill procedure. 
Basically, the cycle average luminosity $\bar L$ reads as:
\begin{equation}
\label{eq-AvLumiGeneralDefinition}
\bar L = N_{\bar p,\textsf{\tiny{0}}} \cdot f_0 \cdot n_t \cdot 
{{\tau \left[1-e^{-{t_{\textsf{\tiny{exp}}}\over{\tau}}}\right]}\over t_{\textsf{\tiny{exp}}} + t_{\textsf{\tiny{prep}}}}
\end{equation}
where $N_{\bar p,\textsf{\tiny{0}}}$ corresponds to the number of available particles at the start of the target insertion.

For the calculations, machine cycles and beam preparation times have to be specified.
The maximum cycle average luminosity is achieved by an optimisation of 
the cycle time $t_{\textsf{\tiny{cycle}}}=t_{\textsf{\tiny{exp}}} + t_{\textsf{\tiny{prep}}}$.
Constraints are given by the restricted number antiprotons in the \HESR,
the achievable effective target thickness and the specified antiproton production rate 
of $R_{\bar p} = 2\cdot10^7$~s$^{-1}$ at \FAIR.
%A further limitation occurs if the antiproton production rate is not high enough 
%to compensate beam losses within one mean beam lifetime.

Main results of calculations performed for different hydrogen targets are summarised in \tabref{table_HydroAvLumi}.
The total hadronic cross section, $\sigma_H^{\bar p p}$, 
decreases with higher beam momentum from approximately 100~mbarn at 1.5~\gevc 
to 50~mbarn at 15~\gevc.
With the limited number of 10$^{11}$ antiprotons, as specified for the high-luminosity mode, 
cycle averaged luminosities of up to $1.6 \cdot 10^{32}$~cm$^{-2}$s$^{-1}$ can be achieved at
15~\gevc for cycle times of less than one beam lifetime.
Due to the very short beam lifetimes at lowest beam momenta more than $10^{11}$ 
particles can not be provided in average.
As a consequence, the average luminosity drops below the envisaged design value at around 2.4~\gevc 
to finally roughly \unit[5\,$\,\cdot\,$\,$10^{31}$]{s$^{-1}$cm$^{-2}$} at 1.5~\gevc.
%As a consequence, cycle average luminosities are below 10$^{32}$ cm$^{-2}$ s$^{-1}$.
Due to the lower assumed target density the achievable luminosity of the cluster-jet target 
is smaller compared to the pellet operation.

\begin{figure}[!b]
\begin{center}
\includegraphics[width=7.5 cm]{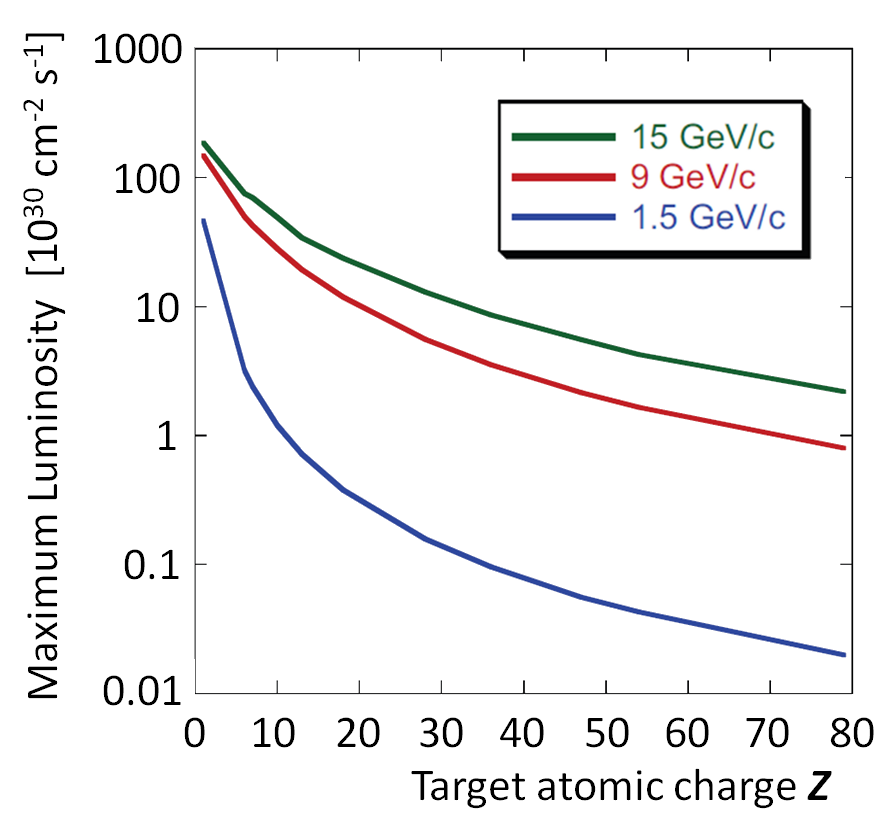}
% Pic3-01_MVD-BTS-Geometry.png: 1239x931 pixel, 125dpi, 25.18x18.92 cm, bb=0 0 714 536 
\caption[Maximum average luminosity vs.~atomic charge $Z$ of the target 
for three different beam momenta]
{Maximum average luminosity vs. atomic charge, $Z$, of the target
for three different beam momenta.}
\label{Pic-HESR_NuclearAvLumi}
\end{center}
\end{figure}

In case of nuclear targets the total hadronic cross section 
for the interaction of antiprotons with target nucleons 
can be estimated from geometric considerations taking into account
the proton radius of \unit[$r_p =0.9$]{fm} and the radius of a spherical nucleus $R_A$,
which can be roughly approximated as  \unit[$R_A =\,$]{$r_0 A^{1/3}$}, 
where \unit[$r_0  = 1.2$]{fm} and $A$ is the mass number.
With the assumption that $\sigma_H^{\bar p p}= \pi r_p^2$, 
the required total hadronic cross section, $\sigma_H^{\bar p A}$, 
for a nucleus of mass number $A$ can be extracted from the given values of 
$\sigma_H^{\bar p p}$ for antiproton-proton collisions as follows:

\begin{equation}
\label{eq-pbarNucleonXsection}
\sigma_H^{\bar p A} = \pi (R_A + r_p)^2 = \sigma_H^{\bar p p} \cdot \left({R_A\over r_p}+1\right)^2 
\end{equation}

Simulation results on maximum average luminosities based 
on equation~\ref{eq-pbarNucleonXsection} are shown 
in \figref{Pic-HESR_NuclearAvLumi}. 
They include adapted beam losses in the target due to 
single Coulomb scattering and energy straggling. 
Compared to antiproton-proton experiments, the maximum average luminosity for nuclear targets decreases rapidly 
with both, higher atomic charge $Z$ and lower beam momenta, by up to three orders of magnitude.  
Specific values for selected nuclear targets are given in \tabref{table_NuclearAvLumi} 
with the effective target thickness required to reach these numbers.

\begin{table*}[t]
\begin{center}
\small
\begin{tabular}{|c|c|c|c|c|c|c|}
\hline
\vspace{-3mm}
\hspace{2.0cm}
  & %\hspace{1.5cm}
    & \hspace{2.0cm}
      & %\hspace{1.5cm} 
	& \hspace{2.0cm}
	  & \hspace{2.0cm}
	    & \hspace{2.0cm} \\
\vspace{-3mm}&&&&&&\\
Target
  & $p_{\textsf{\tiny{beam}}}$
    & $\bar L_{\textsf{\tiny{exp}}}$
      &  $L_{\textsf{\tiny{inst}}}$
	  & $\sigma_H$
	    &  $\bar R_{\textsf{\tiny{exp}}}$
	      &  $\bar L_{\textsf{\tiny{peak}}}/\bar L_{\textsf{\tiny{exp}}}$\\
material
  & [\gevc]
    & [cm$^{-2}$s$^{-1}$]
      & [cm$^{-2}$s$^{-1}$] 
	  & [mbarn]
	    & [s$^{-1}$]
	      & \footnotesize $(R_{\textsf{\tiny{nom}}})$
\parbox[0pt][6mm][t]{0cm}{}\\
\hline
\vspace{-3mm}&&&&&&\\
\multirow{2}{*}{hydrogen}
  & 1.5
    & $5.4\cdot 10^{31}$
      & ($5.9 \pm 0.6$)$\,\cdot \,10^{31}$
	& 100
	  & $5.4\cdot 10^{6}$
	    & 3.7 \\
  & 15
    & $1.8\cdot 10^{32}$
      & ($2.0 \pm 0.2$)$\,\cdot \, 10^{32}$
	& 51
	  & $9.7\cdot 10^{6}$
	    & 2.1 \\
\hline
\vspace{-3mm}&&&&&&\\
\multirow{2}{*}{argon}
  & 1.5
    & $4.0\cdot 10^{29}$
      & ($4.4 \pm 0.4$)$\,\cdot \, 10^{29}$
	& 2020
	  & $8.1\cdot 10^{5}$
	    & \multirow{2}{*}{--} \\
  & 15
    & $2.4\cdot 10^{31}$
      & ($2.6 \pm 0.3$)$\,\cdot \, 10^{31}$
	& 1030
	  & $2.5\cdot 10^{7}$
	    &  \\
\vspace{-3mm}&&&&&&\\
\multirow{2}{*}{gold}
  & 1.5
    & $4.0\cdot 10^{28}$
      & ($4.4 \pm 0.4$)$\,\cdot \, 10^{28}$
	& 7670
	  & $3.1\cdot 10^{6}$
	    & \multirow{2}{*}{--} \\
  & 15
    & $2.2\cdot 10^{30}$
      & ($2.6 \pm 0.3$)$\,\cdot \, 10^{30}$
	& 3911
	  & $8.6\cdot 10^{6}$
	    & 
\parbox[0pt][4.5mm][t]{0cm}{}\\
\hline
\end{tabular}
\caption[Estimate on the expected event rates at \panda.]
{Summary of expected event rates at \panda. 
Numbers for the hydrogen target correspond to the pellet system (see \tabref{table_HydroAvLumi}).
The given ratio $\bar L_{\textsf{\tiny{peak}}}/\bar L_{\textsf{\tiny{exp}}}$ 
corresponds to the maximum value to achieve the nominal interaction rate of $R_{\textsf{\tiny{nom}}} = 2 \cdot10^7$~s$^{-1}$.
Rough estimates for nuclear targets are based on the numbers given in \tabref{table_NuclearAvLumi},
with $\bar L = \bar L_{\textsf{\tiny{exp}}}$, and  $\sigma_H$ calculated according to equation~\ref{eq-pbarNucleonXsection}.
}
%\vspace{-6mm}
\label{table_EventRates}
\end{center}
\end{table*} 

\subsubsection*{Event Rates}
\label{event-rates}

Besides the cycle-averaged luminosity an evaluation 
of the instantaneous luminosity during the data taking 
is indispensable for performance studies of the \panda detector.
Associated event rates define the maximum data load
to be handled at different timescales by the individual subsystems.
The discussions in this section are based on the following assumptions:
\begin{itemize}
 \item Nominal antiproton production rate at \FAIR:  \\ $R_{\bar p}=2\cdot10^7$~s$^{-1}$
 \item Effective target density: \\ $n_t=4\cdot10^{15}$~atoms/cm$^2$
 \item Maximum number of antiprotons in the \HESR: \\ $N_{\bar p,\textsf{\tiny{max}}}$~=~10$^{11}$
 \item Recycling of residual antiprotons at the end of each cycle
\end{itemize}

As indicated in \figref{fig:lumcycle}
the instantaneous luminosity during the cycle changes on a macroscopic timescale.
One elegant way to provide constant event rates in case of a cluster-jet target 
is given by the possibility to compensate the antiproton consumption during an accelerator cycle 
by the increase of the effective target density. 
Alternatively, using a constant target beam density the beam-target overlap might be increased
adequately to the beam consumption.  
With these modifications the instantaneous luminosity during the cycle 
is expected to be kept constant to a level of 10\%.

The values for the luminosity as given in \tabref{table_HydroAvLumi} 
are averaged over the full cycle time. 
However, to extract the luminosity during data taking, $\bar L_{\textsf{\tiny{exp}}}$,
these numbers must be rescaled to consider the time average over the experimental time:
\begin{equation}
\label{eq-LumiExp}
\bar L_{\textsf{\tiny{exp}}} = (t_{\textsf{\tiny{cycle}}}/t_{\textsf{\tiny{exp}}}) \cdot \bar L
\end{equation}
In addition to the fluctuation of the instantaneous luminosity during the operation cycle as dicussed above
($\Delta L_{\textsf{\tiny{inst}}}/L_{\textsf{\tiny{inst}}}\leq 10\%$),
it must be considered that the \HESR will be only filled by 90\% in case of using a barrier-bucket system.
As a consequence, values for $L_{\textsf{\tiny{inst}}}$ during data taking 
are 10\% higher than the ones for $\bar L_{\textsf{\tiny{exp}}}$.

An estimate of peak luminosities, $L_{\textsf{\tiny{peak}}}> L_{\textsf{\tiny{inst}}}$, 
must further include possible effects on a short timescale.
Contrary to homogeneous cluster beams, 
a distinct time structure is expected 
for the granular volume density distribution of a pellet beam.
Such time structure depends on the transverse and longitudinal overlap between 
single pellets and the circulating antiproton beam in the interaction region. 
Deviations of the instantaneous luminosity on a microsecond timescale 
are caused by variations of the pellet size, the pellet trajectory 
and the interspacing between consecutive pellets. 
The latter must be well controlled to avoid the possible presence of 
more than one pellet in the beam at the same instant. 
The resulting ratio $L_{\textsf{\tiny{peak}}}/L_{\textsf{\tiny{exp}}}$ depends on the pellet size.
First studies on the expected peak values for the \panda pellet target have been performed~\cite{Smirnov:2009}.
Results indicate that the peak luminosity stays below $10^{33}$~cm$^{-2}$s$^{-1}$
if the pellet size is not bigger than 20~$\tcmu$m.

Finally, for the extraction of event rates the obtained luminosities are multiplied with the hadronic cross section.
\Tabref{table_EventRates} summarises the main results for a hydrogen target based on a pellet system,
which is expected to deliver upper limits for the occuring event rates.
In addition, a rough estimate for nuclear targets based on the input of \tabref{table_NuclearAvLumi} 
and equation~\ref{eq-pbarNucleonXsection} is given.
Even though these values still must be verified by detailed studies, 
it can be seen that the reduced average luminosity for heavier nuclear targets 
is counter-balanced by an increased cross-section that results in comparable event rates.

Based on the given assumptions and caveats, as discussed in this section, 
a nominal interaction rate of $R_{\textsf{\tiny{nom}}} = 2 \cdot10^7$~s$^{-1}$
can be defined that all detector systems have to be able to handle.
This specification includes the requirement that density fluctuations of the beam-target overlap
have to be smaller than a factor of two ($\bar L_{\textsf{\tiny{peak}}}/\bar L_{\textsf{\tiny{exp}}}$).
However, in order to avoid data loss it might be important to introduce a generic safety factor 
that depends on special features of the individual detector subsystems and their position 
with respect to the interaction region.

%% file: introduction/MainIntro/detector.tex
\section{The \PANDA Detector}
%\section*{I.5 \hspace{4mm} The \PANDA Detector}
%\addcontentsline{toc}{subsection}{I.5 \hspace{4mm} The \PANDA Detector}
\label{s:over:panda}

\begin{figure}[!b]
\begin{center}
\includegraphics[width=0.49\textwidth]{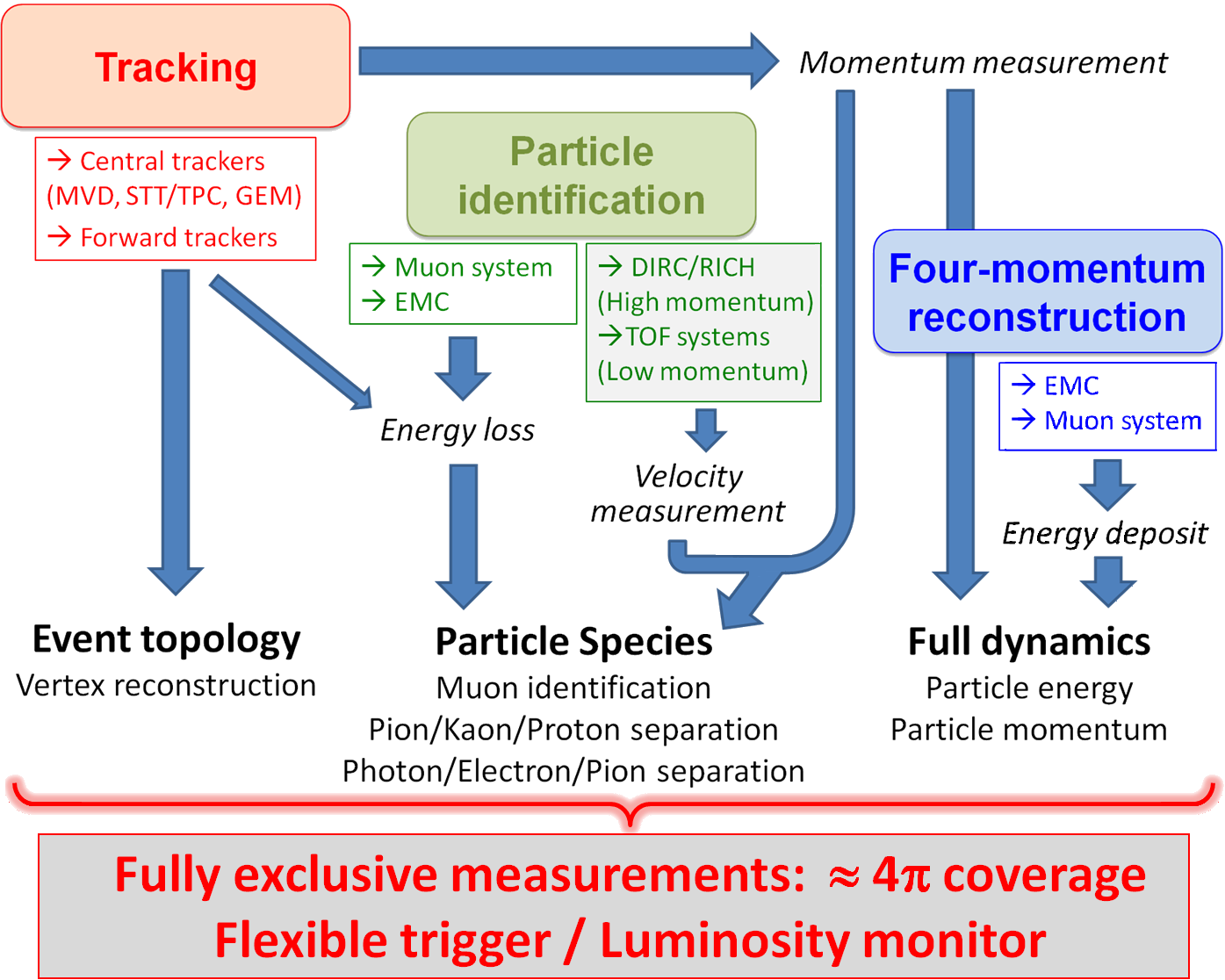}
\end{center}
\caption[Basic detection concept]
{Basic detection concept. The main components will be described in 
chapter~\ref{sec:det:ts} and \ref{sec:det:fs}.
}
\label{pic-DetConcept}       
\end{figure}

The main objectives of the design of the \PANDA experiment 
are to achieve $4\pi$ acceptance, high resolution
for tracking, particle identification and calorimetry, high rate
capabilities and a versatile readout and event selection. 
To obtain a good momentum resolution the detector will be composed of 
two magnetic spectrometers: 
the {\em Target Spectrometer (TS)}, based on a superconducting solenoid magnet surrounding
the interaction point, which will be used to measure at large polar angles 
and the {\em Forward Spectrometer (FS)}, based on a dipole magnet, for small angle tracks. 
An overview of the detection concept is shown in \figref{pic-DetConcept}.

It is based on a complex setup of modular subsystems including tracking detectors
(MVD, STT, GEM), electromagnetic calorimeters (EMC), a muon system,
Cherenkov detectors (DIRC and RICH) and a time-of-flight (TOF) system.
A sophisticated concept for the data acquisition with a flexible trigger is planned 
in order to exploit at best the set of final states relevant for the \PANDA physics objectives.

\subsection{Target Spectrometer}
\label{sec:det:ts}
%\addcontentsline{toc}{subsection}{\hspace{4mm}...  Target Spectrometer}

\begin{figure*}
\begin{center}
\includegraphics[width=0.85\dwidth]{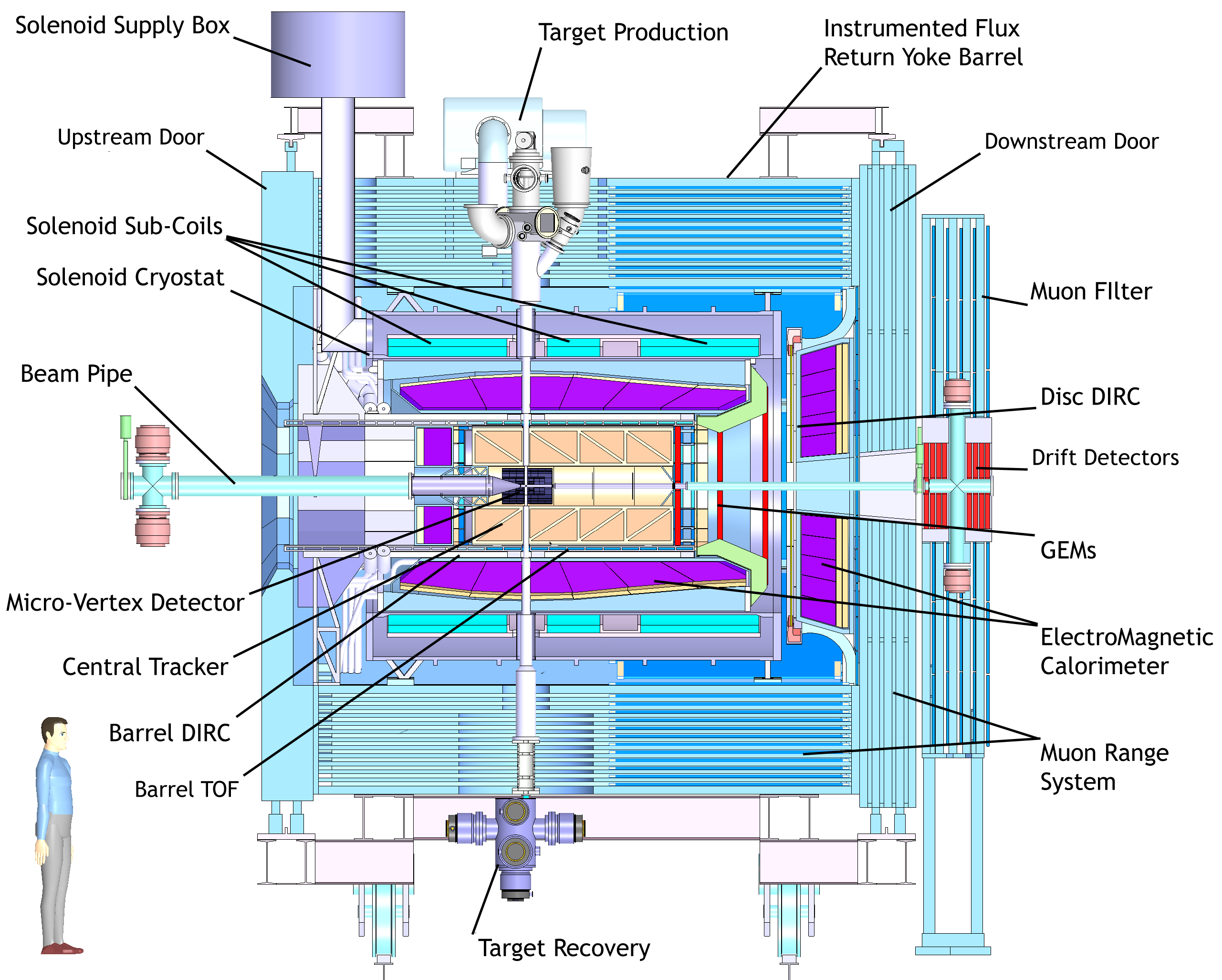}
\caption[Artistic side view of the Target Spectrometer (TS) of \PANDA]
{Artistic side view of the Target Spectrometer (TS) of \PANDA. 
  To the right of this the Forward Spectrometer (FS) follows, which is
illustrated in \figref{f:over:fs_view}.}
\label{f:over:ts_view}
\end{center}
\end{figure*}

The Target Spectrometer will surround the interaction point and measure
charged tracks in a highly homogeneous solenoidal field. 
In the manner of a collider detector it will contain
detectors in an onion shell like configuration. Pipes for the
injection of target material will have to cross the spectrometer
perpendicular to the beam pipe.

The Target Spectrometer will be arranged in three parts: the barrel
covering angles between 22$\degrees$ and 140$\degrees$, the forward
end cap extending the angles down to 5$\degrees$ and 10$\degrees$ in
the vertical and horizontal planes, respectively, and the backward end
cap covering the region between about 145$\degrees$ and 170$\degrees$.
Please refer to \figref{f:over:ts_view} for an overview.

\subsubsection*{Beam-Target System}
\label{sec:det:ts:tgt}

The beam-target system consists of the apparatus for the target production 
and the corresponding vacuum system for the interaction region.  
The beam and target pipe cross sections inside the target spectrometer are decreased 
to an inner diameter of 20~mm close to the interaction region.  
%A beam-target cross is currently foreseen for intersection of the target cross section 
%with the beam pipe. 
The innermost parts are planned to be made of beryllium, titanium or a suited alloy 
which can be thinned to wall thicknesses of 200~$\tcmu$m. 
Due to the limited space and the constraints on the material budget close to the IP, 
vacuum pumps along the beam pipe can only be placed outside the target spectrometer. 
Insections are foreseen in the iron yoke of the magnet which allow the integration 
of either a pellet or a cluster-jet target. 
The target material will be injected from the top. 
Dumping of the target residuals after beam crossing 
is mandatory to prevent backscattering into 
the interaction region.  
The entire vacuum system is kept variable and allows an operation of both target types. 
Moreover, an adaptation to non-gaseous nuclear wire targets is possible. 
For the targets of the planned hypernuclear experiment the whole upstream end cap 
and parts of the inner detector geometry will be modified.
A detailed discussion of the different target options can be found in chapter~\ref{intro:target}.

\subsubsection*{Solenoid Magnet}

The solenoid magnet of the TS will deliver a very homogeneous solenoid field 
of 2~T with fluctuations of less than $\pm$2\%.
In addition, a limit of $\mathrm{\int} B_r/B_z \mathrm{d}z<2$~mm is specified for
the normalised integral of the radial field component.
The superconducting coil of the magnet has a length of 2.8~m and an inner diameter of 90~cm,
using a laminated iron yoke for the flux return.
The cryostat for the solenoid coils is required to have two warm bores of 
100 mm diameter, one above and one below the target position, to
allow for insertion of internal targets.
The load of the integrated inner subsystems can be picked up at defined fixation points.
A precise description of the magnet system and detailed field strength calculations 
can be found in~\cite{PANDA-MagnetTDR}. 

\subsubsection*{Micro Vertex Detector}

\begin{figure}[!b]
\begin{center}
\includegraphics[trim=0 0.1cm 0 0.3cm, clip, width=7. cm]{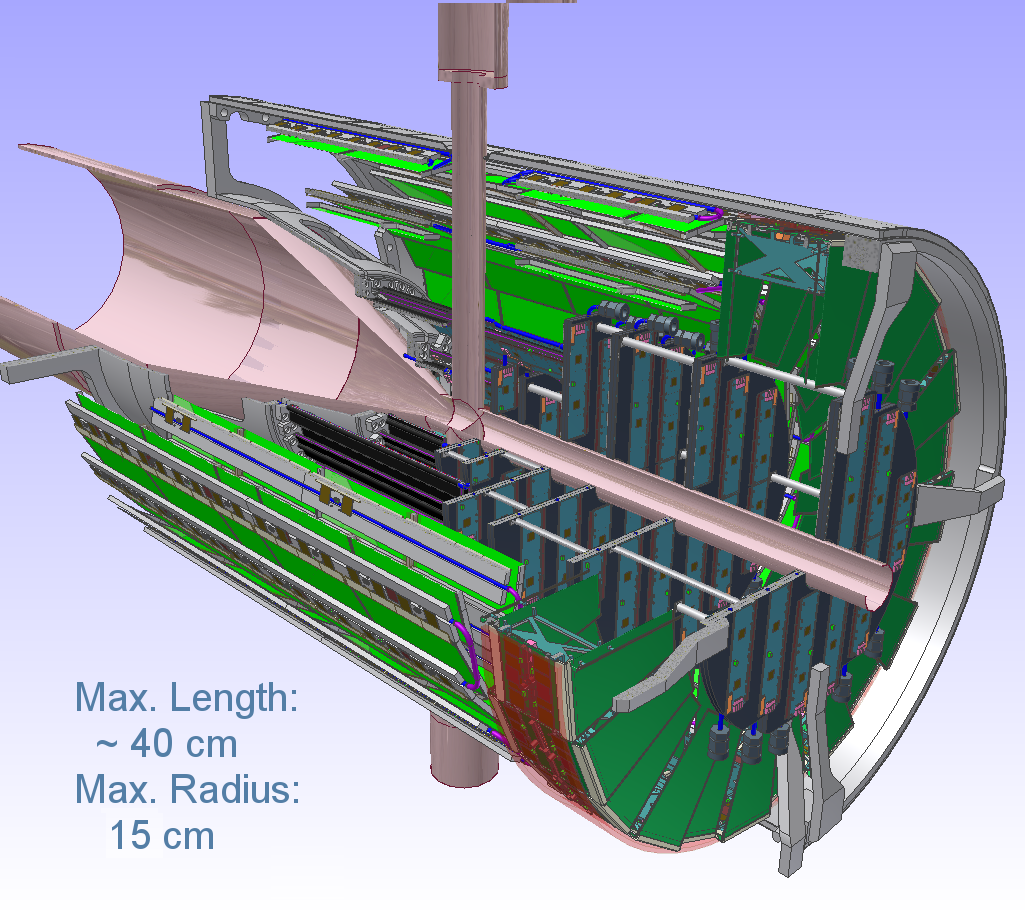}
\caption[The Micro Vertex Detector (MVD) of the Target Spectrometer]
  {The Micro Vertex Detector (MVD) of the Target Spectrometer
  surrounding the beam and target pipes seen from downstream.  
  To allow a look inside the detector a three-quarters portraits is chosen.}
\label{fig:det:mvd}
\end{center}
\end{figure}

The design of the Micro Vertex Detector (\Mvd) for the Target
Spectrometer is optimised for the detection of secondary decay vertices from
charmed and strange hadrons and for a maximum acceptance close to the interaction
point. It will also strongly improve the transverse momentum
resolution. The setup is depicted in \figref{fig:det:mvd}.

The concept of the \Mvd is based on radiation hard silicon pixel
detectors with fast individual pixel readout circuits and silicon
strip detectors. The layout foresees a four layer barrel detector with
an inner radius of 2.5~cm and an outer radius of 13~cm. The two
innermost layers will consist of pixel detectors and the outer two
layers will be equipped with double-sided silicon strip
detectors.

Six detector wheels arranged perpendicular to the beam will achieve
the best acceptance for the forward part of the particle spectrum.
While the inner four layers will be made entirely of pixel detectors,
the following two will be a combination of strip detectors on the outer
radius and pixel detectors closer to the beam pipe. 

\subsubsection*{Additional Forward Disks}

Two additional silicon disk layers are considered further downstream
at around 40~cm and 60~cm 
to achieve a better acceptance of hyperon cascades.
They are intended to be made entirely of silicon strip detectors.
Even though they are not part of the central MVD 
it is planned, as a first approach, to follow 
the basic design as defined for the strip disks of the MVD.
However, an explicit design optimisation still has to be performed.
Two of the critical points to be checked are related 
to the increased material budget caused by these layers 
and the needed routing of cables and supplies for these additional disks
inside the very restricted space left by the adjacent detector systems.

\subsubsection*{Straw Tube Tracker (\Stt)}
\label{sec:det:ts:stt}

\begin{figure}[htb]
\begin{center}
\includegraphics[width=\swidth]{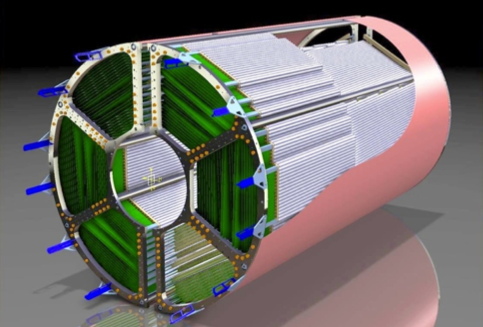} 
\caption[Straw Tube Tracker (STT) of the Target Spectrometer]
  {Straw Tube Tracker (STT) of the Target Spectrometer %with beam and target pipes  
  seen from upstream.}
\label{fig:exp:ts:stt}
\end{center}
\end{figure}

This detector will consist of aluminised Mylar tubes called {\em straws}. 
These will be stiffened by operating them at an overpressure of 1~bar 
which makes them self-supporting.
The straws are to be arranged in planar
layers which are mounted in a hexagonal shape around the \Mvd as shown
in \figref{fig:exp:ts:stt}. In total there are 27 layers of which
the 8 central ones are skewed, to achieve an acceptable resolution of
3 mm also in $z$ (parallel to the beam). The gap to the
surrounding detectors will be filled with further individual
straws. In total there will be 4636 straws around the beam pipe at
radial distances between 15~cm and 41.8~cm with an overall length
of 150~cm. All straws have a diameter of 10~mm 
and are made of a 27~$\tcmu$m thick Mylar foil.
Each straw tube is constructed with a single anode wire in the centre
that is made of 20~$\tcmu$m thick gold plated tungsten 
The gas mixture used will be Argon based with CO$_2$ as quencher. 
It is foreseen to have a gas gain not greater than 10$^5$ in order to
warrant long term operation. With these parameters, a resolution in
$x$ and $y$ coordinates of less than 150~$\tcmu$m is expected.
A thin and light space frame will hold the straws in place, 
the force of the wire however is kept solely by the straw itself. 
This overall design results in a material budget of 1.2\% 
of one radiation length.

\subsubsection*{Forward GEM Detectors}

Particles emitted at angles below 22\degrees{} which are not covered
fully by the STT will be tracked by three planar stations placed approximately 
1.1~m, 1.4~m and 1.9~m downstream of the target.  
Each of the station consists of double planes with two
projections per plane.
The stations will be equipped with Gaseous micro-pattern detectors 
based on Gas Electron Multiplier (GEM) foils as amplification stages. 
The chambers have to sustain a high counting rate of particles peaked 
at the most forward angles due to the relativistic boost of the reaction products 
as well as due to the small angle \pbarp elastic scattering. 
The maximum expected particle flux in the first chamber in the vicinity 
of the 5~cm diameter beam pipe will be about 
$3 \cdot 10^{4}$~cm$^{-2}$s$^{-1}$.

\subsubsection*{Barrel DIRC}

At polar angles between 22\degrees{} and 140\degrees{}, particle
identification will be performed by the Detection of Internally
Reflected Cherenkov (\Dirc) light as realised in the {\INST{BaBar}}
detector~\cite{Staengle:1997xp}.  It will consist of 1.7~cm thick
fused silica (artificial quartz) slabs surrounding the beam line
at a radial distance of 45~cm to 54~cm. At {\INST{BaBar}} the light was imaged across a large
stand-off volume filled with water onto 11,000 photomultiplier
tubes. At \Panda, it is intended to focus the images by lenses onto
Micro-Channel Plate PhotoMultiplier Tubes (MCP PMTs) which are
insensitive to magnet fields. This fast light detector type allows a
more compact design and the readout of two spatial coordinates. 

\subsubsection*{Forward End-Cap DIRC} 

A similar concept is considered to be employed in the forward direction for
particles at polar angles between 5\degrees{} and 22\degrees{}. The same radiator,
fused silica, is to be employed, however in shape of a disk. The
radiator disk will be 2~cm thick and will have a radius of
110~cm. It will be placed directly upstream of the forward end cap
calorimeter. At the rim around the disk the Cherenkov light will be
measured by focusing elements. In addition measuring the time of
propagation the expected light pattern can be distinguished in a
3-dimensional parameter space. Dispersion correction is achieved by
the use of alternating dichroic mirrors transmitting and reflecting
different parts of the light spectrum. As photon detectors either
silicon photomultipliers or microchannel plate PMTs are considered.

\subsubsection*{Scintillator Tile Barrel (Time-of-Flight)}

For slow particles at large polar angles, particle identification will
be provided by a time-of-flight (TOF) detector positioned 
just outside the Barrel DIRC,
where it can be also used to detect photon conversions in the DIRC radiator.
The detector is based on scintillator tiles of $\mathrm{28.5 \times 28.5~mm^2}$ size, 
individually read out by two Silicon PhotoMultipliers per tile.
The full system consists of 5,760 tiles in the barrel part 
and can be augmented also by approximately 1,000 tiles 
in forward direction just in front of the endcap disc DIRC.
Material budget and the dimension of this system are optimised such  
that a value of less than 2\% of one radiation length, including readout
and mechanics and less than 2~cm radial thickness will be reached, respectively.
The expected time resolution of 100~ps will allow precision timing of tracks 
for event building and fast software triggers.
The detector also provides well timed input with a good spatial resolution
for online pattern recognition.

\subsubsection*{Electromagnetic Calorimeters}

Expected high count rates and a geometrically compact design of the
Target Spectrometer require a fast scintillator material with a short
radiation length and Moli\`ere radius for the construction of the
electromagnetic calorimeter (\Emc). Lead tungsten (PbWO$_4$) is a
high density inorganic scintillator with sufficient energy and time
resolution for photon, electron, and hadron detection even at
intermediate energies~\cite{Mengel:1998si,Novotny:2000zg,Hoek:2002ss}.

The crystals will be 20~cm long, i.e.~approximately 22~$X_0$, in
order to achieve an energy resolution below 2\percent{} at
1~$\gev$~\cite{Mengel:1998si,Novotny:2000zg,Hoek:2002ss} at a
tolerable energy loss due to longitudinal leakage of the shower.
Tapered crystals with a front size of  $\mathrm{2.1 \times 2.1~cm^2}$ will be
mounted in the barrel EMC with an inner radius of 57~cm. This
implies 11,360 crystals for the barrel part of the calorimeter.  The
forward end cap EMC will be a planar arrangement of 3,600 tapered
crystals with roughly the same dimensions as in the barrel part, and
the backward end cap EMC comprises of 592 crystals. The readout of
the crystals will be accomplished by large area avalanche photo diodes
in the barrel and in the backward end cap, vacuum photo-triodes will be used 
in the forward end cap. The light yield can be increased by a factor of 
about 4 compared to room temperature by cooling the crystals down to 
$-25$~$\degC$. 
%The arrangement of the barrel and forward end cap calorimeters 
%is shown in \figref{fig:emc}.

The EMC will allow to achieve an $e$/$\pi$ ratio of 10$^3$ for momenta above 0.5~\gevc.  
Therefore, $e$/$\pi$-separation will not require an additional gas Cherenkov detector 
in favour of a very compact geometry of the EMC. 
A detailed description of the detector system can be found in~\cite{PANDA:TDR:EMC}.

\subsubsection*{Muon Detectors}

The laminated yoke of the solenoid magnet acts as a range system for the detection of muons. 
There are 13 sensitive layers, each 3~cm thick (layer ``zero" is a double-layer). They alternate with 3~cm thick iron absorber layers (first and last iron layers are 6~cm thick), introducing enough material for the absorption of pions in the \PANDA momentum range and angles.
In the forward end cap more material is needed due to the higher momenta of the occurring particles. 
Therefore, six detection layers will be placed around five iron layers of 6~cm each 
within the downstream door of the return yoke, 
and a removable muon filter with additional five layers of 6~cm iron and corresponding detection layers
will be moved in the space between the solenoid and the dipole. 

As sensors between the absorber layers, rectangular aluminum Mini Drift Tubes (MDT) are foreseen.
%which are similar to the ones applied at the COMPASS experiment~\cite{COMPASS}.
Basically, these are drift tubes with additional capacitive
coupled strips, read out on both ends to obtain the longitudinal coordinate.
All together, the laminated yoke of the solenoid magnet and the additional muon filters 
will be instrumented with 2,600 MDTs and 700 MDTs, respectively.

\begin{figure*}[htb]
\begin{center}
\includegraphics[width=0.9\dwidth]{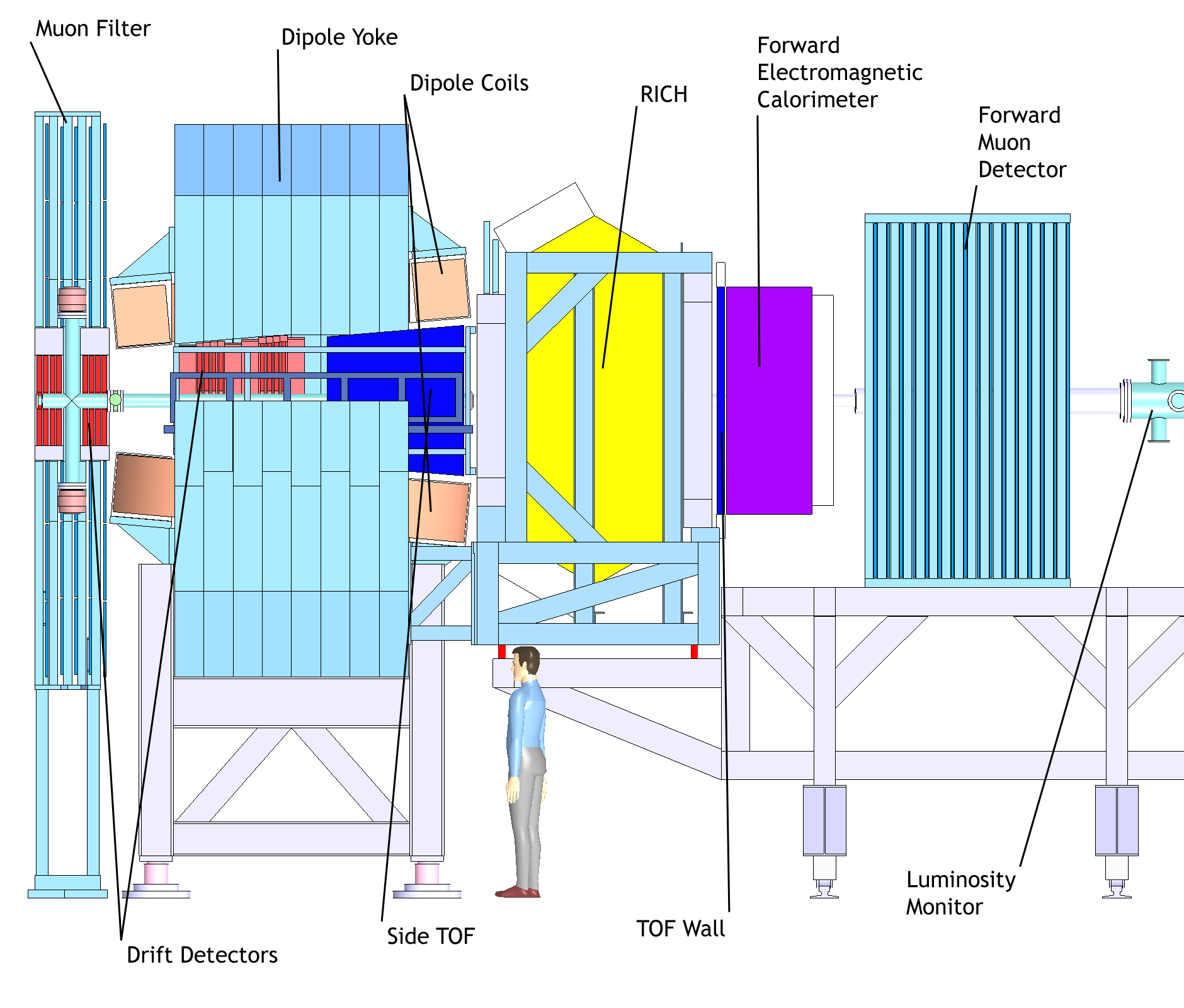}
\caption[Artistic side view of the Forward Spectrometer (FS) of \PANDA]
{Artistic side view of the Forward Spectrometer (FS) of \PANDA. 
  It is preceded on the left by the Target Spectrometer (TS),
which is illustrated in \figref{f:over:ts_view}.}
\label{f:over:fs_view}
\end{center}
\end{figure*}

\subsubsection*{Hypernuclear Detector}

The hypernuclei study will make use of the modular structure of
\Panda. Removing the backward end cap calorimeter and the \Mvd will allow to
add a dedicated nuclear target station and the required additional
detectors for $\gamma$ spectroscopy close to the entrance of
\Panda. While the detection of hyperons and low momentum $K^{\pm}$
can be ensured by the universal detector and its PID system, a
specific target system and a $\gamma$-detector are additional
components required for the hypernuclear studies.
%\paragraph*{Active Secondary Target}
%compact high resolution solid state tracker} 

The production of hypernuclei proceeds as a two-stage process. 
First hyperons, in particular \CascadeCascadebar, are produced on a nuclear target. 
%In some cases the $\Xi$ will be slow enough to be captured in
In addition, a secondary target is needed for the formation of a double hypernucleus.
%a secondary target, where it reacts in a nucleus to form a .
The geometry of this secondary target is determined 
by the short mean life of the \Cascademinus of only 0.164~ns. %$\Xi^-$
This limits the required thickness of the active secondary target to about 25~mm to 30~mm. 
It will consist of a compact sandwich structure of silicon micro-strip detectors and absorbing material. 
In this way the weak decay cascade of the hypernucleus can be detected in the sandwich structure.
%\paragraph*{Germanium Array} 

An existing germanium-array with refurbished readout will be used for
the $\gamma$-spectroscopy of the nuclear decay cascades of
hypernuclei. The main limitation will be the load due to neutral or
charged particles traversing the germanium detectors. Therefore,
readout schemes and tracking algorithms are presently being developed
which will enable high resolution $\gamma$-spectroscopy in an
environment of high particle flux.

%%%%%%%%%%%%%  

\subsection{Forward Spectrometer}
%\addcontentsline{toc}{subsection}{\hspace{4mm}...  Forward Spectrometer}
\label{sec:det:fs}

The Forward Spectrometer (FS) will cover all particles emitted in
vertical and horizontal angles below $\pm \mathrm{5}^\circ$ and $\pm \mathrm{10}^\circ$,
respectively.  Charged particles will be deflected by an integral
dipole field.  Cherenkov detectors, calorimeters and
muon counters ensure the detection of all particle types.  
\Figref{f:over:fs_view} gives an overview to the instrumentation 
of the FS.

\subsubsection*{Dipole Magnet}
A 2~Tm dipole magnet with a window frame, a 1~m gap, and more
than 2~m aperture will be used for the momentum analysis of charged
particles in the FS.  In the current planning, the
magnet yoke will occupy about 1.6~m in beam direction starting from
3.9~m downstream of the target.  Thus, it covers the entire angular
acceptance of the TS of $\pm$10\degrees{} and
$\pm$5\degrees{} in the horizontal and in the vertical direction,
respectively.  The bending power of the dipole on the beam line causes
a deflection of the antiproton beam at the maximum momentum of
15~$\gevc$ of 2.2\degrees{}. 
%/The designed acceptance for charged
%/particles covers a dynamic range of a factor 15 with the detectors
%/downstream of the magnet. 
For particles with lower momenta, detectors
will be placed inside the yoke opening. The beam deflection will be
compensated by two correcting dipole magnets, placed around
the \Panda detection system.
The dipole field will be ramped during acceleration in the \HESR 
and the final ramp maximum scales with the selected beam momentum.

%For more details on the dipole magnet see~\cite{PANDA-MagnetTDR}.

\subsubsection*{Forward Trackers}
\label{sec:det:fs:trk}

%\COM[Inti]{I am not sure if my additions are correct...}

\begin{figure}[htb]
\begin{center}
\includegraphics[width=0.90\swidth]{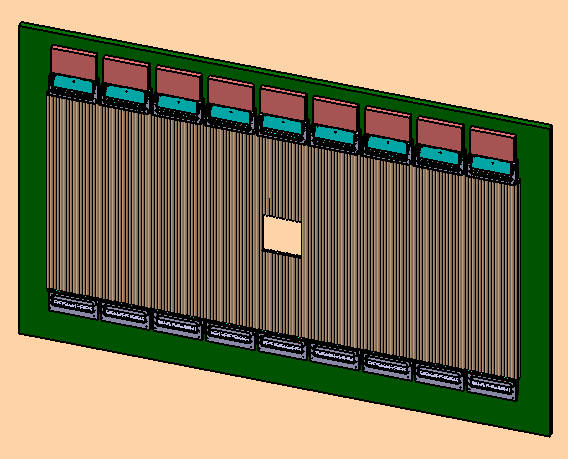} 
\caption[Double layer of straw tubes in the tracker of the Forward Spectrometer]
{Double layer of straw tubes with preamplifier cards and gas manifolds
mounted on rectangular support frame.
The opening in the middle of the detector is foreseen for the beam pipe.}
\label{fig:det:fs:trk:dc1}
\end{center}
\end{figure}

The deflection of particle trajectories in the field of the dipole
magnet will be measured with three pairs of tracking drift detectors.
The first pair will be placed in front, the second within and the
third behind the dipole magnet.  Each pair will contain two autonomous
detectors, thus, in total, 6 independent detectors
will be mounted.  Each tracking detector will consist of four
double-layers of straw tubes (see \figref{fig:det:fs:trk:dc1}), two
with vertical wires and two with wires inclined by a few degrees.  The
optimal angle of inclination with respect to vertical direction will
be chosen on the basis of ongoing simulations.  The planned
configuration of double-layers of straws will allow to reconstruct
tracks in each pair of tracking detectors separately, also in case of
multi-track events.
 
\subsubsection*{Forward Particle Identification}
%\paragraph*{RICH Detector}
To enable the $\pi$/$K$ and $K$/$p$ separation also at the highest
momenta a \Rich detector is proposed. The favoured design is a dual
radiator \Rich detector similar to the one used at
\INST{HERMES}~\cite{Akopov:2000qi}. Using two radiators, silica
aerogel and C$_4$F$_{10}$ gas, provides $\pi$/$K$/$p$ separation in a
broad momentum range from 2 to 15~$\gevc$.  The two different indices
of refraction are 1.0304 and 1.00137, respectively.  The total
thickness of the detector is reduced to the freon gas radiator
\mbox{(5\%$X_0$),} the aerogel radiator (2.8\%$X_0$), and the
aluminum window (3\%$X_0$) by using a lightweight mirror focusing
the Cherenkov light on an array of photo-tubes placed outside the
active volume.

%\paragraph*{Time-of-Flight Wall}
A wall of slabs made of plastic scintillator and read out on both ends
by fast photo-tubes will serve as time-of-flight stop counter placed at
about 7~m from the target. 
Similar detectors will be placed inside the dipole magnet opening 
to detect low momentum particles which do not exit the dipole magnet.
The time resolution is expected to be in the order of 50~ps
thus allowing a good $\pi$/$K$ and  $K$/$p$ separation 
up to momenta of 2.8~\gevc and 4.7~\gevc, respectively.

\subsubsection*{Forward Electromagnetic Calorimeter}

For the detection of photons and electrons a {Shashlyk}-type
calorimeter with high resolution and efficiency will be employed. 
The detection is based on lead-scintillator sandwiches read out with
wave-length shifting fibres passing through the block and coupled to
photo-multipliers. 
The lateral size of one module is $\mathrm{110~mm \times 110~mm}$ and a length of 680~mm ($=20X_0$).
A higher spatial resolution will be achieved by sub-dividing each module 
into 4 channels of 55~mm$\times$55~mm size coupled to 4 PMTs.
To cover the forward acceptance, 351
such modules, arranged in 13 rows and 27 columns at a distance of
7.5~m from the target, are required.
With similar modules, based on the same technique as proposed for \PANDA,
an energy resolution of $4\%/\sqrt{E}$~\cite{bib:emc:KOP99} has been achieved.

\subsubsection*{Forward Muon Detectors}

For the very forward part of the muon spectrum, a further range
tracking system consisting of interleaved absorber layers and
rectangular aluminium drift-tubes is being designed, similar to the
muon system of the TS, but laid out for higher
momenta. The system allows discrimination of pions from muons,
detection of pion decays and, with moderate resolution, also the
energy determination of neutrons and anti-neutrons.
The forward muon system will be placed at about 9~m from the target.

\subsubsection*{Luminosity Monitor}
%\addcontentsline{toc}{subsection}{\hspace{4mm}...  Luminosity monitor}

The basic concept of the luminosity monitor is to reconstruct 
the angle of elastically scattered antiprotons in the polar angle range 
from 3~mrad to 8~mrad with respect to the beam axis
corresponding to the Coulomb-nuclear interference region.
The luminosity monitor will consist of a sequence
of four planes of double-sided silicon strip detectors
located in the space between the downstream side of the forward muon system 
and the \HESR dipole needed to redirect the antiproton beam out of the \PANDA chicane back
into the direction of the \HESR straight stretch 
(i.e.~between $z=+\,11$~m and $z=+\,13$~m downstream of the target).  
The planes are positioned  as close to the beam axis as possible 
and are separated by 10~cm along the beam direction. 
Each plane consists of four wafers 
arranged perpendicular surrounding to the beam axis placed at top, down, right and left. 
In this way, systematic errors can be strongly suppressed. 
The silicon wafers will be located inside a vacuum chamber to minimise
scattering of the antiprotons before traversing the tracking planes. 
With the proposed detector setup 
an absolute precision of about 3\% on the time integrated luminosity 
is considered feasible for this detector concept at \PANDA.

\subsection{Data Acquisition}
%\addcontentsline{toc}{subsection}{\hspace{4mm}...  Data Acquisition}

In \PANDA, a data acquisition concept is being developed to be
as much as possible matched to the complexity of the
experiment and the diversity of physics objectives and the rate
capability of at least $2\cdot10^{7}$ events/s.
Therefore, every sub-detector system is a self-triggering entity.
Signals are detected autonomously by the sub-systems and are preprocessed.
Only the physically relevant information is extracted and transmitted.
This requires hit-detection, noise-suppression and clusterisation at
the readout level. 
The data related to a particle hit, with a substantially reduced rate
in the preprocessing step, is marked by a precise time stamp and
buffered for further processing.
The trigger selection finally occurs in computing nodes which
access the buffers via a high-bandwidth network fabric. The new
concept provides a high degree of flexibility in the choice of trigger
algorithms. It makes trigger conditions available which are
outside the capabilities of the standard approach.

\subsection{Infrastructure}
%\addcontentsline{toc}{subsection}{\hspace{4mm}...  Infrastructure}

\label{s:over:infra}

The experimental hall for \PANDA with a planned floor space of $\mathrm{43~m \times 29~m}$
will be located in the straight section at the east side of the \HESR.  
Within the elongated concrete cave the \Panda detector 
together with auxiliary equipment, beam steering, and focusing elements will be housed.  
Moreover, the experimental hall will provide additional space 
for delivery of components and assembly of the detector parts. 
To allow for access during \HESR operation the beam line is shielded.
%by a concrete radiation shield of 2~m thickness on both sides and 
%is covered on top by concrete bars of 1$\,$m thickness.  
The floor space for the \Panda experiment including dipoles and focusing elements 
is foreseen to have an area of $\mathrm{37~m \times 9.4~m}$
and a height of 8.5~m with the beam line at a height of 3.5~m.  
The general floor level of the \HESR is planned to be 2~m higher
%This level will be kept for a length of 4~m in the north as well
%as the south of the hall (right part in~\Reffig{f:over:Panda_Hall}),
to facilitate transport of heavy equipment into the \HESR tunnel.
A crane spans the whole area with a hook at a height of about 10~m.  
%It is planned that it will be possible to open the roof of the
%cave as well as its wall, so that heavy components can be hoisted in
%by crane.
The TS with electronics and supplies will be mounted
on rails which makes it retractable to parking positions outside of
the \HESR beam line area. 
In the south corner of the hall, a counting house complex with five floors is foreseen. 
It will contain supplies e.g.~for power, high voltage, cooling water and gases (1$^{\mathrm{st}}$ floor),
provide space for the readout electronics and data processing (2$^{\mathrm{nd}}$ floor) 
as well as the online computing farm (3$^{\mathrm{rd}}$ floor)
and house the hall electricity supplies and ventilation (5$^{\mathrm{th}}$ floor).
The fourth floor with the control room, a meeting room and social rooms for the shift crew
will be at level with the surrounding ground. 
The supply point will be at the north-east area of the building.  
All other cabling, which will be routed starting at the counting house, 
will join these supply lines at the end of the rails system of 
the Target Spectrometer at the eastern wall.  
The temperature of the building will be moderately controlled.  
More stringent requirements with respect to temperature and humidity for the detectors 
have to be maintained locally.  
To facilitate cooling and avoid condensation, the Target
Spectrometer will be kept in a tent with dry air at a controlled temperature.

%% file: introduction/MainIntro/req.tex
The different topics of the \PANDA physics program
%each of which is scientifically well justified, 
will impose specific optimisation criteria and
requirements to design and performance of the tracking system.
The optimum design thus depends on the relative weight 
which is given to the different physics aspects. 
Main criteria for the optimisation will be discussed in the following.
\subsubsection*{Acceptance}

Full $2\pi$ azimuthal coverage is mandatory in order to allow
identification of multi-particle final states and studies of
correlations within the produced particles. 
In particular, the spectroscopy program of charmed and strange hadrons 
relies on the measurement of Dalitz plot distributions of three-body final states, 
which requires a smooth acceptance function across the full phase space. 
Particular care has to be taken to avoid gaps in
the acceptance function and to minimise the effect of
discontinuities induced by the transition between adjacent
sub-detector components, by detector frames or by mechanical
support structures.

The fixed-target setup at \PANDA implies a Lorentz boost $\gamma_{CM}$ 
of the centre of mass ranging from 1.20 to 2.92. 
This large dynamic range in the Lorentz boost corresponds 
to a large difference in the typical event topologies
at low and at high antiproton momenta. 
At higher antiproton beam momenta 
the vast majority of the produced particles in the final state 
will be emitted into the forward hemisphere. 
However, light particles like $e^{\pm}$, ${\mu^{\pm}}$ or ${\pi^{\pm}}$ 
may well be emitted into the backward hemisphere 
even at highest beam momentum.
%The critical value of the center of mass momentum 
%that can result in backward emission in the laboratory frame
%is given by 
%$p_{\rm{}crit}=\beta_{CM}\cdot\gamma_{CM}\cdot{}m_x=
%\left(p_{\bar{p}}/\sqrt{s}\right)\cdot{}m_x$. 
As an example, pion backward emission is possible for 
a centre of mass momentum $p_{cm}>93$~\mevc at $p_{\bar{p}}=1.5$~\gevc, 
and for $p_{cm}>380$~\mevc at $p_{\bar{p}}=15$~\gevc. 

%Since muon-pion separation requires
%passage through a large amount of material and thus high momenta
%to efficiently suppress the much more abundant pions, muon
%detection at backward angles doesn't seem to be feasible. 

%However, in the tracking of electrons and pions coverage of a significant
%fraction of the backward hemisphere is required.

Backward charged particle tracking is needed 
for various measurements foreseen at \PANDA.
For instance, for the independent determination of 
the electric and magnetic parts of the time-like proton form factor 
in the reaction \pbarp~$\to$~\ee 
the full angular distribution has to be measured. 
At $q^2=14$~\gevcsq, that is at $p_{\bar{p}}=6.45$~\gevc, 
a polar angle of $160^{\circ}$ in the centre of mass frame 
corresponds to electrons with a momentum of 0.70~\gevc at $\theta_{lab}=113^{\circ}$. 
Detection of pions in the backward hemisphere is important 
in studies of strange, multi-strange and charmed baryon resonances in
\pbarp~$\to$~$Y^{\star}\bar{Y}'$ (+c.c.) reactions where
the excited hyperon $Y^{\star}$ decays by single or double pion
emission. 
Also higher charmonium states may emit pions with decay
energies above the critical value for backward emission in the
laboratory. 
The \PANDA tracking detectors therefore have to cover
the full range of polar angles between $0^{\circ}$ 
%(for particles with $p/q$ different from that of the antiproton beam) 
and about $150^{\circ}$.

Besides the solid angle of the detector 
also the acceptance in momentum space has to be considered. 
Often the final state contains charged particles 
with very large and with very small transverse momentum components 
which need to be reconstructed at the same time. 
Given the strength of the solenoid field of 2~T 
%(at $p_{\bar{p}}\ge{}3.8\,{\rm{}GeV}/c$) 
required to determine the momentum vector of the high transverse momentum particle, 
the radius of the transverse motion of the low transverse momentum particle may be small. 
Sufficient tracking capability already at small distance from the beam axis 
is therefore mandatory. 
As an example one may consider the reaction 
$\bar{p}p\rightarrow{}D^{*+}D^{*-}$ 
close to threshold with $D^{*+}\rightarrow{}D^0\pi^+$ (\& c.c.).
Assuming 39~\mevc momentum of the decay particles in the $D^{*\pm}$ rest frame,
%%and 39~MeV/c momentum of the decay particles in the $D^{*\pm}$ rest frame. 
%The $D^0\rightarrow{}K^-\pi^+$ (\& c.c.) decay particles have 61~MeV/c momentum 
%in the $D^0$/$\bar{D}^0$ rest frame. 
particles of the subsequent decay $D^0\rightarrow{}K^-\pi^+$ (\& c.c.) 
have 61~\mevc momentum in the $D^0$/$\bar{D}^0$ rest frame.
In the solenoid field of the TS,
the charged pions and kaons from the $D^0$/$\bar{D}^0$ decay 
may have helix diameters up to almost 1.5~m.
The transverse motion of the charged pion from the  $D^{*\pm}$ decay
stays within a distance of almost 7~cm from the beam axis
and therefore need to be reconstructed based on the
track information from the MVD only.

\subsubsection*{Delayed Decay Vertex Detection}

An important part of the \PANDA physics program involves final
states consisting of hadrons with open charm or strangeness which
decay by weak interaction and thus have macroscopic decay lengths.
The decay length of charmed hadrons is of the order of 100~$\tcmu$m
($\approx$~310~$\tcmu$m for $D^{\pm}$, $\approx$~150~$\tcmu$m for
$D_s^{\pm}$, $\approx$~120~$\tcmu$m for $D^0$, $\approx$~130~$\tcmu$m for
$\Cascade_c^+$, $\approx$~60~$\tcmu$m for $\Lambdaplain_c^+$, and
$\approx$~30~$\tcmu$m for $\Cascade_c^0$). 
Therefore, the design of the tracking system aims on a detection
%\PANDA detector is designed
%according to the goal to detect 
of decay vertices of particles with
decay lengths above 100~$\tcmu$m. 
In order to achieve sufficient separation of the reconstructed decay vertex, 
the inner part of the tracking system has to be located very close to the
interaction point, both in longitudinal and in radial direction.
This requirement is fulfilled in the design of the MVD.

The identification of hyperons and $K_S$ mesons requires the
reconstruction of delayed decay vertices at much larger distances.
$\Lambdaplain$ and $\Cascade$ hyperons have comparatively large decay
lengths of about $8$~cm and $5$~cm, respectively. 
Due to the Lorentz boost this may result in vertices 
which are displaced by tens of centimetres 
from the interaction point mostly in the downstream direction. 
The considerations in the previous section concerning
the required acceptance thus apply with respect to the shifted
emission points of charged particles. 
The inner part of the \PANDA tracking system,
%in particular the Micro-Vertex-Detector (MVD) as the inner part,
therefore needs sufficient extension to the downstream direction
in order to deliver sufficient track information for charged
particle tracks originating from these displaced vertices.

\subsubsection*{Momentum and Spatial Resolution}

The spatial resolution of the tracking detectors is important in
two aspects. In the vicinity of the interaction point it directly
determines the precision to which primary and displaced decay
vertices can be reconstructed. Further on, based on the deflection
of charged particles in both solenoid and dipole magnetic fields,
it is an essential contribution to the momentum resolution of
charged particles in all three coordinates.

The detection of displaced vertices of charmed hadrons imposes
particular requirements to the spatial resolution close the interaction point. 
With a typical Lorentz boost $\beta\gamma\simeq{}2$, $D$ meson decay vertices 
have a displacement of the order of a few hundreds micrometres 
from the primary production point. 
Hence, to distinguish charged daughter particles of $D$ mesons from prompt particles 
a vertex resolution of 100~$\tcmu$m is required. 
The position resolution is less demanding for the reconstruction of strange hadrons 
having decay lengths on the scale of centimeters. 
In this case a vertex resolution of a few millimetres is sufficient.
Due to the significant Lorentz boost and the small opening angle between the decay particles of hyperons 
the resolution in transverse direction is required to be much better 
than the one for the longitudinal component.

The achievable momentum resolution is a complex function of the
spatial resolution of the tracking sub-detectors, the number of
track-points, the material budget of active and passive components
resulting in multiple scattering, the strength and homogeneity of
the magnetic field, and of the particle species, its momentum and
its emission angle. Due to the respective momentum dependence, it
is generally expected that multiple scattering limits the momentum
resolution of low energy particles, whereas for high energy
particles the smaller curvature of the tracks is the dominant
contribution to the resolution. 

The resolution in the determination of the momentum vectors of the
final state particles directly determines the invariant 
or missing mass resolution of the particles that are to be reconstructed.
Typically, the width of hadrons unstable with respect to strong
interaction (except for certain narrow states like e.g.~charmonium
below the $D\bar{D}$ threshold) is of the order of 10~\mevcc to 100~\mevcc.
As an instrumental mass resolution much below the natural width is
without effect, a value of a few 10~\mevcc seems to be acceptable for
the identification of known states or for the mass measurement of new states. 
With a typical scale of \gevcc for the kinematic particle energy
this translates to a relative momentum resolution $\sigma_p/p$ of
the order of 1\% as design parameter for the \PANDA tracking detectors.

%Simulation results presented in
%this report demonstrate that with the chosen design of the
%tracking detectors the achieved mass and momentum resolution
%fulfills the requirements given above.

%
\subsubsection*{Count Rate Capability}
The expected count rates depend on the event rate as discussed 
in chapter~\ref{lumi-considerations}
and the multiplicity of charged particles produced in the events.
While the total rate is of importance for DAQ design and online
event filtering, the relevant quantity for detector design and
performance is the rate per channel, which is a function of the
granularity per detector layer and of the angular distribution of
the emitted particles. 
The latter depends on the beam momentum and the target material.

The nominal event rate at \PANDA is given by $2 \cdot 10^7$ interactions per second.
In case of $\bar{p}p$ annihilations 
typically only a few charged particles are produced. 
Even if secondary particles are taken into account,
the number of charged particles per event will not be much larger
than 10 in most cases. 
Thus the detector must able to cope with a rate of $2 \cdot 10^8$ particles
per second within the full solid angle.
Particular attention has to be paid to elastic $\bar{p}p$
scattering since this process contributes significantly to the
particle load in two regions of the detector: scattering of
antiprotons at small forward angles and the corresponding emission
of recoil protons at large angles close to $90^{\circ}$. This affects
primarily the inner region of the MVD disc layers and the forward
tracking detector as well as the MVD barrel part and the central
tracker. 

The use of nuclear targets will not create significantly 
higher count rates than obtained with a hydrogen or deuterium target. 
This is due to single Coulomb scattering 
which dramatically increases with the nuclear charge
($\propto{}Z^4$) and results in $\bar{p}$ losses 
with no related signals in the detector. 
In contrast to $\bar{p}p$ collisions in $\bar{p}A$ collisions no
high rate of recoil particles close to $90^{\circ}$ is expected. 
The emission angles of recoil protons from quasi-free $\bar{p}p$
scattering are smeared by Fermi momentum and rescattering, while
recoil nuclei, if they at all survive the momentum transfer, are
too low energetic to pass through the beam pipe.

\subsubsection*{Particle Identification}

Charged particle identification over a wide range of momentum and
emission angle is an essential prerequisite for the capability of
\PANDA to accomplish the envisaged physics program. 
Charged particles with higher momenta will be identified via Cherenkov
radiation by the DIRC detector in the Target Spectrometer  and by
the forward RICH detector in the Forward Spectrometer. 
For positive charged kaon-pion separation in the DIRC 
about 800~\mevc momentum is required. 
While almost all particles emitted within the acceptance of the Forward
Spectrometer are above the Cherenkov threshold due to the forward
Lorentz boost, a number of interesting reaction channels have
final states with heavier charged particles ($K^{\pm},p,\bar{p}$)
at larger angles with momenta below the DIRC threshold. 
In order to separate these low energy kaons
from the much more abundant pions, particle identification
capability based on energy loss information has to be supplied by
the central tracking detector. 
%The required separation power and
%thus quality of the $dE/dx$ measurement in the central tracker
%depends on whether or not a TOF barrel, delivering independent
%information on the particle velocity, will be included in the
%Target Spectrometer.

\subsubsection*{Material Budget}

Any active or passive material inside the detector volume
contributes to multiple scattering of charged particles, electron
bremsstrahlung and photon conversion, and thus reduces the
momentum resolution for charged particles in the tracking
detectors, and detection efficiency and energy resolution for
photons in the EMC. Therefore the material budget has to be kept
as low as possible. Following the more demanding requirements to
meet the performance criteria of the EMC, a total material budget
of MVD and Central Tracker below 10\% is still considered to be
acceptable~\cite{PANDA:TDR:EMC}.

%% file: introduction/introduction.tex
\chapter{The Micro Vertex Detector -- MVD}

\input{introduction/GeneralOverview}

\input{introduction/Specifics}

\input{introduction/MVD-layout}

%\section*{Authors}
%\begin{tabbing}
%\hspace{3cm} \=T. W\"urschig \\
%                       \> ...
%\end{tabbing}

\putbib[lit_intro]

%% file: introduction/GeneralOverview.tex
\section{General Overview}
%\alert{Author: Thomas W\"{u}rschig, Contact: t.wuerschig$\mathrm @$hiskp.uni-bonn.de}

This volume illustrates the technical layout and the
expected performance of the Micro Vertex Detector
(MVD) of the \PANDA experiment.
The introductory part contains a physics motivation that underlines the importance 
of this tracking system for key experiments at \PANDA.
Basic detector requirements are defined taking into account both 
the demands of the scientific program
and the experimental conditions of the 
antiproton accelerator \HESR as well as the \Panda target and detector setup.
Moreover, the overall layout of the MVD is described. 

Silicon detectors will be used as sensitive elements inside the MVD.
The default option foresees hybrid pixel detectors 
and double-sided microstrip detectors in the inner and outer parts, respectively.
Both of them are based on different detector technologies
and thus impose specific requirements for the sensor production and characterisation,
the connected \frontend electronics and the hybridisation.
Hence, all these topics will be discussed separately 
for the silicon pixel and the silicon strip part
in chapter~\ref{pixelpart} and~\ref{strippart}, respectively.
Further aspects related to the MVD infrastructure, which are relevant for both parts,
will be outlined in chapter~\ref{infrastructure}.
The performance of the MVD has been studied in extensive Monte Carlo simulations,
which will be summarised in chapter~\ref{simulations}.

%% file: introduction/Specifics.tex
\section{Physics with the MVD}

\begin{table*}
\begin{center}
\begin{tabular}{lccl}
\hline\hline
particle		& lifetime	& decay length $c \tau$	& decay channel (fraction) \\
\hline
\vspace{-3.85mm}&&&\\
$\Ks$ 		& $89.53(5)\,\ps$  	& 2.6842$\,\cm$	& $\pip \pim  \,((69.20 \pm 0.05)\%)$\\
\vspace{-3.85mm}&&&\\
$D^{\pm}$	& $1.040(7)\,\ps$  	& 311.8$\,\tcmu $m	& $e^+  \mbox{semileptonic} + \mbox{c.c.}  \,((16.07 \pm 0.30)\%)$\\
		&		&	     	& $\Km \mbox{anything} + \mbox{c.c.}  \,((25.7 \pm 1.4)\%)$\\
		&		&   		& $\Kp \mbox{anything} + \mbox{c.c.}  \,((5.9 \pm 0.8)\%)$\\
		&		&   		& $\Kzbar \mbox{anything} + \Kz \mbox{anything}  \,((61 \pm 5)\%)$\\
		&		&   		& e.g.~ $\Km \pip \pip + \mbox{c.c.}  \,((9.4 \pm 0.4)\%)$\\
		&		&   		&         $\Kzbar \pip \pip \pim + \mbox{c.c.}  \,((6.8 \pm 0.29)\%)$\\
\vspace{-3.85mm}&&&\\
$\Dz$		& $410.1(15)\,\fs$	& 122.9$\,\tcmu $m	& $e^+  \mbox{anything} + \mbox{c.c.}  \,((6.49 \pm 0.11)\%)$\\
		&		&   		& $\mu^+ \mbox{anything} + \mbox{c.c.}  \,((6.7 \pm 0.6)\%)$\\
		&		&   		& $\Km \mbox{anything} + \mbox{c.c.}  \,((54.7 \pm 2.8)\%)$\\
		&		&   		& $\Kp \mbox{anything} + \mbox{c.c.}  \,((3.4 \pm 0.4)\%)$\\
		&		&   		& $\Kzbar \mbox{anything} + \Kz \mbox{anything}  \,((47 \pm 4)\%)$\\
		&		&   		& e.g.~ $\Ks \Kp \Km   \,((4.65 \pm 0.30)\%)$\\
		&		&   		&         $\Km \pip \pip \pim + \mbox{c.c.}  \,((8.09 ^{+ 0.21}_{- 0.18})\%)$\\
		&		&   		&         $\Ks \pip \pim \piz + \mbox{c.c.}  \,((5.4 \pm 0.6)\%)$\\
\vspace{-3.85mm}&&&\\
$\Dspm$		& $500(7)\,\fs$  	& 149.9 $\,\tcmu $m 	& $e^+  \mbox{semileptonic} / \mbox{c.c.}  \,((6.5 \pm 0.4)\%)$\\
		&		&  	 	& $\Km \mbox{anything} + \mbox{c.c.}  \,((18.7 \pm 0.5)\%)$\\
		&		&  	 	& $\Kp \mbox{anything} + \mbox{c.c.}  \,((28.8 \pm 0.7)\%)$\\
		&		&   		&  e.g.~ $\Kp \Km \pip + \mbox{c.c.}  \,((5.50 \pm 0.27)\%)$\\
\hline
\vspace{-3.85mm}&&&\\
$\Lambdaplain$	& $263.1(20)\,\ps$  	& 7.89$\,\cm$  	& $p \pim  \,((63.9  \pm 0.5)\%)$\\
\vspace{-3.85mm}&&&\\
$\Sigmaplain^+$	& $80.18(26)\,\ps$	& 2.404$\,\cm$	& $p \piz  \,((51.57 \pm 0.30)\%)$\\
		&		&		& $n \pip  \,((48.31 \pm 0.30)\%)$\\
\vspace{-3.85mm}&&&\\
$\Sigmaplain^-$	& $147.9(11)\,\ps$	& 4.434$\,\cm$	& $n \pim  \,((99.848 \pm 0.005)\%)$\\
\vspace{-3.85mm}&&&\\
$\Cascade^-$		& $163.9(15)\,\ps$	& 4.91$\,\cm$	& $\Lambdaplain \pim  \,((99.887 \pm 0.035)\%)$\\
\vspace{-3.85mm}&&&\\
$\Omegaplain^-$	& $82.1(11)\,\ps$	& 2.461$\,\cm$	& $\Lambdaplain \Km  \,((67.8 \pm 0.7)\%)$\\
		&		&		& $\Cascade^0 \pim  \,((23.6 \pm 0.7)\%)$\\
		&		&		& $\Cascade^- \piz  \,((8.6\pm0.4)\%)$\\
\vspace{-3.85mm}&&&\\
$\Lambdaplain_c^+$	& $200(6)\,\fs$	& 59.9$\,\tcmu $m  	& $p \Kzbar  \,((2.3 \pm 0.6\%)$\\
		&		&		& $p \Km \pip  \,((5.0 \pm 1.3\%)$\\
		&		&		& $\Lambdaplain \pip \pip \pim  \,((2.6 \pm 0.7\%)$\\
		&		&		& $\Sigmaplain^+ \pip \pim  \,((3.6 \pm 1.0\%)$\\
\vspace{-3.85mm}&&&\\
$\Cascade_c^0$	& $112(13)\,\fs$  	& 33.6$\,\tcmu $m  	& $\Cascade^- \pip(not\,known)$\\
\hline\hline
\end{tabular} 
\caption[Strange and charmed candidates for identification by means of their delayed decay]
{Strange and charmed candidates for identification by means of their delayed decay, as listed in \cite{PDG:2010}.}
\label{tab:vtx:vtxtab1}
\end{center}
\end{table*}
\begin{table*}
\begin{center}
\begin{tabular}{lll}
\hline\hline
reaction channel			& detected particle 	& tracking used for ...\\
\hline
\vspace{-3.85mm}&&\\
$\pbarp\,\rightarrow \,\phi\phi$	& 2\Kp 2\Km		& momentum measurement (PID) \\
\vspace{-3.85mm}&&\\
$\pbarp\,\rightarrow \,\etac$	& \Kpm \pimp \Ks		& momentum measurement (PID) \\
\vspace{-3.85mm}&&\\
\multirow{2}{*}{$\pbarp\,\rightarrow \,$\DDbar}	
      & \multirow{2}{*}{ \K's and $\pi$'s}	& charm detection online \\
				& 				&  momentum measurement (PID) \\
\vspace{-3.85mm}&&\\
$\pbar A\,\rightarrow$\,\DDbar $X$	& \D (\Dbar)			& inclusive charm ID online \\
\vspace{-3.85mm}&&\\
$\pbarp\,\rightarrow \,\psi(2S)$ & \pip\pim $J/\psi$ ($\rightarrow e^+ e^-$ / $\mu^+\mu^-$)& vertex constraint \\
\hline\hline
\end{tabular}
\caption[Selection of benchmark channels requiring an optimum performance of the vertex 
tracking] 
{
Selection of benchmark channels requiring an optimum performance of the vertex 
tracking. 
The most serious challenge will be \D identification due to the very short decay lengths of these mesons.}
\label{tab:vtx:vtxtab2}
\end{center}
\end{table*}

Charged particle tracking is one of the essential parts in the overall detection concept at \PANDA,
which aims at fully exclusive measurements with a flexible trigger.
The MVD delivers 3D hit information close to the interaction point,
which in addition defines the earliest possible measurement of particles in the reaction channel.  
Track and time information both provide important input for the subsequent event reconstruction. 
Precise timing is essential for an accurate assignment of individual tracks to the appropriate event.  
In addition, a fast detection of particles inside the MVD can be used 
as time reference for outer detector systems.  
The track information in the magnetic field of the solenoid gives access 
to the particle momentum and the event topology.  

The identification of open charm and strangeness 
is one of the major tasks for the experiment.
and hinges on the capability to resolve secondary decays of 
short-lived particles 
in displaced vertices other than the primary reaction point. 
%Energy loss measurement may provide additional input for a particle hypothesis.
%While particle identification above about $700\,\mevc$ will be facilitated by the 
In addition, the energy loss of slower protons, kaons and pions in the silicon detectors 
may be used as well to contribute to the global particle identification. 
Some examples of decay channels 
that hold the potential to be identified by purely geometrical means, 
are listed in \tabref{tab:vtx:vtxtab1}.
Basically, strange and charmed hadrons exhibit two different decay length scales 
in the order of several centimetres and a few hundred micrometres, respectively.
While \Panda will focus on charm production, 
the identification of kaons will greatly enhance the efficiency also for $\D$ mesons 
since they show large branchings into channels accompanied by kaons. 

Key points of the \PANDA physics program 
presented in  chapter~\ref{sec:sciprog}
can only be met with a suitable vertex resolution. 
In this context, the MVD will be of central importance 
for the associated physics goals.
An identification of ground state $D$ mesons via their displaced vertex 
is of particular interest because they can be produced in associated decays 
of excited $D$ mesons as well as charmonia or charmonium-like states 
above the \DDbar threshold.
The applicability of a precise $D$ meson tagging is thus 
essential for the spectroscopy in the open charm sector and the charmonium mass region.

In addition, the experimental program at \Panda 
puts emphasis on the high-resolution spectroscopy of electrons. 
Electrons are excellent tags for \D meson spectroscopy (cf. \tabref{tab:vtx:vtxtab1}).
They carry information on the flavour of the decaying charm.
With heavy targets leptons can also be used to reconstruct mesonic properties inside the nuclear medium. 

Some of the envisaged physics benchmark channels, 
which all require the use of the vertex detector for quite different reasons, 
are listed in \tabref{tab:vtx:vtxtab2}. 
The most stringent requirements are given by those reaction channels 
where the \D decay will be used to select certain event patterns 
from the data stream during data acquisition, 
as indicated in the last column of the table.

\newpage
\mbox{\hspace{5 mm}}
\newpage
\section{Basic Detector Requirements}
\label{Intro-Specifics}
\authalert{Author: Thomas W\"{u}rschig, Contact: t.wuerschig$\mathrm @$hiskp.uni-bonn.de}

\begin{figure}[]
\begin{center}
\includegraphics[width=8.5 cm]{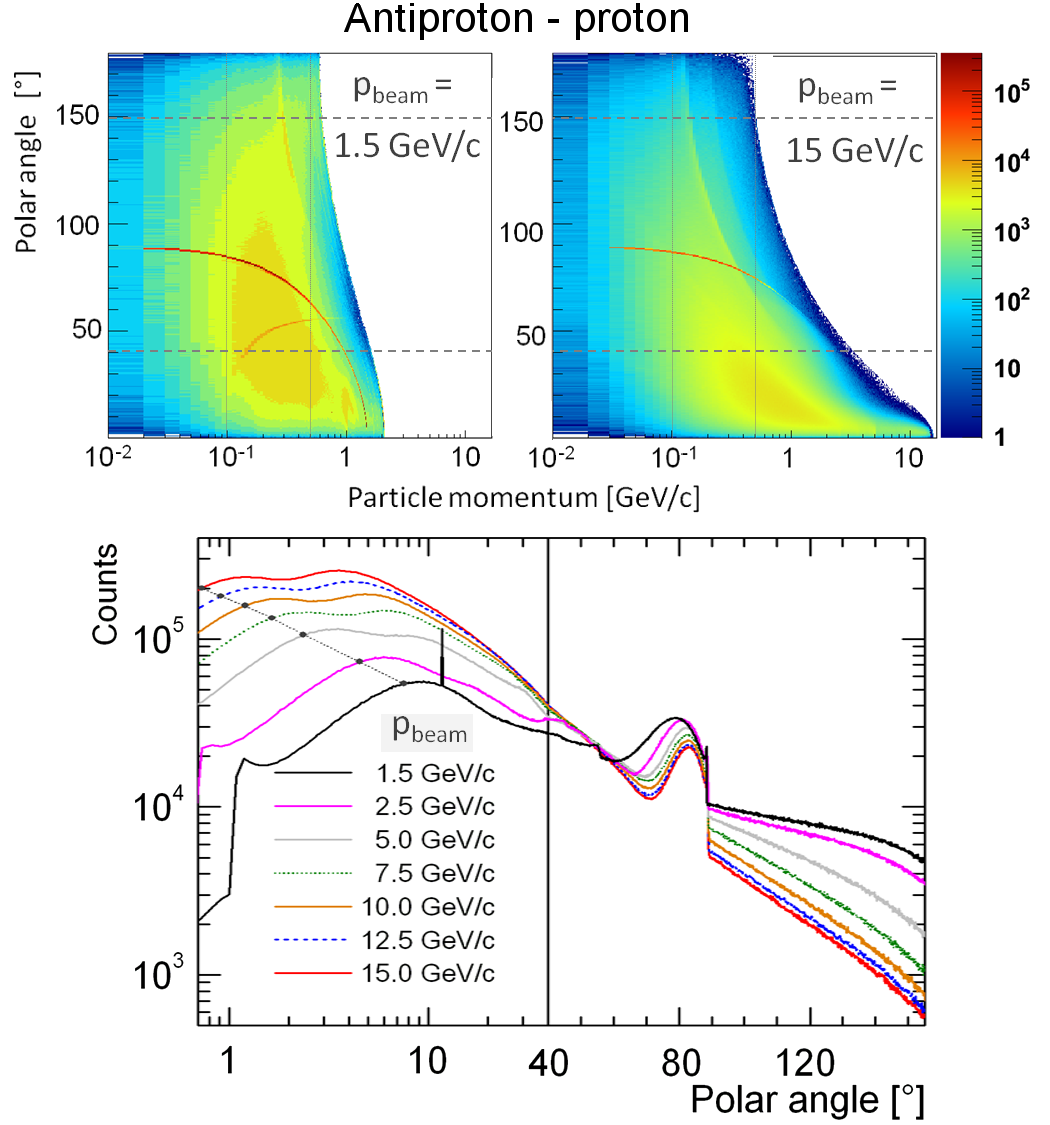}
\caption
[Expected particle distribution in antiproton-proton collisions at different beam momenta]
{Expected particle distribution in antiproton-proton collisions
obtained with  10$^7$ DPM~\cite{DPM} events
at different beam momenta ($p_{\textsf{\tiny{beam}}}$).
Top: Particle distributions as a function of particle momentum and polar angle. 
The red line visible in both plots results from the elastic scattering process
that is implemented in the DPM generator.
Bottom: 1D profile projected onto the polar angle.
%Results were obtained with 10$^7$ DPM~\cite{DPM} 
%and 2\,$\times$\,10$^6$ UrQMD~\cite{UrQMD2} events, respectively.
%A lower statistical sample of 40,000 events is plotted in 
%the 2D maps of the antiproton-nucleus reactions. 
}
\label{pic-ParticleDistr_DPM}
\end{center}
\end{figure}

\begin{figure}[]
\begin{center}
\includegraphics[width=6.5 cm]{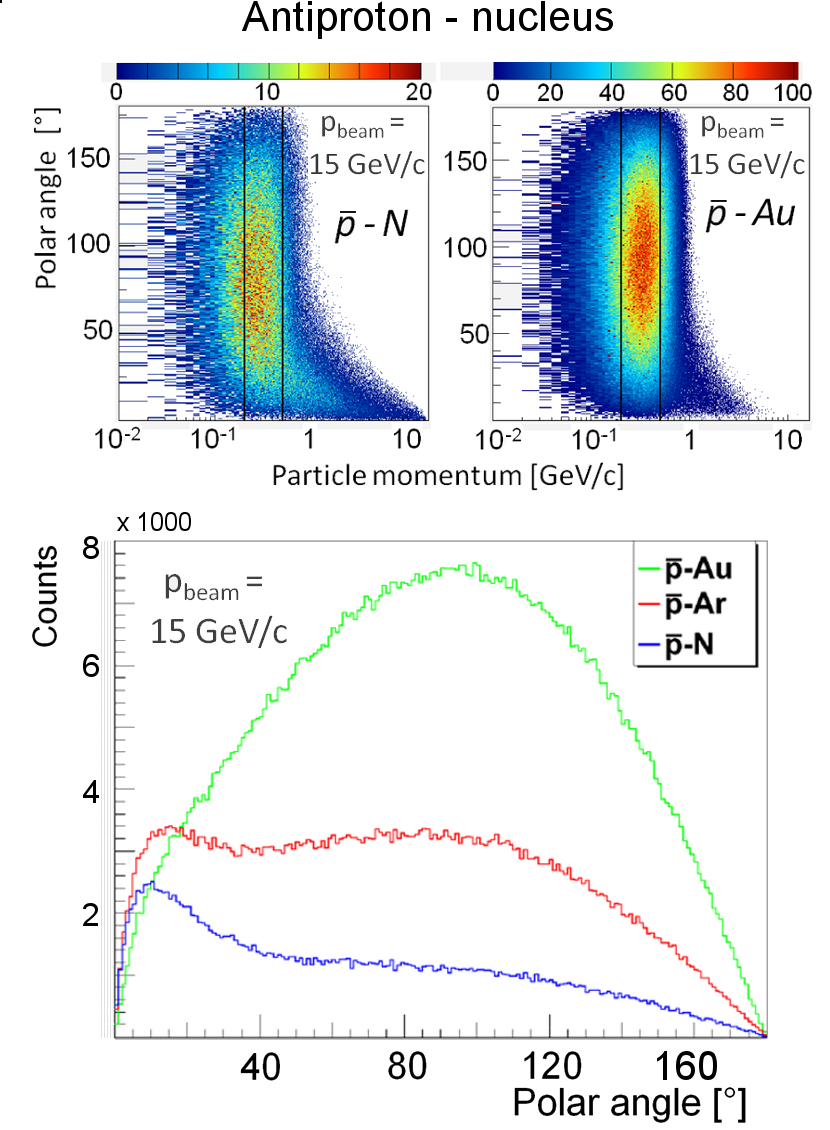}
\caption
[Expected particle distribution in antiproton-nucleus collisions at maximum beam momenta]
{Expected particle distribution in antiproton-nucleus collisions obtained 
at maximum beam momenta ($p_{\textsf{\tiny{beam}}}$)
Top: Particle distributions as a function of particle momentum and polar angle
obtained with 40,000 UrQMD~\cite{UrQMD2} events.
Bottom: 1D profile projected onto the polar angle
obtained with a higher statistical sample of $2\times10^6$.
%Results were obtained with 10$^7$ DPM~\cite{DPM} 
%and 2\,$\times$\,10$^6$ UrQMD~\cite{UrQMD2} events, respectively.
%A lower statistical sample of 40,000 events is plotted in 
%the 2D maps of the antiproton-nucleus reactions. 
}
\label{pic-ParticleDistr_UrQMD}
\end{center}
\end{figure}

Basic detector requirements are mandated by the physics goals of the \Panda experiment
and the experimental conditions the detector will meet during operation.
The MVD has to cope with different experimental setups 
of the target region.
In any case, before reaching the MVD particles must traverse
at least one wall of the beam or target pipe.
The minimum particle momentum for different particle species 
needed to reach the inner layer of the MVD is listed in \tabref{tab-MinMomentumBeamPipe}.
\Figref{pic-ParticleDistr_DPM} illustrates
the distribution of final state particles
in antiproton-proton collisions at different beam momenta,
which is expected to be obtained with a hydrogen target
intended to be used in the first stage of \Panda.

\begin{table}
\centering
\small
\begin{tabular}{|c|c|c|}
\hline
\parbox[0pt][6mm][c]{0cm}{}
Particle 
& $|\vec{p}_{\textsf{\tiny{min}}}|$ , $\alpha_{\textsf{\tiny{inc}}}=90^{\circ}$ 
  & $|\vec{p}_{\textsf{\tiny{min}}}|$ , $\alpha_{\textsf{\tiny{inc}}}=45^{\circ}$  \\ 
\vspace{-3.5mm}&&\\
\hline
\vspace{-3.5mm}&&\\
$p, \bar p$ & $\sim$89 MeV/c & $\sim$98 MeV/c \\ \hline
\vspace{-3.5mm}&&\\
$\pi^{\pm}$ & $\sim$23 MeV/c & $\sim$25 MeV/c \\ \hline
\vspace{-3.5mm}&&\\
%$\pi^{0}$ & $\sim$22 MeV/c & $\sim$25 MeV/c \\ \hline
%\vspace{-3.5mm}\\
$K^{\pm}$ & $\sim$56 MeV/c & $\sim$62 MeV/c \\ \hline
\vspace{-3.5mm}&&\\
$\mu^{\pm}$ & $\sim$19 MeV/c & $\sim$21 MeV/c \\ \hline
\vspace{-3.5mm}&&\\
$e^{\pm}$ & $\sim$1 MeV/c & $\sim$1 MeV/c \\ \hline
\end{tabular}
\caption[Minimum momentum $|\vec{p}_{\textsf{\tiny{min}}}|$
for different particle species needed to traverse 
the beryllium pipe]
{Minimum momentum $|\vec{p}_{\textsf{\tiny{min}}}|$
for different particle species needed to traverse 
the 200~$\tcmu$m thick beryllium pipe intended to be used 
at the beam-target region assuming two diffrent 
incident angles $\alpha_{\textsf{\tiny{inc}}}$.
Calculations are based on the Bethe-Bloch energy loss.}
\label {tab-MinMomentumBeamPipe}
\end{table} 

The fixed-target setup is reflected by 
a Lorentz boost of particles in the forward direction,
which increases with higher beam momentum.
Most of these particles carry away 
a significant part of the initial beam momentum
and thus the momentum range spans more than two orders of magnitude. 
Slower particles in a range of 100~\mevc up to 500~\mevc
are emitted more or less isotropic over nearly the full solid angle. 
The elastic scattering process in antiproton-proton reactions 
delivers an enhanced emission of slow recoil protons 
at polar angles above 80$^{\circ}$. 
Results on the particle distribution in the final state 
to be expected with heavier targets
are shown in \figref{pic-ParticleDistr_UrQMD}.
%The study of antiproton-nucleus reactions 
%is foreseen at a later stage of the \Panda experiment. 
In comparison with antiproton-proton collisions
the Lorentz boost in forward direction in antiproton-nucleus reactions 
is less distinct and vanishes with increasing atomic number.

Central task of the MVD is a precise measurement of both 
the primary interaction vertex 
and secondary decay vertices of short-lived particles.  
For this purpose the detection of a first track point 
very close to the nominal vertex is mandatory.
Moreover, the obtained track information from the vertex detector 
also results in an improved momentum resolution.
The general detection concept of \Panda requests coverage 
over nearly the full solid angle.
The boundary conditions of adjacent detector subsystems
define a range of approximately 3$^{\circ}$ to 150$^{\circ}$ 
in polar angle to be covered by the MVD.

Due to the fixed-target setup, a very good spatial resolution 
is particularly needed in forward direction.
Therefore, a high granularity and intrinsic detector resolution 
are a prerequisite.
Slightly lower quality requirements result for the backward hemisphere.
A sufficient number of track points must be measured inside the MVD 
to achieve an optimum tracking performance.
In the best case, an independent pre-fit of MVD tracklets can be accomplished,
for which a minimum of four track points is necessary.
Further improvements can be achieved by an analogue readout of the signal amplitudes.
It facilitates a cluster reconstruction over detector elements sharing the signal
and additional energy loss measurements,
which would be of advantage as they could be used in the particle identification at low energy.

Another main issue for the MVD is the minimisation of the material budget. 
\Panda will have to cope with a large amount of low-energy particles.
Scattering in the silicon layers is thus a major concern 
since the small-angle scattering drastically increases with decreasing particle energy.
Moreover, any material near the primary target may cause background 
due to bremsstrahlung and pair production. 
The loss of photons related to the conversion or absorption in the detector material 
significantly affects the overall performance of the \Panda apparatus, 

As a consequence of the close position with respect to the interaction region
all detector components must withstand an adequate radiation dose.  
With an overall lifetime of \Panda of about 10 years and an assumed duty cycle of 50\%
the requested radiation tolerance given in terms of 1~\mev neutron equivalent particles per square centimetre,
1~\neueq, 
is in the order of $10^{14}$~\neueq~\cite{MVD-TIDstudy}
and thus calls for dedicated production techniques.
The achievable time resolution of the MVD must be in the order of 10~ns 
in order to cope with the high interaction rate. 
%and to be operable at the standard reference clock of 155 MHz.
Particular challenges arise from the specific trigger and readout concept of the \Panda experiment
that is based on an autonomous readout scheme for each of the individual detector sub-systems.   
Therefore, an internal trigger and a first data concentration must be integrated
for each channel and on \frontend level, respectively.   
This approach exceeds common state-of-the-art solutions and required an extended research and development program.  

In the following the most important detector requirements 
will be listed to conclude this section:

\begin{itemize}

\item{\bf Optimum detector coverage}\\
Within the given boundary conditions a full detector coverage must be envisaged.
The number of layers is defined by the design goal, 
which foresees a minimum of four MVD hit points per track.
An increased number of layers is necessary in the forward part 
in order to fulfill the required tracking performance.

\item{\bf High spatial resolution}\\
Tracks of charged hadrons have to be measured 
with a spatial resolution no worse than 100~$\tcmu$m 
in $z$ and a few tens of $\tcmu$m in $xy$. 
The achievable vertex resolution shall be in the same range 
of 100~$\tcmu$m in order to resolve displaced decay vertices of open-charm states 
such as the $D_0$ and the $D^\pm$
with mean decay lengths of 123~$\tcmu$m and 312~$\tcmu$m, respectively 
(cf. \tabref{tab:vtx:vtxtab1}).
One of the most demanding tasks is the recognition of \DDbar events 
as decay products of charmonium states via their displaced vertices.
At energies not much above the threshold, 
these mesons will be strongly forward focused.
In this case, the best resolution is required along the beam ($z$-direction).

\item{\bf Low material budget}\\
The impact of the MVD on the outer detector systems must be kept as low as possible.
Photon conversion, in particular close to the interaction point, has to be minimised 
for the efficient operation of the electromagnetic calorimeter.
Moreover, scattering deteriorates the overall tracking performance. 
In view of both deteriorating effects,
the total material budget in units of radiation length, $X/X_0$,  
shall stay below a value of $X/X_0 = 10\%$.
Besides the sole reduction of the total material amount,
lightweight and low-Z materials must be used where ever possible.

\item{\bf Improved momentum resolution}\\
The track information of the MVD defines important constraints 
for the applied tracking algorithms
and thus facilitates the determination of the particle momentum.
In this context, it is of special importance for small-angle tracks.
With the minimised material budget, matter effects will be reduced 
to such a level that the MVD can improve the overall momentum resolution
by roughly a factor of two.

\item{\bf Additional input to the global particle identification}\\
The lower momenta for most of the emitted particles 
facilitate an improved particle identification 
based on energy loss measurements,
which can provide additional to sustain
different particles hypothesis.

\item{\bf Fast and flexible readout}\\
The readout has to be designed in a different way as compared to other experiments 
in order to allow continuous data collection from
the detector without external triggering. 
As multiplicities are expected to be rather small 
(a maximum of 16 charged particles was estimated for the studies envisaged), 
the more serious aspect of data extraction from the 
vertex detector is caused by the rather high interaction rates, 
which will be in excess of 10$^7$ annihilations per second. 
The occupancy in different detector regions is very anisotropic and 
will also change due the use of different targets.
This requires an efficient and flexible readout concept.

\end{itemize}

%% file: introduction/MVD-layout.tex
\section{MVD Layout}
\label{Intro-MVD-Layout}
\authalert{Author: Thomas W\"{u}rschig, Contact: t.wuerschig$\mathrm @$hiskp.uni-bonn.de}

\begin{figure}[!b]
\begin{center}
 \includegraphics[width=7.5 cm]{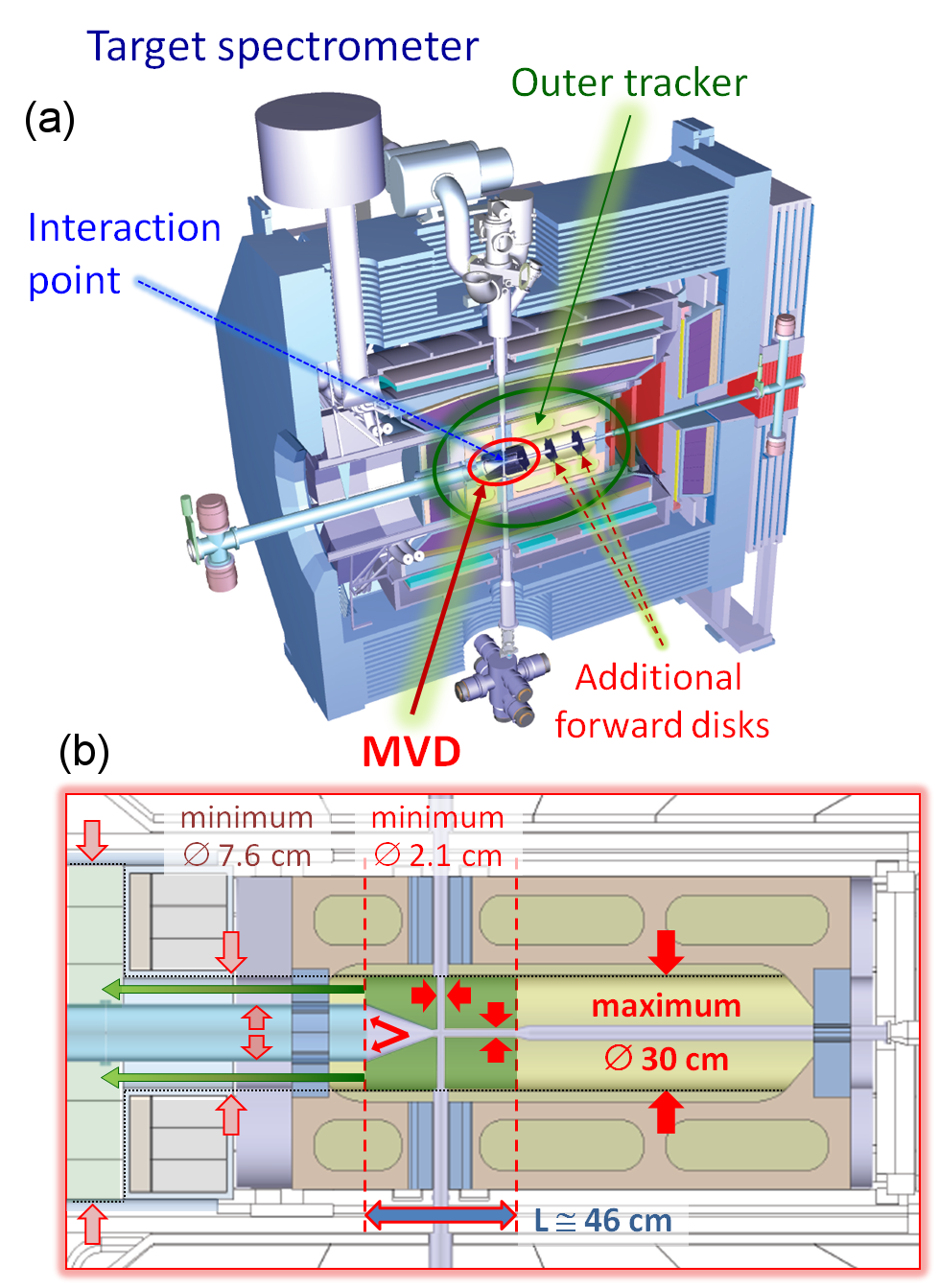}
% Pic3-01_MVD-BTS-Geometry.png: 1239x931 pixel, 125dpi, 25.18x18.92 cm, bb=0 0 714 536 
\caption[Positioning of the MVD inside the Target Spectrometer
and zoom of the beam-target geometry]
{Position of the MVD inside the target spectrometer (a)
and zoom of the beam-target geometry (b).
The MVD volume is highlighted in green, 
the associated readout in upstream direction 
is indicated by green arrows.
Boundary conditions given by adjacent sub-systems,
i.e.~the beam and target pipes as well as the outer tracker,
are visualised by red arrows.
All measures refer to a maximum extension of the MVD volume
without any safety margins.}
\label{pic-BTS-MVD-geometry}
\end{center}
\end{figure}

The MVD is the innermost part of the central tracking system.
Its position inside the \panda target spectrometer, 
the beam-target geometry and the resulting boundary conditions
are shown in \figref{pic-BTS-MVD-geometry}.
In the following subsections, 
the basic geometry and the conceptual design 
of the detector will be discussed.

\subsection{Basic Detector Geometry}

\begin{figure}[!t]
\begin{center}
\includegraphics[width=7.0 cm]{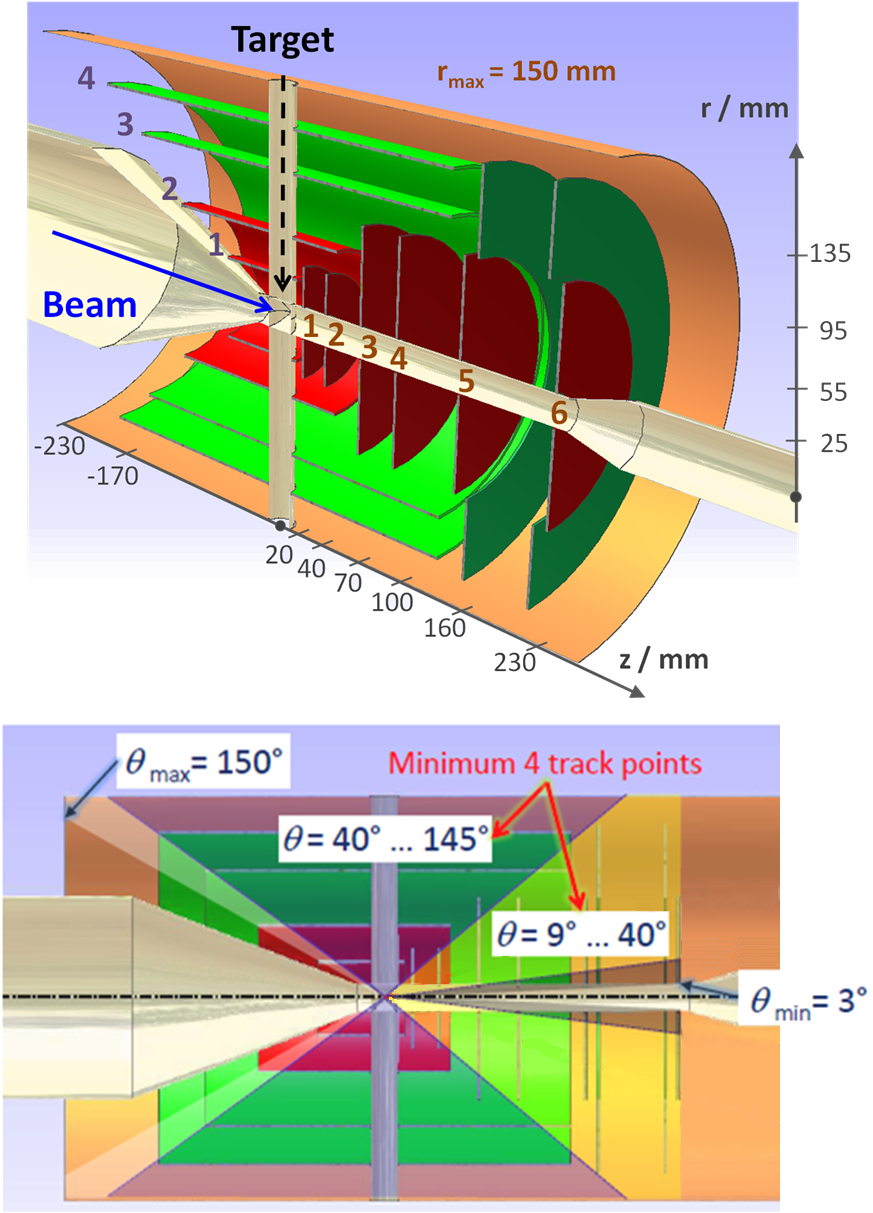}
% Pic3-02_MvdLayout.png: 787x873 pixel, 80dpi, 24.98x27.71 cm, bb=0 0 708 786
\caption[Basic layout of the MVD]
{
Basic layout of the MVD (\textit{top}). 
The red inner parts are equipped with silicon hybrid pixel sensors. 
Double-sided silicon micro-strip detectors utilised in the outer layers 
are indicated in green. 
\textit{Bottom}: 
Sideview along the beam axis illustrating 
the polar angle coverage. 
The barrel and the forward part meet at a polar angle of
\unit[$\theta=$]{40$^{\circ}$}.
}
\label{pic-MVD-GeneralLayout}
\end{center}
\end{figure}

The MVD is divided into a barrel and a forward part. 
The outer limits are given by the detector position inside the central tracker, 
which defines a maximum radius of 15~cm. 
The extension of the MVD along the beam axis is roughly
$z=\pm23$~cm with respect to the nominal interaction point.  
A schematic picture of the MVD layers is shown in \figref{pic-MVD-GeneralLayout}.  
There are four barrel layers and six forward disks.
Silicon hybrid pixel detectors are foreseen in the two innermost barrel layers 
and for all disks.
Double-sided silicon micro-strip detectors will be used in the two outer barrel layers 
and as radial complements in the last two disk layers.
The basic detector layout results in a detector coverage 
with a minimum of four track points in a polar angle interval 
from 9$^{\circ}$ to 145$^{\circ}$.

The barrel part of the MVD covers polar angles 
between 40$^{\circ}$ and 150$^{\circ}$. 
The maximum downstream extension is given by both outer strip layers. 
The two pixel layers end at polar angles around 40$^{\circ}$
thus avoiding shallow crossing angles of particles.
The upstream extension of all barrel layers is chosen 
to fit with the opening cone of the beam pipe. 
The radii of the innermost and the outermost barrel layer 
are set to 2.5 cm and 13.5 cm, respectively. 
The two intermediate barrel layers are arranged in increasing order. 

The disk layers in forward direction enable a measurement at small polar angles 
between 3$^{\circ}$ and 40$^{\circ}$. 
The innermost disk located at $z=2$~cm is the closest 
of all detector layers with respect to the nominal interaction point. 
It has an interspacing of 2~cm to the second pixel disk. 
Both of these small pixel disks are located inside the outer pixel barrel layer.
Further downstream there are four large pixel disks. 
While the first two of them are positioned inside the strip barrel layers, 
the ones still further downstream are outside the barrel layers 
and extended to larger radii by additional strip disks. 

In addition to the MVD, there are two extra disks envisaged 
in forward direction that would
fill the long detector-free gap to the forward GEM tracking station. 
They are intended to contribute to the vertex reconstruction 
of hyperons, which have much longer lifetimes and consequently 
a larger displacement of the secondary vertex than $D$ mesons.
The conceptual design of these additional disks is similar 
to that of the strip disks of the MVD.

\subsection{Conceptual Design}

\subsubsection*{Choice of Detector Technology}

Silicon detectors excel in a fast response and a low material budget. 
Moreover, they allow a high degree of miniaturisation and they can be produced 
in big quantities with very good reproducibility. 
Due to these properties they perfectly meet the requirements imposed on the MVD. 

Silicon hybrid pixel detectors deliver discrete 2D information 
with a high granularity and allow very precise space point measurements. 
They are intended to be used in the innermost MVD layers 
in order to cope with the high occupancy close to the interaction region.
However, the total number of channels required to cover larger surfaces 
increases rapidly.

The material budget of the detector scales with the number of readout channels. 
It must be minimised in order to fulfill the requirements of the \panda experiment. 
For this reason, double-sided silicon micro-strip detectors are foreseen 
in the outer parts of the MVD. 
They facilitate the readout of a much larger area with significantly fewer channels. 
However, an utilisation very close to the primary vertex is disfavored because of
the high probability of multiple hits in the detector, which lead to ambiguities,
so-called ghost hits, 
and the increased probability for hit loss due to pile-up.

In the following, all individual silicon detectors will be denoted 
as either pixel or strip sensors in order to avoid misinterpretations.

\subsubsection*{Hierarchical Structure}

\begin{figure}[!t]
\begin{center}
 \includegraphics[width=7.5 cm]{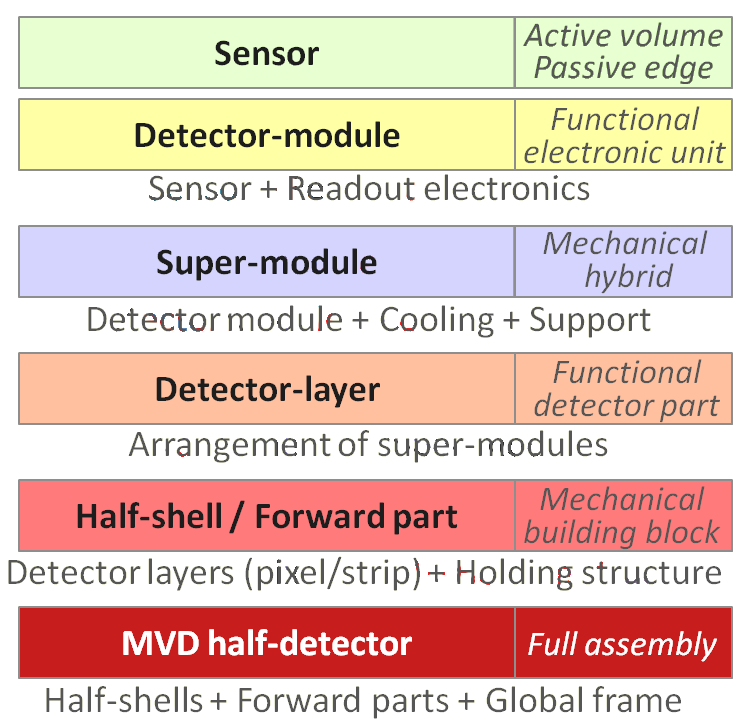}
% Pic3-01_MVD-BTS-Geometry.png: 1239x931 pixel, 125dpi, 25.18x18.92 cm, bb=0 0 714 536 
\caption
[Hierarchical structure of the MVD]
{Hierarchical structure of the MVD.}
\label{pic-ConventionsHierarchy}
\end{center}
\end{figure}

The hierarchical structure of the MVD is based on a modular concept following 
the future test and installation sequence of the detector. 
The silicon sensors represent the lowest level therein. 
A detector module is defined as the smallest functional unit, 
which is electronically independent. 
It is formed by all hard-wired connections between individual sensors 
and assigned components of the readout electronics. 

The finalised hybrid is formed by a super-module including several detector modules 
and the associated cooling and support structure. 
It corresponds to the smallest mechanically independent unit for the detector assembly.  
Different detector layers are then composed of super-modules, 
which are attached to the respective mechanical holding structure. 
In this way four main building blocks are created.
They are given by the individual pixel and strip layers, 
which form two half-shells in the barrel part
and two half-disk structures in the forward part, respectively.  
The half-detector concept is required due to the target pipe, 
which crosses the entire MVD volume from the top to the bottom
and thus breaks rotational symmetry around the beam axis.
As a consequence, the detector will be left-right separated
along its middle plane.
A schematic picture of the hierarchy is given 
in \figref{pic-ConventionsHierarchy}.

\subsubsection*{Sensor Geometry}

\begin{figure}
\begin{center}
\includegraphics[width=7.25 cm]{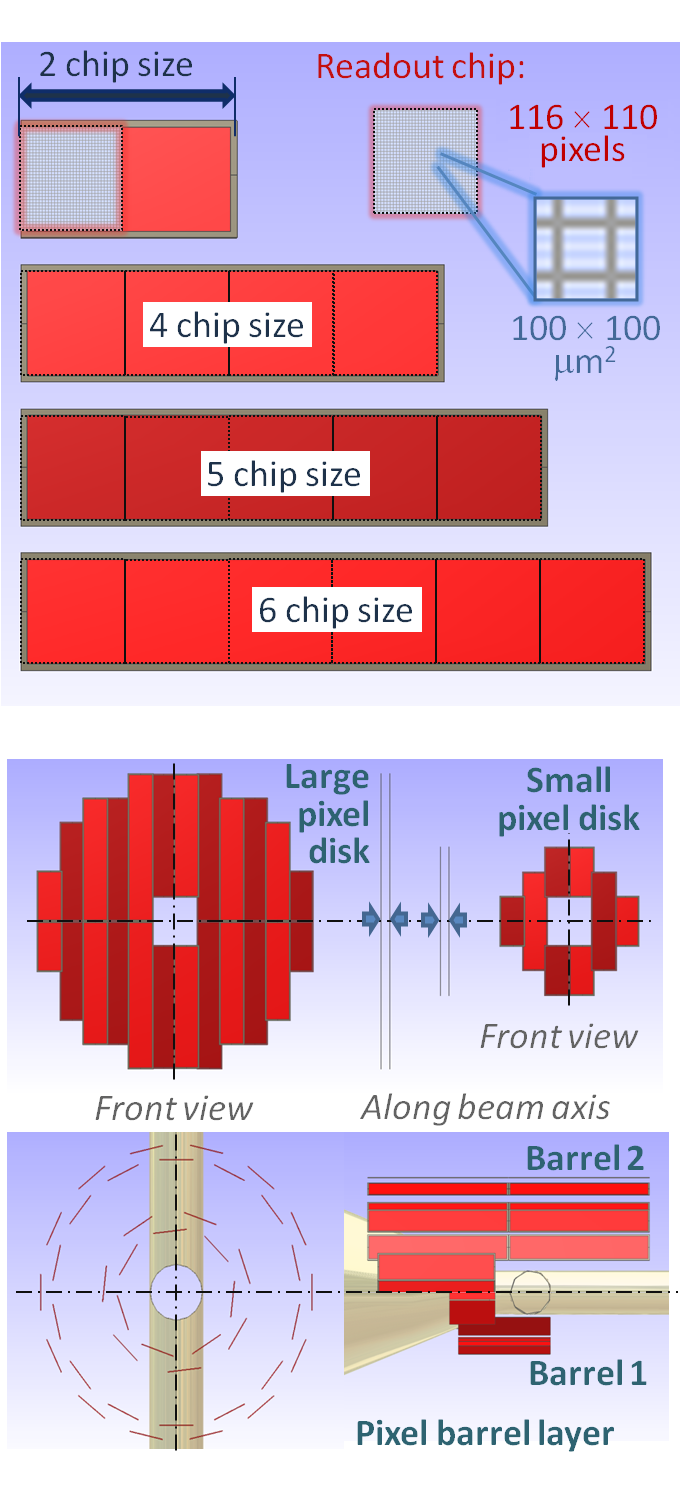}
% Pic3-02_MvdLayout.png: 787x873 pixel, 80dpi, 24.98x27.71 cm, bb=0 0 708 786
\caption[Schematics of the basic pixel sensor geometry]
{
Schematics of the basic pixel sensor geometry. 
\textit{Top:} 
Main pixel sensor types. 
The design is based on a quadratic pixel cell size of 
$100\times100$~$\tcmu$m$^2$ and an effective readout area 
of approximately 1~cm$^2$ per individual readout chip.
\textit{Bottom:} 
Pixel sensor arrangement in different detector layers.
}
\label{pic-PixelGeometry}
\end{center}
\end{figure}

The schematic detector layers as shown in \figref{pic-MVD-GeneralLayout}
must be approximated by an appropriate arrangement of individual detector elements. 
The precise sensor dimensions and the exact sensor positioning result from 
an explicit design optimisation~\cite{PhD-Tesis_Wuerschig}.
It has been accomplished to facilitate compromise between 
good spatial coverage and minimum material on the one hand 
and the introduction of sufficient space needed for 
passive elements and appropriate safety margins for the detector integration on the other.
Therefore, technical aspects have been taken into account 
alongside with purely geometric considerations.
The final number of different sensor types is kept small in order to facilitate 
the modular concept and to obtain a high compatibility.
The specific size of the readout structure has direct impact on the sensor design. 
For the two different detector types it is given by the pixel cell size 
and the strip interspacing, respectively. 

In case of pixel sensors a rectangular shape is favored 
due to technical reasons. 
The \Panda pixel design is based on a quadratic pixel cell size 
with a side length of 100~$\tcmu$m. 
There are four different pixel sensors, 
each of them has the same width but they differ in lengths. 
The dimensions correspond to multiples of the readout matrix 
of the connected readout chip, 
which serves $116\times110$ cells (see chapter~\ref{Section-PixelReadout})
and thus features an area of approximately 1.3~cm$^2$.
The different pixel sensor types are shown schematically 
in \figref{pic-PixelGeometry}, on the top. 

%\clearpage

\begin{figure*}[t]
\begin{center}
\includegraphics[width=14. cm]{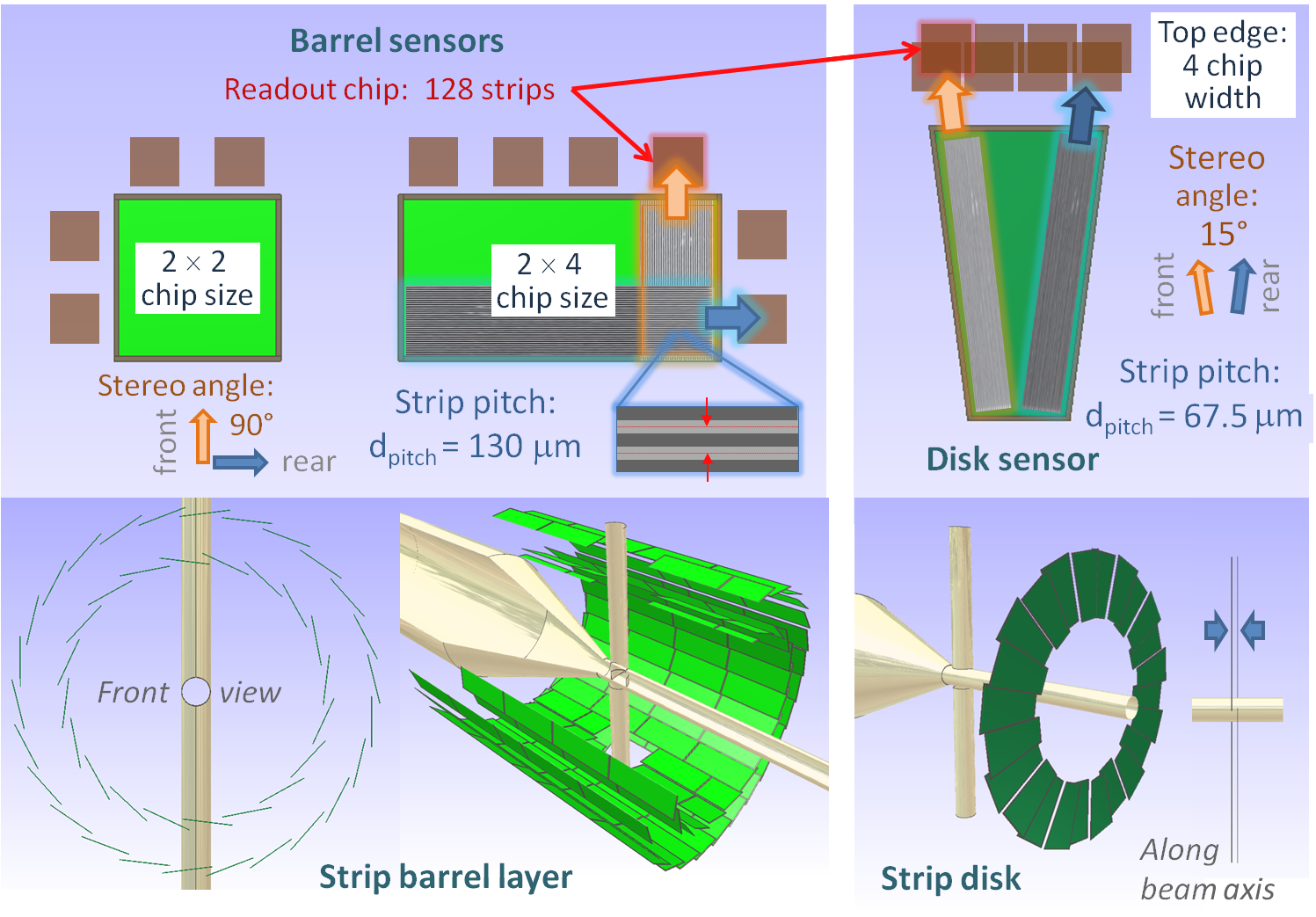}
% Pic3-04_PixelArrangement.png: 1240x910 pixel, 125dpi, 25.20x18.49 cm, bb=0 0 714 524
\caption[Basic sensor geometry for the MVD strip part]
{
Basic sensor geometry for the MVD strip part: Sensor types (\textit{top}) 
and sensor arrangement in the different detector layers (\textit{bottom}). 
The underlying segmentation of the sensors is given by the stereo angle 
and the strip pitch assuming 128 channels per readout chip.
}
\label{pic-StripGeometry}
\end{center}
\end{figure*}

The basic geometry for the pixel sensor arrangement is illustrated below.  
A pixel barrel layer is made of a double-ring arrangement in order 
to achieve sufficient radial overlap. 
Shorter sensors are used to obtain 
sufficient space for the target pipe lead-through. 
The forward disk layers consist of an appropriate configuration of pixel sensors,
which are aligned in different rows.
Alternating rows are positioned with a small interspacing along the beam axis 
thus allowing the compensation of the passive edge around the active area. 

\begin{figure}[htb]
\begin{center}
 \includegraphics[width=7.25 cm]{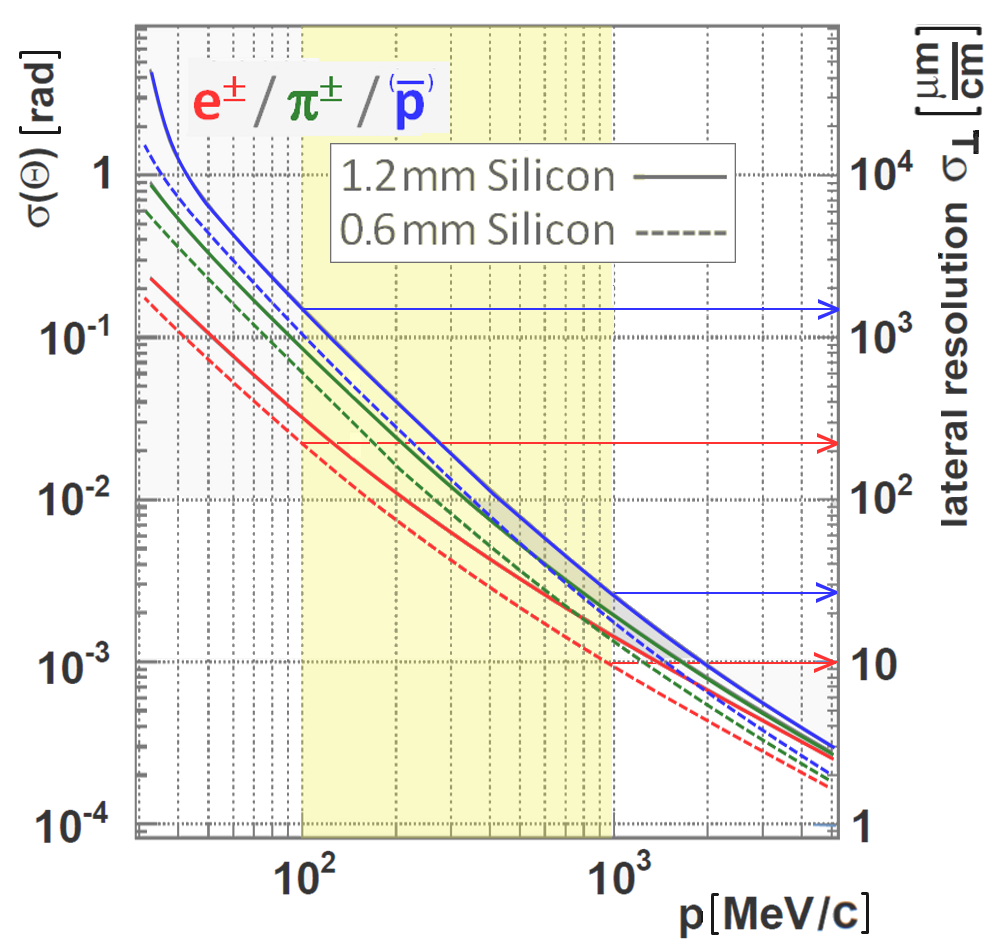}
% Pic3-01_MVD-BTS-Geometry.png: 1239x931 pixel, 125dpi, 25.18x18.92 cm, bb=0 0 714 536 
\caption[Scattering effects in silicon and optimisation of the stereo angle for the trapezoidal sensors]
{
Mean scattering angle, $\sigma(\Theta)$, 
and the associated particle deflection, $\sigma_\bot$, 
as a function of the particle momentum, $p$, 
for two effective silicon layer thicknesses. 
Values for electrons, pions and protons 
are calculated with a Gaussian approximation~\cite{ScatteringDahl}. 
The relevant momentum range for the strip barrel layer of the MVD 
is highlighted in yellow. 
}
\label{pic-StripPitch}
\end{center}
\end{figure}

For silicon strip detectors there is no direct impact 
of the readout chip size on the dimension of the sensor segmentation.
Hence, the distance between neighbouring strips,
which is commonly denoted as 
\textquotedblleft strip pitch\textquotedblright, 
is one of the main parameters for the detector optimisation.
An approximate range is given by the average track deviation 
related to scattering effects in the previous pixel layers.
Calculated values for the mean scattering angle 
and the associated deflection of tracks 
are shown in \figref{pic-StripPitch}. 
They are obtained with an empirical formula for the multiple scattering 
in a Gaussian approximation given by Lynch and Dahl~\cite{ScatteringDahl}. 
\clearpage

\onecolumn
\begin{table}[p]
\begin{center}
\small
\scalebox{1.0}{
\begin{tabular}{|c|c|c|}
\hline
\vspace{-2mm}
\hspace{5cm}
  & \hspace{5cm} 
    & \hspace{5cm}\\
\textbf{Basic parameter}
  & \textbf{Pixel part}
    &  \textbf{Strip part} \\
\vspace{-3mm}&&\\
\hline
\vspace{-2mm}&&\\
Number of super-modules 
  & 66
    & 70 \\
\vspace{-3mm}&&\\
Number of detector modules 
  & 176
    & 140 \\
\vspace{-3mm}&&\\
 \multirow{4}{*}{Number of sensors}
  & 34 (2 chips size)
    & 172 (rectangular) \\
  & 28 (4 chips size)
    & 34 (squared)\,\,\,\,\,\,\,\, \\ 
  & 54 (5 chips size)
    & 48 (trapezoidal) \\
  & 60 (6 chips size)
    & \\
\multicolumn{1}{|r|}{\textit{Total:}} 
  & 176
    & 254 \\
\vspace{-3mm}&&\\
Active silicon area / [m$^2$]
  & 0.106
    & 0.494\\
\vspace{-3mm}&&\\
\multirow{2}{*}{Number of \frontend chips}
  & 338 (barrels)
    & 940 (barrels)\\
  & 472 (disks)
    & 384 (disks)\\
\multicolumn{1}{|r|}{\textit{Total:}} 
  & 810
    & 1324 \\
\vspace{-3mm}&&\\
Number of readout channels
  & $\approx10.3\cdot10^6$
    & $\approx1.7\cdot10^5$
 \parbox[0pt][7mm][t]{0cm}{} \\
\hline
\end{tabular}
}
\caption
{Compilation of design parameters for the MVD.}
%\vspace{-6mm}
\label{tab-BasicInfo-CentralMVD}
%\end{center}
%\end{table*} 
\vspace{6mm}
%\begin{table*}[hp]
%\begin{center}
\small
\scalebox{0.9}{
\begin{tabular}{|c|c|p{7.5mm}|p{7.5mm}|c|c|p{7.5mm}|p{7.5mm}|p{10mm}|c|}
\hline
\vspace{-3mm}&&&&&&&&&\\
\parbox[0pt][5mm][c]{1.5cm}{\centering Main}
& \multirow{2}{*}{Sub-layer}
  & ($r_{\textsf{\tiny{def}}}$)
    & $\langle r_{\textsf{\tiny{opt}}} \rangle$ 
      & $r_{\textsf{\tiny{min}}}$ 
	& $r_{\textsf{\tiny{max}}}$ 
	  &  ($z_{\textsf{\tiny{def}}}$)
	    & $\langle z_{\textsf{\tiny{opt}}} \rangle$ 
	      & \centering $z_{\textsf{\tiny{min}}}$
		& $z_{\textsf{\tiny{max}}}$ \\
\parbox[0pt][5mm][t]{1.5cm}{\centering layer}
&
  & [mm]
    & [mm]
      & [mm]
	& [mm]
	  & [mm]
	    & [mm]
	      & \centering [mm]
		& [mm]  \parbox[0pt][6mm][t]{0cm}{}\\  
\hline
\vspace{-3mm}&&&&&&&&&\\
%%%%%%%%%%%%%%%%%%%%%%%%%%%%%%%%%%%%%%%%%%%%%%%%%% BL1
%%%%%%%%%%%%%%%%%%%%%%%%%%%%%%%%%%%%%%%%%%%%%%%%%%%%
\parbox[0pt][5mm][b]{1.5cm}{\centering \textbf{Barrel}}
& Inner ring
  & \multirow{3}{*}{\parbox[0pt][5mm][t]{7.5mm}{\centering 25}}
    & \centering 22
      & 21.80
	& 22.75
	  & \multirow{3}{*}{\parbox[0pt][5mm][t]{7.5mm}{\centering -}}
	    &  \multirow{3}{*}{\parbox[0pt][5mm][t]{7.5mm}{\centering -}}
	      & \multirow{3}{*}{\parbox[0pt][5mm][t]{1.0cm}{\centering  $-$39.8}}
		& \multirow{3}{*}{\parbox[0pt][5mm][t]{1cm}{\centering 9.8}}  \\  
\vspace{-3.75mm}&&&&&&&&&\\
\parbox[0pt][5mm][t]{1.5cm}{\centering \textbf{layer 1}}
& Outer ring
  & 
    & \centering 28
      & 27.80
	& 28.58
	  &  
	    &  
	      & 
		& \\  
\hline
%%%%%%%%%%%%%%%%%%%%%%%%%%%%%%%%%%%%%%%%%%%%%%%%%% BL2
%%%%%%%%%%%%%%%%%%%%%%%%%%%%%%%%%%%%%%%%%%%%%%%%%%%%
\vspace{-3mm}&&&&&&&&&\\
\parbox[0pt][5mm][b]{1.5cm}{\centering \textbf{Barrel}}
& Inner ring
  & \multirow{3}{*}{\parbox[0pt][5mm][t]{7.5mm}{\centering 50}}
    & \centering 47.5
      & 47.30
	& 47.85
	  & \multirow{3}{*}{\parbox[0pt][5mm][t]{7.5mm}{\centering -}}
	    &  \multirow{3}{*}{\parbox[0pt][5mm][t]{7.5mm}{\centering -}}
	      & \multirow{3}{*}{\parbox[0pt][5mm][t]{10mm}{\centering  $-$79.8}}
		& \multirow{3}{*}{\parbox[0pt][5mm][t]{10mm}{\centering 57.8}}  \\  
\vspace{-3.75mm}&&&&&&&&&\\
\parbox[0pt][5mm][t]{1.5cm}{\centering \textbf{layer 2}}
& Outer ring
  & 
    & \centering 52.5
      & 52.30
	& 52.82
	  &  
	    &  
	      & 
		& \\  
\hline
%%%%%%%%%%%%%%%%%%%%%%%%%%%%%%%%%%%%%%%%%%%%%%%%%% BL3
%%%%%%%%%%%%%%%%%%%%%%%%%%%%%%%%%%%%%%%%%%%%%%%%%%%%
\vspace{-3mm}&&&&&&&&&\\
\parbox[0pt][5mm][b]{1.5cm}{\centering \textbf{Barrel}}
& 
  & \multirow{3}{*}{\parbox[0pt][5mm][t]{7.5mm}{\centering 95}}
    & \multirow{3}{*}{\parbox[0pt][5mm][t]{7.5mm}{\centering 92}}
      &\multirow{3}{*}{\parbox[0pt][5mm][t]{1cm}{\centering 89.72}}
	& \multirow{3}{*}{\parbox[0pt][5mm][t]{1cm}{\centering 96.86}}
	  & \multirow{3}{*}{\parbox[0pt][5mm][t]{7.5mm}{\centering -}}
	    &  \multirow{3}{*}{\parbox[0pt][5mm][t]{7.5mm}{\centering -}}
	      & \multirow{3}{*}{\parbox[0pt][5mm][t]{1.0cm}{\centering  $-$133.8}}
		& \multirow{3}{*}{\parbox[0pt][5mm][t]{1cm}{\centering 139.0}}  \\  
\vspace{-3.75mm}&&&&&&&&&\\
\parbox[0pt][5mm][t]{1.5cm}{\centering \textbf{layer 3}}
&&&&&&&&& \\  
\hline
%%%%%%%%%%%%%%%%%%%%%%%%%%%%%%%%%%%%%%%%%%%%%%%%%% BL4
%%%%%%%%%%%%%%%%%%%%%%%%%%%%%%%%%%%%%%%%%%%%%%%%%%%%
\vspace{-3mm}&&&&&&&&&\\
\parbox[0pt][5mm][b]{1.5cm}{\centering \textbf{Barrel}}
& 
  & \multirow{3}{*}{\parbox[0pt][5mm][t]{7.5mm}{\centering 135}}
    & \multirow{3}{*}{\parbox[0pt][5mm][t]{7.5mm}{\centering 125}}
      &\multirow{3}{*}{\parbox[0pt][5mm][t]{10mm}{\centering 123.20}}
	& \multirow{3}{*}{\parbox[0pt][5mm][t]{10mm}{\centering 129.24}}
	  & \multirow{3}{*}{\parbox[0pt][5mm][t]{7.5mm}{\centering -}}
	    &  \multirow{3}{*}{\parbox[0pt][5mm][t]{7.5mm}{\centering -}}
	      & \multirow{3}{*}{\parbox[0pt][5mm][t]{1cm}{\centering  $-$169.2}}
		& \multirow{3}{*}{\parbox[0pt][5mm][t]{1cm}{\centering 139.0}}  \\  
\vspace{-3.75mm}&&&&&&&&&\\
\parbox[0pt][5mm][t]{1.5cm}{\centering \textbf{layer 4}}
&&&&&&&&& \\  
\hline
%%%%%%%%%%%%%%%%%%%%%%%%%%%%%%%%%%%%%%%%%%%%%%%%%% Dk1
%%%%%%%%%%%%%%%%%%%%%%%%%%%%%%%%%%%%%%%%%%%%%%%%%%%%
\vspace{-3mm}&&&&&&&&&\\
\parbox[0pt][5mm][b]{1.5cm}{\centering \textbf{Disk}}
& Sdk\,1,\,front
  & \multirow{3}{*}{\parbox[0pt][5mm][t]{7.5mm}{\centering $<$\,50}}
    & \multirow{3}{*}{\parbox[0pt][5mm][t]{7.5mm}{\centering -}}
      & \multirow{3}{*}{\parbox[0pt][5mm][t]{7.5mm}{\centering 11.70}}
	&  \multirow{3}{*}{\parbox[0pt][5mm][t]{7.5mm}{\centering 36.56}}
	  & \multirow{3}{*}{\parbox[0pt][5mm][t]{7.5mm}{\centering 20}}
	    &  \multirow{3}{*}{\parbox[0pt][5mm][t]{7.5mm}{\centering 22}}
	      & \centering 19.7
		& 19.9  \\  
\vspace{-3.75mm}&&&&&&&&&\\
\parbox[0pt][5mm][t]{1.5cm}{\centering \textbf{layer 1}}
& Sdk\,1,\,rear
  & & & & & & & \centering 24.1
		& 24.3\\ 
\hline 
%%%%%%%%%%%%%%%%%%%%%%%%%%%%%%%%%%%%%%%%%%%%%%%%%% Dk2
%%%%%%%%%%%%%%%%%%%%%%%%%%%%%%%%%%%%%%%%%%%%%%%%%%%%
\vspace{-3mm}&&&&&&&&&\\
\parbox[0pt][5mm][b]{1.5cm}{\centering \textbf{Disk}}
& Sdk\,1,\,front
  & \multirow{3}{*}{\parbox[0pt][5mm][t]{7.5mm}{\centering $<$\,50}}
    & \multirow{3}{*}{\parbox[0pt][5mm][t]{7.5mm}{\centering -}}
      & \multirow{3}{*}{\parbox[0pt][5mm][t]{7.5mm}{\centering 11.70}}
	&  \multirow{3}{*}{\parbox[0pt][5mm][t]{7.5mm}{\centering 36.56}}
	  & \multirow{3}{*}{\parbox[0pt][5mm][t]{7.5mm}{\centering 40}}
	    &  \multirow{3}{*}{\parbox[0pt][5mm][t]{7.5mm}{\centering 42}}
	      & \centering 39.7
		& 39.9  \\  
\vspace{-3.75mm}&&&&&&&&&\\
\parbox[0pt][5mm][t]{1.5cm}{\centering \textbf{layer 2}}
& Sdk\,2,\,rear
  & & & & & & & \centering 44.1
		& 44.3\\ 
\hline 
%%%%%%%%%%%%%%%%%%%%%%%%%%%%%%%%%%%%%%%%%%%%%%%%%% Dk3
%%%%%%%%%%%%%%%%%%%%%%%%%%%%%%%%%%%%%%%%%%%%%%%%%%%%
\vspace{-3mm}&&&&&&&&&\\
\parbox[0pt][5mm][b]{1.5cm}{\centering \textbf{Disk}}
& Ldk\,1,\,front
  & \multirow{3}{*}{\parbox[0pt][5mm][t]{7.5mm}{\centering $<$\,95}}
    & \multirow{3}{*}{\parbox[0pt][5mm][t]{7.5mm}{\centering -}}
      & \multirow{3}{*}{\parbox[0pt][5mm][t]{7.5mm}{\centering 11.70}}
	&  \multirow{3}{*}{\parbox[0pt][5mm][t]{7.5mm}{\centering 73.96}}
	  & \multirow{3}{*}{\parbox[0pt][5mm][t]{7.5mm}{\centering 70}}
	    &  \multirow{3}{*}{\parbox[0pt][5mm][t]{7.5mm}{\centering 72}}
	      & \centering 69.7
		& 69.9  \\  
\vspace{-3.75mm}&&&&&&&&&\\
\parbox[0pt][5mm][t]{1.5cm}{\centering \textbf{layer 3}}
& Ldk\,1,\,rear
  & & & & & & & \centering 74.1
		& 74.3\\ 
\hline 
%%%%%%%%%%%%%%%%%%%%%%%%%%%%%%%%%%%%%%%%%%%%%%%%%% Dk4
%%%%%%%%%%%%%%%%%%%%%%%%%%%%%%%%%%%%%%%%%%%%%%%%%%%%
\vspace{-3mm}&&&&&&&&&\\
\parbox[0pt][5mm][b]{1.5cm}{\centering \textbf{Disk}}
& Ldk\,2,\,front
  & \multirow{3}{*}{\parbox[0pt][5mm][t]{7.5mm}{\centering $<$\,95}}
    & \multirow{3}{*}{\parbox[0pt][5mm][t]{7.5mm}{\centering -}}
      & \multirow{3}{*}{\parbox[0pt][5mm][t]{7.5mm}{\centering 11.70}}
	&  \multirow{3}{*}{\parbox[0pt][5mm][t]{7.5mm}{\centering 73.96}}
	  & \multirow{3}{*}{\parbox[0pt][5mm][t]{7.5mm}{\centering 100}}
	    &  \multirow{3}{*}{\parbox[0pt][5mm][t]{7.5mm}{\centering 102}}
	      & \centering 99.7
		& 99.9  \\  
\vspace{-3.75mm}&&&&&&&&&\\
\parbox[0pt][5mm][t]{1.5cm}{\centering \textbf{layer 4}}
& Ldk\,2,\,rear
  & & & & & & & \centering 104.1
		& 104.3\\ 
\hline
%%%%%%%%%%%%%%%%%%%%%%%%%%%%%%%%%%%%%%%%%%%%%%%%%% Dk5
%%%%%%%%%%%%%%%%%%%%%%%%%%%%%%%%%%%%%%%%%%%%%%%%%%%%
\vspace{-3mm}&&&&&&&&&\\
\multirow{3}{*}{\parbox[0pt][5mm][b]{1.5cm}{\centering \textbf{Disk}}}
& Ldk\,3,\,front
  & \multirow{3}{*}{\parbox[0pt][5mm][t]{7.5mm}{\centering -}}
    & \multirow{3}{*}{\parbox[0pt][5mm][t]{7.5mm}{\centering -}}
      & \multirow{3}{*}{\parbox[0pt][5mm][t]{7.5mm}{\centering 11.70}}
	&  \multirow{3}{*}{\parbox[0pt][5mm][t]{7.5mm}{\centering 73.96}}
	  & \multirow{5}{*}{\parbox[0pt][5mm][t]{7.5mm}{\centering 160}}
	    &  \multirow{3}{*}{\parbox[0pt][5mm][t]{7.5mm}{\centering 150}}
	      & \centering 147.7
		& 147.9  \\  
\vspace{-3.75mm}&&&&&&&&&\\
\multirow{3}{*}{\parbox[0pt][5mm][c]{1.5cm}{\centering \textbf{layer 5}}}
& Ldk\,3,\,rear
  & & & & & & & \centering 152.1
		& 152.3\\ 
\vspace{-3mm}&&&&&&&&&\\
& StripDk\,1,\,L
  & \multirow{3}{*}{\parbox[0pt][5mm][t]{7.5mm}{\centering -}}
    & \multirow{3}{*}{\parbox[0pt][5mm][t]{7.5mm}{\centering -}}
      & \multirow{3}{*}{\parbox[0pt][5mm][t]{7.5mm}{\centering 74.33}}
	&  \multirow{3}{*}{\parbox[0pt][5mm][t]{10mm}{\centering 131.15}}
	  & 
	    &  \multirow{3}{*}{\parbox[0pt][5mm][t]{7.5mm}{\centering 162.5}}
	      & \centering 160.0
		& 160.3  \\  
\vspace{-3.75mm}&&&&&&&&&\\
&  StripDk\,1,\,S
  & & & & & & & \centering 165.0
		& 165.3\\ 
\hline
%%%%%%%%%%%%%%%%%%%%%%%%%%%%%%%%%%%%%%%%%%%%%%%%%% Dk6
%%%%%%%%%%%%%%%%%%%%%%%%%%%%%%%%%%%%%%%%%%%%%%%%%%%%
\vspace{-3mm}&&&&&&&&&\\
\multirow{3}{*}{\parbox[0pt][5mm][b]{1.5cm}{\centering \textbf{Disk}}}
& Ldk\,4,\,front
  & \multirow{3}{*}{\parbox[0pt][5mm][t]{7.5mm}{\centering -}}
    & \multirow{3}{*}{\parbox[0pt][5mm][t]{7.5mm}{\centering -}}
      & \multirow{3}{*}{\parbox[0pt][5mm][t]{7.5mm}{\centering 11.70}}
	&  \multirow{3}{*}{\parbox[0pt][5mm][t]{7.5mm}{\centering 73.96}}
	  & \multirow{5}{*}{\parbox[0pt][5mm][t]{7.5mm}{\centering 230}}
	    &  \multirow{3}{*}{\parbox[0pt][5mm][t]{7.5mm}{\centering 220}}
	      & \centering 217.7
		& 217.9  \\  
\vspace{-3.75mm}&&&&&&&&&\\
\multirow{3}{*}{\parbox[0pt][5mm][c]{1.5cm}{\centering \textbf{layer 6}}}
& Ldk\,4,\,rear
  & & & & & & & \centering 222.1
		& 222.3\\ 
\vspace{-3mm}&&&&&&&&&\\
& StripDk\,2,\,S
  & \multirow{3}{*}{\parbox[0pt][5mm][t]{7.5mm}{\centering -}}
    & \multirow{3}{*}{\parbox[0pt][5mm][t]{7.5mm}{\centering -}}
      & \multirow{3}{*}{\parbox[0pt][5mm][t]{7.5mm}{\centering 74.33}}
	&  \multirow{3}{*}{\parbox[0pt][5mm][t]{10mm}{\centering 131.15}}
	  & 
	    &  \multirow{3}{*}{\parbox[0pt][5mm][t]{7.5mm}{\centering 207.5}}
	      & \centering 204.7
		& 205.0  \\  
\vspace{-3.75mm}&&&&&&&&&\\
&  StripDk\,2,\,L
  & & & & & & & \centering  209.7
		& 210.0\\ 
\hline
\multicolumn{9}{c}{\footnotesize Sdk ... small pixel disk, Ldk ... large pixel disk, StripDk ... strip disk}
\end{tabular}
} %end scalebox
\caption[Positions of the active sensor volumes within the different detector layers]
{Positions of the active sensor volumes within the different detector layers. 
$r_{\textsf{\tiny{def}}}$ and $z_{\textsf{\tiny{def}}}$ are the predefined values 
for the radius and the axial extension as given in \figref{pic-MVD-GeneralLayout}, 
$\langle r_{\textsf{\tiny{opt}}} \rangle$ and $\langle z_{\textsf{\tiny{opt}}} \rangle$ 
are the corresponding mean values in the sub-layers after optimisation. 
$r_{\textsf{\tiny{min/max}}}$ and $z_{\textsf{\tiny{min/max}}}$ refer to the outer 
limits of the radial and axial extension in the respective layers.
}
\label{Table-SensPOSlayer}
\end{center}
\end{table}
\twocolumn

Plotted results for a silicon thickness of 0.6~mm and 1.2~mm represent the minimum effective path lengths of particles 
crossing two and four pixel layers, respectively. 
With a minimum flight path of roughly 10~cm between the pixel and the barrel layers, 
the lateral track resolution due to scattering effects
is restricted to a few hundred micrometers
for particles with a momentum well below 1~\gevc.

In total there are three different strip sensor types,
which have different shapes.
They are shown in \figref{pic-StripGeometry}, on the top.
Quadratic and rectangular sensors are chosen for the barrel part,
while a trapezoidal design is used for the forward disks. 
The respective stereo angle is set to 90$^{\circ}$ in case of barrel type sensors 
and to a reduced angle of 15$^{\circ}$ for the trapezoids. 
In this way strips always run in parallel to one sensor edge. 
The strip pitch for the rectangular barrel sensors is set to 130~$\tcmu$m.
A smaller value of 67.5~$\tcmu$m for the trapezoids in the disk part
is used to recover the worsened resolution induced by the smaller stereo angle.

The strip barrel layers are formed by a paddle-wheel sensor arrangement. 
Neighbouring trapezoidal sensors of the forward disks are located at two positions 
slightly shifted along the beam axis in order to make up for the passive sensor edge. 
To accommodate the target pipe crossing, 
some barrel strip sensors are left out at the top and bottom. 
A schematic picture of the strip sensor geometry is given 
in the bottom part of \figref{pic-StripGeometry}.

Finally, the defined sensor geometry results in approximately 10.3 million pixel
and roughly 200,000 strip readout channels.
Design parameters for the MVD are summarised in \tabref{tab-BasicInfo-CentralMVD}.
A compilation of the sensor positions in all detector layers 
can be found in \tabref{Table-SensPOSlayer}.

\begin{figure}[b]
\begin{center}
\includegraphics[trim=0.2cm 0  0.1cm 0, clip, width=7.5 cm]{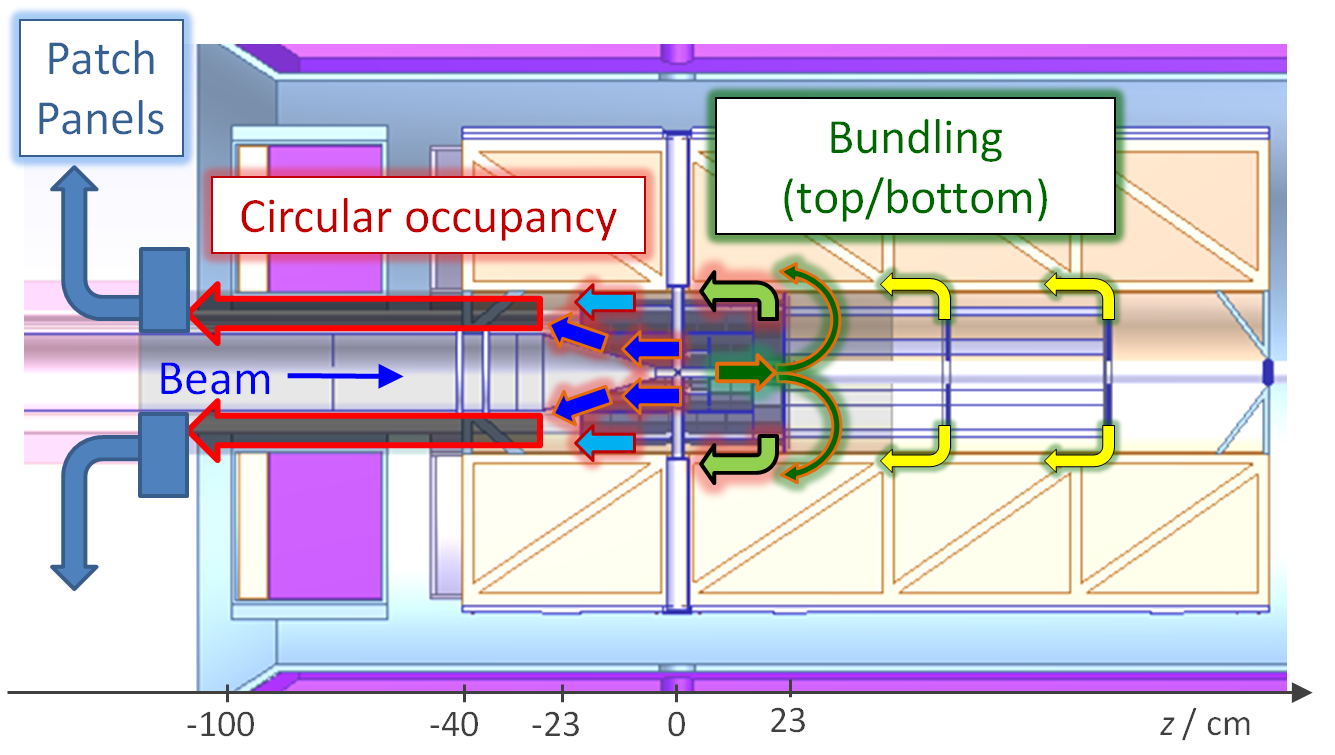}
% Pic3-04_PixelArrangement.png: 1240x910 pixel, 125dpi, 25.20x18.49 cm, bb=0 0 714 524
\caption[Overall routing concept for the MVD]
{
Schematic routing scheme for the MVD. 
Blue and green arrows illustrate the concept for the barrel and the forward part, respectively.
A probable routing of the additional forward disks indicated with yellow arrows 
potentially interferes with the MVD volume.
}
\label{pic-OverallRouting}
\end{center}
\end{figure}

\begin{figure}[t]
\begin{center}
\includegraphics[width=0.45\textwidth]{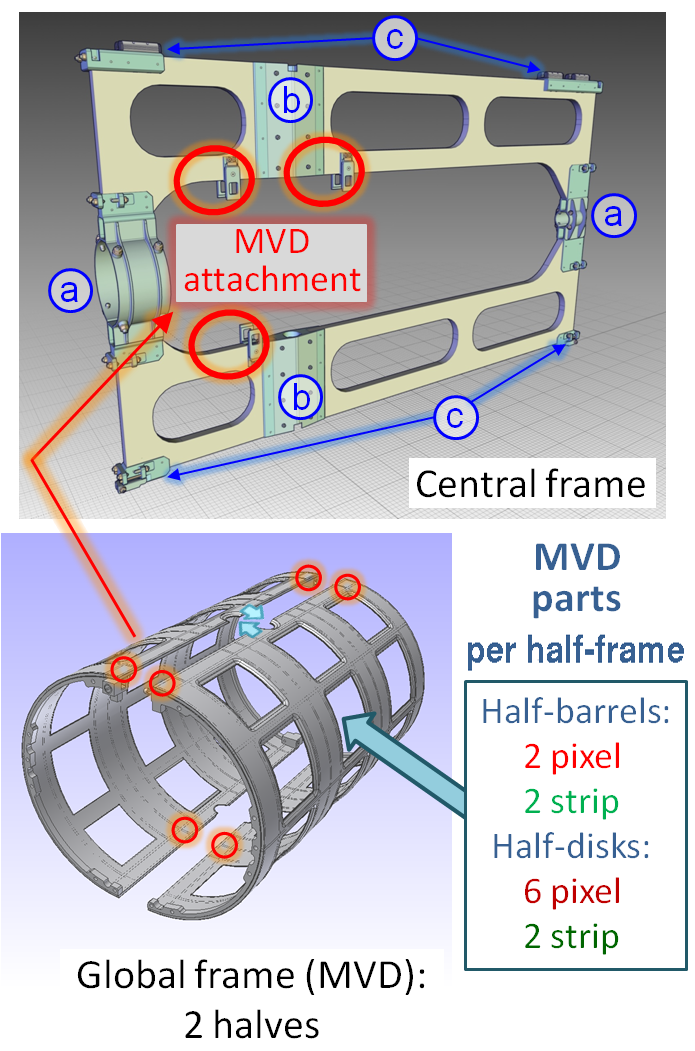}
% Pic3-04_PixelArrangement.png: 1240x910 pixel, 125dpi, 25.20x18.49 cm, bb=0 0 714 524
\caption[Overall detector integration of the MVD]
{Detector integration of the MVD inside the target spectrometer.
\textit{Left:} Three-point fixation to the central frame,
which also provides fixation points for the beam pipe (a), 
the target pipe (b) and the connection to a rail system (c). 
The main MVD parts are integrated in two half-frames (\textit{right}), 
which are connected separately to the central frame.}
\label{pic-MvdOverallIntegration}
\end{center}
\end{figure}

\subsubsection*{Mechanics and Detector Integration}

The MVD will operate under normal pressure conditions 
at room temperature of approximately +25~$^{\circ}$C. 
An active cooling of both pixel and strip readout electronics 
is needed for a long term stability of the detector. 
A detailed description of the cooling system can be found in 
chapter~\ref{Cooling-system}.
Besides the cooling pipes, other service structures 
must be brought from the outside close to the silicon detectors
in order to supply sensors and associated electronics components 
as well as to transmit signals for data and slow control of the readout chips. 
A schematic routing scheme for all MVD services is given in \figref{pic-OverallRouting}.

Limitations are given by the stringent boundary conditions.
Due to the fixed target setup, 
which defines the physically most interesting region 
in the forward part,
a guiding to the outside of the target spectrometer 
is only possible in upstream direction.
In this region the opening cone of the beam pipe reduces 
the available cross section at inner radii.
The restricted radial space requires a circular occupancy 
of all services around the beam pipe.
For the pixel disk inside the barrel layers 
a bundling at the top and bottom is foreseen. 
Patch panels, which allow a connection to peripheral systems, 
are intended to be placed further upstream.

Basic ideas of the overall detector integration for the MVD 
are shown in \figref{pic-MvdOverallIntegration}. 
The mechanical interface to external parts of the \panda apparatus 
is defined by a three point fixation to the central frame, which serves as 
reference frame and global support for the entire central tracking system. 
It also facilitates a consecutive assembly of the beam-target system, 
the MVD and the outer tracker outside of the solenoid magnet. 
A safe insertion of all attached sub-systems to their nominal position 
is realised via a rail system.
The global support structure for the MVD is composed of two half-frames. 
Both halves are mechanically independent in order to avoid any introduction of stress 
to either half-detector. 
They allow the internal mounting of all detector half-layers 
and the attachment to the central frame. 
All support structures inside the MVD will be composed of light-weight 
carbon materials. Details can be found in section~\ref{mechanics}.
Additional support structures are also needed for MVD services.

%% file: pixelpart/pixelpart.tex
\chapter {Silicon Pixel Part}
\label{pixelpart}

\authalert{Author: D. Calvo, calvo@to.infn.it}

This chapter describes the silicon 
pixel part of the Micro Vertex Detector. 
The innermost layers and disks of the MVD are 
the most important ones for determining primary  
and secondary vertices of charm mesons. They need a very
high precision and granularity to cope with the required spatial
resolution.
In addition they have to handle high track densities 
in the forward region for proton targets, they are 
exposed to relatively high radiation levels and they 
have to process and transmit very high data rates due 
to the untriggered readout of \PANDA.  
To cope with these requirements a hybrid pixel detector 
with its two-dimensional segmentation 
featuring a 2D readout without ambiguity, 
 double hit resolution, excellent signal to noise 
ratio at high speed, and reliable sensor technology appears 
to be the most interesting choice for 
equipping the innermost part of the MVD.

	\section{Hybrid Pixel Assembly Concept}

\begin{figure}[htb]
\vspace{1pt}
\centering
\includegraphics[width=0.45\textwidth]{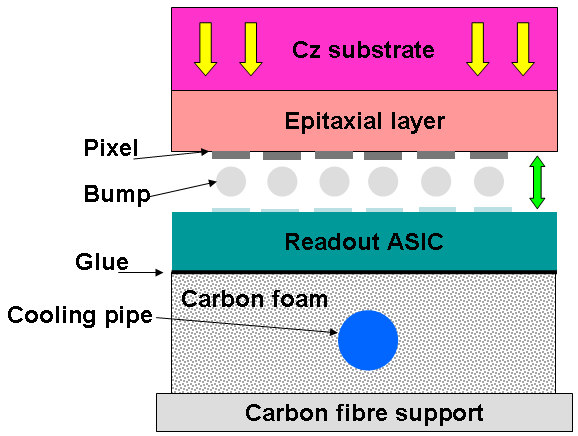}
\caption[Scheme of the hybrid pixel for \PANDA]{Scheme of the hybrid pixel for \PANDA.}
\label{fig:hybrid}
\end{figure}

Hybrid pixel detectors have been studied and developed
for the vertex trackers of the LHC experiment. 
The core issue is a two-dimensional matrix, the sensor
(rectangular shape of few tens of mm by a few 
hundreds $\tcmu$m thick), of reverse biased silicon diodes 
flip-chip bonded to several readout chips. 
Each cell (pixel) on the sensor matrix is connected 
by solder-lead bumps (alternatively indium is used) to a 
frontal cell on a readout ASIC developed in CMOS technology. 
The readout cell must fit into the same area as 
the pixel sensing element. 
In particular, the ASICs used in the LHC experiments 
have been developed 
in  250~nm CMOS technology, whereas the sensor has 
been made of Floating Zone (FZ) silicon with a 
minimum thickness of 200~$\tcmu$m. 
The request of high granularity and the simulation results 
of benchmark channels for the MVD  
indicate a pixel size of $\mathrm{100~\tcmu m \times 100~\tcmu m}$
as the most suitable. However
the reduction of the surface of the pixel increases 
the perimeter-to-area ratio 
and hence the input capacitance, which is usually 
dominated by the inter-cell capacitance. 
To cope with the limited material budget 
an R$\&$D effort to investigate a lower sensor thickness 
has been made 
and the results are reported in chapter~\ref{Sen}. 
Concerning the sensor material, references \cite{semi0}, \cite{semi1} 
indicate that silicon epitaxial material could be 
an interesting material to be applied in \PANDA, 
in particular for radiation hardness, up to the radiation 
level expected.
Taking into account the configuration of the wafer made of an epitaxial silicon layer 
grown on top of a Czochralski substrate, 
thinned pixel sensors could be obtained by removing most 
of the Cz substrate. 
Furthermore, 
an R$\&$D project has started to develop a readout ASIC
in  130~nm CMOS technology equipped with 
pixel cells of the required size. 
In addition to the two-dimensional spatial information, 
 time and energy loss of the hit will be measured.
Besides the feature to obtain by this technology 
the same electrical functionality using a quarter of 
area than older technology with a power reduced 
by a factor 4 appears a challenge towards a 
simplification of the cooling system. 
The readout architecture is described in the section \ref{ToPiX_architecture} 
and the ASIC prototypes in the 
section~\ref{ASIC_prototypes}.   
A schematic view of the custom hybrid pixel assembly 
for the \PANDA experiment is shown in \figref{fig:hybrid}.

The epitaxial silicon layer with a thickness up to 150~$\tcmu$m 
and a high 
resistivity of few kOhm cm is the active region 
of the pixel matrix. The Cz substrate 
features lower resistivity ($0.01-0.02$~$\Omega\cdot \mathrm{cm}$) 
and a thickness of a few hundreds of $\tcmu$m. 
Most of this substrate is removed starting 
from the side opposite to the epitaxial layer.
Each pixel is connected by the bump bonding 
technique to the corresponding readout cell of 
a custom ASIC that will be developed in 130~nm CMOS technology. Then this assembly is 
glued to a carbon foam layer to improve heat 
dissipation towards the cooling pipe as a 
part of a cooling system working in depression 
mode and based on water as cooling fluid. 
A structure made of carbon fibre with suitable 
shape is the mechanical support.

	\section{Sensor}
\label{Sen}
			
\authalert{Author: D. Calvo, calvo@to.infn.it}

		\subsection{First Thinned Prototypes}
To study the epitaxial silicon material, 
first prototypes  of  hybrid pixel  
assemblies using thinned epitaxial silicon 
pixel structures as sensor element have been 
produced and tested in 2008 \cite{epi02}. 
Results were obtained from 
assemblies with three different epitaxial thicknesses 
of 50, 75 and 100~$\tcmu$m. 
The \frontend chip for \PANDA was still under development 
at that time and therefore 
 ALICE pixel \frontend chips and readout chain were used. 
Basic detector characteristics, bump bond yields were measured, and tests with various 
radioactive sources were performed. 

			\subsubsection*{Design and Production}

For the sensors, 4-inch wafers with three different n-epitaxial layer 
thicknesses deposited on standard Czochralski (Cz) substrate wafers 
were provided by ITME (Warsaw). The different epitaxial layer 
characteristics are shown in \tabref{tab:firstprototypes}. 
The Cz substrate parameters 
were equal for all wafers: $n^+$ conductivity type, 
Sb dopant, 525~$\tcmu$m thickness, 
$0.01-0.02$~$\Omega\cdot\mathrm{cm}$ resistivity.

\begin{table}[width=0.4\textwidth]
\centering
\begin{tabular}{|l|c|c|l}
\hline
  % after \\: \hline or \cline{col1-col2} \cline{col3-col4}
 Epi Layer &  Thickness & Resistivity \\ \hline
 Epi-50 & $49\pm0.5$~$\tcmu$m & 4060~$\Omega\cdot\mathrm{cm}$ \\ \hline
 Epi-75 & $73.5\pm1$~$\tcmu$m & 4570~$\Omega\cdot\mathrm{cm}$ \\ \hline
 Epi-100 & $98\pm2$~$\tcmu$m & 4900~$\Omega\cdot\mathrm{cm}$ \\ \hline
\end{tabular}
\caption[Thickness and resistivity of the epitaxial layers of the first assemblies]{Thickness and resistivity of the epitaxial layers of the first assemblies.}
\label{tab:firstprototypes}
\end{table}

The wafers were processed by FBK-ITC (Trento) 
using ALICE pixel sensor masks provided 
by INFN-Ferrara which had already previously 
been used to produce high quality ALICE 
pixel assemblies from 200~$\tcmu$m Float Zone (FZ) 
material \cite{alice0}. The sensor masks include 
full-size and single-chip sensors (the ALICE 
pixel size is $\mathrm{50~\tcmu m \times 425~\tcmu m}$) together 
with the usual range of diagnostic structures including 
simple diodes. 
Six epitaxial wafers were processed, two of each thickness. 
The thinning of the sensor wafers and the bump-bonding 
to ALICE pixel 
readout chips were performed by VTT (Finland). 
The target thicknesses 
for the sensor wafers were chosen to be 100, 120, and 150~$\tcmu$m, respectively, 
for the 50, 75, and 100~$\tcmu$m epi layers. Single chip sensors were bump-bonded 
to chip-sensor assemblies (Epi-50, Epi-75, and Epi-100, 
respectively, corresponding to the 50, 75, and 100~$\tcmu$m thick epitaxial layers).
      
	\subsubsection*{Results}

\begin{figure}[htb]
\vspace{1pt}
\centering
\includegraphics[width=0.45\textwidth]{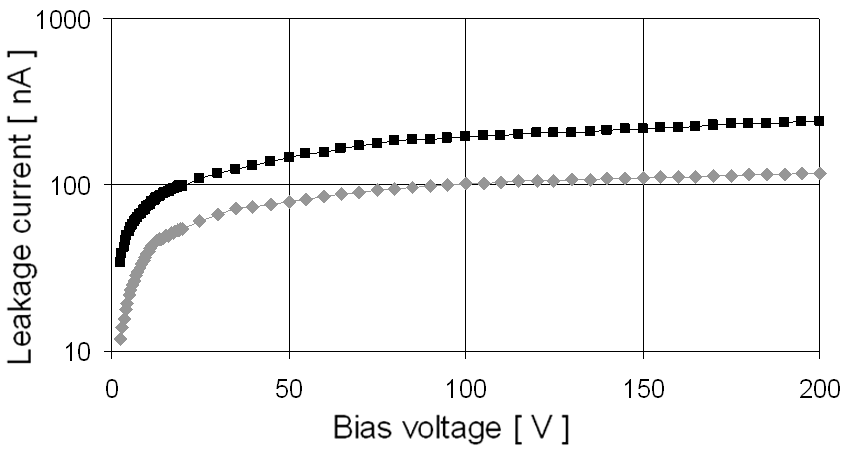}
\caption[Leakage current as a function of the bias voltage for assemblies Epi-50 and Epi-75]{Leakage current as a function of the bias voltage for assemblies Epi-50 (lower curve) and Epi-75 (upper curve).}
\label{fig:lcurrent}
\end{figure}

The full-depletion voltage values for all three epitaxial wafer thicknesses 
were less than 6~V. \Figref{fig:lcurrent} 
shows the total sensor leakage current 
as a function of bias voltage up to 200~V for two assemblies. 

No sign of breakdown is observed, and the thicker sensor has 
a higher current for any given voltage. Good long-term stability was also 
noted.
Total leakage current as a function of the bias voltage was 
measured at different temperatures up to 50~$^\circ$C for one 
assembly (Epi-50).  
A leakage current of about 3~nA, corresponding
to a bias voltage of 100~V, (the assembly has 8192 pixels) 
at 21~$^\circ$C increases up to about 23~nA at 43~$^\circ$C. 
This corresponds to a leakage current per pixel of less than about 10~fA,
which is a value that can be accepted and 
controlled by the readout circuit.

\begin{figure}[htb]
\vspace{1pt}
\centering
\includegraphics[width=0.45\textwidth]{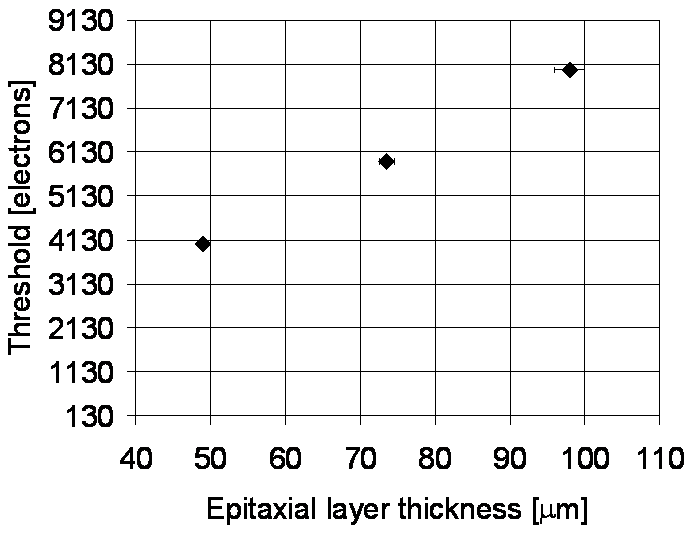}
\caption[Threshold values corresponding to the Landau most probable value for the different epitaxial layer thicknesses]{Threshold values corresponding to the Landau most probable value for the different epitaxial layer thicknesses.}
\label{fig:threshold}
\end{figure}

Three assemblies, S15 (Epi-50), S8 (Epi-75), and S6 (Epi-100), 
were instead mounted on testing boards enabling their complete 
electrical characterisation using the ALICE Pixel DAQ system \cite{alice1}.
Beta particles from a $^{90}$Sr source,  which pass through the 
assemblies and their associated mountings to subsequently 
trigger a plastic scintillator with photomultiplier readout 
should be sufficiently similar in their behaviour to 
minimum-ionising particles (MIP)  to enable a relative 
comparison of the active thickness of the epitaxial sensors. 
Calibration of the threshold DAC setting in electrons is 
obtained by making similar, but self$\textendash$triggered, threshold 
scans while exposing the assemblies to $^{55}$Fe and $^{109}$Cd 
X$\textendash$ray sources. 
\Figref{fig:threshold} shows the threshold value 
in electrons corresponding 
to the most probable value of the Landau distribution plotted 
with respect to the nominal epitaxial active thickness. 

It should be noted, as is confirmed by the linearity of the result, 
that charge sharing is not a major issue despite the 
50~$\tcmu$m pixel width due to the fact that the active 
thickness is equally limited. Taking 22.500 electrons for 
a true MIP in 300~$\tcmu$m of silicon one would expect 7500 electrons 
from 100~$\tcmu$m of silicon, and given the use of the beta 
source which should raise that value somewhat the agreement 
seems more than satisfactory.

		\subsection{Radiation Damage Study}

Epitaxial material has been studied for radiation damage by 
many experimental groups (RD50 collaboration). In particular, at the 
beginning of the R$\&$D phase for the \PANDA pixel detector, results on 
thin epitaxial diodeds (about 25~$\tcmu$m) and lower epitaxial resistivity
(about $100-150$~$\Omega\cdot\mathrm{cm}$)  were available.
However the application of epitaxial material in the \PANDA experiment asked for 
higher thickness and higher resistivity to obtain more electron production 
from MIP and a lower full depletion voltage, respectively.
For these reasons systematic tests for radiation damage were planned.
The most important parameters as full depletion voltage and  leakage current
have been studied as a function of fluences and annealing phases.
To study the behaviour of these parameters, displacement damage tests 
were performed using neutrons from the nuclear reactor (TRIGA MARK II) 
of Pavia (Italy) at the LENA laboratory. 
Diodes from wafers, featuring different epitaxial thicknesses and resisitivity 
as reported in \Tabref{tab:radiation},  
have been characterised for full depletion voltage 
and leakage current  using a probe station with bias voltage provided by a 
Keithley 237 High Voltage Source Measure Unit.

\begin{table}[width=0.4\textwidth]
\centering
\begin{tabular}{|l|c|c|c|l}
\hline
  % after \\: \hline or \cline{col1-col2} \cline{col3-col4}
 Diodes &  Thickness & Resistivity \\ \hline
 Epi-50, HR  & $49\pm0.5$~$\tcmu$m & 4060 $\Omega\cdot\mathrm{cm}$ \\ \hline
 Epi-75, HR & $73.5\pm1$~$\tcmu$m & 4570 $\Omega\cdot\mathrm{cm}$ \\ \hline
 Epi-100, HR & $98\pm2$~$\tcmu$m & 4900 $\Omega\cdot\mathrm{cm}$ \\ \hline
 Epi-50, MR  & $50\pm1$~$\tcmu$m & 3100 $\Omega\cdot\mathrm{cm}$ \\ \hline
 Epi-75, MR & $74.6\pm1$~$\tcmu$m & 3200 $\Omega\cdot\mathrm{cm}$ \\ \hline
 Epi-100, MR & $100\pm1$~$\tcmu$m & 3610 $\Omega\cdot\mathrm{cm}$ \\ \hline
 Epi-75, LR & $74.5\pm1$~$\tcmu$m & 460 $\Omega\cdot\mathrm{cm}$ \\ \hline
\end{tabular}
\caption[Thickness and resistivity featuring the epitaxial layer of the
wafers used for the radiation damage tests]{Thickness and resistivity featuring the epitaxial layer of the
wafers used for the radiation damage tests. HR, MR and LR
mean High resistivity, Middle Resistivity and Low Resistivity, respectively.}
\label{tab:radiation}
\end{table}

I-V measurements have been performed using an HP4145B Semiconductor Parameter Analyser and C-V
mesurements have been performed using an HP4248A Precision Meter.
The measurents of I-V and C-V trends for different diodes have 
been carried out before and after the irradiation as a function of 
the equivalent fluence and specifically with the following values: 
$1.5\cdot10^{14}$ and $5.3\cdot10^{14}$~\neueq, respectively,
corresponding to about 10 years of \PANDA lifetime with antiproton 
proton annihilation and antiproton Xe interactions (50\% duty cycle). 

Then, a long annealing phase at 60~$^\circ$C was
started and full depletion bias voltage and leakage current 
characterisation were performed many times during this period. 
The full depletion voltage is evaluated from the intersection of 
the two extrapolated 
lines in the capacitance-voltage characteristic. 
In \Tabref{tab:fdv} the full depletion voltage of diodes under test are 
reported before the irradiation.

\begin{table}[width=0.4\textwidth]
\centering
\begin{tabular}{|l|c|c|c|l}
\hline
  % after \\: \hline or \cline{col1-col2} \cline{col3-col4}
 Diodes &  Full Dep.~Voltage [V] & $\sigma$ [V] \\ \hline
 Epi-50, HR  & 4.35 & 0.06 \\ \hline
 Epi-75, HR & 5.6 & 0.1 \\ \hline
 Epi-100, HR & 5.9 & 0.2 \\ \hline
 Epi-50, MR  & 4.91 & 0.07 \\ \hline
 Epi-75, MR & 8.2 & 0.3  \\ \hline
 Epi-100, MR & 10.4 & 0.6 \\ \hline
 Epi-75, LR & 42.6 & 1.3 \\ \hline
\end{tabular}
\caption[Pre irradiation full depletion voltage values as measured
for epitaxial diodes]{Pre irradiation full depletion voltage values as measured
for epitaxial diodes under test for radiation damage. HR, MR and LR
mean High resistivity, Middle Resistivity and Low Resistivity, respectively. $\sigma$ is the standard deviation.}
\label{tab:fdv}
\end{table}

\begin{figure*}[t]
\vspace{1pt}
\centering
\includegraphics[width=0.84\textwidth]{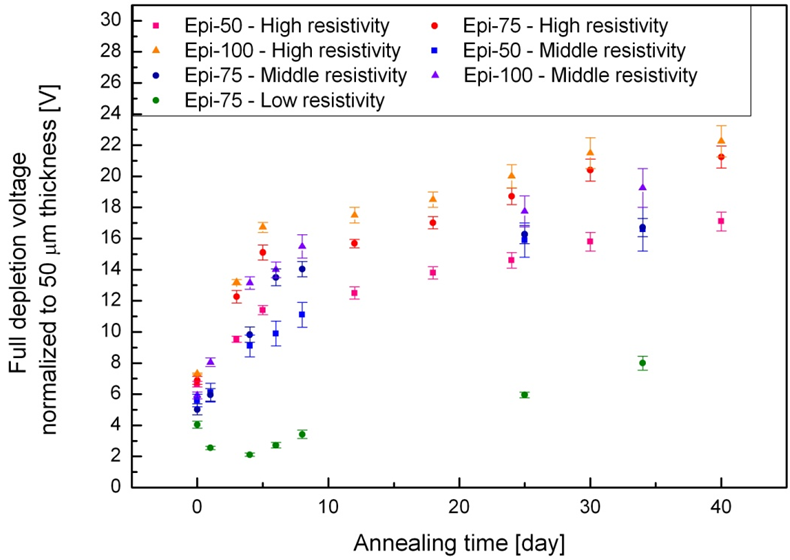}
\caption[Annealing behaviour of the full depletion voltages as a 
function of the 
annealing time for the irradiated diodes]{ Annealing behaviour of the full depletion voltages as a 
function of the 
annealing time for the diodes irradiated to 
$1.5\cdot10^{14}$~\neueq.
All bias voltage values are normalised to the 50~$\tcmu$m thickness. 
The values reported at time zero are the voltages of irradiated diodes, 
measured after the irradiation, but before the annealing phase. 
In this plot, the square corresponds to diodes Epi-50, the circle to 
diodes Epi-75, 
and the triangle to diodes Epi-100; 
the colours in red scale correspond to 
the high resistivity devices, 
the colours in blue scale to the middle resistivity diodes, 
and the green to the low resistivity one.}
\label{fig:3year}
\end{figure*} 

As expected the full depletion voltage values decrease as the resistivity 
of the epitaxial part increases. 
In \figref{fig:3year} the full depletion bias voltages measured 
during the annealing 
phase at 60~$^\circ$C following the irradiation corresponding to 
$1.5 \cdot 10^{14}$~\neueq.
All voltage values are normalised to the 50~$\tcmu$m epitaxial thickness. 
The bias voltages reported at time zero are the values 
measured after the neutron 
irradiation and before the annealing phase \cite{re:ciao}.

The measured values of the full depletion voltages for high and middle 
resistivity 
diodes have increasing trends showing a reverse annealing effect. 
Practically the devices have undergone the type inversion phenomenon immediately 
after the neutron irradiation.
Besides, the larger epitaxial thickness needs higher full depletion voltages 
that could be explained both by higher resistivity values 
of the epitaxial material or by a non-uniform concentration of oxygen 
diffusing 
from the Czochralski substrate, limiting its beneficial 
action against the effects induced by radiation.
The  full depletion voltage values for the lowest resistivity diodes 
as a function of the annealing time show a different trend in comparison 
to the others. 
The evidence for type inversion is found after few days of the annealing phase.

The diode volume currents corresponding to the full depletion bias voltage 
of the 
devices under study increase to values of about $10^{-2}$~A/cm$^3$ for
all tested devices starting from $10^{-8}-10^{-7}$~A/cm$^3$.
A leakage current of about 13~nA/pixel, corresponding  to the maximum 
measured volume current, assuming a volume 
of $\mathrm{100~\tcmu m \times 100~\tcmu m \times 100~\tcmu m}$, 
 the \PANDA choice, is reached.
Since the \frontend electronics has a leakage compensation scheme designed to 
withstand a leakage current of 60~nA with a baseline shift smaller than 2~mV, 
this increase of leakage current is already acceptable for 
the \PANDA experiment.
The leakage current decreases to 50\% after some days of annealing.
The damage constant $\alpha$ ranges between $5-8\cdot10^{-17}$~A/cm. 
This constant is estimated as $\Delta J$/$\phi$ where $\Delta J$ is the difference between the diode volume current before and 
after the irradiation, 
both measured once the full depletion bias voltage has been reached, 
and $\phi$ is the corresponding equivalent fluence.

Results from a test at $5.3\cdot10^{14}$~\neueq in particular for 
the annealing phase
at  60~$^\circ$C confirm the trends already discussed, but 
with higher full depletion voltage values corresponding to the same 
time of annealing.
The diode volume currents after the irradiation are of the same 
order of magnitude.

Comparison of full depletion voltage values as a function of
 two different 
temperatures in the 
annealing phase (40~$^\circ$C and 60~$^\circ$C) shows different trends: evidence 
of inversion type is not limited to the lowest resistivity under test 
but even in the 
higher resistivity samples. Apart from that, the full depletion voltages are 
systematically 
lower at lower temperature as expected.     

Looking at these results the lower thickness appears better for an application 
as sensor, but this parameter has to be a compromise with the  $\text{d}E/\text{d}x$
information and the signal to noise ratio. Besides a higher thickness entails 
larger stiffness. The 100~$\tcmu$m thickness appears to be the best compromise.\\
The resistivity choice has to be a compromise between the initial 
full depletion voltage
values and the values measured during the annealing phase.
The resistivity, for the epitaxial layer, can be adjusted in a certain range
between 1~k$\Omega\cdot\mathrm{cm}$ and 3~k$\Omega\cdot\mathrm{cm}$.

		\subsection{Fullsize Prototype Sensors}
					
A first design of the full size sensors for the pixel detector 
has been carried out, taking into account that sensors of 
different sizes are needed for optimising the disk coverage.

      \subsubsection*{Design and Production}

\begin{figure}[htb]
\vspace{1pt}
\centering
\includegraphics[width=0.45\textwidth]{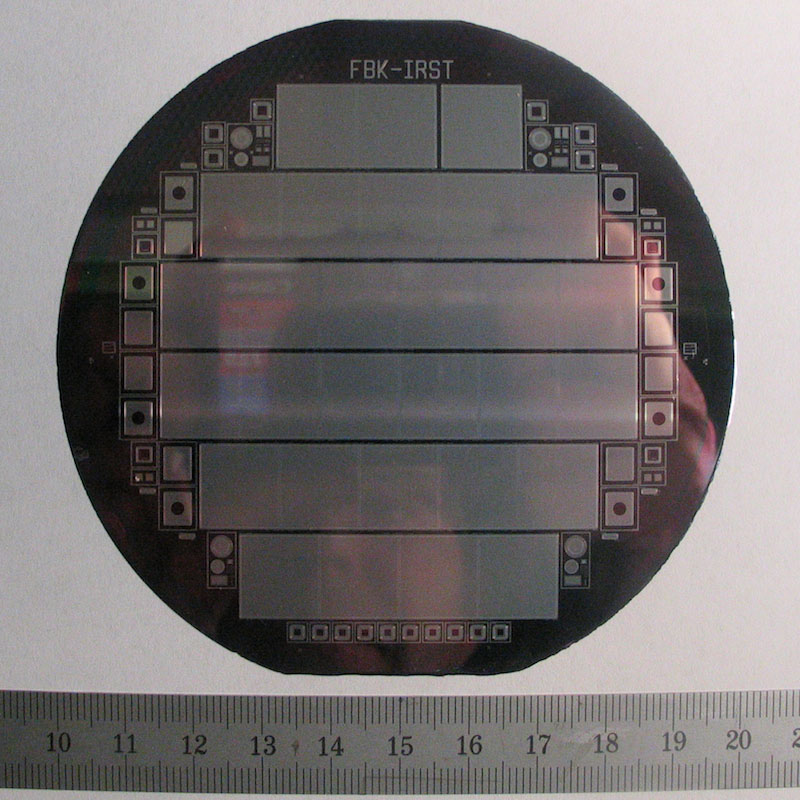}
\caption[Picture of a \PANDA wafer]{Picture of a \PANDA wafer.}
\label{fig:wafer}
\end{figure}

The basic idea for the \PANDA pixel detector is a modular concept based on the readout 
chip size. Sensors housing 2, 4, 5 or 6 readout chips have 
been arranged in the epitaxial silicon wafer to allow a 
first production for thinning and dicing studies. 
In \figref{fig:wafer} the picture of a whole wafer with pixel sensors 
and diagnostic structures on the edge can be seen.

\begin{figure}[htb]
\vspace{1pt}
\centering
\includegraphics[width=0.45\textwidth]{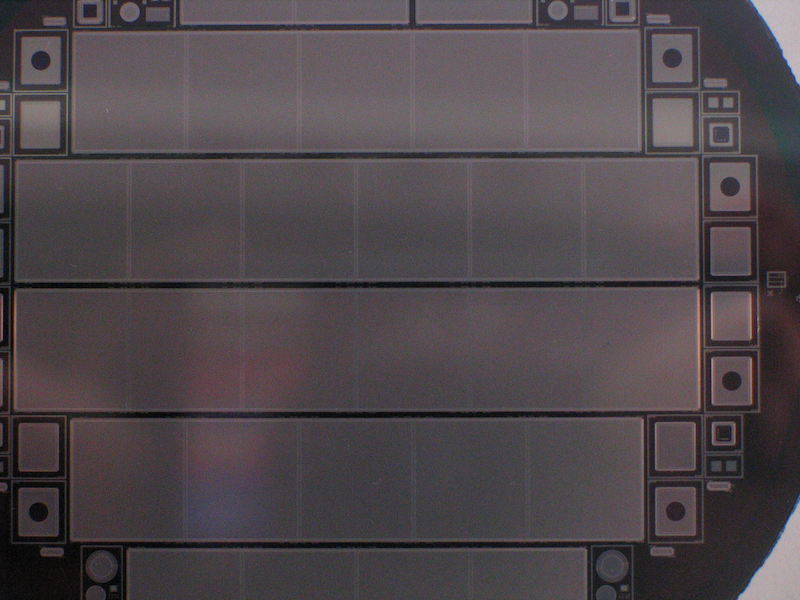}
\caption[Picture showing a partial view of the sensor wafer]{Picture showing a partial view of the sensor wafer.}
\label{fig:pwafer}
\end{figure}

\Figref{fig:pwafer} shows a partial view of the wafer. 
In the middle, two sensors for arranging six readout chips 
are visible. The vertical white lines correspond 
to two larger pixel ($\mathrm{100~\tcmu m \times 300~\tcmu m}$) columns in the 
region where two readout ASICs will be arranged side by side.

\begin{figure}[htb]
\vspace{1pt}
\centering
\includegraphics[width=0.3\textwidth]{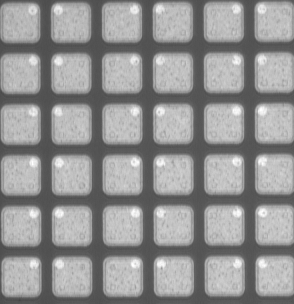}
\caption[Photograph of a part of the pixel matrix]{Photograph of a part of the pixel matrix. Pads for
bumps are the white circles at one corner of the gray pixel.}
\label{fig:pad}
\end{figure}

\Figref{fig:pad} represents a partial view of a sensor made 
with $\mathrm{100~\tcmu m \times 100~\tcmu m}$ pixels.
At one corner of each pixel the pad for 
the bump bonding is visible. These are arranged in a mirror configuration 
with respect to a bus serving two pixel columns.
The pitch along the column is 100~$\tcmu$m, and along 
the rows is alternatively 50~$\tcmu$m and 150~$\tcmu$m.  
      
\clearpage

In these sensors the passive silicon edge housing guard 
rings is 700~$\tcmu$m wide (about 10\% of the sensor size). The guard ring closest to 
the sensor has a width of 100~$\tcmu$m and it is connected to ground, 
additional guard rings are floating. 
This configuration has been applied as a preliminary 
layout similar to the previous sensors designed for 
LHC experiments. In our case the use of epitaxial silicon 
sensors gives the possibility of a reduced bias voltage 
during the sensor lifetime as a consequence of a lower full 
depletion voltage. A limited guard ring 
number reducing the previous passive silicon edge of 50\% 
could be possible. This evaluation has been obtained from
simulations carried out in collaboration with FBK, showing 
the break down voltage as a function of the floating guard ring number.
Experimental measurement using the single chip assembly based on ToPix~v3
will be performed to validate the simulation results.\\
Middle resistivity wafers used during the radiation 
damage test have been investigated for 
these tests. 
Three epitaxial 
thicknesses were chosen to better investigate the thickness 
dependence during the thinning process. Epi-75 
low resistivity wafers were added for comparison purposes.
Sixteen  wafers were thinned at the Wafer Solution company: 
the Epi-50 and Epi-75 wafers to the target of 100~$\tcmu$m thicknesses 
and the Epi-100 wafer to 120~$\tcmu$m thickness, 
in addition, half of these wafers were diced to obtain sensors. 
No mechanical breakage occurred.

			\subsubsection*{Tests}
			
In addition to the radiation damage test of the epitaxial material used 
to make the first sensor prototypes with the results already described in the 
previous section, I-V and C-V measurements and planarity and thickness tests
 were performed. 

In particular I-V and C-V measurements were made on some test structures 
before and after thinning and dicing of wafers to investigate potential 
mechanical damage in the bulk of the wafer. Variations within measurement
errors have been reported  in the I-V and C-V trends and no changes in the 
leakage current and full depletion voltage values were observed.

Concerning the planarity evaluation, two techniques were used compatible 
with the equipment at disposal at INFN-Torino: a Mitutoyo measurement machine
equipped with both a precision CCD camera (magnifier 1:10) and a stylus probe
(3~$\tcmu$m over 1~m precision).
Simply to put down the 4 inch thinned wafers on the measurement base
and using a CCD camera, a total unplanarity of about 40\% was measured by a grid of $8\times9$
measurement positions and a preferential 
cup shape was observed. This result can be explained 
both with the elasticity of the wafer now thinned and a measurement 
imprecision due to the reflection of the light on the mirror like wafer 
surface. 

\begin{figure}[htb]
\vspace{1pt}
\centering
\includegraphics[width=0.48\textwidth]{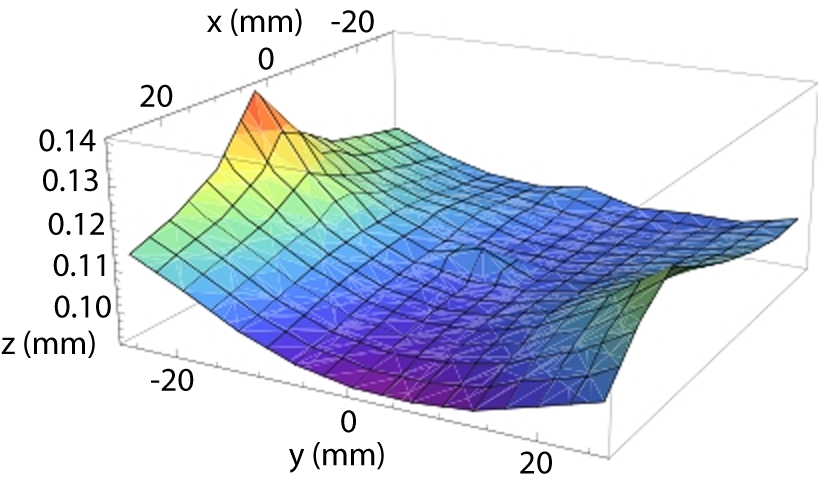}
\caption[Plot showing the unplanarity of a thinned wafer prototype measured with a 
stylus probe]{Plot showing the unplanarity of a thinned wafer prototype measured with a 
stylus probe using a 72 position grid.}
\label{fig:thickness}
\end{figure}

By using the probe the elasticity of the wafer was immediately detected 
by observing the  movement of the wafer under the probe pressure, without any
breakage.
The thickness measurement has been obtained using the same grid of
 72 measurement positions as previously discussed. 
\Figref{fig:thickness} shows the result obtained for a 100~$\tcmu$m 
thick nominal wafer.

		\subsection{Technology Choice Epi vs. Oxygen}

\begin{figure}[htb]
\vspace{1pt}
\centering
\includegraphics[width=0.45\textwidth]{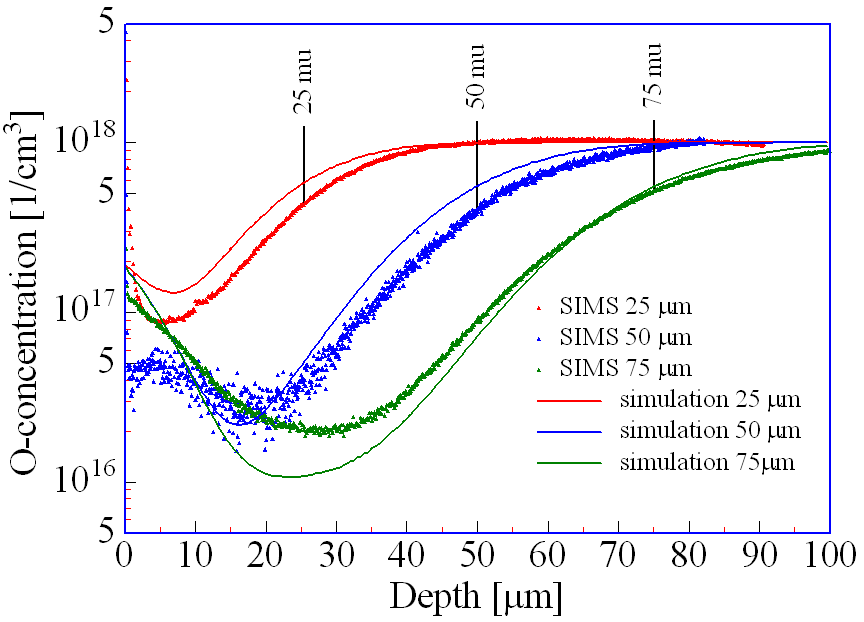}
\caption[Oxygen concentration as a function of wafer depth for different 
epitaxial thicknesses]{Oxygen concentration as a function of wafer depth for different 
epitaxial thicknesses \cite{oxygen0}.}
\label{fig:oxidationbib2}
\end{figure}

The higher radiation tolerance of the epitaxial devices in contrast to 
FZ ones can be explained 
taking into account the oxygen diffusion from the Cz substrate 
\cite{oxygen0}, as shown 
in \figref{fig:oxidationbib2}.

\begin{figure}[htb]
\vspace{1pt}
\centering
\includegraphics[width=0.45\textwidth]{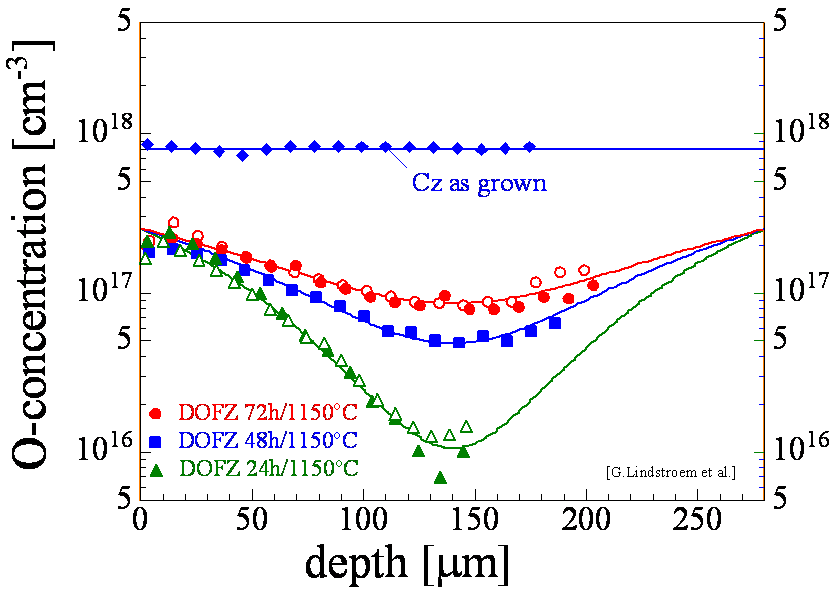}
\caption[Oxygen concentration as a function of FZ wafer depth for
some oxidation processes]{Oxygen concentration as a function of FZ wafer depth for
some oxidation processes \cite{oxygen0}.}
\label{fig:oxidationbib}
\end{figure}

According to the oxidation process used to increase oxygen concentration 
in the FZ wafers \cite{oxygen0} as shown in \figref{fig:oxidationbib}, 
some epitaxial wafers were submitted to an oxygen process at the FBK.
In particular a  1100~$^\circ$C phase for 12 hours with wafers 
in oxygen atmosphere was followed by a phase of 53.6 hours at 1100~$^\circ$C temperature in a nitrogen atmosphere.
Then SIMS measurements of the oxygen concentration as a funtion of depth  
were performed at ITME. 

\begin{figure}[htb]
\vspace{1pt}
\centering
\includegraphics[width=0.43\textwidth]{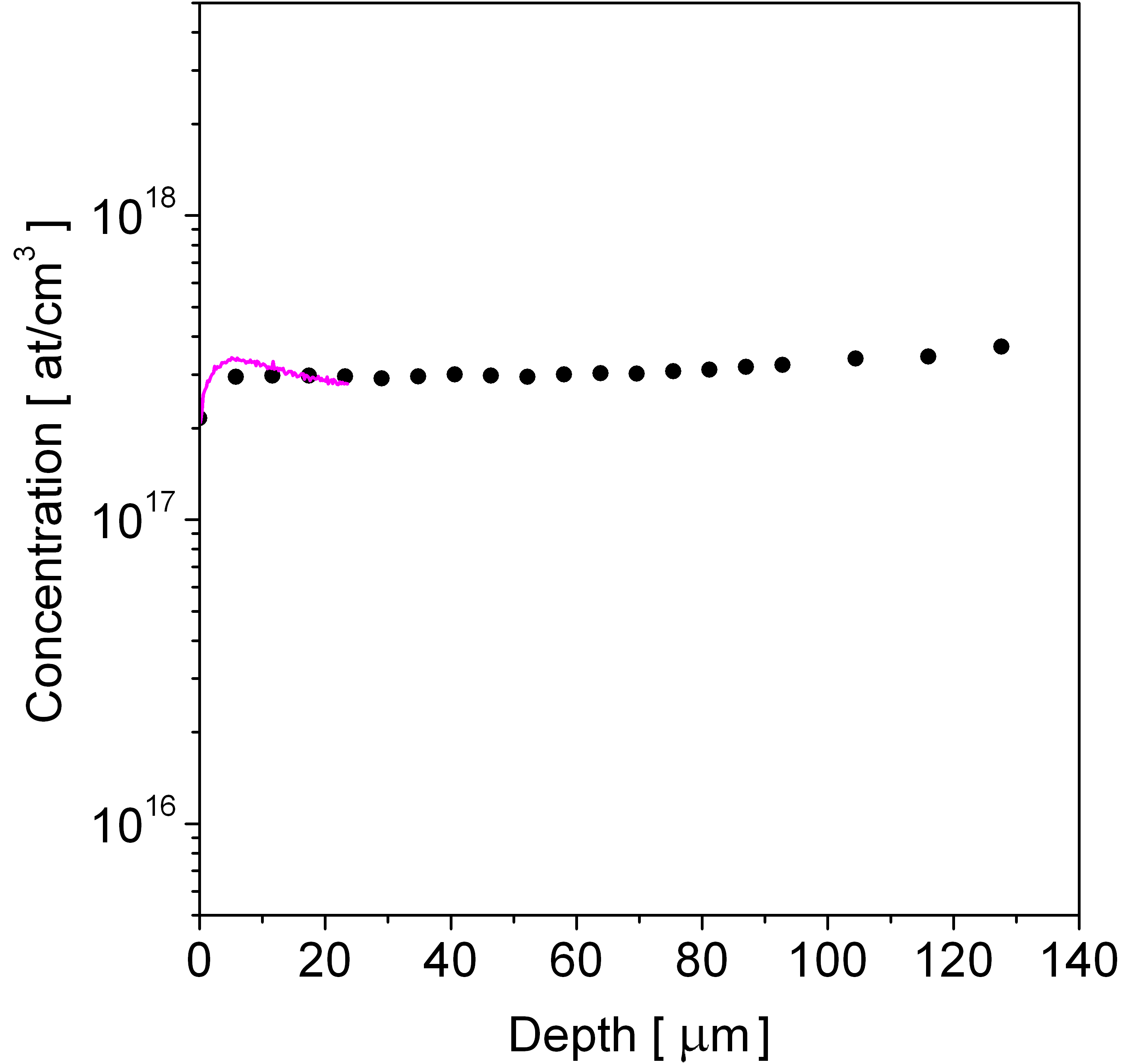}
\caption[Oxygen concentration as a function of epitaxial wafer depth]{Oxygen concentration as a function of epitaxial wafer depth. The scatter graph represents the value of oxygen concentration measured on beveled surface. 
The magenta line shows the concentration of oxygen in shallow depths at sample ion etching.}
\label{fig:oxidation100}
\end{figure}

\Figref{fig:oxidation100} reports the result for an epitaxial wafer      
100~$\tcmu$m thick. The flat trend can be explained with both contributions 
coming from the Cz substrate oxygen diffusion and from  diffusion 
from oxygen atmosphere during the high temperature phase. Such a result 
shows the possibility to have a uniform tolerance to the radiation 
along the sensor bulk.

		\subsection{Production}

\begin{figure}[htb]
\vspace{1pt}
\centering
\includegraphics[width=0.42\textwidth]{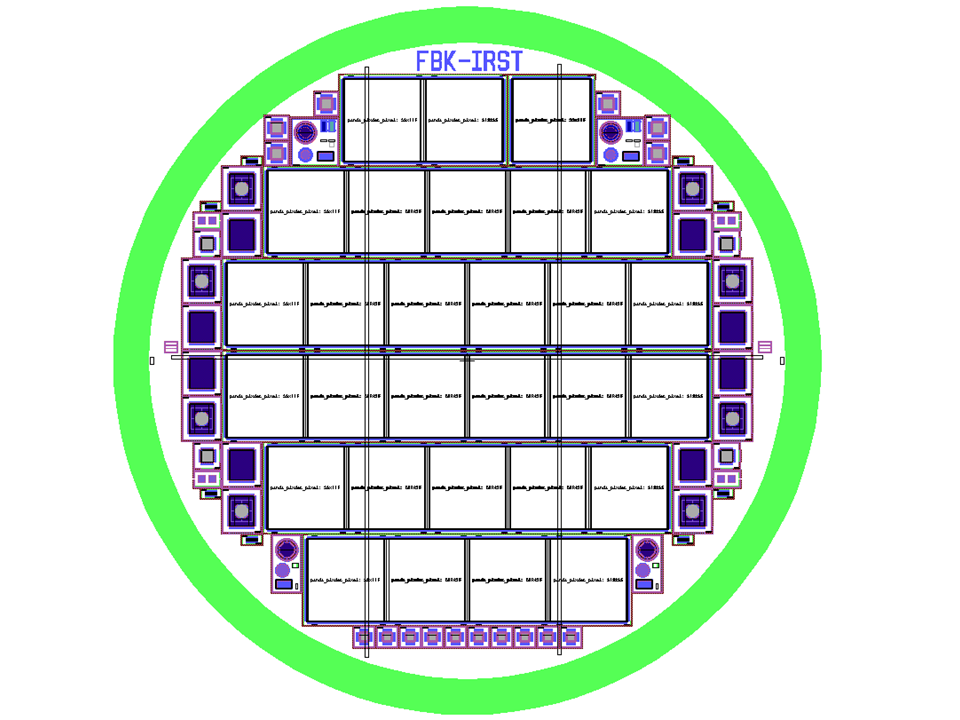}
\caption[Scheme of the wafer with the first design of the full size sensors]{Scheme of the wafer with the first design of the full size sensors.}
\label{fig:waferscheme}
\end{figure}

Taking into account the first design of the full sensors for the pixel 
detector as reported in the drawing of figure  
\ref{fig:waferscheme}
and the needed sensor number for making the whole pixel detector as reported in table
 \ref{tab:sensornumber}
a minimum number of 34 wafers has to be produced.
More details will be explained in the section \ref{module}.

\begin{table}[htb]%[width=0.3\textwidth]
\centering
\begin{tabular}{|l|c|c|c|c|c|c|l}
\hline
  % after \\: \hline or \cline{col1-col2} \cline{col3-col4}
   &  S2 & S4 & S5 & S6 & Total \\ \hline
 Barrel1 & 6 & 8 & 0 & 0 & 14 \\ \hline
 Barrel2 & 0 & 0 & 6 & 44 & 50 \\ \hline
   Disk1 & 6 & 2  & 0 & 0 & 8 \\ \hline
   Disk2 & 6 & 2 & 0 & 0 & 8 \\ \hline
   Disk3 & 4 & 4 & 12 & 4 & 24 \\ \hline
   Disk4 & 4 & 4 & 12 & 4 & 24 \\ \hline
   Disk5 & 4 & 4 & 12 & 4 & 24 \\ \hline
   Disk6 & 4 & 4 & 12 & 4 & 24 \\ \hline
   Total & 34 & 28 & 54 & 60 & 176 \\ \hline
   Chip Number & 68 & 112 & 270 & 360 & 810 \\ \hline
\end{tabular}
\caption[Sensor numbers]{Sensor numbers. S2 is the sensor with two readout chips, S4 with 
four, S5 with 5 and S6 with 6 sensors, respectively.}
\label{tab:sensornumber}
\end{table}

In particular, quality control of each patterned wafer will be performed 
on test structures adjacent to the sensors: C-V, I-V trends on test diodes, 
C-V on MOS structure, I-V trend on a gated diode, breakdown voltage for 
several GR (Guard Ring) configurations, C-V interpixel measurement.
Visual inspection of metals and bonding pads of each sensor can be planned for 
checking for example the etching uniformity.
Doping profiles as a function of epitaxial depth and bias voltage 
are additional measurements useful to compare a wafer before and
after the oxygen process.
Besides, wafer samples from production processes will be used to 
obtain diodes for radiation damage tests to monitor the radiation 
resistance uniformity.

ITME has been selected as the company able to provide the epitaxial wafers 
useful for our project by checking samples during the R\&D. 
FBK has shown the capabilities for sensor production.

In \tabref{tab:pixelproperties} the most probable properties of the 
epitaxial wafer for the production are summarised.

\begin{table}[width=0.4\textwidth]
\centering
\begin{tabular}{|l|c|c|l}
\hline
  % after \\: \hline or \cline{col1-col2} \cline{col3-col4}
 \multicolumn{2}{|l|}{\textbf{Cz substrate}}  \\ \hline
%\textbf{Cz substrate} & \\ \hline
 Diameter & $100\pm0.5$~mm  \\ \hline
 Orientation & $<100>$ \\ \hline
 Conductivity type & $n^+$ \\ \hline
 Dopant & Sb \\ \hline
 Thickness & $525\pm20$~$\tcmu$m \\ \hline
 Resistivity & $0.01-0.02$~$\Omega\cdot\mathrm{cm}$ \\ \hline
 \multicolumn{2}{|l|}{\textbf{Epitaxial Layer}}  \\ \hline
%\textbf{Epitaxial Layer} & \\ \hline
 Conductivity type & n \\ \hline
 Dopant & P \\ \hline
 Thickness(centre) & $100\pm2$~$\tcmu$m \\ \hline
 Resistivity & $1000-2000$~$\Omega\cdot\mathrm{cm}$ \\ \hline
\end{tabular}
\caption[Epitaxial wafer properties for sensor production]{Epitaxial wafer properties for sensor production.}
\label{tab:pixelproperties}
\end{table}

\input{./pixelpart/pixel_electronics.tex}  % text by Gianni Mazza

%	\section{ToPix Architecture} 
				
%	\subsubsection*{Design}

	\section{Hybridisation} 

\authalert {Author: D. Calvo, calvo@to.infn.it}

Both the electrical and mechanical connections between the sensor and the 
readout chip are made by solder balls in a flip--chip solder--bonding process.
This technique is now well developed for the application in the \PANDA 
experiment not only for the daily connection process 
established in the FPGA production but also  for the strong experience acquired
in the LHC experiment realisation compared to the first solder flip--chip 
process developed by IBM in 1969 \cite{ibm01}. 

	    	\subsection{Bump Bonding}

Many companies (IZM, VTT, SELEX) have the capacity to perform the 
bump bonding process. 
Solder balls made of Sn--Pb or In have been used with good results in the 
LHC experiments. Balls sizes up to 20~$\tcmu$m have been deposited on 
pads corresponding on one side to the pixel sensor 
and on the other size to the 
single readout cell. This technique can be regarded as well known. 
Pads on new epitaxial material and circuits made by the new 130~nm CMOS 
technology have been checked. The test of the first one is reported in the 
following subsection and test of the second one has been performed in an other 
experiment \cite{Na62}.    
%A drawback could be  the close position of pads in this pixel
%project. With the single--chip assembly prototype, based on ToPix3, a complete
%study will be possible.
       
      \subsubsection*{First Thinned Prototypes}
      
\begin{figure}[htb]
\vspace{1pt}
\centering
\includegraphics[width=0.45\textwidth]{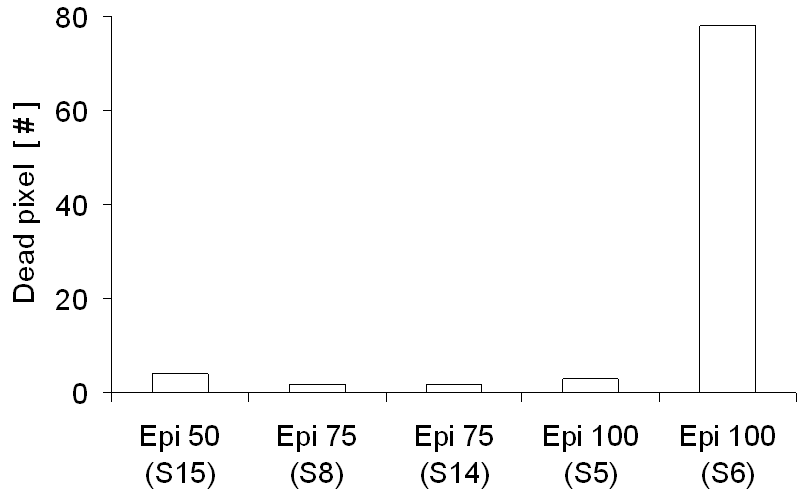}
\caption[Dead pixel counts in the tested assemblies]{Dead pixel counts in the tested assemblies.}
\label{fig:bump}
\end{figure}

\begin{figure}[htb]
\vspace{1pt}
\centering
\includegraphics[width=0.45\textwidth]{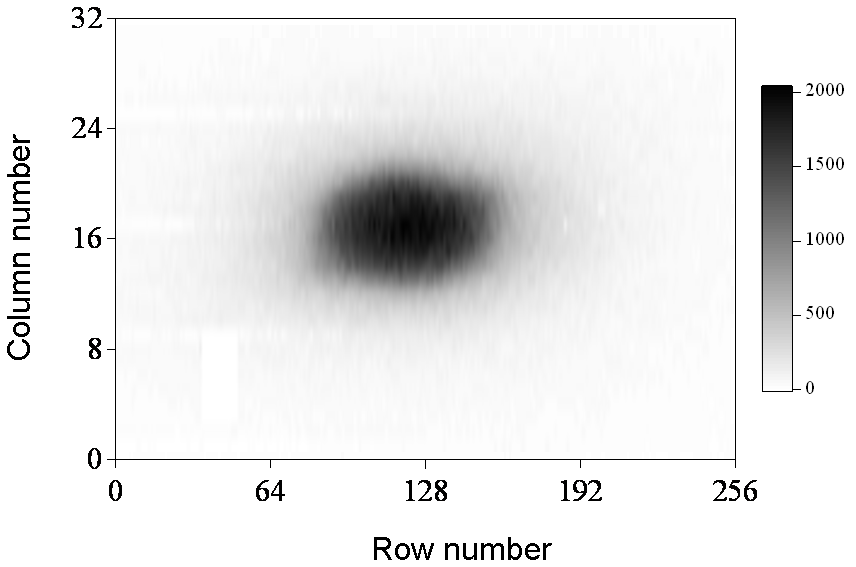}
\caption[Source profile obtained from S8 (Epi-75) assembly 
using a $^{90}$Sr source]{Source profile obtained from S8 (Epi-75) assembly 
using a $^{90}$Sr source, the x-axis corresponding to the pixel 
column number (32 in the ALICE readout) and the y-axis to the row number (256 in the ALICE readout).}
\label{fig:source}
\end{figure}

With the first thinned prototypes described at the beginning of Section 
\ref{Sen}, 
the quality of the bump-bonding process was measured using 
a $^{90}$Sr source in self-triggered mode.
\Figref{fig:bump} shows the number of dead pixels for 
the tested assemblies (8192 pixels per assembly) \cite{epi02}. 
In this case the term ``dead" 
is used to cover several possible reasons for a pixel 
not being usable for signal acquisition, electronics 
channels with zero-gain, high channel noise due to high 
individual pixel leakage current, and failed bump connection. 
Apart from the S6 assembly which had an obvious localised 
problem covering a region of pixels, the connection yield 
and overall efficiency is very high, better than 99$\%$. 
An example of the source profile for the S8 (Epi-75) assembly is shown in 
\figref{fig:source} .

During the realisation of the first thinned sensor, unfortunately, 
all the epi wafers broke in the final stages of the thinning process, 
either during cleaning or during removal from the protective tape 
used to hold the wafers during thinning. One possible explanation 
is stress caused by significant differences in the lattice 
constants of the high-purity epitaxial layer and 
the heavily doped substrate. 
Then development work in collaboration with the VTT company  
was concentrated on 
understanding this problem better with the aim of resolving it.

In particular, a set of 12 patterned wafers, featuring 
50, 75 and 100~$\tcmu$m epitaxial thickness,
with deposited bumps were thinned on the Cz side up to the target thickness of 
200~$\tcmu$m. This final thickness was selected with the aim to evaluate 
a possible contribution to the wafer breakage by the last 
passivation processes on the top of the patterned wafer, optimised for 
200~$\tcmu$m thickness as in the ALICE application.Only two wafers were 
destroyed during the thinning process, giving a yield of 83\%.

Then 9 blank wafers with 100~$\tcmu$m epitaxial layer were thinned in groups of 
three to the final thicknesses of 130, 120 and 110~$\tcmu$m 
by removing most of the Cz substrate. 
The first two wafers broke during the CMP process used to remove the 
defect layers immediately after the backgrinding. The etching technique 
was planned at this time and no other breakage was detected with this 
last technique.

%      \subsection{FE Wafer Thinning}

	\section{Single Chip Assembly Prototype}

\authalert {Author: D. Calvo, calvo@to.infn.it}

The first prototype of single chip assembly made of an ASIC developed 
in 130~nm CMOS technology, ToPix~v3, and thinned epitaxial silicon sensor
will be produced as the final step of the R\&D phase for the hybrid 
pixel detector development. 

% no other subsection, yet - it wouldn't make sense
%			\subsection{Description of the Prototype}
	
\begin{figure}[htb]
\vspace{1pt}
\centering
\includegraphics[width=0.45\textwidth]{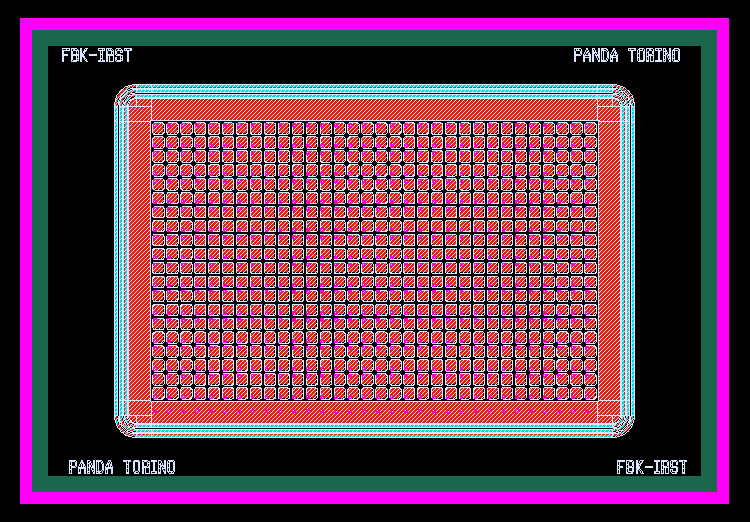}
\caption[Pixel matrix layout]{Pixel matrix layout.}
\label{fig:s100}
\end{figure}

The assembly is based on ToPix3, already described in the readout 
prototype section. The reduced scale of the pixel sensor matrix asks for
a dedicated sensor layout.
\Figref{fig:s100} shows the pixel matrix layout made of 640 $\mathrm{100~\tcmu m \times 100~\tcmu m}$
pixels to overlap the readout cell matrix of ToPix~v3.  
On the perimeter the guard ring sequence, the first one connected to ground 
and the others in a floating configuration, is visible.

A second layout has been foreseen taking into account the larger pixels needed 
across the sensor region where two readout chips will be arranged side by side 
in the final project.
This last prototype is made of 600 $\mathrm{100~\tcmu m \times 100~\tcmu m}$
 pixels and 40 
$\mathrm{100~\tcmu m \times 300~\tcmu m}$
pixels.

\begin{figure}[htb]
\vspace{1pt}
\centering
\includegraphics[width=0.45\textwidth]{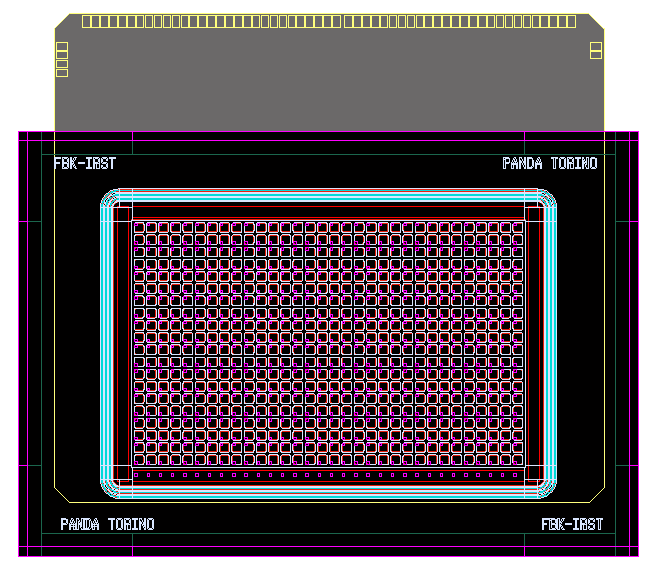}
\caption[Scheme of the assembly]{Scheme of the assembly. The sensor (in black) is placed above the chip (in gray).}
\label{fig:stop}
\end{figure}

\Figref{fig:stop} shows the schematic assembly between ToPix~v3 and the
sensor.

\begin{figure}[htb]
\vspace{1pt}
\centering
\includegraphics[width=0.45\textwidth]{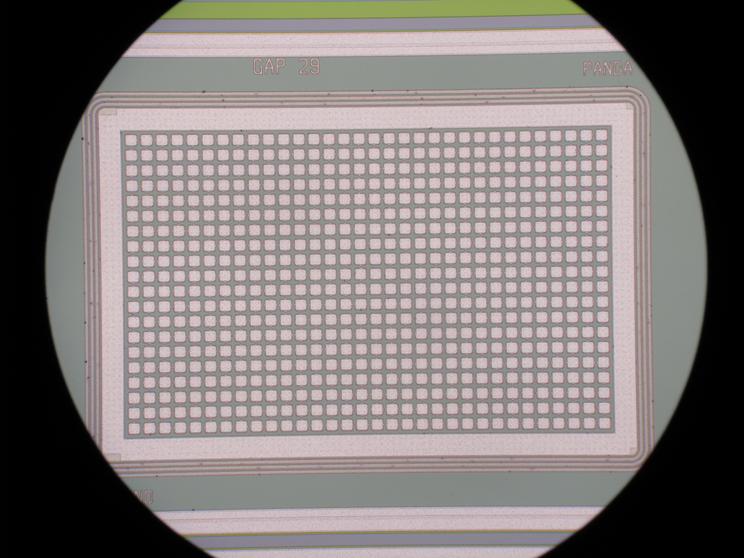}
\caption[Photograph of epitaxial sensor produced at FBK]{Photograph of epitaxial sensor produced at FBK. Pixel matrix matching ToPix~v3.}
\label{fig:sensortopix3}
\end{figure}

The pixel sensors have been arranged in 4~inch epitaxial wafers and produced 
at FBK. 
In particular MR and LR wafers as described in the Radiation Damage section
have been used. Besides a  wafer of each type was oxygen enriched.
\Figref{fig:sensortopix3} shows a sensor view by means of a microscope.

\section{Module}
\label{module}
\authalert {Author: D. Calvo, calvo@to.infn.it}

The philosophy behind the design process of the pixel modules is 
to first optimise forward coverage, then adjust barrel design as a result. 
Only one variable is to take into account to perform this study, the chip size.
Nevertheless a small production of silicon sensors as foreseen must fit 
on 4~inch wafers.

A hybrid pixel module is composed of several readout chips bump bonded to 
a unique silicon sensor, on the top of the sensor a multilayer bus 
routing both power supply lines and signal lines is foreseen  
equipped with minimal SMD components. Bus and readout chips are 
connected by wire bonding technique. Potentially a chip controller 
of this module could be arranged on the top of the bus to drive Gbit links 
and serving two or three readout chips. 

\begin{figure}[htb]
\vspace{1pt}
\centering
\includegraphics[width=0.45\textwidth]{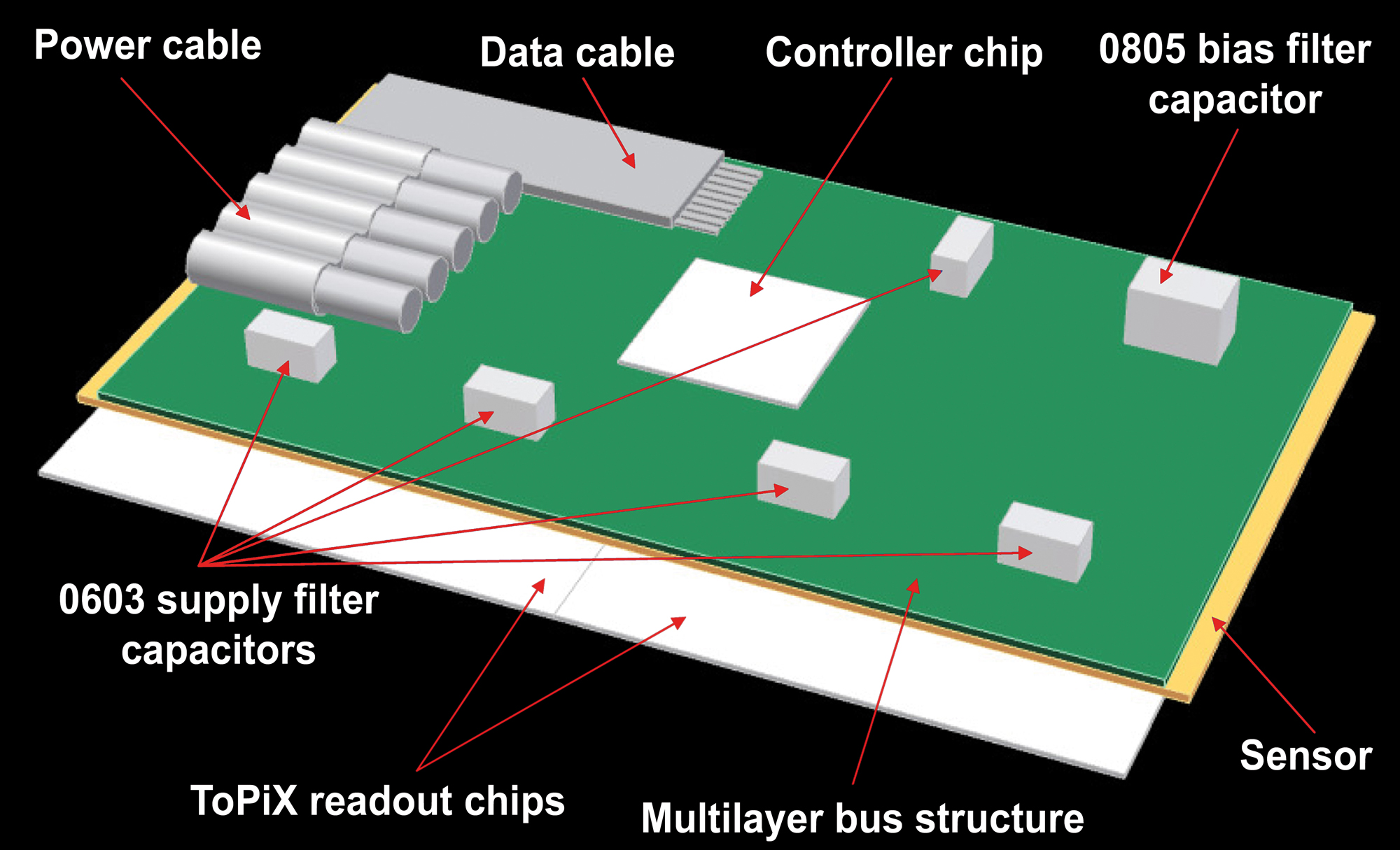}
\caption[S2 module: two readout chips are bump bonded to a single sensor]{S2 module: two readout chips are bump bonded to a single sensor.}
\label{fig:S2module}
\end{figure}

Starting from the disks, the most useful sensor layout appears to be an 
alternating front and back configuration with respect to the carbon 
foam layer used to increase the heat transport to the cooling pipes 
embedded in the middle of this layer and acting as a mechanical support, too.
In particular, a pixel active area of $\mathrm{11.6~mm \times 11.4~mm}$
per chip is foreseen 
and 2, 4, 5, 6 readout chips sensors (S2, S4, S5, S6 respectively) 
are planned, with vertical gaps bridged with longer pixels as 
already explained in \ref{Sen}.
The nominal readout chip size is $\mathrm{14.8~mm \times 11.4~mm}$
or slightly less 
to include dicing tolerances.
\Figref{fig:S2module}  shows an S2 module scheme.

\begin{figure}[htb]
\vspace{1pt}
\centering
\includegraphics[width=0.3\textwidth]{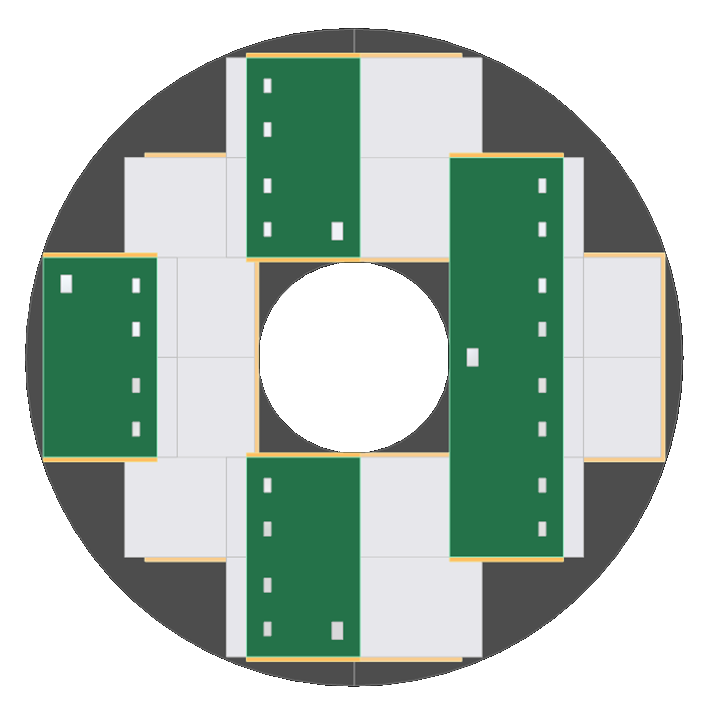}
\caption[Module configuration of a small disk]{Module configuration of a small disk. The radius is 36.56~mm.}
\label{fig:smalldisk}
\end{figure}

The small disks are equipped with six S2 modules and two S4 modules, 
with horizontal gap, and as starting point a sensor overlap 
due to the dead space driven by the conventional guard ring design. 
A geometrical total coverage of 65\% is reached, but a systematic small 
shift will be studied to investigate the spatial distribution 
of the emitted particle tracks, to maximise the reconstructed hits.
\Figref{fig:smalldisk} shows the module configuration of a small disk.

\begin{figure}[htb]
\vspace{1pt}
\centering
\includegraphics[width=0.3\textwidth]{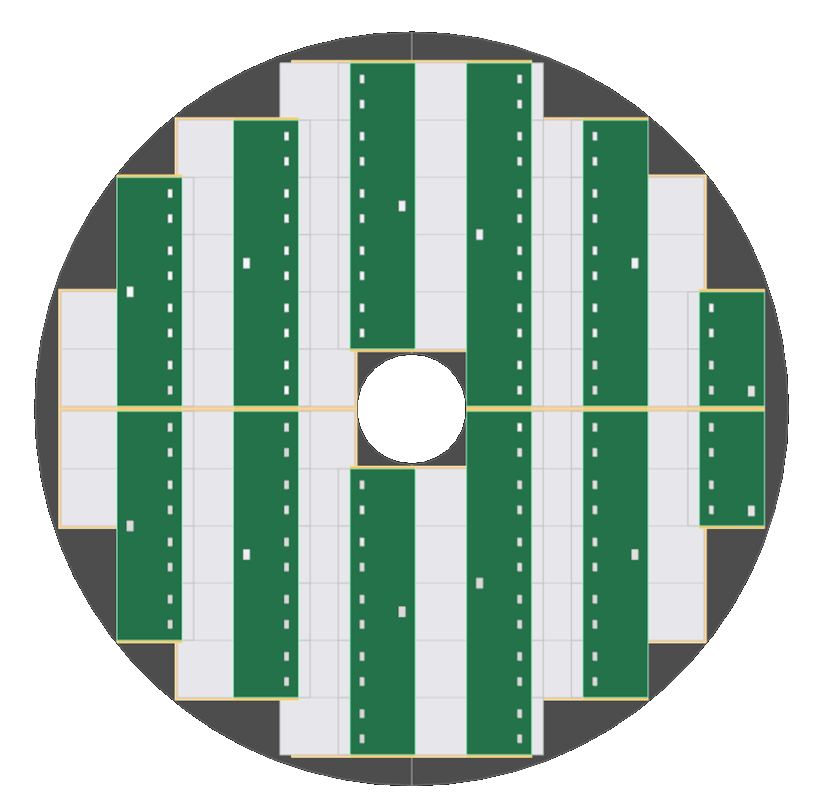}
\caption[Module configuration of a big disk]{Module configuration of a big disk. The radius is 73.96~mm.}
\label{fig:bigdisk}
\end{figure}

The big disks will be equipped with four S2, four S4, twelve S5  
and four S6 modules.
In this case a small dead zone (of about 1~mm) will result 
where two sensors meet in the horizontal plane. 
\Figref{fig:bigdisk} shows the module configuration of a big disk.

In general, keeping cables out of the active region will mean that 
some modules may require two designs according to which end 
the cables have to be connected.

\begin{figure}[htb]
\vspace{1pt}
\centering
\includegraphics[width=0.45\textwidth]{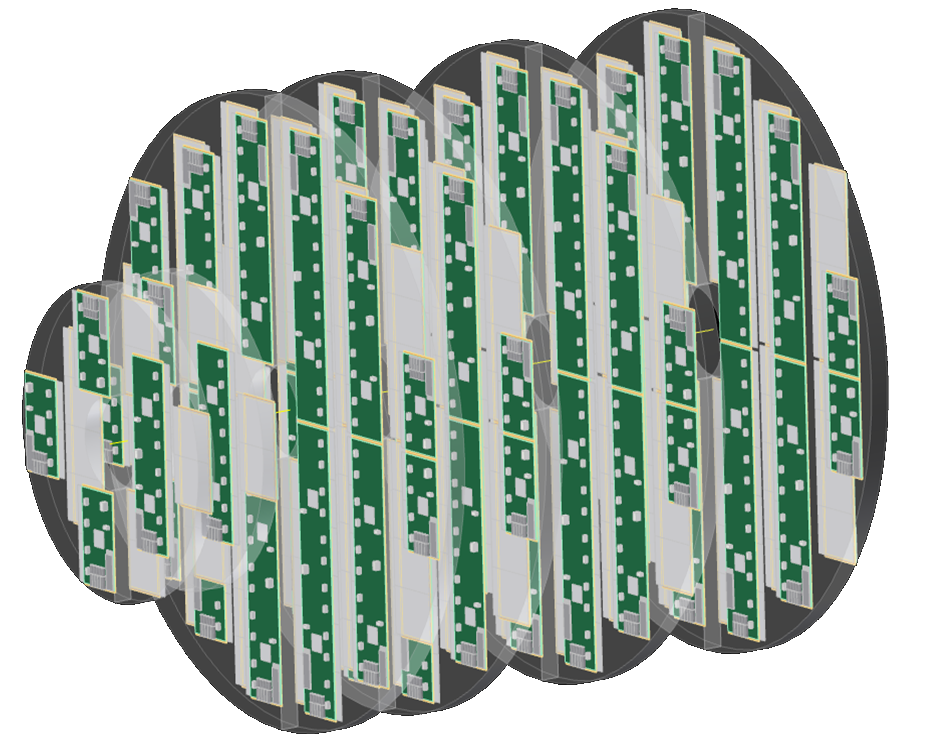}
\caption[The complete forward pixel assembly scheme]{The complete forward pixel assembly scheme.}
\label{fig:diskassembly}
\end{figure}

In \figref{fig:diskassembly} the complete forward assembly is shown.

The pixel barrels are made of two layers of staves with modules 
featuring  differently sized sensors to leave out the target pipe. 
The first layer houses 8 staves made of S4 modules and 
six staves based on S2 module each.
The second layer is assembled  from six staves made of 
S5 modules each and 22 staves housing two S6 modules each.

	\section{Bus-Flex Hybrid }	
	
\authalert {Author: D. Calvo, calvo@to.infn.it}

The bus on the top of the sensor in a hybrid pixel module routes 
electrical signals and power lines. 
For minimising material and for handling high data transmission, 
different approaches, in particular for the electrical signals, 
have been followed studying this part of the project
taking into account the two different solutions with and without a module 
controller serving two or three readout chips.

A first approach is based on a custom solution where at least two 
aluminium metal layers are interconnected with vias.
The major parameters are: aluminium layer thickness of about 
5~$\tcmu$m, a minimum track pitch $\leq$~100~$\tcmu$m, a minimum 
via diameter  $\leq$~25~$\tcmu$m, lowest possible dielectric constant 
to minimise thickness with the impedance control, the top layer compatible 
with wire bonding and SMD mounting, at least the overall structure size 
has to be compatible with a 4"~Si~wafer.

\begin{figure}[htb]
\vspace{1pt}
\centering
\includegraphics[width=0.5\textwidth]{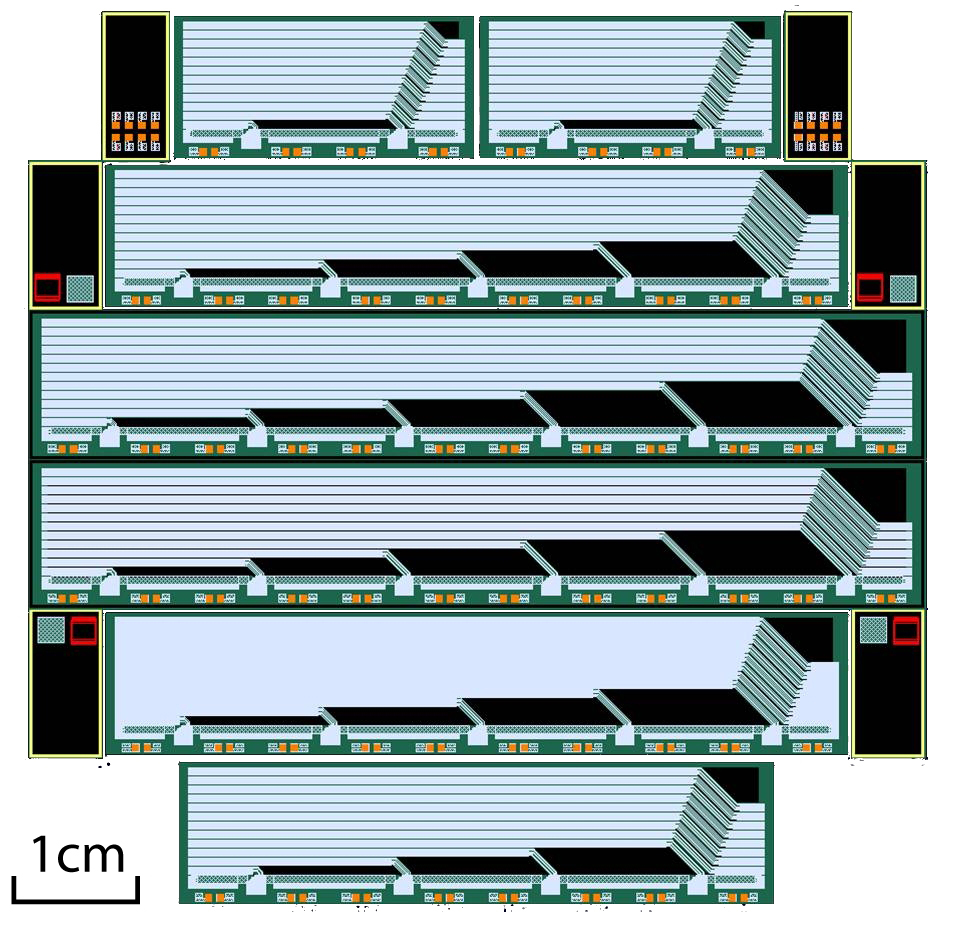}
\caption[Busses for different pixel modules]{Busses for different pixel modules.}
\label{fig:morebus}
\end{figure}

In \figref{fig:morebus} 
different size buses useful 
for different pixel 
modules are arranged 
inside an area suitable for a 4"~Si~wafer. 

The bus layouts are designed 
for the case  the module controller is not foreseen and a 
high density of tracks has to be handled and routed on 
the bus first layer  up to a aluminium strip cable 
(see the Cable section in the Infrastructure chapter, \ref{cable-sec}). 
The second layer avoids the track 
crossing, e.g.~of the clock line in case a single line is used for
the whole module.

\begin{figure}[htb]
\vspace{1pt}
\centering
\includegraphics[width=0.49\textwidth]{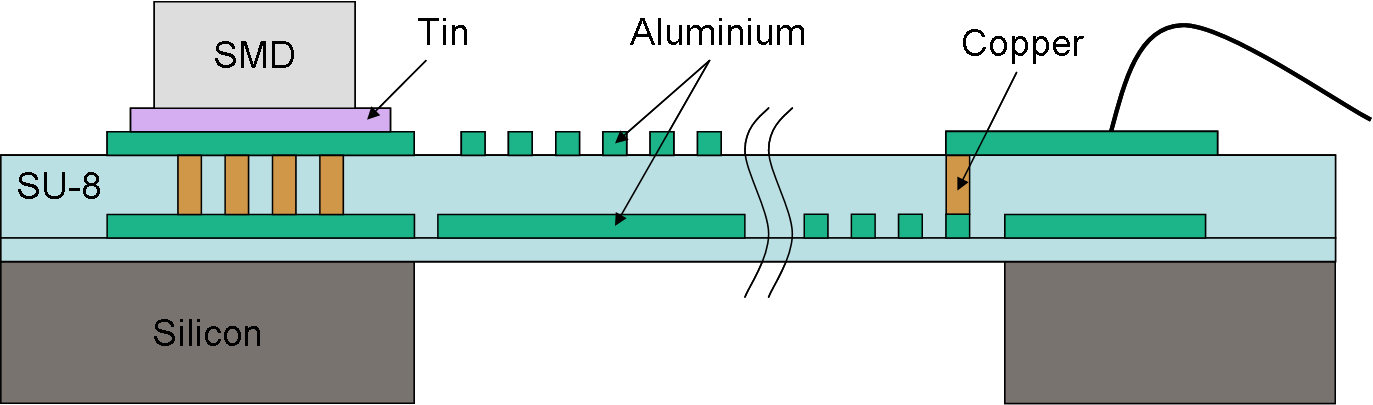}
\caption[Scheme of the SU8 technique]{Scheme of the SU8 technique.}
\label{fig:su8scheme}
\end{figure}

The feasibility of such a solution was investigated 
with a first company (Techfab, Italy).
The base idea was to build the bus structures on silicon wafer as support and
then take advantage of the Si--detector processes and quality 
lithography. SU--8 was identified as the most interesting dielectric 
for its relative dielectric constant (about 3). 
This lower value allows a dielectric thickness of about 15~$\tcmu$m between Al layers.  
The electroless copper deposition was proposed as an useful process for filling the vias 
(\figref{fig:su8scheme}).

\begin{figure}[htb]
\vspace{1pt}
\centering
\includegraphics[width=0.40\textwidth]{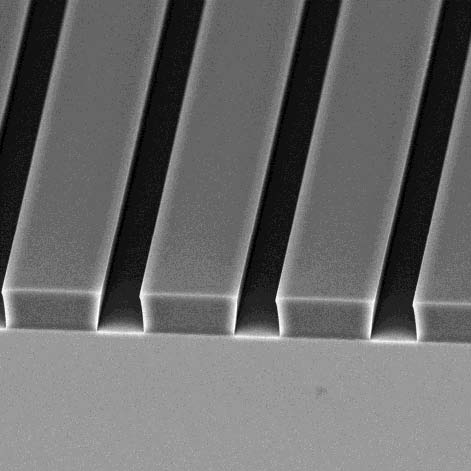}
\caption[SU8 strips]{SU8 strips.}
\label{fig:su8strips}
\end{figure}

\begin{figure}[htb]
\vspace{1pt}
\centering
\includegraphics[width=0.45\textwidth]{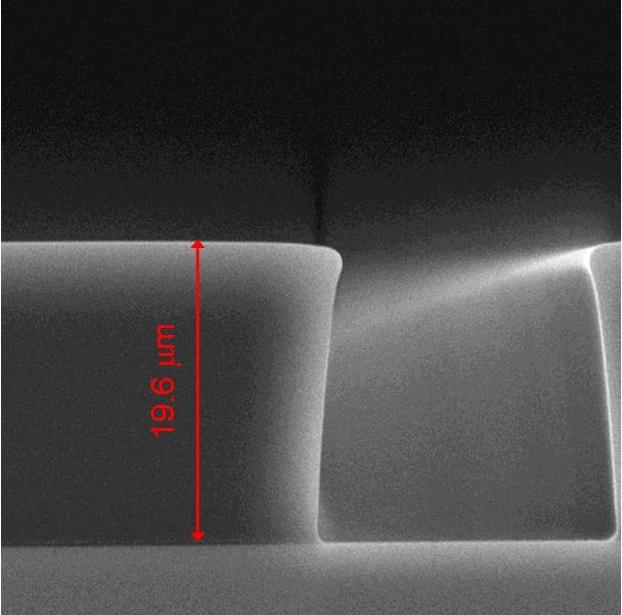}
\caption[Slope of an SU8 strip]{Slope of an SU8 strip.}
\label{fig:su8slope}
\end{figure}

Techfab went out of business after the very first process tests so 
that the project will now be pursued in cooperation with FBK, Trento.
First test of SU--8  3010 was performed by using 
a strip-shaped mask. A layer that was only 19.6~$\tcmu$m thin has been 
obtained as shown in the photograph of 
\figref{fig:su8strips}, 
where the wall of the strip shows a reverse angle 
(\figref{fig:su8slope}) not perfectly suited for aluminium deposition.

Besides an electroless technique with nickel 
appears to be better suited for the cleanliness of the manufacturing equipments.  
 
\begin{figure}[htb]
\vspace{1pt}
\centering
\includegraphics[width=0.5\textwidth]{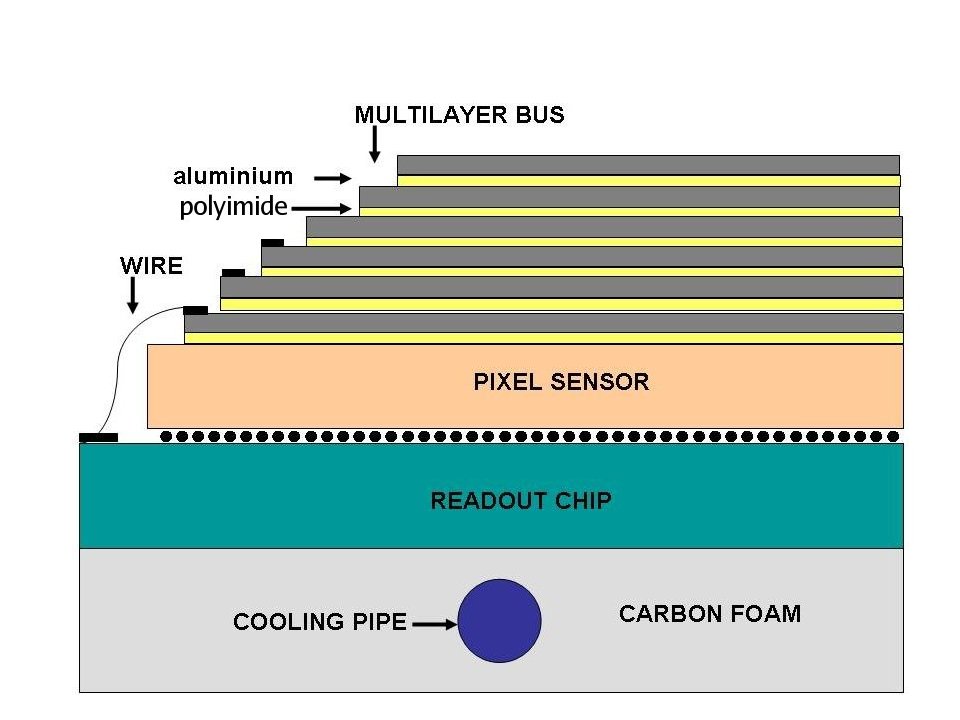}
\caption[Staircase multilayer]{Staircase multilayer.}
\label{fig:stairbus}
\end{figure}

In parallel to this new approach, a base solution has been 
considered using  techniques developed at CERN for the Alice 
experiment.
In this case a multilayer bus with a staircase shape
(\figref{fig:stairbus}) houses 
deposited aluminium tracks featuring 80--90~$\tcmu$m width 
and 150~$\tcmu$m pitch on 
each layer and microvias between two layers have been made.

Alternatively to 10~$\tcmu$m aluminium deposition, 15~$\tcmu$m aluminium foil 
glued to polyimide has been considered. Different etching 
processes have been studied for both solutions \cite{bus03}.

\begin{figure*}[htb]
\vspace{1pt}
\centering
\includegraphics[width=0.65\textwidth]{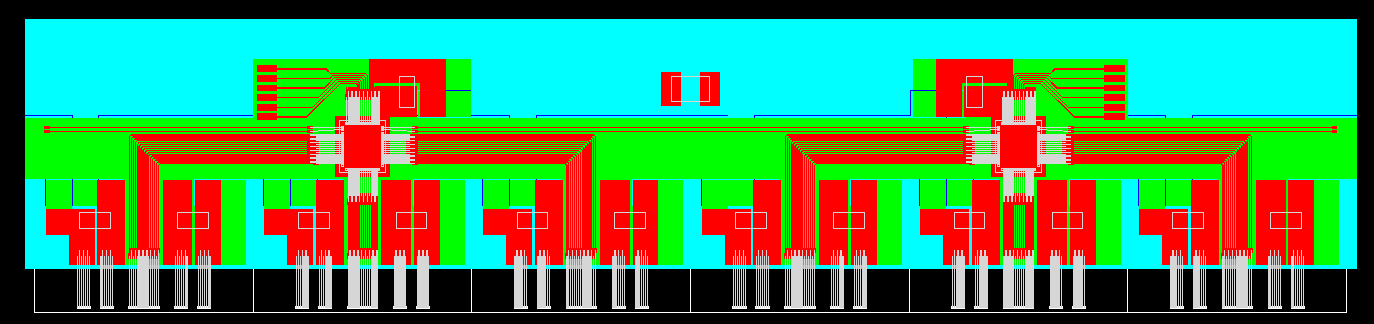}
\caption[Bus with module controllers]{Bus with module controllers.}
\label{fig:modulebus}
\end{figure*}

As an example, in \figref{fig:modulebus} a layout of a bus, 
foreseeing a module controller for the readout chips, is shown.

The power supply routing bus can be made of aluminium strips glued to
polyimide, using the same technique that will be explained  
in the Infrastructure chapter,
Cable section (\ref{cable-sec}) for data transmission. 
Analogue and digital power supply 
lines planned for the readout chips and the 
bias voltage for the sensor with  corresponding ground 
planes will be the first layers of the staircase bus 
on the top of the sensor.

\putbib[lit_pixel]

%% file: pixelpart/pixel_electronics.tex
\section{Front-End Electronics}

\authalert{Author: G. Mazza, mazza@to.infn.it}

\label{Section-PixelReadout}
	\subsection{Requirements}

The pixel readout ASIC has to provide for each hit a simultaneous position, 
time, and energy loss measurement. The specifications for the readout 
electronics depend on the requirements of the MVD. The fundamental parameters 
to take into account are the \PANDA radiation environment and the close proximity 
of the MVD to the interaction point.\\
A pixel size of $\mathrm{100~\tcmu m \times 100~\tcmu m}$ has been chosen based on 
the study of track performance. Since each readout cell has to be bump bonded to 
the sensor, the readout cell matrix has to respect the same geometry as the pixel 
sensor matrix. Therefore the readout cell has to fit in an area of 
$\mathrm{100~\tcmu m \times 100~\tcmu m}$. 
To guarantee the reliability of the system, time and charge 
digitisation has to be performed at the pixel cell level in 
order to transmit only digital information 
out of the pixel cell. For a Minimum Ionising Particle (MIP) 
in a 100~$\tcmu$m thick epitaxial sensor an average ionisation charge of 
$\sim$ 1.3~fC  is expected. The upper limit of the dynamic range is given 
by the measurement of protons with a momentum down to 
$\sim$~200~\mevc, which can deposit ionisation charges up to 50~fC. 
A power density of up to $\mathrm{W/cm^2}$ can be tolerated. 
However, a strong effort is going into the reduction of the power consumption 
of the \frontend electronics and hence the material budget for the cooling
system.\\
It is foreseen to employ a p-in-n epitaxial sensor, which should 
give an adequate radiation hardness with a simple geometry. However, 
the possibility of using the already well known n-in-n sensor employed in 
LHC is considered as a backup solution. Therefore, the \frontend 
has to be designed to be compatible with sensors of either polarity. 
The noise level is mainly limited by the input transistor thermal 
noise, and therefore
is a function of its bias current. An Equivalent Noise Charge (ENC) 
of $\sim$ 200 electrons is a good trade-off between power consumption 
and minimum detectable charge. The readout electronics has to work 
in the \PANDA radiation environment, with an accumulated Total Ionising 
Dose (TID) of 100~kGy in 10 years. Hit rate simulation studies estimate 
an average rate of maximal $\sim 10^3$ events per second on a single pixel 
cell. The system will be self-triggered and all the events above a given 
threshold have to be read out, leading to a high data rate. A time 
resolution lower than the average event rate of $2 \times 10^7$ events/s is 
required. \Tabref{tab:ToPiX_specs} summarises the specifications for 
the ASIC design.

\begin{table}[h]
\begin{center}
\begin{tabular}{|l|c|}
\hline
Pixel size & $\mathrm{100~\tcmu m \times 100~\tcmu m}$\\
\hline
\multicolumn{2}{|c|}{Self trigger capability}\\
\hline
Chip active area & $\mathrm{11.4~mm \times 11.6~mm}$\\
~ & 116 rows, 110 columns\\
\hline
Chip size & $\mathrm{11.6~mm \times 14.8~mm}$\\
\hline
 $\text{d}E/\text{d}x$ measurement & ToT\\
Energy loss & 12 bits resolution\\
measurement  & 7 bits linearity\\
\hline
Input range & up to 50~fC\\
Noise floor & $\le 0.032$~fC\\
\hline
Clock frequency & 155.52~MHz\\
\hline
Time resolution & 6.45~ns (1.9~ns r.m.s.) \\
\hline
Power budget & $<$ 800~mW/cm$^2$\\
\hline
Max event rate/cm$^2$ & $6.1\times 10^6$~Hits/s\\
Max data rate/chip & $\sim250$~ Mbit/s\\
\hline
Total ionising dose & $\le 100$~kGy\\
\hline
\end{tabular}
\end{center}
\caption[Specification summary for the \frontend ASIC]{Specification summary for the \frontend ASIC. ToT: Time over Threshold.}
\label{tab:ToPiX_specs}
\end{table}

	\subsection{Readout Architecture}

\begin{figure*}[h]
\centering
\includegraphics[width=0.7\textwidth]{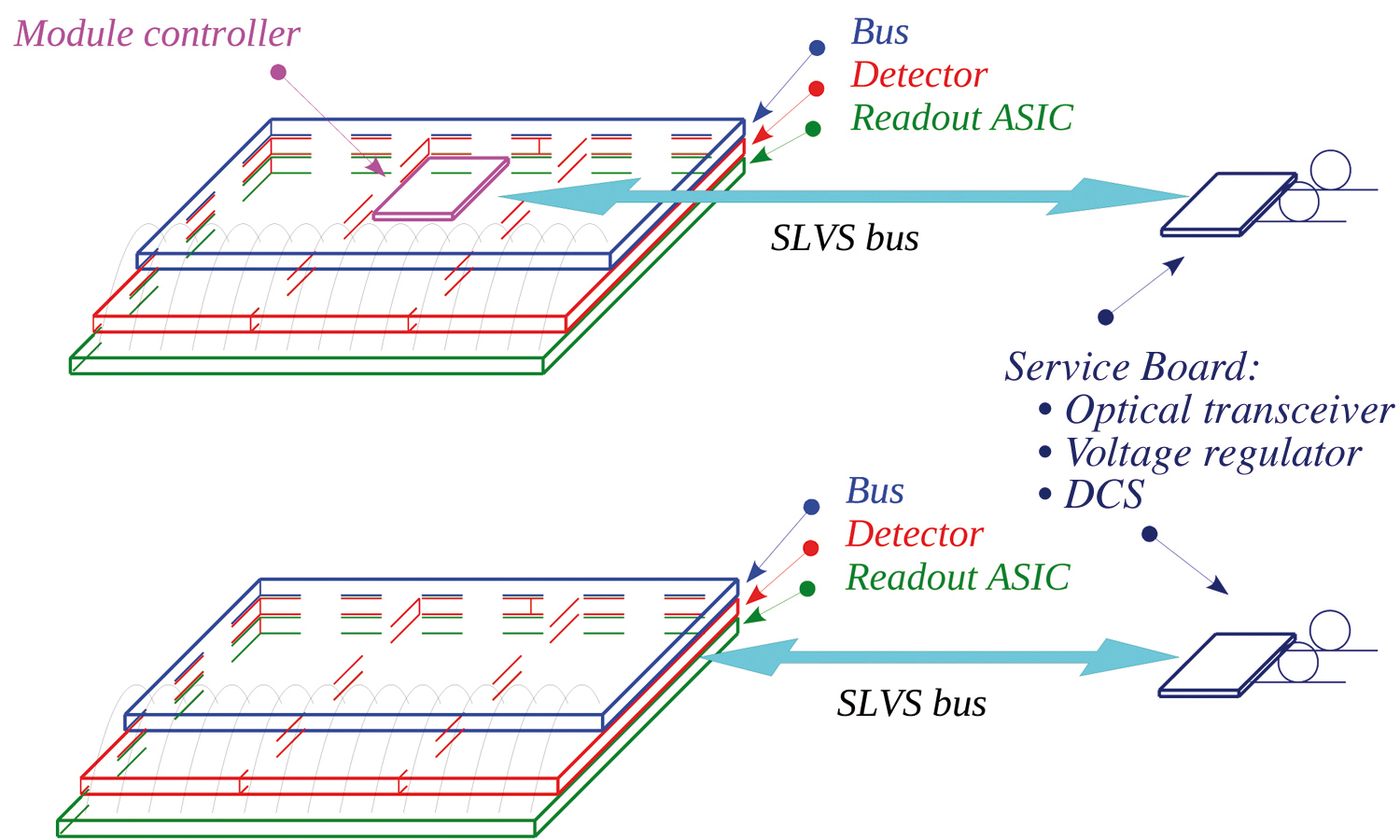}
\caption[Pixel readout scheme]{Pixel readout scheme.}
\label{f:readout_scheme}
\end{figure*}	

The readout architecture, depicted in \figref{f:readout_scheme}, 
foresees a module structure formed by a three-layer structure: the
sensor, the \frontend chip and the transmission bus. The sensor is 
bump-bonded to the readout chips which, in turn are wire-bonded to the 
transmission bus.\\
The module is connected to a service board, located outside the sensitive
area, which performs the data multiplexing and the electrical-to-optical
conversion to decrease the amount of cabling. The link is bi-directional
and allows to send the clock and to upload configuration parameters to 
the \frontend. A second service board hosts the voltage regulators. 
Due to the low power supply voltage of the \frontend (1.2~V), a 
DC-DC down-conversion will be implemented at this level.

As shown in \figref{f:readout_scheme}, two possible solutions are 
under evaluation: in the first one, a module controller chip is used 
on top of the transmission bus to receive, multiplex and re-transmit 
the data from the \frontend chips over high speed serial links. In the second 
solution the module is directly connected to the service board. The advantage 
of the first solution is the reduction of the number of wires coming out the 
module; moreover, by multiplexing the data from various \frontend chips, a 
better bandwidth allocation is possible. On the other hand, when compared to 
the second solution, it requires an extra ASIC which increases the power 
consumption and becomes a critical point of failure.\\

	\subsection{Front-End ASIC}

A new \frontend ASIC, named ToPix, is under development in order to
cope with the unique requirements of the \PANDA experiment.
The circuit is developed in a commercial CMOS 0.13~$\tcmu$m technology, in
order to profit from the high level of integration and the intrinsic radiation
tolerance of deep submicron technologies. Indeed, it has been shown 
\cite{Faccio05} that deep  submicron technologies with very thin gate oxide and 
Shallow Trench Insulation (STI) have the potential to stand high levels of 
total ionising dose with standard layout techniques. In order to cope with the 
Single Event Upset (SEU) effect, all flip flops and latches in the pixel 
cell are fully static and are based either on Hamming encoding or triple 
redundancy.\\

		\subsubsection*{Pixel Cell}
\label{section:front_end_electronics:pixel_cell}
In the proposed design, the Time over Threshold (ToT) technique has been adopted
for the amplitude measurement. The ToT allows to achieve good 
linearity and excellent resolution even when the preamplifier is 
saturated, thus making room for a high dynamic range. Owing to 
the saturation of the input stage, a cross talk of the order of few percent is expected
for large input charges. The ToT is obtained by 
taking advantage of
the time stamp which is already distributed to all pixels in order 
to obtain the time information.

\begin{figure*}
\begin{center}
\includegraphics[width=0.7\textwidth]{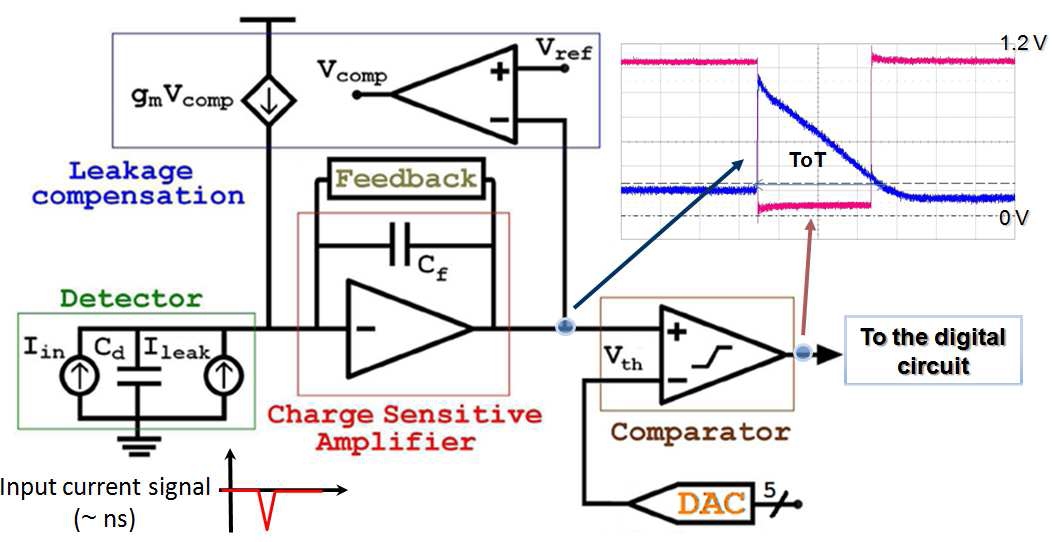}
\end{center}
\caption[Analog readout channel]{Analog readout channel.}
\label{f:px_cell_analog}
\end{figure*}

\begin{figure*}
\begin{center}
\includegraphics[width=0.9\textwidth]{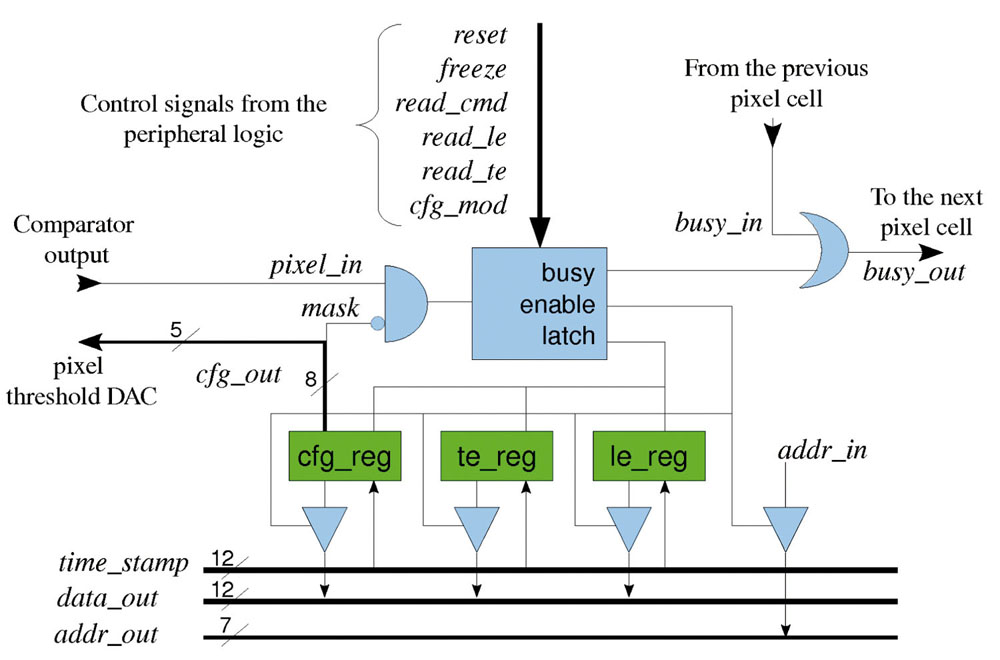}
\end{center}
\caption[Pixel control logic]{Pixel control logic.}
\label{f:px_cell_digital}
\end{figure*}

\Figsref{f:px_cell_analog} and \ref{f:px_cell_digital} show the schematic 
of the analogue and digital parts of the pixel cell, respectively. 
The charge preamplifier is a gain-enhanced direct cascode amplifier with a 
feedback capacitor and a constant current discharge circuit. A low frequency 
differential amplifier in feedback with the input stage allows for 
leakage current compensation by forcing the stage DC output value to 
a reference voltage. The input amplifier can be configured to accept 
signals of both polarities via control bits. A calibration signal can 
be applied to the input of each preamplifier via a 24 fF integrated 
capacitor. Each pixel can be independently enabled to receive the 
calibration signal via the configuration register. The comparator is 
based on a folded cascode architecture. The threshold voltage is set 
by global DAC and can be fine adjusted via a 4 bits local DAC.\\
When the preamplifier output crosses a threshold, the comparator switches 
to 1 and the time stamp value is loaded into the leading edge register. The 
integrating capacitor is then slowly discharged and when the amplifier 
output goes 
below the threshold, the comparator switches back to 0 and the time 
stamp value 
is loaded into the trailing edge register. Therefore the time stamp 
value in the leading edge register gives the timing information, 
while the difference between leading and trailing edge time stamps 
gives the amplitude information.\\

\begin{figure*}[h]
\includegraphics[width=0.95\textwidth]{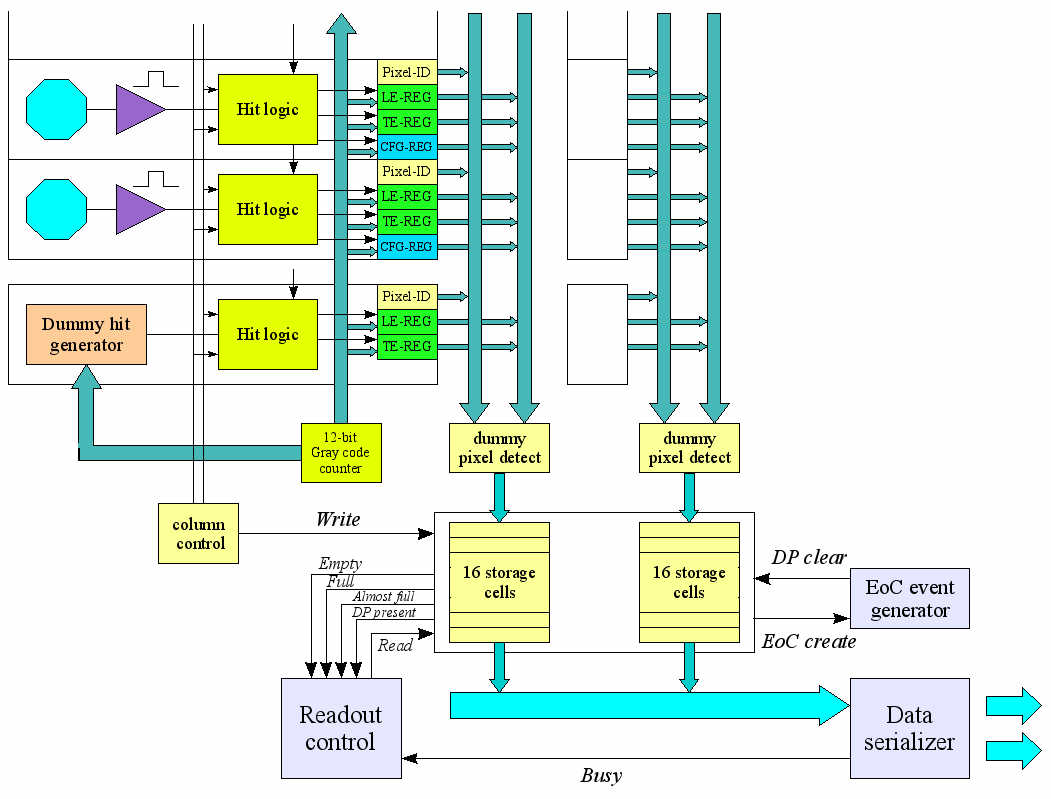}
\caption[Column readout schematic]{Column readout schematic.}
\label{f:column_readout}
\end{figure*} 

		\subsubsection*{ToPix Architecture}
		\label{ToPiX_architecture}

The pixel cells are organised in double columns which share the same 
time stamp and readout buses. Each double column is composed of 
$2\times116$ cells, a 12 bit time stamp bus, a 7 bit address bus 
and a 12 bit data bus. Each cell in the column has
a unique cell address which is hardwired.
The readout buses are differential. The time stamp bus uses a reduced 
swing pseudo differential logic with pre-emphasis to limit the power 
consumption without degrading the speed performance. The time information 
is sent on the bus with a Gray encoding to reduce the switching noise and 
to avoid synchronisation problems with the asynchronous pixel logic. 
The address and data readout buses are CMOS pseudo differential lines and 
are read out at the end of the column via sense amplifier in order to keep 
the cell output drivers to a manageable size and power consumption.\\
The column readout scheme is based  on a concept similar to the one used 
in other pixel readout designs \cite{Peric06}. When a pixel cell receives 
a hit it generates a busy signal which is  put in logical OR with the busy 
of the previous cell, and then sent to the next cell. Therefore at the end 
of the column the busy signal is high if at least one pixel contains data.\\
The readout is made in a fixed priority scheme: the read commands are 
sent to all pixels, but only the one with its busy input signal at zero 
(i.e.~with no pixel at higher priority with data) sends its data and address 
to the output bus. A freeze signal is used to block 
the generation of new busy signals (but not the possibility to 
store an event into the pixel) 
during readout in order to make sure that all the pixels are addressed 
in a readout cycle.\\
\begin{figure*}[h]
\begin{center}
\includegraphics[width=0.7\textwidth]{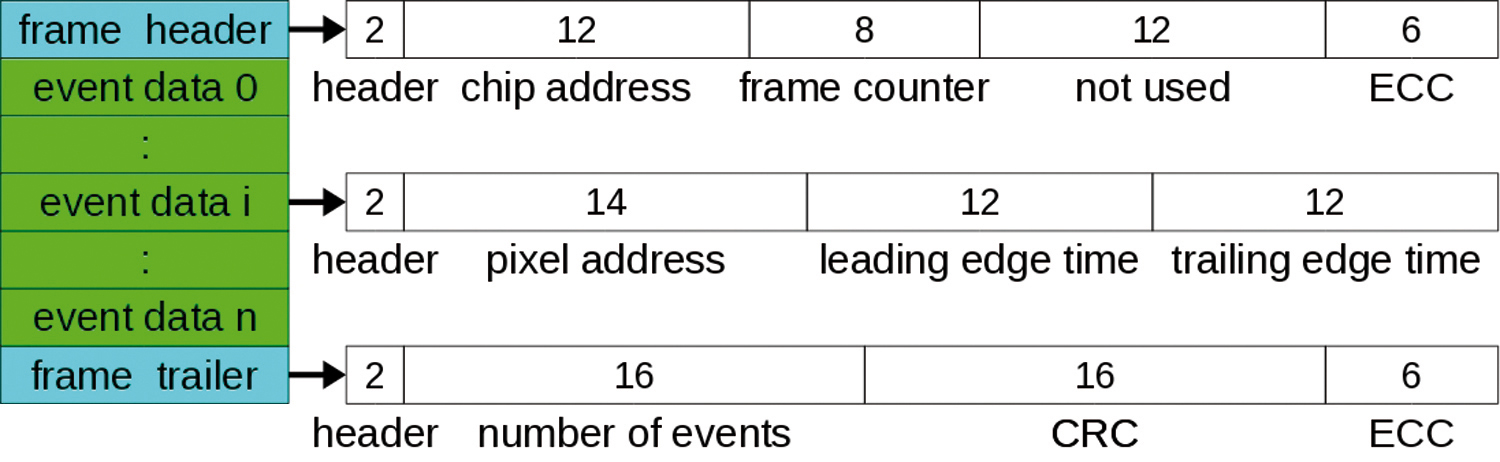}
\caption[Frame structure]{Frame structure. CRC: Cyclic Redundant Check, ECC: Error Correction Code.}
\label{f:frame}
\end{center}
\end{figure*}
The configuration data upload uses the same scheme: the config mode 
signal puts all pixels in 
the busy state. Successive commands address the pixels from the one 
with highest priority  
to the one with the lowest. During each address phase, it is possible 
to write the data on 
the time stamp bus (now used to send configuration information and thus not connected to the 
counter) in the configuration register or to put the data stored in 
the configuration register 
on the data bus.\\

Each double column is controlled by a Column Readout Control Unit (CRCU). The
CRCU consists of four main components: a bank of differential drivers for the
time stamp propagation, a bank of sense amplifiers for the data readout, a 
32 cells output FIFO and the control logic. The CRCU sends the readout 
signals on the base of the state of the busy from the pixel column and the state 
of the output FIFO.\\
The ToPix readout is controlled by a Chip Controller Unit (CCU), which reads
the data from the 55 CRCUs and multiplexes them to up to 3 double data rate 
serial links. The multiplexing operation is made in a round robin scheme with 
two queues, a fast one for the CRCU FIFOs with less than 5 free cells and a 
slow one for the others.\\
Data are transmitted in frames. A frame contains all the data received by 
the chip during a single time stamp counter cycle. Each frame is divided from 
the previous and the next ones by a header and a trailer, respectively. \Figref{f:frame} 
shows the frame structure.

Frames are composed by 40 bits wide words with a 2 bit header to identify a
data word from a header or a trailer word. The data word contains 14 bits for
the pixel address and 24 bits for the leading and trailing edge time 
information.
The header word contains the chip address, the frame counter and error 
detection and correction bits. The trailer word contains the number of words 
present in the frame, the Cyclic Redundant Check (CRC) of the entire frame and the error detection and 
correction bits.

	\subsection{Module Controller}

One of the two readout options for the pixel sensors shown in \figref{f:readout_scheme} 
foresees a module controller ASIC located on the module structure. The use of 
a module controller has been a common choice in the electronic readout of the LHC 
pixel detectors \cite{Beccherle02}\cite{Schnetzer03}\cite{Kluge07}, because it gives 
a greater degree of flexibility.\\
In the \PANDA MVD readout scheme the use of a module controller would allow a
reduction of the transmission lines by using a Phase Locked Loop (PLL) and therefore the possibility
to transmit data at a bit rate faster than the clock frequency. Moreover, it would
also allow a better load balancing between \frontend chips with high and 
low event rates.\\
However, the disadvantages in terms of higher power consumption, increased number of
cooling pipes and increased system fragility due to the introduction of a critical 
single point of failure makes the solution without module controller more attractive.

	\subsection{ASIC Prototypes}
         \label{ASIC_prototypes}
An R\&D activity for the design of the ToPix ASIC is ongoing. In the following
the three ToPix prototypes will be described, together with the test results.

		\subsubsection*{ToPix Version 1}

\begin{figure*}[h]
\centering
\includegraphics[width=0.95\textwidth]{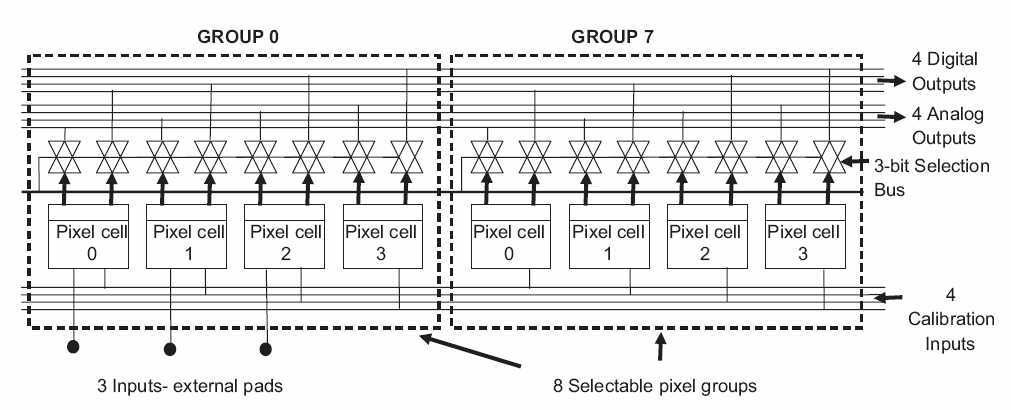}
\caption[ToPix~v1 simplified schematic]{ToPix~v1 simplified schematic.}
\label{f:ToPiX_v1}
\end{figure*}

A small prototype with the basic pixel cell building blocks was designed in order 
to evaluate the analogue performances and gain experience with the technology.\\
This first prototype \cite{Calvo08} chip contains 32 readout cells, 
each one equipped with a preamplifier and a discriminator. The preamplifier consist 
of a Charge Sensitive Amplifier (CSA) with a feedback capacitance of 10~fF, a 
constant current feedback and a baseline restorer. The amplifier size is $\mathrm{37~\tcmu m \times 51~\tcmu m}$
 and the discriminator size is $\mathrm{12.8~\tcmu m \times 48~\tcmu m}$. \Figref{f:ToPiX_v1} shows the ToPix~v1 
architecture. The ASIC size is $\mathrm{2~mm \times 1~mm}$. The cells are arranged in 
8 groups of 4 cells. Each group has 4 calibration input lines, 4 analogue output 
lines, and 4 digital output lines. 
Each input calibration line is in common for one of the cells in each of 
the 8 groups. In each cell, a calibration capacitor of 30 fF shunts the 
calibration line to the input node. Therefore, sending a voltage step on 
the calibration line a delta shaped current pulse is injected at the input node. 
The output lines are multiplexed in order to be shared between the cells belonging 
to different groups. In order to drive the capacitive load of output pads and 
measurement probes, which can be up to 10 pF, the analogue signals are driven by 
a source follower and the digital signals are driven by a buffer. Six cells 
belonging to two different groups have direct connection to dedicated input pads. 
This configuration allows the observation of the cross-channel effects.

\begin{figure}[h]
\centering
\includegraphics[width=0.48\textwidth]{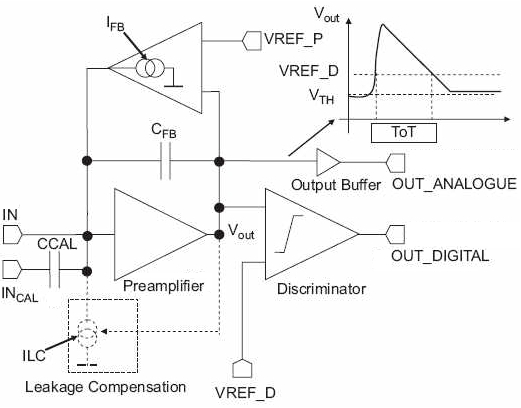}
\caption[ToPix~v1 cell schematic]{ToPix~v1 cell schematic.}
\label{f:ToPiX_v1_cell}
\end{figure}

\Figref{f:ToPiX_v1_cell} shows a simplified schematic view 
of ToPix~v1 readout, 
which is designed to work with an n-type sensor (negative current pulses). 
The CSA core amplifier has three feedback components:

\begin{itemize}

\item The feedback capacitance ${C_{\text{FB}}}$ fixes the charge gain (1/${C_{\text{FB}}}$). 
On the one hand a high value of ${C_{\text{FB}}}$ decreases the CSA charge gain, on the other 
hand a low value of ${C_{\text{FB}}}$ decreases the CSA loop stability. In this prototype a 
nominal value of the feedback capacitance of 10~fF has been chosen as the 
optimal compromise between gain and loop stability.

\item The baseline restorer is implemented with a $gm/C$  low pass filter 
stage which computes the difference between the baseline reference voltage and 
the CSA output. The resulting low frequency voltage signal controls the gate 
of a PMOS which generates the leakage compensation current at the input node.

\item The discharging current generator is based on a differential pair. The two
inputs are connected to the CSA output and to the baseline reference voltage, respectively.
The differential pair is biased by a current source. When ${V_{\text{OUT}}} ={V_{\text{REF}}}$ the 
current is equally shared in the two branches of the differential pair. When a 
negative current signal is present at the CSA input, ${V_{\text{OUT}}}$ increases, thus switching 
off one of the two branches. Consequently the current flows in the input node 
and ${C_{\text{FB}}}$ is discharged. A good constant discharge is achieved only when ${V_{\text{OUT}}}$ 
is sufficiently large to fully steer  the current from one branch to the other 
of the differential pair. Since the two transistors at the differential pair input
work in weak inversion this voltage is about 40~mV.

\end{itemize}

The constant current discharge provides a constant slope trailing edge
of the preamplifier output signal which, in turn, provides a linear ToT 
measurement of the sensor charge signal.The discharge current source 
(${I_{\text{FB}}}$) is normally turned off. When an input signal is applied, 
the output voltage of the preamplifier increases very fast, exceeding 
the voltage VREF\_P, the current source is turned on and the feedback 
capacitor is discharged with constant current until equilibrium is restored,
and the output level of the amplifier turns back to VREF\_P.
The dynamic range is extended by discharging the feedback capacitor at the 
input node, preserving the linearity of the ToT measurement even for higher 
input signals, when the preamplifier output is saturated. The feedback 
topology considered is similar to the one employed in \cite{Peric06}.
The discriminator design follows a low-voltage approach employing a single
stage folded-cascode topology followed by two inverters. For test purposes 
each channel is provided with an internal calibration capacitor ${C_{\text{CAL}}}$ driven by 
an external signal ${IN_{\text{CAL}}}$, with amplitude ${V_{\text{CAL}}}$. The injected charge is given 
by the product ${V_{\text{CAL}}} \cdot{C_{\text{CAL}}}$. The absolute value of ${C_{\text{CAL}}}$ is known with 
an accuracy of 10\%.

		\subsubsection*{ToPix Version 1 Test Results}

\begin{figure}[h]
\includegraphics[width=2.8in]{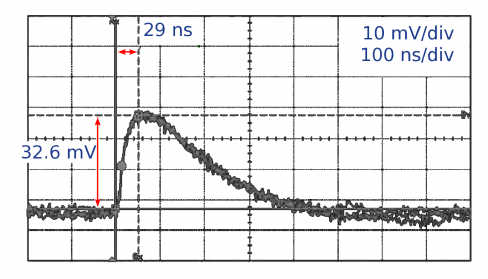}
\caption[Cell output for an input charge of 0.5~fC]{Cell output for an input charge of 0.5~fC.}
\label{f:ToPiX_v1_pulse}
\end{figure}

\Figref{f:ToPiX_v1_pulse} shows a typical pulse at the output of the preamplifier 
for an input charge signal of 0.5~fC, measured with an oscilloscope and a differential 
probe, for best common mode noise rejection. The preamplifier gain in the 
linear region is 60~mV/fC. This value was obtained fitting 5 
calibration points 
with nominal values of 1, 3, 4, 8, 12~fC. The r.m.s. output noise is 1~mV. 
Assuming a gain of 60~mV/fC this translates to an equivalent input noise of 100~e$^-$ rms. 
The power consumption is 12~$\tcmu$W per pixel cell using a 1.2~V power supply.
The test results are summarised in \tabref{tab:ToPiXv1test}.

\begin{table}[!ht]
\begin{center}
\begin{tabular}{|l|c|}
\hline
analogue gain					& 60~mV/fC\\
\hline
Linear input gain			& 100~fC\\
\hline
$V_{\text{noise,rms}}$ (${C_{\text{d}}}=0$)	& 1~mV\\
\hline
ENC  (${C_{\text{d}}}=0$)			& 104~e$^-$\\
\hline
Power consumption			& 12~$\tcmu$W/cell\\
\hline
ToT dispersion (rms)		& 20\%\\
\hline
Baseline dispersion (rms)	& 13~mV\\
\hline
\end{tabular}
\end{center}
\caption[ToPix~v1 electrical tests result]{ToPix~v1 electrical tests result.}
\label{tab:ToPiXv1test}
\end{table}

		\subsubsection*{ToPix Version 2}
A second prototype of the ToPix ASIC has been designed and tested in order to
prove the pixel cells and column functionality. This chip is designed for a 
clock frequency of 50~MHz as in the original specifications.

\begin{figure}[h]
\includegraphics[width=2.8in]{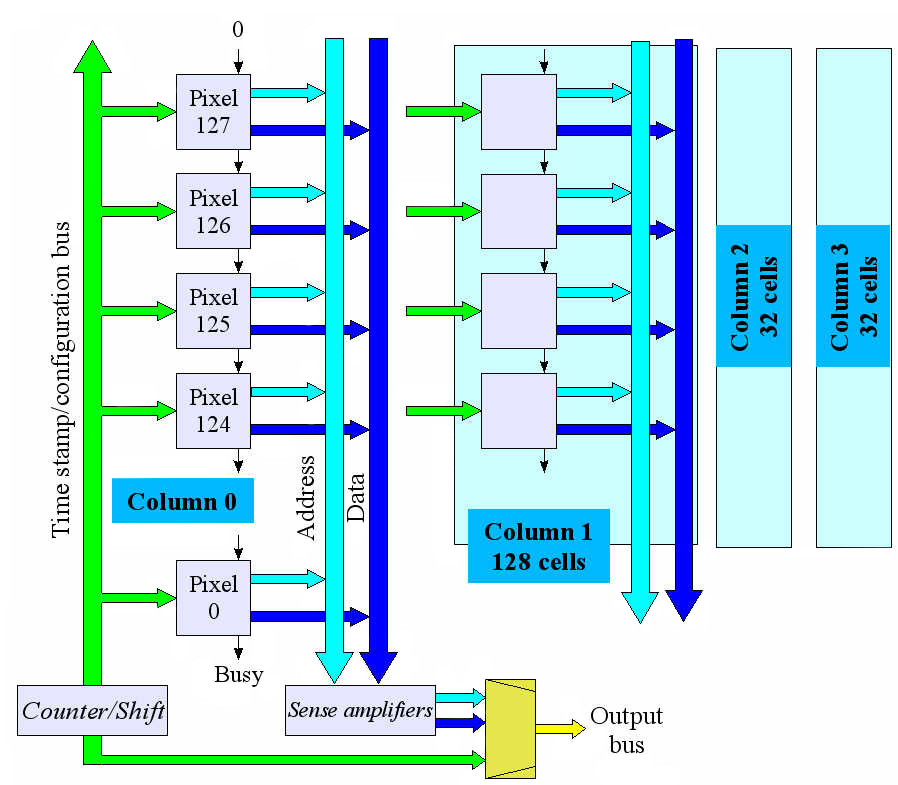}
\caption[ToPix~v2 simplified schematic]{ToPix~v2 simplified schematic.}
\label{f:ToPiX_v2_scheme}
\end{figure}

The prototype includes two 128-cell and two 32-cell columns. The pixel cells 
include all the functional blocks required for the final ASIC except for the 
bonding pad opening. The end-of-column logic has been greatly simplified for 
test purposes and includes only the counter and the bus readout sense amplifiers. 
The prototype block diagram is shown in \figref{f:ToPiX_v2_scheme}.

\begin{figure}[h]
\begin{center}
\includegraphics[width=0.45\textwidth]{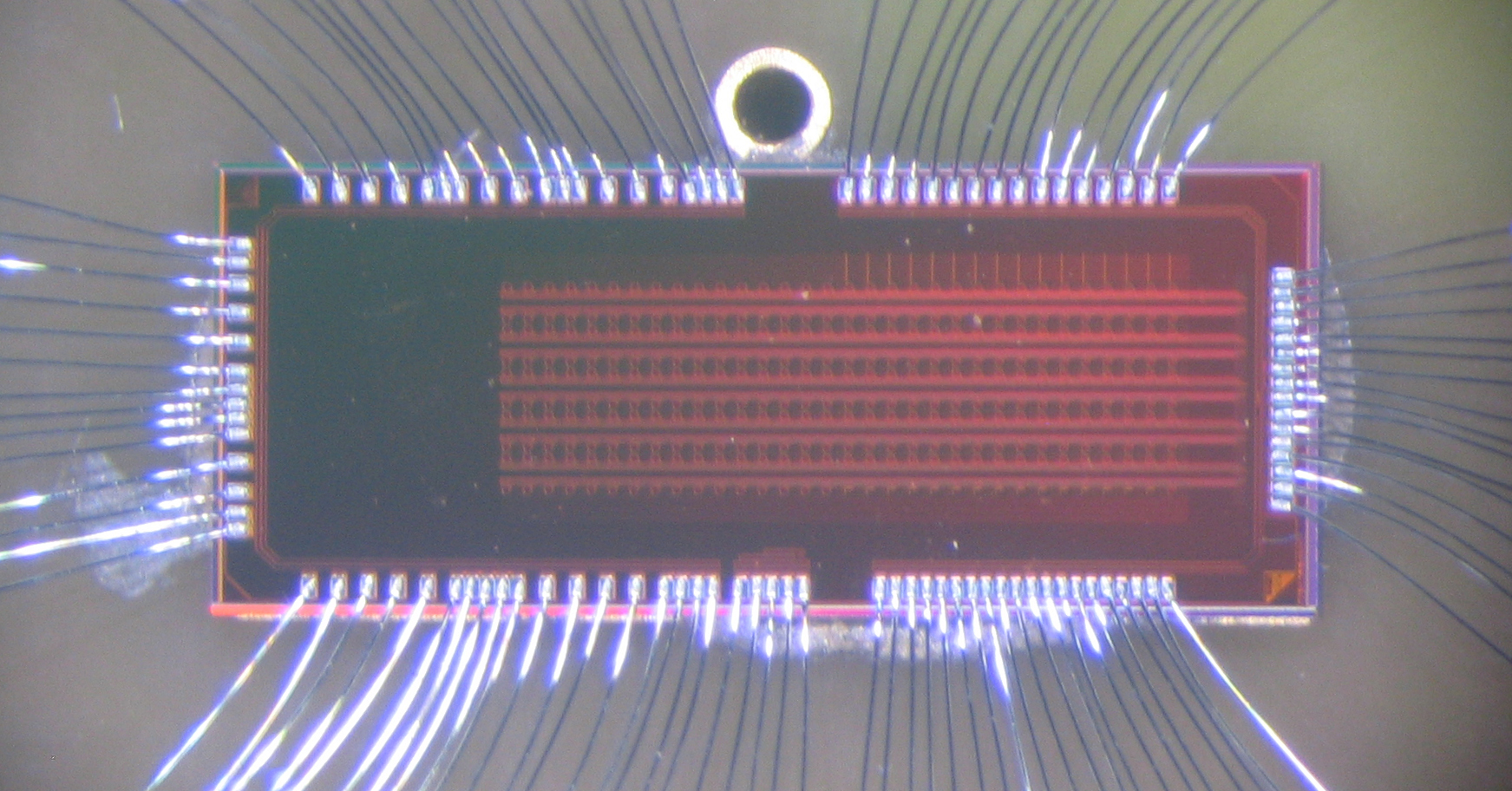}
\end{center}
\caption[ToPix~v2 die]{ToPix~v2 die.}
\label{f:ToPiX_v2_photo}
\end{figure}

The chip has been designed in a commercial CMOS 0.13~$\tcmu$m technology with 8 metal
layers. The two 128-cell columns are folded in four 32-cell slices in order to 
have an acceptable form factor. This arrangement represents a worst case with respect
to the final floorplan because of the longer interconnections. The die size is 
$\mathrm{5~mm \times 2~mm}$. \Figref{f:ToPiX_v2_photo} shows the chip die.

\begin{figure}[h]
\includegraphics[width=2.8in]{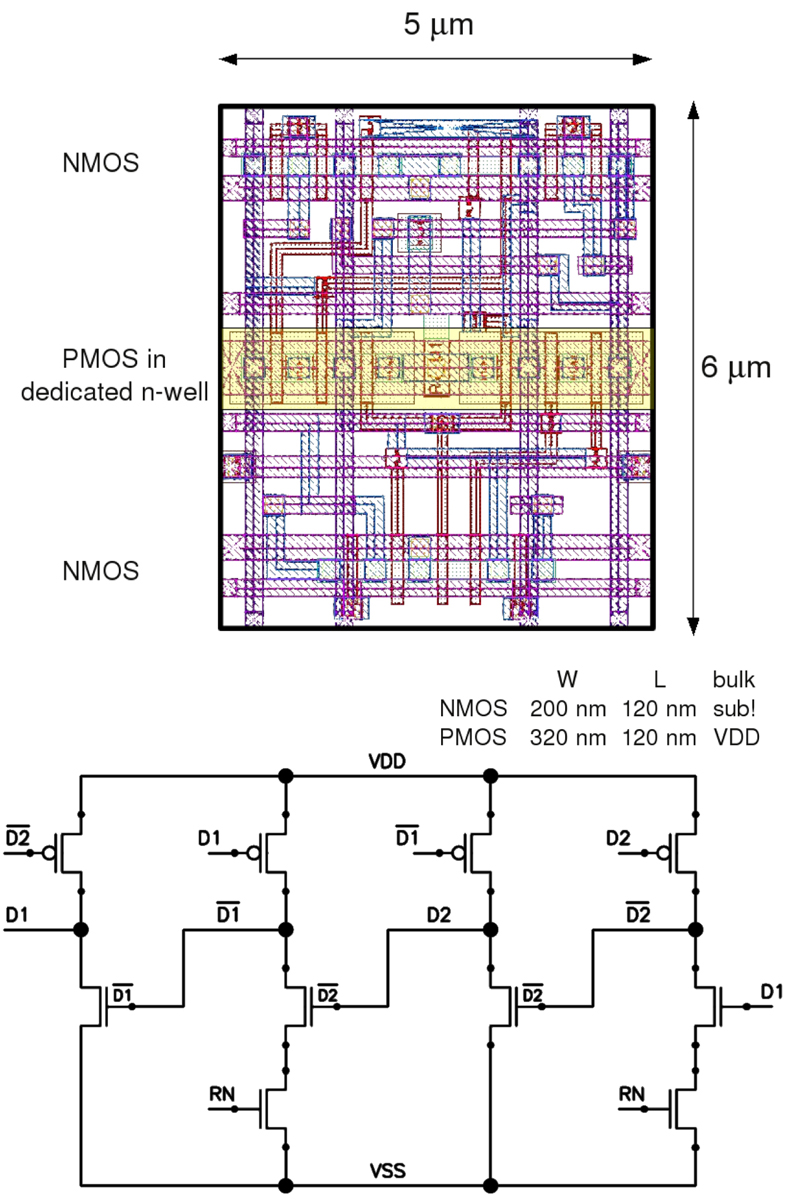}
\caption[Dice cell schematic]{Dice cell schematic.}
\label{f:dice_cell}
\end{figure}

The Single Event Upset (SEU) effect has been mitigated by designing 
all flip flops and latches in the pixel cell as fully static and based on 
the DICE architecture \cite{Calin96}. This architecture, shown in 
\figref{f:dice_cell}, is insensitive to SEU as far as only one circuit
node received the voltage spike due to the incoming particle. The area
penalty is about a factor of 2, compared with the factor $3.5-4$ of a triple 
redundancy circuit.\\

A number of features has been added in order to improve testability. 
In addition 
to the internal charge injection circuit, the inputs of 16 cells of 
the first column 
are connected to an external pad. It is therefore possible to 
connect the chip to 
a detector in a simple way. The drawback of such an arrangement 
is that it leads 
to a significant increase of the input capacitance and therefore of the noise.
All digital control signals are provided externally, in order to allow testing 
and optimisation of the speed and synchronisation of the readout sequence. It 
should be noted that the folding configuration is a slightly worse case with 
respect to the timing because of the longer bus lines.

\subsubsection*{ToPix Version 2 Test Results}
\begin{figure}[h]
\begin{center}
\includegraphics[width=2.8in]{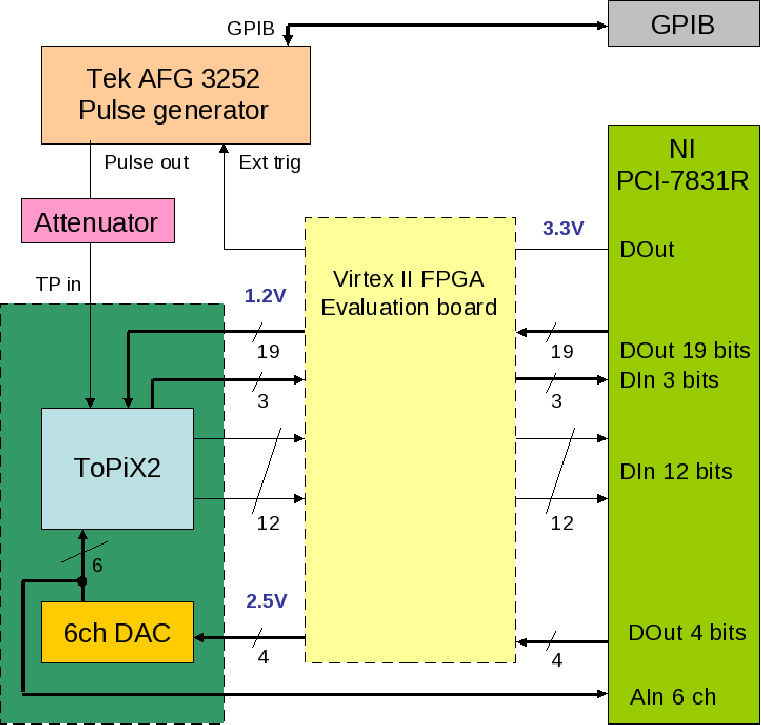}
\caption[Diagram of the data acquisition system  for  testing ToPix~v2]{Diagram of the data acquisition system  for  testing ToPix~v2.}
\label{f:torino_scheme}
\end{center}
\end{figure}

Test of the ASIC was performed with a DAQ chain based on a NI-PCI (National Instruments) system using the PCI-7831R 
board and LabVIEW software.  The NI-FPGA generates pattern and works as a triggered memory. 
A Virtex II FPGA shifts voltage levels. 
To allow the test of ToPix~v2 with 
a 50~MHz clock, a setup with a bidirectional FIFO has 
been implemented on the FPGA board, see \figref{f:torino_scheme}.

\begin{figure}[h]
\includegraphics[width=0.96\columnwidth]{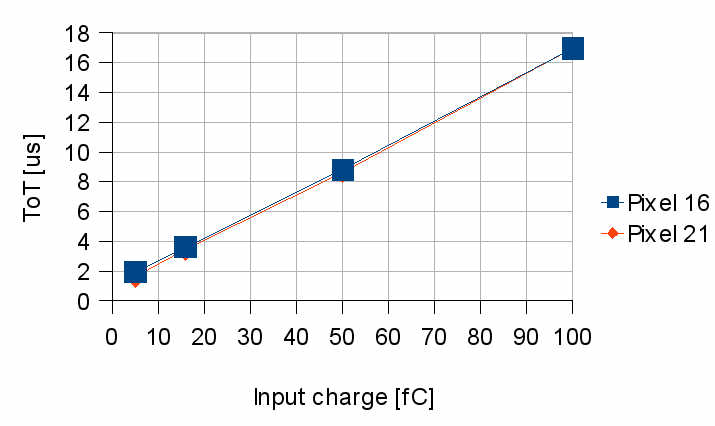}
\caption[Input-output transfer function]{Input-output transfer function.}
\label{f:ToPiX_v2_io}
\end{figure}

\begin{figure}[h]
\includegraphics[width=0.96\columnwidth]{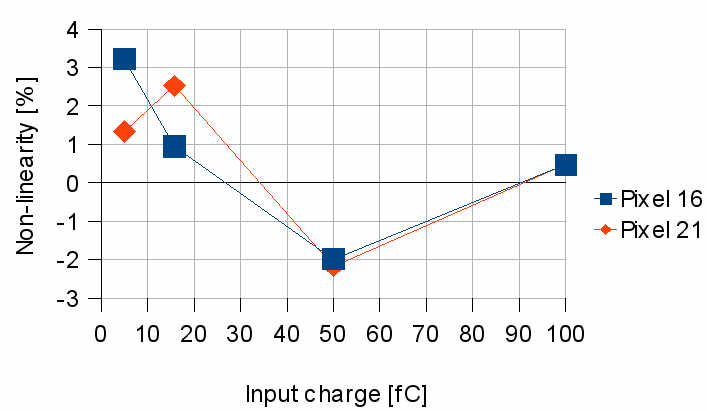}
\caption[Deviation from linear fit]{Deviation from linear fit.}
\label{f:ToPiX_v2_lin}
\end{figure}

The prototype has been tested both by injecting charge through the 
inputs connected  to the pads and through the internal calibration circuit. 
\Figsref{f:ToPiX_v2_io} and \ref{f:ToPiX_v2_lin} show the input-output transfer 
function and the linearity over the whole dynamic range for two typical pixels.

\begin{figure}[h]
\includegraphics[width=0.88\columnwidth]{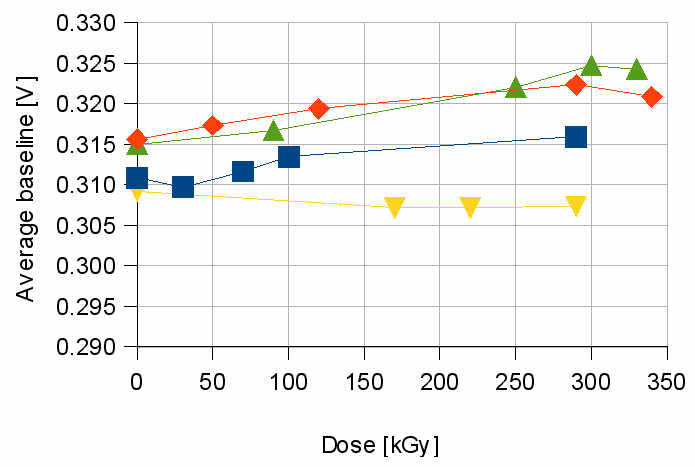}
\caption[Average baseline during irradiation]{Average baseline during irradiation.}
\label{f:ToPiX_v2_irr_bsline}
\end{figure}

\begin{figure}[h]
\includegraphics[width=0.88\columnwidth]{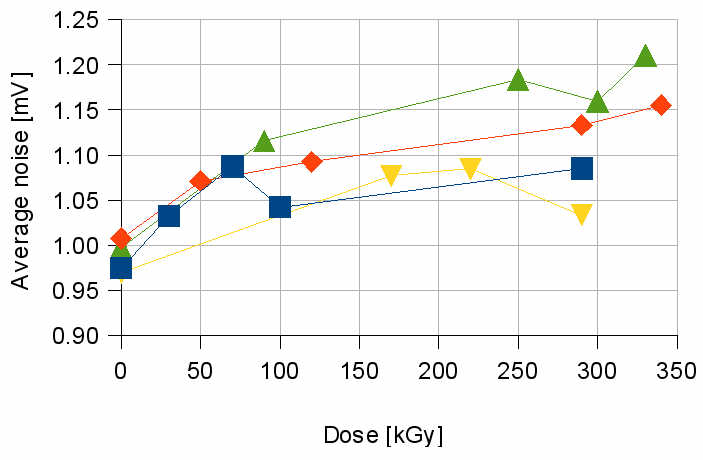}
\caption[Average noise during irradiation]{Average noise during irradiation.}
\label{f:ToPiX_v2_irr_noise}
\end{figure}

Total dose irradiation tests have been performed on four chips  
with a rate of 4.07~Gy/s. \Figsref{f:ToPiX_v2_irr_bsline} and \ref{f:ToPiX_v2_irr_noise} 
show the average of the baseline and the noise over some of the 320 cells of 
4 chips during irradiation, obtained with a threshold scan in order to 
remove the quantisation error. The baseline variation is below 3\% 
and is comparable with the dispersion due to the process variation.
The noise level is around 1~mV r.m.s. before irradiation, which corresponds 
to 0.025~fC. An increase of 20\%, 
to 0.03~fC,  has been observed after irradiation. The measured noise value if higher than the 0.017~fC
measured for the first ToPix prototype. It should be considered that in the ToPix~v2 a large metal plate
has been connected to the pixel input to emulate the bonding pad capacitance. Simulations show that the
capacitance added by the metal plate justify the noise increase.

\begin{figure}[h]
\includegraphics[width=2.9in]{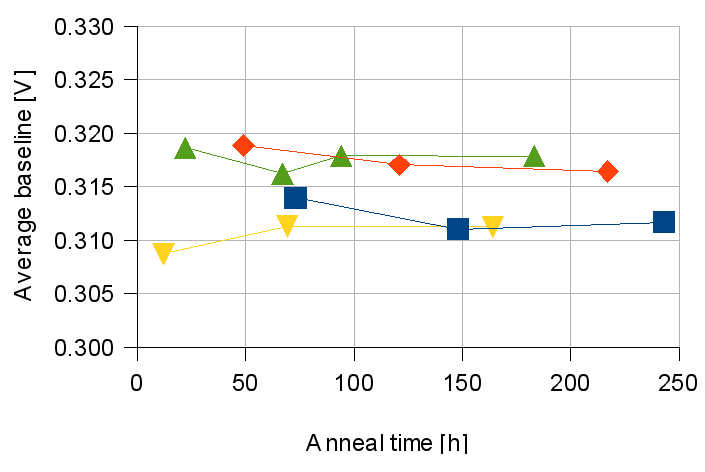}
\caption[Average baseline during annealing]{Average baseline during annealing.}
\label{f:ToPiX_v2_ann_bsline}
\end{figure}

\begin{figure}[h]
\includegraphics[width=2.9in]{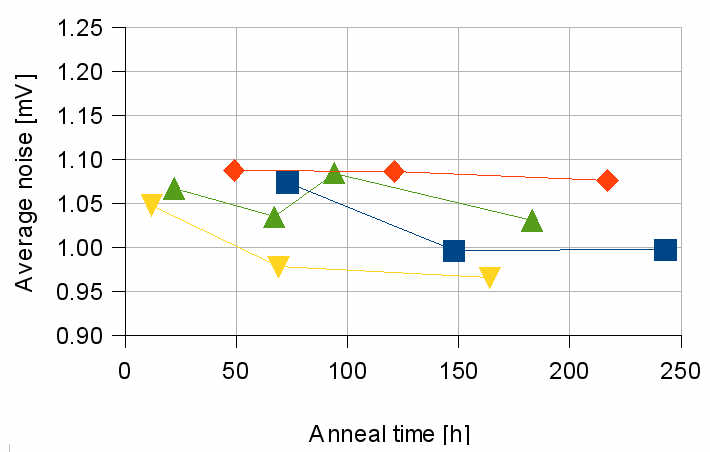}
\caption[Average noise during annealing]{Average noise during annealing.}
\label{f:ToPiX_v2_ann_noise}
\end{figure}

After irradiation the chips have been annealed at 100~$^{\circ}$C. 
\Figsref{f:ToPiX_v2_ann_bsline} and \ref{f:ToPiX_v2_ann_noise} show a 
partial recovery in the baseline value and an almost complete recovery of 
the noise level to the pre-irradiation values.

\begin{figure}[h]
\includegraphics[width=2.8in]{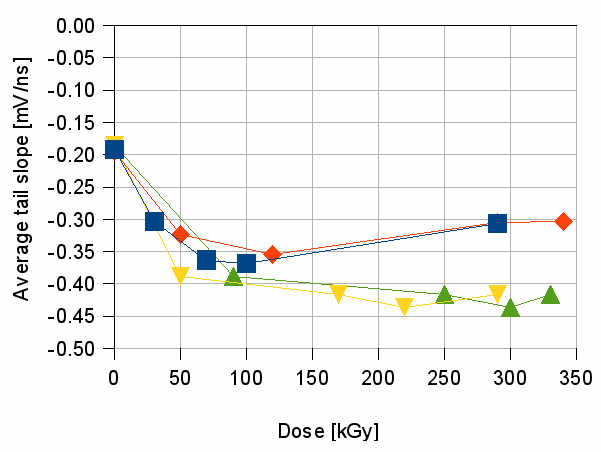}
\caption[Average slope during irradiation]{Average slope during irradiation.}
\label{f:ToPiX_v2_irr_slope}
\end{figure}

\begin{figure}[h]
\includegraphics[width=2.8in]{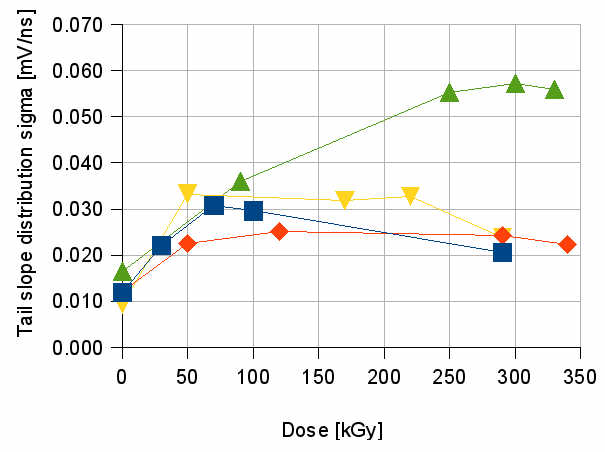}
\caption[Tail slope sigma during irradiation]{Tail slope sigma during irradiation.}
\label{f:ToPiX_v2_irr_sigma}
\end{figure}

\Figsref{f:ToPiX_v2_irr_slope} and \ref{f:ToPiX_v2_irr_sigma}
show the average and $\sigma$ tail slope during irradiation. In this case 
the variation is much more relevant, up to 2 times in case of the average 
and up to 6 times for the $\sigma$. The problem lies in the very small
current (around 6~nA) required to discharge the 
integrating capacitor to 
obtain the required resolution. With such low currents the leakage current 
induced by radiation plays an important role.

A possible solution is to use an enclosed gate layout for the critical 
transistors of the discharge circuit. Such a layout technique has been proven to be
effective in preventing radiation-induced current leakage \cite{Anelli99}.
However, irradiation tests made on a previous prototype of the ToPix 
analogue part \cite{Calvo08} showed that at dose rate of 0.011~Gy/s the 
relative ToT variation decreases by an amount between 22\% and 54\% 
in comparison with the variation after a dose rate of 4~Gy/s.

\input{pixelpart/v2_seu.tex}  % text by laura

\subsubsection*{ToPix Version 3}

\authalert{Author: G. Mazza, mazza@to.infn.it}

\begin{figure}[h]
\begin{center}
\includegraphics[width=2.8in]{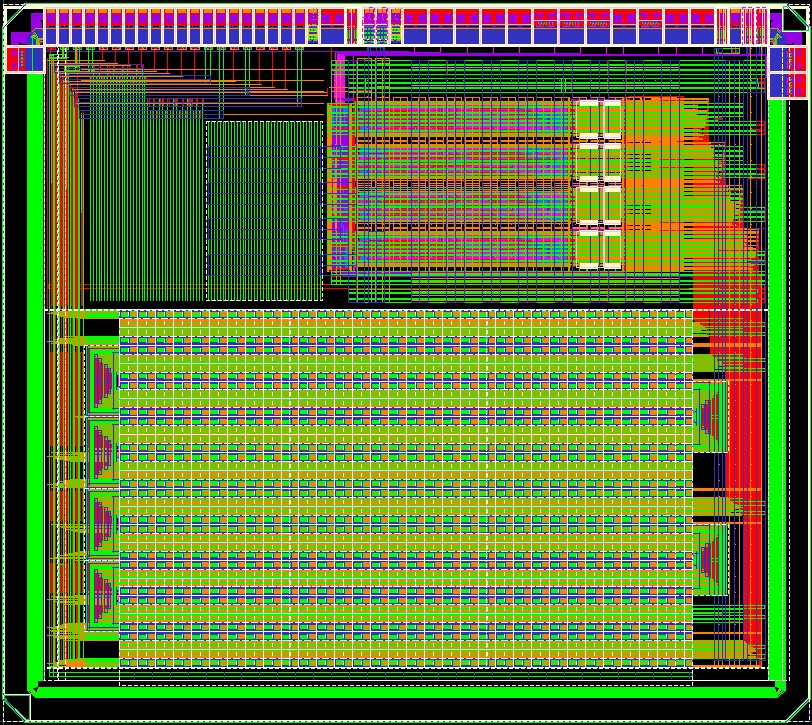}
\caption[ToPix~v3 layout,  $\mathrm{4.5~mm \times 4~mm}$ die]{ToPix~v3 layout,  $\mathrm{4.5~mm \times 4~mm}$  die. The pixel matrix is visible in the bottom part and 
the peripheral logic in the upper part. Pads for wire connections are in the upper edge.}
\label{f:ToPiX_v3_layout}
\end{center}
\end{figure}

A third prototype of the ToPix chip, designed to work with 
the 155~MHz clock
updated to  the new specifications, has been designed and tested.
The prototype includes two $2 \times 128$ cell and two 
$2 \times 32$ cell  double 
columns. The die size is $\mathrm{4.5~mm \times 4~mm}$. Bump bonding pads 
have been used in 
the pixel cells for direct connections to the detector. The 
128 cells columns have been folded in four 32-cells sub-columns in order 
to have an almost square form factor, thus simplifing the die 
handling. The complete pixel cells and the end of column 
control logic and buffers have been implemented, while the output multiplexer and 
serialiser is still in a simplified revision. The chip layout is shown in 
\figref{f:ToPiX_v3_layout}.

In this prototype a clipping circuit has been implemented in 
order to avoid long dead-times for large input charges. The circuit forces a ToT 
saturation at $\sim$16.5~$\tcmu$s 
while keeping a good linearity for charges up to 50~fC.\\
All flip flops in the pixel cell are fully static and use 
the triple redundancy technique 
in order to make them more SEU tolerant. The configuration 
register implements an asynchronous 
error detection and self-correction circuitry while the 
leading and trailing edge registers 
implement only the majority voter to save area. Therefore 
the error correction works only 
in the readout phase. In order to fit in the pixel area 
the configuration register has been
reduced from 12 to 8 bits.\\
The chip is powered by a single 1.2~V power supply. 
Digital input and output drivers 
use the differential SLVS standard \cite{SLVS} in 
order to reduce the digital noise 
contribution and to be compatible with the 1.2~V power supply.\\
Simulations with layout parasitic extraction and process 
and mismatch variations show a gain 
of $200.022 \pm 0.046$~ns/fC at ${V_{\text{TH}}}=30$~mV and a 
channel to channel ToT spread of 7\%. 
The simulated noise level is 155~e$^-$ at ${I_{\text{LEAK}}}=0$ and 
209~e$^-$ at ${I_{\text{LEAK}}}=10$~nA. 

\input{pixelpart/v3_analog.tex}  % text by laura for thanu

		\subsubsection*{ToPix Version 3 Test Results}

\begin{figure}[!h]
\begin{center}
\includegraphics[width=0.5\textwidth]{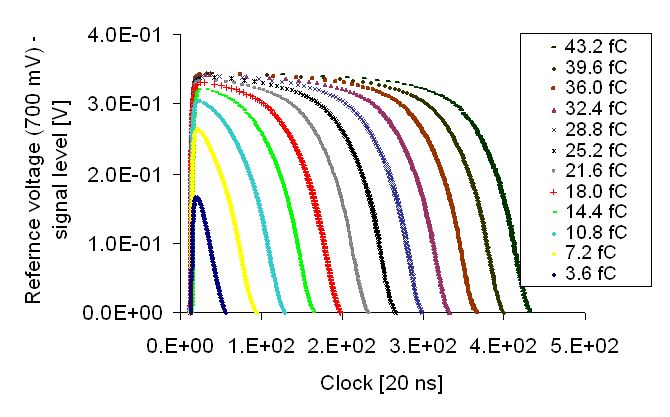}
\caption[Reconstructed signal shapes with threshold scan for different input charge values]{Reconstructed signal shapes with threshold scan for different input charge values.}
\label{f:ToPiX_v3_signal_th}
\end{center}
\end{figure}

The ToPix~v3 prototype has been tested on a bench to verify the performances.
The experimental setup foresees the testing board housing ToPix~v3 and a Xilinx evaluation 
board equipped with the Virtex 6 FPGA. The acquisition system is based on the LabView software. 
Due to a problem in the column buses only the 32-cells column can work
at 160~MHz; therefore all the tests have been performed at 50~MHz.
The problem has been investigated and is partially due to the folded structure
of the column, which makes the bus significantly longer (by a factor of 1.3),
and partially from the fact that the 3 latches of each bit (due to the triple
redundancy) are directly connected to the bus without any buffer, thus
tripling the capacitive load.\\
\Figref{f:ToPiX_v3_signal_th} shows the preamplifier output for differet input charge values 
obtained with a threshold scan and the internal test signal generator. The obtained 
shapes show the fast rising edge of the charge amplifier and the slow discharge at constant 
current, even in the saturation case.

\begin{figure*}[!h]
\begin{center}
\includegraphics[width=0.99\textwidth]{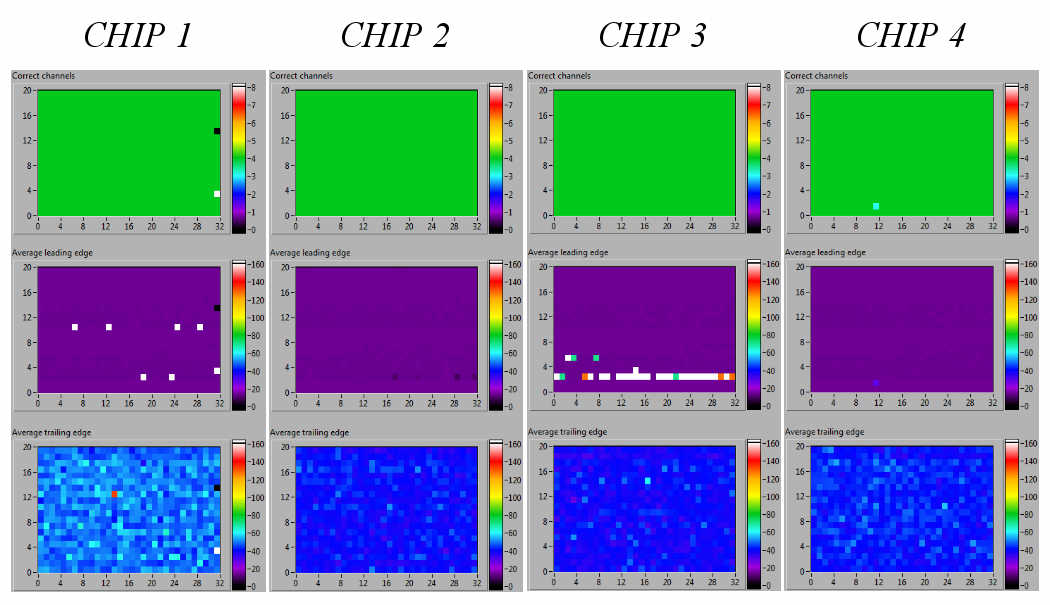}
\caption[ToPix~v3 pixel map for the four tested samples]{ToPix~v3 pixel map for the four tested samples.}
\label{f:ToPiX_v3_pixelmap}
\end{center}
\end{figure*}

At present four ToPix chips have been tested, showing similar performances.
\Figref{f:ToPiX_v3_pixelmap} shows the pixel response of the four tested
samples. In the first row the chips which respond correctly to a sequence of
4 test pulses are shown in green. Pixels transmitting twice the correct number
are marked in white, while pixels which do not respond at all are in black.
Intermediate numbers of responses are shown accordingly with the colour scale.
In the second and third rows, the leading and trailing edge values 
are represented, respectively. The time stamp values are represented
in a colour scale. All the test signals are sent at the same time and with
the same charge, therefore a uniform colour pattern is expected. Here
pixels in black and in white correspond to a all 0's and all 1's 
output pattern, respectively. It can be seen that the number of defective 
pixels is slightly higher, apart from chip 2 where a column shows significant 
problems. It can also be observed that the variation of the trailing edge value
is higher than the leading edge. This variation is expected due to the higher
variation in the charge injection system and to the higher noise of the
trailing edge.

\begin{figure}[!ht]
\begin{center}
\includegraphics[width=0.5\textwidth]{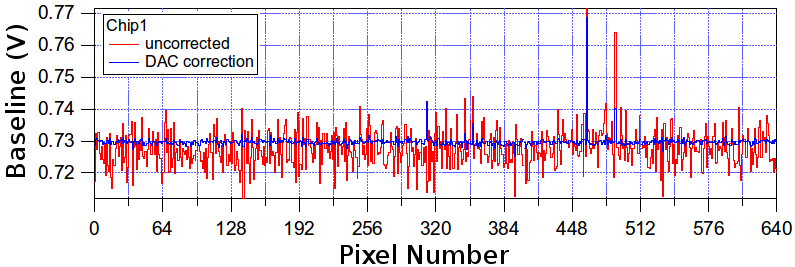}
\caption[ToPix~v3 extrapolated baseline before and after equalisation]{ToPix~v3 extrapolated baseline before and after equalisation.}
\label{f:ToPiX_v3_bsline}
\end{center}
\end{figure}

\Figref{f:ToPiX_v3_bsline} shows the baseline level for all 640 pixels
of a chip, before and after baseline equalisation.

\begin{figure}[!ht]
\begin{center}
\includegraphics[width=0.5\textwidth]{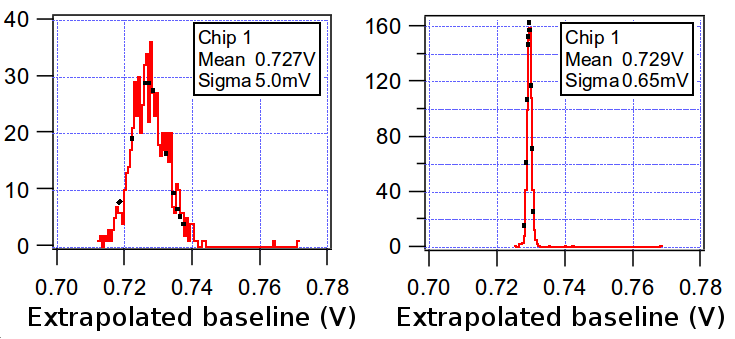}
\caption[ToPix~v3 baseline distribution before and after correction]{ToPix~v3 baseline distribution before and after correction.}
\label{f:ToPiX_v3_bsline_sigma}
\end{center}
\end{figure}

\Figsref{f:ToPiX_v3_bsline_sigma} shows the baseline distribution before (left) 
and after (right) correction. A baseline sigma of 5~mV has been measured before correction, 
which correspond to an equivalent charge dispersion of 473~e$^-$. After the correction with 
the on-pixel 5 bit DACs a  dispersion of 650~$\tcmu$V, corresponding to less than 62~e$^-$, 
can be obtained. The measurements confirm the effectiveness of the correction DAC to reduce the
threshold dispersion.

\begin{figure}[!ht]
\begin{center}
\includegraphics[width=0.5\textwidth]{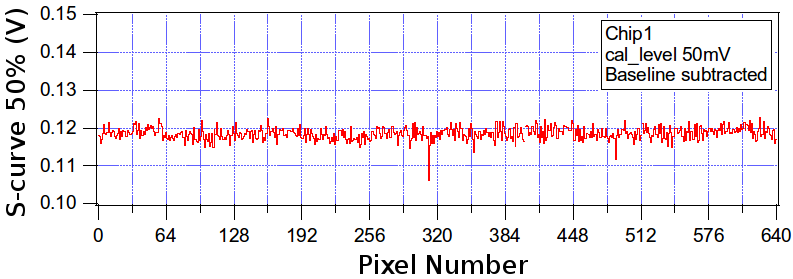}
\caption[ToPix~v3 measured analogue gain]{ToPix~v3 measured analogue gain.}
\label{f:ToPiX_v3_gain}
\end{center}
\end{figure}

\Figref{f:ToPiX_v3_gain} shows the measured gain. The measured value of 66~mV/fC is
slightly lower than the simulated one (75~mV/fC) but compatible with the process variations
and with the uncertainties of the charge injection mechanism.

\begin{figure}[!ht]
\begin{center}
\includegraphics[width=0.5\textwidth]{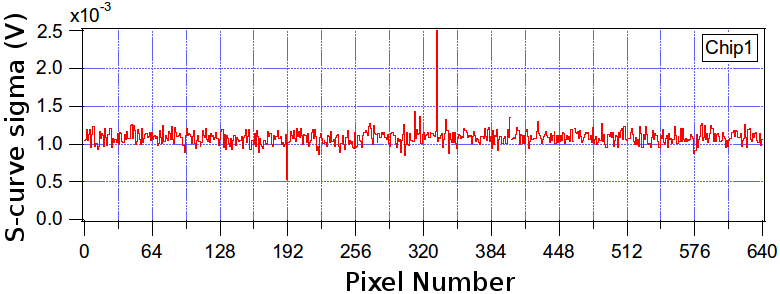}
\caption[ToPix~v3 measured noise]{ToPix~v3 measured noise.}
\label{f:ToPiX_v3_noise}
\end{center}
\end{figure}

\newpage
\Figref{f:ToPiX_v3_noise} shows the measured noise. The average value of 100~e$^-$ obtained 
without the detector connected to the preamplifier input is consistent with a simulated value 
of 150~e$^-$ with the detector capacitance included.

\begin{figure}[!ht]
\begin{center}
\includegraphics[width=0.5\textwidth]{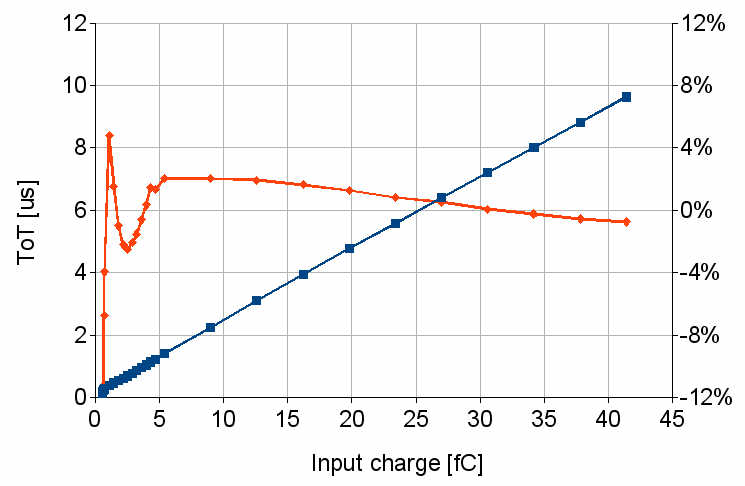}
\caption[ToPix~v3 transfer function for nominal feedback current - full range]{ToPix~v3 transfer function for nominal feedback current - full range.}
\label{f:ToPiX_v3_ToT_5nA_full}
\end{center}
\end{figure}

\begin{figure}[!ht]
\begin{center}
\includegraphics[width=2.8in]{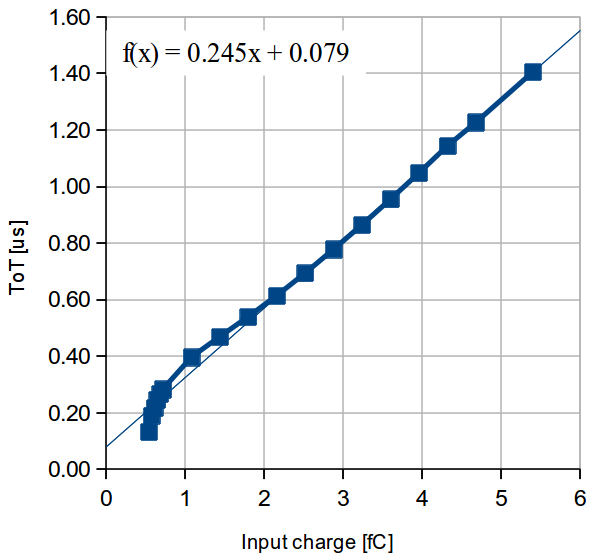}
\caption[ToPix~v3 transfer function for nominal feedback current - 0.6~fC range]{ToPix~v3 transfer function for nominal feedback current - 0.6~fC range.}
\label{f:ToPiX_v3_ToT_5nA_6fC}
\end{center}
\end{figure}

\vspace{1cm}
The charge to time overall transfer function is shown in \figref{f:ToPiX_v3_ToT_5nA_full}
for the full charge input range and in \figref{f:ToPiX_v3_ToT_5nA_6fC} for charges up
to 6~fC. Also shown in \figref{f:ToPiX_v3_ToT_5nA_full} is the deviation from the linear fit.

\begin{figure}[!ht]
\begin{center}
\includegraphics[width=0.9\columnwidth]{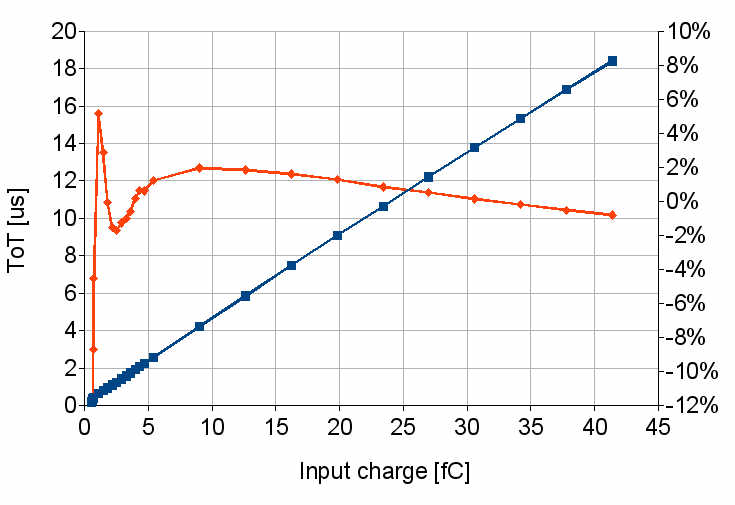}
\caption[ToPix~v3 transfer function for halved feedback current - full range]{ToPix~v3 transfer function for halved feedback current - full range.}
\label{f:ToPiX_v3_ToT_2nA5_full}
\end{center}
\end{figure}

\begin{figure}[!ht]
\begin{center}
\includegraphics[width=2.8in]{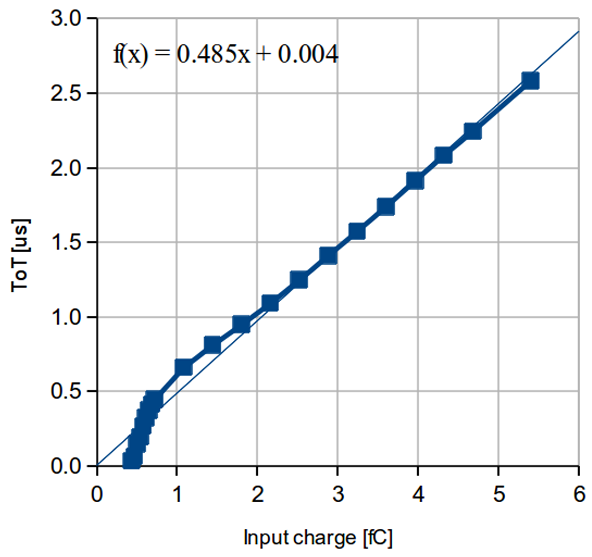}
\caption[ToPix~v3 transfer function for halved feedback current - 0.6~fC range]{ToPix~v3 transfer function for halved feedback current - 0.6~fC range.}
\label{f:ToPiX_v3_ToT_2nA5_6fC}
\end{center}
\end{figure}

Due to the reduced clock frequency, the chip transfer function has been measured
also for a feedback current which is half of the nominal value, in order to 
partially compensate for the reduction in charge resolution. The results are shown
in \figref{f:ToPiX_v3_ToT_2nA5_full} for the full range and in \figref{f:ToPiX_v3_ToT_2nA5_6fC} for the first part of the dynamic range.

%% file: pixelpart/v2_seu.tex
\subsubsection*{SEU Study}

A dedicated study of the SEU in ToPix~v2 has been performed in order to 
 verify if the DICE architecture was sufficient in the expected \PANDA environment.
To this aim  the device has been tested with heavy ions at SIRAD (LNL-INFN)\cite{sirad}, and the convolution of
the experimental results with simulation studies allowed to estimate the maximum SEU rate of ToPix~v2  in \PANDA (see \cite{PANDAnote8}).\\
More ions of different energies, see \tabref{tab:ioni_seu}, have been used in order to have a wide range of 
deposited energy.
 Moreover it was
 possible to investigate the deposited energy due
 to different incident angles, rotating the support of ToPix~v2 \cite{angolo}. 
 The irradiation runs lasted about 20 minutes and before and after
 each run dosimetry measurements were carried out.

\begin{table}[!ht]
\begin{center}
\begin{tabular}{|c|c|c|c|}
\hline \begin{footnotesize}Ion \end{footnotesize}& \begin{footnotesize}Beam Angle\end{footnotesize} &\begin{footnotesize} ${E_{\text{cin}}}$  [MeV]  \end{footnotesize} &\begin{footnotesize} ${E_{\text{dep}}}$  [MeV]\end{footnotesize}\\
\hline $^{16}$O & $0^{\circ}$  & 101 & 0.70\\
\hline $^{16}$O & $30^{\circ}$ & 101 & 0.81\\
\hline $^{19}$F & $0^{\circ}$  & 111 & 0.91\\
\hline $^{19}$F & $30^{\circ}$ & 111 & 1.07\\
\hline $^{28}$Si & $30^{\circ}$& 146 & 2.40\\
\hline $^{35}$Cl & $0^{\circ}$ & 159 & 3.01\\
\hline $^{58}$Ni & $0^{\circ}$ & 223 & 6.47\\
\hline $^{79}$Br & $0^{\circ}$ & 215 & 8.96\\
\hline
\end{tabular}
\end{center}
\caption[Ions used in the beam test]{Ions used in the beam test with the corresponding kinetic energy, incident beam angle and deposited energy in a sensitive volume of 1~$\tcmu$m$^3$.}
\label{tab:ioni_seu}
\end{table}

\begin{figure}[!ht]
\begin{center}
 \includegraphics[width=0.5\textwidth]{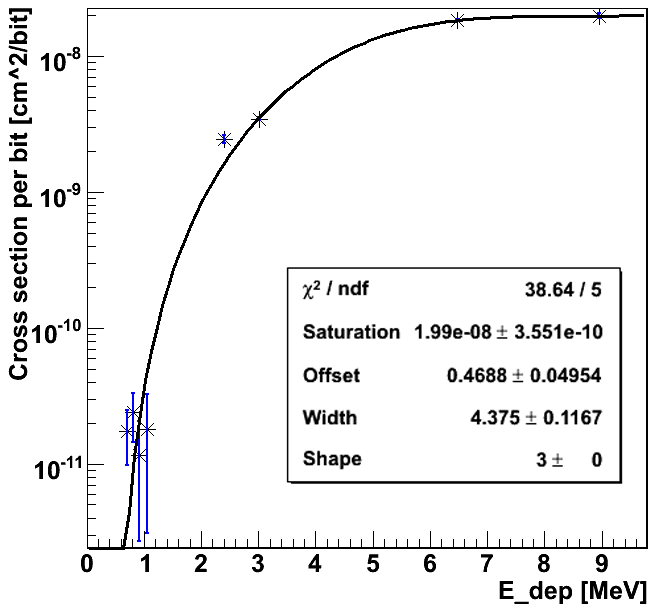}
\end{center}
\caption[Weibull fit to the experimental ion data of the ToPix~v2 configuration register]{Weibull fit to the experimental ion data of the ToPix~v2 configuration register. A sensitive volume of 1~$\tcmu$m$^3$ is assumed.}
\label{fig:weibull}
\end{figure}

The study of the configuration register was performed using a control program
developed with the LabView software, that wrote a 12 bit sequence of
alternated 0 and 1 every 2 seconds and counted
whenever some bits were changed. 
Knowing the number of SEUs (${N_{\text{errors}}}$) and the total incident particle fluence ($\Phi$), it is possible to calculate 
the probability to have an upset, usually called SEU cross section:
\begin{equation}
\sigma_{\text{SEU}}=\frac{N_{\text{errors}}}{\Phi \cdot N_{\text{bit}}}
\end{equation}
The $\sigma_{\text{SEU}}$ has been calculated as the average of the cross sections coming from the several irradiation runs 
with the same ion species,
and the standard deviation was used  as an estimation of the error.\\
The experimental cross sections obtained have been fitted by a Weibull function (see \figref{fig:weibull})
 It has the form:
\begin{equation}
\sigma=\sigma_{0} \left[1-e^{-\left(\frac{E_{\text{dep}}-E_{0}}{W}\right)^s} \right]
\end{equation}
where W and s are shape parameters. Here the s parameter was
set to 3. 

Our fit parameters for the configuration register, see \figref{fig:weibull}, are:
\begin{itemize}
\item ${E_{\text{th}}}$=0.4688~\mev
\item $\sigma_0 =1.99\cdot 10^{-8}$ [$\mathrm{cm}^2$ $\mathrm{bit}^{-1}$]
\item $W = 4.375$~\mev
\end{itemize}

To evaluate the SEU rate of electronics developed in a standard layout without DICE architecture, two other components of the end  column circuit
 of ToPix~v2 were tested for 
irradiation. \\
The 12 bit shift register was tested using a method similar to the previous one (for the configuration register).\\
 The 12 bit counter was tested with
 a software comparing at  fixed time intervals its value with the one of a ``mirror" counter. 
All the SEU tests were performed with a 10~MHz clock, although the nominal clock 
of ToPix~v2 is 50~MHz, because of the long cable of the setup.\\
The obtained fit parameters, see \tabref{tab:sr_cnt}, in these cases are very different from the Weibull parameters obtained from the 
test of the configuration register. In particular the energy threshold is practically zero, this means that even a small deposited energy can 
trigger an upset. The saturation cross section is bigger by a factor 2, in agreement with the fact that a higher SEU rate is expected in 
 electronics
 designed in this standard layout. 
\begin{table}[!ht]
\begin{center}
\resizebox{\columnwidth}{!}{
\begin{tabular}{|c|c|c|c|}
\hline  Device  &  $E_{\mathrm{th}}$ [MeV] &   $\sigma_0$ [$\mathrm{cm^2}$ $\mathrm{bit^{-1}}$] &  W [MeV] \\
\hline SR  & $1.58 \cdot 10^{-15}$ & $2.6 \cdot10^{-8}$ & 1.03 \\ 
\hline CNT &  $7.31 \cdot 10^{-14}$ &  $2.6 \cdot 10^{-8}$&  1.46\\
\hline
\end{tabular}
}
\end{center}
\caption[Fit parameters of the shift register and the 12b counter in standard electronics]{Fit parameters of the shift register (SR) and the 12b counter (CNT) in standard electronics.}
\label{tab:sr_cnt}
\end{table}

In order to obtain the SEU rate, it is necessary to know the probability for a certain particle to deposit a certain energy in 
the sensitive volume. 
Following the procedure explained in \cite{faccio_seu}, the rate of SEU for the \PANDA
 hadron flux has been calculated, taking the lepton contribution to be negligible.
The evaluated SEU cross section $\Sigma$ in the hadron environment for the three tested elements of the ToPix~v2 prototype, 
has to be corrected with a 30\% safety factor 
to take into account 
the under-estimation of the sensitive volume of 1~$\tcmu$m$^3$, as explained in \cite{faccio_seu}. 
The calculated  $\Sigma$s are reported in \tabref{tab:sigma_seu}.\\

   \begin{table}[!ht]
   \begin{center}
   \begin{tabular}{|c|c|c|}
   \hline  \multicolumn{3}{|c|}{$\Sigma$ [$\mathrm{cm^2}$/bit]}\\
   \hline  CR & SR & CNT \\
   \hline  $11.65\cdot 10^{-16}$ & $3.94 \cdot10^{-14}$ & $6.25 \cdot10^{-14}$\\
   \hline
   \end{tabular}
   \end{center}
      \caption[SEU cross section in the hadron environment, corrected for the
   under-estimation of the SV]{SEU cross section in the hadron environment, corrected for the
   under-estimation of the SV: configuration register (CR), shift register (SR) and counter (CNT). }
      \label{tab:sigma_seu}
   \end{table}
   
   Multiplying the $\Sigma$ cross section by the hadron flux, obtained from a 
   dedicated particle rate study \cite{PANDAnote7}, 
   the SEU rates for annihilations on protons and on nuclei are obtained. The results are reported in \tabref{tab:cross_seu}.
   
   \begin{table}[!ht]
   \begin{center}
   \resizebox{\columnwidth}{!}{
   \begin{tabular}{|c|c|c|c|}
   \hline  \begin{footnotesize}Annihilation \end{footnotesize}&CR & SR & CNT \\
   \hline  $\mathsf{\overline{p}-p}$  & $6.99 \cdot 10^{-9}$ & $23.64 \cdot 10^{-8}$ & $37.5 \cdot 10^{-8}$\\
   \hline  $\mathsf{\overline{p}-N}$  & $1.43 \cdot10^{-9}$  &$ 4.85 \cdot 10^{-8}$ & $7.69 \cdot10^{-8}$\\
   \hline
   \end{tabular}
   }
   \end{center}
    \caption[SEU (bit$^{-1}$ s$^{-1}$) in the ToPix~v2 prototype]{SEU (bit$^{-1}$ s$^{-1}$) in the ToPix~v2 prototype: configuration register (CR), shift register (SR) and counter (CNT).}
   \label{tab:cross_seu}
   \end{table}

Taking into account the final design of ToPix, the SEU rate /h can be derived, taking into account 
that there are 
12760 pixels, the 8 bits for the configuration registers and the 12 bits for the shift register and the counter. 
The estimated rates are listed in \tabref{tab:conclusione_seu}. 
Considering the total number of chips (810) in the MVD pixel part, we can estimate about 2430 upsets per hour. The 
SEU rates foreseen for the
shift register and the counter are higher in comparison to the configuration register as expected.
\begin{table}[!ht]
\begin{center}
\begin{tabular}{|c|c|c|c|}
     \hline \begin{footnotesize}Annihilation \end{footnotesize} & CR & SR & CNT \\
     \hline  $\mathsf{\overline{p}-p}$ &3 & 130 &207\\
     \hline $\mathsf{\overline{p}-N}$ & 1 & 27 & 42 \\
     \hline
     \end{tabular}
     \end{center}
          \caption[SEU (chip$^{-1}$ h$^{-1}$) for the \PANDA environment, in the final ToPix prototype]{SEU (chip$^{-1}$ h$^{-1}$) for the \PANDA environment, in the final ToPix prototype: configuration register (CR), shift register (SR) and counter (CNT).}
          \label{tab:conclusione_seu}
     \end{table}

The variation of one bit in the digital circuitry (SEU), following the passage
of a particle, can be problematic in the \PANDA environment. 
The corruption of the data is not a dangerous effect in itself, 
because the corrupted data can be easily identified and discarded. Instead a
more dangerous effect is the loss of the detector control function that can be
restored resetting or rewriting the chip. As a consequence all the data stored from the
upset event to the reset operation have to be discarded. 
As shown  there is a relevant different SEU rate between a radiation tolerant 
architecture such as the configuration register of ToPix~v2 and the non-radiation
tolerant architecture, e.g.~the shift register and the counter.
Hence triple redundancy architecture application has been designed for the ToPix~v3.

%% file: pixelpart/v3_analog.tex
\begin{figure}[!h]
\begin{center}
 \includegraphics[width=0.45\textwidth]{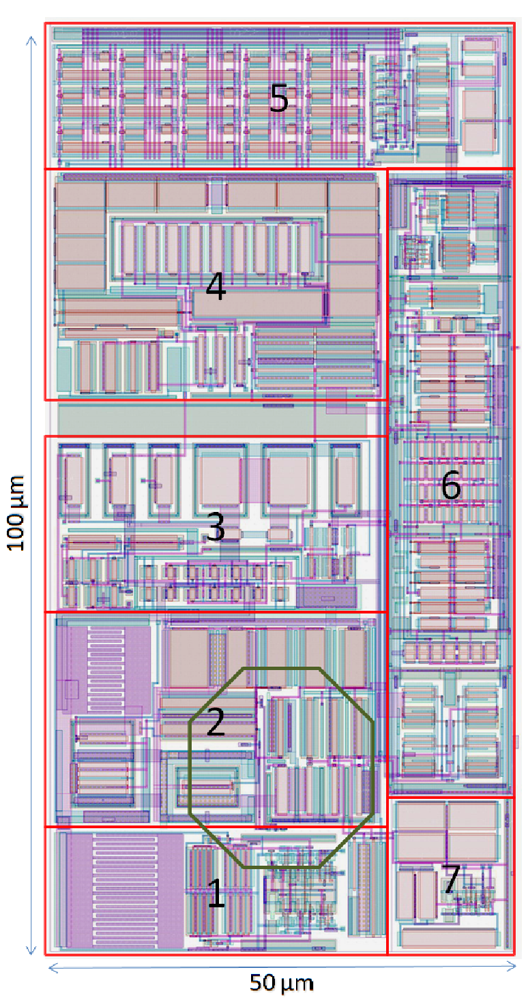}
\end{center}
\caption[Analog Cell Layout]{Analog Cell Layout. 1: Calibration circuit, 2: Charge Sensitive Amplifier, 
3: Constant Current Feedback, 4: Baseline Holder, 5: 5 bit DAC for the threshold correction, 
6: Comparator, 7: Large charge cut-off circuit. The octagonal shaped bump bonding pad area is shown in green.}
\label{fig:alayout}
\end{figure}

The Analogue Cell Layout is implemented in an area of $\mathrm{50~\tcmu m \times 100~\tcmu m}$.
The floorplan has been optimised in order to place the comparator between the most sensible analogue blocks 
(preamplifier, constant current feedback and baseline holder) and the digital cell.
In \figref{fig:alayout} the different components of the analogue cell are shown.

\begin{figure}[!h]
\begin{center}
 \includegraphics[width=0.50\textwidth]{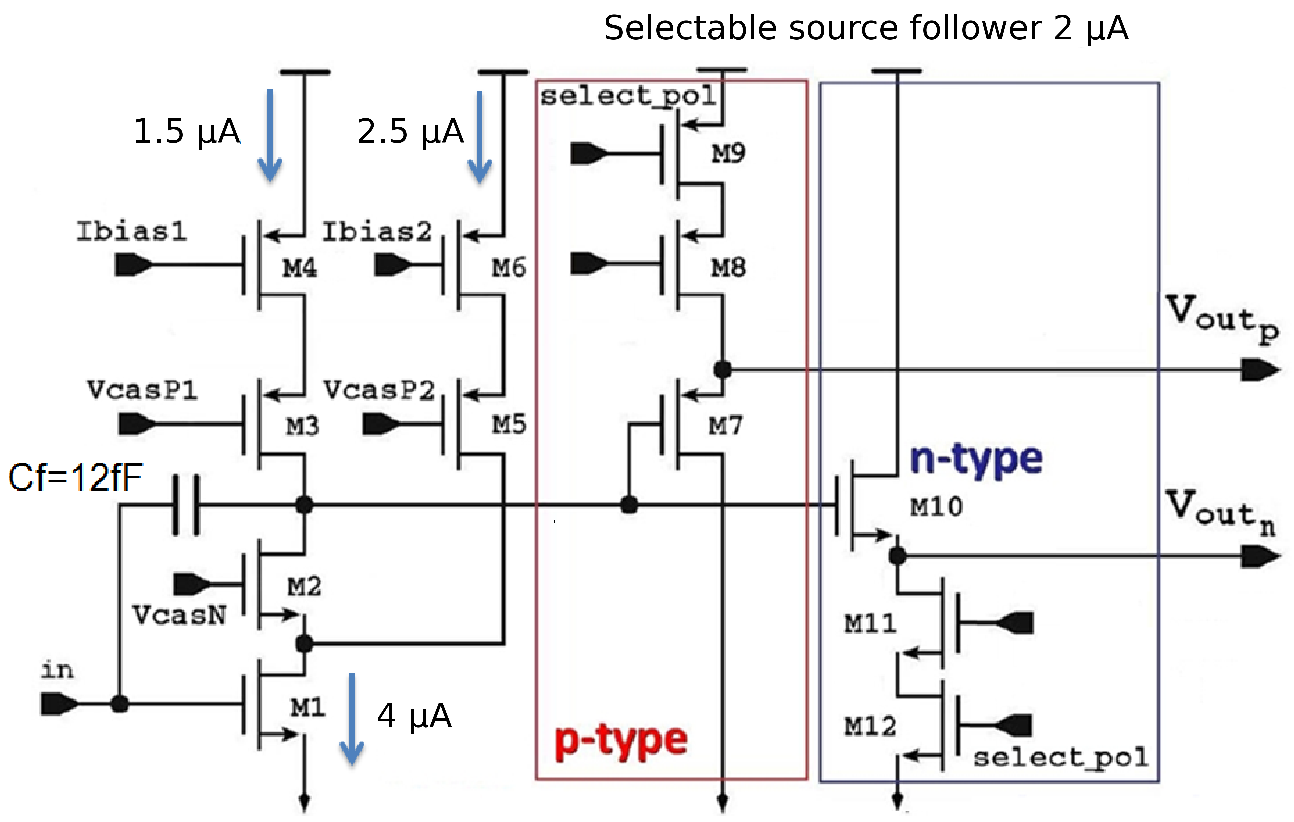}
\end{center}
\caption[Charge Sensitive Amplifier schematic]{Charge Sensitive Amplifier schematic.}
\label{fig:CSA_sch}
\end{figure}

\textbf{Charge Sensitive Amplifier}\\
The bias lines are propagated vertically, low levels metal are employed for the bias line and high levels metal for the power supply lines.
In order to avoid the propagation of spurious signals through the power supply, 7 separate lines have been employed.
The input cascode stage of the CSA has a dedicated ground (gnd\_pre).
The triple well transistors of the input stage have a dedicated supply net (vdd\_pre) as shield.
The leakage compensation stage bias is given by vdd\_ilc.
The comparator has a dedicated supply line (vdd\_d) and a dedicated ground (gnd\_d).
All the other analogue blocks are supplied by vdd\_a and gnd\_a.
The input stage is a gain enhanced cascode amplifier with capacitive feedback $\mathrm{C_f}$.
A feedback capacitance of 12~fF, which is half of the one employed in ToPix~v2, has been chosen as the best compromise between the need of maximising the gain for low signals and the one of having an adequate phase margin in the feed-back loop. The output stage is a source follower which is employed as a voltage buffer to drive the output capacitive load. The schematic view of this stage is shown in \figref{fig:CSA_sch}.

\begin{figure}
\begin{center}
  \includegraphics[width=0.45\textwidth]{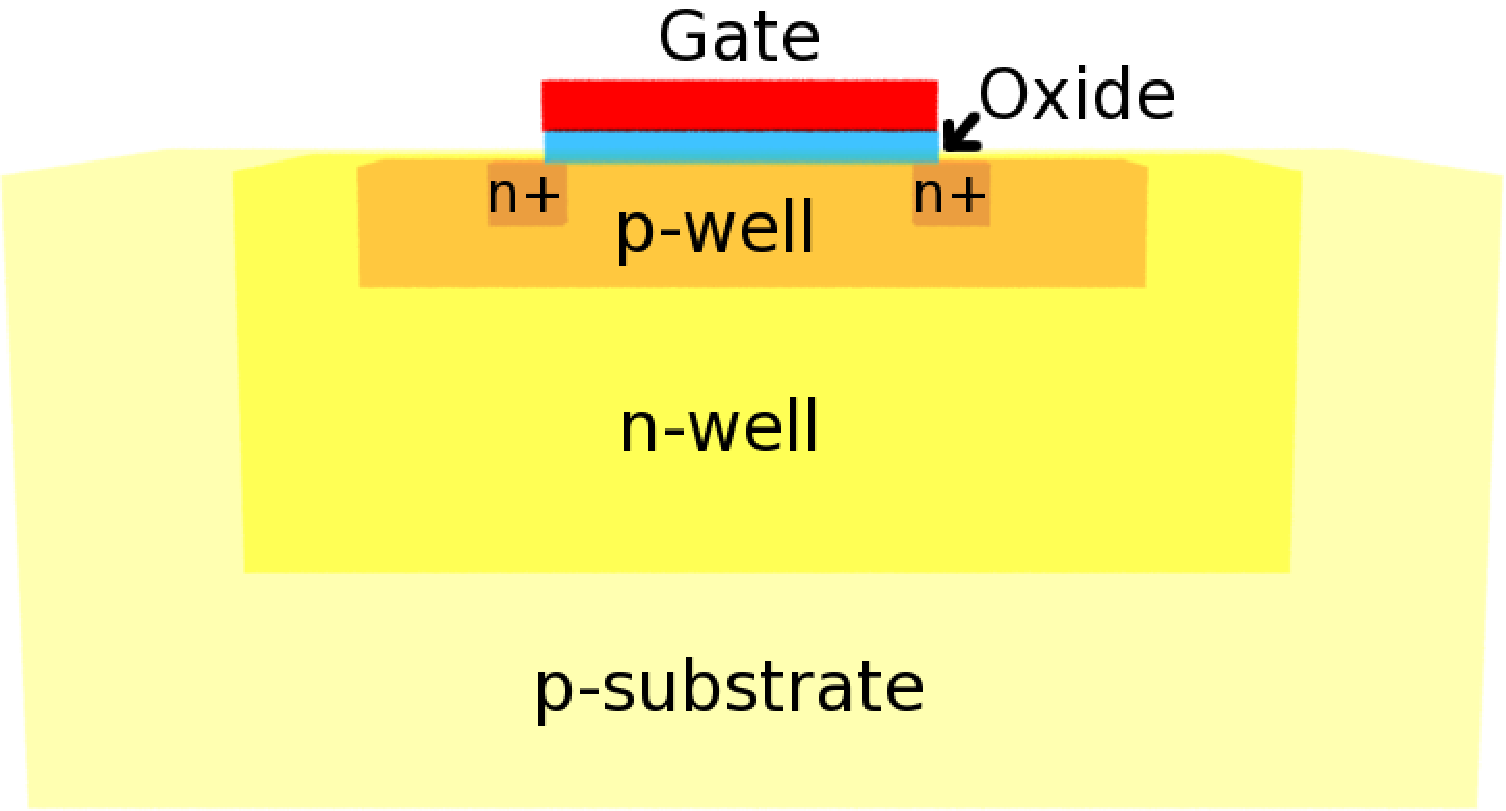}
\end{center}
\caption[Triple well graphical representation]{Triple well graphical representation (not to scale).}
\label{fig:triplewell}
\end{figure}

The chip will be produced in a p-substrate process, therefore the standard NMOS transistors are 
fabricated directly on the chip substrate. The process offers also the possibility of implementing 
the NMOS devices in dedicated wells (triple well option) as sketched in \figref{fig:triplewell}.
In this case a deep n-well is fabricated in which the p-well hosting the NMOS transistors is put. 
This solution allows a good shielding of the transistor from the common chip substrate at the expense 
of an increased area.
In a mixed mode design it is very important to protect the sensitive
nodes from the potential interference of the digital switching
circuitry. The triple well solution has hence been used for the input
transistor M1 and its cascode device M2.

The CSA output stage is a source follower, which is employed as a voltage buffer to drive the loading capacitances at the input of the feedback stages and of the comparator.
Since the source follower has asymmetric rise and fall times and the polarity of the injected charge determines the direction of the leading edge, a NMOS source follower is needed for n-type sensors and a PMOS one for p-type sensors, hence maximising the output swing. The selection between the two is done via the select\_pol status bit.
In p-type configuration select\_pol is low, M9 is switched on and M12 is switched off.
In n-type configuration select\_pol is high, M12 is switched on and M9 is switched off.
This stage is biased with a nominal current of 2~$\tcmu$A.
The output DC voltage is regulated by the baseline holder. The value of the baseline voltage V\_ref must be set according to the sensor type. Typical values are 300~mV for n-type sensors and 700~mV for p-type ones.
The buffers have the source and bulk short circuited in order to
suppress the bulk effect that would decrease their voltage gain. For
this reason transistor M11 in \figref{fig:CSA_sch} is implemented
as a triple well device in order to decouple its bulk from the global
substrate and to allow the source-bulk connection.\\

\begin{figure}[!h]
\begin{center}
 \includegraphics[width=0.5\textwidth]{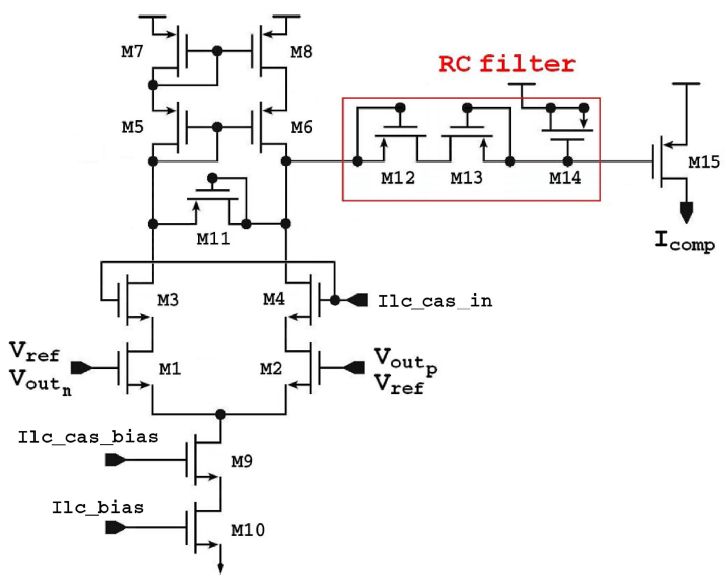}
\end{center}
\caption[CSA with a feedback circuitry which automatically compensates for the detector leakage current]{CSA with a feedback circuitry which automatically compensates for the detector leakage current proposed in \cite{Krummenacher91}.}
\label{fig:bslinerest}
\end{figure}

\textbf{Baseline Holder}\\
The baseline holder is a specific feedback network that controls the DC output of the preamplifier, making it insensitive to the sensor 
leakage current.
As a first step the solution presented in \cite{Krummenacher91} has been considered for the leakage compensation.
A similar circuit has been designed in the current technology with a filtering capacitance of 10~pF in order to 
evaluate
its leakage compensation capabilities. This filtering capacitance requires an area of about 500~$\tcmu$m$^2$, 
which is close
to the maximum area that can be allocated for that component. The simulation results gave a ToT compression of 
13\%
for a 10~nA leakage current and of 38\% for a 50~nA current. Therefore a different approach, based on a dual 
feedback
loop has been preferred. This novel approach, shown in \figref{fig:bslinerest}, is based on the fact that 
in deep submicron
technologies MOS transistors with $V_{\text{GS}}=0$ can provide a resistance in the G$\Omega$ range.
Thanks to the high value of the equivalent resistance, a capacitance in the order of 5~pF is sufficient
to achieve a filter cut off frequency in the order of 10~Hz. The filtering capacitor is implemented
through MOS devices exploiting the gate oxide, and it occupies an area of 172~$\tcmu \mathrm{m^2}$.

The output of the CSA is sensed by a differential pair and compared to the preset reference voltage.
The resulting error signal is used to drive a PMOS, which acts as a
voltage controlled current source and injects the compensation current
into the input node.\\

\begin{figure}[!h]
\centering
 \includegraphics[width=0.5\textwidth]{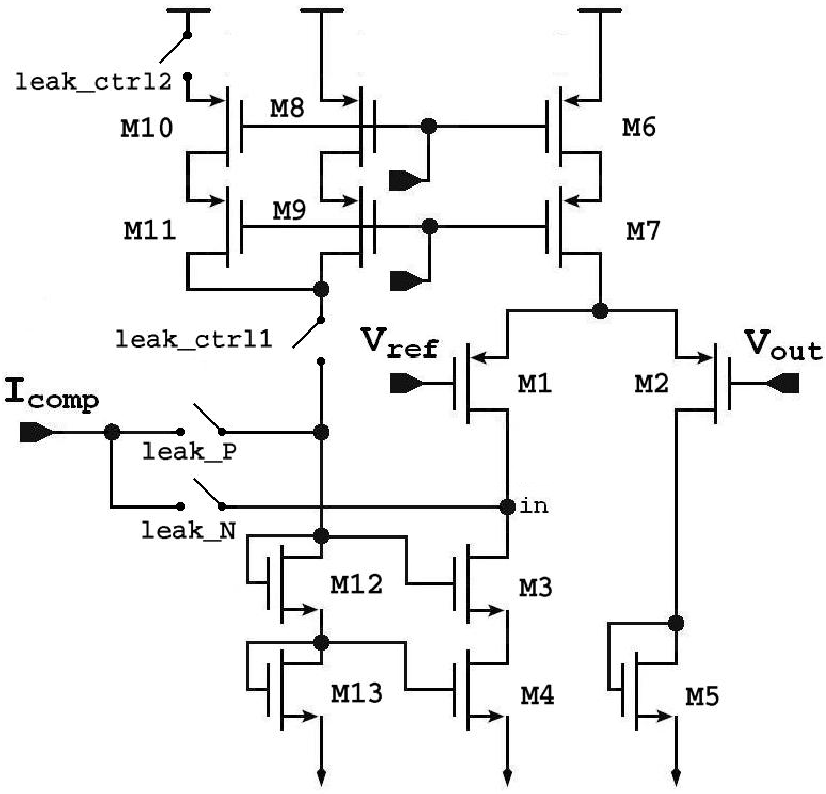}
\caption[Schematic of the constant feedback current generator]{Schematic of the constant feedback current generator.}
\label{fig:IFB}
\end{figure}

\textbf{Constant Feedback Current Generator}\\
The constant  current feedback generator is a network that provides
the CSA discharging current \cite{PericThesis}.
\Figref{fig:IFB} shows the schematic view of this stage.
The CSA output signal is presented to the gate of M2, while M1 receives the baseline reference voltage.
With a p-type sensor, $V_{\text{out}}=0$ will have a negative polarity, steering all the current from M1 to M2. In this case the current in M3 is used to recharge the feed-back capacitor.
With n-type sensors, $V_{\text{out}}=0$ will be positive, steering the current from M2 to M1. M1 will have an excess current that will discharge the feedback capacitor.
At the equilibrium this stage provides an equivalent small signal feed-back resistance of 6.7~M$\Omega$  with $I_{\text{FB}}= 5$~nA.

This stage provides also the injection of the leakage compensation current at the input node.
Using four switches it is possible to change the topology of the
circuit in order to choose the polarity of the current to be
compensated.\\

%\newpage
\textbf{ToT Linear Range Limitation}\\
When the CSA saturates the input node rises (p-type case) or drops (n-type case).
 In both cases when the transistors connected at the input node are pushed out from the saturation region, 
 the discharging current increases leading to a compression of the ToT signal. This phenomenon depends on the 
 input capacitance and the input DC voltage.
Since the input DC voltage is at $\sim 200$~mV, assuming a detector capacitance of 200~fF, for an n-type sensor 
the compression occurs from $Q_{\text{in}} \approx 90$~fC.
For a p-type sensor a much larger current ($\sim 200$~fC) could be
collected on the input capacitance while ensuring the proper bias of
the constant current feedback.\\

\begin{figure}
\begin{center}
  \includegraphics[width=0.5\textwidth]{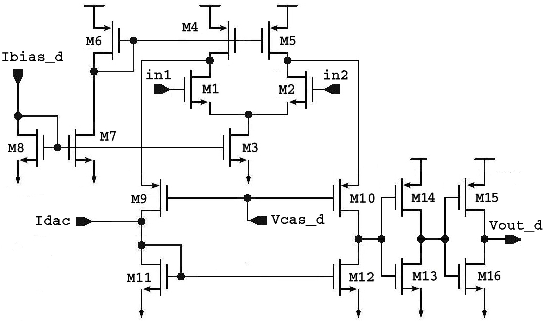}
\end{center}
\caption[Comparator schematic]{Comparator schematic.}
\label{fig:ToPix_comp}
\end{figure}

\textbf{Comparator}\\
The comparator has a folded cascode input stage, and two CMOS inverters in cascade as driver in order to have a fast transition between the two logic states.
Since the comparator performance in ToPix~v2 was satisfactory, the
transistor level implementation has not been changed. However the
layout has been redesigned in order to fit in the available cell area.
\Figref{fig:ToPix_comp} shows the transistor level implementation. The comparator gives a high level output when $V_{\text{in2}} > V_{\text{in1}}$.
Therefore in n-type configuration, where the CSA output is positive, in1 is connected to the comparator reference voltage and the CSA output to in2. In p-type configuration where the CSA output is negative the signals are exchanged using CMOS switches.\\

\textbf{DAC Current Source}\\
Since Vref\_d  is common to all pixels, in order to mitigate the
threshold dispersion a local 5 bit DAC is added in each pixel, to
allow a fine tuning of the threshold on a pixel by pixel basis.

\begin{figure}[!h]
\begin{center}
  \includegraphics[width=0.5\textwidth]{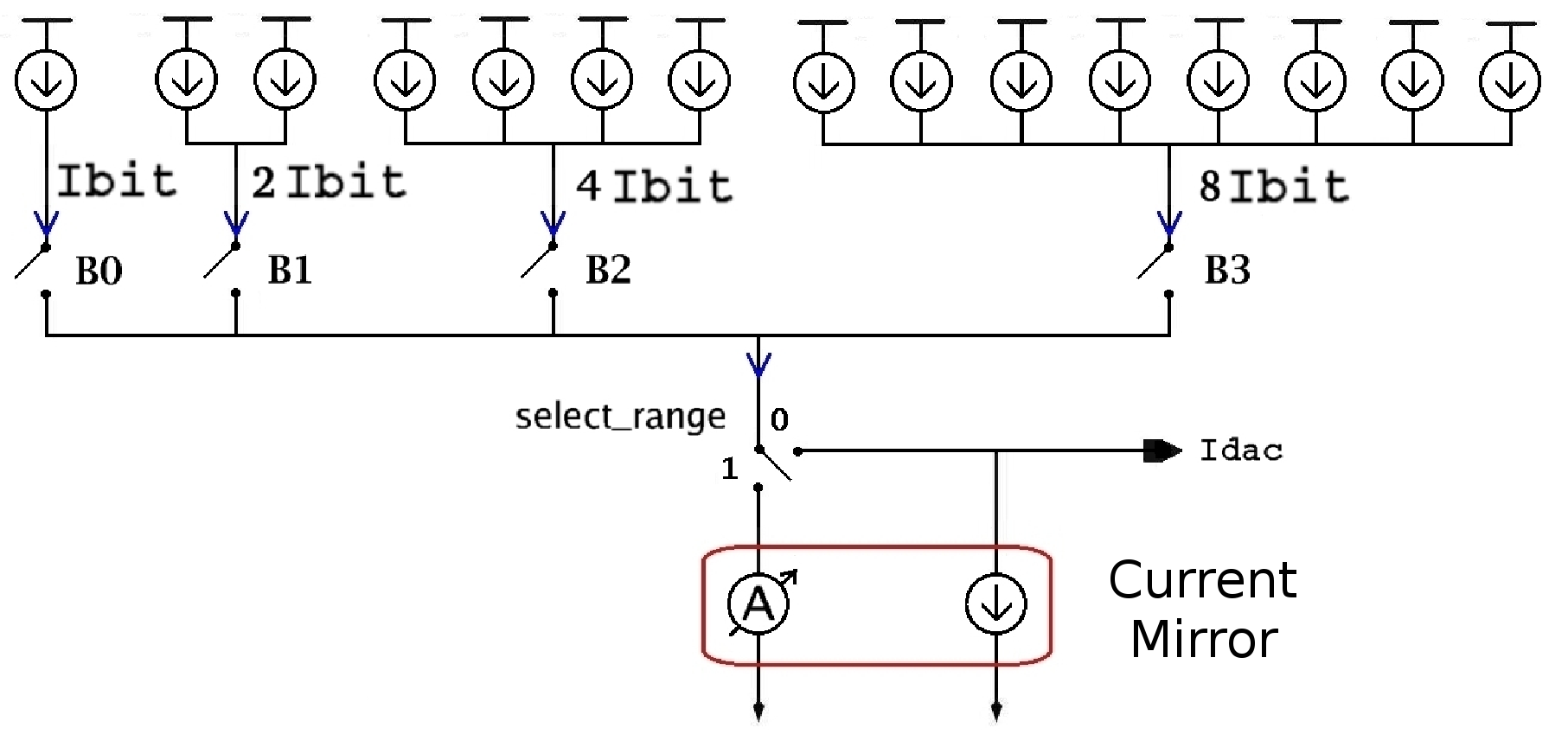}
\end{center}
\caption[Simplified model of the 5 bit DAC]{Simplified model of the 5 bit DAC.}
\label{fig:DAC}
\end{figure}

A simplified scheme of the DAC is shown in \figref{fig:DAC}. For simplicity the switches are represented in common for each group of Ibit current generators.
Each of the bit {B0, B1, B2, B3} controls a switch that activates the relative current generator.\\

\textbf{Clipping Circuit}\\
A clipping circuit has been designed in order to protect the \frontend when a large charge is presented at the input node, thus avoiding long dead times.
The \frontend requires to measure charges up to 50~fC, so larger charges have to be discharged with an extra current.
The clipping circuit exploits the fact that the input node rises
(p-type case) when the CSA saturates. Connecting the input node at the
gate and drain of a low power NMOS and its source at ground, it is
possible to clip the voltage signals larger than the threshold voltage
of the transistor ($V_{\text{th}} \approx 600$~mV).\\

\begin{figure}[!h]
\begin{center}
 \includegraphics[width=0.5\textwidth]{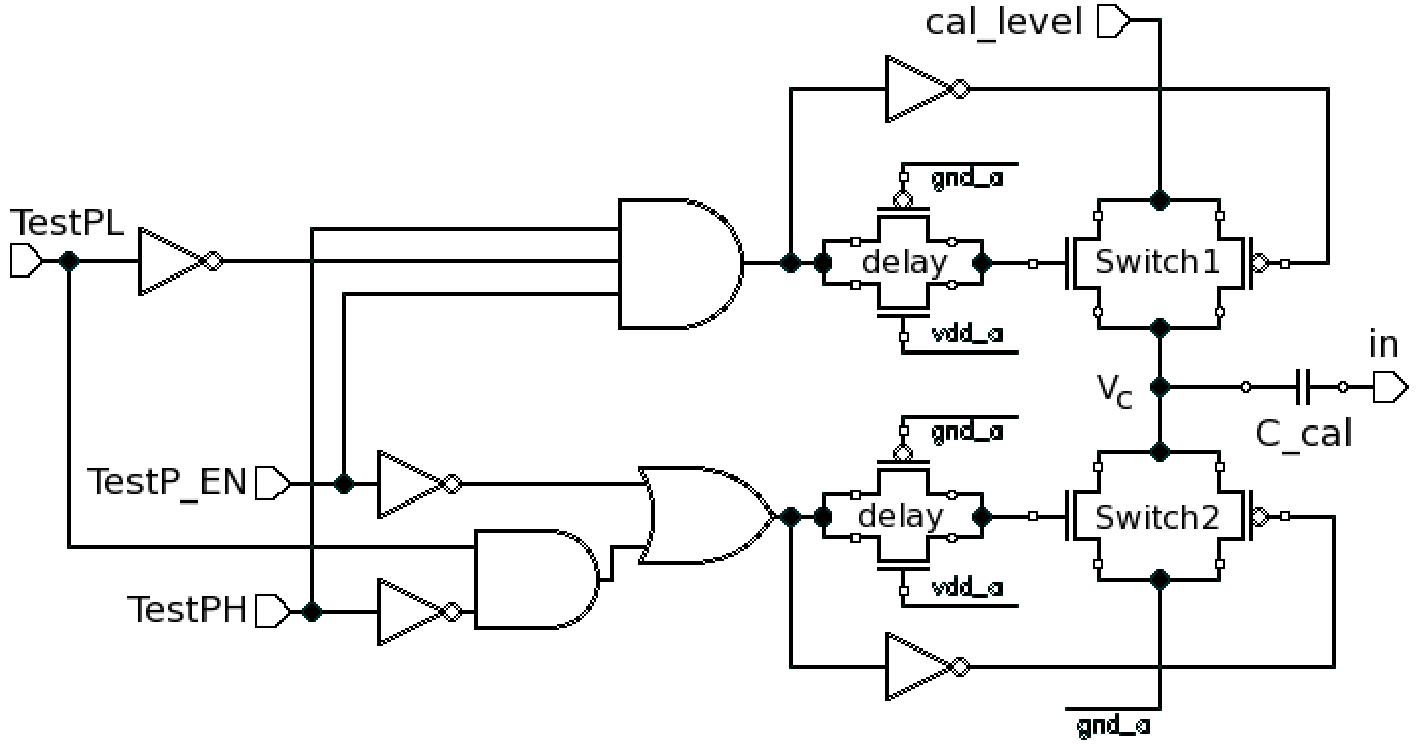}
\end{center}
\caption[Calibration circuit schematic]{Calibration circuit schematic.}
\label{fig:cal_sch}
\end{figure}

\textbf{Calibration Circuit}\\
The calibration circuit allows to inject a current pulse ($I_{\text{inj}}$)
into the preamplifier to test the circuit response.
\Figref{fig:cal_sch} shows the schematic of the calibration circuit.
The TestP signal is a CMOS differential signal common to all the pixels. The positive line is TestPH and the negative one is TestPL.
TestP\_EN is a status bit of the pixel configuration register, and
activates the injection subcircuit.

%% file: strippart/strippart.tex
%\begin{bibunit}[plain]

\chapter{Silicon Strip Part} %Bonn and J\"ulich
\label{strippart}

This chapter describes the silicon strip part of the MVD,
which consists of a barrel and a forward disk section as outlined in section~\ref{Intro-MVD-Layout}.
The basic design foresees double-sided strip detectors 
of rectangular and trapezoidal shape, respectively.
The overall concept of the MVD is based on a standard process for radiation hard sensors with a target thickness of 200 to 300~$\tcmu$m.
It follows approved solutions of other tracking systems 
already installed at HEP facilities.
As for the entire apparatus the trigger-less readout concept represents one of the major technical challenges 
and requires the implementation of new technologies. 
Moreover, the readout on both sensor sides requires sophisticated 
technical solutions for the hybridisation of the detector modules.

%	\section{Overview}
	\section{Double-Sided Silicon Strip Detectors (DSSD)}
	\label{sec:dssd}
	Presently silicon strip detectors are deployed to a great extent as tracking detectors in particle physics experiment worldwide. The division of the semiconducting detector by means of strips enables a one-dimensional spatial resolution of particle tracks traversing the detector. The utilisation of double-sided strip detectors offers the ability of precise point reconstruction with the advantage of less detector channels compared to pixel detectors. However, ambiguities occur in case of multiple particles crossing the detector in the same time frame. Therefore the \panda MVD utilises silicon strip detectors at the outer layers of the MVD where the particle flux is reduced in contrast to the inner layers.

		\subsection{Barrel Sensors}
		\authalert{author: Hans-Georg Zaunick, zaunick@hiskp.uni-bonn.de}

\begin{figure}[]
				\centering
				%\fbox{
				  \includegraphics[width=0.475\textwidth]{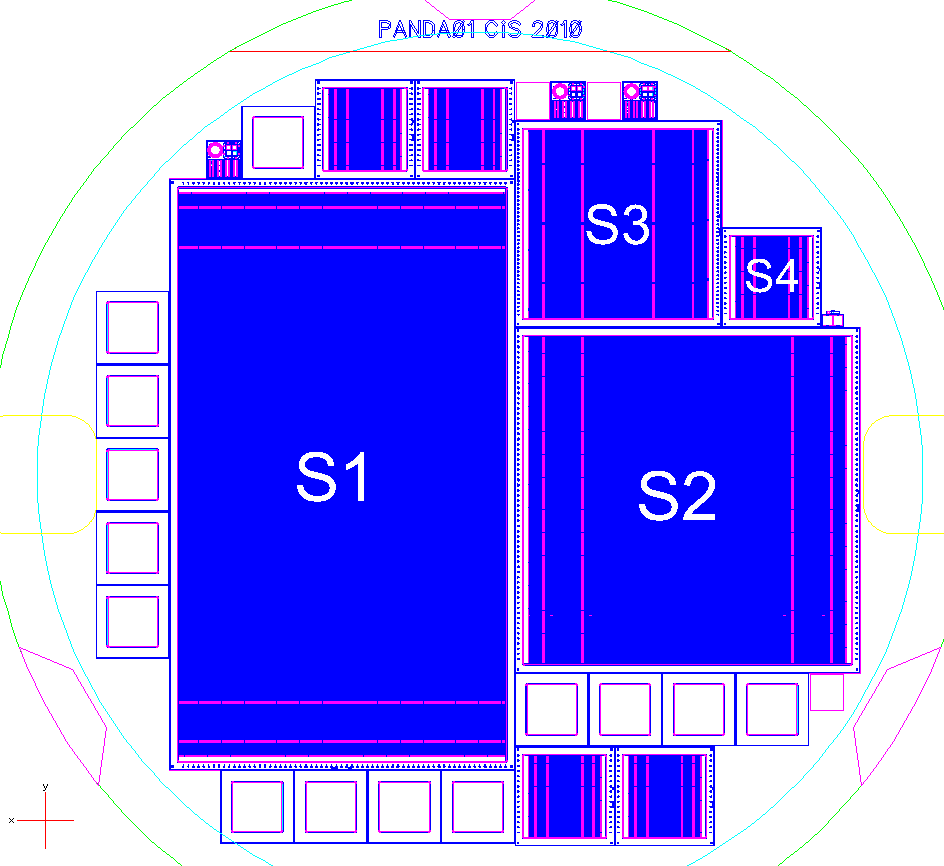}
				%}
				\caption[Floorplan of wafer with full size prototype sensors]{Floorplan of wafer with full size prototype sensors. The Sensor properties are comprehended in \tabref{tab:strip:sensor-params}.
				}
				\label{fig:strip:cis-wafer-floorplan}
			\end{figure}

			First \panda full size prototype sensors have been produced in the first half of 2011 by the company CiS GmbH in Erfurt (Germany).
			The production process utilises a 10+2-layer process with a single metal layer laid out on 4$''$-wafers consisting of \textlangle111\textrangle~-~cut 
			material. A floorplan of the designed layout of the implemented structures can be seen in \figref{fig:strip:cis-wafer-floorplan}.
			
			\Tabref{tab:strip:sensor-params} gives an overview of the specifications of the wafer and of the implemented sensors. As starting material, a monocrystalline Floating Zone (FZ) silicon wafer was chosen. For the implemented sensor structures front side strips are realised as p$^+$ in n doping, whereas the back side contains n$^+$-strips embedded in the n-substrate. All sensors are double-sided with strip structures oriented orthogonal to each other on either side. The depletion is achieved by punch-through biassing from the bias ring towards the strips on both front and back side and the n-side charge separation is realised by p-spray implants. The active area is protected by eight guard rings to assure a stable electric field within the active area. In this scheme, front side is considered the junction contact, or p-side, while the ohmic, or n-side, is always referred to as back side.

			\begin{table}[width=0.4\textwidth]
			\centering
				\begin{tabular}{|l|c|}
				% after \\: \hline or \cline{col1-col2} \cline{col3-col4}
				\hline
				\multicolumn{2}{|l|}{\textbf{General}}  \\ \hline
				wafer material & FZ Si, 4$''$, n/P   \\ \hline
				thickness & 285~$\pm$~10~$\tcmu$m  \\ \hline
				resistivity & 2.3\,$\dots$\,5.0~k$\Ohm\cdot\mathrm{cm}$  \\ \hline
				n-side isolation & p-spray  \\ \hline
				guard rings & 8  \\ \hline
				stereo angle & 90\textdegree \\ \hline
				passive rim & 860~$\tcmu$m \\ \hline
				\multicolumn{2}{|l|}{\textbf{S1}}  \\ \hline
				n-side strips & 896  \\ \hline
				p-side strips & 512  \\ \hline
				pitch & 65~$\tcmu$m \\ \hline
				active area & 58.275~$\times$~33.315~mm$^2$\\ \hline
				\multicolumn{2}{|l|}{\textbf{S2}}  \\ \hline
				n-side strips & 512  \\ \hline
				p-side strips & 512  \\ \hline
				pitch & 65~$\tcmu$m \\ \hline
				active area & 33.315~$\times$~33.315~mm$^2$\\ \hline
				\multicolumn{2}{|l|}{\textbf{S3}}  \\ \hline
				n-side strips & 384  \\ \hline
				p-side strips & 384  \\ \hline
				pitch & 50~$\tcmu$m \\ \hline
				active area & 19.230~$\times$~19.230~mm$^2$\\ \hline
				\multicolumn{2}{|l|}{\textbf{S4}}  \\ \hline
				n-side strips & 128  \\ \hline
				p-side strips & 128  \\ \hline
				pitch & 65~$\tcmu$m \\ \hline
				active area & 8.355~$\times$~8.355~mm$^2$\\ \hline
				\end{tabular}
			\caption[Specifications of the prototype strip sensors]{Specifications of the prototype strip sensors.}
			\label{tab:strip:sensor-params}
			\end{table}

\begin{figure}[]
				\centering
				%\fbox{
				  \includegraphics[width=0.475\textwidth]{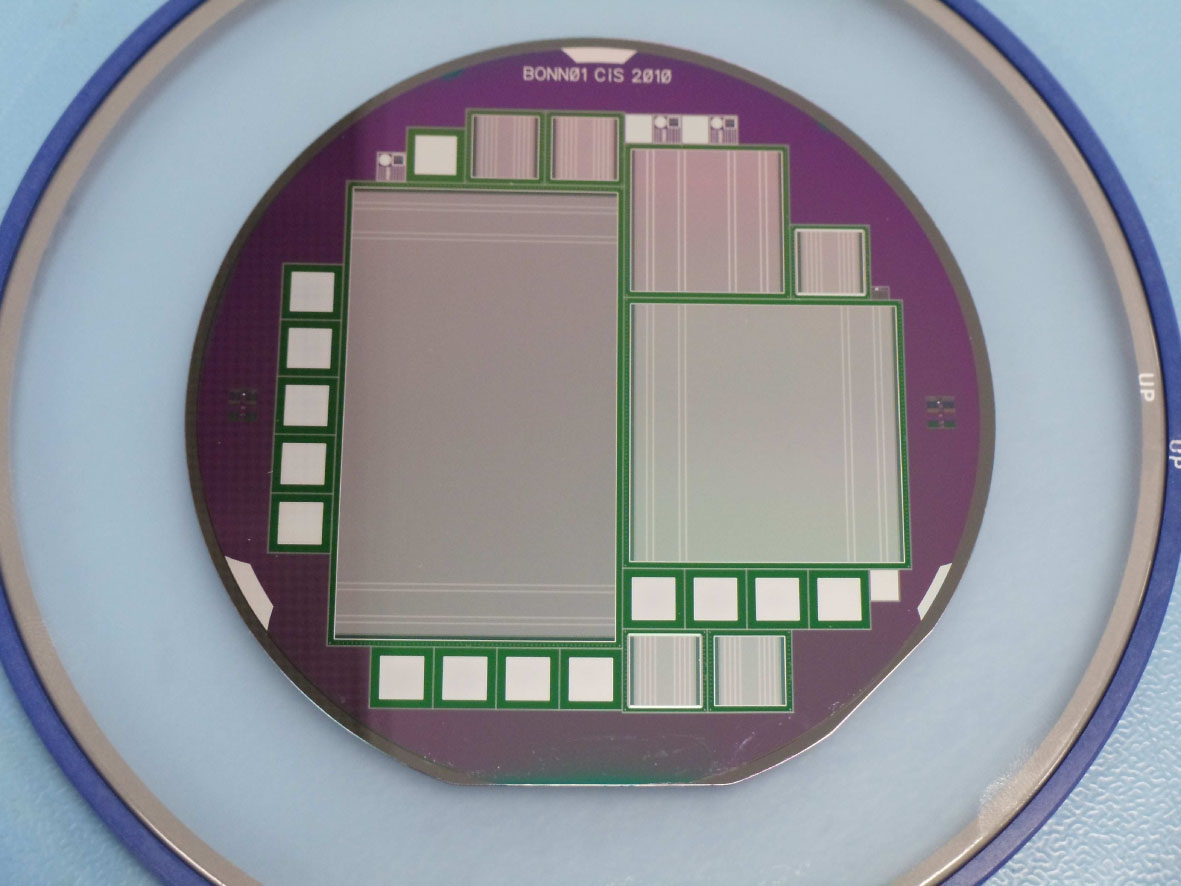}
				%}
				\caption[Wafer including Prototype Sensors]{4$''$-Wafer with prototype sensors.}
				\label{fig:strip:cis-wafer}
			\end{figure}

			Besides the full size \panda prototype sensors, a smaller, square shaped sensor S3 with 50~$\tcmu$m pitch and an active area of roughly 2~$\times$~2~cm$^2$ was realised on the delivered wafer (see \figref{fig:strip:cis-wafer}). This sensor was foreseen to be an equally sized replacement for already existing prototype sensors from earlier prototyping stages in order to benefit from the existing readout infrastructure. Furthermore, five ``Baby''-Sensors (S4) with only 128 strips on either side were placed on the wafer. Additional test structures are implemented as well, serving mainly as markings during wet-processing stages, bonding calibration tags or similar. Finally, fourteen  diodes with different numbers of guard rings are placed around the main elements on the wafer. These diodes may be used to derive the radiation dose during irradiation tests.

\begin{figure}[]
				\centering
				\fbox{
				  \includegraphics[width=0.225\textwidth]{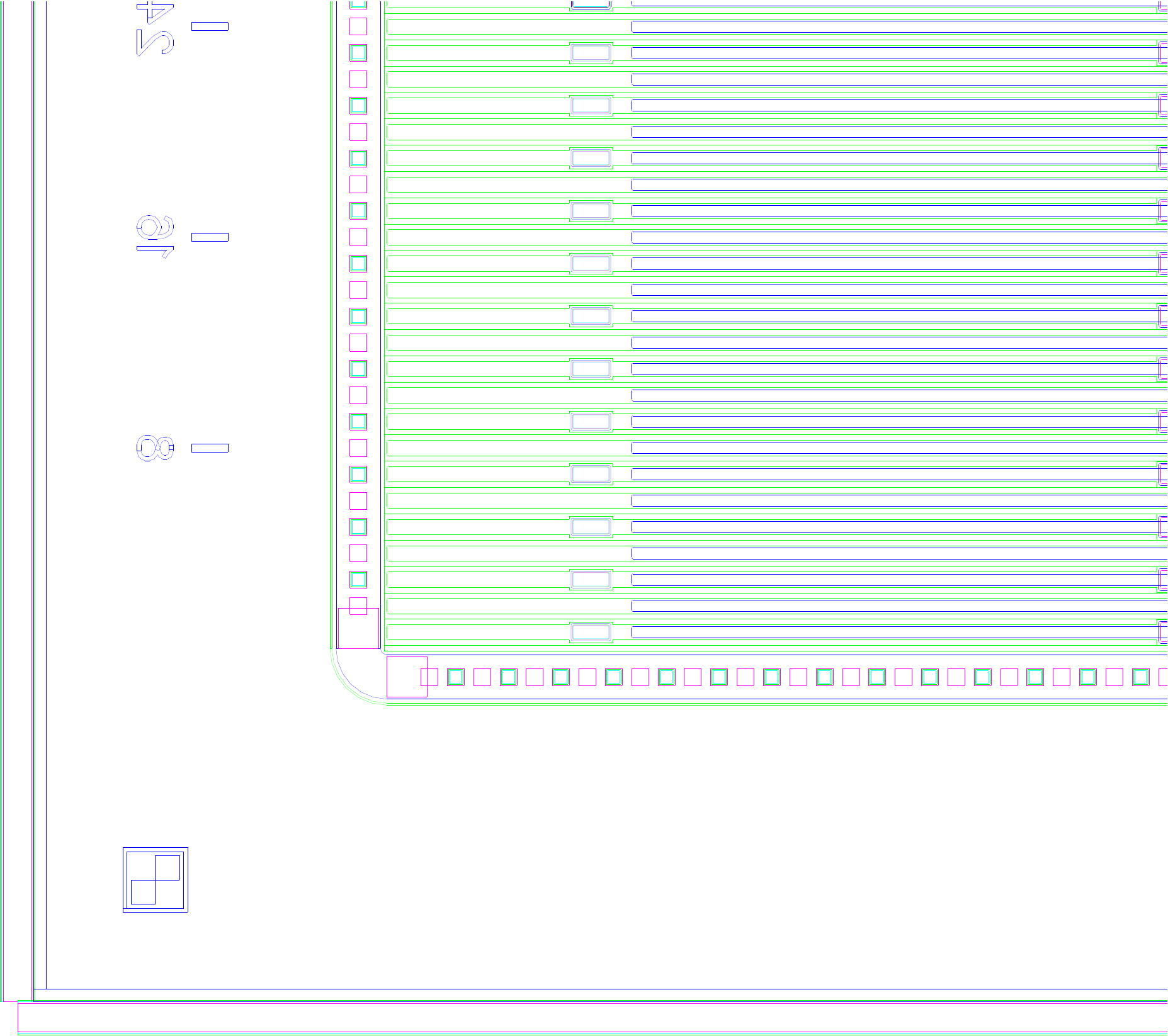}
				  \hfill
				  \includegraphics[width=0.225\textwidth]{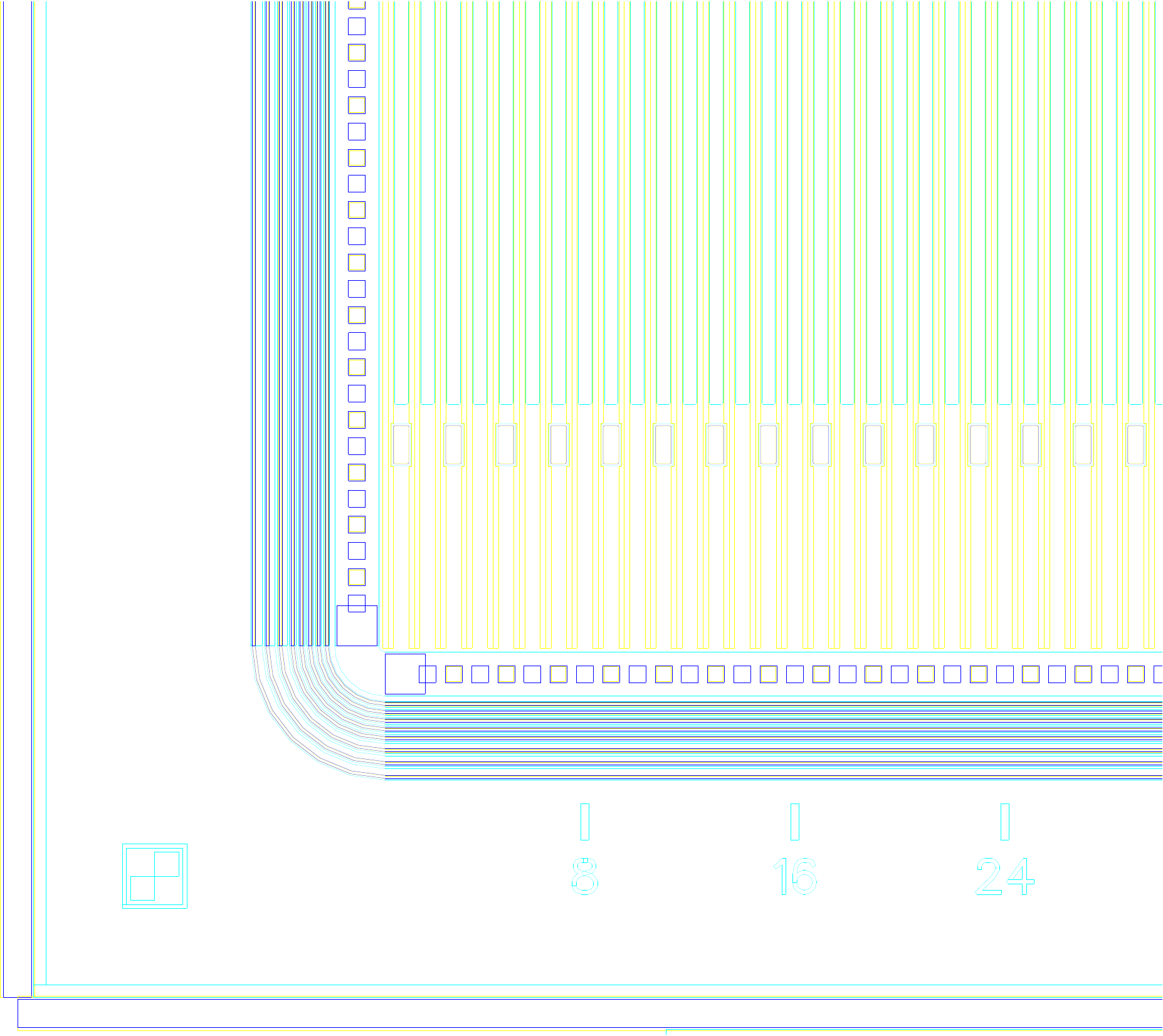}
				}
				\caption[Layout detail of MVD-Barrel DSSD Sensors]{Layout detail of MVD-Barrel Double-Sided Strip Sensors: ohmic- or n-side (left), junction- or p-side (right).}
				\label{fig:strip:cis-sensors1}
			\end{figure}

			In \figref{fig:strip:cis-sensors1} a corner of the sensors on the n-side (left picture) and one on the p-side (right picture) are shown with the first strips and the high potential bias contact ring with square-shaped passivation openings. The pads on the n-side strips are connected to the odd-numbered n$^+$-implants, while the even-numbered pads are located at the opposite edge (not shown). On the junction- or p-side additional guard ring structures for insulation of the high potential difference to the bias line are necessary. The contacts along the inner bias line are used for the supply of the negative bias potential. The DC-pads on the strips are direct connections to the p-side 
			p$^+$-implants. For ease of channel identification, additional orientation numberings and position marks are implanted inside the metal layer.

\begin{figure}[]
				\centering
				%\fbox{
				  \includegraphics[width=0.5\textwidth]{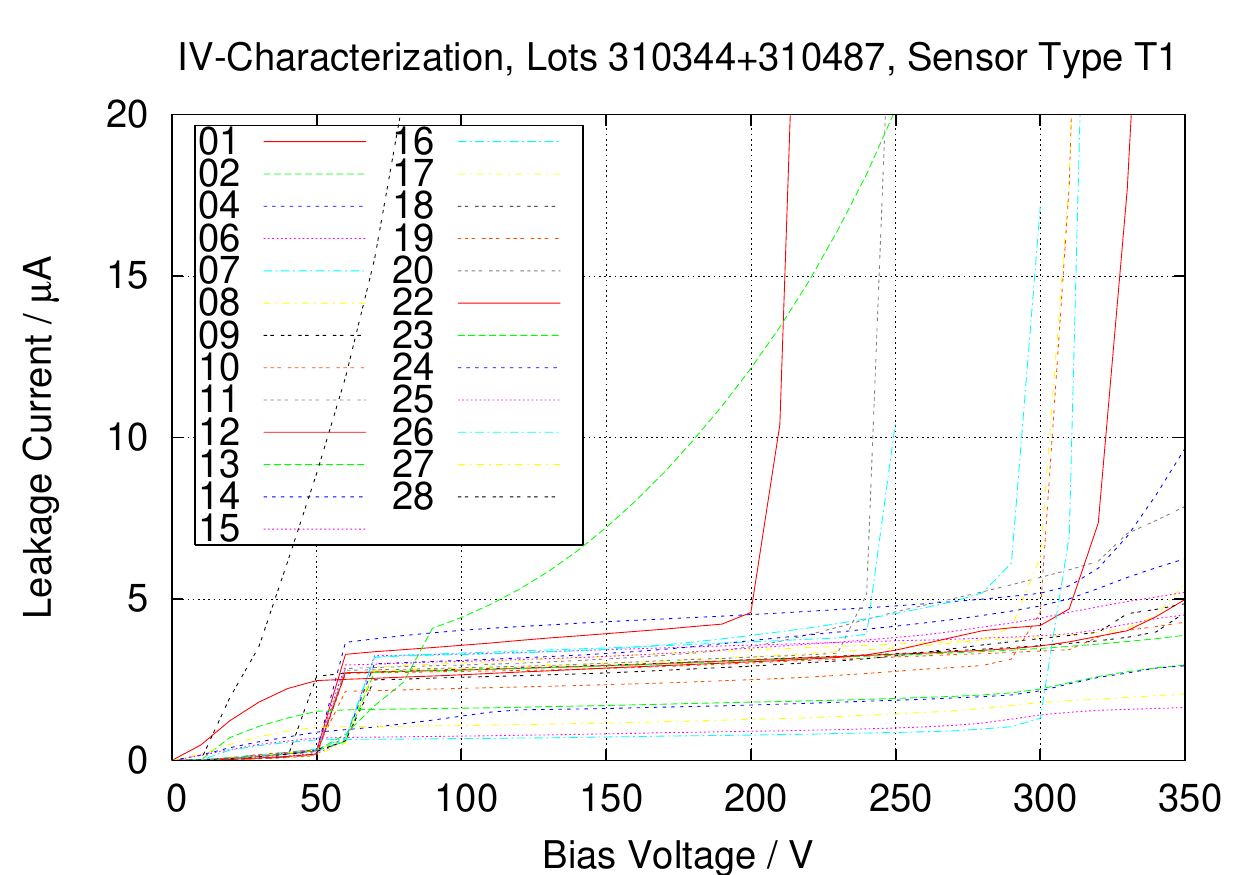}
				%}
				\caption[I-V-Curves of Si-Strip prototype Sensors]{I-V-Curves of full size ($\mathrm{6.0 \times 3.5~cm^2}$) silicon sensor prototypes originating from 25 different wafers. At full depletion the majority of the sensors shows a nearly constant leakage current between $\approx\,1\,\dots\,4~\tcmu$A over a wide reverse voltage range. The steep increase at high voltages is due to the avalanche breakdown and varies significantly from wafer to wafer. The temperature during the measurements was held constant at 20~$^\circ$C.}
				\label{fig:strip:cis-wafer-iv}
			\end{figure}

\begin{figure}[]
				\centering
				%\fbox{
				  \includegraphics[height=0.495\textwidth,angle=90]{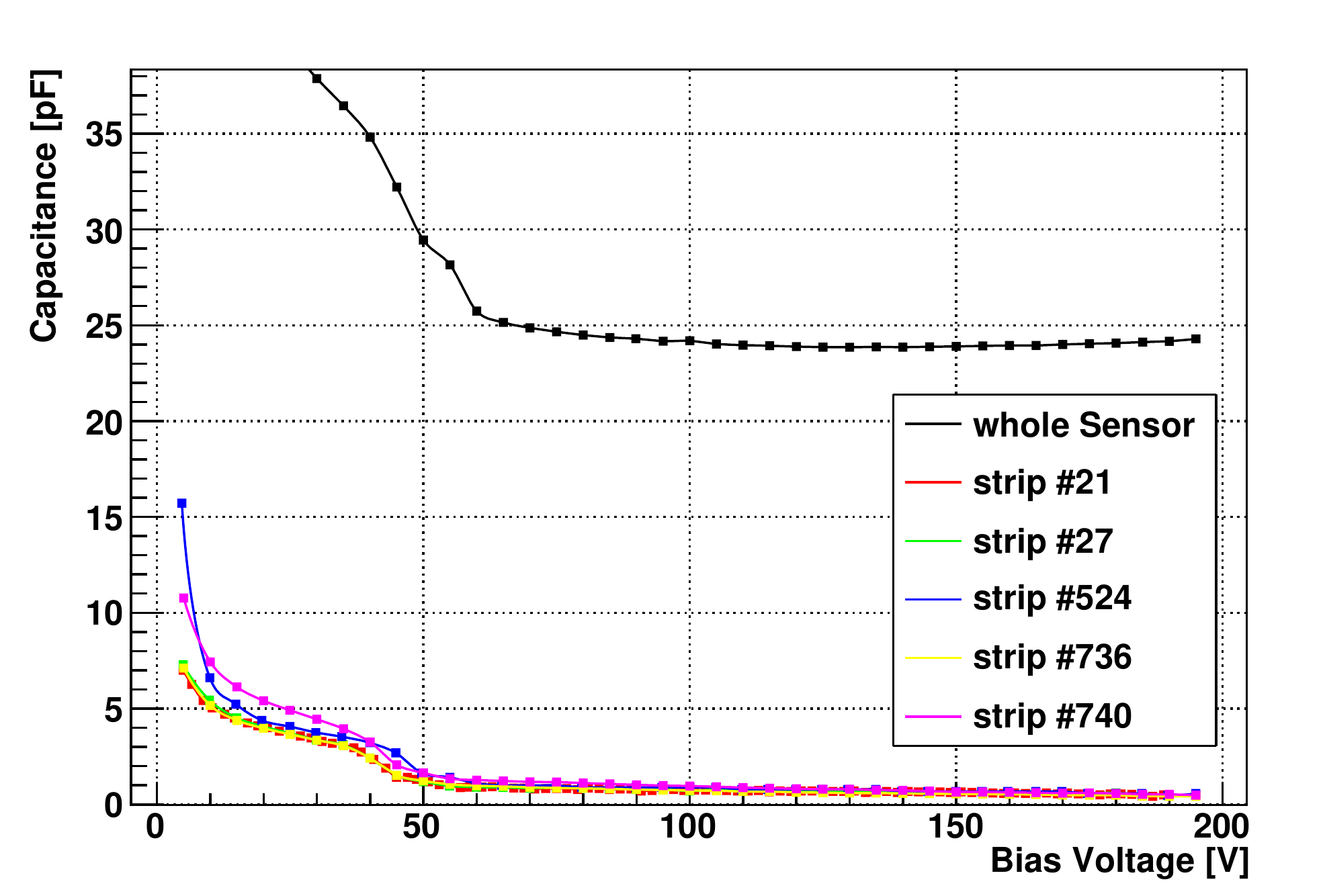}
				%}
				\caption[Capacitance characteristics of large prototype Sensor]{Capacitance characteristics of p-side strips of one large ($\mathrm{6.0 \times 3.5~cm^2}$) prototype sensor. At full depletion the capacitance of single strips lies well below 1~pF. The capacitance between bias and bulk contacts (black) is shown for comparison.}
				\label{fig:strip:cis-wafer-cv}
			\end{figure}

			A measurement of the leakage current flowing through the sensor with applied variable reverse voltage at the bias contacts is shown in \figref{fig:strip:cis-wafer-iv}. The recorded curves are the {I-V}\,-\,characteristics of the large sensors (6.0~$\times$~3.5~cm$^2$) from 25 different wafers out of two lots. Almost all of the sensors show the expected behaviour, 
			i.e.~very low and constant leakage current above full depletion and a steeply increasing current at reverse breakdown voltage. Two of the analysed sensors obviously completely fall off this pattern; one which shows a continuous strong rise of the current even at low voltages and one that appears to become depleted at all but shows excessive leakage at higher voltages. It is already clear from this first picture that sensors past production should be casted into quality categories to ensure equal parameters of the sensors utilised in a collective context (in this case the modules inside the MVD).
			The full depletion voltage for each sensor is extracted before the actual assembly takes place. This is done in a probe station setup by sampling the capacitance vs. bias voltage characteristics as seen in \figref{fig:strip:cis-wafer-cv}. The p-side single strip capacitances are measured with typical values of $\approx$\,450~fF at a depletion voltage of 150~V. For comparison the total capacitance at the bias contacts (black trace) is shown.
			In order to assure error-free operation inside the \panda-detector a procedure is suggested that allows the selection of sensors whose parameters satisfy following criteria:
			\begin{itemize}
			 \item the global leakage current must not exceed 10~$\tcmu$A below breakdown voltage
			 \item the global capacitance must show clear depletion above full depletion voltage
			 \item the breakdown voltage must be at least 50~V higher than the maximum foreseen bias voltage of 200~V
			\end{itemize}
			In the first 2011 prototyping batch 80\% of the delivered sensors (type S1) conformed to these criteria. It is expected that the yield increases once the production parameters are fixed and the producing site imposes post production tests allowing the delivery of sensors according to these criteria \cite{hgz-long-communication}.
			
\begin{figure}[]
				\centering
				\includegraphics[width=0.5\textwidth]{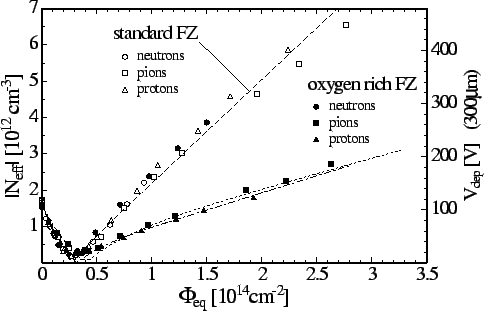}
				\caption[Effective bulk doping concentration vs. 1~\mev equiv. neutron fluence]{Effective bulk doping concentration and required depletion voltage vs. 1~\mev equivalent neutron fluence for standard Floating Zone and oxygen enriched silicon sensors~\cite{rd48-report}.}
				\label{fig:type-inversion}
			\end{figure}

\begin{figure}[]
				\centering
				\includegraphics[height=0.495\textwidth,angle=90]{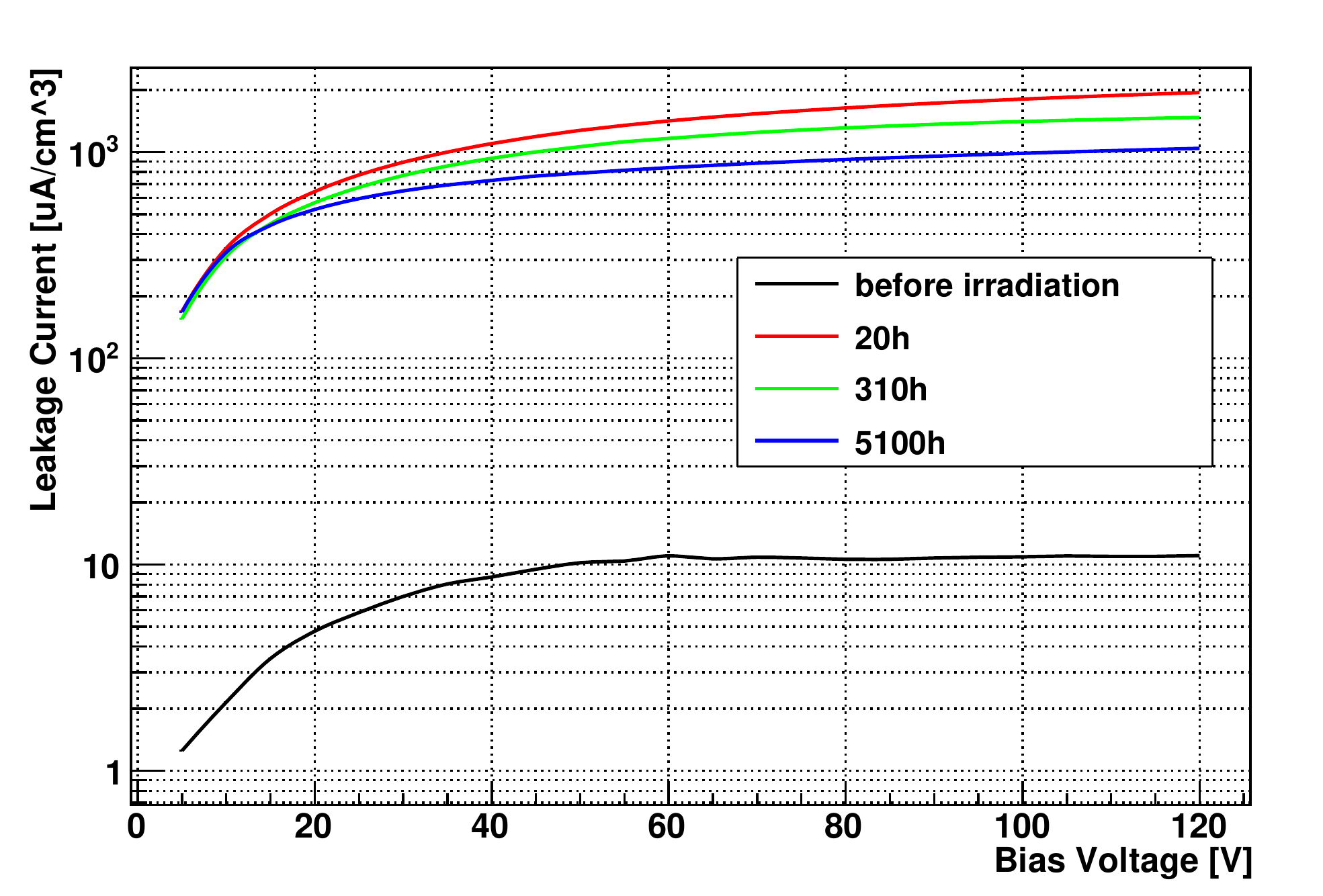}
				\caption[Annealing behaviour of leakage current characteristics]{Annealing behaviour of leakage current characteristics after irradiation of S4 sensors with 10~MRad 14~\mev protons. The {I-V}\,-\,trends of the sensors were recorded after 20 hours (red), 310 hours (green) and 5100 hours (blue) of annealing time normalised to 25~$^\circ$C.}
				\label{fig:annealing-iv}
			\end{figure}

\begin{figure}[]
				\centering
				\includegraphics[height=0.495\textwidth,angle=90]{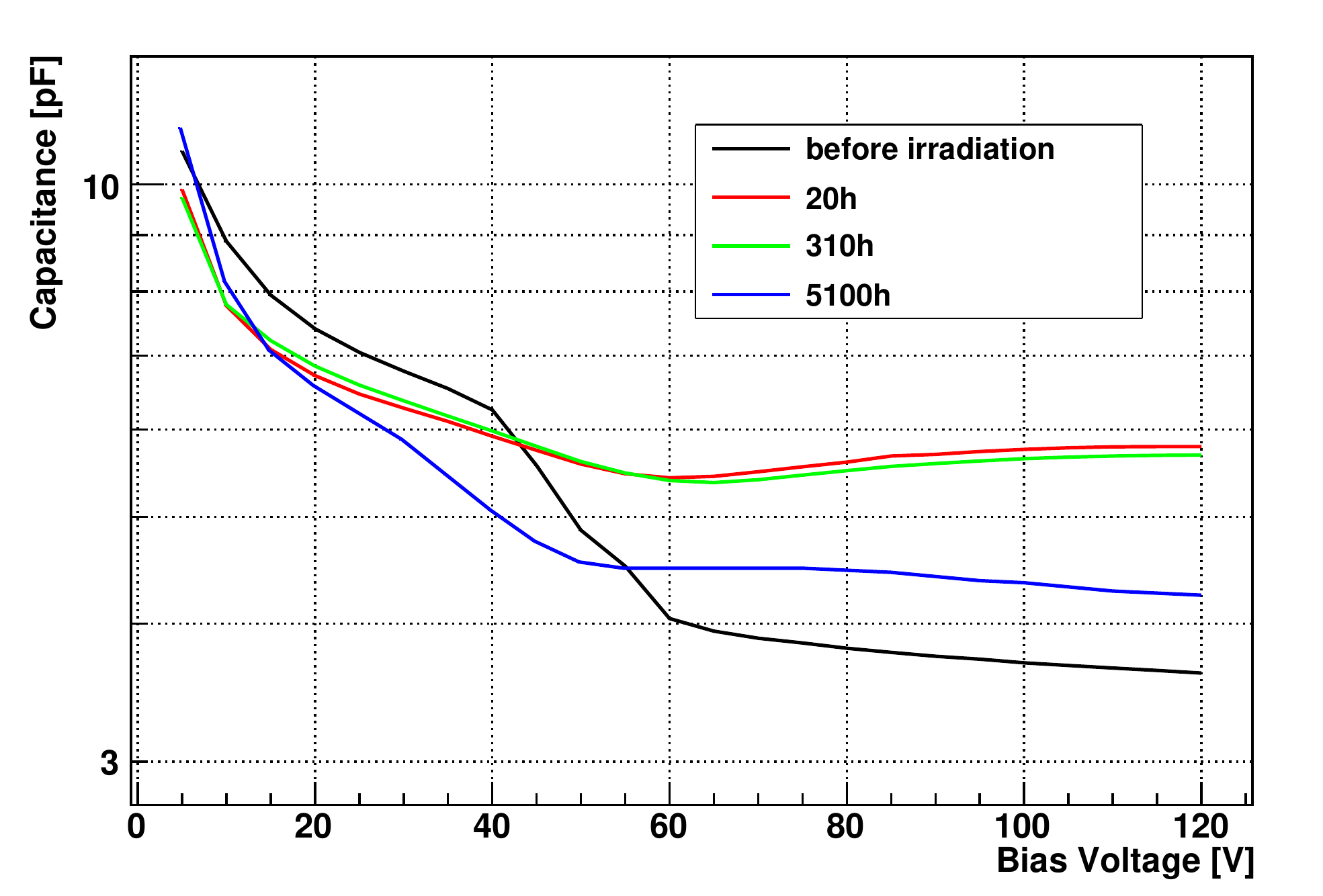}
				\caption[Annealing behaviour of capacitance characteristics]{Annealing behaviour of capacitance characteristics after irradiation of S4 sensors with the annealing times given in \figref{fig:annealing-iv}.}
				\label{fig:annealing-cv}
			\end{figure}

			For the qualification of radiation hardness the S4 sensors (128 strips on either side with a pitch of 65~$\tcmu$m) have been irradiated with protons and neutrons of several fluences in order to categorise the obtained sensors with respect to the systematic radiation damage studies undertaken by the RD48 collaboration~\cite{rd48-report}. Particularly the definition of the maximum depletion voltage of 200~V necessary to operate irradiated sensors at the end of the full \panda lifetime with an applied fluence of $1 \cdot 10^{14}$~\neueq is derived from these collected data (see \figref{fig:type-inversion}). With this choice an additional oxygen enrichment stage is not mandatory but left as optional degree of freedom to compensate for possible yield losses.
			Irradiations with slow (14~\mev) protons were carried out at the cyclotron facility in Bonn with a total fluence of up to $2.2\cdot10^{13}$ p$\cdot$cm$^\text{-2}$ which corresponds to a 1~\mev neutron equivalent fluence of roughly $8\cdot10^\text{13}$~cm$^\text{-2}$ obtained by NIEL scaling. {I-V} and {C-V}\,-\,characteristics of the irradiated sensors (type S4) have been measured in different time intervals after the end of the irradiation (\figsref{fig:annealing-iv} and \ref{fig:annealing-cv}). The measurements took place between annealing intervals of 20 hours (red curves), 310 hours (green curves) and after an annealing at a temperature of 60~$^\circ$C for 24 hours (blue curves). The green curves correspond to the standard annealing interval of 80~min in a 60~$^\circ$C environment as proposed by \cite{rd48-report}. The {I-V}\,-\,trends in \figref{fig:annealing-iv} are normalised to the volume leakage current in order to have a proper comparison to measurements available from other groups. The obtained leakage current of 
			%$\approx$\,1~mA$\cdot$cm$^\text{-3}$
			$\approx$\,1~$\text{mA}\cdot\text{cm}^\text{-3}$ 
			after irradiation meets well with the value to be expected from field studies \cite{Moll-paper}.
			The recorded post-irradiation {C-V}\,-\,trends in \figref{fig:annealing-cv} show nicely the expected beneficial (red and green) and reverse (blue) annealing behaviour as well as the change in the full depletion voltage from 60~V to 50~V due to the change in the doping concentration according to \figref{fig:type-inversion} and thus demonstrate the suitability of the first batch of prototype sensors for the targeted requirements.

		    \subsection{Wedge Sensors} 
		    \authalert{Author: Thomas W\"{u}rschig, Contact: t.wuerschig$\mathrm @$hiskp.uni-bonn.de}

\begin{figure}[htb]
			\begin{center}
			\includegraphics[trim=0 2.1cm 0 0, clip, width= 5.5 cm]{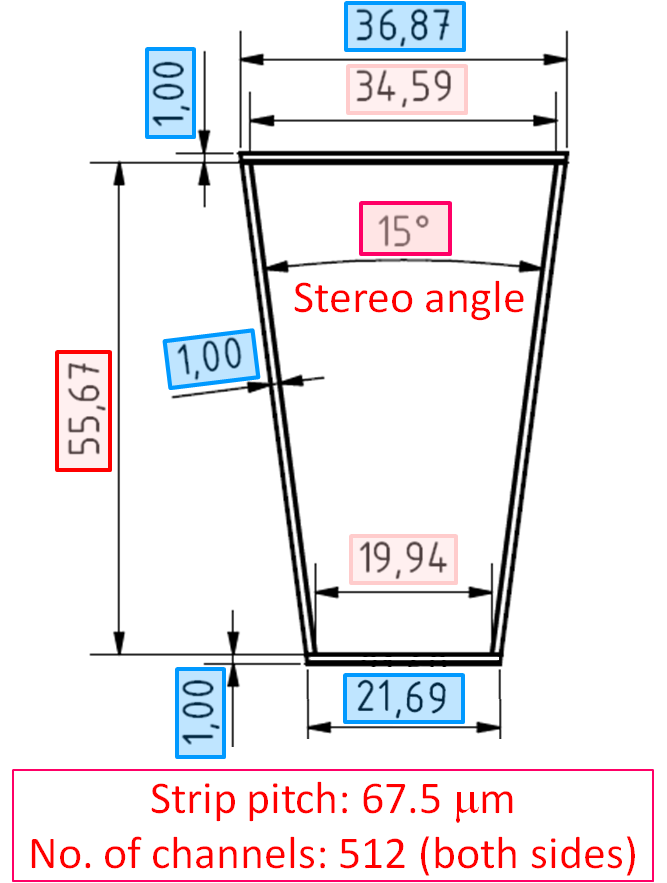}
			\caption[Dimensions of the strip wedge sensor]{Dimensions of the strip wedge sensor.}
			\label{fig-WedgeLayout}
			\end{center}
			\end{figure}

			Basically, the technological option chosen for the wedge sensors will be the same than for the barrel sensors.
			A picture of the main sensor layout is shown in \figref{fig-WedgeLayout}.

			\begin{figure}[b]
			\begin{center}
			\includegraphics[trim=0 0.15cm 0 0.5cm, clip, width= 7. cm]{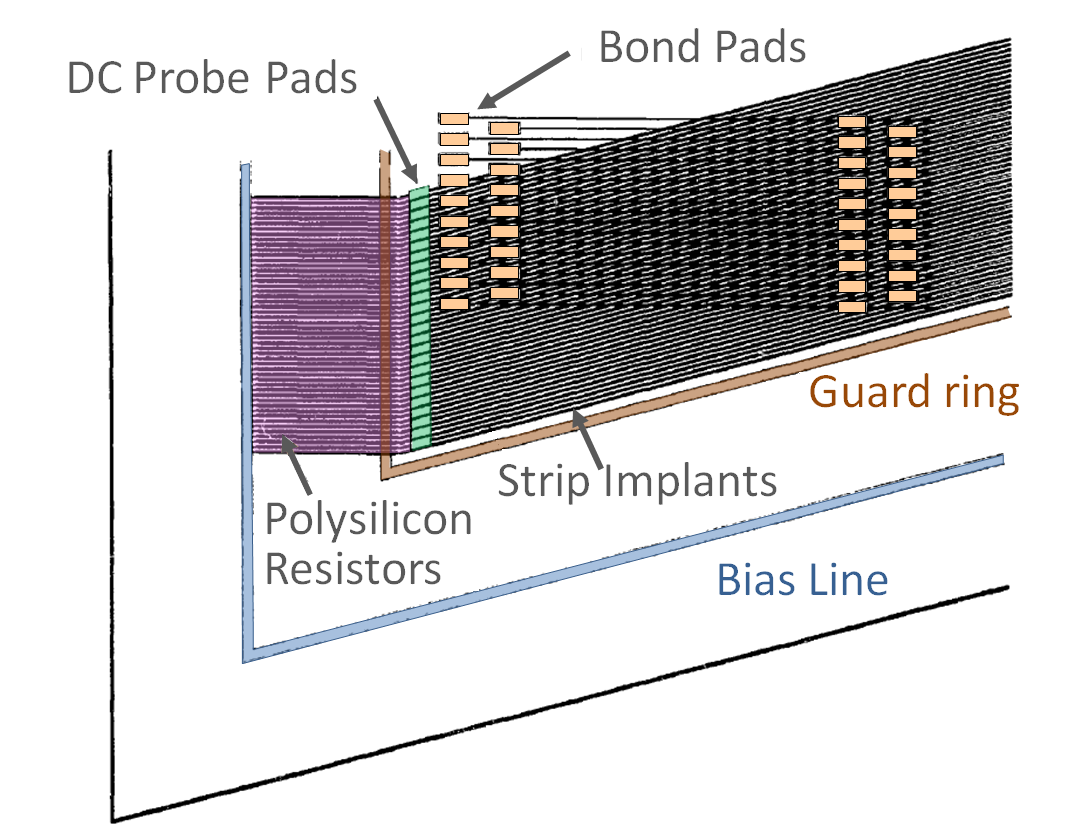}
			\caption[Illustration of the pad geometry at the top edge of the sensor]{Illustration of the pad geometry at the top edge of the sensor (adapted from~\cite{D0-WedgeLayout}).}
			\label{fig-WedgePads}
			\end{center}
			\end{figure}

			The active area of 1517~mm$^2$ stays below the one for the rectangular barrel sensor (1941~mm$^2$) 
			so that major characteristics of the sensors are expected to be in the same range 
			than the ones of the already tested full size barrel prototypes.
			In contrast to the barrel sensor, the readout pads on both sides will be at the same sensor edge.
			Strips run parallel to one of the edges of the trapezoid. 
			Thus, those on the opposite side get shorter when approaching the sensor edge.			
			A connection scheme adapted from~\cite{D0-WedgeLayout} 
			is shown in \figref{fig-WedgePads}.
			Basic design parameters of the sensor are listed in \tabref{tab-wedge}.
 
\begin{table}[!t]
\begin{center}
\small
\begin{tabular}{|c|c|}
\hline
\vspace{-3.9mm}&\\
Strip pitch & 67.5~$\tcmu$m \\
\vspace{-3.9mm}&\\
Strip orientation & $\parallel$ to sensor edge \\
\vspace{-3.9mm}&\\
Number of strips &  512 / side \\
\vspace{-3.9mm}&\\
Stereo angle &  15$^{\circ}$\\
\vspace{-3.9mm}&\\
Sensor height & 57.67~mm\\
\vspace{-3.9mm}&\\
Total area & 1688~mm$^{2}$\\
(Active area) &  (89.9\%)\\
\hline
\end{tabular}
\caption[Design parameters of the wedge sensor]{Design parameters of the wedge sensor.}
\label{tab-wedge}
\end{center}
 \end{table}

	\section{Front-End Electronics ASIC} %Bonn/J\"ulich \\
	\authalert{main author: Robert Schnell, schnell@hiskp.uni-bonn.de}
	\label{strip:ToT}
		\subsection{Requirements}
		\label{stripASICreqs}
			The data acquisition concept of the \panda detector demands \frontend electronics that is able to run without the need of an external trigger. Therefore the \frontend must be able to distinguish detector noise from physical events and send this hit information together with a precise time stamp to the data acquisition.\par
			
			A decision towards the \frontend to be deployed for the strip part of the \panda MVD has not yet been made. %is presently not achieved.
			The \frontend should feature a measurement of the deposited charge of a particle crossing the detector in order to achieve a spatial resolution through clustering algorithms. This energy loss measurement may also be used for a particle identification hypothesis (PID).
			Since only digital information about the charge measurement should be transfered from module level, the digitisation should be done on the front-end. %please don't change to \frontend, it doesn't work fine here 
			Digitisation can be achieved via the Time-over-Threshold method (ToT) or sampling-ADCs. Regardless whether ToT or ADCs will be utilised a low power consumption should be aspired. The digitisation resolution should be in the order of 8~bit or better to ensure a precise charge measurement and, moreover, a high spatial resolution. Assuming linearity of the digitiser within a dynamical range of 10 minimum ionising particles, 8~bit are required to assure that the least significant bit will be comparable with the one sigma detector noise.
			In order to save board space and number of data cables the \frontend should feature highspeed serial data links.
			The \frontend shall be radiation hard to a level of up to 10~MRad.
			The time-stamping of observed hits will be done with the \panda timing clock of 155~MHz.
			\Tabref{tab:strip:fe-requirements} summarises the requirements for the strip readout \frontend.

			\begin{table*}[width=0.4\textwidth]
			\centering
			%	\small
			\begin{tabular}{|p{5cm}|c|p{5cm}|}
			\hline
			\large{\bfseries{Parameter}} & \large{\bfseries{Value}} & \large{\bfseries{Remarks}} \\
			\hline
			\hline
			\multicolumn{3}{|c|}{\bfseries{Geometry}} \\
			\hline
			width	& $\le$~8~mm & \\
			depth	& $\le$~8~mm & \\
			input pad pitch & $\approx$~50~$\tcmu$m & \\
			\hline
			pad configuration & \multicolumn{2}{p{9cm}|}{lateral pads occupied only for diagnostic functions, should be left unconnected for final setup} \\
			\hline
			channels per \frontend & $2^6\dots2^8$ & default: 128 channels \\
			\hline
			\multicolumn{3}{|c|}{\bfseries{Input Compliance}} \\
			\hline
			sensor capacitances,& & \\
			fully depleted sensor & & \\
			& $<$~10~pF & rect. short strips \\
			& $<$~50~pF & rect. long strips + ganging \\
			& $<$~20~pF & trapezoidal sensor \\
			input polarity & either & selectable via slow control \\
			input ENC & $<$~800~e$^-$ & ${C}_\text{sensor}$\,=~10~pF \\
			& $<$~1,000~e$^-$ & ${C}_\text{sensor}$\,=~25~pF \\
			\hline
			\multicolumn{3}{|c|}{\bfseries{Signal}} \\
			\hline
			dynamic range & 240~ke$^-$ ($\approx$\,38.5~fC) & \\
			min. SNR for MIPS & 12 & 22,500~e$^-$ for MIPs in 300~$\tcmu$m silicon, guaranteed within lifetime\\
			peaking time & $\approx 5 \dots 25$~ns & typical Si drift times \\
			digitisation resolution & $\ge$~8~bit & \\
			\hline
			\multicolumn{3}{|c|}{\bfseries{Power}} \\
			\hline
			overall power dissipation & $<$~1~W & assuming 128 channels per \frontend \\
			\hline
			\multicolumn{3}{|c|}{\bfseries{Dynamical}} \\
			\hline
			trigger & internally generated & when charge pulse exceeds adjustable threshold value \\
			time stamp resolution & $<$~20~ns & \\																			%should be lower ?!
			dead time / ch & $<$~6~$\tcmu$s & baseline restored to 1\% \\
			overshoot recovery time / ch & $<$~25~$\tcmu$s & \\
			\hline
			average hit rates / ch & & derived from simulations at \\													%new count rates from Thomas?
			(poissonian mean) & & a beam momentum of 15~\gevc \\
			hot spots & 9,000~s$^{-1}$ & \pbarp \\
			& 40,000~s$^{-1}$ & \pbarAu \\
			average occupancy & 6,000~s$^{-1}$ & \pbarp \\
			& 30,000~s$^{-1}$ & \pbarAu \\
			\hline
			\multicolumn{3}{|c|}{\bfseries{Interface}} \\
			\hline
			slow control & any & low pincount, e.g.~I$^2$C \\
			data & sparsified data & \\
			\hline
			\end{tabular}
			\vspace{0.1cm}

			\caption[Requirements for the strip \frontend ASIC]{Requirements for the strip \frontend ASIC.}
			\label{tab:strip:fe-requirements}
			\end{table*}

		\subsection{Options}
			Currently there are no \frontend ASICs available to fully comply with the demanded requirements. However, there are developments for self-triggering \frontends which are mostly in an early prototyping state. The available and future developments will have to be tested with respect to the demands of the MVD. Some of the \frontend options are presented below.

			\subsubsection*{Modified ToPix Version 3}
				The use of very similar \frontend architectures for both pixel and strip sensors in the MVD opens interesting perspectives. In this case, in fact, the back-end electronics for the two sub-systems could share the same hardware design, with minimal customisation being done at the firmware and software level. Such an  approach would allow important synergies inside the MVD community, reducing the development time and costs of the overall project. However, the two sub-detectors differ significantly in key specifications, such as the sensor capacitance (few hundreds of femtofarad for the pixels and several picofarad for the strips), data rate (few kilohertz per pixel versus about ten kilohertz per strip) and power consumption per channel (few microwatts in case of pixels and few milliwatts for the strips). From these numbers, it is apparent that the two parts of the MVD demand a detailed customisation of the very \frontend electronics. 
				
				In ToPix, the \frontend ASIC for the pixel sensors described in chapter~\ref{ToPiX_architecture}, the Time over Threshold technique (ToT) has been adopted to digitise the charge information. In this approach the feed-back capacitor of a high gain amplifier is discharged with a constant current. The time during which the amplifier's output remains above a given threshold is measured by counting the clock pulses. One of the interesting features of the method is that the linear dynamic range is extended well beyond the saturation point of the amplifier. This combines low power operation with the capability of measuring energy losses much greater than those ones of minimum ionising particles.
				
				In the present design, the input amplifier employs a single stage topology. This choice was dictated by the limited space available in the pixel cell for the implementation of the analogue part (50~$\tcmu$m~$\times$~100~$\tcmu$m). One of the drawbacks of this configuration is the increased sensitivity to cross-talk for large input signals that saturate the amplifier. When this happens, the amplifier's open-loop gain drops, thereby determining an enhancement of the cross-talk between adjacent channels through the inter-electrode capacitance. For pixel sensors this capacitance is small and such an effect can be tolerated. However, the saturation of the input stage would be a serious issue for strip sensors, where the inter-strip capacitance may reach several picofarads.
				
				The measure of the charge with the ToT gives a variable dead-time. The high granularity of the pixels (cell size of 100~$\tcmu$m~$\times$~100~$\tcmu$m) reduces the event rate per channel to the kHz level and dead times of several microseconds can hence be tolerated. For strip sensors, on the other hand, the event rate per channel is significantly higher.

				\begin{figure*}[t]
					\centering
					%\fbox{
					\includegraphics[width=0.9\textwidth]{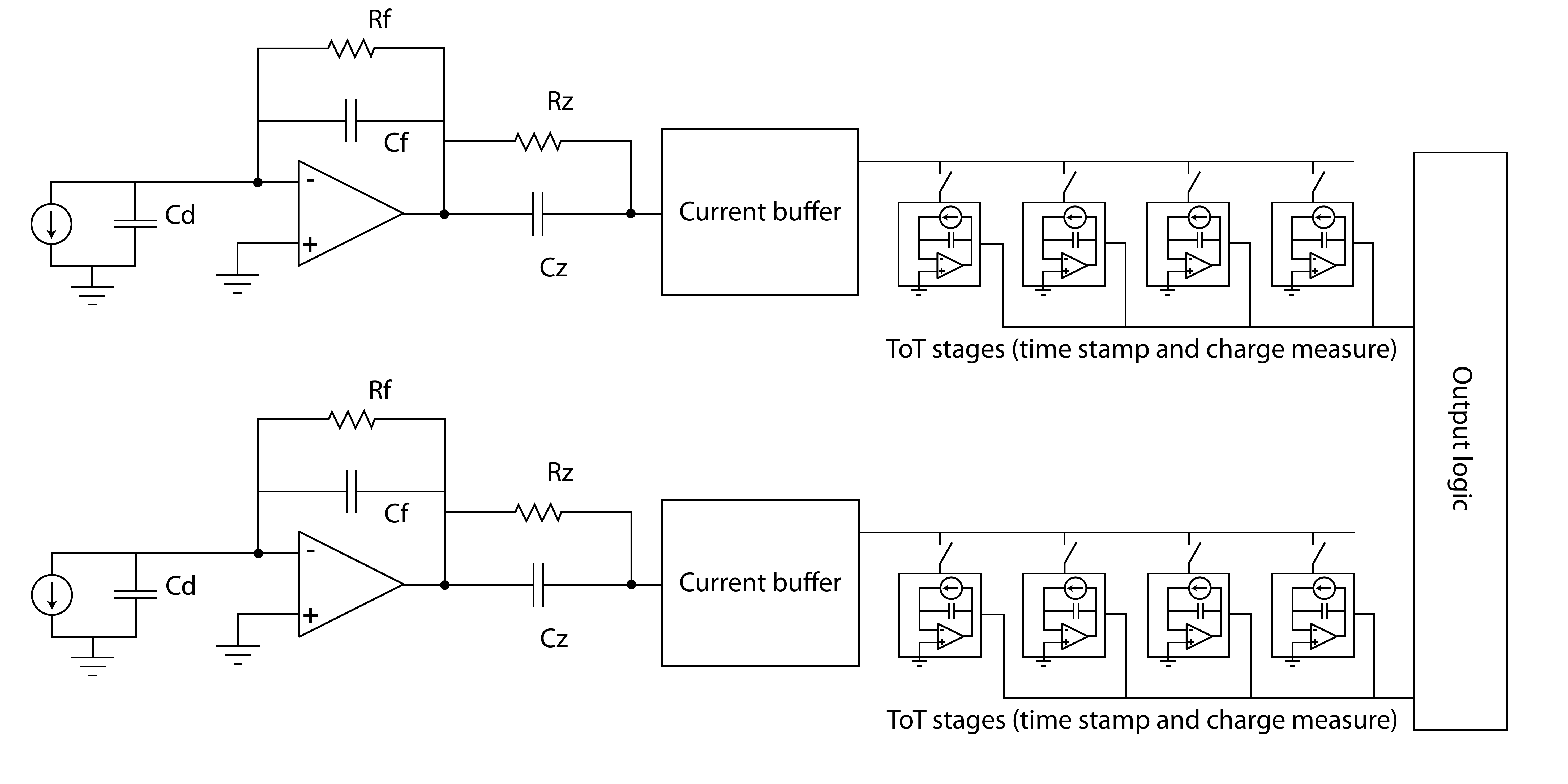}
					%}
					\caption[Schematic of modified ToPix ASIC for strip readout]{Schematic of modified ToPix ASIC for strip readout.}
					\label{fig:strip:topix-schematic}
				\end{figure*}

\vspace{1cm}				
				\Figref{fig:strip:topix-schematic} shows a possible adaption of the ToPix architecture that would take into account the aforementioned issues. The input stage is a charge sensitive amplifier followed by a pole-zero cancellation network. This block can be seen as a current amplifier with Gain $C_\text{z}/C_\text{f}$. By properly sizing $R_\text{f}$ and $C_\text{f}$ one can limit the signal swing at the preamplifier output, so the saturation is avoided. In this way, an adequate loop gain is always maintained and the cross-talk is minimised. The input stage is followed by a current buffer, which provides additional gain and drives the ToT block with proper impedance. The ToT stage has the same topology of the pixel cell in ToPix. A constant current discharges the feed-back capacitance of an integrator. The start and stop time of the discharge are measured by latching into local registers Gray-encoded words provided by a counter which is common to all channels. Assuming the standard reference clock of \panda of 155~MHz a 10~bit resolution translates into a maximum dead-time of 6.4~$\tcmu$s.
				The average dead-time will be significantly smaller, since higher input charges which leads to longer signals are less probable. 
				By using more ToT cells per channel a derandomisation of the arrival time is achieved. Assuming an average dead time of 6.4~$\tcmu$s (which is very conservative for 10~bit resolution) and four digitising cells one can accommodate a rate of 100~kHz per channel with an efficiency better than 99.5\%.    

				Preliminary simulations of the architecture were done using the parameters of the same CMOS process chosen for the prototypes of the pixel \frontend  electronics. The strip-optimised front-end shows a noise of 1000 electrons rms for a power consumption of 1~mW per channel.
				From the readout point of view each channel, which contains four independent digitising elements, can be seen a ``short" pixel column with four pixels. Such similarities would of course favour the development of a common readout framework and the reuseability of different modules and concepts between the two subsystems.

			\subsubsection*{STS-XYTER}
				The STS-XYTER is a development based on the n-XYTER~\cite{NXYTER-paper}. It will make use of the same token-ring-architecture common to the n-XYTER. It is supposed to be a dedicated \frontend ASIC for the readout of silicon strip detectors to be used in several \Fair experiments.
				The \frontend will feature a self-triggering architecture and is designed for low power consumption. Therefore it will make use of a ToT-based digitisation. In order to handle the data rate in high occupancy environments a fast digital link with a data rate up to 2.5~Gbps is envisaged. Two prototype ASICs - TOT01~\cite{STS-XYTER-Test1} and TOT02~\cite{STS-XYTER-Test2} - were produced and tested.
				Further specifications and parameters measured with first prototypes can be extracted from \tabref{tab:strip:Front-end_compar}.

			\begin{table*}[h]
				\centering
				\begin{tabular}{|c|c|c|c|c|}
					\hline
					\bfseries{Parameter} & \bfseries{adapted ToPix} & \bfseries{STS-XYTER} & \bfseries{FSSR2} \\
					\hline
					input pad pitch &  $\approx$~50~$\tcmu$m & 50~$\tcmu$m & 50~$\tcmu$m \\% & 44~$\tcmu$m \\
					\hline
					channels per \frontend & 128 & 128 & 128 \\% & 128  \\
					\hline
					dynamical range & 100~fC & 15~fC & 25~fC \\
					\hline
					ENC & 1,000~e$^-$~@~20~pF & 700~e$^-$~@~28~pF & 240~e$^-$\,+\,35~e$^-$/pF \\
					\hline
					peaking time & 6~ns & 80~ns & 65~ns \\% & 50~ns \\
					\hline
					power consumption & 0.8~mW/ch & 1.2~mW/ch & 4.0~mW/ch \\% & 3.5~mW/ch \\
					\hline
					trigger & self-triggering & self-triggering & self-triggering \\% & external \\
					\hline
					digitisation technique & ToT & ToT & Flash-ADC \\% & none \\
					\hline
					digitisation resolution & 10~bit & $4-6$~bit & 3~bit Flash-ADC \\% & analogue \\
					\hline
					time resolution & 1.85~ns~@~155~MHz & 1.85~ns~@~155~MHz & 132~ns time stamp \\
					\hline
					data interface & e-link - SLVS & up to 2.5~Gbps & LVDS \\% & analogue differential \\
					\hline
					number of data lines & 1 pair & 1 pair & 1 to 6 pairs \\% & 1 pair \\
					\hline
					slow control & custom (serial) & custom & custom LVDS \\% & I$^2$C \\
					\hline
					process & 0.13~$\tcmu$m CMOS & UMC 0.18~$\tcmu$m & 0.25~$\tcmu$m CMOS \\% & 0.25~$\tcmu$m CMOS \\
					\hline
					radiation hardness & $>$~10~MRad & 10~MRad & up to 20~MRad \\
					\hline
				\end{tabular}
			\vspace{0.1cm}
			\caption[Comparison of different \frontends for strip detectors]{Comparison of different front-ends for strip detectors. The specifications for the STS-XYTER are taken from prototype evaluations~\cite{STS-XYTER-Test1,STS-XYTER-Test2}.}
			\label{tab:strip:Front-end_compar}
			\end{table*}

			\subsubsection*{FSSR2}
				The FSSR2~\cite{fssr-paper1} is a 128-channel ASIC for silicon strip readout. It features a fast, self-triggered readout architecture with no analogue storage, similar to the FPIX2 chip (the pixel \frontend for the BTeV experiment) where it was derived from. 
				Each analogue channel contains a programmable charge sensitivity preamplifier, a $CR-(RC)^2$ shaper with selectable shaping time, a selectable base line restorer and a discriminator. The discriminator with an adjustable threshold simply yields a binary hit information. Moreover each channel contains a 3-bit flash ADC with selectable thresholds to achieve an amplitude information.

	\section{Module Data Concentrator ASIC}
	\authalert{main author: Hans-Georg Zaunick, zaunick@hiskp.uni-bonn.de}
		\subsection{Architecture}
			%\Figref{fig:strip:mod-controller-architecture}
			Each strip sensor module will have the \frontend chips as well as a module controller on board. The module controller serves as link between the \frontend chips and the MVD detector data acquisition-system. The module controller has to be able to handle sensor modules of different sizes up to the maximum of $896 \times 512$ strips corresponding to the large double-sided sensors outlined in \tabref{tab:strip:sensor-params}. The main tasks to be performed are:
			\begin{itemize}
 			\item readout, decoding and multiplexing of \frontend chip data;
			\item buffering, pedestal subtraction and optional one-dimensional or two dimensional clustering;
			\item interfacing to DAQ via serial GBT~\cite{gbt} E-links;
			\item slow control functions (parameter transfer, monitoring).
			\end{itemize}

			The module controller has to be finally realised as a fully digital ASIC in radiation-hard technology. Since just one module controller is needed for each sensor module,  the additional material budget, the current consumption and the thermal load of this component will be neglegible compared to that of the \frontend chips. 
			The design of the module controller is currently in progress~\cite{PANDA-KolMet}. To have a starting point, it is assumed to use the n-XYTER plus a separate ADC as \frontend hardware. In this case, 11 n-XYTERs and ADC's are needed for the largest strip sensor module. The design is fully VHDL-based using Xilinx FPGA as technological platform. The modular structure of the design guarantees a minimum effort to incorporate different \frontend solutions or additional features in future versions.   

			\begin{figure}[h]
				\centering
				\includegraphics[width=0.495\textwidth]{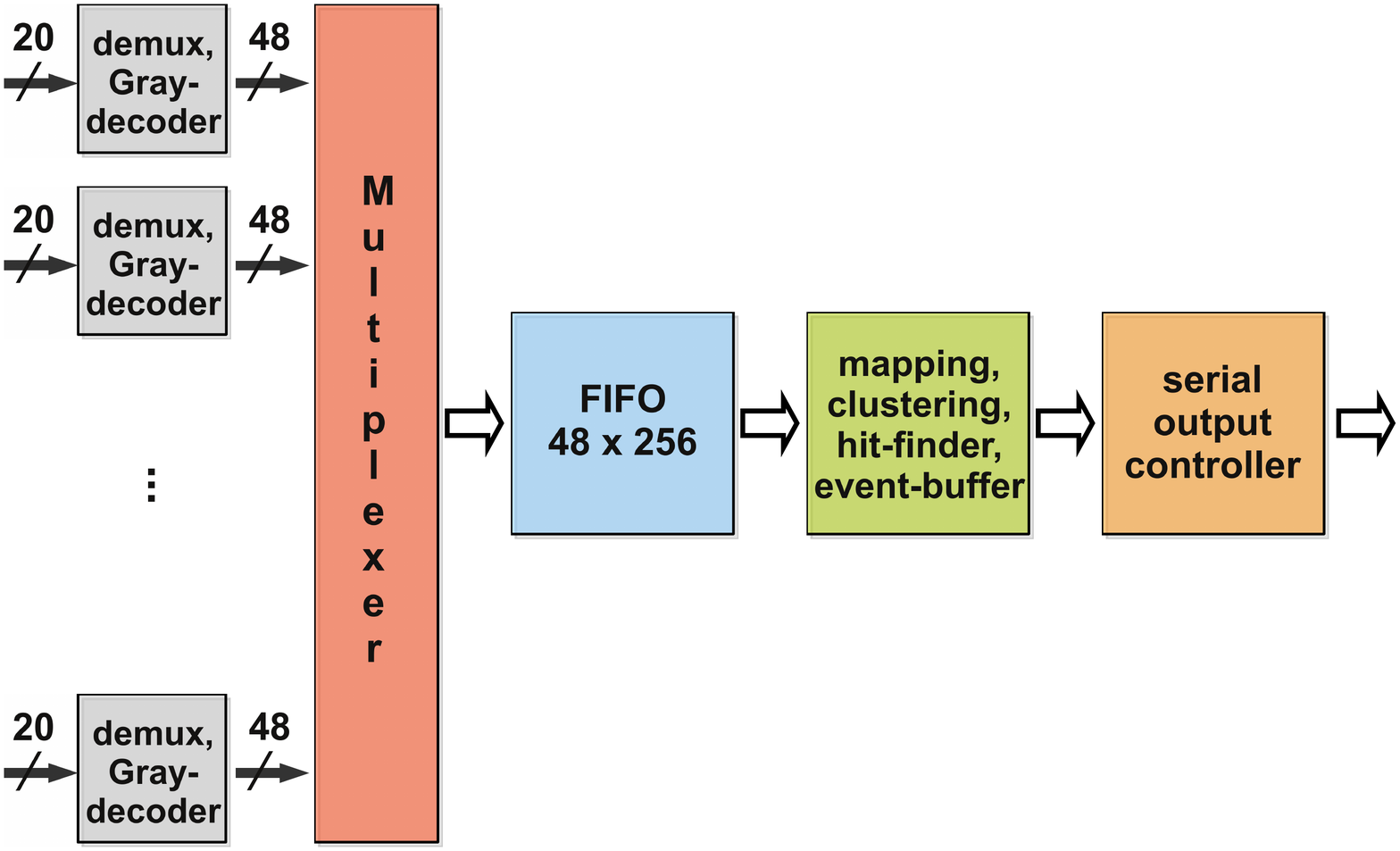}
				\caption[Schematic of the Module Data Concentrator]{Schematic of the Module Data Concentrator.}
				\label{fig:strip:mod-controller-architecture}
			\end{figure}
			
\clearpage
		\subsection{Implemented Feature Extraction Algorithms}
		The data frames from the different \frontend chips contain the time stamp, strip-identifier and charge information.
		After subtracting the pedestal from the charge the frames are stored in a FIFO, from which they may be extracted in the raw format or a feature extraction may be performed by the module controller.
		With the foreseen feature extraction stage enabled, the module controller will continuously scan the FIFO output and collect frames of a certain time interval $\Delta t$. $\Delta t$ is determined by bit-truncation of a parameter-based number n of least significant bits from the time stamp. The corresponding frames are buffered and analysed for clusters with a minimal multiplicity m of next and next but one hits in the $x$- and $y$-strips of one sensor module~\cite{Koop}. If hits are detected, the time stamp, the centroid, the width in the $x$- and $y$-plane, the sum of charges and the multiplicity will be stored in the hit-register. The contents of the hit register may be transferred to the DAQ system or, if enabled, passed to a 2-dimensional correlation stage that generates a list of all possible x-y-combinations of clusters. For each possible combination a combinatorial and a charge-correlation probability is calculated based on the utilisation of fast look up tables, particularly of error functions. The resulting 2d-clusters then are buffered in the output FIFO. Important clustering parameters, like the minimal multiplicity m, the number n of truncated bits in the time stamps, the cluster thresholds or the cutoff threshold of the combinatorial probability are adjustable via the slow control interface.

				\subsection{Implementation Status}
			The VHDL-design is expected to be finished in 2011. First test were performed using device simulations. The design is able to buffer up to 5 simultaneous hits at the n-XYTER clock speed of 128 MHz. Based on a Xilinx Spartan 6-device timing simulations resulted in approximately 150~ns for the multiplexing of the 11 \frontend devices (n-XYTER plus ADC). The current consumption and thermal load strongly depend on the technology chosen for the final ASIC design and have to be evaluated later.
			Since the final decision concerning the \frontend chip for the strip sensors has not been reached yet, the module controller design still has to be adapted to the final \frontend and transfered to the appropriate ASIC technology basis.

	\section{Hybridisation}
		\subsection{Overview}
\authalert{Author: Thomas W\"{u}rschig, Contact: t.wuerschig$\mathrm @$hiskp.uni-bonn.de}

\begin{figure}
\begin{center}
\includegraphics[width=6.8 cm]{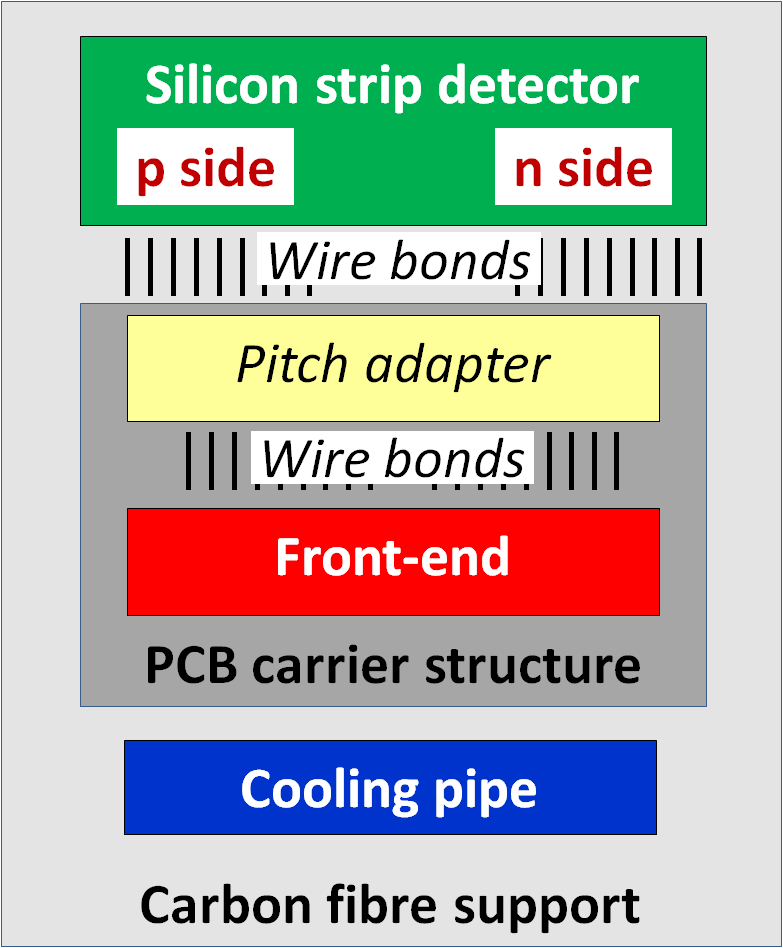}
\caption[Basic concept for the hybridisation of double-sided strip detectors]{Basic concept for the hybridisation of double-sided strip detectors.}
\label{pic-SchematicsHybrid}
\end{center}
\end{figure}

			The hybridisation of double-sided strip detectors 
is one of the technically most demanding tasks for the strip part of the MVD. 
Particular challenges arise from the electrical connection on both sensor sides, as described in chapter~\ref{sec:strip_power}.
\Figref{pic-SchematicsHybrid} gives a schematic illustration of the overall concept. 
In contrast to the hybrid pixel design, readout electronics can be placed 
outside the acceptance of the sensors. 
Therefore, appropriate adapter components are introduced to interconnect to sensor 
and the \frontend electronics. 
Wire bonding is the default option for all electrical connections from the sensor to the \frontend chip. 
Basically, the layout of the readout pads on the senor differs from the one on the \frontend. 
This accounts for both the effective contact area as well as the pitch, i.e.~the interspacing of two neighbouring pads. 
Technical limitations of the wire-bonding process must be considered if the pitch size is smaller than \unit[100]{$\tcmu$}m 
as it is the case for the wedge sensors and the input channels of the \frontend. 
These require parallel straight-line connections between corresponding pads, which can be achieved by the introduction of a fan-like structure. 
Such pitch adapter can be either integrated on the hybrid carrier PCB or designed as an individual component. 
While the first option represents an advanced technical solution, 
the other one follows the solution of other HEP experiments recently installed. 
Finally, the detector module must be integrated onto a local support structure in order to achieve the required mechanical stability. 
The local support structure includes a cooling pipe needed to prevent a heating of the readout electronics or the sensors.

		\subsection{Basic Approach for the MVD}

			\begin{figure}
				\begin{center}
					\includegraphics[width=7.5 cm]{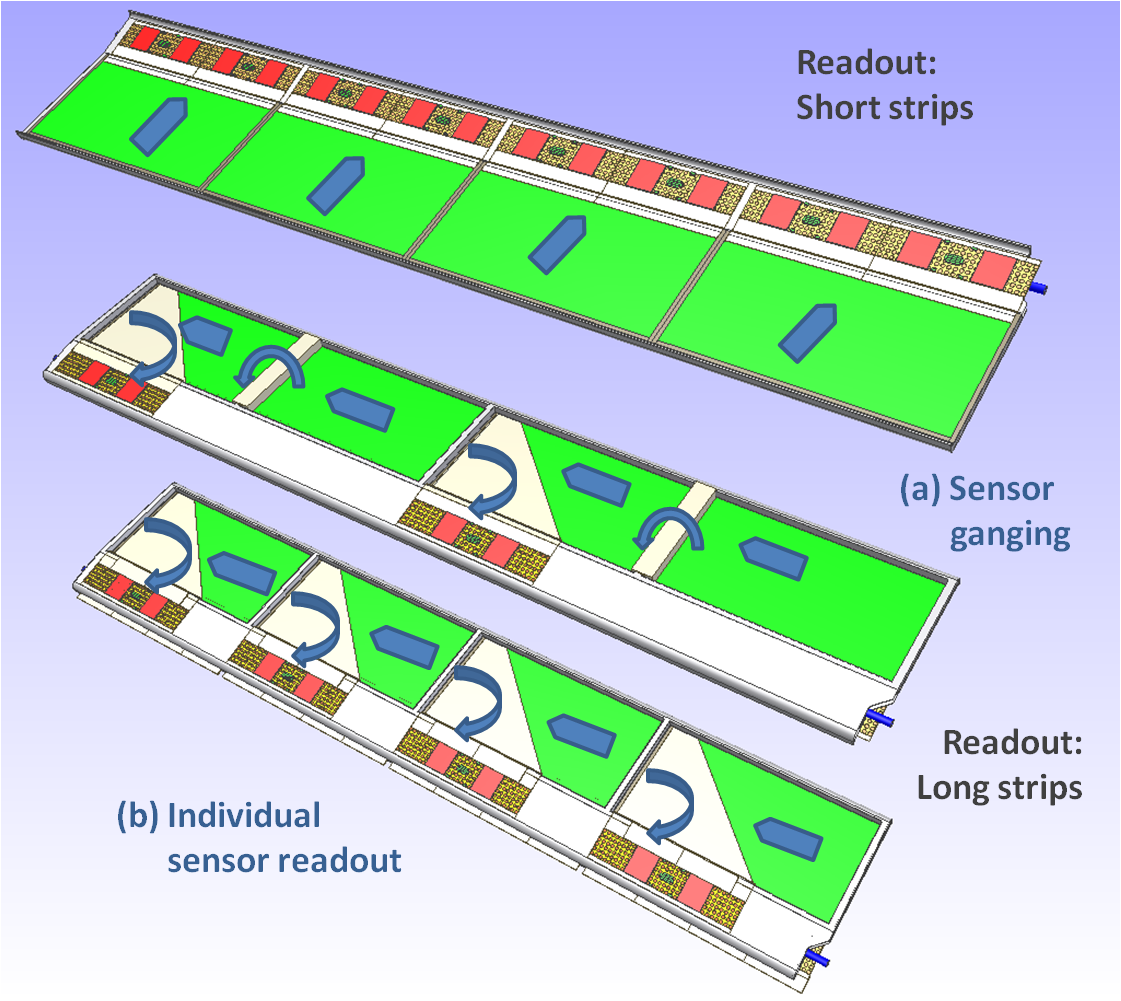}
					\caption[Illustration of the hybridisation concept for the strip barrel part]{Illustration of the hybridisation concept for the strip barrel part. For the readout of the long strips there are two major options: The ganging of long strips of adjacent sensors (a) or an individual readout of all sensors (b).}
				\label{pic-SchematicsHybridStripBarrel}
				\end{center}
			\end{figure}

			\begin{figure*}[h]
				\centering
				\includegraphics[width=15cm]{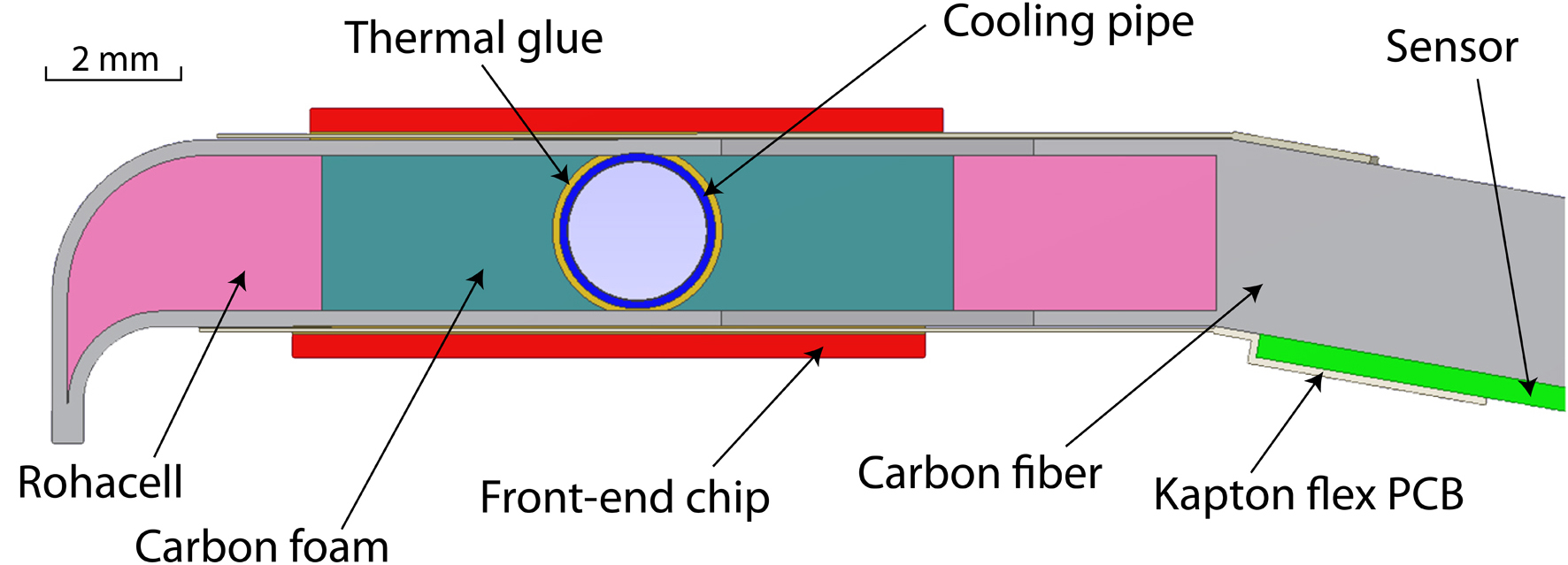}
				\caption[Schematic cross section of a strip hybrid structure in the barrel part]{Schematic cross section of a strip hybrid structure in the barrel part. The carbon frame carries the sensors (green), the flex-PCB with electrical structures and \frontend electronics (red). The Kapton-Flex-PCB is glued with an overlap of ca. 6~mm onto the sensor. Connection to the latter is realised with aluminum wire-bonds. The Flex-PCB also serves as a fan-in adaptor to the pitch of the \frontend ASIC which is mounted above the cooling pipe.}
				\label{fig:strip:hybrid-cross-section}
			\end{figure*}

			The different sensor geometry as well as the changed sensor orientation with respect to the beam pipe lead to individual hybridisation concepts in the barrel and the forward part. \Figref{pic-SchematicsHybridStripBarrel} illustrates the basic approach for the strip modules in the barrel layers.
			A complete super-module contains several sensors, which are aligned in one row. The carbon fibre support structure is extended at the long edge thus leaving space for the positioning of the \frontend electronics. Due to the stereo angle of 90$^{\circ}$ the sensor is read out at two edges. A flexible carrier structure is introduced to flip the \frontends at the short side by 90$^{\circ}$ to the support frame. 
The overall number of readout channels can be further decreased by connecting strips of neighbouring sensors. 
However, this option is technically more challenging and may cause deteriorations of the overall performance. 
Therefore, a fall-back solution is given by an individual readout of all sensors. 
A schematic cross section through a hybridised strip module is shown in \figref{fig:strip:hybrid-cross-section}.

			\begin{figure}
				\begin{center}
					\includegraphics[width=7.5 cm]{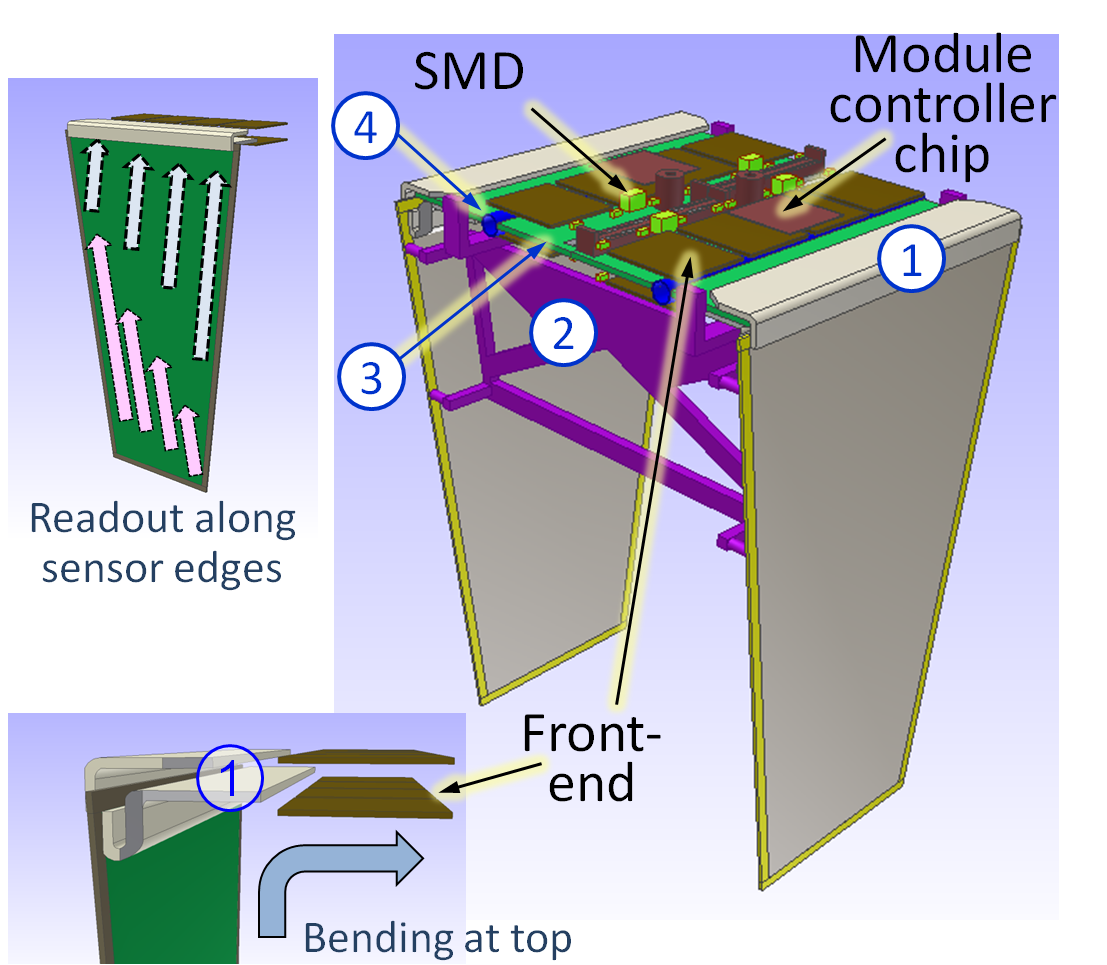}
					\caption[Illustration of the hybridisation concept for the strip disks]{Illustration of the hybridisation concept for the strip disks. A flexible adapter (1) allows a bending of the readout electronics at the top. The super-module is supported by a sophisticated carbon structure (2) and a flat structure at the top plane in between the sensors (3). Cooling pipes (4) are placed below the \frontend.}
					\label{pic-SchematicsHybridStripDisk}
				\end{center}
			\end{figure}

			The hybridisation concept of the strip disk part follows 
the basic idea presented in~\cite{HermesLambdaWheels}. 
A disk super-module combines two successive sensors of the two layers. 
In contrast to the barrel sensors the readout of both sensor sides is performed 
at the same sensor edge at the top. 
Flexible adapter pieces facilitate a bending by 90$^{\circ}$ 
so that the \frontend electronics can be placed at the top plane in the gap between 
the two sensors. 
In this way a minimum radial occupancy of the passive elements can be achieved. 
While a flat support serves as contact surface for the PCB carrier board, 
a more sophisticated carbon support structure is needed to combine the readout plane 
at the top and the two sensors to a compact and robust object. 
An illustration of the hybridisation concept for the strip disks is given in \figref{pic-SchematicsHybridStripDisk}.

		\subsection{Layout of Hybrid Carrier PCB}
			%\alert{layout schematics coming soon}
			The material of the \frontend electronics carrier structure should satisfy the following requirements:
			\begin{enumerate}
			 \item low thickness at high structural integrity, low-Z compound;
			 \item well defined dielectric constant up to frequencies of at least 1~GHz and;
			  low dielectric loss at the same time;
			  \item radiation tolerance, i.e.~insignificant change of properties under irradiation in ionising as 
			  well as non-ionising fields.
			\end{enumerate}
			From readily available materials only few fulfill all of these demands. An optimal choice seems to
			be Kapton\footnote{developed and patented by duPont}-Polyimide (chemical composition: Poly-(4,4$'$-oxy\-di\-phenylene-pyro\-mellitimide)) since beside
			the mentioned requirements additional advantages, e.g.~easy bondability with metal-films and wet-processing 
			with standard-PCB-structuring techniques are applicable~\cite{kapton-datasheet}.

			\begin{table}[width=0.4\textwidth]
			\centering
			\resizebox{\columnwidth}{!}{
			\begin{tabular}{|l|c|}
				% after \\: \hline or \cline{col1-col2} \cline{col3-col4}
				\hline
				material & kapton polyimide   \\
				dielectric thickness & 25.4~$\tcmu$m  \\
				dielectric constant & 3.4 \\
				dielectric loss tangent & 0.004  \\
				conductor type & Cu  \\
				conductor thickness & 18~$\tcmu$m  \\
				radiation length (min-max) & $0.01-0.26$\%~$X/X_0$ \\
				\hline
				\end{tabular}
				}
			\caption[Specifications of the hybrid carrier material]{Specifications of the hybrid carrier material.}
			\label{tab:strip:kapton-flex-props}
			\end{table}			

			For the hybrid carrier PCB a two layer layout is foreseen with the parameters given in \tabref{tab:strip:kapton-flex-props}. The value for radiation length for this material is in general very low but may peak up to ca. 0.25\%~$X/X_0$ locally, when copper traces or larger areas on both sides overlap.

			\begin{figure}[htb]
				\centering
%				\fbox{
%				\includegraphics[trim=0 0 1.3cm 2.5cm,  clip, height=0.8\columnwidth, angle=-90]{strippart/pictures/nxyter_hybrid_4fe_white}
%				\includegraphics[trim=0 0.8cm 3.5cm 0, clip, width=7.cm]{strippart/pictures/nxyter_hybrid_4fe_white}
				\includegraphics[width=7.cm]{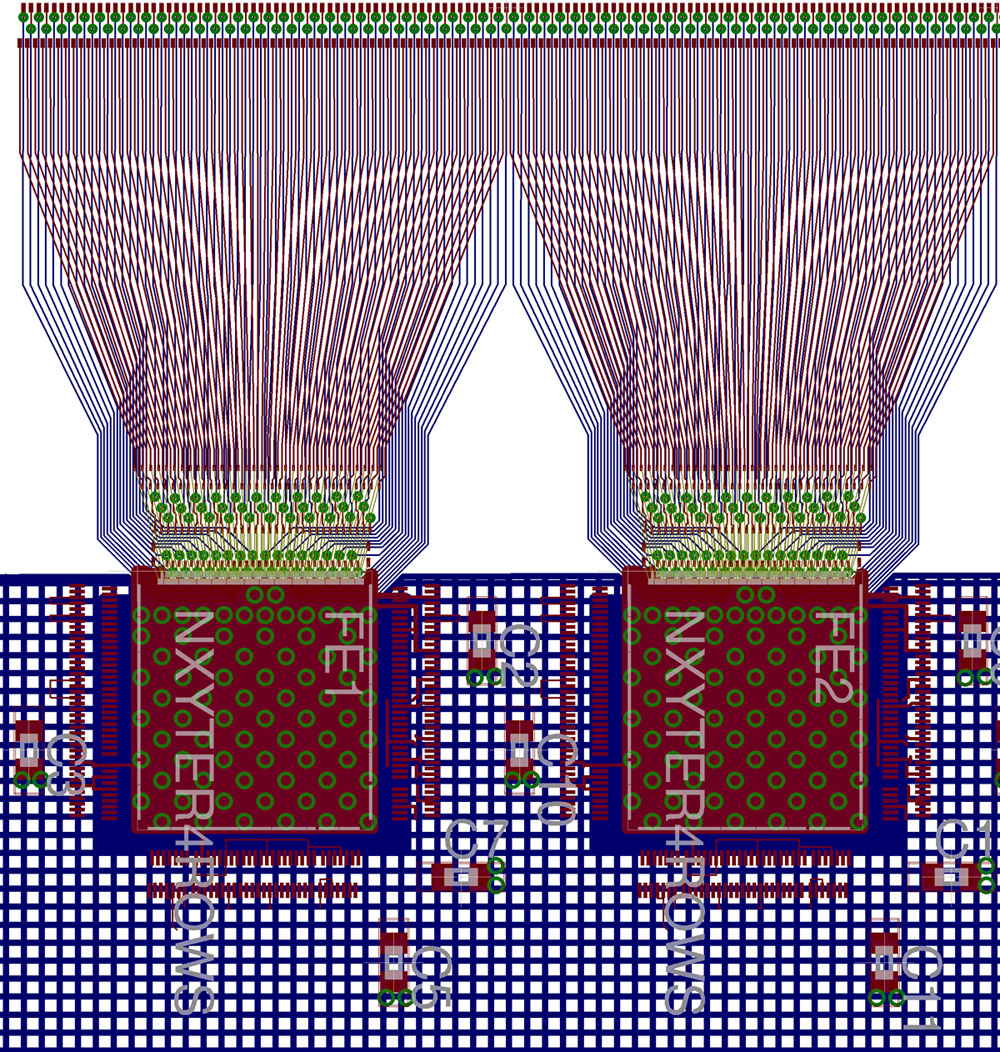}
%				}
				\caption[Layout of pitch adaptor hybrid]{Layout detail of pitch adaptor hybrid.}
				\label{fig:strip:hybrid-routing}
			\end{figure}

			The fan-in routing of the default 130~$\tcmu$m pitch structure to the \frontend pitch can be processed by standard PCB manufacturing techniques. \Figref{fig:strip:hybrid-routing} shows a layout of a prototype for the connection to already existing n-XYTER \frontend ASICs which is being produced and evaluated at the moment. The bonding pads at the top edge are placed in two staggered rows at the sensor pitch. The next step is the interconnection of this hybrid prototype to the sensor by directly glueing the PCB with an overlap of 6.7~mm onto the sensor such that the innermost AC-pads of the sensor and the bonding pads on the flex-PCB face each other and can be connected via bonding wires.

\clearpage
		\subsection{Interconnections}

\authalert{Author: Thomas W\"{u}rschig, Contact: t.wuerschig$\mathrm @$hiskp.uni-bonn.de}

Basically, there are three different subjects to be considered as interconnection between different components of a super-module:

\begin{itemize}
\item{{Electrical connections}\\
Electrical connections are needed 
to supply the sensor and the electronics, 
to lead the charge signal of individual strips 
to an input channel of the \frontend electronics 
as well as to transmit data signals on the super-module.
Therefore, wire bonds with a thickness of a few tens of micrometres are foreseen.
With respect to a minimised effective radiation length,
aluminium bonds will be preferably used. 
However, there are technical challenges in particular for very thin wires, 
which are taken for the charge signals.
Therefore, gold wires are the defined standard for some of the wire-bonding technologies.
}
\item{{Mechanical connections}\\
All the components of a super-module will be glued together.
The choice of the glue depends on kind of interconnection
and thus must be in accordance with given specifications 
on the electrical and thermal conductivity, the expansion coefficient
and the effective material budget.
In any case it must withstand the expected radiation dose.
For the interconnection of the \frontend to the carrier boards a high thermal conductivity of the glue is needed.
On the other hand, a too high material load must be prevented.
}
\item{{External connections}\\
Connectors are needed for the overall integration of a super-module.
Cooling connectors must allow for the change from a stable pipe to a flexible tube.
Moreover, separate connectors for the power supply and the data transmission have to be inserted.
Finally, the mechanical connection to external holding structure 
must be ensured with a high reproducibility. 
}
\end{itemize}

Most of the parts mentioned above related to integration aspects  
are very similar for the pixel and the strip part.
Therefore, more details on specific materials, e.g.~glues, 
can be found in chapter~\ref{infrastructure}.

\subsection{Test Assembly}

First evaluation studies of the hybridisation concept for the strip part
have been performed with a laboratory test setup~\cite{DTS-PANDAnote}, 
for the characterisation of double-sided silicon strip detectors
and connected readout electronics.
Therefore, square sensors with a side length of \unit[2]{cm} 
and a thickness of \unit[300]{$\tcmu$m}~\cite{DTS-sensor} 
were used in combination with the APV25-S1 readout chip~\cite{APV25}. 
Specifications of both components differ partly from \Panda requirements.
However, shape and dimensions of the implemented test sensor are similar 
to the one defined for the MVD barrel sensors. 
Moreover, the stereo angle of 90$^{\circ}$ between the strips 
of both sensor sides is compliant.
The strip pitch of \unit[50]{$\tcmu$m} falls short 
of the smallest value of \unit[67.5]{$\tcmu$m} defined for the wedge sensors.
Therefore, the hybridisation of appropriate detector modules 
facilitates an evaluation of technical details
such as mounting procedures and the electrical connection
via wire bonds at very small dimensions.

\begin{figure}[htb]
\begin{center}
 \includegraphics[width=7.5 cm]{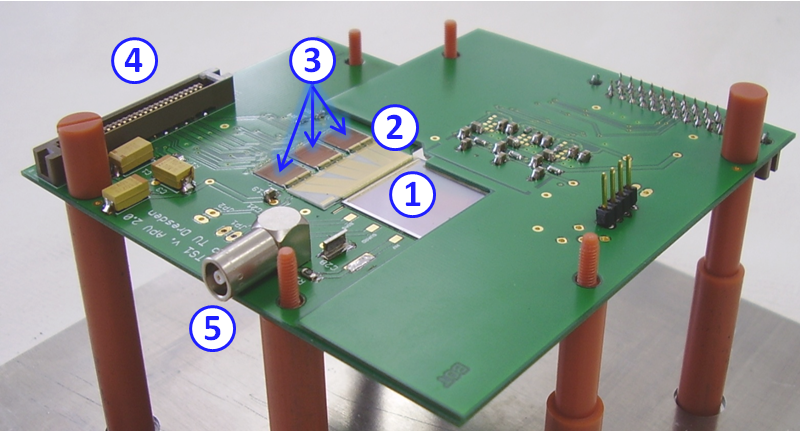}
% Pic3-01_MVD-BTS-Geometry.png: 1239x931 pixel, 125dpi, 25.18x18.92 cm, bb=0 0 714 536 
\caption[Photograph of a finally assembled detector module for the laboratory test setup]{Photograph of a finally assembled detector module for the laboratory test setup. Main components are: (1) Sensor; (2) Pitch-adapter; (3) \Frontend chips; (4) High-density connector for power supply, slow control, data output; (5) HV connection for sensor supply.}
\label{pic-DTS_DetectorModule}
\end{center}
\end{figure}

\begin{figure}[b]
\begin{center}
 \includegraphics[width=7.5 cm]{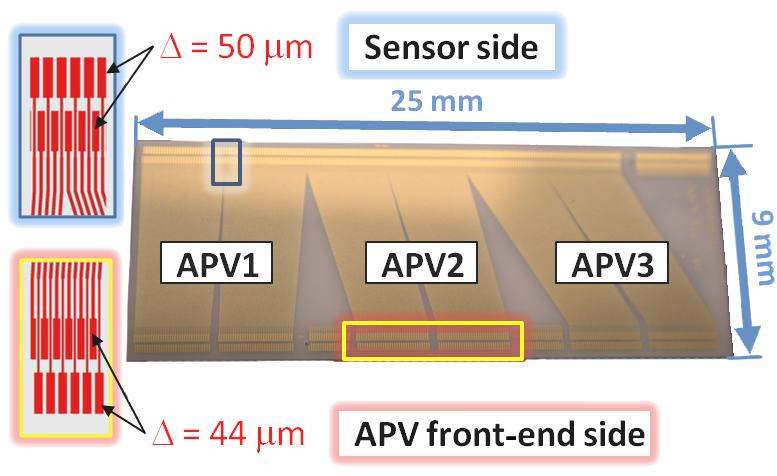}
% Pic3-01_MVD-BTS-Geometry.png: 1239x931 pixel, 125dpi, 25.18x18.92 cm, bb=0 0 714 536 
\caption[Photograph of the fabricated pitch adapter and schematics of the pad geometry]{Photograph of the fabricated pitch adapter and schematics of the pad geometry.}
\label{pic-DTS_PA}
\end{center}
\end{figure}

\begin{figure*}[!t]
\begin{center}
 \includegraphics[width=15 cm]{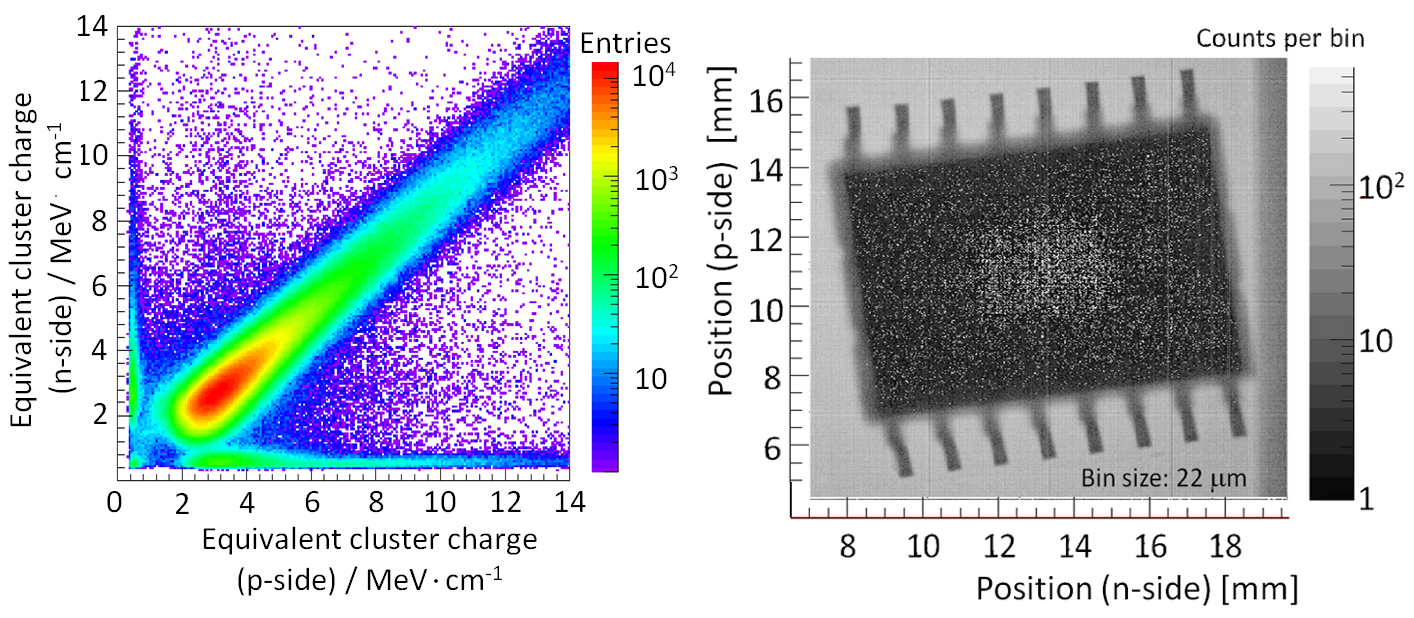}
\caption[Results obtained with a laboratory setup for double-sided strip detectors]{Results of long-term measurements with a $^{90}$Sr source on the laboratory setup: Correlation of deposited energy (\textit{left}) for both sensor sides ($p$-side and $n^+$-side) and 2D imaging of a SMD device placed on top of the sensor and acting as an absorber for the $\beta$-electrons (\textit{right}).
}
\label{pic-DTS_Measurement}
\end{center}
\end{figure*}

A photograph of the completed detector module is shown 
in \figref{pic-DTS_DetectorModule}. 
The sensor is glued to L-shaped circuit boards 
thus allowing a double-sided readout with one common design. 
A ceramic pitch adapter has been fabricated,
which  leads from the readout pitch of the sensor 
to the needed structure of the APV25 \frontend chip with 
a pitch of just \unit[44]{$\tcmu$m}.
One photograph of the fabricated pitch adapter 
can be found in \figref{pic-DTS_PA} along with
the schematics of the pad geometry. 

Extended studies with a series of detector modules
have been performed proving the full functionality of the setup.
Moreover, there are no indications of an impact of 
the assembly procedure on the final detector performance.
Selected results of the measurements are collected 
in \figref{pic-DTS_Measurement}.

%	\addcontentsline{toc}{chapter}{\bibname}

%	\newpage
\vspace{1cm}
\putbib[lit_strip]
		
%\end{bibunit}

%% file: integration/offdetector.tex
		
\chapter{Infrastructure }
\label{infrastructure}
%\section{Optical Data Transmission }(Gianni)
\input{integration/odt}

\section{Off-Detector Electronics}

The \panda DAQ System architecture defines 3 multiplexing and aggregation layers, before the detector data are sent to the Level 1 Trigger farm implemented by the Compute Nodes in ATCA crates. The first layer does the multiplexing of several \frontend input links into one outgoing link. In addition to this, information on beam structure (2~$\tcmu s$ bursts and 500~$\tcmu s$ super bursts) and global clock is received by the SODA system and propagated to the \frontend modules in order to provide proper time-stamping of the detector data. In two further aggregation stages data are grouped into data blocks corresponding to one burst and super blocks corresponding to one super burst.

\begin{figure}[ht]
\centering
\includegraphics[width=0.49\textwidth]{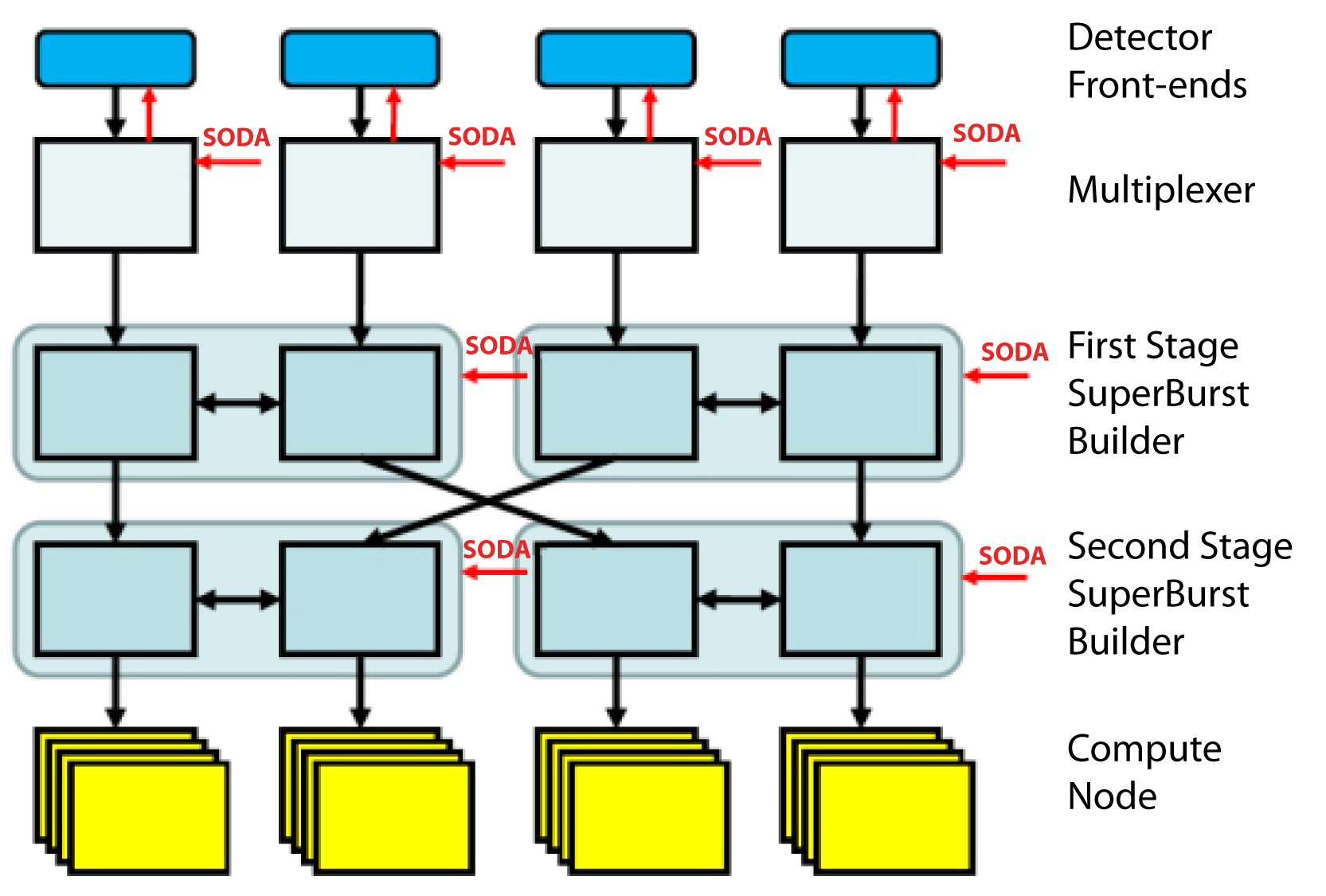}
\caption[Architecture of the lower levels of the \panda DAQ system]{Architecture of the lower levels of the \panda DAQ system.}
\label{fig:pandaDAQ}
\end{figure}

Due to its extreme data rates the MVD provides high challenges for this architecture. Assuming that the MVD module controllers and service boards (with the GBT) are combined such that the GBT links are almost fully loaded, the MVD Multiplexer Boards (MMBs)will do a 3-to-1 multiplexing with 3 optical links from the GBTs on the service boards and one optical uplink with 10~Gb/s. Since the maximum user data rate for the GBT is 3.28~Gb/s, all three links from the service boards can be fully loaded, simultaneously. Due to the high output rate the MMBs will be designed as MicroTCA-Boards. This allows direct insertion into the Compute Node version 3 and skipping of the SuperBurst building stages. Implementation of test systems is supported, as well.

\begin{figure}[ht]
\centering
\includegraphics[width=0.45\textwidth]{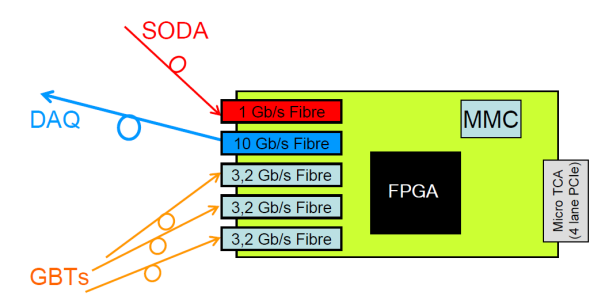}
\caption[Main components of the MMB]{Main components of the MMB (showing netto data rates of the link without GBT protocol or 8b/10b overhead).}
\label{fig:pandaMMB}
\end{figure}

The MMB will not use the GBT, but will implement the GBT protocol with a commercial FPGA. The Xilinx Virtex 6 is foreseen for this purpose, but the final decision will be done during the hardware design phase. \Figref{fig:pandaMMB} shows the main components of the MMB. 
SODA provides global clock and timing as well as information on bursts and super bursts, which will be propagated by the FPGA via the optical links to the service modules using the trigger\&timing interface component of the GBT. This information in combination with the reference clocks and reset pulses generated by the GBT allows proper time stamping of the FE ASIC data. Via the FPGA the GBT-SCA is accessible in order to do environmental monitoring. Setting the parameters of the FE ASIC (e.g.~thresholds) is done via the DAQ interface of the GBT and not with the GBT-SCA. On the MicroTCA interface the PCIe protocol (with 4 lanes) is used in the fat pipe. The PCIe protocol is implemented by the FPGA and gives access to the above-mentioned management and control functions. Also the detector data can be sent via the MicroTCA interface instead of the optical uplink. In this case, the average load over all 3 links from the GBTs must not overcome 80\%. The module management controller (MMC) is implemented by a Microcontroller, which communicates with the shelf manager via I2C. 

In normal operation it is foreseen that the MMBs will reside in MicroTCA crates. Each MicroTCA crate will contain one commercial off-the-shelf CPU-board for management and control purposes. Linux will run on the CPU-board enabling comfortable local and remote access to the MMBs via the PCIe link on the MicroTCA backplane. The shelf-monitoring functions of MicroTCA support additional remote monitoring and control operations.

%\section{Power supply system} (Bonn-Hans)

\input{integration/ps}

%	\subsection{Powering Concept}
%				Request for strips and pixel-integration
			
%		\subsubsection{Design}
												
%			Power Distribution 					
%			Grounding/Shielding (just to remember us)								

%	Power regulators							 
%		Design						

\input{integration/cables}   % part by Paolo de Remigis			
							
%		Requirements 			
%		signal cable
%		Al prototype/Results					
%		power cable/Selection				

%		routing scheme (Beppe + others)

%\\\\\\\\\\\\\\\\\\\\\\\\\\\\\\\\\\

%\newpage

\input{integration/mechanics}

%\\\\\\\\\\\\\\\\\\\\\\\\\\\\
		
\input{integration/cooling}

\section{DCS}

	For a reliable and safe detector operation the monitoring of operating parameters is crucial.
	Therefore a Detector Control System (DCS) will be deployed to permanently monitor parameter of importance for the MVD.
	This data will be collected and provided to the shift crew while running the experiment.
	The occurrence of critical values outside predefined parameter ranges need to create an alert or even cause an interrupt so that measures can be taken to prevent disturbance in operation or even a damage to the detector.
	The data shall be archived in order to evaluate the detector performance over time and identify possible problems.
	The \panda Experiment Control System (ECS) will be based on EPICS\footnote{{\bfseries E}xperimental {\bfseries P}hysics and {\bfseries I}ndustrial {\bfseries C}ontrol {\bfseries S}ystem}. It allows to collect, display, and archive sensor readings from various devices. 
	The following parameters should be kept under surveillance:
	\begin{itemize}
		\item voltage and current of the \frontend electronic
		\item voltage and leakage current of the sensors
		\item temperature of electronic and coolant
		\item coolant flow
		\item temperature and humidity in the detector volume
	\end{itemize}
	
The supply voltages and the current consumptions of the \frontend electronics will be measured at the power supplies and the voltage regulators close to the detector. 

%The \frontend electronics used in the MVD with its low feature sizes is very sensitive to voltage spikes. Therefore a constant monitoring of the supply voltages and current consumptions is important to protect the electronics. These values will be measured at the power supplies and the voltage regulators close to the detector.

	For the silicon sensors the depletion voltage and the leakage current have to be supervised. They give an indication of the condition of the sensors and can be used to check the radiation damage the sensors have taken. Here a measurement at the power supplies will be performed.
	
	The sensors and, to a lesser extend, the electronics are  very sensitive to the operation temperature. A close surveillance of the temperature and the cooling system are therefore very important for a safe and stable operation of the MVD.
	
	Thus each pixel module and each strip super module will be equipped with an NTC thermistor to measure the temperature on the module. The readout cables of the resistors are routed along with the signal bus to the patch panel. Here the resistivity is measured and the data is fed into the DCS system.
	
The cooling system is monitored via temperature and pressure sensors on each cooling pipe and the correct temperature of the water in each  pipe can be kept by a dedicated heater.
 In addition the overall temperature and humidity in the MVD and the temperature of the off-detector electronics will be measured and controlled.

%\section*{Authors}
%\begin{tabbing}
%\hspace{3cm} \=S. Coli \\
%                       \>P. De~Remigis \\
%                       \>G. Giraudo \\
%                       \>G. Mazza \\
%               		\>T. Stockmanns\\
%\end{tabbing}

\putbib[lit_intg]

%% file: integration/odt.tex
\section{Optical Data Transmission }
\authalert{Author: G. Mazza; mazza@to.infn.it}

\begin{figure*}[h]
\begin{center}
\includegraphics[width=0.9\textwidth]{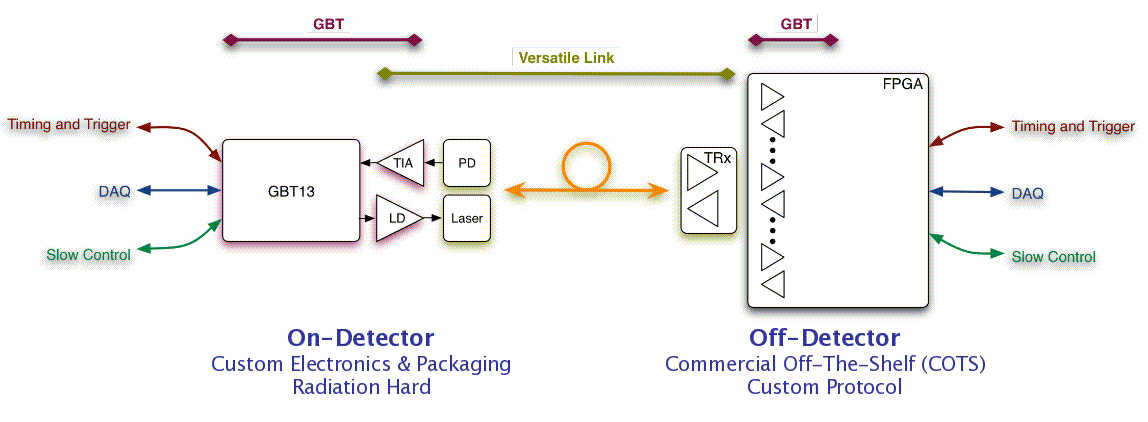}
\caption[GBT and VL scheme]{GBT and VL scheme. LD: Laser Driver, TIA: TransImpedance Amplifier, PD: Photo Diode, TRx: TRansceiver.}
\label{f:GBT_scheme}
\end{center}
\end{figure*}

The maximum expected data rate for the pixel MVD is estimated to
be 450~Mbit/chip. The  8b/10b encoding on the output data could increase the number to $\sim$~600~Mbit/chip.\\
Due to the absence of a trigger signal, no trigger matching logic can be
used to reduce the amount of data already at the \frontend chip level. Moreover,
the track distribution over the barrel and the disks can vary significantly
depending on the type of the target, thus not allowing an optimisation based
on the detector position. These constraints make the pixel MVD data transmission
extremely demanding in terms of required bandwidth.\\
Another stringent constraint is due to the reduced dimensions of the MVD and its
closeness to the interaction point, which limits the number of cables that can
be accommodated and tolerated (for material budget issues).\\
These constraints lead to the choice of using high speed serial links and to 
convert the data transmission from electrical to optical as close as possible 
to the detector, in order to profit from the high data rates allowed by the
optical fibers.\\
Unfortunately the use of off-the-shelves components for this application is
not straightforward because of the radiation environment where the components
will be placed. However, the problem is common to most of the HEP community
and in particular for the upgrades of the LHC experiments. In this context
two research projects, named GigaBit Transceiver (GBT) \cite{Moreira09} and 
Versatile Link (VL) \cite{Troska09} are ongoing in order to develop a radiation 
tolerant transceiver (the GBT) and to qualify optical components for the radiation 
environment (the VL).The connection between the MVD and the DAQ is going to make 
profit of this development and therefore to design a \frontend electronic compatible 
with the GBT interface. \Figref{f:GBT_scheme} shows the scheme proposed by the 
GBT and VL projects.

	\subsection{GigaBit Transceiver}

The goal of the GBT project \cite{Moreira09} is to produce the electrical 
components of a radiation hard optical link, as shown in \figref{f:GBT_scheme}. 
One half of the system resides on the detector and hence in a radiation environment, 
therefore requiring custom electronics. The other half of the system is free from 
radiation and can use commercially-available components. Optical data transmission is 
via a system of opto-electronics components produced by the Versatile Link project. 
The architecture incorporates timing and trigger signals, detector data and slow controls 
all into one physical link, hence providing an economic solution for all data 
transmission in a particle physics experiment.\\
The on-detector part of the system consists of the following components.

\begin{figure}[h]
\includegraphics[width=0.5\textwidth]{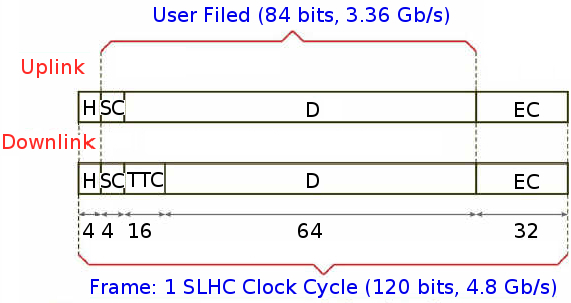}
\caption[GBT data packet format]{GBT data packet format.}
\label{f:GBT_packet}
\end{figure}

\begin{itemize}

\item{GBTX \cite{Moreira11} \cite{Cobanoglu09}: 
a serialiser-de-serialiser chip receiving and transmitting serial data at 4.8~Gb/s . 
It encodes and decodes the data into the GBT protocol and provides the interface to 
the detector \frontend electronics.}

\item{GBTIA \cite{Menouni09}: a trans-impedance amplifier receiving the 4.8~Gb/s serial 
input data from a photodiode. This device was specially designed to cope with the 
performance degradation of PIN-diodes under radiation. In particular the GBTIA can handle 
very large photodiode leakage currents (a condition that is typical for PIN-diodes 
subjected to high radiation doses) with only a moderate degradation of the sensitivity. 
The device integrates in the same die the transimpedance pre-amplifier, limiting amplifier 
and 50~$\Ohm$ line driver.}

\item{GBLD \cite{Mazza09}: a laser-driver ASIC to modulate 4.8~Gb/s serial data on a laser. 
At present it is not yet clear which type of laser diodes, edge-emitters or VCSELs, 
will offer the best tolerance to radiation \cite{Troska09}. The GBLD was thus conceived to drive 
both types of lasers. These devices have very different characteristics with the former type 
requiring high modulation and bias currents while the latter need low bias and
modulation currents. The GBLD is thus a programmable device that can handle both types 
of lasers. Additionally, the GBLD implements programmable pre- and de-emphasis equalisation, 
a feature that allows its optimisation for different laser responses.}

\item{GBT-SCA \cite{Gabrielli09}: 
a chip to provide the slow-controls interface to the \frontend electronics. This device is 
optional in the GBT system. Its main functions are to adapt the GBT to the most commonly 
used control buses used HEP as well as the monitoring of detector environmental 
quantities such as temperatures and voltages.}

\end{itemize}
The off-detector part of the GBT system consists of a
Field-Programmable-Gate-Array (FPGA), programmed to be
compatible with the GBT protocol and to provide the interface
to off-detector systems.\\
To implement reliable links the on-detector components
have to be tolerant to total radiation doses and to single event
effects (SEE), for example transient pulses in the photodiodes
and bit flips in the digital logic. The chips will therefore be
implemented in commercial 130~nm CMOS to benefit from its
inherent resistance to ionizing radiation. Tolerance to SEE is
achieved by triple modular redundancy (TMR) and other
architectural choices. One such measure is forward error correction (FEC), 
where the data is transmitted together with a Reed-Solomon code which allows
both error detection and correction in the receiver.
The format of the GBT data packet is shown in \figref{f:GBT_packet}. A
fixed header (H) is followed by 4 bits of slow control data
(SC), 80 bits of user data (D) and the Reed-Solomon FEC
code of 32 bits. The coding efficiency is therefore 88/120~=~73\%, 
and the available user bandwidth is 3.2~Gb/s.
FPGA designs have been successfully implemented in
both Altera and Xilinx devices, and reference firmware is
available to users. Details on the FPGA design can be found
in \cite{Baron09}.

%	\subsection{Optoboards}

%% file: integration/ps.tex
%File: powerspl.tex .
%Comment: Power supply for the low and the high voltage system, by De Remigis.
%Date: 21.09.11.
\section{Power Supply System}
\authalert{deremigi@to.infn.it}
\subsection{Introduction}
The power supply system for the whole MVD has to provide low voltage for a total
number of 2130 readout circuits, including pixel and strip electronics, and 346
controller or transceiver circuits to transmit the data towards the DAQ.

Besides the power supply system has to provide high voltage for a total number
of 136 super modules, including both pixel and strip sensors.

\subsection{Powering Concept for the Pixel Part}
For the hybrid pixel part the smallest unit considered for the segmentation of
the power supply system is a super module, that can include up to 6 readout
chips (ToPix) and a transceiver chip (GBT).
For each basic unit, the design of the power supply system has to provide two
independent floating channels, for the analogue and digital sections, with
sensing feedback, for the voltage compensation due to the cable voltage drop.

With respect to the power supply system, the requirements for the ToPix readout
chip are: a channel for the analog circuits with a nominal voltage $V_{\text{a}}=1.2$~V,
an upper voltage limit $V_{\text{al}} =1.4$~V, a nominal current $I_{\text{a}}=200$~mA, an upper
current limit $I_{\text{al}}=220$~mA; and a channel for the digital circuits with a
nominal voltage $V_{\text{d}}=1.2$~V, an upper voltage limit $V_{\text{dl}}=1.4$~V, a nominal
current $I_{\text{d}}=1.2$~A, an upper current limit $I_{\text{dl}}=1.4$~A.
These numbers are evaluated after the measurement on the ASIC prototype ToPix~v3,
and extrapolating the result to a full size chip with the complete array of
cells, that affects mainly the analogue current, and the entire set of columns,
that affects the digital current.
A significant amount of the digital current is due to the differential drivers
(7~outputs) and receivers (9~inputs) constituting the interface for the chip
that, at present, is considered completed and does not require to be increased
as in the case of the current for the end of column blocks.
Currently, the expected currents are evaluated considering a readout circuit
running with a system clock 160~MHz fast and a pre emphasis capability to drive
the long transmission lines, while working at 80~MHz or releasing the boosting
could decrease the total current consumption of the 20\% or 5\% respectively.
Therefore, sensible current limits for the power supply have to be selected to
protect the readout, without requiring too much power that could lead to an
oversised system; for this purpose, since the present figures constitute the
worst case, an increment of just 10\% for all the parameters has been chosen.

\subsubsection*{Power Regulators}
As power regulator the linear regulator solution has been excluded due to the
low efficiency in the power conversion, while a switching DC-DC solution has been
chosen because it has an higher efficiency and can be based on the project
currently under development at CERN for the upgrade of the LHC experiments.

The switching converter needs to store the energy either by the magnetic field
(using inductors) or by the electrical field (using capacitors); at present the
main project is addressing the architecture with inductors that allows larger
efficiency, although the architecture with capacitor looks very attractive due
to the small size of the whole circuit.
The full DC-DC converter is a kind of hybrid based on a small PCB containing few
passive components, mainly the inductor, and an ASIC providing the switching
transistors and the control logic to monitor the correct behaviour
\cite{ref:dcdc}.
Currently there are different prototypes featuring integrated circuits made with
different technology to test their hardness with respect to the radiation dose,
but in any case all of them foresee an input voltage around $V_{\text{i}}=10$~V, an
output voltage $V_{\text{o}}=1.2$~V compatible with the new readout circuits and an
output current of roughly $I_{\text{o}}=3$~A.

This type of inductor based converter, also called ``buck converter'', requires
particular care when working in a high magnetic field and for this reason it
cannot make use of ferrite core coils since it risks the saturation, but an air
core coil is preferred although an increment in the size is expected.
Another issue related to the inductor is the radiated switching noise and his
electromagnetic compatibility with the system, for this purpose the coil is
shielded by a small box with the conductive surface and the geometrical
dimensions become in the order of $\mathrm{30 \times 10 \times 10~mm^3}$.

At present the CERN development is selecting the best technology for the ASIC
production from the point of view of the radiation hardness, but a backup
solution has been found in the 350~nm On Semiconductor that has been already
qualified for this application.
The other technology option comes from the 250~nm IHP, but since it is a more
advanced solution it has not yet reached the necessary level of radiation
hardness and it isn't enough tolerant against the single event burnout or the
displacement damage.
Specifically these DC-DC converters have been tested to cope with the environment
of an experiment for high energy physics considering a total integrated dose of
around 1~MGy, a particle fluence of roughly $10^{15}$~\neueq and a
magnetic field of about 4~T.

\subsection{Powering Concept for the Strip Part}
\label{sec:strip_power}
For the strip part the foreseen powering scheme for one super module contains separate low voltage supply paths for the n-side and the p-side readout electronics, respectively since the n-side \frontend electronics is operated without galvanic reference to a common ground in a floating regime. This avoids high parasitic currents in the case of breakdown of the thin insulation barrier between n-side implants and metallisation (pinholes) of the sensors, which would otherwise result in the loss of the complete sensor. Thus a total of 4 independent low voltages (one analogue and one digital for either sensor readout side) and 4 high voltages for each sensor bias have to be supplied. The low voltage paths are connected with sense feedback to compensate for the voltage drop on the cables. Assuming a maximum of 1~W power dissipation per \frontend at an operation voltage of 3.3~V a value of 5.3~A can be estimated for the maximum supply current on the low voltage lines. The current drawn on the bias voltage lines is negligible even with the maximum applied irradiation dose at the end of the \panda lifetime and amounts to at most 2~mA per sensor. Sensing lines are thus not required for this supply but a good filtering with very low cutoff frequencies is mandatory on the module level.

%% file: integration/cables.tex
%File: cables.tex .
%Comment: Low mass cables in aluminium for the data connection, by De Remigis.
%Date: 21.09.11.
\section{Cables}
\authalert{deremigi@to.infn.it}

\label{Cables}
\label{cable-sec}

A special feature of the \PANDA experiment comes from the
triggerless DAQ, that requires a very high output bandwidth because to the
continuous flux of data that cannot exploit the de-randomisation coming from an
external trigger at a lower rate.
Besides, since at present the only way out from the MVD is in the backward
region, the cabling becomes a serious issue because of the amount of material
needed to take out all the data from the sensitive volume.

\subsection{Requirements}
\label{cable-reqs}

Due to the average annihilation rate of 20~MHz a flux of $3\cdot 10^6$ particles per 
second and chip is expected in the hottest region, the forward disks.
The current design for the pixels considers a ToPix readout circuit managing a
 100~$\tcmu$m thick sensor, a data word of 40~bit per event, a variable luminosity
structure due to the peak in the beam profile and the crossing of the pellet
target, therefore the readout chips in the middle of the disks have to deal with
a data rate of 450~Mbit/s each.

The present solution for the architecture of the serial link comes from the Giga
Bit Transceiver (GBT) project under development at CERN, for the upgrade of the
experiments that will be running at the Super LHC (SLHC).
The GBT is foreseen to be interconnected to the readout circuits by an Elink
that is an electrical interface implementing a serial connection.

\subsection{Signal Cable}
To deal with the low material budget for the vertex region, the use of copper
cables is excluded and one has to employ aluminium cables.
In this case considering that aluminium has a lower atomic number than copper,
even if it has a higher resistivity, an overall gain is obtained in the material
budget because the aluminium radiation length is 88.9~mm instead of copper
radiation length that is 14.4~mm.

From the physical point of view each logical signal going to, or coming from,
ToPix chip is implemented by a differential pair to be less sensitive against
the noise that can be induced at the same time on each single line, and that
will be canceled after the subtraction between the two phases.
In particular these connections are implemented as a transmission lines using the
microstrip technique, and are composed as a sandwich of two aluminium conductive
layers separated by a polyimide insulator layer \cite{ref:cable}.
On one side the two tracks composing the differential pair are laid down, while
on the other side the metal plane constitutes the reference ground.

\subsubsection*{Al Prototypes}
Several kinds of cable prototypes were made and tested, differing either in the
manufacturing technology or in the physical layout.
For the first choice the standard solution is a laminated sheet over an
polyimide layer, that is worked in a way similar to a PCB etching the metal from
the area where it must be removed.

The other manufacturing solution is an aluminium deposition on top of an
insulator film, that is obtained dispensing the metal just in the required area
defined by the layout pattern, by the Physical Vapour Deposition (PVD).

\subsubsection*{Folded Layout}
The first samples were produced to test both the technologies, by using the
metal lamination from the Technology Department at CERN and successively the
aluminium deposition from the Techfab company near Torino.
The CERN prototype has a stack of up to two aluminium foils with a thickness of 15~$\tcmu$m, on both sides of the plastic film 51~$\tcmu$m thick produced by DuPont
under the commercial name of Pyralux.
Various layout prototypes were produced with different widths and spacing of 100~$\tcmu$m + 100~$\tcmu$m, 150~$\tcmu$m + 150~$\tcmu$m and 200~$\tcmu$m + 200~$\tcmu$m
respectively.

The Techfab prototype features a cross section of two aluminium depositions with
a thickness of roughly 7~$\tcmu$m, around the  50~$\tcmu$m thick plastic sheet
produced by DuPont as common kapton.
In this case, a sample with a unique layout showing 150~$\tcmu$m width and 150~$\tcmu$m spacing has been obtained.
Due to some issues with the wire bonding procedure, a try has been performed
with an aluminium/silicon alloy to improve the bond strength.
The samples with the folded layout have been used to perform an irradiation test
by neutrons at the LENA nuclear reactor in Pavia with a total fluence greater
than what is expected for the \PANDA lifetime, and no significant
variations have been observed.

\subsubsection*{Straight Layout}
\begin{figure}[ht]
	\center{\includegraphics[width=.45\textwidth]
	{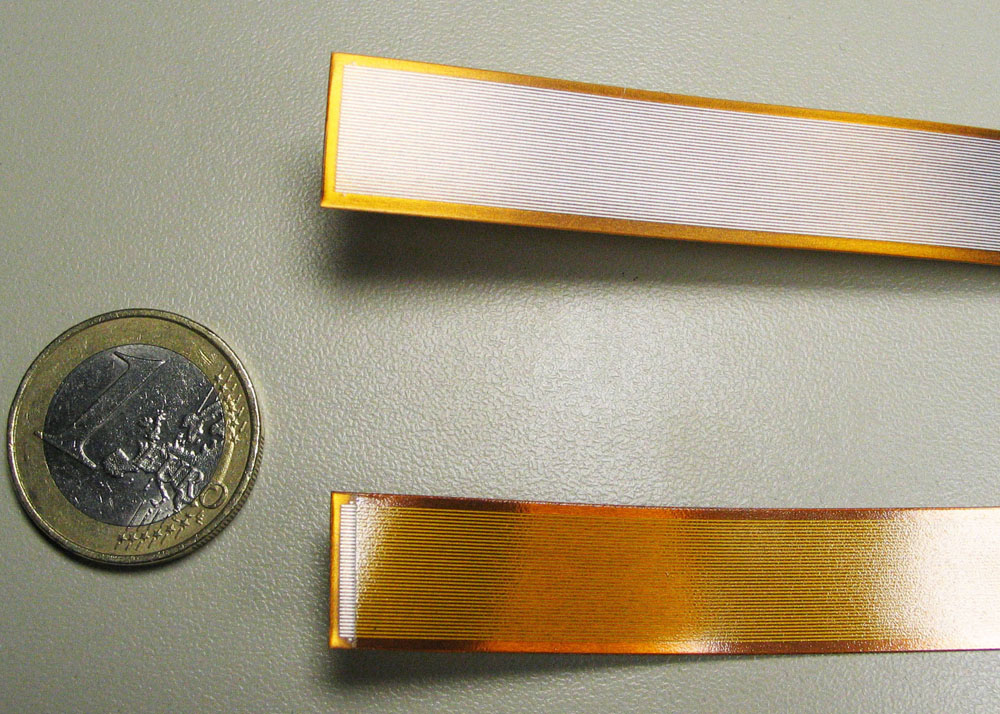}}
	\caption{Low mass cables implementing aluminium
	microstrips\label{fig:cable}.}
\end{figure}

The low mass cables have been designed implementing the connections as a
transmission line, using the microstrip technique on DuPont materials and
manufacturing them in a way similar to a PCB.
Currently there are aluminium microstrips, produced by the Technology
Department at CERN, with a layout including 18 differential pairs for a total
number of 36 tracks; at present there are two versions differing for the nominal width
that is 100~$\tcmu$m or 150~$\tcmu$m and the spacing that is 100~$\tcmu$m or 150~$\tcmu$m respectively.
All of them have the insulator cover on the bottom side to protect the ground
plane, while one more parameter introducing another difference on the available
samples is the presence or the absence of the insulator cover laid on the top
side where the differential strips are implemented (\figref{fig:cable}).
Summarising, at present, there are 4 different kinds of aluminium microstrips 1~m
long under study differing for the pitch size or for the top covering.

\subsubsection*{Results}
\begin{figure}[t]
	\center{\includegraphics[width=.45\textwidth]
	{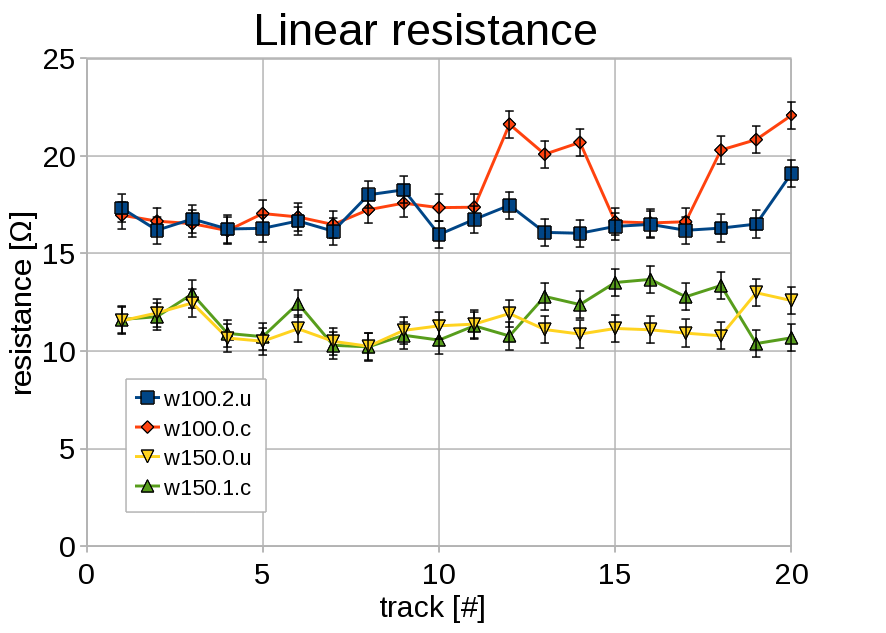}}
	\caption{Linear resistance measured for the first 20 tracks of each
	sample\label{fig:resistance}.}
\end{figure}

\begin{figure}[ht]
	\center{\includegraphics[width=.5\textwidth]
	{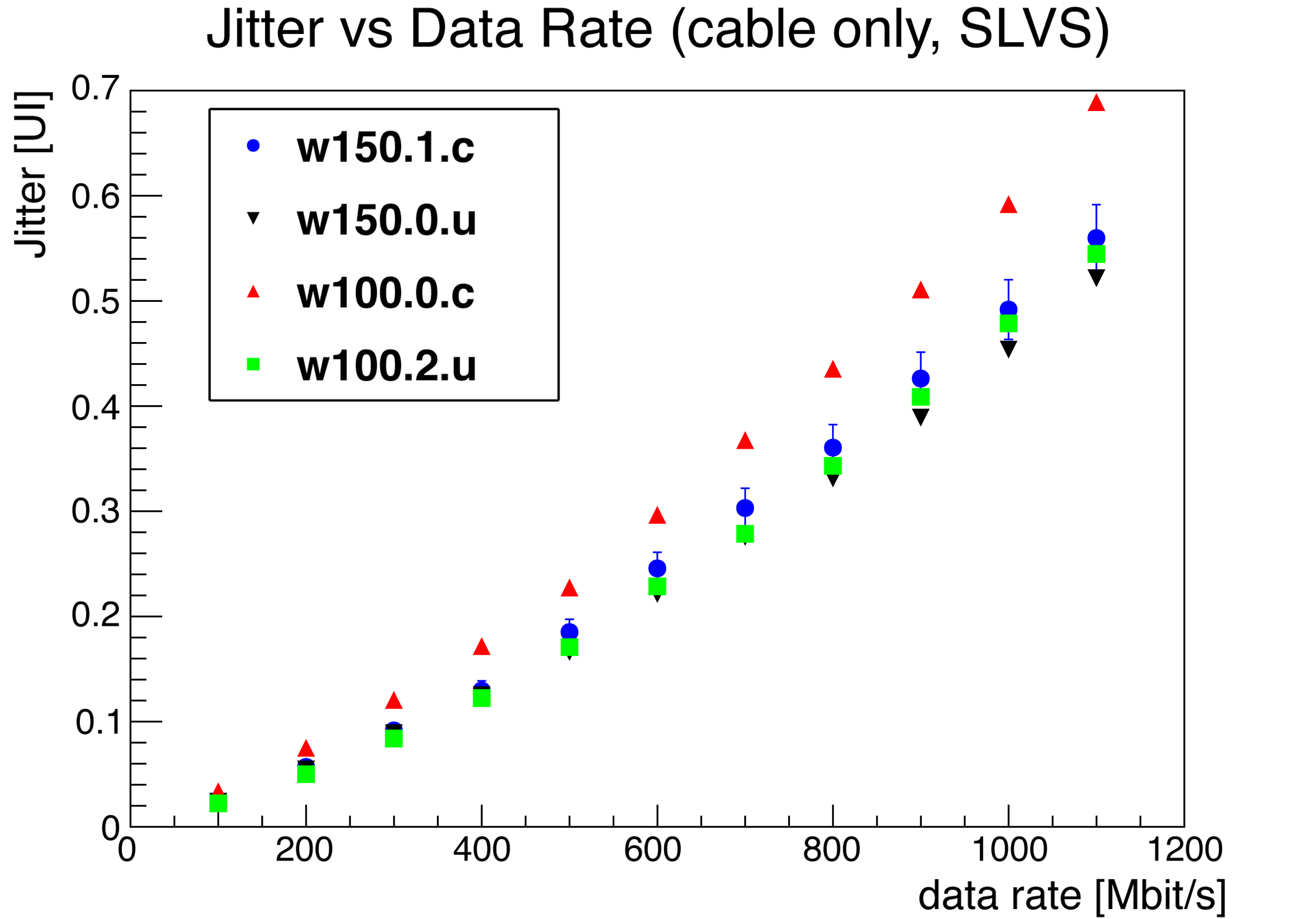}}
	\caption{Total jitter evaluated for just the microstrips without any
	transceivers.}
	\label{fig:jitter}
\end{figure}

From a static point of view the aluminium microstrips were tested to measure the
resistance of each track between its ends, and the capacitance referenced to the
ground plane; moreover the uniformity for these parameters was evaluated (\figref{fig:resistance}).
The measurement has been performed bonding both cable ends to the test station
to reduce the contact problems, and no dependence between static parameters and
track position has been observed.
Instead, as expected, the results show that there is a clear relation between
the linear resistance and the track width, that is around 12~$\Omega$ for the
large strips (150~$\tcmu$m) and about 17~$\Omega$ for the narrow ones (100~$\tcmu$m).
Regarding the dynamic behavior, a test bench has been setup with some prototype
circuits implementing the Scalable Low Voltage Signaling (SLVS) standard, to
measure the total jitter and its main components like the deterministic jitter
or to perform a Bit Error Ratio Test (BERT) as a function of the data rate
\cite{ref:slvs}.
Other tests were performed to evaluate the cross talk effect with the
transceiver prototypes, and to measure the total jitter produced by just the
aluminium cables without introducing the component due to the SLVS circuits.
The present solution considers that signals running on the low mass cables
comply with the SLVS differential standard which foresees for each line a
$V_{\text{low}}=100$~mV and a $V_{\text{high}}=300$~mV, against the LVDS differential standard
which typically foresees for each line a $V_{\text{low}}=1.05$~V and a $V_{\text{high}}=1.45$~V.
The setup for the measurement is composed of the aluminium cable under test that
is glued by the shield on two boards specifically designed, while the
interconnections are made with a standard wire bonding.
The test was performed by a pulse generator configured to produce a Pseudo
Random Bit Sequence (PRBS) with a pattern length of $p=2^{23}-1$ bits, and output levels
according the SLVS standard.
In this way the performance of the aluminium microstrip can be evaluated
studying the analog shape of the signals coming out of cable, without the
insertion of any distortion attributable to the transceivers.
The 4 samples measured so far are called w100.0.c, w100.2.u, w150.0.u and
w150.1.c where the first number represents the track width in micrometer, and
the last letter represent the condition to be covered or uncovered with the
insulator film.
The total jitter has been measured and evaluated in Unit Interval (UI), that
represents the ideal period of a bit irrespective of the link speed, allowing
the relative comparison between different data rate (\figref{fig:jitter}).
Since the commercial serialiser and deserialiser are considered safe when their
total jitter came up to a threshold of ${T_\text{j}}=0.3$~UI, it is possible to conclude
that the aluminium microstrips with track width ${w}=100$~$\tcmu$m are able to run up
to a data rate ${f}=600$~Mbit/s while the differential cables with track width ${w}=150$~$\tcmu$m can work up to a data rate ${f}=800$~Mbit/s.

\begin{figure}[ht]
	\centering{\includegraphics[width=.45\textwidth]
	{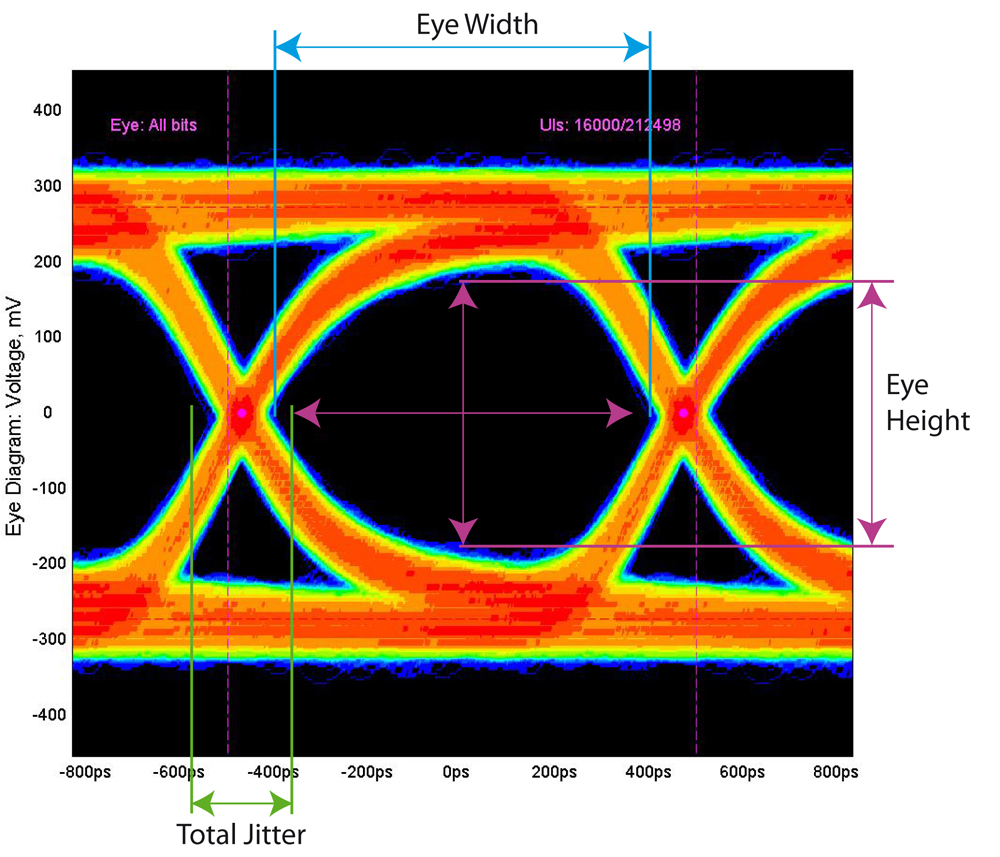}}
	\caption{Example of an eye diagram.}
	\label{fig:eyediagram}
\end{figure}

An useful tool to evaluate the total jitter with respect to the unit interval is
the eye diagram, see \figref{fig:eyediagram}, that is a composite view of the signal made by the overlapping
of many acquired waveforms upon each other corresponding to bit periods
representing any possible logic states and transitions.

Moreover, analysing the eye diagram height at the end of the flexcable relative
to the input PRBS, the differential amplitude can be evaluated and it results in
final value of roughly 300~mV instead of the nominal value of about 400~mV, with
an attenuation that is due to the tracks linear resistance.
The test setup has allowed an evaluation of the cross talk keeping a victim
differential pair biased in a steady state and driving the aggressor
differential lines, on the left and right sides of the first one, by the same
signal, but a very light effect has been reported and probably this is due to
the little swing of the SLVS standard and the relatively low transition time of
the signals connected to the high capacitive load.

\subsection{Power Cable}
Each ToPix circuit is expected to consume around 1.4~A and since it requires a
supply voltage at 1.2~V, that means a total power of 1.68~W typically,
corresponding to a relative power of 979~mW/cm$^2$.
However two independent lines have to be distributed from the power converter to
separate the analog part from the digital part, to reduce the risk of mutual
interference.

Also in this case the material budget requirement represents a constraint that
prevents the use of copper conductors inside the active volume of the MVD, 
then a
solution with aluminium was investigated.
At present a sample of enameled wire made of copper clad aluminium has been
evaluated to verify that it is suitable for the routing in a very 
limited space, reducing the voltage drop to roughly 100~mV.

%% file: integration/mechanics.tex
\section{Mechanical Structures}

\label{mechanics}
\authalert{Author: G. Giraudo, giraudo@to.infn.it'}

\begin{figure*}[!ht]
\begin{center}
\includegraphics[width=0.67\textwidth]{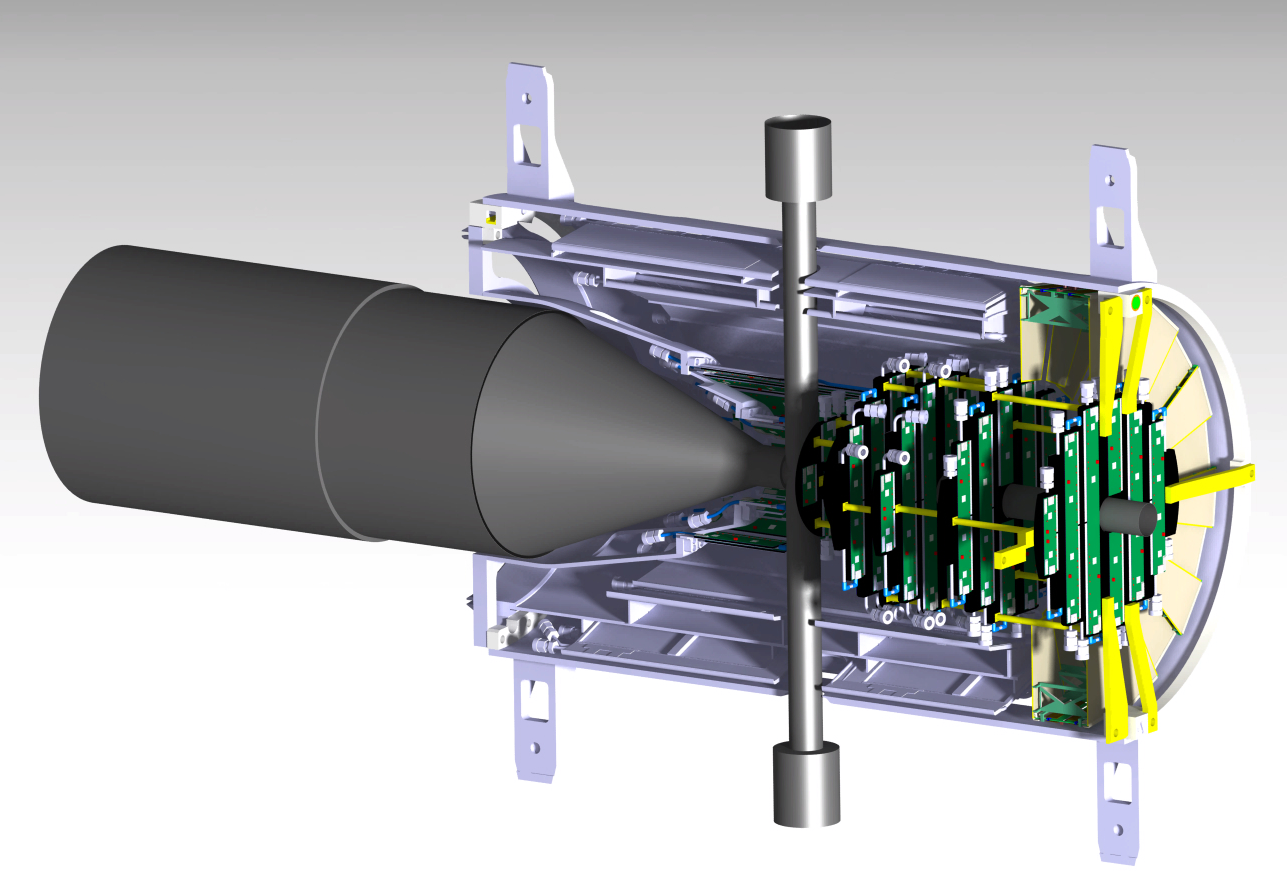}
\caption[Longitudinal cross-section of the MVD and the cross-pipe sector]{Longitudinal cross-section of the MVD and the cross-pipe sector.}
\label{fig:mechFig_01}
\end{center}
\end{figure*}

The whole MVD is composed by four mechanically independent
sub-structures:
\begin{itemize}
\item pixel barrels;
\item pixel disks;
\item strip barrels;
\item strip disks.
\end{itemize}
Each of them is assembled independently, then
gathered on a stand and positioned on a holding frame.
 \Figref{fig:mechFig_01} shows a longitudinal
cross section of the MVD together with the cross-pipe sector.

\subsection{Global Support: The Frame}

 The frame is
the main structural element of the whole MVD. Its purpose is to
suspend and keep in position all the sub-structures, with a
relative positioning error of less than 100~$\tcmu$m as well as to connect the
MVD to the external world (i.e.~the central tracker). The
frame also acts  as support for the services from the inner
part of the MVD.  The frame has a sandwich
structure, made by a high modulus composite of unidirectional carbon fiber
cyanate ester resin and foam as core material. The
cyanate ester resin systems feature good behavior and high
dimensional stability when immersed into a temperature-humidity varying
field. The use of a conventional epoxy system would have drawbacks,
due to excessive moisture absorption and residual stress. As
compared to epoxies, cyanate ester resins are inherently tougher, and they
have significantly better electrical properties and lower moisture
uptake \cite{re:mechRef_1} \cite{re:mechRef_2}. The main properties of the
M55J/LTM110-5 composite material are reported in \tabref{tab:mechTab_1}.

\begin{table}[b]
\centering
\resizebox{\columnwidth}{!}{
\begin{tabular}{|l|c|}
  \hline
  Tensile Modulus $E_{11}$~(GPa) &  310 \\ \hline
  Tensile Strength $\sigma_{11}$~(MPa) & 2000 \\ \hline
  Poisson's ratio $\nu_{12}$ & 310 \\ \hline
  CTE fibre (ppm/$^\circ$C) & $-1.1$ \\ \hline
  CTE resin (ppm/$^\circ$C) & 60 \\ \hline
  Moisture uptake (155\,d~-~22$^\circ$C~-~75\%\,RH) & 1.4\% \\
  \hline
\end{tabular}
}
\caption[M55J/LTM110-5 Composite properties]{M55J/LTM110-5 Composite properties.}
\label{tab:mechTab_1}
\end{table}

A finite elements analysis has been performed in order to define
the minimum number of plies necessary to achieve the requested stiffness.
According to the simulation, the frame has been designed 
including  two skins,
each composed of four plies with a quasi-symmetric stacking sequence
of unidirectional M55J carbon fiber prepreg, and  3~mm of
Rohacell\textsuperscript{\textregistered}51IG as core material. A set of eight 
semi-circular, 1~mm thick, reinforcing
ribs is embedded into the structure. These  ribs will be realised with the
same composite material 
used for the external skins. The final thickness
of the frame is 4~mm.

\begin{figure}[htb]
\begin{center}
\includegraphics[width=0.42\textwidth]{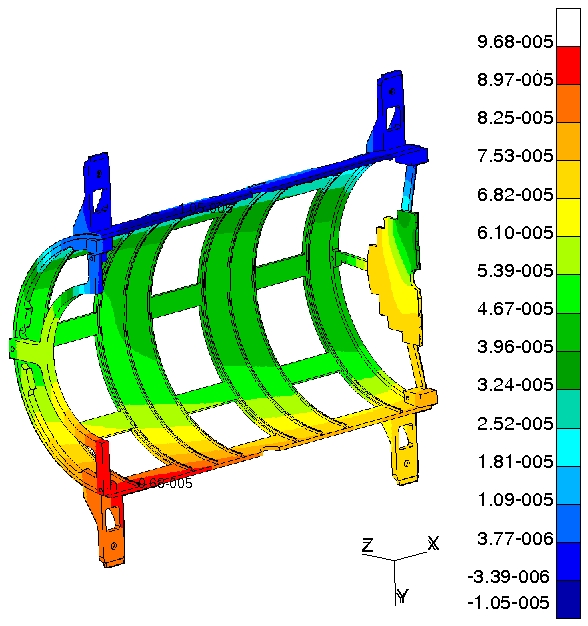}
\caption[Maximum vertical displacement of the frame under static load]{Maximum vertical displacement of the frame under static load. A safety
factor of 2 is applied.}
\label{fig:mechFig_02}
\end{center}
\end{figure}

\Figref{fig:mechFig_02} shows the behavior of the frame under 
gravitational load. Extra loads are applied, 
distributed on the end-rings to take into 
account a safety factor of 2. The simulation shows a maximum 
displacement in vertical direction of less than 100~$\tcmu$m.

The frame holds the four sub-structures containing the
silicon devices in their place. Two rings glued at the ends of the frame act as
connectors between the sub-structures and the frame itself. The rings are
made of mattglass/epoxy laminate composite, Durostone\textsuperscript{\textregistered}EPM203,
pre-machined and finished after gluing in order to achieve the final
dimensions and the requested accuracy. A set of proper tools is
necessary for these operations.

\begin{figure}[!ht]
\centering
\includegraphics[width=0.45\textwidth]{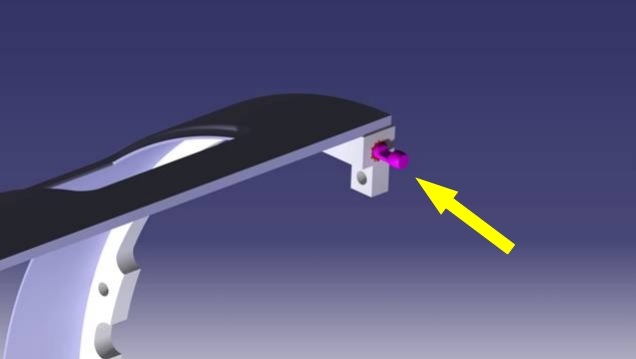}
\caption[Cams system. Cylindrical cam on upper end for 
fixing the half frame]{Cams system. Cylindrical cam on upper end for 
fixing the half frame.}
\label{fig:mechFig_04a}
\end{figure}

\begin{figure}[!ht]
\centering
\includegraphics[width=0.40\textwidth]{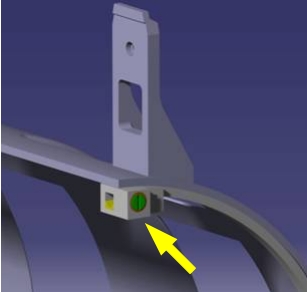}
\caption[Cams system. The V insert]{Cams system. The V insert.}
\label{fig:mechFig_04b}
\end{figure}

\begin{figure}[!ht]
\centering
\includegraphics[width=0.40\textwidth]{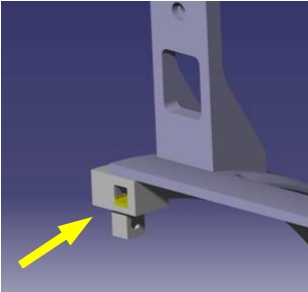}
\caption[Cams system. The U insert]{Cams system. The U insert.}
\label{fig:mechFig_04c}
\end{figure}

The relative position of the two halves is ensured by means of a three
cam system. \Figsref{fig:mechFig_04a},
\ref{fig:mechFig_04b}  and \ref{fig:mechFig_04c} show the ancillary
components of the system. 
Two cylindrical 
cams are located on the upper ends
of one half frame, whereas a third one is
located at the lower end. A sphere is positioned on top of the cams. 
On the other half frame there are inserts, which are
embedded into the rings. One insert, on the top, has a ``V" shape
while a second one, also on the top but at the opposite end,  is ``U"-shaped.

The upper cams pass through the inserts and fix the position
along the Z axis. The third cam, located on the bottom of the half frame,
fixes the radial position.  All the ancillaries are made by hardened aluminium
alloy (Hotokol).  
A similar system allows the MVD to be suspended from the Central Frame
and to be fixed with three springs. The springs are shaped so to get a
two-components locking force: one component (X) is on the horizontal
plane fixing the radial position, while the second is on the vertical plane,
downward oriented, and defines the Y and Z components.
Finally, the position is achieved by means of the V and U shaped inserts
embedded into the Central Frame structure.

This system allows the control of the position to be achieved, so
that re-positioning is possible with a high degree of accuracy,
that we can evaluate to be less than 10~$\tcmu$m \cite{re:mechRef_3}.

\subsection{The Pixel Support Structure}

\subsubsection*{Pixel Barrels}
The pixel barrels are a collection of two cylindrical layers of
detectors placed around the interaction point and coaxial with the
beam pipe axis.

The first, innermost, layer is composed by 14 super-modules, while
the second layer by 28. Each super-module is a linear structure of
pixel sensors, hosting from two to twelve elements, glued on a
mechanical support made of a carbon foam (POCO HTC and POCO FOAM) thin layer and a
$\Omega$-shaped substrate realised with a  M55J carbon fibre cyanate ester
resin composite. The cooling tube is embedded between the carbon
foam and the $\Omega$ substrate.

\begin{figure*}[!ht]
\centering
\includegraphics[width=0.85\textwidth]{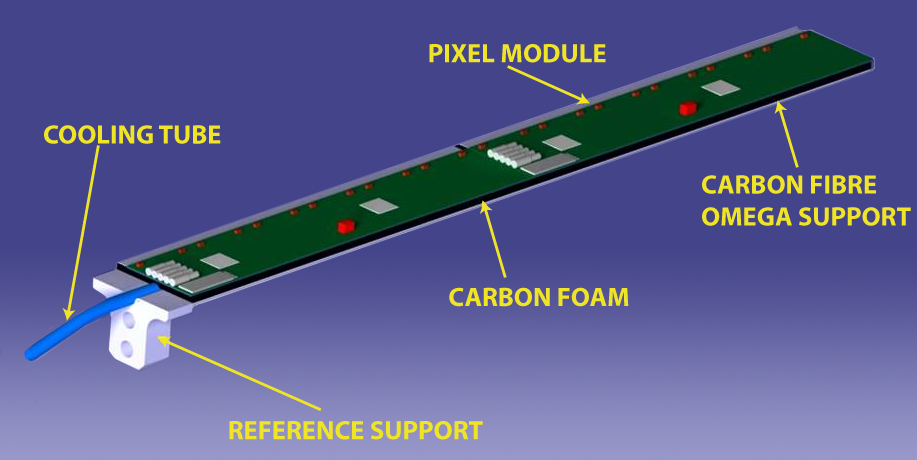}
\caption[Structure of a pixel barrel super module]{Structure of a pixel barrel super module.}
\label{fig:mechFig_05}
\end{figure*}

\begin{figure}[htb]
\centering
\includegraphics[width=0.45\textwidth]{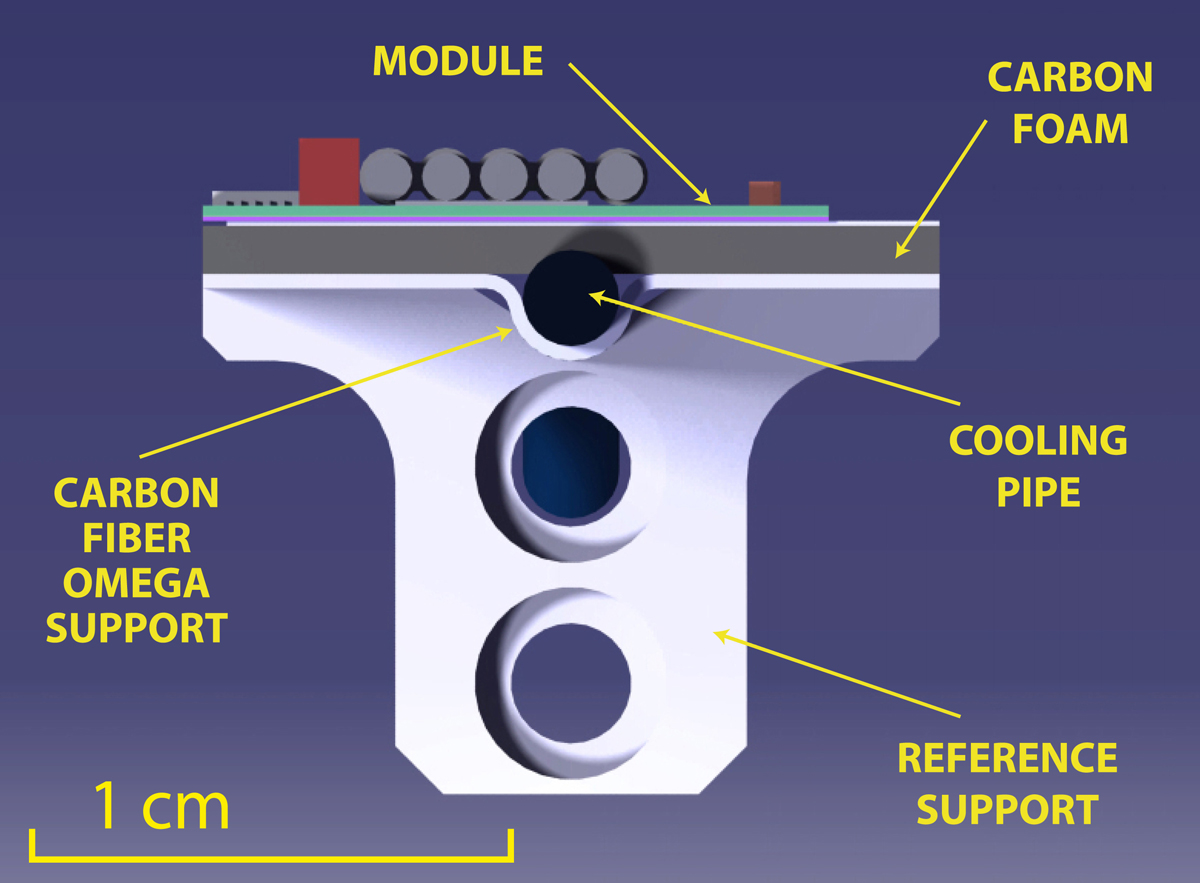}
\caption[Transverse section of a pixel barrel super module]{Transverse section of a pixel barrel super module.}
\label{fig:section_module}
\end{figure}

\vspace{1cm}
\Figref{fig:mechFig_05} sketches
the structure of a super module and \figref{fig:section_module} shows 
its transverse section.

A plastic block, glued on one end of the $\Omega$ substrate, acts
as reference. The block has two precision holes through which two
precision screws fasten the super module to the support structure.

The support structure is an assembled set of sub-structures. Due to
the strong geometrical constraints, the shape of the support
structure is formed by the surfaces of two semi-truncated cones interconnected
by two reference rings. The reference rings contain the positioning
holes for the super-modules and allow the precise positioning
of the detectors.

%\clearpage

\begin{figure}[htb]
\centering
\includegraphics[width=7.5cm]{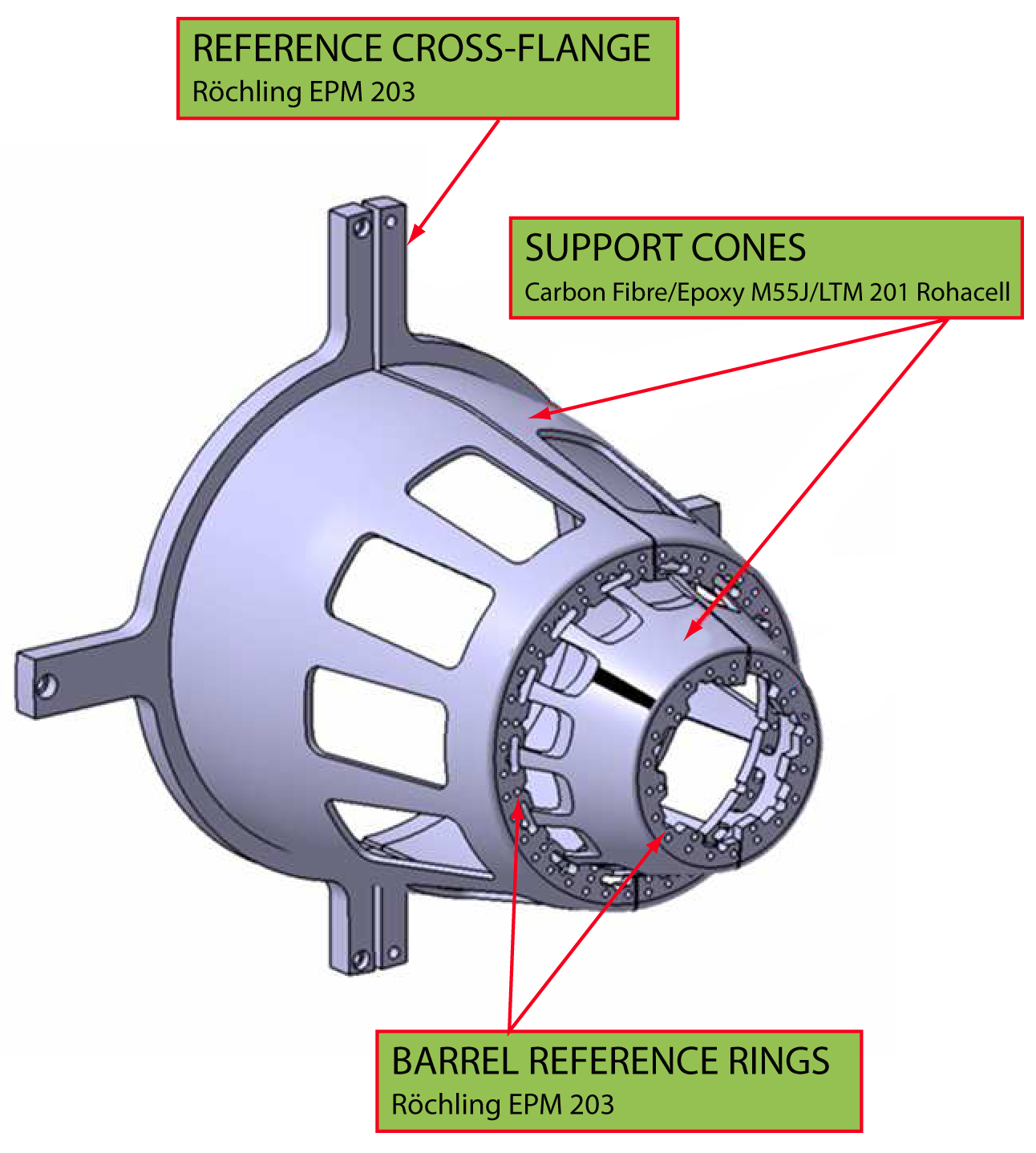}
\caption[Support structure of the pixel barrel]{Support structure of the pixel barrel.}
\label{fig:mechFig_06}
\end{figure}

\begin{figure*}[htb]
\centering
\includegraphics[width=0.65\textwidth]{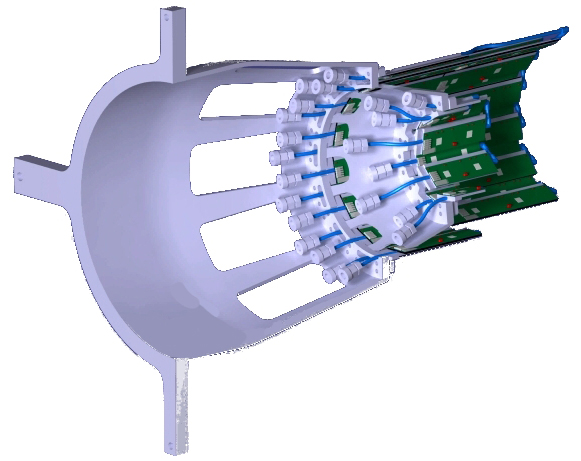}
\caption[Half barrel assembled on the cone support]{Half barrel assembled on the cone support.}
\label{fig:halfbarrel}
\end{figure*}

%\clearpage
The support structure, shown in \figref{fig:mechFig_06} 
is suspended from the frame by a reference
cross flange. In \figref{fig:halfbarrel} the half barrel of the
second layer of pixel is shown assembled on the cone support.
In each barrel the staves are arranged at two different radius and 
the overlap of the active areas is $\sim$~0.3~mm.

\subsubsection*{Pixel Disks}

The disks are six planar structures arranged at different distances 
from the interaction point. 
Each disk is divided vertically in two parts, called half disks.
In this way it is possible to assemble the whole set around the pipe. Each
disk is composed by a planar support structure, by detectors
modules and by cooling tubes. The carbon foam planar structures also act 
as cooling bridges. The cooling tubes are embedded into
the disks while the detectors modules are glued on both sides of
each disk in order to get the largest possible coverage.

The carbon foams, both the POCO HTC and the POCO FOAM, are supplied in
rectangular plates. 
Typical dimensions are $\mathrm{300~mm \times 300~mm \times 13~mm}$. With
the goal of cutting the shapes for both, the barrel elements and the disks
elements, in the most efficient
 way, with minor possible machine rejection in mind, different production
 methods have been investigated. 
The most efficient way is to cut the shapes from the carbon foam
plates with wire EDM.

\begin{figure}[h]
\centering
\includegraphics[width=0.43\textwidth]{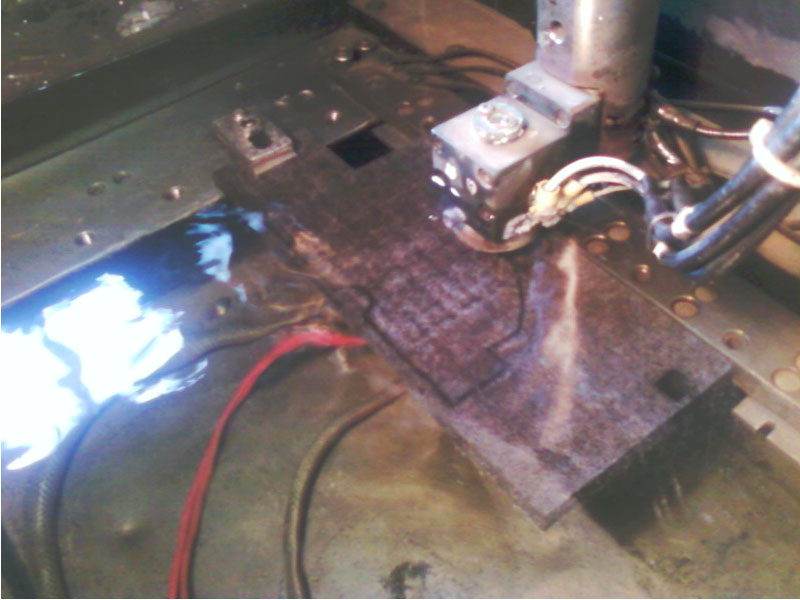}
\caption[The cut of the shape of the disk from the plate with wire EDM]{The cut of the shape of the disk from the plate with wire EDM.}
\label{fig:disk_cut}
\end{figure}

\begin{figure}[htb]
\centering
\includegraphics[width=0.43\textwidth]{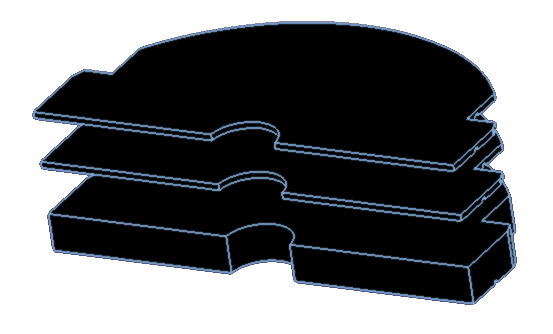}
\caption[Machining of the disks elements]{Machining of the disks elements: the half disk shapes are cut with wire EDM and then sliced-up.}
\label{fig:machining}
\end{figure}

As general
 philosophy the shapes are cut from the plate (\figref{fig:disk_cut}) and
 subsequently 
sliced-up (\figref{fig:machining}).

The wire EDM operates with the material to be cut immersed into
water. The 
foam acts as a ``sponge''. After the wire cut a big quantity of water
was embedded
into the foam element. 
In order to remove the water absorbed, the element was ´baked¡ in an
oven for 48 
hours at 60~$^{\circ}$C. Before the baking the foam element weight was 15.12~g,
while after 
48 hours of baking the weight became 13.12~g, as expected for the dry element.

\begin{figure}[htb]
\centering
\includegraphics[width=0.45\textwidth]{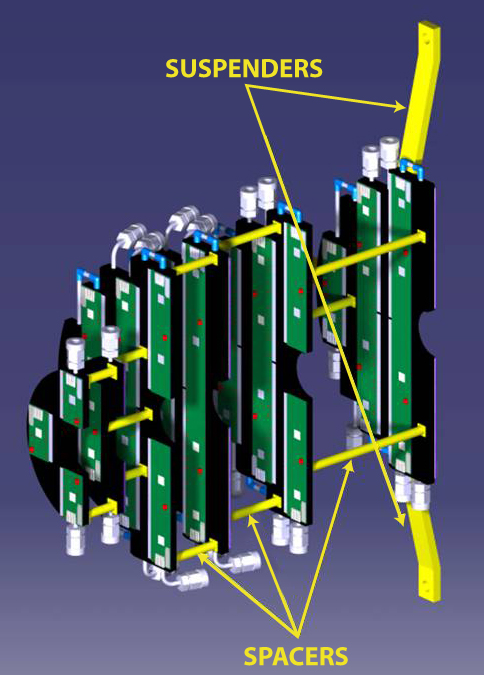}
\caption[Sketch of the set of half disks composing half of the pixel 
forward detector]{Sketch of the set of half disks composing half of the pixel 
forward detector.}
\label{fig:mechFig_08b}
\end{figure}

%\clearpage
The set of six half disks is assembled to obtain a half forward
pixel detector (see \figref{fig:mechFig_08b}). The relative distances 
between disks are ensured by
spacers, glued on the disk surfaces. The final position of each
half disk is surveyed and referred to a set of external benchmarks
fixed on a suspender system, which fastens the half forward detectors
to the frame as well.

\subsection{Support Structures of the Strip Part} 

\subsubsection*{Strip Barrels}
\authalert{Author: D. Grunwald, T. Stockmanns}

\begin{figure}[t]
\centering
\includegraphics[width=0.9\textwidth]{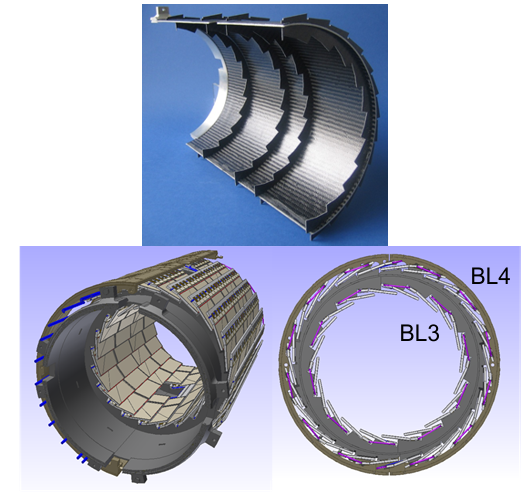}
\caption[Sketch of the common support barrel for BL3 and BL4]{Sketch of the common support barrel for BL3 and BL4 with the sawtooth connections for the super modules. 
The picture at the top shows a first prototype of the support barrel with the sawtooth connectors.}
\label{fig:strip_supportBarrel}
\end{figure}

\begin{figure}[t]
\centering
\includegraphics[width=0.8\textwidth]{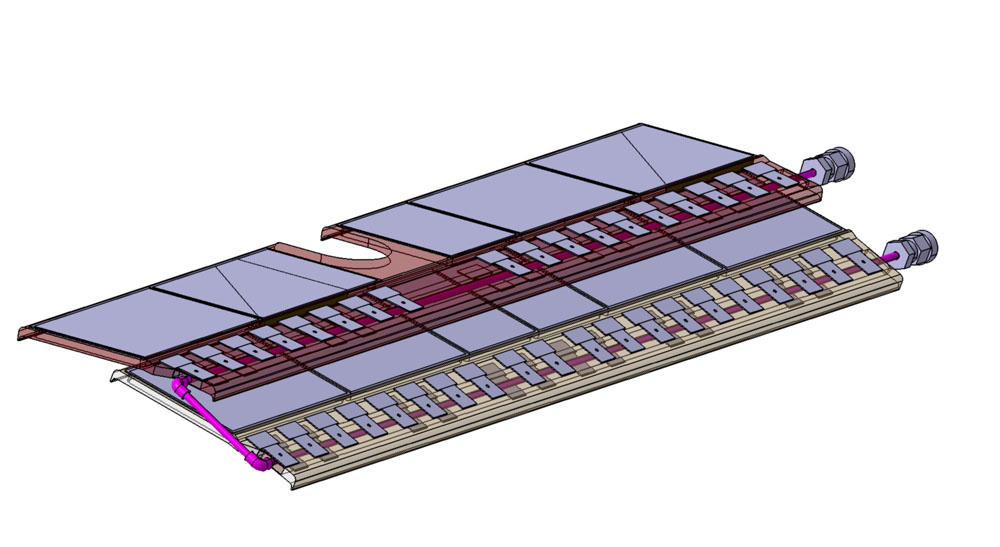}
\caption[Sketch of the set of two strip super modules with a common cooling pipe]{Sketch of the set of two strip super modules with a common cooling pipe.}
\label{fig:BL4_coolingLoop}
\end{figure}

The main holding structure of the strip barrel part consists of two half-cylinders with a radius of 11.4~cm and an overall length of 49.3~cm. They are made out of a sandwich structure of two layers of carbon fibers (M55J) with a thickness of 0.2~mm each and a 2~mm core of Rohacell foam. These half-cylinders are attached to the global support structure of the MVD via 4 connectors at the upstream end of the detector (\figref{fig:strip_supportBarrel}).  
Sawtooth shaped structures on the inside and the outside form a support surface for the sensor elements sitting on so called super-modules. A super module consists of four 6~cm long sensor modules and one or two short modules with a length of 3.5~cm. In addition a super module houses the readout electronics, interconnections,  and the cooling system. The support structure is made out of the same sandwich structure used already for the support cylinders. Underneath the readout electronics the Rohacell core is replaced by a carbon foam (POCO HTC) with high thermal conductivity and an embedded cooling pipe of 2~mm diameter. 
The edges of the structure are bent to increase the stiffness of the structure (\figref{fig:BL4_coolingLoop}).  
The fixation of a super module on the sawtooth rings is done via special bearing points which allow a precise positioning and a compensation for different thermal expansion coefficients.
The overall concept is based on the ALICE ladder repositioning system with a precision better than 
\unit[6]{$\tcmu$m}~\cite{re:mechRef_3}.

\subsubsection*{Strip Disks}

\begin{figure}[!ht]
 \centering
 \includegraphics[width=0.4\textwidth]{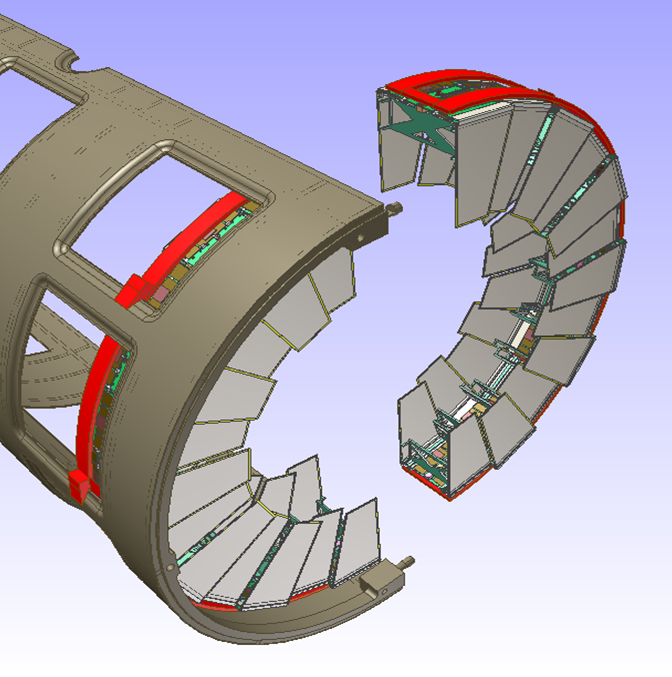}
 \caption[Sketch of the support structure for the strip disks]{Sketch of the support structure for the strip disks.}
 \label{fig:strip_supportDisks}
 \end{figure}

The two forward disks are made of two double layer half-disks, which are separated along the vertical axis. 
They have an outer radius of 137~mm, an inner radius of 75~mm and a spacing of 70~mm between them. The basic 
building block of the double-disk is a module which consists of two wedge shaped sensors which share a common 
PCB for the readout electronics and the cooling pipes. They are connected to a carbon foam sandwich structure 
which holds the sensors, the PCB, the two cooling pipes and does the connection to a support ring which sits at 
the outer rim of the disks. This support ring is then connected to the global support barrel of the MVD 
(\figref{fig:strip_supportDisks}).

\onecolumn

\twocolumn

%

%% file: integration/cooling.tex
\section{The Cooling System}
\label{Cooling-system}

\authalert{Author: S. Coli, coli@to.infn.it'}

The cooling system project provides to operate close to room
temperature, with the temperature of the cooling fluid above the dew
point. The difficulty in retrieval of the most appropriate coolant  for
the required operating temperatures and at moderate pressures, led to
discard evaporative solutions, in favor of a depression  system, 
using water as cooling fluid at 16~$^\circ$C. 
A depression system should avoid leaks
and reduce vibrations and stress on plastic fittings and manifolds,
because of moderate pressure.

For geometry constraints and installation procedure, the MVD structure is
divided in two halves, which will be connected together after target
pipe installation. In addition, due to the MVD collocation relatively to the
beam and target pipes, all the services and cooling pipes 
are routed from the upstream side.
 These constraints have conditioned the cooling
design, which, always with the view to saving material, has made use of
U-shape tubes, to avoid aimless material for the return lines and to
reduce the number of cooling tubes.
For an efficient cooling system, it is advisable to work with a turbulent 
regime flow, which means, considering an inner diameter of 1.84 mm for the 
cooling pipes and water as cooling fluid, with flow rates higher than 0,25 l/min. 
In a depression mode system, the pressure drops in the pipes are
limited, confining cooling mass flows in reduced adjusting
ranges. In fact the part provided with the pressure starts just near the
MVD, in the inlet pipe, till the end of the cooling line (at the cooling
plant, about $20-30$~m far from MVD). Pressure sensors on inlet (and
return) lines ensure the depression  mode in the MVD volume.

\subsection{The Pixel Cooling System}
\label{Cooling-system-Pixel}

\begin{figure*}[ht]
\centering
\includegraphics[width=0.63\textwidth]{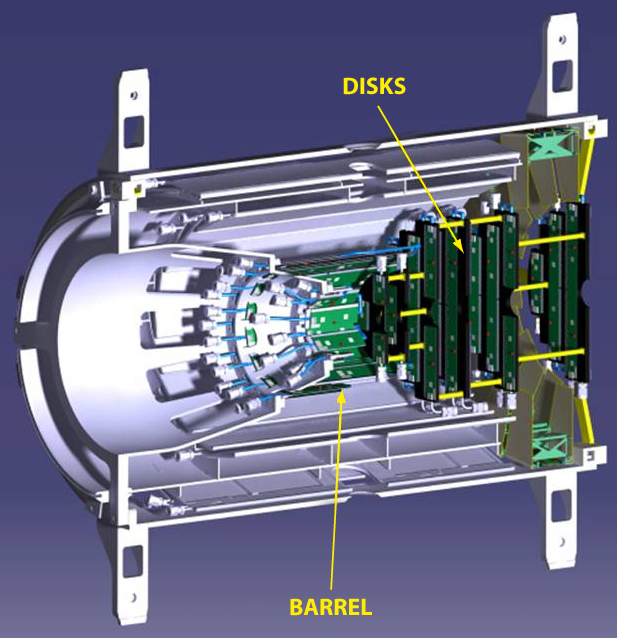}
\caption[The Pixel volume, with 2 barrels and 6 disks]{The Pixel volume, with 2 barrels and 6 disks.}
\label{fig:coolFig_01}
\end{figure*}

The pixel volume of the Micro Vertex Detector (see \figref{fig:coolFig_01} 
consists of a cylinder 
of 300~mm in diameter and 460~mm in length, in which 810 readout
chips  dissipate 1~W/cm$^2$ for a total power 
of about 1.4~KW.

The main purpose of the cooling system is to guarantee the maximum 
temperature of the assembly under 35~$^\circ$C. A dry air flow (nitrogen could be better) 
has to be foreseen in the MVD volume.
For the cooling system design a material with low density, high thermal 
conductivity, low thermal expansion coefficient, machinable, gluable, 
stable at different temperatures and radiation resistant has been 
searched. The material which complied to all these requirements was the
carbon foam: his open pore structure graphite
produces a material with good thermal properties and low 
density. 
In \tabref{tab:coolTab_1} and \ref{tab:coolTab_2}, density and thermal 
conductivity of two carbon foams, POCO HTC and POCO FOAM, are presented.

\begin{table}[width=0.4\textwidth]
\centering
\begin{tabular}{|l|c|}
  \hline
  {\bf POCO HTC} &  \\ \hline
  {\bf Density} & 0.9~g/cm$^3$ \\ \hline
  {\bf Thermal conductivity} &  \\
  Out of plane & 245~W/($\mathrm{m\cdot K}$) \\
  In plane & 70~W/($\mathrm{m\cdot K}$) \\ \hline
\end{tabular}
\caption[POCO HTC properties]{POCO HTC properties.}
\label{tab:coolTab_1}
\end{table}

\begin{table}[width=0.4\textwidth]
\centering
\begin{tabular}{|l|c|}
  \hline
  {\bf POCO FOAM} &  \\ \hline
  {\bf Density} & 0.55~g/cm$^3$ \\ \hline
  {\bf Thermal conductivity} &  \\
  Out of plane & 135~W/($\mathrm{m\cdot K}$) \\
  In plane & 45~W/($\mathrm{m\cdot K}$) \\ \hline
  \hline
\end{tabular}
\caption[POCO FOAM properties]{POCO FOAM properties.}
\label{tab:coolTab_2}
\end{table}

\begin{figure}[!ht]
\centering
\includegraphics[width=0.5\textwidth]{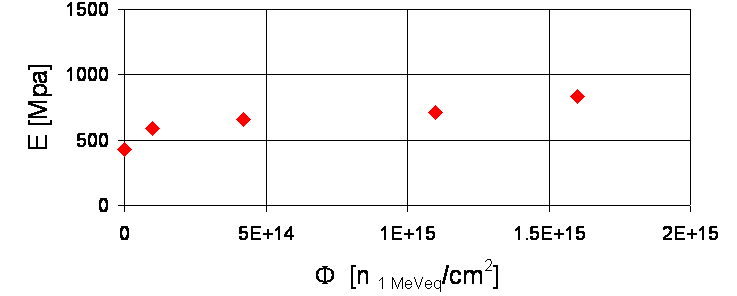}
\caption[Test results: Young's modulus {[}MPa{]}
vs radiation levels for POCO FOAM]{Test results: Young's modulus [MPa]
vs radiation levels for POCO FOAM.}
\label{fig:coolFig_02}
\end{figure}

\begin{figure}[!ht]
\centering
\includegraphics[width=0.5\textwidth]{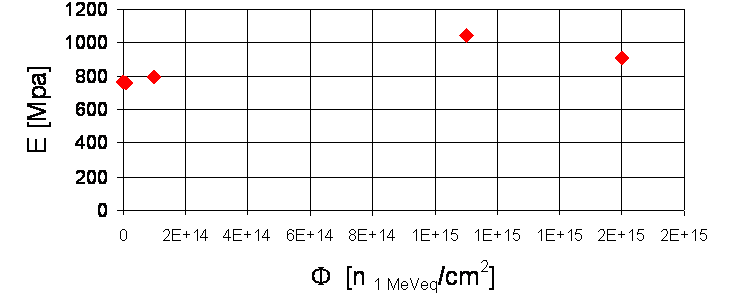}
\caption[Test results: Young's modulus {[}Mpa{]}
vs radiation levels for POCO HTC]{Test results: Young's modulus [Mpa]
vs radiation levels for POCO HTC.}
\label{fig:coolFig_03}
\end{figure}

Tests have been made on POCO HTC and on POCO FOAM to verify carbon 
foam thermal conductivity and mechanical stability at the radiation 
levels expected in the experiment. Using the TRIGA MARK II reactor in Pavia,  
five samples per foam type have been immersed in the reactor central channel  
at different reactor power, for a time of $1000-1500$~s, 
while one sample for foam type, not irradiated, was used as reference.
The tests for the Young's modulus estimation consisted on a carbon foam 
specimens ($\mathrm{15 \times 50~mm^2}$ and 5~mm of thickness), irradiated at 
different radiation levels, 
%{\bf (from 0 to 250 KW, 15 minutes for each radiation 
%load)}, 
and then tensioned with a set of loads. For the evaluation of the 
deformation, strain gauges have been glued on the specimens and read by a  
HBM-MGCplus acquisition system.
Test results are shown in \figref{fig:coolFig_02} and \ref{fig:coolFig_03}.
The carbon foam stiffness increases with the neutron fluence, the POCO FOAM Young's modulus doubles 
in the considered irradiation range, 
while a 23\% variation of the nominal value is observed for the POCO HTC.

\begin{figure}[!ht]
\centering
\includegraphics[width=0.5\textwidth]{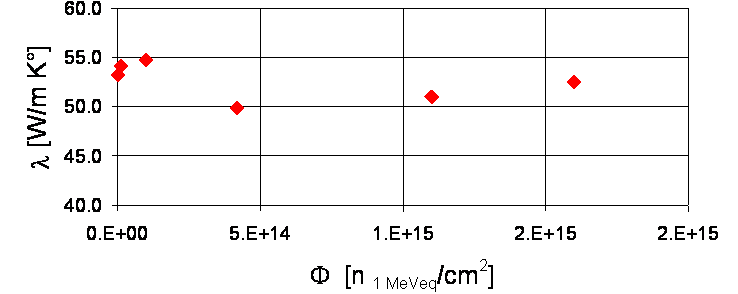}
\caption[Test results: thermal conductivity {[}W/($\mathrm{m\cdot K}$){]} vs radiation levels  
for POCO FOAM]{Test results: thermal conductivity [W/($\mathrm{m\cdot K}$)] vs radiation levels  
for POCO FOAM.}
\label{fig:coolFig_04}
\end{figure}

\begin{figure}[!ht]
\centering
\includegraphics[width=0.5\textwidth]{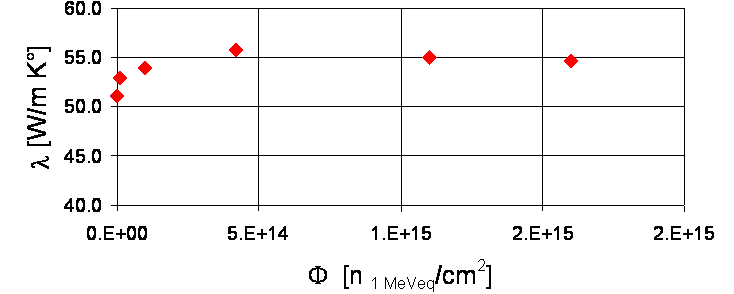}
\caption[Test results: thermal conductivity {[}W/(m$\cdot$K){]} vs radiation levels  
for POCO HTC]{Test results: thermal conductivity [W/($\mathrm{m\cdot K}$)] vs radiation levels  
for POCO HTC.}
\label{fig:coolFig_05}
\end{figure}

The tests for the thermal conductivity estimation consisted on the same 
carbon foam specimens  heated on one side with a resistance and cooled 
on the other side. For the evaluation of the thermal conductivity, two 
thermocouples, arranged at a well known distance, have been glued on the 
specimens and read by the HBM-MGCplus acquisition system.
The results of the thermal tests are reported in \figref{fig:coolFig_04} and \ref{fig:coolFig_05}.
the thermal conductivity variations are within 10\% for both carbon foams. 

%\newpage

The nominal values pre-irradiation have been used in the FEM simulations.
Test results show variation of the thermal conductivity values within 10\%.\\
Because of different geometries, different chips concentrations, different 
mechanical structure solutions, the cooling system design can be distinguished 
in two sub-groups, the disks and the barrels.

\subsubsection*{Disk Cooling System}
\begin{figure}[ht]
\centering
\includegraphics[width=0.4\textwidth]{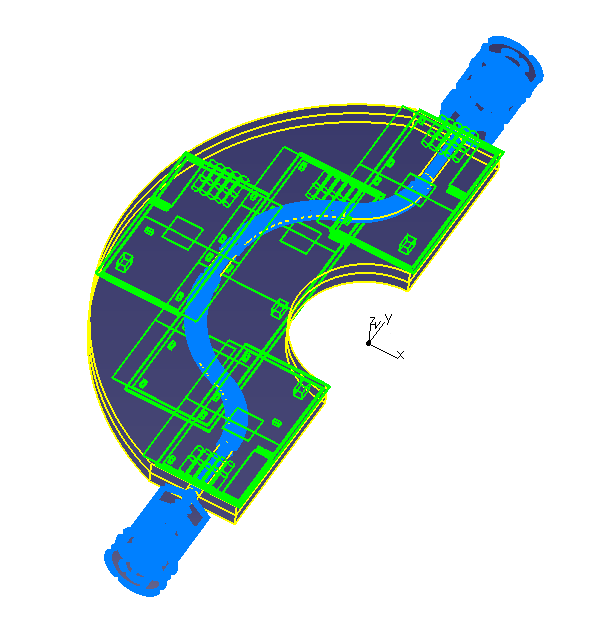}
\caption[Half small disk in carbon foam completed of chips, detectors, 
cooling tube and fitting]{Half small disk ($\mathrm{r}=36.56$~mm) in carbon foam completed of chips, detectors, 
cooling tube and fitting.}
\label{fig:coolFig_06}
\end{figure}

\begin{figure}[ht]
\centering
\includegraphics[width=0.45\textwidth]{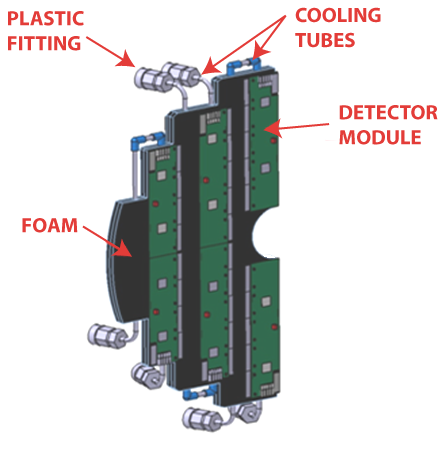}
\caption[Half big disk in carbon foam completed of chips, detectors, cooling tube and plastic fittings]{Half big disk ($\mathrm{r}=73.96$~mm) in carbon foam completed of chips, detectors, cooling tube and plastic fittings.}
\label{fig:coolFig_07}
\end{figure}

The cooling system for the disks consists of tubes of 2~mm external diameter, 
inserted in a carbon foam disk with a thickness of 4 mm, 
on which the detectors are glued on both sides. The tubes, with a 
wall thickness of 80~$\tcmu$m, are in MP35N, a Nickel-Cobalt 
chromium-molybdenum alloy (by Minitubes) with excellent corrosion 
resistance and good ductility and toughness. 
In order to save material, to reduce pressure drops and to allow irregular 
paths to reach the manifolds of the first patch panel, the metal pipes are 
connected to flexible polyurethane tubes, \cite{re:poli_tubes}, through small plastic 
fittings (in ryton R4, custom made). The plastic fittings are glued on 
the metal cooling tubes with an epoxy glue (LOCTITE 3425), and all the 
connections are individually tested to ensure  the pressure tightness.
The disks group consists of two disks (4 halves) with a diameter of 75~mm, 
dissipating about 70~W (35~W each disk) and of four disks (8 halves), 
with a diameter of 150~mm, dissipating about 740~W (about 185~W each disk).
On the small disks, due to space constraints, only one tube for each half 
disk is foreseen, while, for the bigger disks, three U-shaped tubes for 
each half, has been arranged. The U-tube is obtained with plastic elbows 
(in ryton R4, custom made) glued on the metal cooling pipes.
In \figref{fig:coolFig_06} and \ref{fig:coolFig_07} the two kinds of 
half disks in carbon foam, 
completed of detector modules, cooling tubes and plastic fittings are shown.

Finite Element Method (FEM) analyses, through temperature maps, allow 
fast evaluations of heat conduction, identifying the most effective 
solutions for the future prototype constructions. After a first prototype, 
useful to estimate the errors due to contact surface and to manual gluing, 
many FEM simulations have been performed to find out the thermally most efficient 
configuration of carbon foam and cooling pipes, considering different 
carbon foam thicknesses, tube diameters, tube numbers and cooling flow rates.
 
From FEM analyses it appears that the most effective operation to be applied is the doubling 
of cooling pipes. 

\begin{figure}[ht]
\centering
\includegraphics[width=0.45\textwidth]{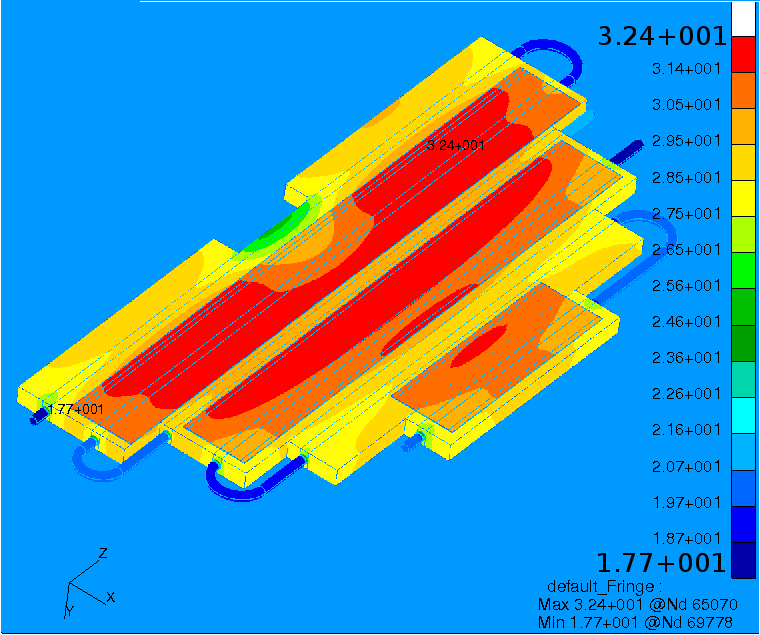}
\caption[FEM analysis results with real 
configuration]{FEM analysis results with the following real 
configuration: total power 94~W 
removed by 6 cooling pipes (2~mm external diameter),
cooling flow 0.3~l/min at 18.5~$^\circ$C, carbon foam thickness 4~mm.}
\label{fig:coolFig_08}
\end{figure}

\begin{figure}[ht]
\centering
\includegraphics[width=0.45\textwidth]{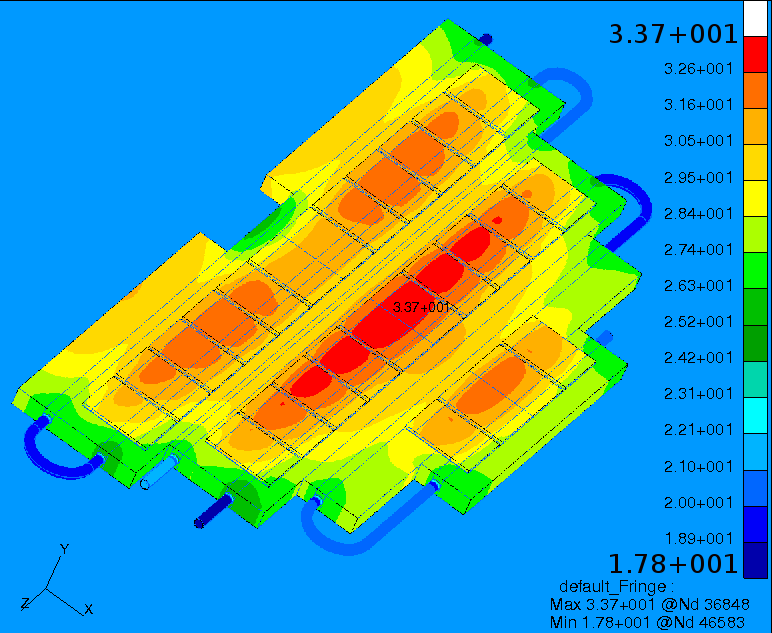}
\caption[FEM analysis results with test 
configuration]{FEM analysis results with the following test 
configuration: total power 94~W 
removed by 6 cooling pipes (2~mm external diameter),
cooling flow 0.3~l/min at 18.5~$^\circ$C, carbon foam thickness 4~mm.}
\label{fig:coolFig_09}
\end{figure}

The test on a second prototype validates FEM results.
In \figsref{fig:coolFig_08} and \ref{fig:coolFig_09} FEM analyses compared to test results are shown.
 
In \figref{fig:coolFig_08} a FEM model with the real 
configuration 
of half disks, in carbon foam with a thickness of 4~mm, six cooling pipes, 
with pixel chips on both sides and a total power of 94~W, was used. 

In \figref{fig:coolFig_09}, a FEM model with the dummy chips 
layout, simulating the pixel chips, dissipating 94~W, on the carbon foam 
disk with a thickness of 4~mm and six cooling pipes, was used.

\begin{figure}[ht]
\centering
\includegraphics[width=0.4\textwidth]{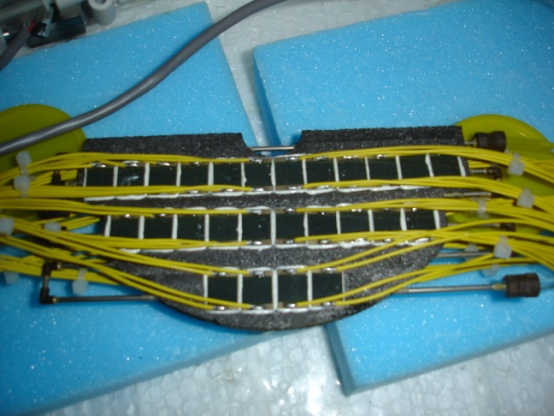}
\caption[Carbon foam disk with 6 inserted cooling pipes and 54 resistors 
(dummy chips)]{Carbon foam disk with 6 inserted cooling pipes and 54 resistors 
(dummy chips).}
\label{fig:coolFig_10}
\end{figure}

For both these FEM analyses, cooling water at 18.5~$^\circ$C and 
flow rate of 0.3~l/min in each tube have been adopted. \Figref{fig:coolFig_10} shows 
the disk prototype, in carbon foam, with six cooling tubes inside and 
54 dummy chips (resistors). 

\begin{figure}[ht]
\centering
\includegraphics[width=0.45\textwidth]{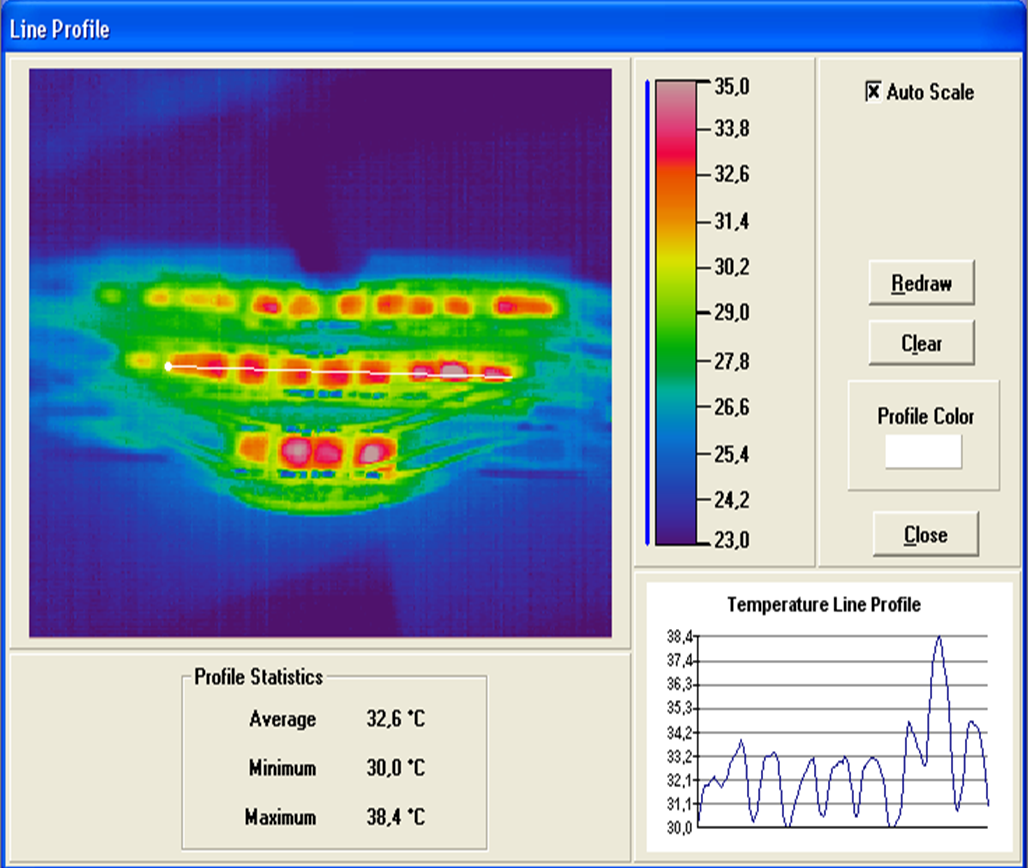}
\caption[Test results: Thermal map of disk prototype, dissipating 94~W]{Test results: Thermal map of disk prototype, dissipating 94~W.}
\label{fig:coolFig_11}
\end{figure}

\Figref{fig:coolFig_11} shows the test results, in line with FEM 
predictions. In this figure some hot spots are visible, due to manual 
gluing imperfections.

To guarantee a depression system, considering the pressure drops 
for the return lines, the pressure drops in the MVD line between the 
patch panel, in which pressure sensors can take place, must be limited 
and below $450-500$~mbar. 

In the disks, to reduce the number of cooling circuits, it was 
thought to use S-shaped 
tubes, but the high pressure drop did not allow this solution. 

\begin{figure}[ht]
\centering
\includegraphics[width=0.4\textwidth]{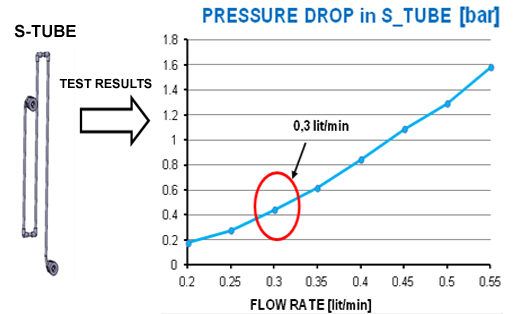}
\caption[Pressure drop {[}bar{]} in a S-shape tube vs mass flow rate {[}l/min{]}]{Pressure drop [bar] in a S-shape tube vs mass flow rate [l/min].}
\label{fig:coolFig_12}
\end{figure}

\begin{figure}[ht]
\centering
\includegraphics[width=0.4\textwidth]{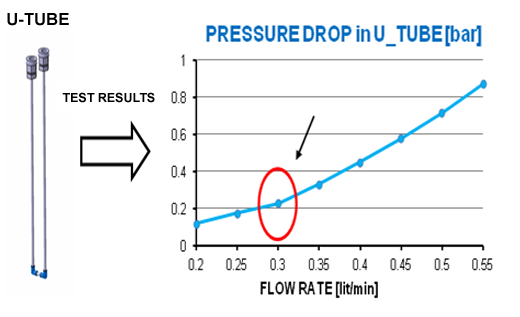}
\caption[Pressure drop {[}bar{]} in a U-shape tube vs mass flow rate {[}l/min{]}]{Pressure drop [bar] in a U-shape tube vs mass flow rate [l/min].}
\label{fig:coolFig_13}
\end{figure}

In \figref{fig:coolFig_12} and \ref{fig:coolFig_13} the pressure drops 
in S-tubes and in U-tubes are shown.

These measured pressure drops are referred to  
the only metal pipe, while the pressure sensors will be positioned 
in the patch panel manifolds, at least 1~m far from MVD and after 
plastic tubes connections. In the first case (S-shaped tube) the pressure 
drop is already close to the limit but all the path through plastic tubes 
and the manifolds is neglected. Therefore, the solution with S-shaped tubes 
cannot be adopted. On the other hand the U-shaped solution, thanks  to its 
limited pressure drops, allows a more extended range of flow rate settings.

\subsubsection*{Barrel Cooling System}
The cooling system is designed to drain the total power, about 75~W on 
the first layer and 500~W on the second, by 21 MP35N U-tubes (2mm of outer 
diameter and 80~$\tcmu$m of thickness). The two halves of the layer 1, 
due to cooling constraints in use U-shaped tubes, cannot be symmetric: 
one half includes 8 staves and the other only 6.

Furthermore the barrel system, besides the drain of the heat dissipation, for the 
material budget limits, has to provide mechanical support. The system 
consists of the cooling tube inserted and glued in a sandwich of carbon 
foam and an Omega made by 3 plies of carbon fiber M55J (properly oriented). 
The high modulus of this carbon foam ensures the rigidity of the structure and, at 
the same time, acts as heat bridge for the cooling. 
The super module structure, which consists of the silicon detector, 
completed of chips, mechanical structure and cooling system, is shown in \figref{fig:mechFig_05}.

\begin{figure}[ht]
\centering
\includegraphics[width=0.45\textwidth]{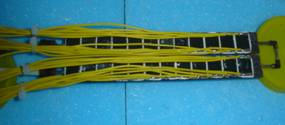}
\caption[Prototype of two staves of the second layer of the barrel, with 
dummy chips (resistors)]{Prototype of two staves of the second layer of the barrel, with 
dummy chips (resistors), with a power of about 21~W for each stave.}
\label{fig:coolFig_14}
\end{figure}

\begin{figure}[ht]
\centering
\includegraphics[width=0.45\textwidth]{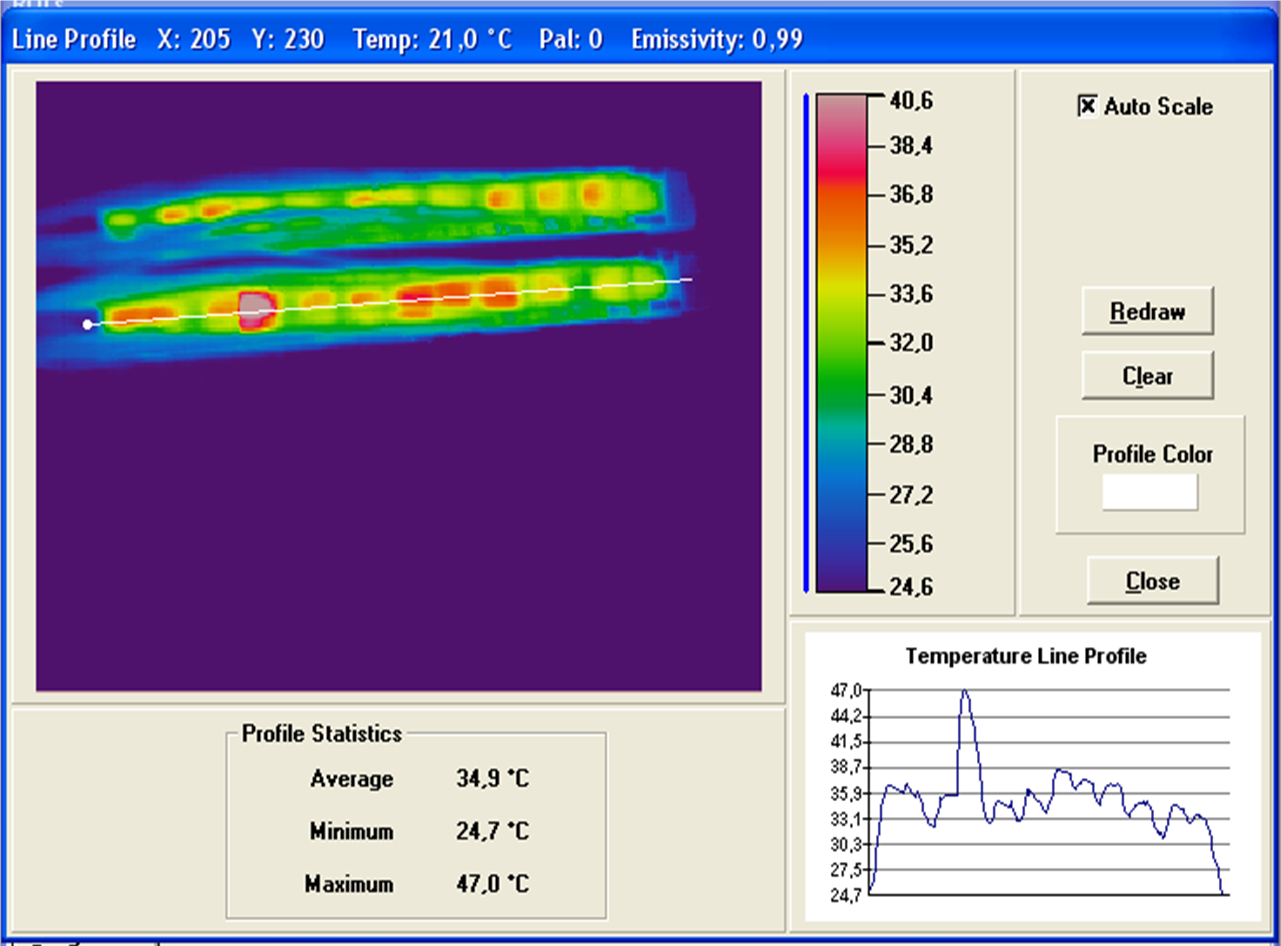}
\caption[Test result: Thermal map of two staves prototype]{Test result: Thermal map of two staves prototype, dissipating 42~W 
in total.}
\label{fig:coolFig_15}
\end{figure}

Even for the barrel part FEM analyses have been useful to find out the most 
efficient solutions and test results are presented below. \Figref{fig:coolFig_14} shows the U-shaped cooling tubes in MP35N, with plastic 
fittings and plastic bends glued to the pipes. The chips are simulated by 
resistors, dissipating about 21 W in each stave. The cooling water for the 
test at 18~$^\circ$C was fed with a flow rate of 0.3~l/min. 
The temperature map, shown in \figref{fig:coolFig_15}, was been obtained 
with a thermo camera (NEC thermo tracer).

\subsection{The Strip Cooling System}

The strip part of the MVD is subdivided into a barrel section and a disk section. 
The barrel section again consists of two barrels with 552 \frontend chips needed 
for the inner one and 824 \frontend chips for the outer one.
Compared to the numbers in \tabref{tab-BasicInfo-CentralMVD}
these higher numbers consider the fall back solution of an individual readout for each sensor 
instead of a ganging of long strips for subsequent barrel sensors. 
The disk section consists of one double disk with 384 \frontend chips. Each of these \frontend chips consumes up 
to 1~W of power which sums up to a total power consumption of the strip part of 1.76~kW.

\begin{figure*}[ht]
\centering
\includegraphics[width=0.65\textwidth]{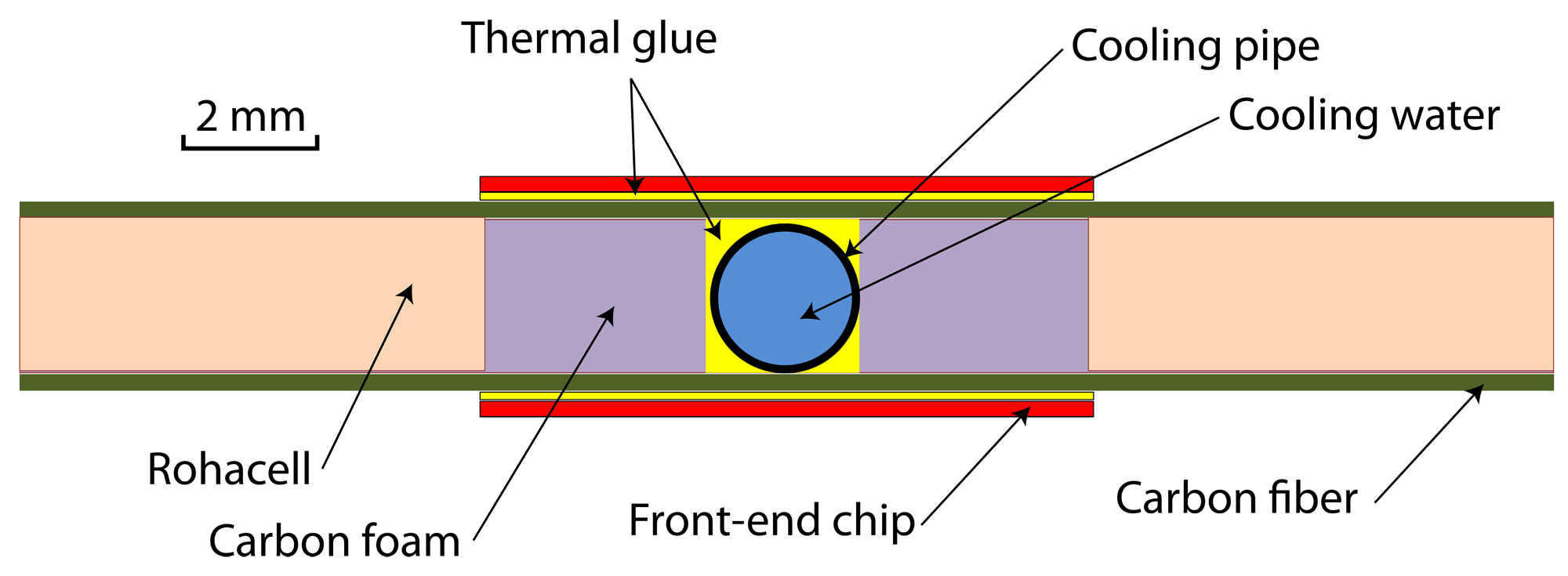}
\caption[Schematic view of the cooling concept of the strip barrel part]{Schematic view of the cooling concept of the strip barrel part. The relevant parts are: \frontend chip (red), thermal glue (yellow), carbon fiber (green), Rohacell (orange), carbon foam (violet), cooling pipe (black), cooling water (blue).}
\label{fig:strip_barrel_cooling}
\end{figure*}

To cool away this power the same concept as in the pixel part is used. A water based system
 will be used which is operated in depression mode. Each support structure is equipped with 
 a cooling tube of 2~mm diameter and a wall thickness of 80~$\tcmu$m made out of MP35N. This 
 cooling tube is glued with thermal conducting glue to a carbon foam material (POCO HTC) 
 with excellent thermal conductivity which sits underneath the \frontend electronics 
 (see \figref{fig:strip_barrel_cooling} for the strip barrel part).

\begin{figure*}[ht]
\centering
\includegraphics[width=0.9\textwidth]{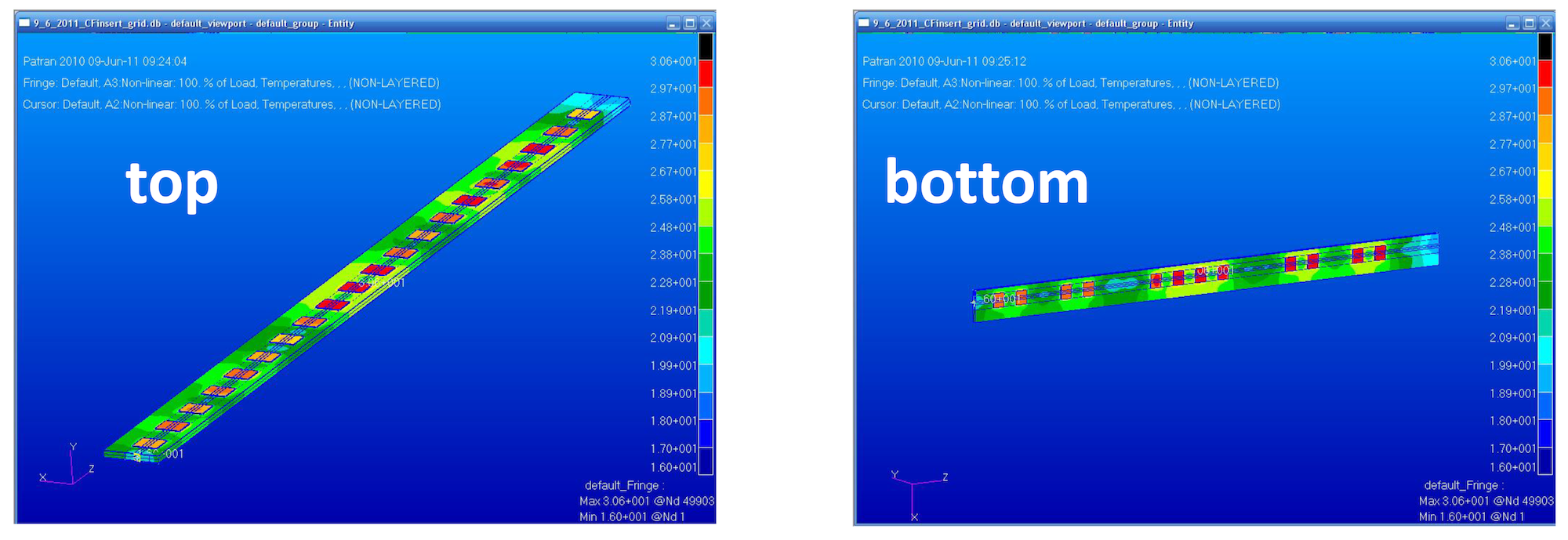}
\caption[FEM analysis result for one barrel section of the strip part]{FEM analysis result for one barrel section of the strip part with 20 strip readout chips on top and 12 chips on the bottom side. The used parameters are: 32~W total power of the chips, the chips are glued to carbon foam with an internal cooling pipe with 2~mm diameter, a water flow of 0.3~l/h and 16~$^{\circ}$C inlet temperature. Left picture shows the top side of the super module, right picture shows the bottom side of the super module.} 
\label{fig:FEM_Strip}
\end{figure*}

Detailed finite element simulations of different setups have been performed for the barrel part of the strip detector. The results of the best setup can be seen in \figref{fig:FEM_Strip}. The maximum temperature the \frontend electronics reaches is 32~$^{\circ}$C in this setup. Almost all heat of the electronics is cooled away.

\begin{figure}[ht]
\centering
\includegraphics[width=0.4\textwidth]{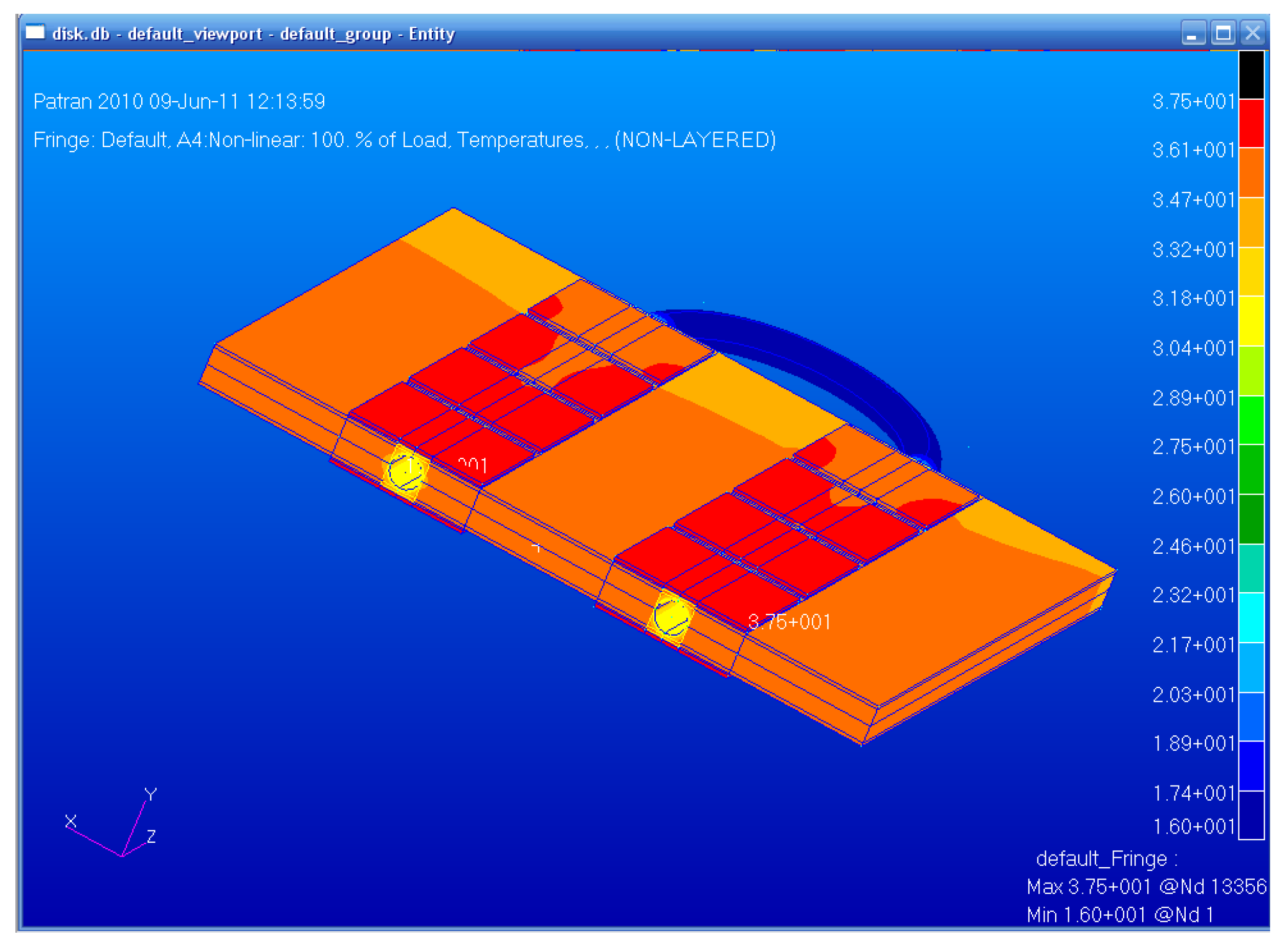}
\caption[FEM analysis result for one disk section of the strip part]{FEM analysis result for one disk section of the strip part of the MVD with 8 strip readout chips on the top and 8 chips on the bottom side. The used parameters are: 32~W total power of the chips, the chips are glued to carbon foam with an internal cooling pipe with 2~mm diameter, a water flow of 0.3~l/h and 16~$^{\circ}$C inlet temperature.} 
\label{fig:FEM_Strips}
\end{figure}

In the strip disk part the situation is more complicated because of a higher density of the readout electronics. Here the achieved temperature of the readout electronics would be 37.5~$^{\circ}$C which is slightly above the required 35~$^{\circ}$C. To solve this problem without changing the cooling concept would require to either reduce the power consumption of the electronics or double the number of cooling pipes which would result in a chip temperature of 28~$^{\circ}$C. While a doubling of the cooling pipes would significantly increase the material budget in this part of the detector the first solution would be more desirable but it is not clear if this goal can be reached.

\subsection{Cooling Plant}

\begin{figure}[ht]
\centering
\includegraphics[width=\textwidth]{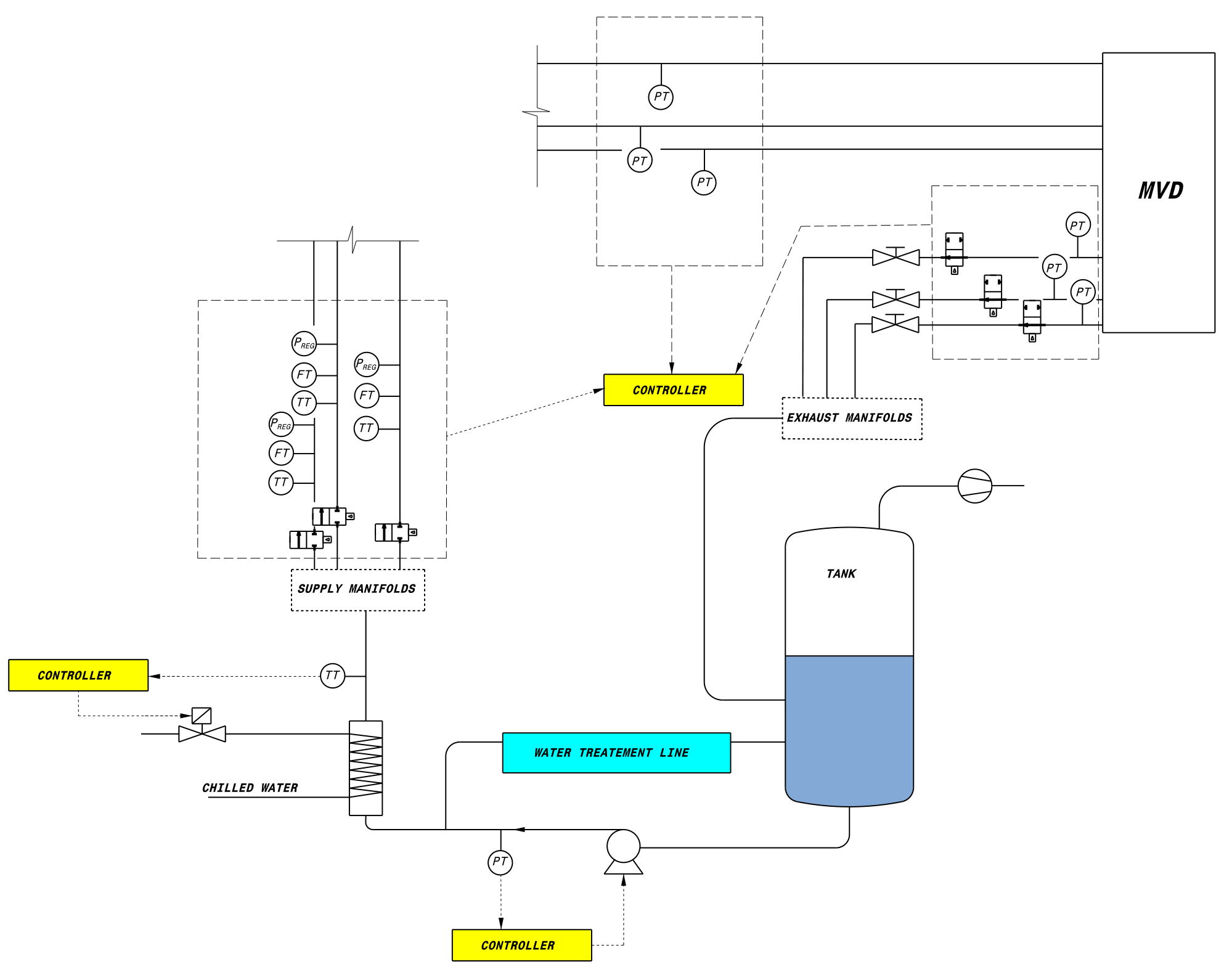}
\caption[Simple hydraulic scheme of the cooling plant]{Simple hydraulic scheme of the cooling plant. TT: Temperature Transmitter, FT: Flow Transmitter, Preg: Pressure Regulator, PT: Pressure Transmitter.}
\label{fig:coolFig_coolcircuit}
\end{figure}

The cooling plant will be placed in a reserved and dedicated area at 
about $20-30$~m far from the MVD final position. In this area the inlet 
pump to feed all the cooling circuits, the tank for the water kept in 
depression  mode by vacuum pump, the heat exchanger connected to a chiller, 
the water cleaning apparatus, the control unit and the manifolds for the 
supply and the return lines of the cooling circuits, each equipped with 
valves, regulators and sensors, will be located. A simple scheme of the 
hydraulic circuit is shown in \figref{fig:coolFig_coolcircuit}.

\onecolumn

\begin{figure}[ht]
\centering
\includegraphics[width=0.49\textwidth]{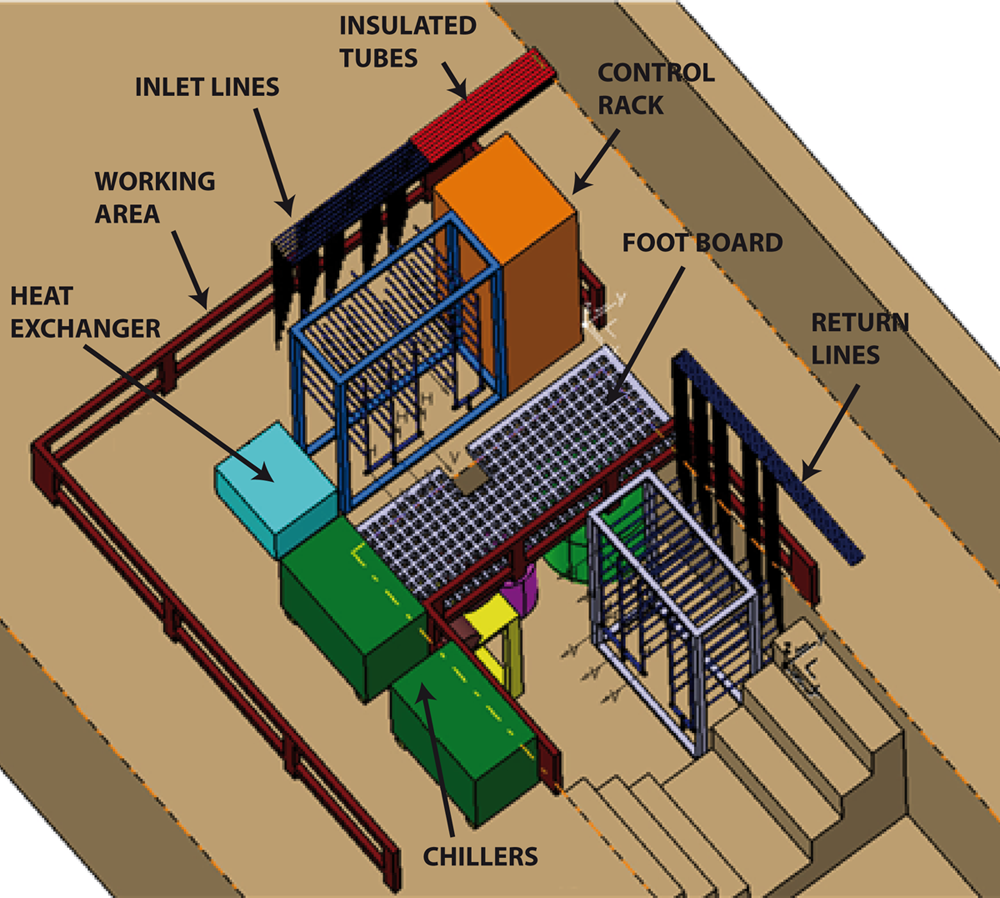}\hspace{0.2cm}
\includegraphics[width=0.49\textwidth]{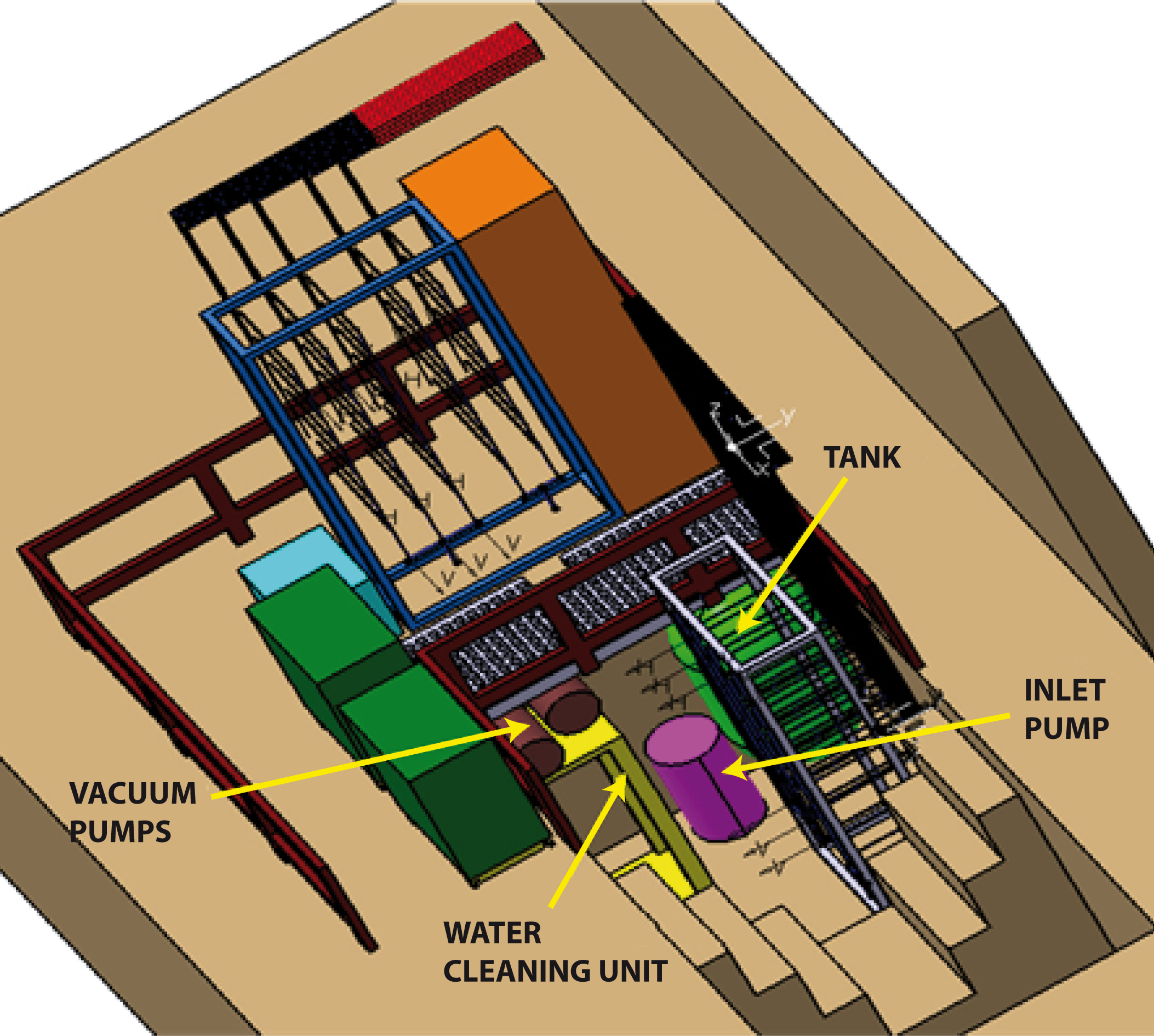}
\caption[Preliminary draft of the cooling plant]{Preliminary draft of the cooling plant.}
\label{fig:coolFig_17}
\end{figure}

\twocolumn

For fault-tolerance reasons, the pixel cooling system is segmented in different 
cooling loops, 9 for the barrels, and 12 for the disks, respectively. 
In the barrel, 3 cooling circuits are foreseen for layer 1 and 6 for 
layer 2. In the disks, 4 cooling circuits are for the two small disks and 8 
cooling circuits for the bigger ones. 
The total number of cooling circuits for the pixel part is 21.
For the strip part there are 32 cooling circuits, 24 for the barrel part and 8 for the strip part.
%For the strip part a detailed layout of the number of cooling pipes is still missing but it will be in the order of 30 lines.
The most important requirement for the hydraulic system is that it works 
at sub-atmospheric pressure, all along the MVD, ensuring a leak-less system.  
To achieve the sub-atmospheric pressure inside the detectors, the pressure 
drops along the detectors and the return lines are limited. 

Pressure sensors on the supply lines, at the entrance of the MVD, ensure 
to operate in depression  mode, while flow meters and pressure regulators 
on each cooling circuit, positioned at the cooling plant, guarantee the 
pressure stability and the nominal operating conditions. 

All the metallic tubes  and the wet parts of all the accessories (sensors, 
valves, flow meters, filters and fitting) will be in stainless steel 
AISI~316. The flexible tubes will be in polyurethane. 

The water purity required for corrosion concerns of thin cooling pipes, 
will be obtained with a cleaning treatment chain which consists of 
deioniser, micro porous membrane to remove the dissolved O$_2$ and sterilisation 
UV lamp \cite{re:coolRef_1}.

In \figref{fig:coolFig_17} preliminary drafts
of the cooling plant are shown.

%\begin{figure}[ht]
%\centering
%\includegraphics[width=0.45\textwidth]{integration/pictures/cool_11b.png}
%\caption{Preliminary draft of the cooling plant.}
%\label{fig:coolFig_18}
%\end{figure}

%% file: simulations/simulations.tex
\chapter{Monte-Carlo Simulations and Performance}
\label{simulations}
% --- Introduction

%\mypicture{simulations/pictures/Pandaroot-Logo1.png}{0.4}{\pndrt Logo}{\pndrt Logo}{fig:pndrtlogo}
A full simulation of the \panda detector components and their answer to the passage of particles has been developed within the \pndrt framework \cite{pandaroot:chep08}.
The simulations for the \Mvd are implemented within that framework. The same reconstruction and analysis algorithms and infrastructure will work on the simulation output as well as on real data.

This chapter will give an overview of the framework, the MVD detector and reconstruction code, as well as the results of performance studies.

% --- Input of sections

\input{simulations/sim-framework}

\input{simulations/sim-sds}

\input{simulations/sim-trkvtx}
\input{simulations/sim-pid}

\input{simulations/sim-studies}
\input{simulations/sim-resolutions}

\input{simulations/sim-channels}
\clearpage

% --- This command adds the bibliography defined in an external file  
\putbib[lit_sim]

% --- End of document

%% file: simulations/sim-framework.tex
% ---
\section{Software Framework Layout}
\authalert{Author: Ralf Kliemt (ralf.kliemt(at)hiskp.uni-bonn.de)}

The software framework is the collection of software and tools for the description of the detectors and the simulation of physics reactions. The layout is organized such as to allow the re-use of well known programs and tools, common to particle and nuclear physics simulations. One basic concept of the framework is its modularity, the possibility to switch algorithms and procedures at mostly any point in the chain of processes. 
Hence a set of interfaces and data input/output is provided to connect all these tools properly.

%\begin{figure*}[]
%  \centering
%  \includegraphics[width=0.8\textwidth]{simulations/pictures/Fairsoft-design.png}
%  \caption[Schematic chain of processing.]{Layout scheme of FairSoft \alert{[]}.}
%  \label{fig:sim:fairsoft}
%\end{figure*}

In practice the frameworks organization features these levels:
\begin{itemize}
\item External Packages containing the main bulk of software:
 Geant3 and Geant4 (Particle propagation through matter in Fortran \cite{Geant3} and C++ \cite{GEANT4}),  VMC (Virtual Monte-Carlo \cite{vmcweb2010}, \cite{Hrivnacova:2003yy}), \rt (Plotting, fitting, graphics etc. \cite{Brun1996}), Pythia \cite{Sjostrand:2007gs}
 %, Pluto \cite{Frohlich:2007bi}  
 and auxiliary tools
\item \frt handling the framework, data I/O, interfaces, infrastructure
\item \pndrt for detector simulations, tracking and reconstruction 
\end{itemize}

To this framework, user processes are added as detector class implementations for the transport model and as tasks, which steer the processing at any stage. The availability of online parameters is provided with the runtime database (RTDB) and data are stored in a \rt file, using \rts objects handling with chains, trees and branches. 
Using VMC, together with the geometry description in \rt format, it is possible to switch between different transport engines. This enables an easy way to compare the outputs of different engines with each other and with real data in order to get a reliable description of the detectors behaviour.

A typical chain of processing is sketched in \figref{fig:sim:chain}. It has the following stages:

\begin{figure}[]
  \centering
  \includegraphics[width=\columnwidth]{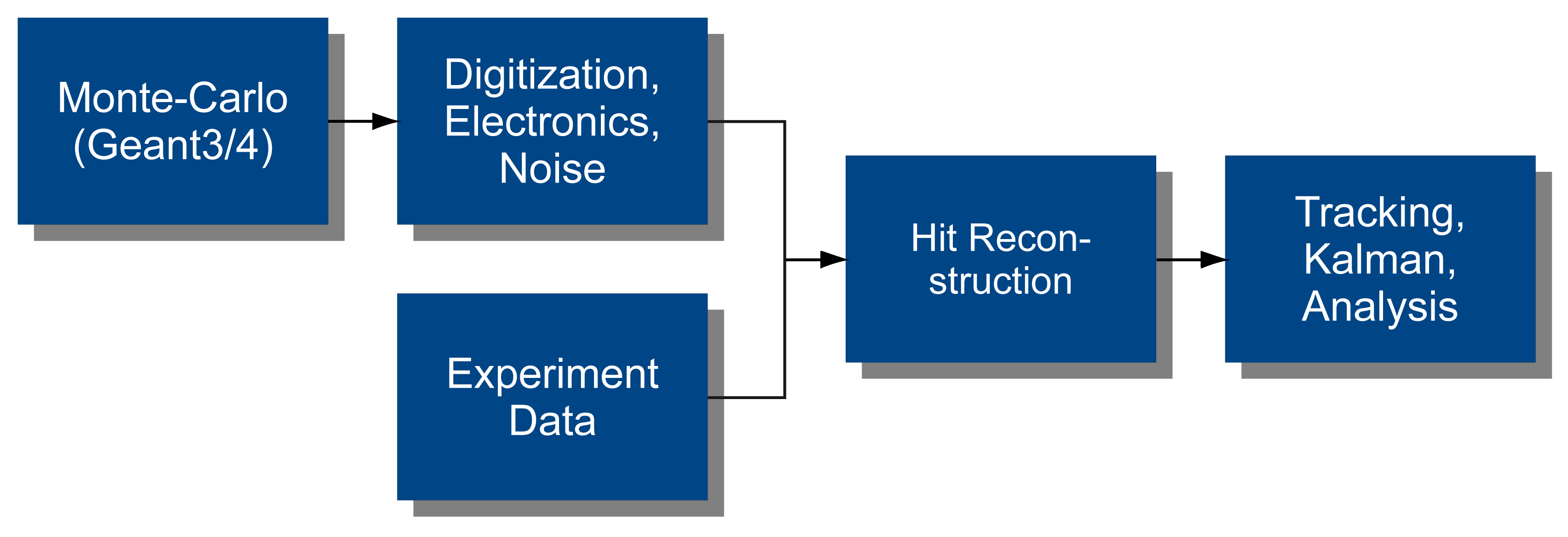}
  \caption[Schematic chain of processing]{Schematic chain of processing.}
  \label{fig:sim:chain}
\end{figure}

{\bf Event Generator:} The event generator produces in each event a set of particles to be processed. Particles are defined by their four-momentum, charge, creation point and time. These properties are randomly distributed by the selected model, such as the technically important ``particle gun'', single-channel generation with EvtGen \cite{Evtgen} or the full description of antiproton-proton or antiproton-nucleus background reactions with DPM\footnote{Dual Parton Model \cite{DPM}} and UrQMD\footnote{UltraRelativistic Quantum molecular Dynamics model for \pbarN \cite{UrQMD},\cite{UrQMD2}} respectively. Also Pythia and Pluto are available.

{\bf Transport Code (VMC):} The particle transport through matter is usually simulated with one version of Geant. The Virtual Monte-Carlo interface (VMC) allows to switch between the different Geant flavors and allows to introduce other packages as well. Each sub-detector has its own geometry description (\rt format) and the and the definition of the active elements in which the interaction of particles can occur according to different, selectable, physical processes (such as energy loss by ionization, showering or Cherenkov emission).

{\bf Digitization:} All Monte-Carlo-Data are processed to model the detectors answer or to produce the effective answer of a readout chip.
The digitized data formats out of the simulation must be the same as those obtained in a real implementation, such as a prototype detector or even the final setup.

{\bf Local Reconstruction:} The digitized data are subject to a local reconstruction procedure which associates them to information with a physical meaning, like a 3D point (from \Mvd, GEM and central tracker), a total energy (from \Emc) or a Cherenkov angle (from \Dirc, \Rich). 

{\bf Tracking:} The tracking procedure in \pndrt proceeds through two steps: track finding (pattern recognition and track prefits) and track fitting (Kalman filter). While the track finder algorithms are very different for each sub detector (or combinations of them), the fitting is done by means of the GENFIT package \cite{Genfit2010}, based on the kalman filter technique \cite{R.E.Kalman1961}, \cite{Fruehwirth1987444}, using GEANE (\cite{GEANE1991}, \cite{A.Fontana2007} and \cite{Lavezzi2007}) as propagation tool.

{\bf PID:} Particle Identification is performed globally, taking information from all detectors into account. Each detector provides probability density functions (p.d.f.) for each particle kind, merged together to provide a global identification probability using Bayes theorem. In addition, different multi-variate algorithms are available, such as a KNN\footnote{K Nearest Neighbors Classifier} Classifier or a Neural Network approach. The tuning of these algorithms still is ongoing.

{\bf Physics Analysis:} The analysis tools have to deal with a collection of information based on four-momenta, positions and the identity of the reconstructed particles in a unified way. Particle combination, selection mechanisms and manipulation tools (like boosting between lab and center of mass frame) are provided. Furthermore a set of fitters is available to fit the four momenta and positions of particles (e.g.~from a decay) under different types of constraints.

\section{Detector Model of the MVD}
\label{MvdCadModel}
\authalert{Author: Thomas W\"{u}rschig, Contact: t.wuerschig$\mathrm @$hiskp.uni-bonn.de}

\begin{figure}[]
\begin{center}
\includegraphics[width=\columnwidth]{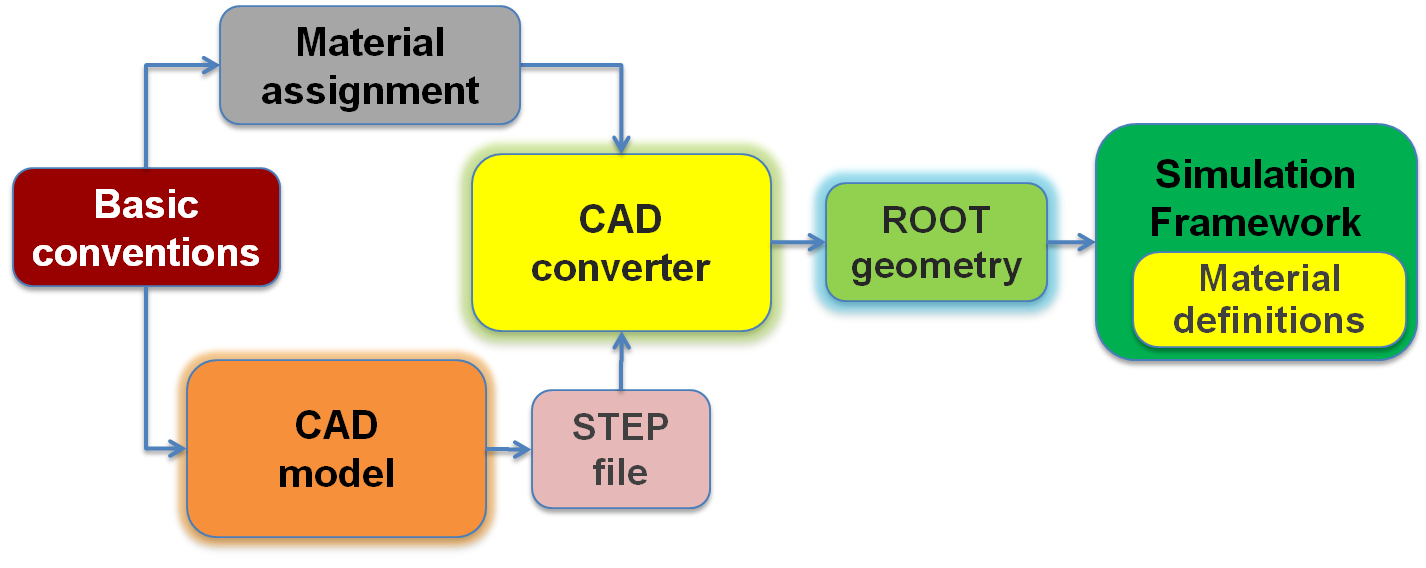}
\caption
[Basic approach for the introduction of CAD models into physics simulations]
{Basic approach for the introduction of CAD models into physics simulations \cite{NIM-MvdCadModel}.}
\label{pic-CADtoROOT}
\end{center}
\end{figure}

\begin{figure*}[]
\begin{center}
\includegraphics[width=15cm]{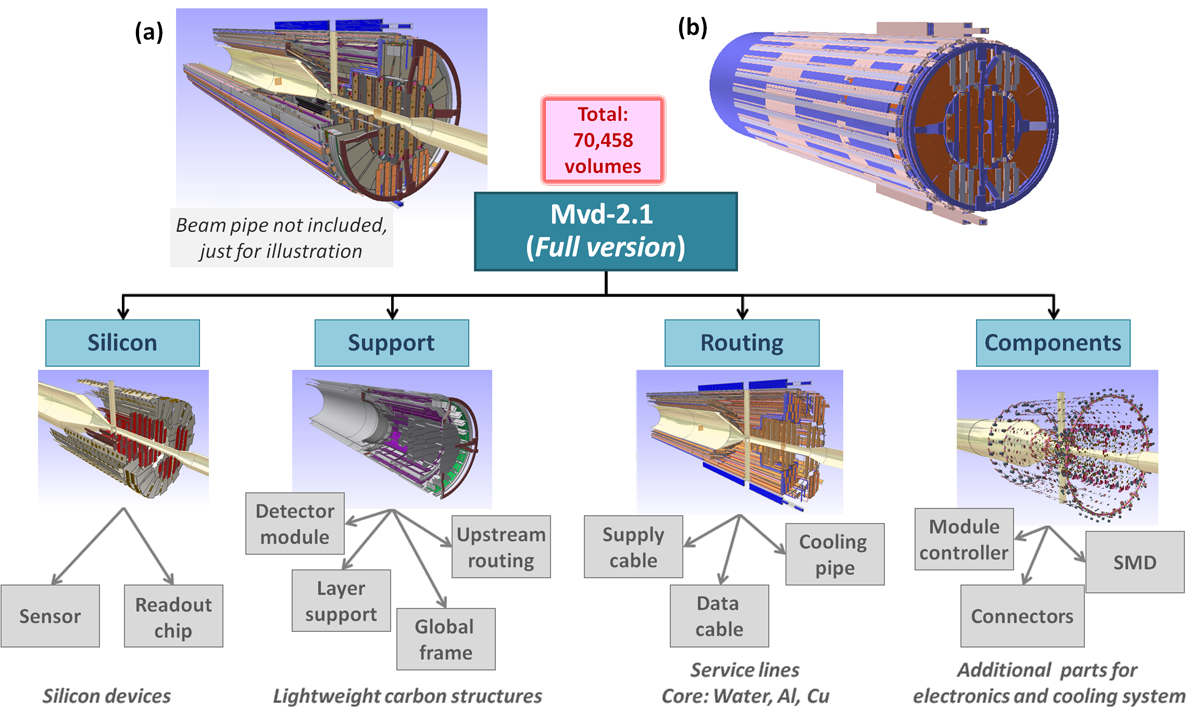}
\caption[Overall view and main structure of the MVD model as introduced into physics simulations]
{
\textit{Top:} View of the detailed MVD model before (a) and after (b) the conversion from a technical CAD project into a ROOT geometry. 
The main structure of the model is shown at the bottom.
}
\label{pic-Mvd-2.1model}
\end{center}
\end{figure*}

Along with the hardware developments a very detailed CAD model of the MVD has been accomplished, 
which delivers a very comprehensive and accurate
detector description~\cite{NIM-MvdCadModel}. 
It is based on a design optimization between a good spatial coverage on one hand 
and the introduction of sufficient space needed for passive elements and the detector integration on the other.
Thereby, a minimum material budget is achieved. 
The model includes active silicon detectors, connected readout electronics, 
local and global support structures, cables, a dedicated cooling system as well as the routing of all services. 
In addition, connectors and electronic components such as surface-mounted devices (SMD) are introduced. 
In total this comprehensive detector description contains approximately 70,500 individual volumes.

A converter~\cite{CAD-converter} has been developed to transform the CAD model 
into a ROOT geometry suited for simulations within the PandaRoot framework. 
It uses a common output format of CAD software (STEP format~\cite{STEP}) as input data. 
Besides the proper definition of the entire CAD in a ROOT geometry,
material properties, i.e.~the density and the radiation length, 
are given explicitly for each of the components.
Therefore, basic conventions have been fixed, 
which allow for an unambiguous and automated allocation of each single volume~\cite{PANDAnoteCAD}.
A schematic illustration of the procedure described is given in \figref{pic-CADtoROOT}.

The main structure of the available detector geometry is shown in \figref{pic-Mvd-2.1model}.
At the top, an illustration of the model before and after the conversion is shown.
Basically there are four functional parts, which are introduced as independent objects.
The silicon part contains all silicon detectors, i.e.~the pixel and the strip sensors, 
along with all associated readout chips. 
The second group is formed by the support structures of the detector.
It includes the local support for detector modules and the individual detector layers 
as well as the global frame for the overall detector integration of the MVD.
Moreover, a schematical support structure needed for the routing of services 
in upstream direction is added.
All support components are defined as lightweight carbon composites of different
effective density and radiation length. 

The entire model of the routing part scales with the number of readout channels 
and is based on low-mass cables and the overall cooling concept outlined in
chapter~\ref{Cables} and~\ref{Cooling-system}, respectively. 
Different cable types are introduced for the high voltage supply of the sensors, 
the power supply of the readout electronics 
as well the data transmission and slow control of the readout electronics. 
All cables are split into a conductive core and an insulation layer. 
The cooling lines consist of a water core and a casing material, 
which is defined either as steel or PVC. 
The barrel layer services are guided straightforward in upstream direction 
following the shape of the target pipe. 
The disk services are lead out to the top from where 
they are further extended in upstream direction.
The concept for the pixel disks is more complex
because the inner layers are inserted inside the barrel part.
Therefore, they must be routed at first in forward direction 
until a position from where they can be brought to outer radii.
While for all other parts a circular arrangement is chosen, 
the routing of the pixel disks is performed at the top and bottom. 

The last part of the MVD model includes additional components 
of the electronics  and cooling system. 
In the latter case these are given by connectors needed for the intersection 
from steel pipes used on the detector modules
to flexible pipes at the outer part of the cooling circuits.
The electronics part includes capacitors and
SMD components, i.e.~condensators and resistors, 
in order to deliver a complete description of fully equipped detector modules.

% - EOF

%% file: simulations/sim-sds.tex
% ---
\section{Silicon Detector Software}
\label{sim-sds}

%Representing the \Mvd in the simulations, 
A software package for strip and pixel silicon sensors, called Silicon Detector Software (SDS), has been developed to reproduce both the \Mvd and the Luminosity monitor in the \pndrt framework.

  \subsection{Monte-Carlo Particle Transport}
  Particles produced in the event generation phase are transported through the detector geometry and the magnetic field by the selected Monte-Carlo engine. When a particle crosses an active volume, a Monte-Carlo object is stored, containing the entry and exit coordinates, the total energy loss inside the volume, as well as the exact time the interaction in the detector occurs..

  Passive detector components, e.g.~holding structures, contribute with their material budget to scattering, energy loss, secondary particle production (delta-electrons) etc. 
  
  \subsection{Digitization}
  \label{Sim-sds_digi} 
  Digitization is the emulation of the detector answer before using reconstruction algorithms. The entry and exit coordinates and the charge information are used to determine which pixels or strips fired. The output data is the same format like the real data will have. Each fired strip or pixel is called a ``digi'', identifying the strip/pixel itself as well as the involved \frontend chip and the sensor. Furthermore the measured charge and the determined time stamp are stored. The digitization is a process in two stages, geometric and electronic digitization, described below.

  \subsubsection*{Geometric Digitization}
  %\pict{simulations/pictures/particlecrossingsensor.png}
\begin{figure}[]
\begin{center}
\includegraphics[width=\columnwidth]{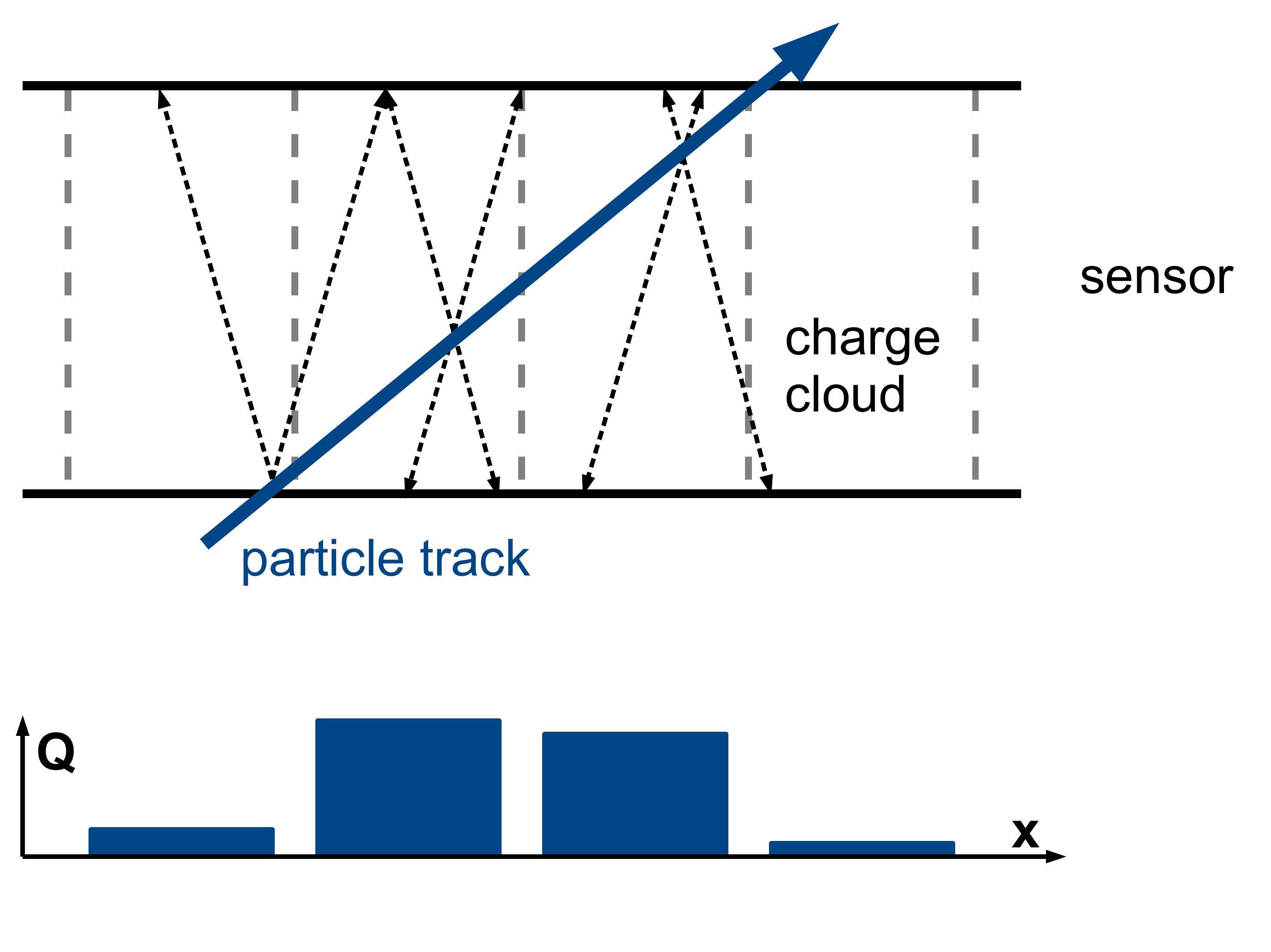}
\caption[Schematic sensor profile view and the linear digitization model with a charge cloud]
{Schematic sensor profile ($\approx 100 - 200$~$\tcmu$m thick) view and the linear digitization model with a charge cloud. Below the charge distribution on the readout cells is illustrated.}
\label{pic:sim:sdsdigiview}
\end{center}
\end{figure}
Geometrically one can project the particle path through the sensor to its surface. As the path crosses the readout structure the discretization into pixels or strips is done. The charge collected in each of the channels is calculated by a gaussian cloud distribution around the path (illustrated in \figref{pic:sim:sdsdigiview}). The charge $Q_i$ for one strip is given by
\begin{equation}\begin{aligned}
Q_i &= \frac{Q}{|x_{\text{out}}-x_{\text{in}}|}\\
& \cdot \left(F_i(x_{\text{out}})-F_i(x_{\text{in}})-F_{i+1}(x_{\text{out}})+F_{i+1}(x_{\text{in}}) \right)
\end{aligned} \end{equation}
and the charge $Q_{i,j}$ in one pixel is
\begin{equation}\begin{aligned}
Q_{i,j} &= \frac{Q}{4\cdot|x_{\text{out}}-x_{\text{in}}|\cdot|y_{\text{out}}-y_{\text{in}}|}\\
&  \cdot \left(F_i(x_{\text{out}})-F_i(x_{\text{in}})-F_{i+1}(x_{\text{out}})+F_{i+1}(x_{\text{in}}) \right) \\
&  \cdot \left(F_j(y_{\text{out}})-F_j(y_{\text{in}})-F_{j+1}(y_{\text{out}})+F_{j+1}(y_{\text{in}}) \right)
\end{aligned} \end{equation}
Q being the total number of electrons produced by the particle, $x_{\text{in}}$ and $x_{\text{out}}$ (y respectively) the entry and exit coordinates of its track and $F_k(x)$ being:
\begin{equation}
F_k(x) = (x_k - x)\cdot \text{erf}\left(\frac{x_k-x}{\sqrt{2}\sigma_\text{C}}\right) + \sqrt{\frac{2\sigma_\text{C}^2}{\pi}}\cdot e^{-\frac{(x_k-x)^2}{2\sigma_\text{C}^2}}.
\end{equation}
% and $\text{erf}(x)$ is the error function
The width $\sigma_\text{C}$ of this cloud is a parameter which is assumed to be in the order of 8~$\tcmu$m.

  \subsubsection*{Electronic Digitization}
  \authalert{Author: Simone Esch (s.esch(at)fz-juelich.de}
Charge from the sensors is measured in the \frontend electronics with the Time-over-Threshold method (ToT) (See also section~\ref{section:front_end_electronics:pixel_cell}). 
The charge is integrated in a charge sensitive amplifier (CSA), at the same time the capacitor of the CSA is discharged by a constant current source. 
The output of the CSA then is compared with a threshold voltage. 
The time the amplifier output is above this threshold is proportional to the deposited charge. 

%The electronic digitization simulates the work of the \frontend electronics like measuring the deposit charge and setting the TimeStamp of a hit.
%The deposit charge is measured with the Time-over-Threshold method (See chapter~\ref{front_end_lectronics:pixel_cell}).
%To measure the signal amplitude in the \frontend electronics the Time-over-Threshold (ToT) method is used (See chapter~\ref{front_end_lectronics:pixel_cell}).
%To obtain the the same data as the \frontend chip would deliver the number of electrons needs to be converted to the corresponding Time-over-Threshold value (ToT value) (See chapter \alert{frontend electronics}.)

%The SDS framework provides the possibility to include different models for this converstion of the digi charge into the corresponding ToT value.

\begin{figure}[]
\begin{center}
\includegraphics[width=0.5\textwidth]{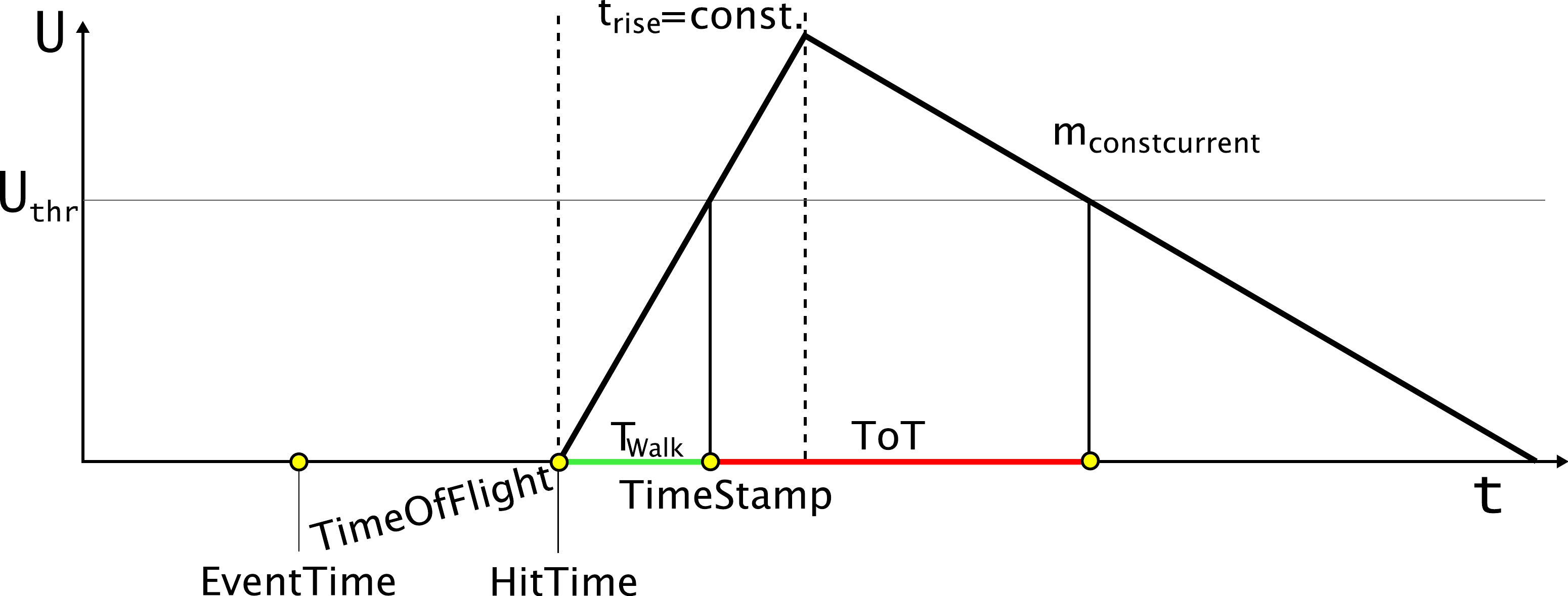}
\caption
[Simple model of the signal shape at the amplifier output]{Simple model of the signal shape at the amplifier output.}

\label{fig:triangular_signal_model}
\end{center}
\end{figure}
  
The SDS package provides the possibility to include different models for the ToT calculation.   

The simplest model for the \frontend chip behavior is the triangular one (See \figref{fig:triangular_signal_model}). It assumes a linear rising and falling of the signal with a constant peaking time: The characteristic parameters are the peaking time, the constant discharge current and the discriminator threshold. 

%As simulation parameters the peaking time, the constant current and the threshold are necessary. 

For the event building it is necessary to mark every digi with precise time information, the Time Stamp.
In reality the Time Stamp is set by the \frontend electronics when the rising signal crosses the threshold voltage. 
Due to the not negligible rising time of the preamplifier this time is not the same instant the particle hits the sensor. 
This delay between the physical hit and the recognition by the electronics is called Time Walk (\figref{fig:triangular_signal_model}). 
  
The simulated Time Stamp of a digi is the sum of the Event Time (MC information), the flight time of the particle till it hits the frontside of the sensor and the Time Walk of the Signal. 
  
  %The Time Stamp of a digi is the sum of the Event Time (MC information), the flight time of the particle till it hits the frontside of the sensor and the Time Walk of the Signal.
  \subsection{Noise Emulation}
\begin{figure}[]
\begin{center}
\includegraphics[width=0.7\columnwidth]{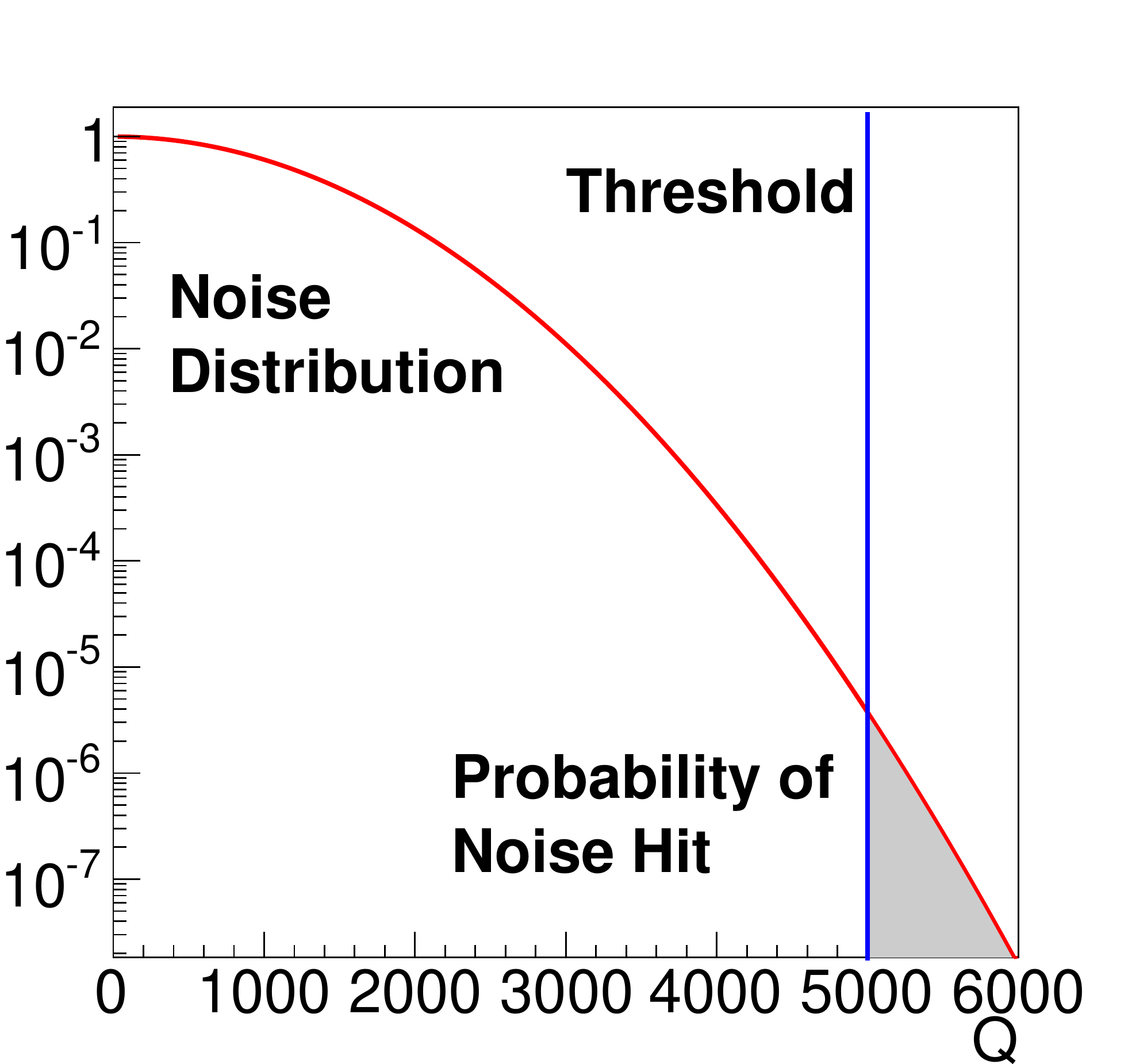}
\caption[Illustration of the probability for a pixel, or strip, to produce a noise hit (grey area)]{Illustration of the probability for a pixel, or strip, to produce a noise hit (grey area).}
\label{fig:noisegaus}
\end{center}
\end{figure}

  Electronic noise has the character of a gaussian distribution, centered around the baseline with the width of $\sigma_{\text{noise}}$. The noise baseline is set as the nominal zero in the simulations.
  To deal with the electronics noise each collected charge in a pixel and strip channel is redistributed by the noise gaussian with $\sigma_{\text{noise}}=200$ electrons for pixel sensors and $\sigma_{\text{noise}}=1000$ electrons for the strip sensors, as the hardware measurements show (see sections~\ref{Section-PixelReadout} and~\ref{strip:ToT}). A threshold cut is applied to suppress too low charge entries being set to $5\,\sigma_{\text{noise}}$.   
  The gaussian tail over the threshold will produce hits by the noise statistics (see \figref{fig:noisegaus}). This is taken into account by adding randomly selected channels. The number of noise hits is calculated by a poissonian random distribution with its mean value being
\begin{equation}
\bar{N}_{\text{noise}} = N_{\text{channels}}\cdot t_{\text{evt}}\cdot  f_{\text{clock}}\cdot  \frac{1}{2}\text{erfc}\left(\frac{Q_{\text{thr}}}{\sqrt{2}\cdot \sigma_{\text{noise}}}\right)
\end{equation}
while $t_{evt}$ is the time interval from the previous event and $f_\text{clock}$ the bus readout frequency
and  $\text{erfc}(x)$ being the complementary error function.

%\pict{simulations/pictures/noiseoverthreshold.png}

%  \subsubsection*{\alert{Anisotropic?} readout structures}
%  The geometric digitization model assumes that charge carriers drift inside the sensor volume to the surfaces and are collected at the metallization, However, the metallization's have gaps which deforms the electric field inside the sensor. \alert{What about p-stops?} Furthermore the metallization \alert{sees induced field -> ``crosstalk''}
%  
%  
%  \subsubsection*{Magnetic field effects}
%%See \figref{fig:magfieldart}.
%\alert{...}
%\alert{cite atlas tdr? less imprtant for thinner sensors ...}

%\begin{figure}[]
%\begin{center}
%\includegraphics[width=0.5\textwidth]{simulations/pictures/ArtisticFields.png}
%\caption
%{Artistic view of the magnetic field strength distribution in \panda}
%\label{fig:magfieldart}
%\end{center}
%\end{figure}

  \subsection{Local Reconstruction}

Digitized data are reconstructed by finding clusters and forming hits at the clusters centeroid, using the charge measurements to improve the accuracy. Furthermore the time of the particle passage is reconstructed by the time walk correction.

   \subsubsection*{Cluster Finding}
Cluster finding means to collect all digis belonging to one track passing through the sensor. This is done iteratively by taking all signaling channels around one digi wthin a given tolerance of channels. A minimum charge sum of a clusters digis can be required to reject ghost hits produced by noise only.

   \subsubsection*{Hit Reconstruction}
Having a cluster of digis, calculating the centeroid finds several approaches \cite{V.Radeka1980}.

Simply taking the central position of the channel with the highest charge yield in a cluster defines the binary centeroid. Its uncertainty is $\Delta x = p/\sqrt{12}$. This method is the only option to calculate coordinates with only one entry per cluster. 

A better resolution for clusters with two or more digis is given by the center of gravity, which uses the digis charge measurements as weight. The average position is calculated by 
\begin{equation}
\bar{x} = \frac{\sum_{i}q_{i}x_{i}}{\sum_{i}q_{i}} .
\end{equation}
This approach mirrors the geometric digitization procedure in the sense of isotropy in the readout structures and the lack of magnetic field effects. The precision $\Delta x$ is smaller than $p/\sqrt{12}$, which will be shown in section~\ref{sim:hitres}.

%   {\bf Head-Tail Algorithm:}\\
%For clusters with a large number ($n_{cl} > 5$) of signalling channels in any %coordinate, the central ones do not yield informatoin of the centeroid and thus %do not add to the precision. The Idea is to fit the clusters edges to retrieve %the centeroid. \alert{... }
%
%   {\bf $\eta$-Distribution:}\\
%When one splits each cluster around its highest charge entry into two halves, %called left and right, the centeroid  \alert{... continue, }

%{\bf Other algorithms:}\\
%As presented in \cite{V.Radeka1980}, there are other algorithms possible. Since these are more suited for large clusters their application in the MVD is not done yet.

To get a precise 3D information with the strip sensors, top and bottom side hits have to be correlated. This produces ambiguities when at least one of the sides of a sensor has more then one cluster. Such a situation is created by multiple tracks hitting a sensor in one event or in case of hits generated by electronic noise. Because the charge carriers arriving at either side correspond to  pairs produced by the same particle, a correlation in the measured charge (in form of a cut in the charge difference $|Q_\text{top}-Q_\text{bottom}|<Q_\text{cut}$) can reduce the number of ambiguities.

   \subsubsection*{Time Walk Correction}
\authalert{\ra author Simone Esch s.esch@fz-juelich.de}

To improve the time resolution of the TimeStamp, a correction of the TimeWalk effect can be applied. 
Based on a ToT conversion model the TimeWalk can be calculated from the ToT value and subtracted from the TimeStamp.

%Since the peaking time of the signal can be assumed as constant for different charges the Time Walk is not constant and depends on the deposited charge in the pixel/strip.
%This error for the measured Time Stamp of a digi can be partly corrected by the measured charge via the ToT mechanism.
%A Time Walk correction can be calculated from the charge which corrects the Time Stamp to improve the time resolution.
%Due to this dependency on the charge a Time Walk correction can be calculated by the measured ToT value and the Time Stamp can be corrected to improve the time resolution.

% The Time Walk Correction is calculated in the reconstruction in each amplifier model class individualy.
%In the simple amplifier model class \code{PndSdsFEAmpModelSimple} (See chapter digitization \alert{link}) the Time Walk  $t_{\text{TW}}$ is calculated as:

For the triangular ToT model the TimeWalk is given by:
\begin{equation}
 t_{\text{TW}} = \frac{Q_{\text{Thr}}[e^{-}]}{Q[e^{-}]}\cdot t_{\text{rise}}
\end{equation}

%The charge Q was calculated from the measured ToT value is then substracted from the Time Stamp of the corresponding digi to get the corrected Time Stamp.

%\alert{Put the simulations into the "results" or add all technical results here and put only full sim results below?}

%A simulation has been done to check the influence of the Time Walk effect on the Time Stamp resolution. 
%The simulation was done with the following values:
%\alert{fill in parameters and values}

%The effect is the bigger the less energy is deposit by the passing particle.  

% - EOF

%% file: simulations/sim-trkvtx.tex
% Ralfs LAzy definitions

  \section{Tracking and Vertexing}
\label{sec::trkvtx}

The tracking procedure in \pndrt proceeds through three steps. First, local pattern recognition algorithms find tracklets in different subdetectors. Afterwards a global track finding and tracklet merging is performed. The global tracks are then globally fitted.

   \subsection{\Mvd Tracklet Finding and Fitting}
      \label{sim:trackletfnd}
The track finding and pre-fitting for the hit points inside the MVD is based on a fast circle fit using the x-y-coordinate of a hit point and a second linear fit of the arc length of a tracklet on the fitted circle and the z-coordinate. For a computational fast circle fitting a translation of the hit points to a Riemann surface is performed. Here a plane is fitted to the track points. The parameters of the plane can be than back calculated to the circle parameters mid point and radius. A detailed description of the algorithm can be found in \cite{Fruehwirth2002366}.

\begin{figure}[]
\begin{center}
\includegraphics[width=7.5 cm]{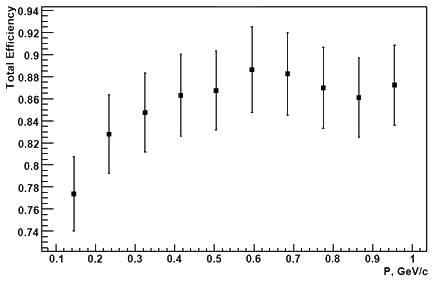}
\caption[Riemann efficiency for different particle momenta]
{Track finding efficiency of the Riemann-Track-Finder for six pion tracks with random momentum and theta angle in the MVD vs. the momentum of a single track. The error bars indicate the range of efficiency for different theta angles.
}
\label{pic-RiemannTrackEfficiencyAll}
\end{center}
\end{figure}

\begin{figure}[]
\begin{center}
\includegraphics[width=7.5 cm]{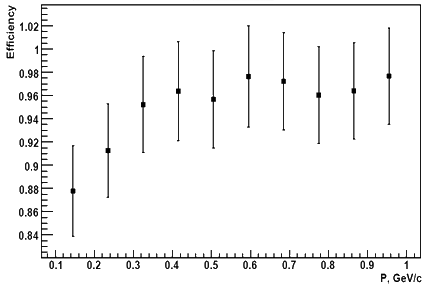}
\caption[Riemann efficiency for different particle momenta]
{Track finding efficiency for tracks with at least four hits in the MVD of the Riemann-Track-Finder for six pion tracks with random momentum and theta angle vs. the momentum of a single track. The error bars indicate the range of efficiency for different theta angles.
}
\label{pic-RiemannTrackEfficiency4Hits}
\end{center}
\end{figure}

The performance of the track finder is shown in the following figures. Six charged tracks per event with a random angular distribution and random momentum between 0.1 \gevc and 1 \gevc have been simulated. In the first diagram (\figref{pic-RiemannTrackEfficiencyAll}) the number of correctly found tracks divided by the number of all simulated tracks is plotted against the momentum of a single track. The error bars indicate the variation of the track efficiency for different theta angles. For tracks with very low momenta of about 100 \mevc the finding efficiency is about 77\% which rises up to 88\% for tracks above 600 \mevc. The increase of track finding efficiency with rising momentum can be explained by the dominance of multiple scattering for low momentum tracks. The low overall track finding performance inside the MVD is coming from the low number of hit points per track. In the second diagram (\figref{pic-RiemannTrackEfficiency4Hits}) the efficiency was calculated for all tracks with at least four hit points in the MVD, which are required for a successful track finding.

If one calculates the tracking efficiency only for those tracks with at least four track points it starts with 88\% at 100 \mevc and rises up to 97\% for tracks above 400 \mevc The error bars are exceeding 100\% due to tracks originating from the primary vertex with only three hit points. They can be found by the Riemann-Track-Finder as well but they are not used for the efficiency evaluation.
%counted as tracks with at least four track points.

To improve the overall track finding efficiency a combination of the MVD hit points with those of the central tracker is necessary, as explained in subsection~\ref{sim:trkfnd}.

   \subsection{Track Finding}
   \label{sim:trkfnd}
   The global track finding operates in a sequence, each step combining several detection systems. 
   First tracklets from the STT pattern recognition are merged with the \Mvd tracklets and unmatched \Mvd hits. 
   Second, any unmatched \Mvd tracklets are merged with remaining STT hits. 
   Then, all remaining \Mvd and STT hits are checked for forming new track candidates.
   Finally, the GEM disk hits are matched to the existing track candidates.
   Each step features an update of the track description parameters, as well as a more stringent condition to clean the candidates of spurious solutions. 
   A detailed description can be found in the STT technical design report \cite{SttTdr}.

      \subsection{Track Fitting}
Track fitting in \Panda is performed globally using all tracking devices. It employs the Kalman Filter (\cite{R.E.Kalman1961}, \cite{Fruehwirth1987444}) technique using the GENFIT package \cite{Genfit2010}. Particle transport is done with the GEANE package (\cite{GEANE1991}, \cite{A.Fontana2007} and \cite{Lavezzi2007}), which backtracks charged particles in a magnetic field, taking care of energy loss and scattering in materials. 
For each track a momentum resolution of $\Delta p/p < 2\%\;(20^\circ<\Theta<140^\circ)$ and $\Delta p/p < 3\%\;(12^\circ<\Theta<20^\circ)$ is achieved \cite{SttTdr}. By default all tracks are propagated to the point of closest approach to the z axis to form particle candidates.

   \subsection{\label{sec::vtxvtx}Vertex Finding Algorithms}
Finding the common starting point (vertex) of several tracks is the central goal of the \Mvd. There are several approaches under development. For primary vertices and short-lived particle decays, all tracks are treated purely as helices, because scattering and energy loss are negligible inside the beam pipe vacuum.

	\subsubsection*{POCA Finder}
%{\bf POCA Finder:}\\
A first approach is to find the point of closest approach (POCA) of two tracks, which is done analytically. 
Both helices are projected to the x-y-plane, assuming a constant magnetic field in z direction, forming circular trajectories. The intersection, or minimal distance, of the two circles defines the desired region. This is not the POCA in a full 3D treatment, but produces a good approximation for tracks originating from the same vertex.
The calculated distance between the tracks gives a measure on how well this assumption holds.

Having more than two tracks combined, the idea is to average the POCAs of each combination of two tracks, weighted by the inverse of the tracks distances $D_{i,j}$. For n tracks one gets the vertex $\vec{V}$:
\begin{equation}
\vec{V} = \frac{\sum^{n}_{i=0, j=i+1}D^{-1}_{i,j}\vec{V_{i,j}}}{\sum^{n}_{i=0, j=i+1}D^{-1}_{i,j}}
\end{equation}

To evaluate the "goodness" of the retrieved vertex the inverse of the weights sum is used: 
\begin{equation}
\tilde{D} = \left(\sum^{n}_{i=0, j=i+1}D^{-1}_{i,j}\right)^{-1}
\end{equation}

This method does not take the tracks covariances and error estimates into account, hence is not able to estimate the vertex covariance and error. However, in order to find a seed value for a fitter or to reject outlying tracks from bad combinations this ansatz is fast and useful.

	\subsubsection*{Linearised Vertex Fitter}
%{\bf Linearized Vertex Fitter:}\\
In this approach \cite{Billoir:1992yq} the tracks are locally parametrized and their propagation close to the expansion point\footnote{Point chosen where to do the parameterization and linearization procedure. It is the Origin for vertices close to the interaction point} is linearized. The fit's basic idea is to reduce the dimensions of the matrices to be inverted. For $n$ tracks one gets $n$ $3\times3$-matrices (5 being the number of parameter necessary to describe a helix, while two of them, the polar angle and the trajectory curvature, remain fixed) to be inverted instead of one 5$n\times$5$n$-matrix in the general case.
This improves the calculation speed of the fitter but lacks precision. The procedure is slightly sensitive to the choice of the expansion point, thus a good estimate on the vertex position is required. 
%In \cite{Billoir:1992yq} it is stated to work within a few cm.

Because of the imposed linearisations the calculation of the track parameters and covariances feature mainly sums over the tracks. Hence it is quite easy and fast to add or subtract tracks from the set without repeating the full calculations. This procedure is very useful to reject outlying tracks or to form mother particle candidates ``on the fly'' by vertex constraint only.

The procedure may be separated into two parts, one being a fast vertex fitter and the other the full fitter, which also fits the particles four-momenta. The calculation of the covariance matrices and the $\chi^2$ gives the estimate of the fit quality.

	\subsubsection*{Kinematic Vertex Fitter}
%{\bf Kinematic Vertex Fitter:}\\
A vertex constrained kinematic fitter is available. It is based on the Paul Avery papers (e.g.~\cite{avery4} and \cite{avery5}) and fits the whole set of momenta and the vertex position. It employs the inversion of the whole 5$n\times$5$n$-matrix for the 5 helix parameters of the $n$ tracks. The fit quality is estimated by the $\chi^2$ value and the covariance matrices. 

% - EOF

%% file: simulations/sim-pid.tex
%laura

  \section{PID Algorithm for the \Mvd}
\authalert{Author L. Zotti}

The passage of a particle through matter is characterised by a deceleration due to the scattering with the 
quasi-free electrons therein. 
The energy loss information can contribute to the global particle identification (PID), especially in the low 
momentum region (below a MIP), where the
Bethe-Bloch distribution is steeper depending on the particle mass. Because the energy loss process has a 
statistical nature, it's necessary to reproduce the behavior of a probability 
density function (p.d.f), of the energy loss, that will depend on the particle momentum and on the measurement 
uncertainty.
 Different statistical approaches have
been investigated in order to obtain a robust PID algorithm.

\subsection{Energy Loss Information}
\begin{figure}[!ht]
\begin{center}
\includegraphics[width=0.45\textwidth]{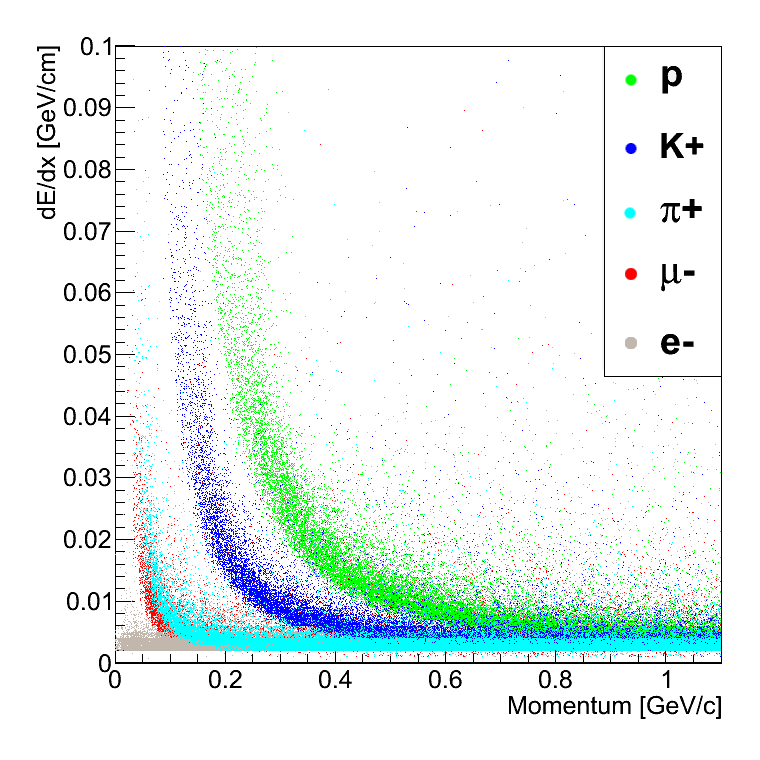}
\caption
[$\text{d}E/\text{d}x$ in the MVD as a function of the momentum]{$\text{d}E/\text{d}x$ in the MVD as a function of the momentum.}
\label{fig:mc_dedx}
\end{center}
\end{figure}

The energy loss information of different particle species have been studied and parametrised. The interaction 
of various particles has
been simulated within \pndrt using Geant4 and the
reconstruction tools available in the framework. In order
to take the uncertainty of the detector and of the full track fit into account, the reconstructed tracks have 
been used. The total energy
loss of a track is considered, which corresponds to the energy loss through few hundreds $\tcmu$m of silicon, 
because the number of contributing hits will be at maximum 6 in the 
forward part and 4 in the barrel part. The energy loss distribution of the reconstructed tracks for 
different particles is shown in \figref{fig:mc_dedx}.

\subsection{Statistical Approach for the Energy Loss Parametrisation}
Different statistical approaches have been studied to assign a $\text{d}E/\text{d}x$ estimator to each track. 
The arithmetic mean value of the energy loss
of course is not a good
estimator due to the long tail of the energy loss distribution. Several estimators have been compared, 
applying different algorithms
to the single-particle tracks. To estimate the quality of the different estimators a comparison in 
terms of standard deviation and of 
efficiency and purity of the samples, selected by different algorithms, has been performed.\\
Each particle species has been simulated at seven different momenta from 400~\mevc up to 1~\gevc in 
step of 100~\mevc,
 then the energy loss of the reconstructed tracks has been studied.
The particle energy lost in each layer $\text{d}E/\text{d}x$ is approximated by the quantity $\Delta E/(D/\cos \theta)$ 
where $\Delta E$ 
is the energy lost in each sensor (measured by the integrated
charge of a ``cluster'', i.e.~a set of neighbouring pixels or strips), D is the active sensor 
thickness, and $\theta$ is the angle between the track direction and the axis normal to the sensor surface.
The distribution of the  track energy loss has been fitted with a Gaussian or with a Landau function in 
order to find the mean
 value, or the most probable value (mpv), and the corresponding sigma for the different particle species 
 and for different momenta.\\

\begin{figure*}[]
\begin{center}
\includegraphics[width=\textwidth]{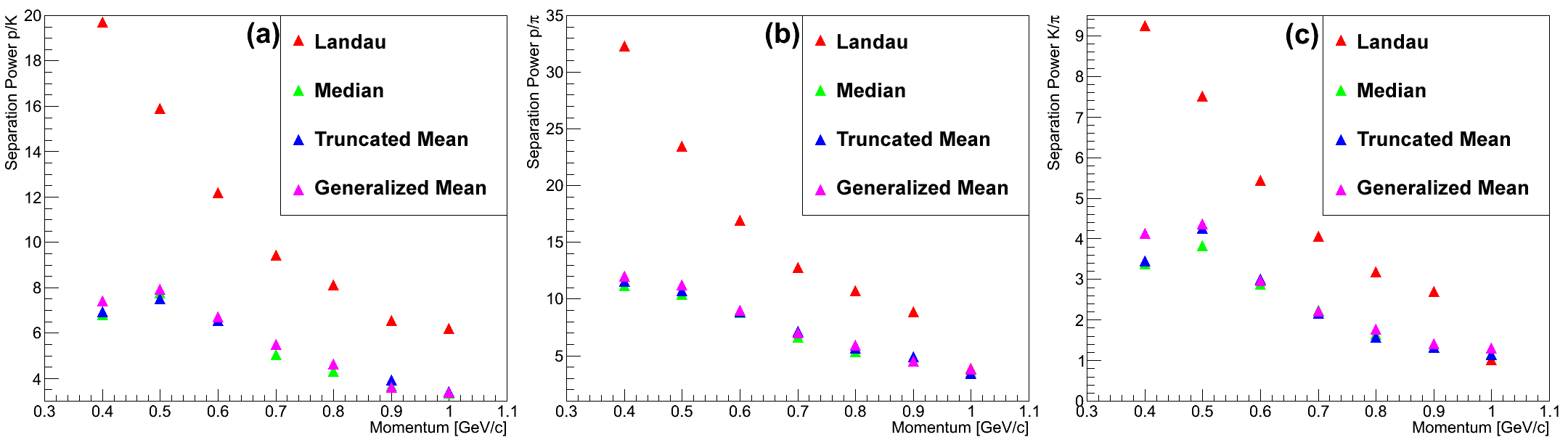}
\caption
[Separation power as a function of the momentum]{Separation power as a function of the momentum: (a) SP for p/K, (b) SP for p/$\pi$, (c) SP for K/$\pi$.}
\label{fig:sep-pk}
\end{center}
\end{figure*}

{\bf Mpv of a Landau Distribution}\\
A unique specific energy loss $\Delta E/\Delta x$
 value along the whole trajectory is estimated, because the loss of energy through each layer is negligible with respect to the considered momenta. Such a value is calculated as 
\begin{equation}
\Delta E/\Delta x=\frac{\sum_{k=1}^n \Delta E_{k}}{\sum_{k=1}^n \Delta x_{k}};
\end{equation}
where
n is the number of track hits in the MVD, $\Delta E_k$ is the energy
lost by the particle in each layer, and $\Delta x_k$ is the crossed thickness.\\
A Landau distribution has been used to fit the total reconstructed energy loss of each track. The most probable values and the 
related sigmas 
have been used for the parametrisation. 

{\bf Median}\\
The median of a distribution is calculated by arranging all the $\Delta E/\Delta x$ of the track on the different detector planes from the lowest to
the highest value and picking out the middle one (or the mean of the two middle value in case of even observations). The mean value and the sigma
of the Gaussian distribution obtained from the different tracks is used as reference value of the energy lost by that particle at a given 
momentum. 

{\bf Truncated Mean}\\
The truncated mean method discards the highest value in a fixed fraction of the total number of observations, 
calculating the mean
of the remainder values. In this study 20\% of
the hits are discarded. This cut is applied only if a track has more then 3 hits.

{\bf Generalised Mean}\\
The ``generalised mean'' of grade $k$ of a variable $x$ is defined as:
\begin{equation}
 M_k(x_1,x_2,....x_n)=\left( \frac{1}{n}\sum_{i=1}^n x_i^k\right)^{\frac{1}{k}};
\end{equation}
In this case the grade $k$ = -2 \cite{CMSnote} has been used. \\

The separation power (SP, \cite{SepPow}) has been calculated as:
\begin{equation}
 \text{SP}=\frac{\left | <\text{d}E/\text{d}x>_\text{p1}-<\text{d}E/\text{d}x>_\text{p2} \right |}{\sigma_\text{p1}/2+\sigma_\text{p2}/2};
\end{equation}
where $<\text{d}E/\text{d}x>$ is the mean value (or the mpv) of the energy lost by a particle at a certain momentum, p1 and p2 are two particle species to be
separated and $\sigma$ is the standard deviation of the distribution. 
The separation power p/K, p/$\pi$ and  K/$\pi$ for the different algorithms 
is shown in \figref{fig:sep-pk}. As expected the separation power decreases with the increasing momentum.

Even if the Landau distribution shows a better separation power
(because of the smaller sigma of the Landau distribution, at
least by a factor 2), it is not the most suitable estimator because
of the long tail, which introduces a significant contamination.

\begin{figure}[htb]
\begin{center}
\includegraphics[width=0.84\columnwidth]{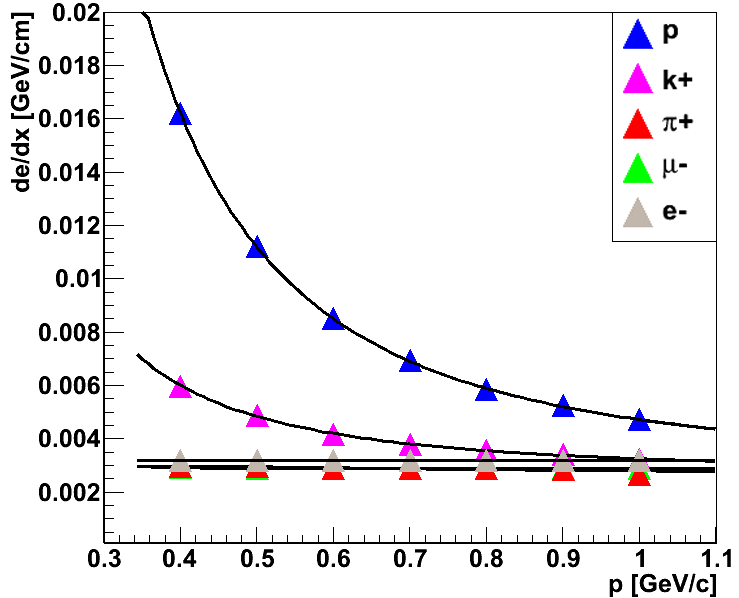}
\caption
[Fit function of the mean values of the median distribution as a function of the momentum for different particle species]
{Fit function of the mean values of the median distribution as a function of the momentum for different particle species.}
\label{fig:med-para}
\end{center}
\end{figure}

In order to build a probability density function, the trend of the mean (or MPV) values and of the sigma
 as a function of the momentum have been fitted by the following function:
\begin{equation}
\frac{\text{d}E}{\text{d}x}=\frac{a}{p^2}\left[b\cdot \ln p^2-p^2-c\right]
\end{equation}
where \textit{a}, \textit{b} and \textit{c} are the parameters of the fit. 
As an example the parametrisation of the mean value of the median as a function of the momentum
 is shown in \figref{fig:med-para}.

\begin{figure*}[!ht]
\begin{center}
\includegraphics[width=0.95\textwidth]{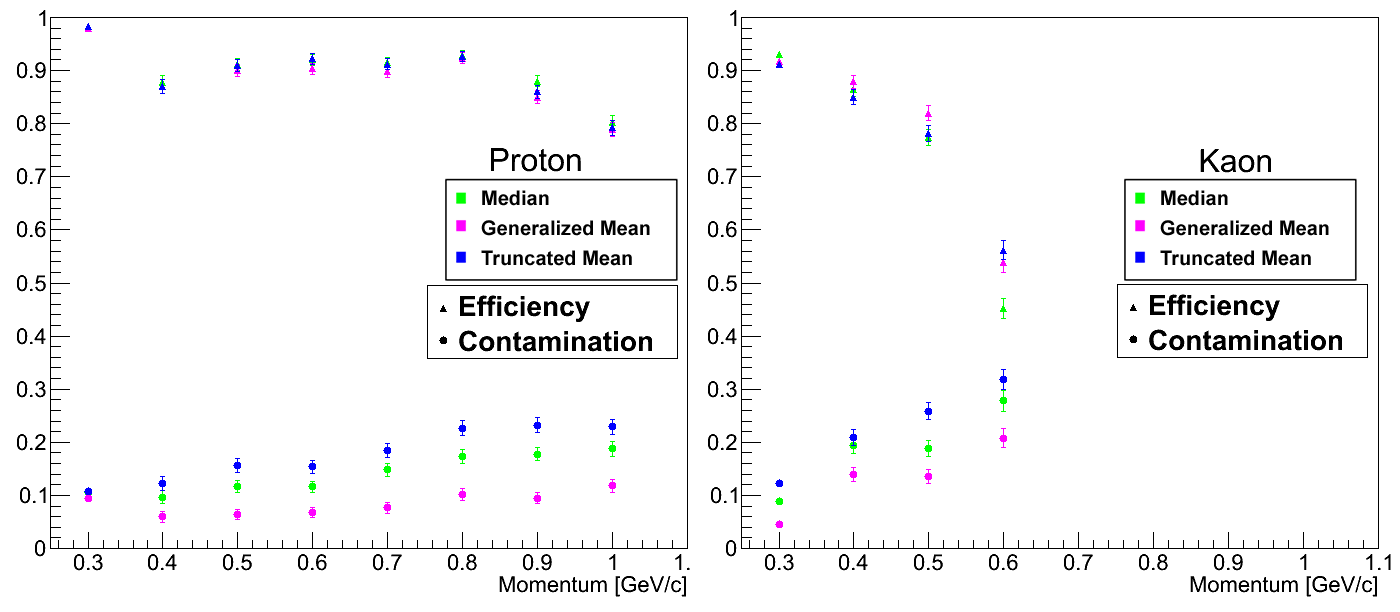}
\caption
[Efficiency and contamination for protons and kaons]{Efficiency and contamination for protons (left) and kaons (right).}
\label{fig:eff-cont_protoni}
\end{center}
\end{figure*}

In this way, the momentum and the energy loss of a given reconstructed track are used to
build the probability density function and to extract the probability to belong to a certain particle species.\\

The efficiency and the contamination of the different algorithms has
been plotted as a function of the momentum for the proton and kaon hypothesis, as shown in \figref{fig:eff-cont_protoni},
using a probability threshold of 90\% (i.e.~the particle hyphotesis should have a 
probability bigger than 90\% to belong to a certain particle species). 
The efficiency of proton identification is almost constant around 90\%
and the contamination in the worst case (at larger momenta), is smaller than 30\%. 
The efficiency of kaon identification is more dependent on the momentum,
but it is larger than 80\%, at least up to 500~\mevc. The 
contamination is higher than in the case of proton, by a factor of 2 at 600~\mevc. The comparison of
the three algorithms shows that the efficiency is almost the same in
the three cases, while the contamination is smaller for the
generalised mean. 
Other particle species are not taken into account because, as shown in \figref{fig:mc_dedx}, the signals of
pions, muons and electrons in the MVD are too overlapped to deliver a good identification 
efficiency.\\
The MVD can contribute to the global PID decision up to a momentum of
500~\mevc for kaons and up to 1~\gevc for protons.

%% file: simulations/sim-studies.tex
% ---

\section{General Simulation Studies}
  \subsection{Radiation Damage}
  \label{sim-RadDam}

\begin{figure}[]
\begin{center}
\includegraphics[width=0.98\columnwidth]{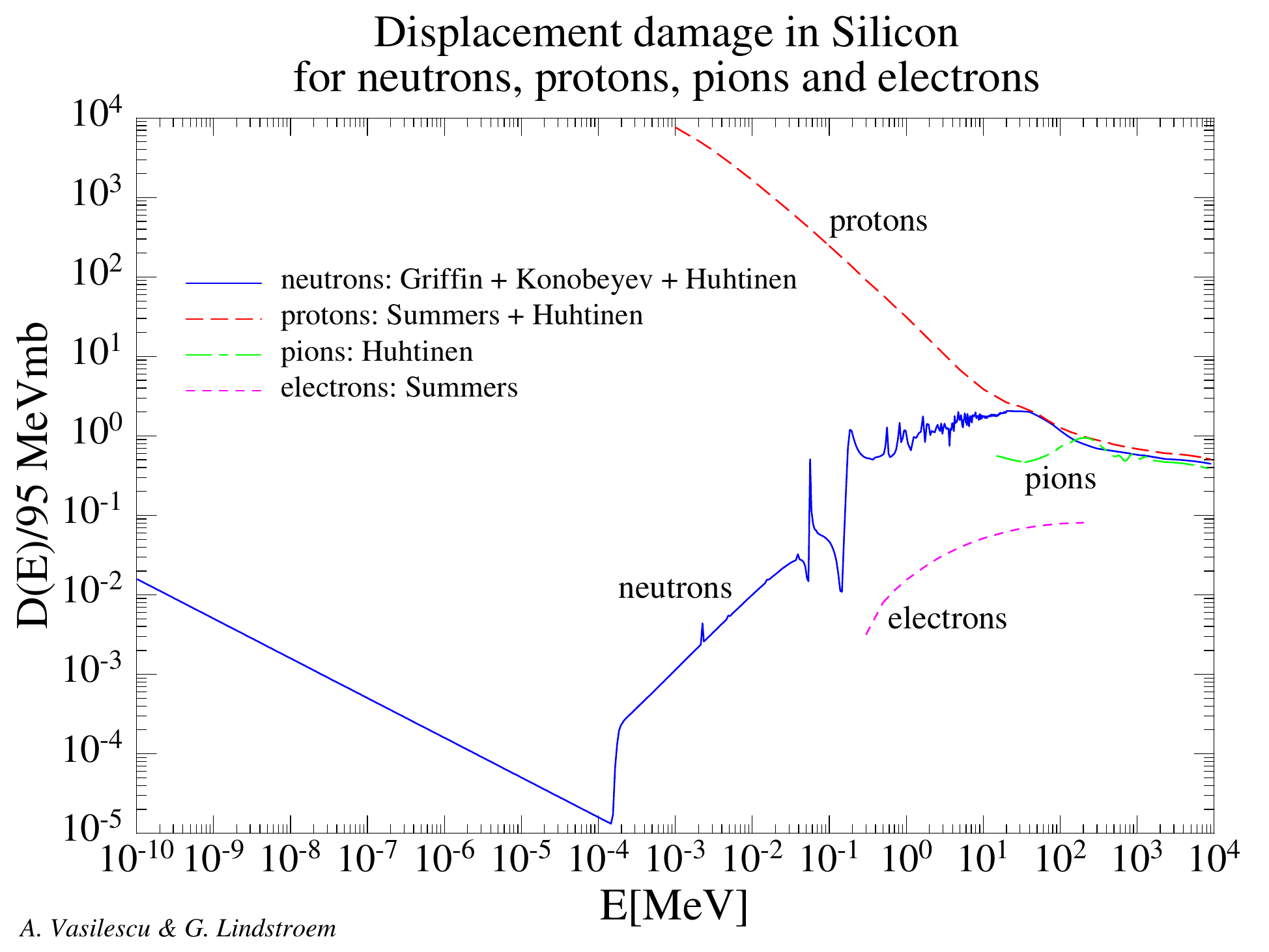}
\caption[Damage factor for different particle types and momenta]
{Damage factor for different particle types and momenta \cite{NIELonline}.}
\label{pic-Hardnessfactor-RadDam}
\end{center}
\end{figure}

Most important for the damage of the silicon sensors is the non-ionising energy loss which causes damages in the lattice structure of the semiconductor material which influences the collection efficiency and the leakage current of the sensor. The damage caused by a particle strongly depends on the particle type and its momentum. To compare the damage caused by different particles one can define a hardening factor $\kappa$ which scales the different particle types to the damage caused by the flux of 1~\mev neutrons according to the NIEL hypotheses \cite{Gunnar200330}. This scaling factor can be seen in \figref{pic-Hardnessfactor-RadDam} for various particle types.

The simulation of the radiation damage was performed with the \pndrt framework. Depending on the target material, two different event generators have been used: DPM\footnote[1]{Including elastic Coulomb scattering with a cut off below a scattering angle of 0.17$^{\circ}$} for hydrogen targets and UrQMDSmm for nuclear targets. The generated particles then were propagated through the full \pnd detector using Geant4 to include also backscattered particles from detectors further outside of \pnd. The neutron equivalent flux of particles then was determined by weighting each intersection of a particle with a silicon sensor with the momentum and particle type dependent scaling factor. This simulation was done for different target materials and beam momenta from 2~\gevc up to 15~\gevc. For each setting more than $10^6$ antiproton-target interactions have been simulated on the pandaGRID.

With the simulated events no time information is given. Therefore it is necessary to estimate the interaction rate R for different target materials and beam momenta to come from a fluence for a number of events A to a fluence for one operation year with $10^7s$. 

\begin{equation}
t=\frac{A}{R}=\frac{A}{\sigma \cdot L}
\end{equation}
The cross section $\sigma$ for hydrogen targets is given by the DPM generator, the values for nuclear targets was taken from \cite{Lehrach06}.

\begin{figure}[]
\begin{center}
\includegraphics[width=0.95\columnwidth]{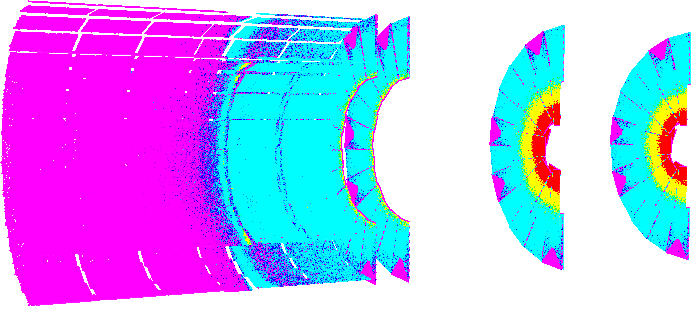}
%\hfill
\includegraphics[width=0.95\columnwidth]{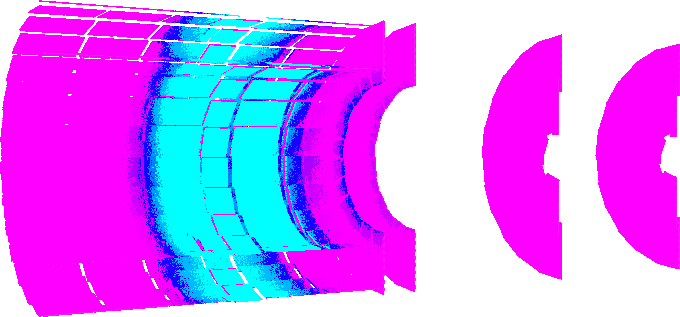}
\caption[Dose distribution in the strip part]
{Relative Dose distribution for a beam momentum of 15~\gevc in the strip part with a hydrogen target (top) and a xenon target (bottom) \cite{Mertens}.}
\label{pic-DoseDistributionDPMStrip-RadDam}
\end{center}
\end{figure}

\begin{figure}[]
\begin{center}
        \subfloat[Hydrogen target, momentum 2~\gevc.]
        {\label{pic-DoseDistributionDPMPixel-RadDam-a}
	    \includegraphics[width=0.95\columnwidth]{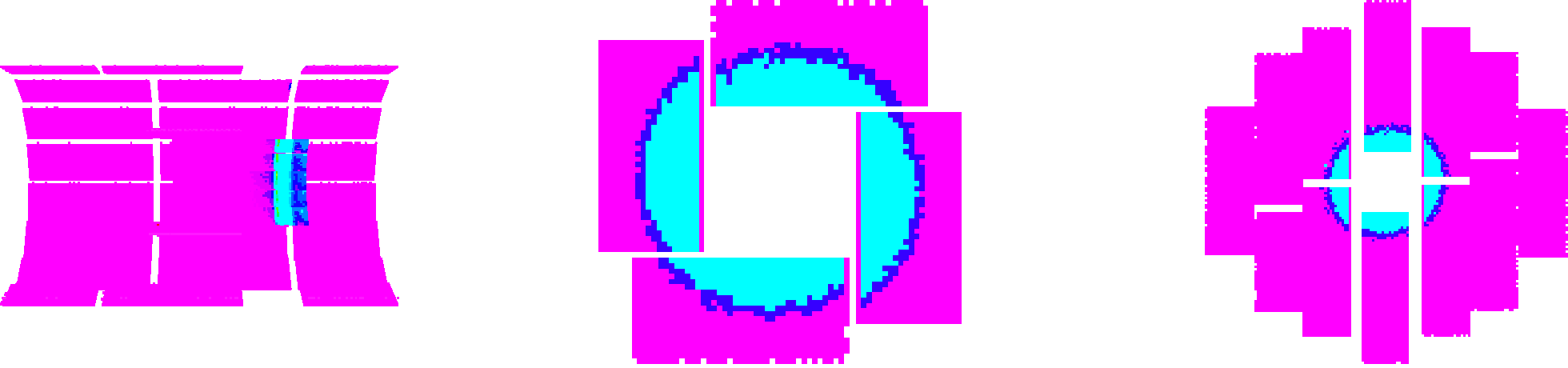}}\\
        \subfloat[Hydrogen target, momentum 15~\gevc.]
        {\label{pic-DoseDistributionDPMPixel-RadDam-b}
	    \includegraphics[width=0.95\columnwidth]{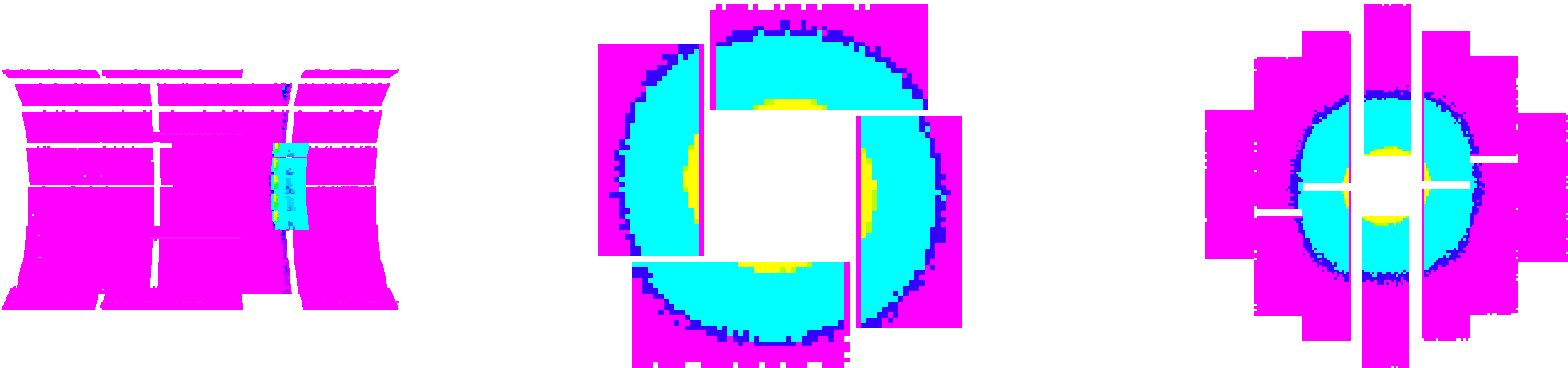}}\\
        \subfloat[Xenon target, momentum 2~\gevc.]
        {\label{pic-DoseDistributionDPMPixel-RadDam-c}
        \includegraphics[width=0.95\columnwidth]{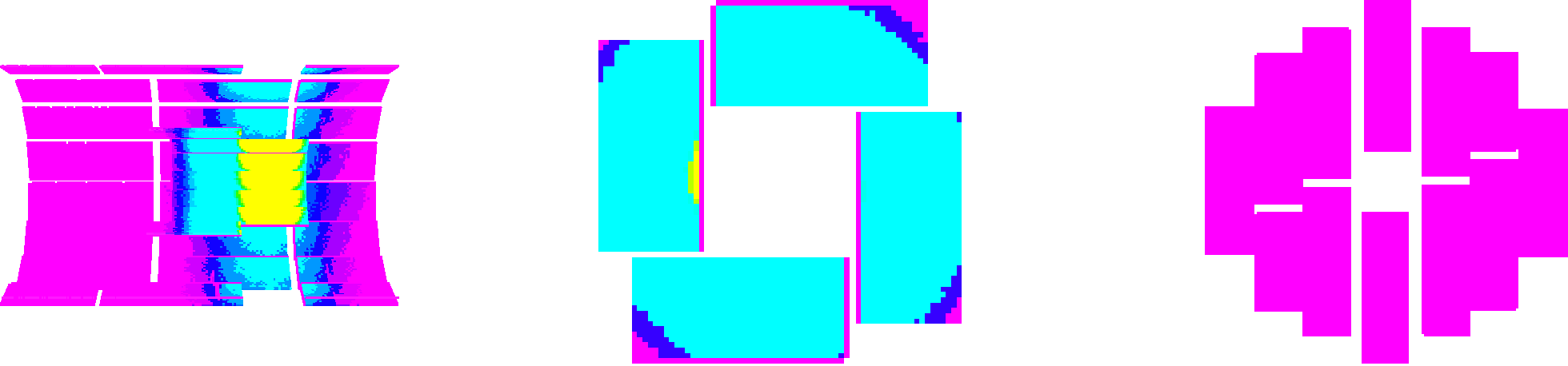}}\\
        \subfloat[Xenon target, momentum 15~\gevc.]
        {\label{pic-DoseDistributionDPMPixel-RadDam-d}
        \includegraphics[width=0.95\columnwidth]{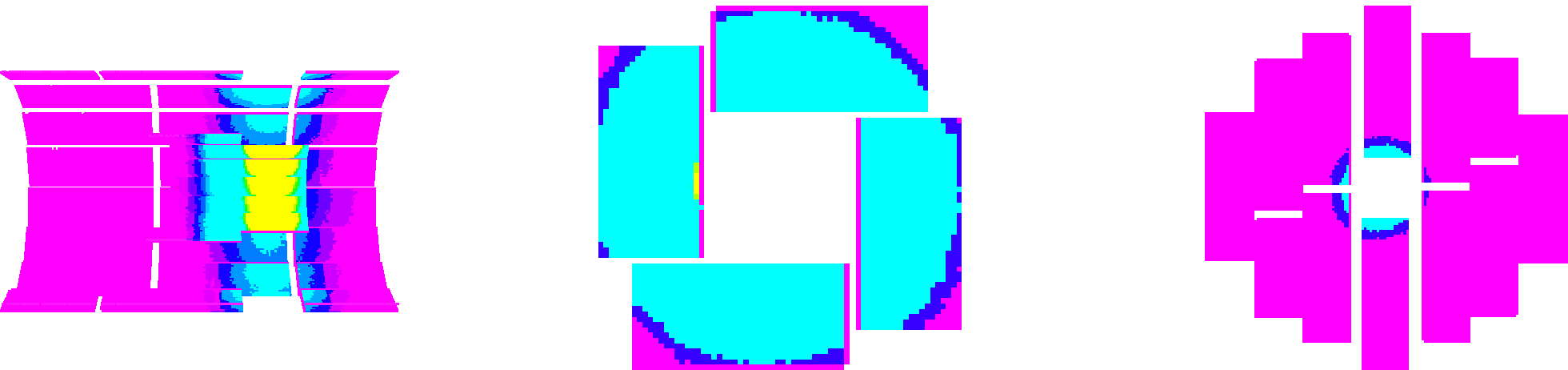}}\\
\caption[Dose distribution in the pixel part]
{Relative dose distribution for a beam momentum of 2~\gevc and 15~\gevc in the pixel part with a hydrogen and a xenon target. From left to right: pixel barrels, small disks and large disks \cite{Mertens}.}
\label{pic-DoseDistributionDPMPixel-RadDam}
\end{center}
\end{figure}

\Figref{pic-DoseDistributionDPMStrip-RadDam} shows the different dose distributions for a light hydrogen target and a heavy xenon target for a beam momentum of 15~\gevc. The same diagrams can be seen for the pixel part in \figref{pic-DoseDistributionDPMPixel-RadDam} for two beam momenta of 2~\gevc and 15~\gevc and for a hydrogen and a xenon target. The colour scale of the diagrams is different for the different setting and does not allow a comparison between the different drawings. For the light target the maximum radiation dose is seen for the inner parts of the forward disks while for the heavy target the distribution of the radiation dose is more isotropic with a maximum at the innermost barrel layers which are closest to the target.

%\begin{figure*}[t]
%\begin{center}
%\includegraphics[width=0.95\columnwidth]{simulations/pictures/dpm15ana2.pdf}
%\includegraphics[width=0.95\columnwidth]{simulations/pictures/xe15ana3.pdf}
%\caption[Dose distribution for Hydrogen and Xenon targets]
%{Dose distribution per mm$^{2}$ and operation year for a hydrogen target (top) and a xenon target (bottom) with 15~\gevc beam momentum \cite{Mertens}.}
%\label{pic-DoseDistributionDPMHisto-RadDam}
%\end{center}
%\end{figure*}

%\begin{figure}[t]
%\begin{center}
%\includegraphics[width=0.95\columnwidth]{simulations/pictures/dpm15ana2.pdf}
%\hfill
%\includegraphics[width=0.95\columnwidth]{simulations/pictures/xe15ana3.pdf}
%\caption[Dose distribution for Hydrogen and Xenon targets]
%{Dose distribution per mm$^{2}$ and operation year for a hydrogen target (top) and a xenon target (bottom) with 15~\gevc beam momentum \cite{Mertens}.}
%\label{pic-DoseDistributionDPMHisto-RadDam}
%\end{center}
%\end{figure}

\begin{figure}[t]
\begin{center}
\includegraphics[width=0.98\columnwidth]{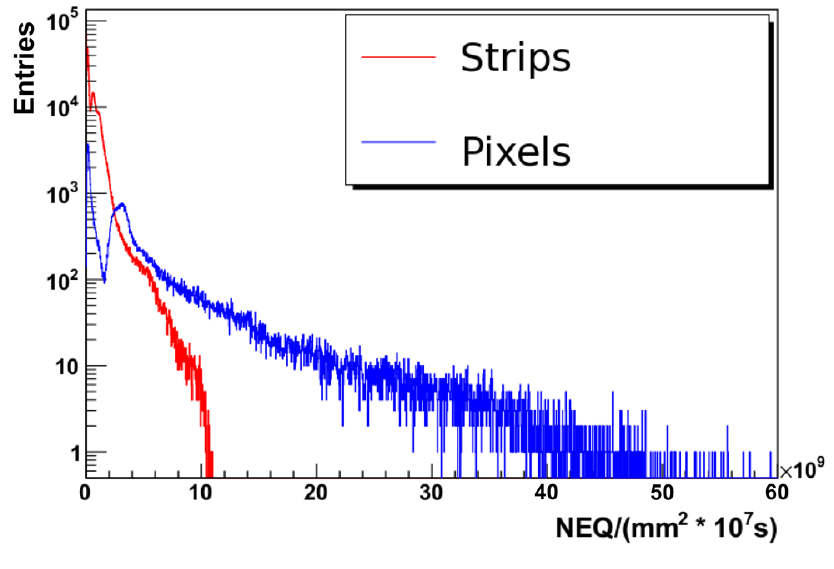}
\hfill
\includegraphics[width=0.98\columnwidth]{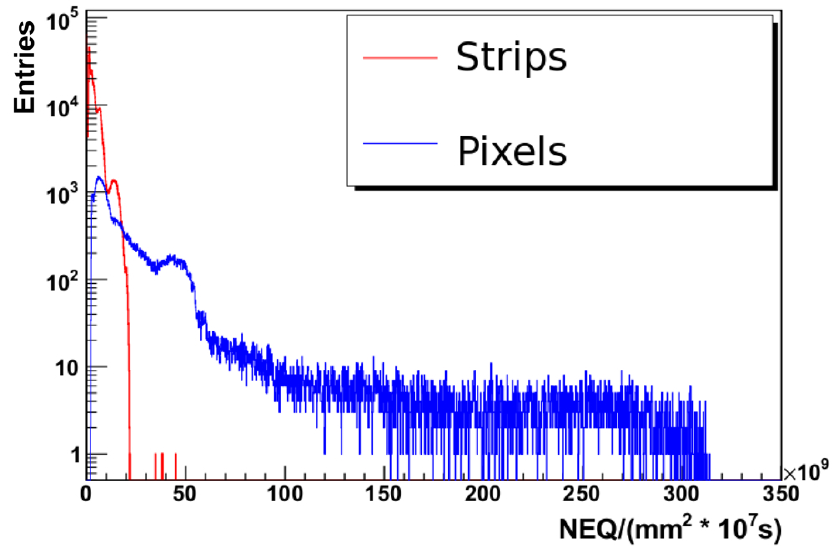}
\caption[Dose distribution for Hydrogen and Xenon targets]
{Dose distribution per mm$^{2}$ and operation year for a hydrogen target (top) and a xenon target (bottom) with 15~\gevc beam momentum \cite{Mertens}.}
\label{pic-DoseDistributionDPMHisto-RadDam}
\end{center}
\end{figure}

The two histograms \ref{pic-DoseDistributionDPMHisto-RadDam} show the overall dose distribution for the two different target materials for one year of \pnd operation. One can see a strong anisotropy of the distributions with few very hot regions and most areas with more than a magnitude less radiation damage than the maximum.

\clearpage

\begin{figure}[htb]
\begin{center}
\includegraphics[width=\columnwidth]{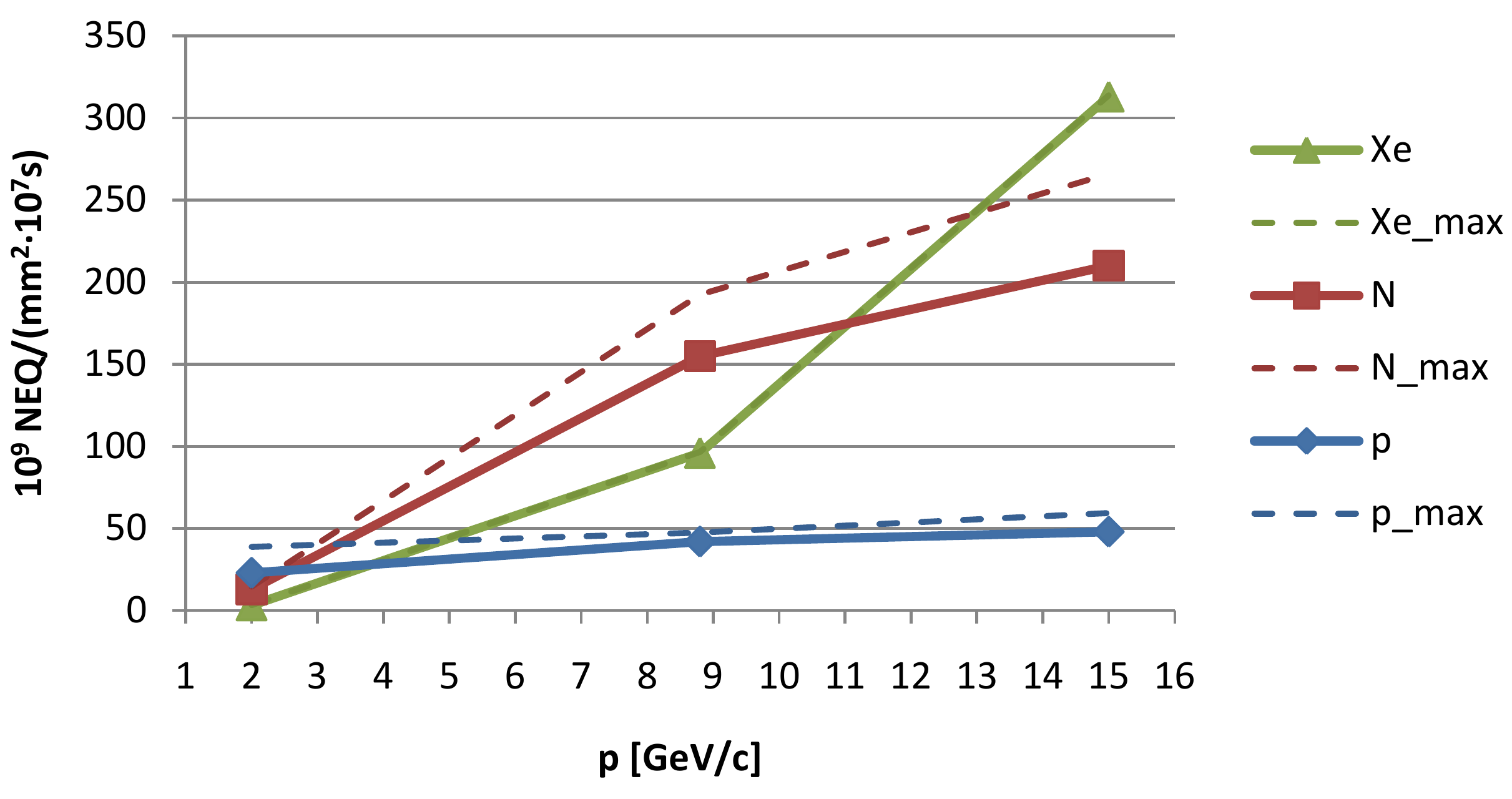}
\caption[Dose distribution for various targets and momenta]
{Maximum radiation dose per mm$^{2}$ and operation year for various target materials for the full MVD \cite{Mertens}.}
\label{pic-DoseDistributionSummary-RadDam}
\end{center}
\end{figure}

A summary of the radiation damage for one operation year for various target materials and beam momenta is shown in \figref{pic-DoseDistributionSummary-RadDam}. This diagram shows the maximum radiation dose a square mm of sensor material sees in one operation year. For a hydrogen target it stays below $1\cdot10^{13}$~\neueq with only a slight increase with the beam momentum while for nuclear targets a strong momentum dependency can be seen with a maximum dose rate of $5\cdot10^{13}$~\neueq for a xenon target with 15~\gevc beam momentum. A duty cycle of 50\% was assumed for these estimates.

%\clearpage
%\newpage

  \subsection{Detector Coverage}
\label{Sim-MvdCoverage}
\authalert{Author: Thomas W\"{u}rschig, Contact: t.wuerschig$\text @$hiskp.uni-bonn.de}

In order to achieve the required tracking performance, 
a high spatial coverage with a sufficient number of track points inside the MVD is needed. 
The implemented optimisation of the detector geometry aims at a minimum of four MVD hit points 
for central tracks at polar angles above 10$^\circ$. 
In this way an independent pre-fit of MVD tracklets can be performed 
by means of fast tracking algorithms (for details see~\ref{sim:trackletfnd}). 
Moreover, for a precise vertex reconstruction 
the first hit point must be detected as close as possible to the origin of the track. 

Extensive studies have been performed to evaluate the detector coverage 
obtained after the optimisation process. 
Therefore, a reduced model of the MVD was used, 
which only includes active sensor elements and no passive material. 
Due to the focus on mere geometrical aspects connected to 
the sensor dimensions and their arrangement in the individual layers, 
all pipes of the beam-target system were omitted thus minimising scattering effects.
Material effects will be explicitly discussed in the next subsection~\ref{Sim-RadLenght}.

In the presented coverage study, particles of fixed momentum  were generated at the nominal interaction point
and isotropically emitted over the polar angle, $\theta$, and the azimuthal angle, $\phi$. 
For the following particle propagation through the detector the magnetic field of the 2~T solenoid was switched on. 
Simulations were carried out in the \pndrt framework~\cite{Group2007}, 
which allows an interface to the Geant4 transport code~\cite{GEANT4}.
\Tabref{tab-SimCoverage-Setup} summarises all basic input parameters. 
The different setups were chosen in order to check a possible influence of 
the resulting bending radius on the effective detector coverage. 

\begin{table}[hb]
\begin{center}
\small
\begin{tabular}{|c|c|}
\hline
\vspace{-3mm}&\\
MVD model & Only active elements \\
\vspace{-3.5mm}&\\
Magnetic field & $|\vec B| = B_z = 2$\ T\\
\vspace{-3mm}&\\
Particle species & $\pi^+, \pi^-,\hspace{1mm} p\,,\hspace{1mm} \overline p$\\
\vspace{-3.5mm}&\\
Momentum /  [\gevc] & 0.15, 0.5, 1.0, 1.5 \\
\vspace{-3.5mm}&\\
Particle distribution & Isotropic in $\theta$ and $\phi$\\
\vspace{-3.5mm}&\\
Start vertex & $(x,y,z)=(0,0,0)$\\
Number of events & 2 million / scan \\
\hline
\end{tabular}
\caption[Main setups for the studies of the detector coverage]
{Basic parameters for the simulations.} 
\label{tab-SimCoverage-Setup}
\end{center}
\end{table}

\begin{figure}[ht]
\begin{center}
\includegraphics[width=\columnwidth]{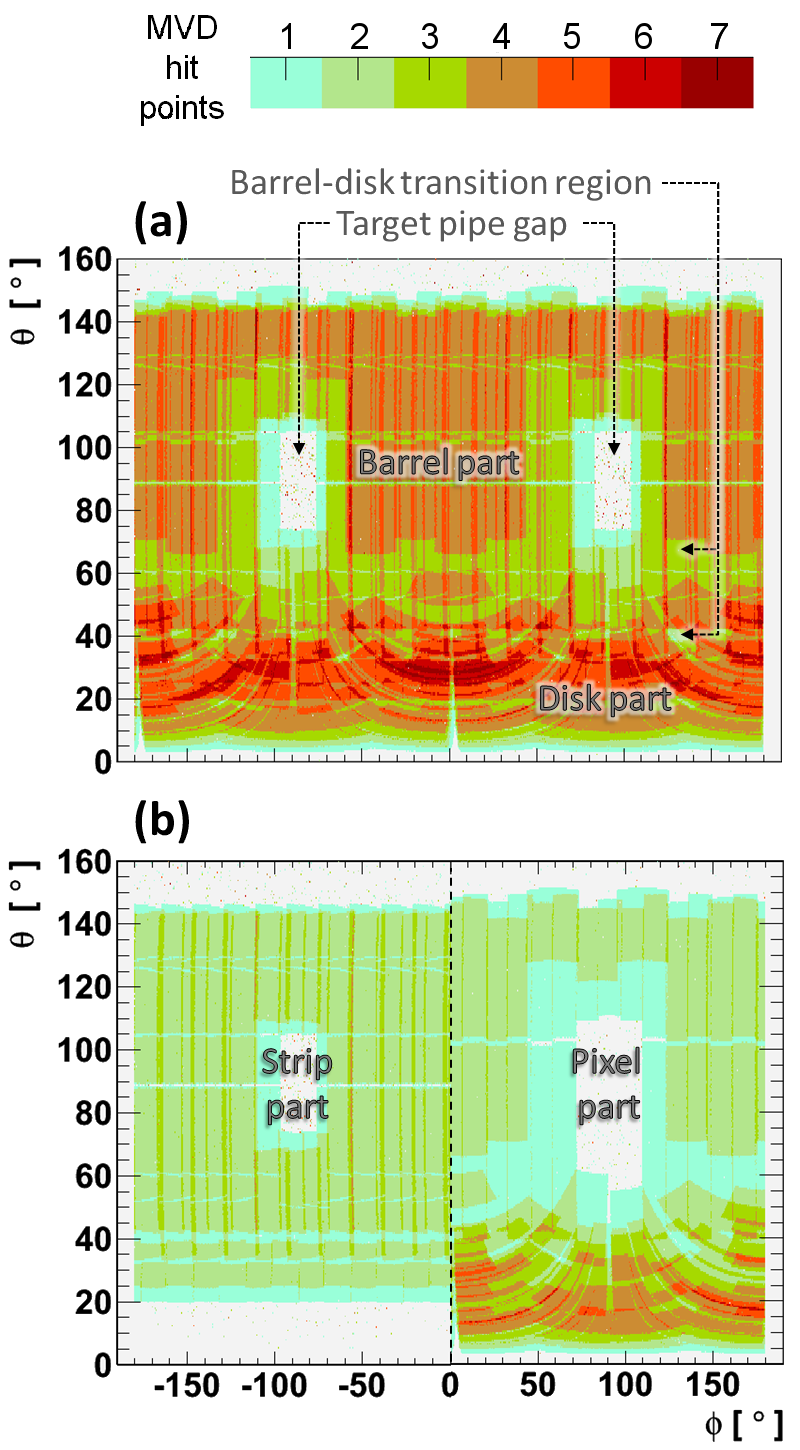}
% Pic3-02_MvdLayout.png: 787x873 pixel, 80dpi, 24.98x27.71 cm, bb=0 0 708 786
\caption[Detector coverage for pions with a momentum of 1~\gevc 
and respective contributions of the pixel and the barrel part]
{
Overall detector coverage for pions with a momentum of 
1~\gevc emitted from the nominal IP~(a)
and respective contributions of the pixel and the barrel part~(b). 
}
\label{pic-Coverage2DHisto-Example}
\end{center}
\end{figure}

Due to the high statistical sample for each of the simulation runs,  
a fine-meshed mapping of the detector coverage becomes possible. 
A corresponding 2D diagram for one selected configuration is shown in \figref{pic-Coverage2DHisto-Example}\,(a).  
In the chosen representation, the number of obtained MVD hit points per track  
is plotted colour-coded against its initial emission angles $\theta$ and $\phi$. 
Contributions of the MVD can be observed in a polar angle interval 
from 3$^{\circ}$ to 150$^{\circ}$, which is in accordance with the basic detector layout 
(cf. \figref{pic-MVD-GeneralLayout}).
The envisaged design goal of at least four hit points is achieved in a large range of the solid angle.
Regions with more than four hit points are related to the radial overlap of adjacent elements within one barrel layer 
and interleaves of different disk layers in the forward direction.
The two gaps around $\theta=90^\circ$ reflect the keep-outs needed for the target pipe. 
Larger areas with a reduced number of hit points around the target pipe gaps 
and in the transition region from the barrel to the disk part 
are related to safety margins required for the overall detector integration. 
They reflect the adjustment to the crossing target pipe in each of the barrel layers  
and the fitting of the inner pixel disks inside the MVD barrel part, respectively. 
Smaller structures with a reduced number of hit points result from passive sensor edges.

In \figref{pic-Coverage2DHisto-Example}\,(b) 
the overall hit distribution is split into the pixel and the strip part of the MVD.
Due to the rotational symmetry of the MVD halves in $\phi$, 
only the region between $\text{0}^\circ \leq \phi \leq \text{180}^\circ$ is shown for each of them.
In both cases the barrel region between $\theta$ = 70$^{\circ}$ and $\theta$ = 140$^{\circ}$ is covered very homogeneously.
Larger areas with a reduced number of hit points result from the adjustment to the crossing target pipe in the different barrel layers.
A significant difference between the two main detector parts occurs in the forward region ($\theta < \text{60}^{\circ}$).
The six double-sided pixel disks deliver a more complex pattern compared to the homogeneous coverage of the two strip disks. 
Due to the approximation of disk layers with rectangular-shaped pixel sensors,
the covered area in the $\theta$-$\phi$ diagram corresponds to a circular arc shaped.

\begin{figure}[]
\begin{center}
\includegraphics[width=\columnwidth]{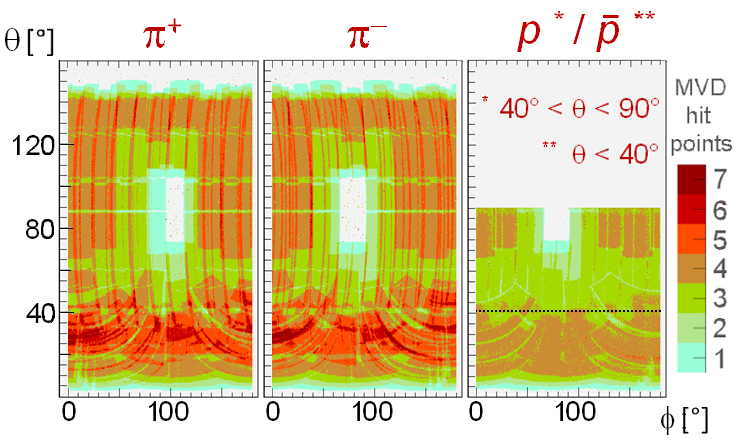}
\caption[Comparison of the detector coverage for low-momentum particles]
{
Comparison of the detector coverage for three different particle species 
with low momentum ($p$=150\,\mevc).
}
\label{pic-Coverage2DHisto-DifferentSetups}
\end{center}
\end{figure}

A comparison between different configurations is shown 
in \figref{pic-Coverage2DHisto-DifferentSetups}. 
The opposed bending radius of charge conjugated particles with low momenta 
is reflected by the different curvatures of the overlaps in the barrel part. 
A comparison between different configurations is shown 
in \figref{pic-Coverage2DHisto-DifferentSetups}. 
The opposed bending radius of charge conjugated particles with low momenta 
is reflected by the different curvatures of the overlaps in the barrel part. 
%Besides, there are no major distinctions proving the hermetic design of the MVD. 
In case of low momentum nucleons strong attenuation effects can be observed. 
They are caused by interactions with the active silicon material during the particle propagation. 
Considering that in a realistic scenario all pipes of the beam-target system  
and the full material budget for the MVD must be included,  
it can be deduced that most of the slow nucleons with momenta below 200~\mevc get stuck in one of the MVD layers.  
In particular, this fact becomes relevant for elastically scattered recoil protons.

Quantitative results on the relative detector coverage in the maximum polar angle range 
from 3$^\circ$ to 150$^\circ$ are summarised in \tabref{tab-CoverageSetups}. 
They are extracted from simulations with positively and negatively charged pions 
in a momentum range from 150~\mevc to 1.5~\gevc. 
The impact of opposed bending radii for the $\pi^+$ and the $\pi^-$ is very small, 
occurring deviations are dominated by systematic errors and statistical fluctuations of the simulation. 
Obtained values for a minimum of one and four hit points are in the order of 95\% and 60\% , respectively.
The missing 5\% in the overall acceptance is related to the target pipe gaps.

\begin{table*}[!t]
\begin{center}
\small
\begin{tabular}{|c|c|c|c|c|c|c|c|c|}
\hline
\multirow{2}{*}{Particle} 
& \multirow{2}{*}{Charge} 
& \multicolumn{7}{c|}{\parbox[0pt][9mm][b]{0cm}{}
Relative coverage within $3^{\circ}\,<\theta\,<150^{\circ}$ / [\%]}\\
\multirow{1}{*}{momentum} 
& \multirow{1}{*}{conjugation} 
& \multicolumn{7}{c|}{Number of MVD hit points}\\
&
  & \hspace{5mm} 1 
    & \hspace{5mm} 2 
      & \hspace{5mm} 3 
	& \hspace{5mm} 4 
	  & \hspace{5mm} 5 
	    & \hspace{5mm} 6 
	      & \hspace{5mm} Total\\
\hline
\vspace{-3mm}&&&&&&&&\\
\multirow{2}{*}{0.15~\gevc} 
& $+$
& 6.3\hspace{1mm} 
& 7.6\hspace{1mm} 
& 22.3\hspace{1mm} 
& 38.5\hspace{1mm} 
& 16.9\hspace{1mm} 
& 3.6\hspace{1mm} 
& 95.0\hspace{1mm} \\
& $-$
& \hspace{1mm} \sl 4.5 
& \hspace{1mm} \sl 9.0 
& \hspace{1mm} \sl 21.0 
& \hspace{1mm} \sl 38.4 
& \hspace{1mm} \sl 17.1
& \hspace{1mm} \sl 3.4 
& \hspace{1mm} \sl 93.2\\
\hline
\vspace{-3mm}&&&&&&&&\\
\multirow{2}{*}{0.5~\gevc}
& $+$
& 6.1\hspace{1mm} 
& 6.4\hspace{1mm} 
& 23.0\hspace{1mm} 
& 39.7\hspace{1mm} 
& 16.9\hspace{1mm} 
& 3.8\hspace{1mm} 
& 94.9\hspace{1mm}\\
& $-$
& \hspace{1mm} \sl 5.4 
& \hspace{1mm} \sl 7.1 
& \hspace{1mm} \sl 22.4 
& \hspace{1mm} \sl 40.9 
& \hspace{1mm} \sl 16.6 
& \hspace{1mm} \sl 3.6 
& \hspace{1mm} \sl 94.5\\
\hline
\vspace{-3mm}&&&&&&&&\\
\multirow{2}{*}{1.0~\gevc}
& $+$
& 6.0\hspace{1mm} 
& 6.4\hspace{1mm} 
& 22.7\hspace{1mm} 
& 40.0\hspace{1mm} 
& 17.1\hspace{1mm} 
& 3.7\hspace{1mm} 
& 95.2\hspace{1mm}\\
& $-$ 
& \hspace{1mm} \sl 5.6 
& \hspace{1mm} \sl 6.9 
& \hspace{1mm} \sl 22.4 
& \hspace{1mm} \sl 40.8 
& \hspace{1mm} \sl 16.6 
& \hspace{1mm} \sl 3.8 
& \hspace{1mm} \sl 95.0\\
\hline
\vspace{-3mm}&&&&&&&&\\
\multirow{2}{*}{1.5~\gevc}
& $+$
& 6.0\hspace{1mm} 
& 6.4\hspace{1mm} 
& 22.6\hspace{1mm} 
& 40.2\hspace{1mm} 
& 17.1\hspace{1mm} 
& 3.8 \hspace{1mm} 
& 95.2\hspace{1mm} \\
& $-$
& \hspace{1mm} \sl 5.7 
& \hspace{1mm} \sl 6.8 
& \hspace{1mm} \sl 22.4 
& \hspace{1mm} \sl 40.8 
& \hspace{1mm} \sl 16.6 
& \hspace{1mm} \sl 3.9
& \hspace{1mm} \sl 95.2
\parbox[0pt][5mm][t]{0cm}{}\\
\hline
\end{tabular}
\caption[Relative detector coverage in the active detector region 
obtained with pions of both charge conjugations]{Relative detector coverage in the active detector region 
obtained with pions of both charge conjugations.}
\label{tab-CoverageSetups}
\end{center}
\end{table*}

\begin{figure}[]
\begin{center}
\includegraphics[width=0.98\columnwidth]{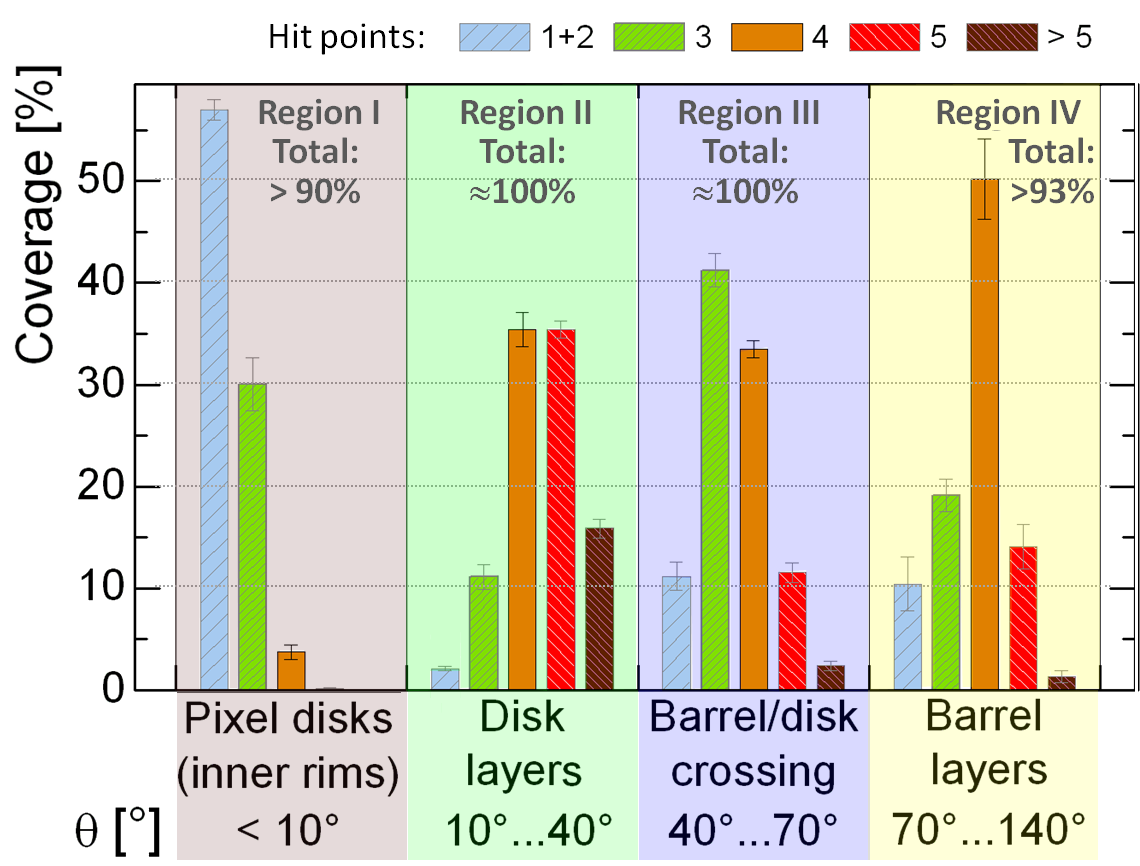}
\caption[Yield of different hit counts in four different detector regions]
{Yield of different hit counts in four detector regions.
Results obtained for charged pions with different momenta
(see \tabref{tab-CoverageSetups}). 
Error bars include deviations of the different setups
and statistical fluctuations.
}
\label{pic-CoverageResult-HitPoints}
\end{center}
\end{figure}

For a more precise evaluation in terms of physics performance,  
the total detector acceptance can be split into four main regions according to the basic design of the detector
(cf. \figref{pic-MVD-GeneralLayout}):

\begin{description}
 \item [Region I:]{Polar angles between 3$^{\circ}$ and 10$^{\circ}$ representing 
the solid angle covered by the inner rims of the pixel disks,}
\item[Region II:]{Polar angles between 10$^{\circ}$ and 40$^{\circ}$ representing 
the solid angle covered by the forward disk layers,}
  \item[Region III:]{Polar angles between 40$^{\circ}$ and 70$^{\circ}$ representing 
the solid angle covered by the strip barrel part, the outer pixel barrel layer 
and the innermost pixel disk delivering the fourth hit point,}
  \item[Region IV:]{Polar angles between 70$^{\circ}$ and 140$^{\circ}$ representing 
the area covered by all four barrel layers.}
\end{description}

The hit information in \textit{Region I} contributes to the forward tracking systems,
which basically use a straight line fit before and after the dipole.
Therefore, one or two hit points from the MVD is already sufficient. 
The design goal of at least four hit points has been applied to the three remaining regions. 
They are connected to the central tracking part for which a helix fit must be performed. 
The most important part for many physics channels is given by \textit{Region II},  
which also includes the kinematic region of charged decay modes of the ground state \D mesons. 
Compromises had to be done in \textit{Region III} in order to obtain a feasible solution for the detector assembly. 
However, the slightly reduced performance can be counter-balanced by the outer tracking system. 
The crossing target pipe affects the main barrel part defined by \textit{Region IV} 
and results in an acceptance cut.
A full coverage of 100\% is reached for the disk layers (\textit{Region II}), 
the crossover between the barrel and the forward part (\textit{Region III}) 
and down to a minimum polar angle of $\theta$ = 4.5$^{\circ}$ in the forward direction (\textit{Region I}). 

\Figref{pic-CoverageResult-HitPoints} illustrates the percentage coverage 
with different hit counts in the four detector regions. 
In \textit{Region I}, a hit count of one and two dominates (56.5\%).
Contributions from the inner rims of the four pixel disks start at 
3$^{\circ}$, 4.5$^{\circ}$, 6.5$^{\circ}$ and 8$^{\circ}$, respectively. 
A maximum number of 14 hit points as well as 
the highest averaged value are obtained in \textit{Region II}.  
In this area four and five hits contribute with one third each   
while the proportion of lower hit counts is only slightly above 10\%.
Within the covered solid angle of \textit{Region III} mostly three hit points (41\%) are obtained. 
However, at least four hit counts are still reached in roughly half of the area.
The coverage in \textit{Region IV} is clearly dominated by four hit points with a contribution of 50\%.
While the percentage of radial sensor overlaps, which deliver a larger hit count, is in the order of 15\%,   
the area around the target pipe with a reduced number of hit points covers roughly 29\% 
of the full acceptance in this region.

\begin{figure}[]
\begin{center}
\includegraphics[width=0.98\columnwidth]{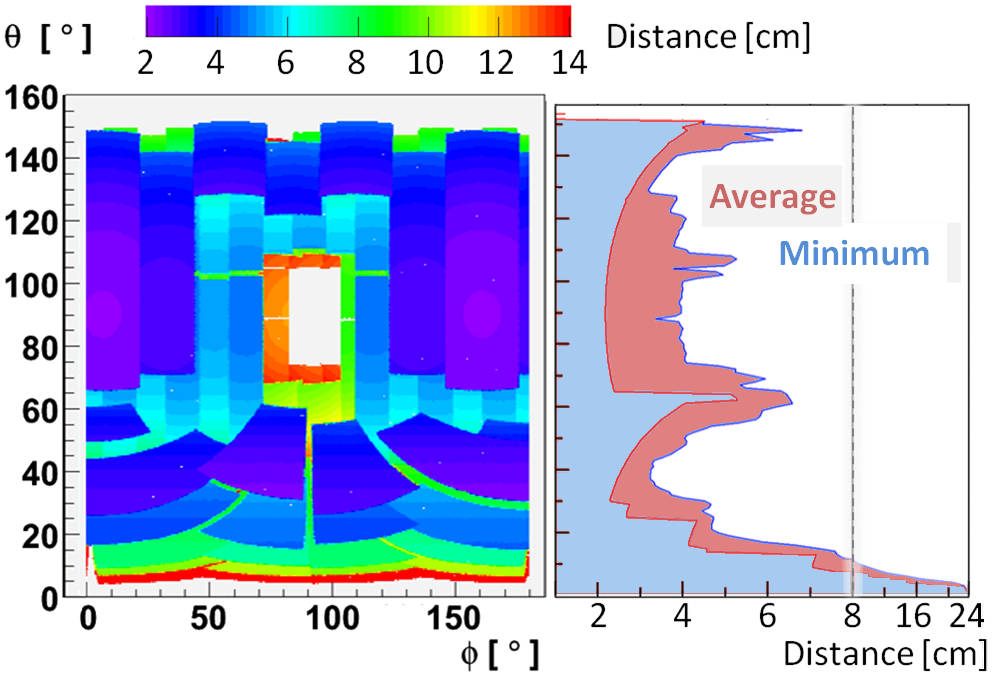}
\caption[Distances of the first MVD hit point to the nominal IP]
{
Distances of the first hit point to the origin (0,0,0)
obtained for slow pions ($p\,=\,$150~\mevc). 
}
\label{pic-CoverageResult2D-Distance}
\end{center}
\end{figure}

Besides a sufficient number of MVD hits for each particle track,
a minimum distance of the first hit point with respect to the nominal IP 
is crucial for the physics performance of the detector. 
A 2D illustration of the absolute distance of the first hit point to the origin 
is given in \figref{pic-CoverageResult2D-Distance}, on the left.
Next to it, on the right, corresponding 1D profiles along the the polar angle are added.
They show the minimum and the averaged value obtained within the full azimuthal angle.
Results indicate that a distance of less than 4~cm can be reached
within a wide range of the covered solid angle. 
Minimum distances to the nominal interaction point are obtained in the barrel part and at polar angles around 30$^\circ$
are in the order of 2~cm.
\vfill

  \subsection{Material Budget}
\label{Sim-RadLenght}
\authalert{Author: Thomas W\"{u}rschig, Contact: t.wuerschig$\text @$hiskp.uni-bonn.de} 

\begin{figure*}[htpb]
\begin{center}
\includegraphics[width=15cm]{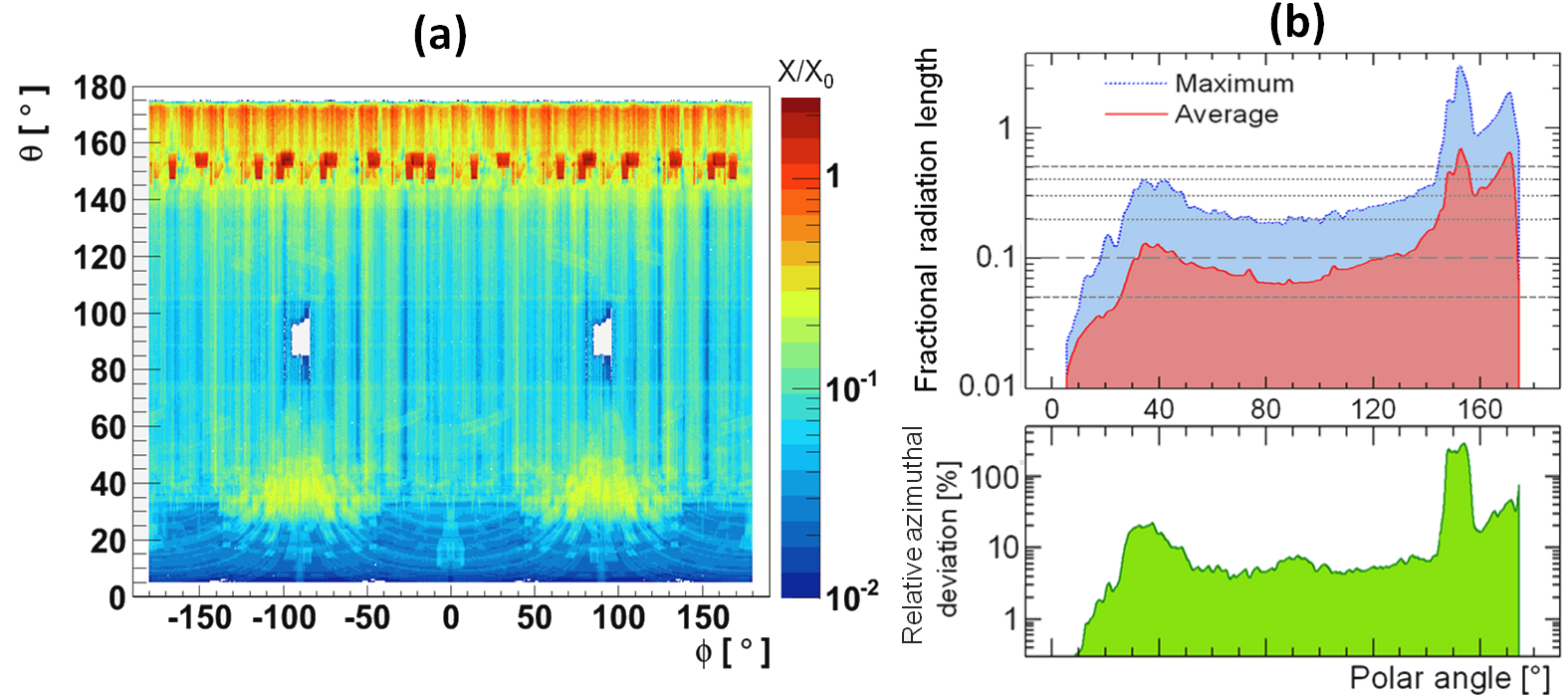}
\caption[2D material map and extracted 1D profile for the full MVD]
{
2D map of the material budget (a) and extracted 1D profiles along the polar angle (b).
}
\label{pic-TotalRadLengthMVD}
\end{center}
\end{figure*}

The introduction of the MVD is inevitably connected to some detrimental effects  
given by scattering and energy straggling of charged particles   
as well as the conversion and attenuation of photons inside the detector material. 
The impact of the introduced material can be quantitatively described by   
a resulting fractional radiation length, $X/X_0$, which is accumulated in all traversed volumes $j$:

\begin{equation}
\label{eq-FractRadLength}
X/X_0 = \sum_{j}{\rho_j\cdot L_j\over  X_{0_j}}
\end{equation}

where $X_{0_j}$ and $\rho_j$ are the specific radiation length 
and the density of the material defined for the volume $j$, respectively,
and $L_j$ corresponds to the traversed path length therein. 
In the further discussion the corresponding value of the material budget 
will be given in a percentage of one full radiation length, \%$X_0$, if $X/X_0<1$ 
or in multiples of $X_0$ if $X/X_0\geq1$. 

The following studies about the material budget are based on the comprehensive MVD model, 
which has been introduced in chapter~\ref{MvdCadModel}.  
%It represents a very detailed detector description and thus facilitates the approach of 
%a realistic estimate on the expected material distribution.  
For the extraction of a material map,  
virtual particles (\textquotedblleft geantinos\textquotedblright) were propagated through the detector. 
The fictitious \textquotedblleft geantino\textquotedblright particle undergoes no physical interactions  
but flags boundary crossings along its straight trajectory.  
Two million events were simulated starting from the origin ($x$;$y$;$z$)~=~]{(0;0;0)}
with an isotropic emission over the polar angle, $\theta$, and the azimuthal angle $\phi$.  
Further on, the effective path lengths in all traversed volumes ($L_j$) 
were identified for each propagated geantino. 
The fractional radiation length ($X/X_0$) was then calculated based on 
the given material definition of the volumes according to equation~\eqref{eq-FractRadLength}.

In \figref{pic-TotalRadLengthMVD}\,(a) the obtained material map of the MVD  
is given in the representation of the fractional radiation length,  
which is plotted colour-coded against both initial emission angles.   
Corresponding profiles along the polar angle are shown in \figref{pic-TotalRadLengthMVD}\,(b).  
They are obtained for a bin size of $\Delta \theta$ = 1$^{\circ}$. 
Resulting numbers for the average and the maximum fractional radiation length  
within each interval are shown at the top. 
The standard deviation of all values normalised to the corresponding mean value is plotted below. 
This ratio is a measure of the isotropy in the material distribution 
along the integrated azimuthal angle.
The obtained material distribution indicates a very low material budget of the MVD 
in the physically most important region  
at polar angles smaller than 140$^{\circ}$. 
Corresponding values stay mainly well below 10\%$X_0$  
and the overall material map shows a rather isotropic occupancy
with relative fluctuations of not more than 10\%.  
Larger deviations are located between $\theta=30^{\circ}$  
and $\theta=50^{\circ}$ 
as well as in the very backward region~($\theta<150^{\circ}$),
which generally exhibits an increased material budget 
with peak values of several radiation lengths.
Further details of the material map will be discussed in the following paragraphs.

\begin{figure}[]
\begin{center}
\includegraphics[width=\columnwidth]{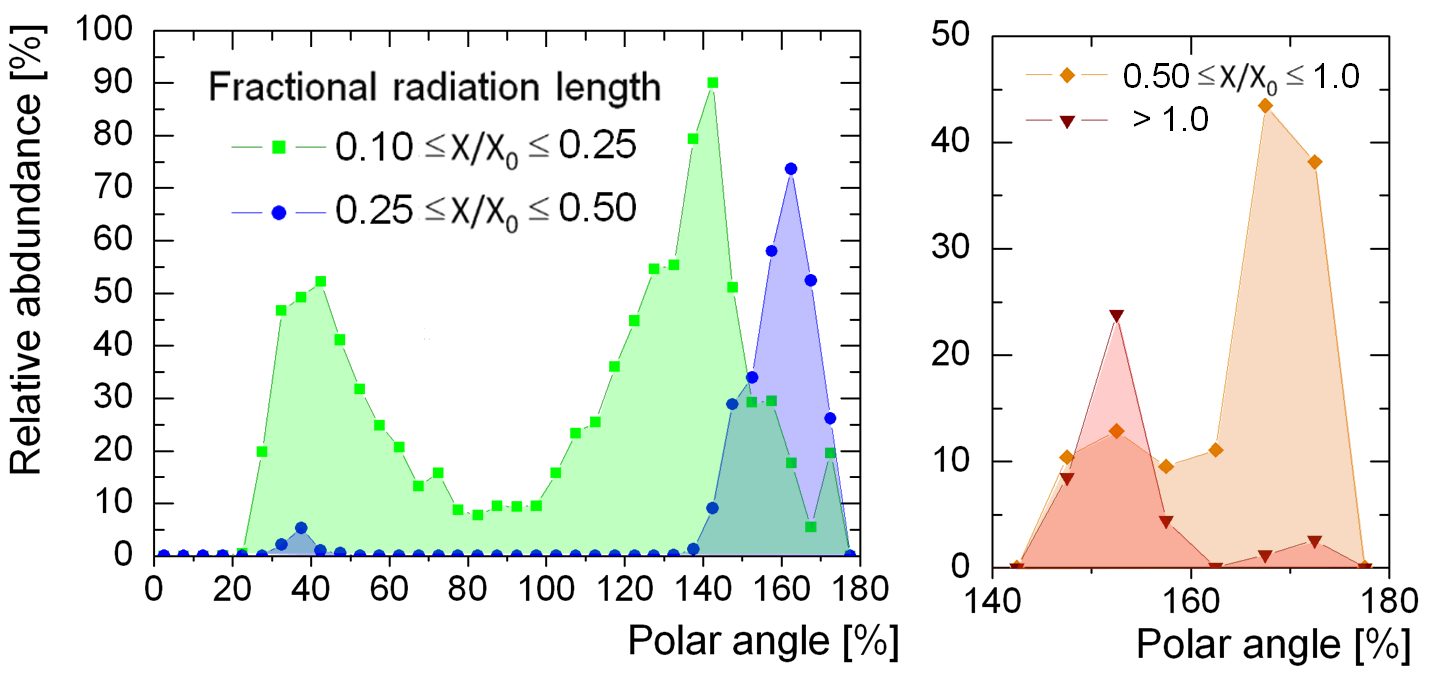}
\caption[Determination of hot spots in the material map]
{
Relative abundance, $N_{\text{spot}}$/$N_{\text{all}}$, of hot spots in the material map
along the polar angle. Specified threshold values are given inside the diagrams. 
}
\label{pic-Hotspots}
\end{center}
\end{figure}

First of all, the focus is given to local hot spots, 
i.e.~regions with an enhanced fractional radiation length.
\Figref{pic-Hotspots} illustrates the azimuthally integrated distribution of such hot spots along the polar angle.
The given relative abundance is defined as the fraction of counted tracks, $N_{\text{all}}$, 
for which the obtained value of $X/X_0$ stays in the specified range,
divided by the overall number of tracks, $N_{\text{all}}$, 
to be expected within an interval of $\Delta\theta = \text{5}^{\circ}$.  
The latter, $N_{\text{all}}$, is given by the percentage 
of the covered area in the $\phi-\theta$ diagram 
multiplied with the total number of simulated events. 

Different intervals for the fractional radiation length were chosen 
to classify hot spots with respect to the average value obtained 
in the sensitive detector region ($\theta<\text{140}^{\circ}$).
A lower limit for the definition of a hot spot is given by

\begin{equation}
\label{eq-HotSpotDefinition} 
\langle X/X_0\rangle+2\cdot\langle \sigma_{\theta_{i}}\rangle = 9.1\%
\end{equation}

where $\langle X/X_0\rangle$ and $\langle \sigma_{\theta_{i}}\rangle$
represent the mean values of the fractional radiation length and 
the standard deviation within the $\Delta\theta$ intervals, respectively.
Both of them are extracted from the 1D profiles shown \figref{pic-TotalRadLengthMVD}.
Finally, four different categories have been specified. 
The first category (I) defined in an interval of $10\% < X/X_0 < 25\%$ 
rather describes larger fluctuations to the averaged value. 
Moderate hot spots of category II are in the order of $25\%X_0$ and $50\%X_0$.
The last two categories III and IV account for hot spots with a very large value 
below and above one full radiation length $X_0$, respectively. 

Compiled results in \figref{pic-Hotspots} indicate that hot spots of category II to IV
are basically %shifted 
outside the sensitive detector region of the MVD.
An exceptional contribution of moderate hot spots (category II) in the order of 5\% can be 
located between $\theta=35^{\circ}$ and $\theta=40^{\circ}$.
%It is correlated to the lead out of the pixel services to outer radii.  
All other hot spots of the second, third and fourth category are located 
in the upstream region between 
$\theta=140^{\circ}$ and $\theta=170^{\circ}$.
Fluctuations of the material map included in category I are mainly spread
over the sensitive detector region.
The maximum of this distribution shows up at $\theta = \text{145}^{\circ}$. 

\begin{figure*}[!htpb]
\begin{center}
\includegraphics[width=\textwidth]{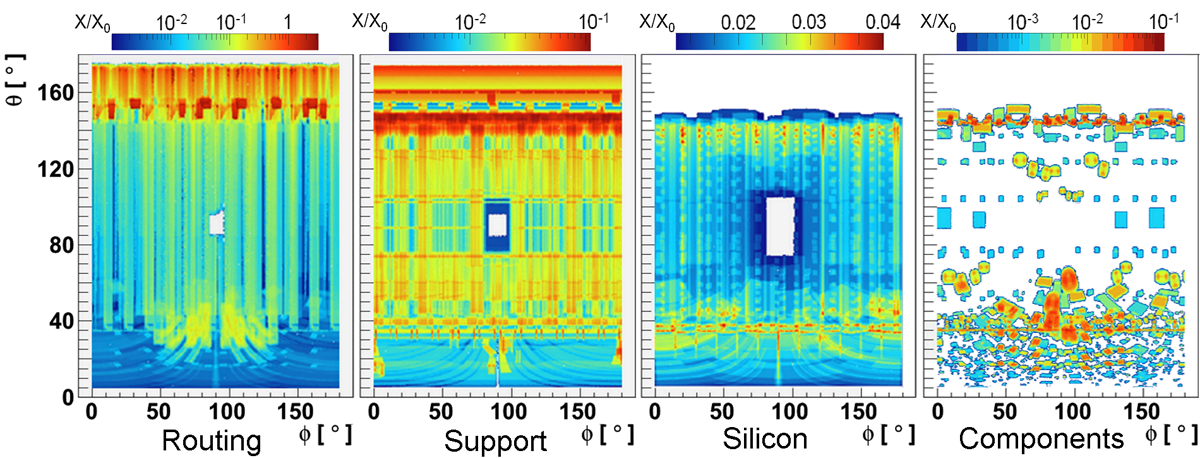}
\caption
[Maps of the contributing material budget in terms of the fractional radiation length ($X/X_0$)]
{Maps of the contributing material budget in terms of the fractional radiation length ($X/X_0$) 
for the four main parts of the model as illustrated in \figref{pic-Mvd-2.1model}. 
Note the different colour scales for $X/X_0$.
}
\label{pic-RadLength-MainComp}
\end{center}
\end{figure*}

In order to obtain a better understanding of the resulting pattern in the material map
it is useful to perform a separate analysis of different detector parts.   
Therefore, only the individual components of interest were switched on during the simulation.  
In a first step, the four main substructures of the MVD model have been chosen. 
These main parts are (cf. \figref{pic-Mvd-2.1model}):  

\begin{description}
 \item [Silicon:]~\\
  Sensor elements and associated readout chips
 \item [Support:]~\\
  Global and local support structures  
 \item [Routing:]~\\
  Service structures for sensors and \frontend chips 
  including supply and data cables as well as cooling pipes
 \item [Components:]~\\
  Electronic and cooling connectors as well as additional SMD components
\end{description}

The individual contributions of these substructures are disentangled 
in \figref{pic-RadLength-MainComp}.
In comparison with \figref{pic-TotalRadLengthMVD} it is evident that 
the most significant impact on the material map originates from the routing part.
It delivers the obvious jump to a larger fractional radiation length 
at polar angles above 140$^{\circ}$ 
and also includes hot spots with maximum values of several $X_0$.
These enhancements are related to the fixed routing scheme of the inner barrel layers 
along the opening beam pipe in upstream direction 
in combination with the generally increased incident angle of tracks traversing 
the circularly arranged service structures. 
In the forward region (25$^{\circ}<\theta<$50$^{\circ}$), 
the routing of the pixel disk services at the top and bottom is clearly visible.
It results in a partial increase above $10\%X_0$ (cf. \figref{pic-Hotspots}) 
thus defining the maximum material budget in the sensitive detector region.
Besides, average values therein level around 5\%$X_0$.

Compared to the routing part, maximum values of all other sub-structures in the 2D distribution 
fall short by more than one order of magnitude.
Support structures contribute with 1\%$X_0$ to 3\%$X_0$ in the active detector region. 
Moreover, higher values reflect the load pick-up of the main building blocks,
which is concentrated in upstream direction at polar angles above 140$^{\circ}$.  
A maximum of approximately 5\%$X_0$ and 10\%$X_0$ is given below (forward part) 
and above (barrel part), respectively.
The most homogeneous material occupancy around 2\%$X_0$ is given by the silicon part. 
It shows a typical $\cos\theta$-modulation reflecting the incident angle of straight lines originating 
from the nominal interaction point. 
A further increase at smaller polar angles is related to the enlarged number 
of individual detector layers in forward direction.
Finally, the resulting pattern of the introduced components defines a very anisotropic distribution. 
While the averaged number remains relatively small, 
maximum values of around 5\%$X_0$ are reached at singular points. 
Most of them are related to cooling connectors 
needed to change from steel pipes to flexible tubes.

\begin{figure}[htb]
\begin{center}
\includegraphics[width=0.93\columnwidth]{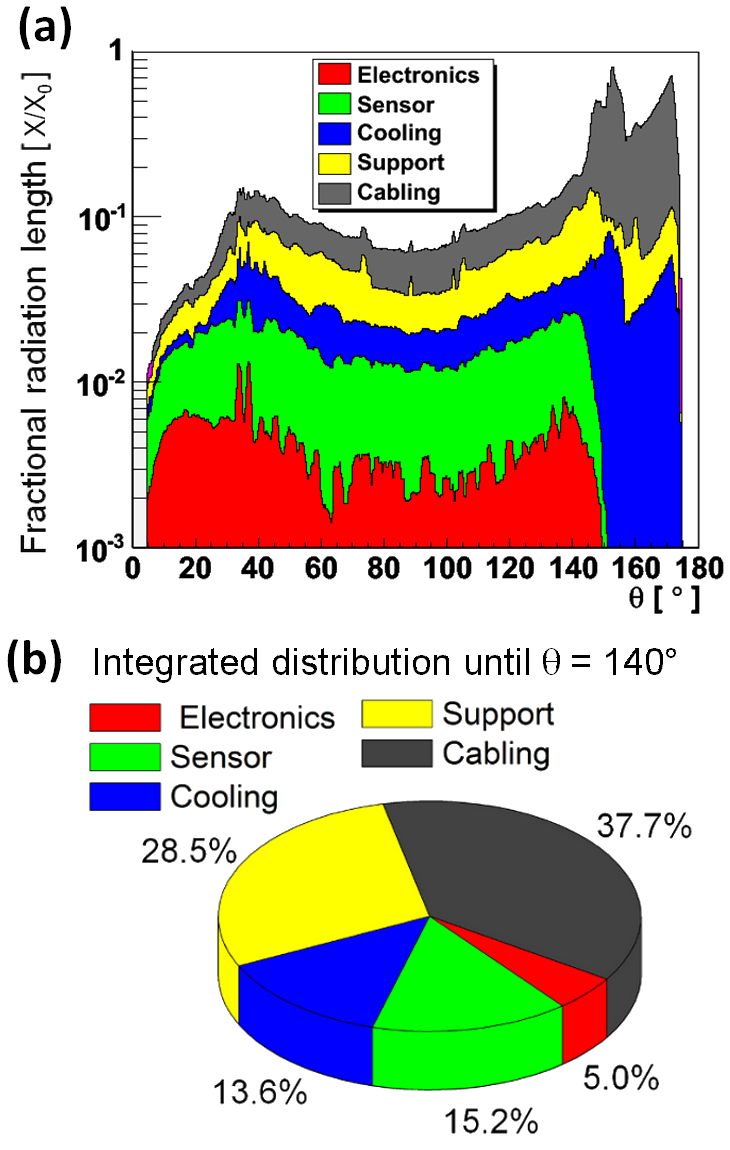}
\caption[Contributions of different functional detector parts to the overall material budget]
{
Contributions of different functional parts of the MVD to the overall material budget. 
Stacked 1D profile along the polar angle (a) and
percentage of the different parts within the sensitive detector region ($\theta < \text(140)^{\circ}$) (b). 
}
\label{pic-RadLength-ReOrderComp-1DProfile}
\end{center}
\end{figure}

For the evaluation of different hardware developments in terms 
of their impact on the overall material budget 
a slightly modified representation is used,
which re-orders the different volumes of the MVD model to functional groups:

\begin{description}
 \item [Sensors:]~\\
  Silicon pixel and strip detectors	
 \item [Electronics:]~\\
  Readout chips, SMD components and electronic connectors  
 \item [Cabling:]~\\
  Routed supply and data cables (core and insulation)
 \item [Cooling:]~\\
  Coolant, pipes and tubes as well as cooling connectors
 \item [Support:]~\\
  All support structures (as defined before)
\end{description}

The stacked diagram in \figref{pic-RadLength-ReOrderComp-1DProfile}\,(a) 
illustrates the individual contributions for each of 
these functional detector parts along the polar angle.
Plotted values represent the azimuthally averaged fractional radiation lengths $X/X_0$ 
within an interval of $\Delta\theta=\text{1}^{\circ}$. 
Besides some accumulation effects in the very forward region,
the proportions between all parts remain relatively constant 
within the sensitive detector region ($\theta\leq$140$^{\circ}$). 
In \figref{pic-RadLength-ReOrderComp-1DProfile}\,(b) 
their percentage on the overall budget is presented in a pie chart.
The biggest amount of material is delivered 
by the cables (38\%) and the support structures (29\%).
The dominating part (70\%) of 
the material budget of the cabling part splits into 78\% and 22\% 
for the supply and the data cables, respectively.
The introduced silicon sensors assuming standard total thicknesses of 
200~$\tcmu$m and 300~$\tcmu$m for the pixel and the strip detectors, respectively,
account for only 15\%.
The cooling system adds another 14\% to the material map 
and the impact of the electronics stays in total rather small (5\%).
However, it should be recalled that the discussion on integrated distributions 
does not supersede a careful evaluation of the 2D map as done before. 

In summary, the obtained composition of the material budget is typical for 
inner silicon tracking systems of hadron physics or high-energy physics experiments.
In case of \pnd, complications for a minimised material budget are given by
the overall routing scheme, which requires the upstream routing of all services 
thus implying e.g.~folded service lines for the pixel disks,
and the DAQ concept, which is based on a triggerless readout thus increasing 
the material due to the higher functionality of the readout chip and the larger amount of data to be handled.
%An exceptional contribution of moderate hot spots (category II) in the order of 5\% can be 
%located between \unit[$\theta=$]{35$^{\circ}$} and \unit[$\theta=$]{40$^{\circ}$}.
%It is correlated to the lead out of the pixel services to outer radii.  
%All other hot spots of the second, third and fourth careadout thus increasing 
%%the material due to the higher functionality of the readout chip 
%and the larger amount of data to be handled.
Even though the total amount of material for all service structures 
can be kept in an acceptable range.
This fact results from an optimised powering scheme for the readout electronics,
which decreases the required core cross section of the supply cables,
and the use of ultra-thin cables based on an Aluminum core.
Both reduce the overall material budget by roughly 45\%.
Moreover, a significant minimisation is given by the use of lightweight carbon support structures
with a tailored design, i.e.~larger cut-outs and sandwiched structures with Rohacell.
Further improvements, which are not included in the presented studies,
are related to thinned silicon detectors, the use of aluminum supply cables
surrounded by just a thin copper layer and the implementation of a higher data concentration.
The maximum achievable reduction in best case is expected to be in the order of 25\%.
A detailed discussion of these point can be found in~\cite{PhD-Tesis_Wuerschig}.

  \subsection{Rate Estimation}

\authalert{Author: Thomas W\"{u}rschig, Contact: t.wuerschig$\text @$hiskp.uni-bonn.de}  

%The spatial distribution and the corresponding kinetic energy of all particles 
%emitted from the primary interaction vertex define  
%the maximum count rates 
%and the radiation load to be expected for different detector parts, respectively. 
%\Figref{pic-ParticleDistr_DPM_UrQMD} illustrates the expected particle distributions 
%in case of antiproton-nulceon and antiproton-proton relations.

\begin{figure*}[]
\begin{center}
\includegraphics[width=15cm]{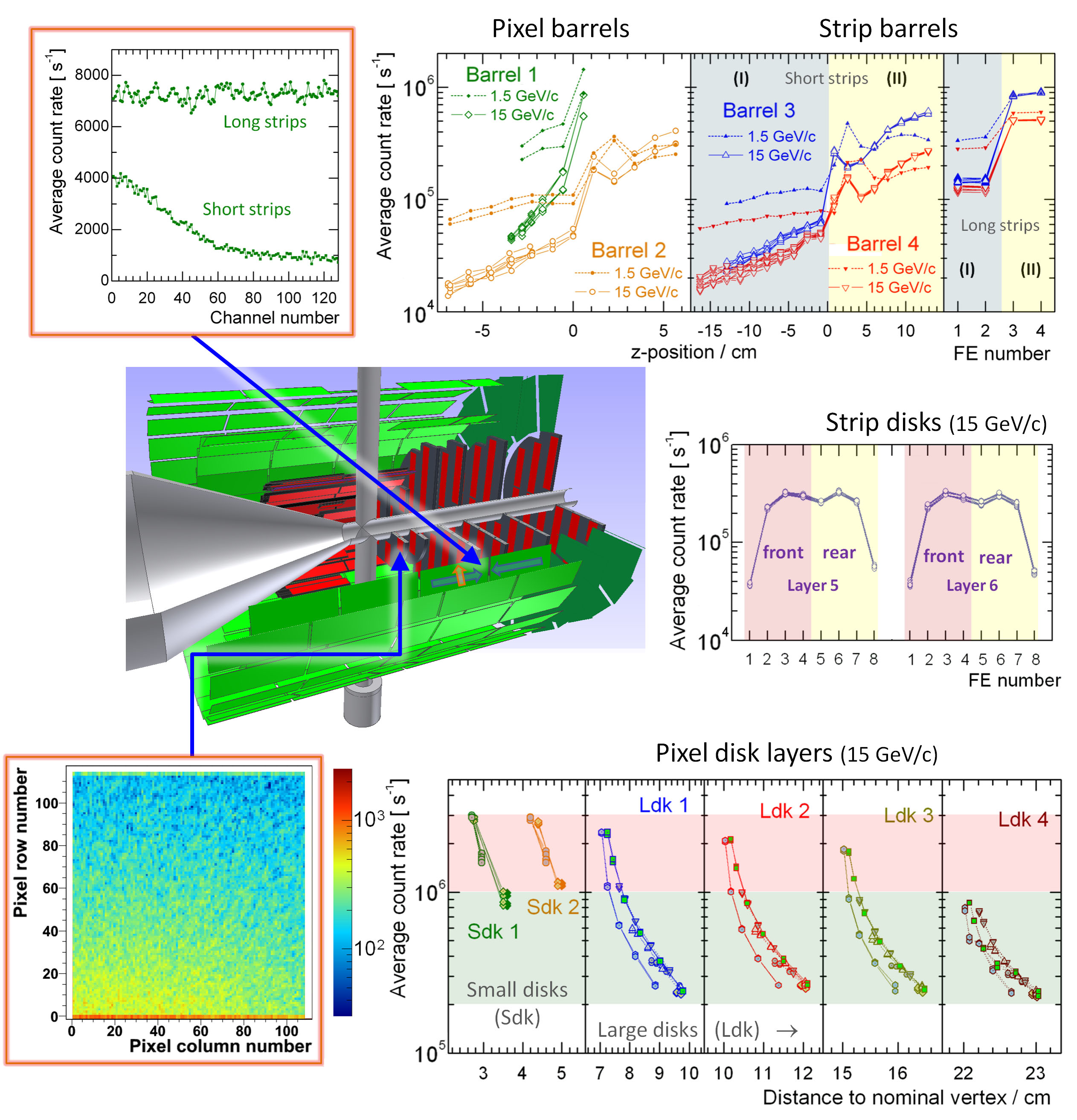}
\caption[Extracted average count rates for antiproton-proton reactions obtained in the different detector layers]
{
Extracted average count rates for antiproton-proton reactions
obtained with the DPM generator. 
The distribution over all individual channels
of the strip and the pixel \frontend with the highest occupancy 
are shown in the frames on the left at top and bottom, respectively. 
Arrows indicate the position inside the MVD. 
Results for all readout chips in the different detector layers are shown on the right.
(All results are based on an assumed interaction rate of $2\cdot 10^7$s$^{-1}$).
}
\label{pic-Rates-DPM-summary}
\end{center}
\end{figure*}

For the MVD as innermost subdetector of \PANDA, 
an evaluation of the expected count rates in different detector regions 
is from utter importance for the development of all readout electronics. 
Extracted numbers result from the initial conditions of the experiment, 
i.e.~the particle distribution in the reaction channel 
(cf. \figref{pic-ParticleDistr_UrQMD}, chapter~\ref{Intro-Specifics})
and the beam-target interaction rate taken as basis for the calculations.
In a first approximation, the latter can be correlated with 
the nominal \PANDA interaction rate, which is given by 2$\cdot 10^7$~s$^{-1}$.
With the time-averaged information it is possible to extract
the hit occupancy in different detector regions
and the overall data load expected to be handled by the MVD.
For a reasonable estimate on peak rates occurring on single readout channels and on \\frontend level,
variations of the beam-target interaction on a shorter time scale must be considered. 

A detailed count rate study for $\bar p p$ reactions has been performed 
at minimum (1.5~\gevc), maximum (15~\gevc) 
and an intermediate beam momentum of 5~\gevc~\cite{PhD-Tesis_Wuerschig}. 
A large statistical sample of two million DPM events 
were simulated for each of the three setups. 
Generated particles were propagated through the detector within the \pndrt framework.
Moreover, the created hit points in the detector were digitised as described in chapter~\ref{Sim-sds_digi}. 
In the further analysis only binary hit information, 
i.e.~the IDs of sensor and \frontend as well as the number of the readout channel, 
was stored thus allowing an unambiguous allocation of each hit point in the detector.
The summed hit counts were then multiplied with the nominal \PANDA interaction rate
in order to obtain the time-averaged count rate.
The given number for an individual readout chip results from 
an integration of all associated readout channels.
In the same way, average count rates for individual sensors 
are defined by the contribution of all connected readout chips. 
The hit occupancy for different detector parts is then given by 
the accumulated count rates of all sensor elements therein.

Obtained results for the count rate distribution on \frontend level 
and the individual contributions of single readout channels
are summarised in \figref{pic-Rates-DPM-summary}.
Due to the distinct emission of particles in forward direction
with increasing beam momentum,
the highest count rates inside the MVD occur 
at maximum beam momentum in the inner pixel disk layers.
Upper limits of three million counts per second (Mcps) 
are obtained for individual pixel readout chips
located at the inner sectors of the small pixel disks.
Also at the inner rims of the subsequent pixel disks, 
average count rates stay at a high level 
between 1.9~Mcps and 2.5~Mcps. 
At outer radii values drop quickly by one order of magnitude.
For the two innermost pixel disks the situation is more challenging  
because all the \frontends have to cope with a high count rate of more than 1~Mcps. 
These special conditions require a sophisticated cooling concept for the readout electronics
(cf. chapter~\ref{Cooling-system-Pixel}), which is crucial for a stable operation.
The highest count rates for readout chips in the strip disks are in the order of 0.3~Mcps. 
The shorter strips at one sensor edge of the trapezoidal sensors
lead to a lower count rate for the \frontend at the outermost position.
This effect is mirrored at the opposite sensor side. 

In contrast to the forward part, the occupancy in the barrel part 
increases with lower beam momentum due to the increased number of particles 
emitted at larger polar angles. 
%Therefore, 
The distributions at highest and lowest beam momentum 
are plotted in \figref{pic-Rates-DPM-summary}.
The impact of the slow elastic recoil protons results in 
a local maximum for \frontends located just behind $z=0$,
which corresponds to the observed peak in the overall particle distribution
shown in~\ref{pic-ParticleDistr_UrQMD}.
Due to the orientation along the beam axis, 
the average count rates of long strips stay for one sensor 
always above the ones of short strips.
Upper limits for the strip part are obtained 
in the third barrel layer at minimum beam momentum.
Assuming an interconnection of the two foremost sensors,
which is one of the possible solutions envisaged 
to reduce the total number of readout channels,
maximum values for the strip \frontends stay slightly below 1~Mcps. 
This value is slightly exceeded for the foremost readout chips of the first pixel layer,
thus defining the maximum average count rates occurring in the overall barrel part. 
In the upstream part ($z<0$) obtained values do not exceed a level of 0.1~Mcps.

Pixel and strip readout chips with the maximum occupancy 
also contain readout channels with the highest individual count rates. 
In case of the strip part, the distribution over all connected readout channels 
ranges between 7000~counts per second (cps) and 8000~cps.
For the pixel part a more anisotropic occupation on the pixel readout chip is obtained. 
Values for single pixel cells differ by roughly one order of magnitude
with an upper limit of roughly 1200~cps. 
The increased count rate at the top and the bottom of the pixel matrix
stems from an enlarged pixel size, 
which is implemented to counter-balance the passive edge of the readout chip.

\begin{table*}
\begin{center}
\small
\begin{tabular}{|c|c|c|c|c|c|}
\hline
\vspace{-3mm}&&&&&\\ 
  & Barrel
    & Forward 
      & Sensor 
	& \Frontend
	  & Readout  \\
  & part
    & part
      & level
	& level
	  & channel
\parbox[0pt][6mm][t]{0cm}{}\\
\hline
\hline
\multicolumn{6}{|c|}{
\parbox[0pt][8mm][c]{0cm}{}\textbf{Pixel part}}\\
\hline  
\hline
\vspace{-3mm}&&&&&\\
Digitised hits 
  & 6.7 $\mapsto$\,4.2
    & 16.6\,$\mapsto$\,32.2
      &	$\leq\,$0.89 
	& $\leq\,$0.29 
	  & $2\cdot 10^{-4}$\\
$N_{\textsf{\tiny{dig}}}$ / [10$^6$]  
  &  \footnotesize (1.5 $\mapsto$ 15)~\gevc
    &  \footnotesize (1.5 $\mapsto$ 15)~\gevc
      & $\langle\,\leq\,0.21\,\rangle$
	&  $\langle\,$\footnotesize$\lesssim$\small\,0.04\,$\rangle$
	  &  $\langle\,3.5\cdot 10^{-6}\rangle$ 
\parbox[0pt][6mm][t]{0cm}{}\\
\hline
\vspace{-3mm}& \multicolumn{2}{c|}{\parbox[0pt][2mm][b]{0cm}{}}  &&&\\
Average count rate 
  & \multicolumn{2}{c|}{233\,$\mapsto$\,364}
    & \footnotesize$\leq$\small\,8.9
      &	\footnotesize$\leq$\small\,2.9
	& \multirow{2}{*}{\footnotesize$\lesssim$\small\,0.002} \\
$\langle\,\dot{N}_{\textsf{\tiny{dig}}}\rangle$ / [Mcps]
  & \multicolumn{2}{c|}{\footnotesize (1.5 $\mapsto$ 15)~\gevc}
    & $\langle\,$\footnotesize$\leq$\small\,2.1$\,\rangle$
      & $\langle\,0.4\,\rangle$
	& 
\parbox[0pt][6mm][t]{0cm}{}\\
\hline
\vspace{-3mm}& \multicolumn{2}{c|}{\parbox[0pt][2mm][b]{0cm}{}}  &&&\\
Expected data rate
  & \multicolumn{2}{c|}{\multirow{2}{*}{$\lesssim\,2200$}}
    & 45
      & 17
	& \multirow{2}{*}{$-$}\\
$\langle\dot{N}_{\textsf{\tiny{dig}}}\rangle$\hspace{-0.1mm}$\cdot$\hspace{-0.1mm}$f_{\textsf{\tiny{DAQ}}}$ / [MB/s]    
   &  \multicolumn{2}{c|}{\parbox[0pt][2mm][b]{0cm}{}}
     & $\langle\,12\,\rangle$
	& $\langle\,2\,\rangle$
	  & 
\parbox[0pt][6mm][t]{0cm}{}\\
\hline
\vspace{-3mm}& \multicolumn{2}{c|}{\parbox[0pt][2mm][b]{0cm}{}}  &&&\\
Estimated peak rate
  & \multicolumn{2}{c|}{\multirow{2}{*}{$-$}}
    & \multirow{2}{*}{$-$}
      &	$>$\,4.0
	& \multirow{2}{*}{\footnotesize$\lesssim$\small\,0.01}\\
$\langle\dot{N}_{\textsf{\tiny{dig}}}\rangle$\hspace{-0.1mm}$\cdot$\hspace{-0.1mm}$f_{\textsf{\tiny{peak}}}$ / [Mcps]
  & \multicolumn{2}{c|}{\parbox[0pt][2mm][b]{0cm}{}}
    &
      & $<$\,14.5
	&
\parbox[0pt][6mm][t]{0cm}{}\\
\hline
\hline
%STRIP
%%%%%%%%%%%%%%%%%%5
%%%%%%%%%%%%%%%%%%%%%%%%%%%%%%%%%
%%%%%%%%%%%%%%%%%%%%%%%%%%%%%%%%%%%%%%%
\multicolumn{6}{|c|}{
\parbox[0pt][8mm][c]{0cm}{}\textbf{Strip part}}\\
\hline  
\hline
\vspace{-3mm}&&&&&\\
Digitised hits 
  & 21.1\,$\mapsto$\,17.6
    & 5.0\,$\mapsto$\,8.4
      &	0.30 
	& 0.09 
	  & $8\cdot 10^{-4}$\\
$N_{\textsf{\tiny{dig}}}$ / [10$^6$]  
  &   \footnotesize (1.5 $\mapsto$ 15)~\gevc
    &    \footnotesize (1.5 $\mapsto$ 15)~\gevc
      & $\langle\,0.10\,\rangle$
	&  $\langle\,0.02\,\rangle$
	  &  $\langle\,1.5\cdot 10^{-4}\rangle$ 
\parbox[0pt][6mm][t]{0cm}{}\\
\hline
\vspace{-3mm}& \multicolumn{2}{c|}{\parbox[0pt][2mm][b]{0cm}{}}  &&&\\
Average count rate 
  & \multicolumn{2}{c|}{418\,$\mapsto$\,416}
    & 4.8
      &	1.5
	& \multirow{2}{*}{\footnotesize$\lesssim$\small\,0.013} \\
$\langle\dot{N}_{\textsf{\tiny{dig}}}\rangle$\hspace{-0.1mm}$\cdot$\hspace{-0.1mm}$f_{\textsf{\tiny{r/o}}}$ / [Mcps]
  & \multicolumn{2}{c|}{  \footnotesize (1.5 $\mapsto$ 15)~\gevc}
    & $\langle\,$\footnotesize$\leq$\small$\,1.6\,\rangle$
      & $\langle\,$\footnotesize$\leq$\small$\,0.3\,\rangle$
	& 
\parbox[0pt][6mm][t]{0cm}{}\\
\hline
\vspace{-3mm}& \multicolumn{2}{c|}{\parbox[0pt][2mm][b]{0cm}{}}  &&&\\
Expected data rate
  & \multicolumn{2}{c|}{\multirow{2}{*}{$\lesssim\,2500$}}
    & 29
      & 9
	& \multirow{2}{*}{$-$} \\
$\langle\dot{N}_{\textsf{\tiny{dig}}}\rangle$\hspace{-0.1mm}$\cdot$\hspace{-0.1mm}$f_{\textsf{\tiny{r/o}}}$
\hspace{-0.6mm}$\cdot$\hspace{-0.1mm}$f_{\textsf{\tiny{DAQ}}}$ / [MB/s]    
   &  \multicolumn{2}{c|}{\parbox[0pt][2mm][b]{0cm}{}}
     & $\langle\,10\,\rangle$
	& $\langle\,2\,\rangle$
	  &
\parbox[0pt][6mm][t]{0cm}{}\\
\hline
\vspace{-3mm}& \multicolumn{2}{c|}{\parbox[0pt][2mm][b]{0cm}{}}  &&&\\
Estimated peak rate
  & \multicolumn{2}{c|}{\multirow{2}{*}{$-$}}
    & \multirow{2}{*}{$-$}
      &	$>$\,2.0
	& $>$\,0.02\\
$\langle\dot{N}_{\textsf{\tiny{dig}}}\rangle$\hspace{-0.1mm}$\cdot$\hspace{-0.1mm}$f_{\textsf{\tiny{r/o}}}$
\hspace{-0.6mm}$\cdot$\hspace{-0.1mm}$f_{\textsf{\tiny{peak}}}$  / [Mcps]
  & \multicolumn{2}{c|}{\parbox[0pt][2mm][b]{0cm}{}}
    &
      & $<$\,5.5
	& $<$\,0.07\hspace{-1mm}
\parbox[0pt][6mm][t]{0cm}{}\\
\hline
\end{tabular}
\caption[Main results of the count rate study for antiproton-proton reactions]
{
Main results of the count rate study performed with 2 million DPM events.
Average count rates are obtained with 
the nominal interaction rate of 2$\cdot 10^7$~s$^{-1}$, 
i.e.~$\langle\dot{N}_{\textsf{\tiny{dig}}}\rangle=10\cdot{N}_{\textsf{\tiny{dig}}}\cdot$~s$^{-1}$.
Given numbers at sensor, \\frontend and channel level 
represent the maximum values obtained for a single element.
Mean values of all elements in the corresponding setup are indicated with $\langle...\rangle$.
}
\label{tab-CountRateStudies}
\end{center}
\end{table*}

The most important results of the presented count rate study 
are compiled in \tabref{tab-CountRateStudies}. 
For the integrated pixel and strip part an upper limit on the average count rate
is given by 360~Mcps and 420~Mcps, respectively.
In case of the strip part final numbers contain a factor 
of $f_{\textsf{\tiny{r/o}}}=1.6$~\cite{PhD-Tesis_Wuerschig},
which considers an induced charge sharing between neighboring channels, 
i.e.~a minimum number of two activated strips per hit.
Such technique based on a smaller substructure with interstitial floating strips
is envisaged for the silicon strip detectors in order to improve the spatial resolution. 
Taking into account a size of 40~bit for the output format of the hit data
(cf. chapter~\ref{cable-reqs}) it is possible to estimate the corresponding data load.
By introducing a scaling factor, $f_{\textsf{\tiny{DAQ}}}=6$~byte
\footnote{Compared to the 5~bit format defined in chapter~\ref{cable-reqs} 
an additional safety bit was added to estimate an upper limit}, 
obtained numbers for the average count rate can be translated into 
a data rate given in megabytes per second (MB/s).  
The maximum data load of roughly 4.7~GB/s is expected at highest beam momentum,
at which the contribution of the pixel and the strip part are in the same order of 
of magnitude.

Fluctuations of the total number of registered hits on a short timescale
deliver increased peak rates,
which have to be buffered by the readout electronics.
As a first approach, the intrinsic change of the instantaneous luminosity profile 
during the operation cycle can be taken to deduce a lower limit 
on the occurring peak rates.
According to model calculations for the \HESR~\cite{HESR-BeamDynHinterberger} 
it is given by an increase of $f_{\textsf{\tiny{peak}}}\approx1.38$,
which results from the ratio of the luminosity at the start 
of the data taking period to the cycle-averaged value.
Additional fluctuations may be introduced 
by variations of the effective target thickness 
and the microscopic structure of the beam.
An initial rudimentary study for the use of a pellet target at \PANDA~\cite{HESR-PelletPeakLumi}
allows for a very rough estimate of an upper limit, 
which results in a scaling factor of $f_{\textsf{\tiny{peak}}}\approx5$.

%Further simulations included in~\cite{PandaNote7-Rate},~\cite{PandaNote4-Rate}
%have been performed to study the possible effects 
%of antiproton-nucleon reactions on the expected count rates inside the MVD.
%Results indicate that the extracted hardware specifications 
%based on the numbers as given in \tabref{tab-CountRateStudies}
%suffice for a compliance of the required detector performance under these conditions.

%ABOVE LINES MOVED TO sim-resolutions.tex FOR EDITING REASONS

% put all remaining  talbes, figures etc. now.
%\clearpage

% - EOF

%% file: simulations/sim-resolutions.tex
\begin{table*}
\begin{center}
%\small
\begin{tabular}{|c|c|c|c|}
\hline  & Pixel & Strip Barrel & Strip Disks  \\
\hline Readout size & 100$\times$100~$\tcmu$m$^{2}$ & 130~$\tcmu$m  & 65~$\tcmu$m \\
\hline Stereo Angle & -- & $90^{\circ}$ & $30^{\circ}$ \\
\hline Noise $\sigma_\text{n}$ & 200~$e^{-}$ & 1000~$e^{-}$ & 1000~$e^{-}$ \\
\hline Threshold & 1000~$e^{-}$ & 5000~$e^{-}$ & 5000~$e^{-}$ \\
\hline Charge Cloud $\sigma_\text{c}$ & 8$~\tcmu$m & 8~$\tcmu$m & 8~$\tcmu$m \\
\hline
\end{tabular}
\caption[Simulation parameters]{Simulation parameters overview.}
\label{tab:sim:params}
\end{center}
\end{table*}

Further simulations included in~\cite{PandaNote7-Rate},~\cite{PandaNote4-Rate}
have been performed to study the possible effects 
of antiproton-nucleon reactions on the expected count rates inside the MVD.
Results indicate that the extracted hardware specifications 
based on the numbers as given in \tabref{tab-CountRateStudies}
suffice for a compliance of the required detector performance under these conditions.

\section{Resolution and Performance Studies}
\label{sim:hitres}

This section contains studies dealing with the performance of the MVD as standalone detector and in the whole \pnd setup. For comparable results, a set of default parameters, see \tabref{tab:sim:params},  has been used.

  \subsection{Hit Resolution}
\authalert{Author: Ralf Kliemt}

\begin{figure*}[]
\begin{center}
\includegraphics[width=0.45\textwidth]{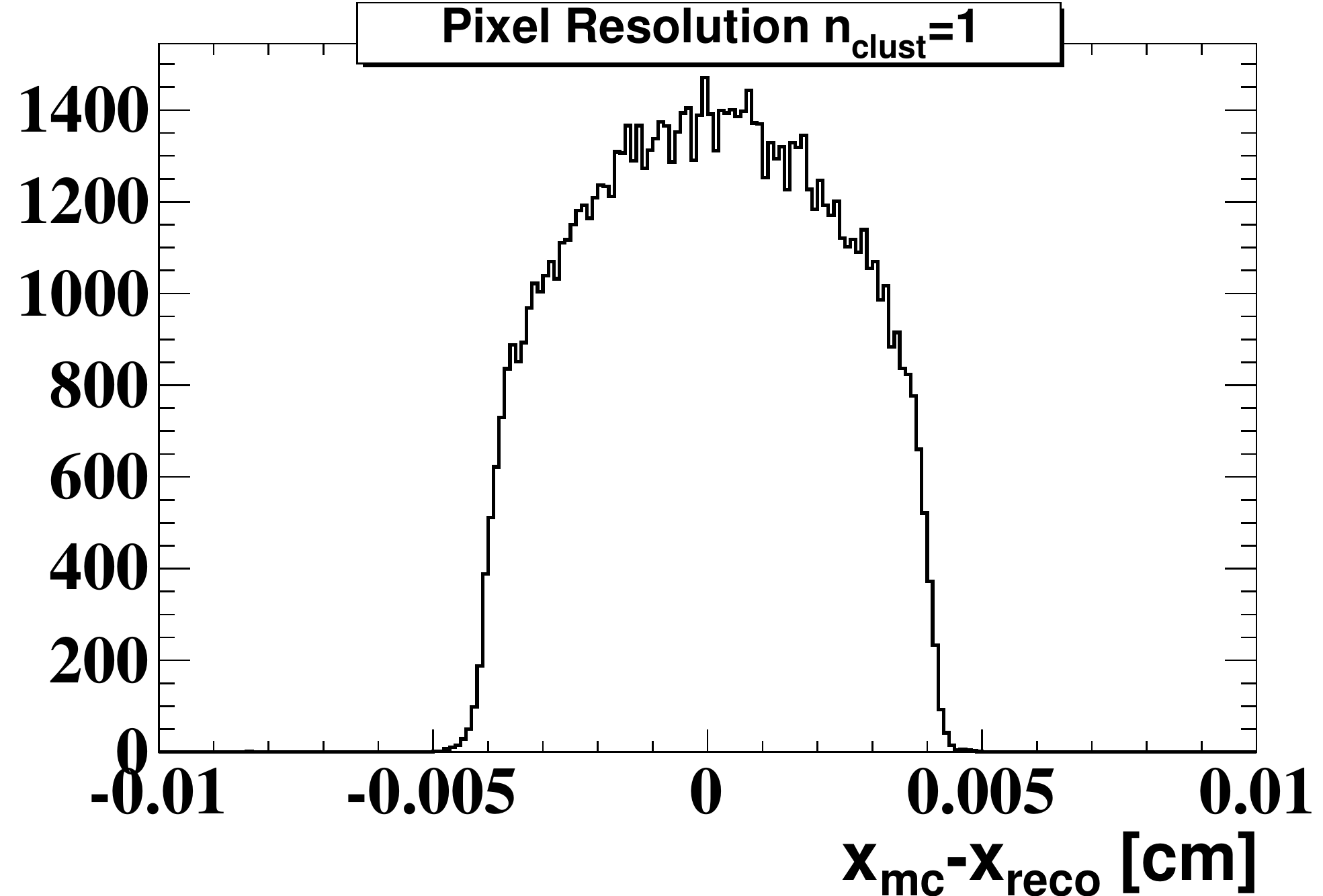}
\includegraphics[width=0.45\textwidth]{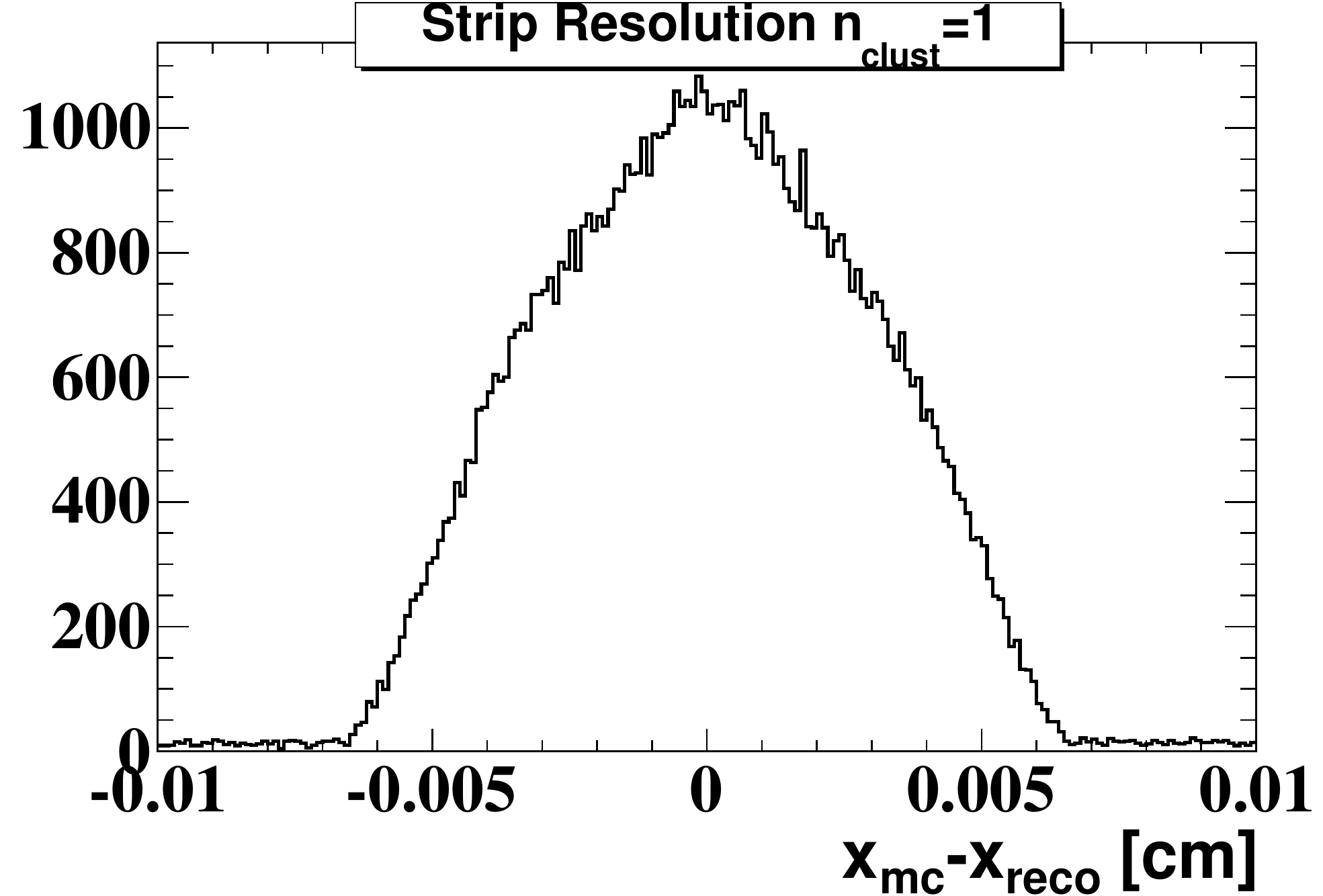}
\includegraphics[width=0.45\textwidth]{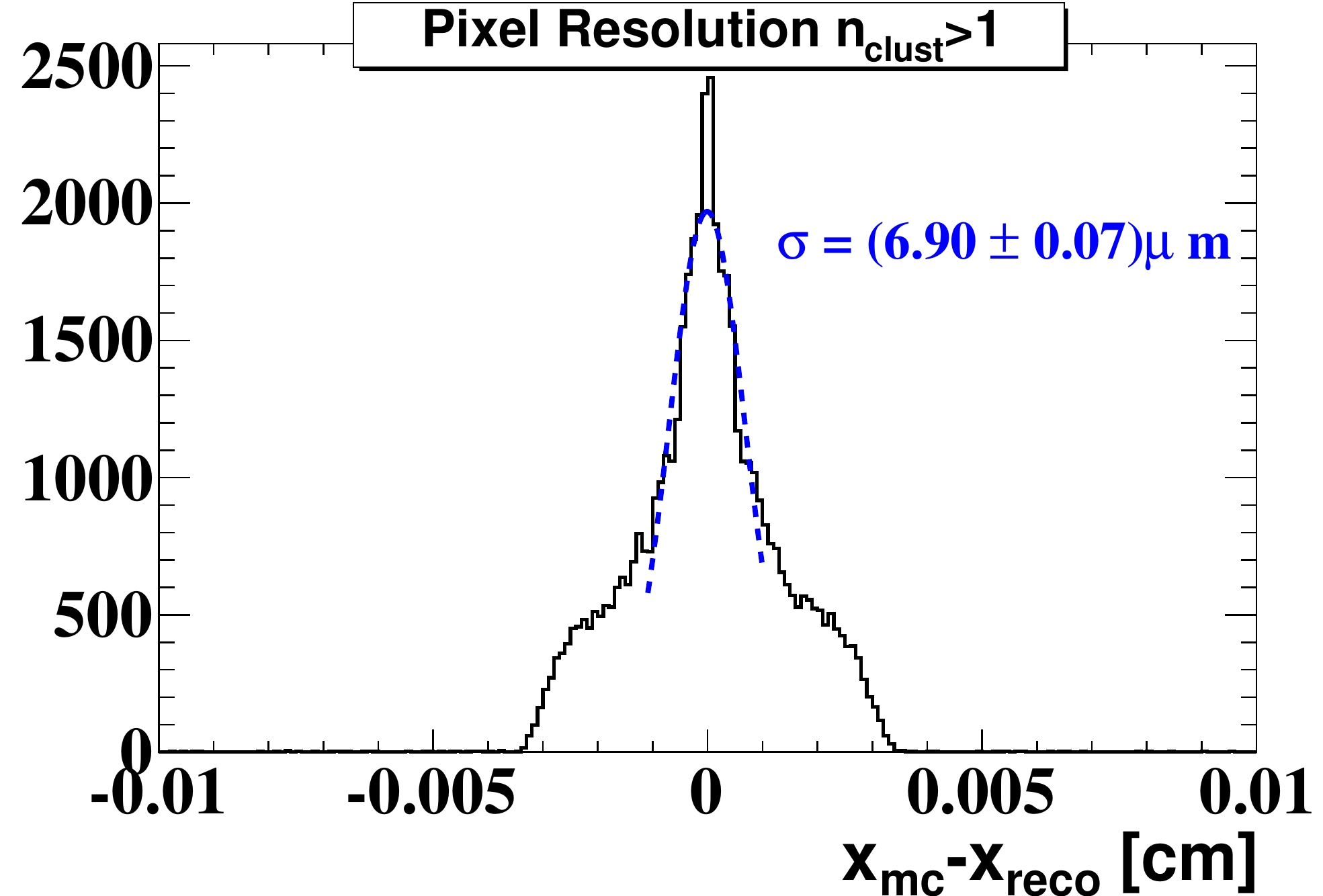}
\includegraphics[width=0.45\textwidth]{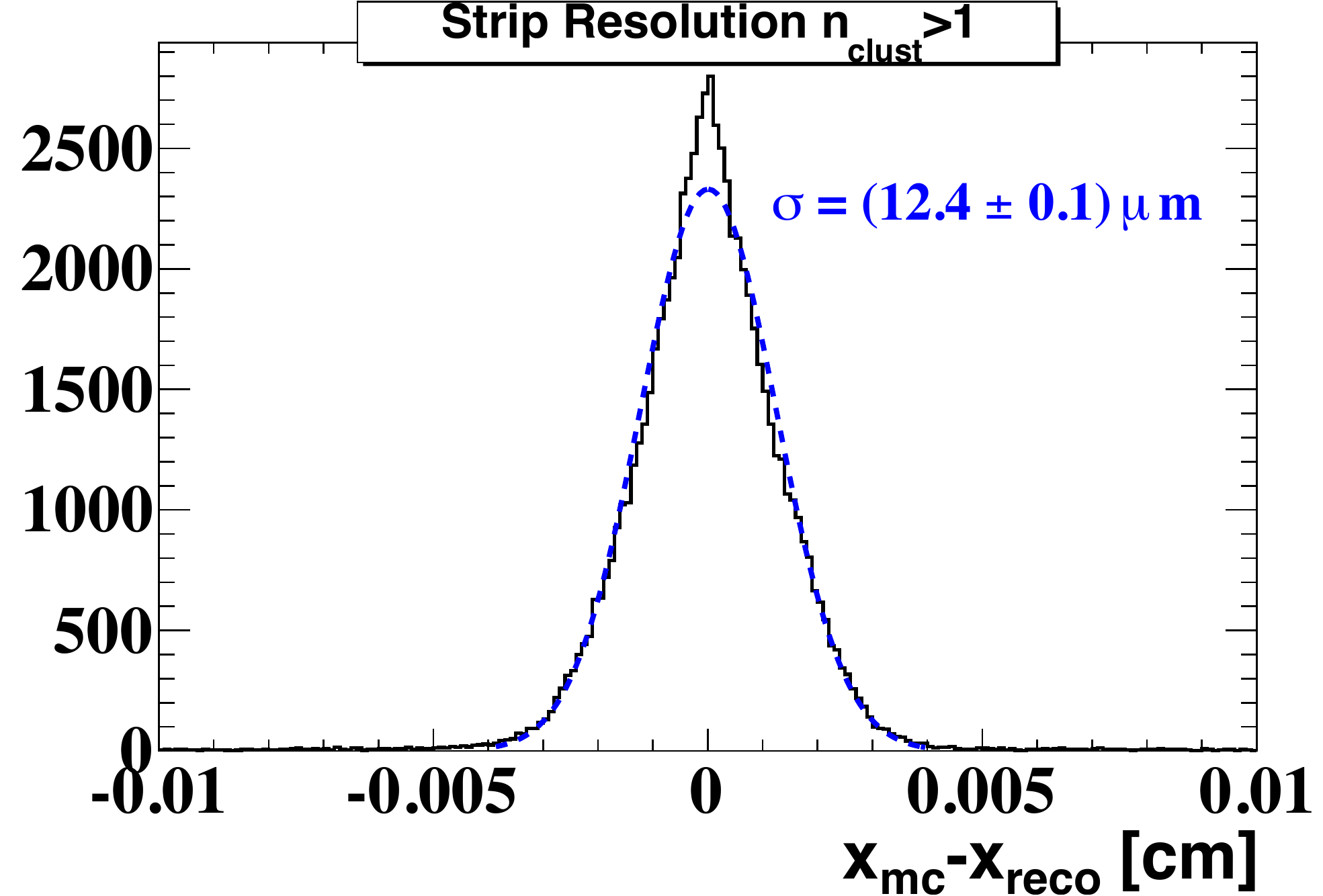}
\caption[Resolution on pixel and strip sensors for 1~\gevc pions]{Resolution on pixel and strip sensors for 1~\gevc pions.}
\label{pointres}
\end{center}
\end{figure*}
A good position resolution of each hit on a sensor is mandatory for a good momentum and vertex reconstruction. To have a clean sample, only charged pions with a momentum of 1~\gevc have been used, covering the whole azimutal angle range while the polar angle covers 7$^{\circ}$ to 140$^{\circ}$. 
The \figref{pointres} shows the resolutions per sensor type and by cluster multiplicity. For the case of one digi per cluster, the resolution is limited by the readout structure size (100~$\tcmu$m for the pixels, 130~$\tcmu$m and 65~$\tcmu$m for the strips). When the charge weighted mean is used, the achieved single hit resolution largely improves ($\sigma_{x} =\text{6.9}$~$\tcmu$m for the pixels and $\sigma_{x} =\text{12.4}$~$\tcmu$m for the strips). 
  
%  \subsection{Tracklet Reconstruction}
%\authalert{Tobias}

%The \Mvd can act as a standalone tracking station, providing there are enough reconstructed hits. An approach with projecting the the hit coordinates on a Riemann surface \alert{[citation needed]} is used. A minimum of 4 hits per tracklet are required. 
%    
  %  \clearpage
    
    \subsection{Vertexing Performance on Pions \label{sec::vertexing}}
    
\authalert{Author: Simone Bianco}

Several scans were performed to characterize the vertexing performance of the MVD. Four pions (two $\pi^+$ and two $\pi^-$) of 1~\gevc were propagated through the detector from a common vertex distributing homogeneously the four particles along the $\theta$ and $\phi$ ranges ($\theta \in [\text{10}^{\circ},\text{140}^{\circ}]$, $\phi \in [\text{0}^{\circ},\text{360}^{\circ}]$).
%The full reconstruction chain has been applied in order to reconstruct the vertex of these four pions. 
The vertices were set in different regions to map the resolution and to perform systematic studies (10.000 simulated events per point):
 \begin{itemize}
   	\item longitudinal scan: moving the vertex along the longitudinal (z) axis in the range [-1,+1]~cm;
	\item circular scan: positioning vertices along a circle with a radius 0.1~cm in the transverse (x-y) plane
	\item radial scan: changing the distance between (0,0,0) and the vertex moving along a radius in the transverse plane.
    \end{itemize}
These studies were realized using the MVD, the STT, the GEMs and the forward tracker. 
%\emph{The STT was chosen since at the time of these studies it was the fastest option}. 
The tracking was performed merging the information provided by the barrel and the forward spectrometers. A Kalman filter was applied to tracks measured in the barrel spectrometer, correcting for energy loss and scattering occurring in the detectors (see section~\ref{sec::trkvtx}). Forward tracks were reconstructed using the ideal pattern recognition and smearing the track parameters according to resolution and efficiency foreseen for the forward tracker.
Finally the POCA vertex finder (see section~\ref{sec::vtxvtx}) was applied to track candidates, obtaining the reconstructed position of the vertices.

All the scans showed a similar behavior: the vertex resolution achieved for the x coordinate is worse than y. This is not expected due to the $\phi$ symmetry of the MVD geometry. The reason of this discrepancy is the design of the inner barrel layers of the MVD, which is limited by mechanical constraints. The top and bottom regions close to the nominal interaction point cannot be covered with pixel modules in the range $\theta\in[\text{50}^{\circ},\text{90}^{\circ}]$. 

\begin{figure}[!ht]
\begin{center}
\includegraphics[width=0.95\columnwidth]{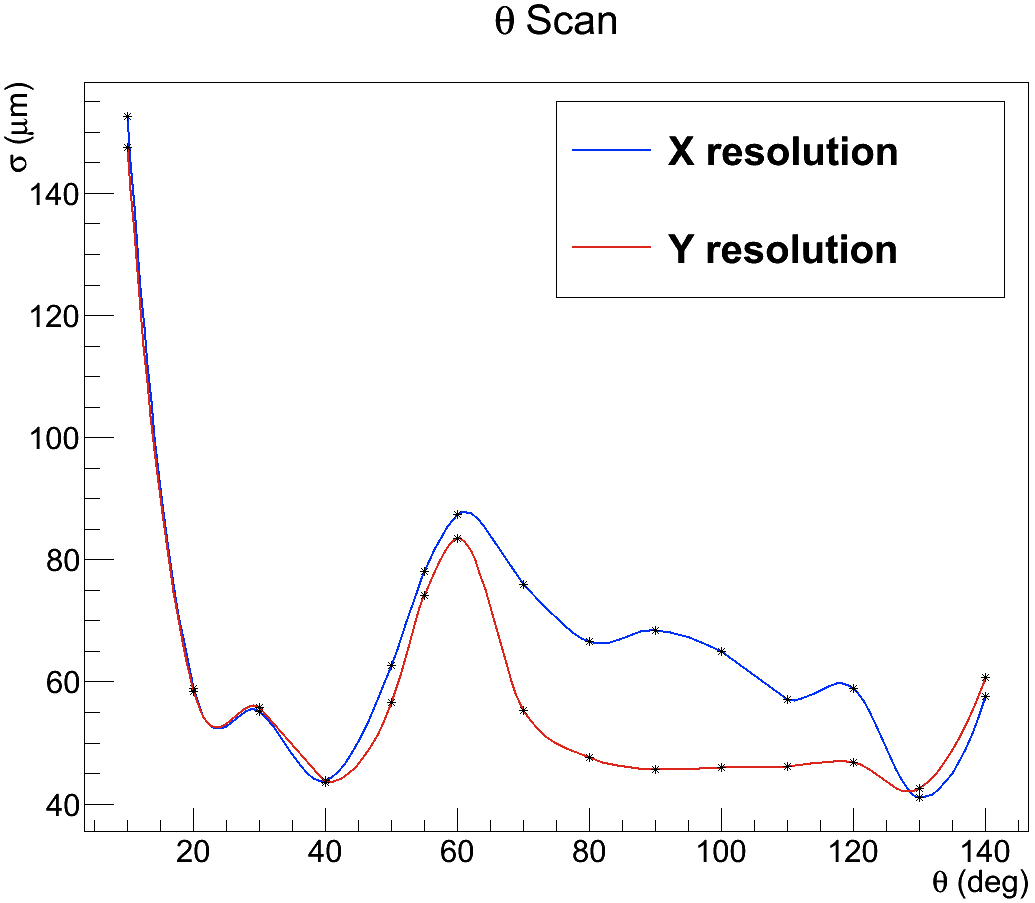}
\caption[Resolutions obtained with a fixed $\theta$ and an uniform $\phi$ distribution]{Resolutions obtained selecting a fixed value for the $\theta$ angle and using an uniform distribution for the $\phi$ angle.}
\label{ThetaScan}
\end{center}
\end{figure}

This results in a different precision in the determination of the coordinates of the reconstructed vertices, which is significant in this portion of solid angle. \Figref{ThetaScan} shows the x and y resolutions obtained with selected $\theta$ values for the four pions. A detailed analysis of the origin of this different behavior can be found in the appendix~\ref{app::vertexing}. 
The studies presented this section were performed covering homogeneously a wide $\theta$ range. This is the reason why in the results a significant difference appears between the x and the y vertexing performances. 
Most of the physics channels of interest for the experiment foresee mainly decays in the forward region of the detector, therefore the x-y different performance at big $\theta$ angles will not influence most of the physics studies. For example all the results of the next section~(\ref{section::channels}) show the same resolution for the x and y vertex reconstruction.

\subsubsection*{Results of the Scans}

\begin{figure}[!htpb]
\begin{center}
\includegraphics[width=1.\columnwidth]{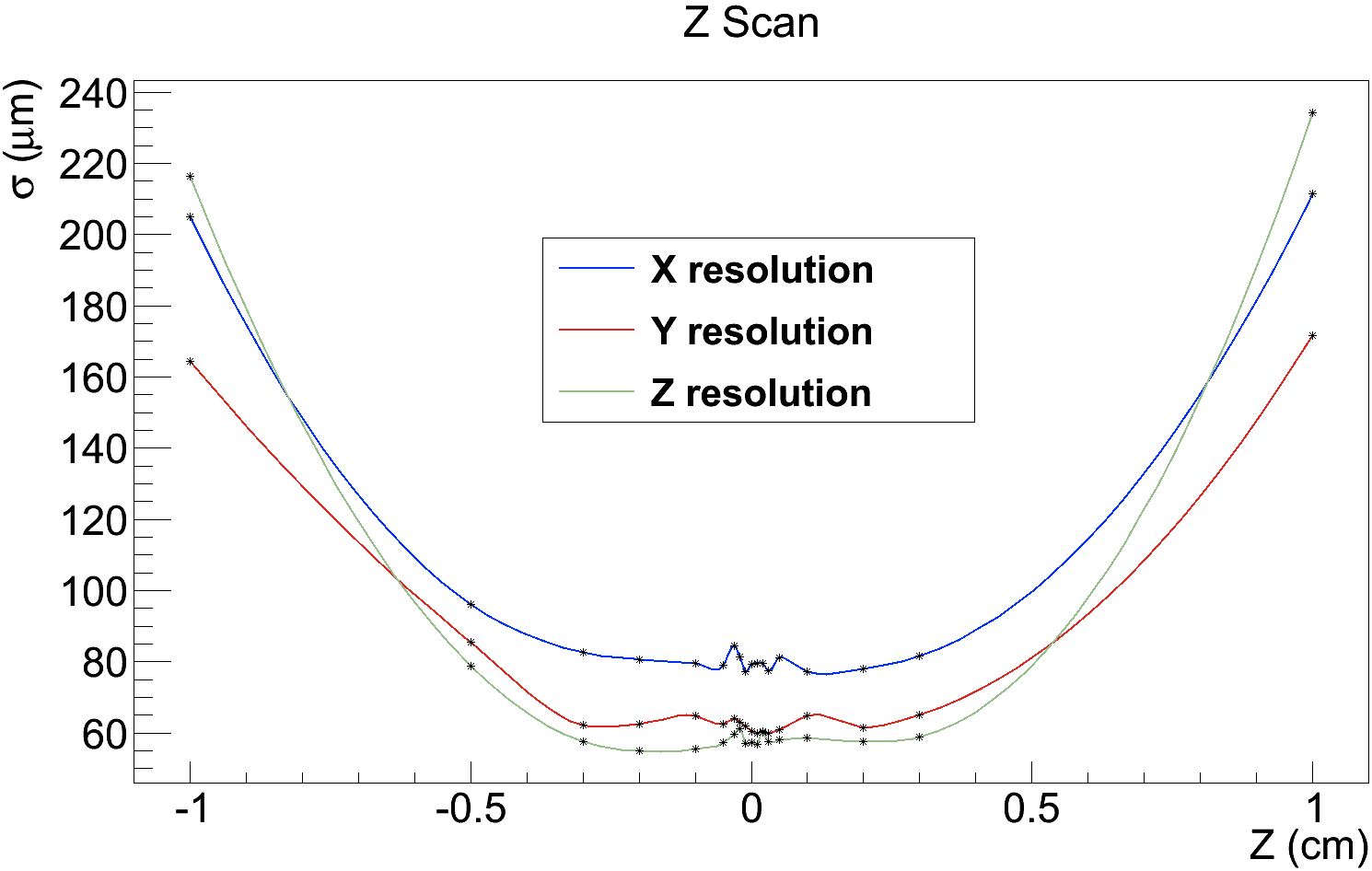}
\caption
[Vertexing results obtained placing defined vertices in different points along the longitudinal axis of the MVD]
{Vertexing results obtained placing defined vertices in different points along the longitudinal axis of the MVD.}
\label{LongScan}
\end{center}
\end{figure}

\begin{figure}[!htpb]
\begin{center}
\includegraphics[width=1.\columnwidth]{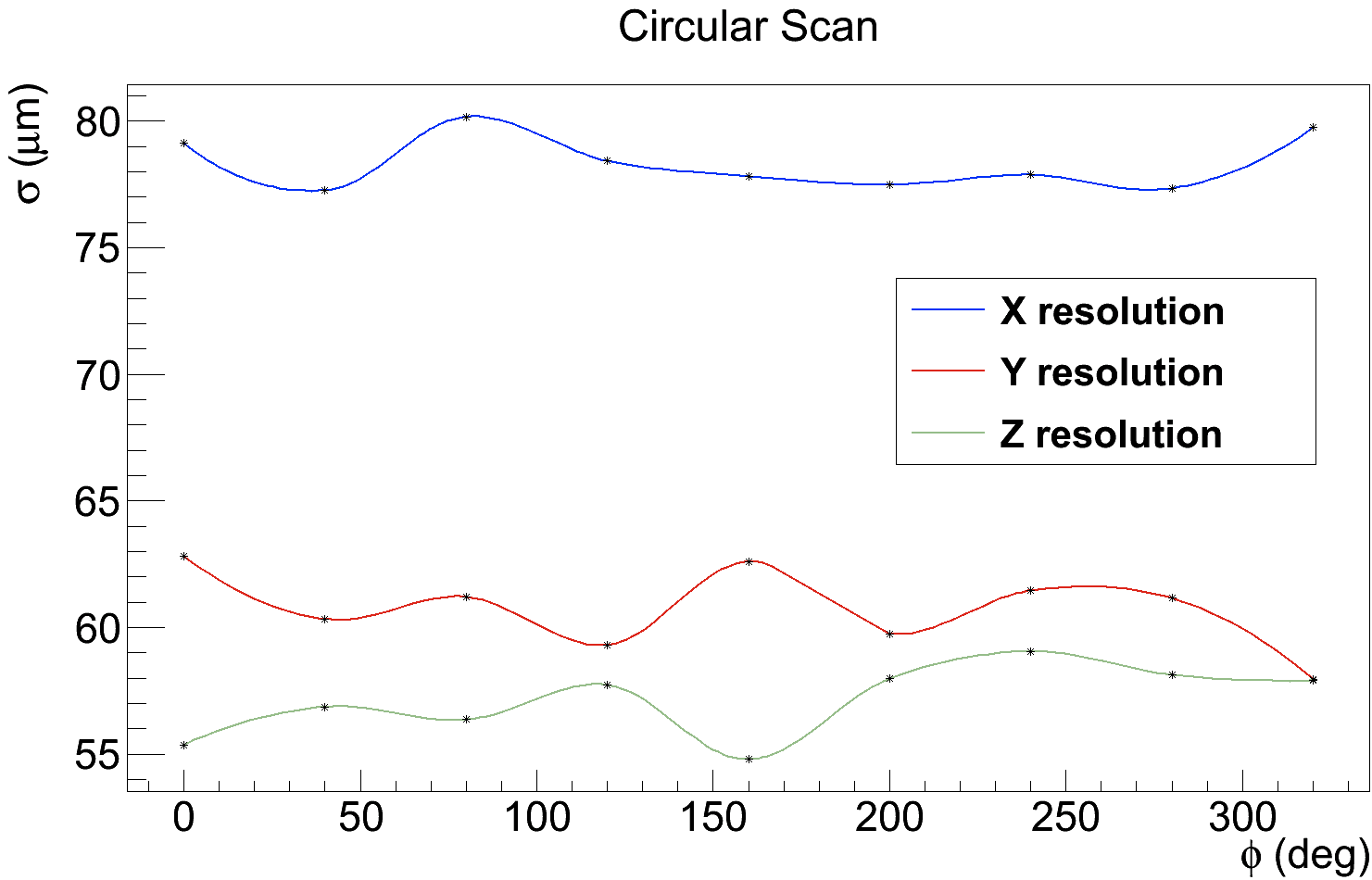}
\caption
[Scan of the vertexing behavior moving vertices on a circle of radius 1~mm laying in the transverse plane]
{Scan of the vertexing behavior moving vertices on a circle of radius 1~mm laying in the transverse plane.}
\label{CircScan}
\end{center}
\end{figure}

\begin{figure}[!htpb]
\begin{center}
\includegraphics[width=1.\columnwidth]{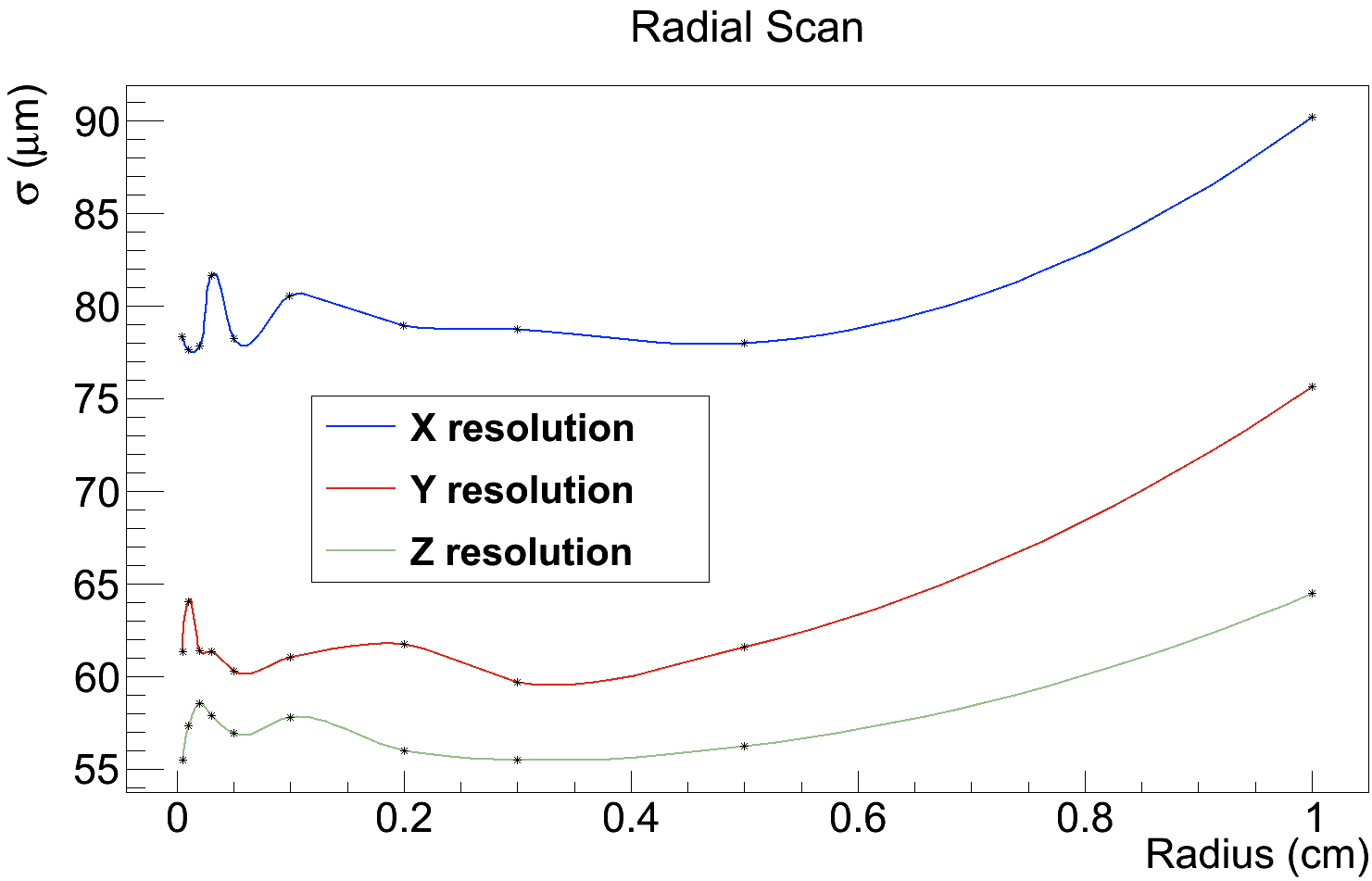}
\caption[Mapping of the vertexing results changing the distance from the nominal interaction point]
{Mapping of the vertexing results changing the distance from the nominal interaction point moving along the radius corresponding to $\phi$=0$^{\circ}$ and z=0~cm.}
\label{RadScan}
\end{center}
\end{figure}

The longitudinal scan (\figref{LongScan}) shows that vertex resolution values are quite stable in the range [-5,+5]~mm, where the pattern recognition for primary particles is foreseen to work efficiently.
The circular scan (see \figref{CircScan}) confirms a good $\phi$-uniformity of the performance expected on the basis of the detector geometry. 
In \figref{RadScan} the effects of a movement along a radius with $\phi=0^{\circ}$ and $z=0$~cm are shown. The resolutions are not severely influenced by this kind of shift as long as one stays at distances below 1~cm from the nominal interaction point. There the vertexing starts to show worse performances.

\begin{figure}[!htpb]
\begin{center}
\includegraphics[width=0.97\columnwidth]{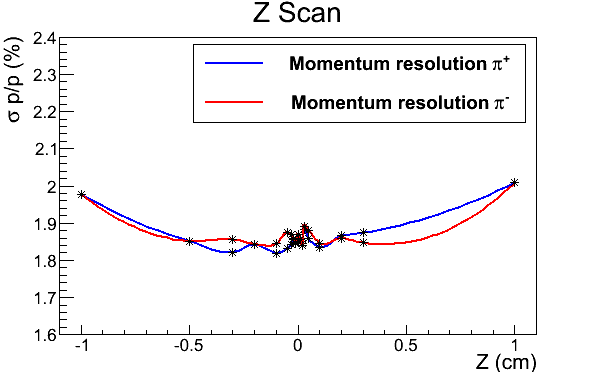}\\
\includegraphics[width=0.97\columnwidth]{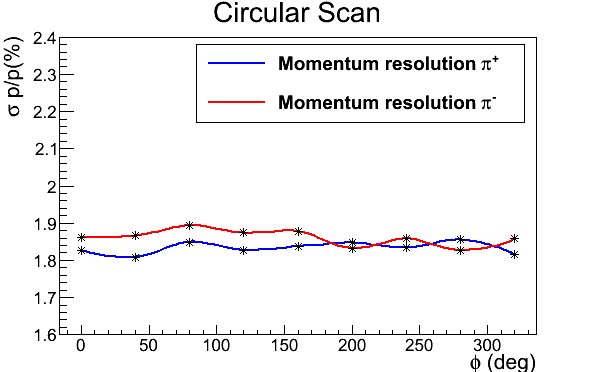}\\
\includegraphics[width=0.97\columnwidth]{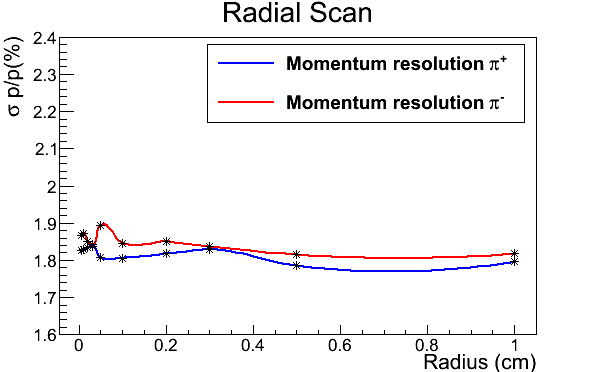}
\caption
[Momentum resolution values obtained for $\pi^+$ and $\pi^-$]
{Momentum resolution values obtained for $\pi^+$ and $\pi^-$ with the three scans.}
\label{MomResScan}
\end{center}
\end{figure}

The momentum resolution is much less effected by changes in the position of the vertex and it stays constant within a few per mill in all the performed scans (see \figref{MomResScan}).

%\begin{figure*}[!htpb]
%\end{figure*}

%% file: simulations/sim-channels.tex
\section{Physics Channels Analysis\label{section::channels}}

The tasks of the \Mvd are mainly to provide high tracking performance close to the interaction zone. We report three physics benchmark cases where the performance of the \Mvd is monitored.
First we have particles decaying directly in the interaction point ($\Psi(2S) \rightarrow J/\psi \pi^+ \pi^-$), 
and in the second case particles which decay further away from the interaction point 
%but still inside the beam pipe 
(\D mesons).
%and third decays which occur even further outside ($K^0$).

  \subsection{Detector Setup}
In the physics channels we chose the following geometric setup. 
%The choice of the central tracker option was settled by the higher speed of the available software code at this time.\\
Active components:
\begin{itemize}
\item \Mvd%\footnote{Geometry file: Mvd-2.1\_FullVersion.root}
: 4 barrel layers and 6 disks
\item STT: 1.5~m long, with skewed layers
\item GEM: 3 stations 
\item FTS: 6 forward tracking straw stations
\end{itemize}
For these elements hit digitization and reconstruction is performed. In the barrel spectrometer part the combined STT+\Mvd+GEM pattern recognition with a Kalman Filter track fitting is used, with the default particle hypothesis being muons. In the forward spectrometer a Monte-Carlo truth based tracking is applied with a 95\% efficiency and gaussian smearing of $\sigma_\text{p}/p=3\%$ in momentum as well as $\sigma_\text{v}=200$~$\tcmu$m in position. 
%Particle identification is omitted (using Monte-Carlo truth) to be independent of it.
Realistic particle identification algorithms are not used, in order to have results independent from them. To identify particles the Monte-Carlo information is used.

The following components were present as passive material, thus contributing to the material being present, excluding the time consuming digitization processes of these components not being relevant in these studies.
\begin{itemize}
\item Pipe Beam-Target cross (passive)
\item Disc DIRC
\item Barrel DIRC
\item EMC crystals
\item Solenoid and forward dipole magnet, including flux return yokes and muon chambers
\end{itemize}

The simulations of physics channels were done with the \pndrt ``trunk'' revision number 12727 from July, 18 2011, using the ``may11'' release of the external packages.

% ------------------
  \subsection{Benchmark Channel: $\overline{p}p \rightarrow \psi(2S) \rightarrow J/\psi \pi^+ \pi^-$}
\authalert{Author: Simone Bianco}

\subsubsection*{$J/\psi \rightarrow \mu^+ \mu^-$}

\begin{figure*}[p]
\begin{center}
\includegraphics[width=0.48\textwidth]{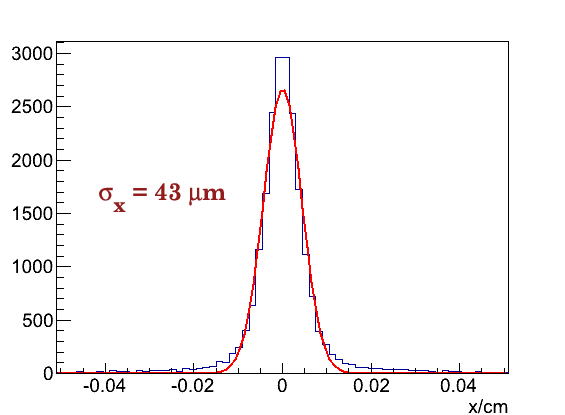}
\includegraphics[width=0.48\textwidth]{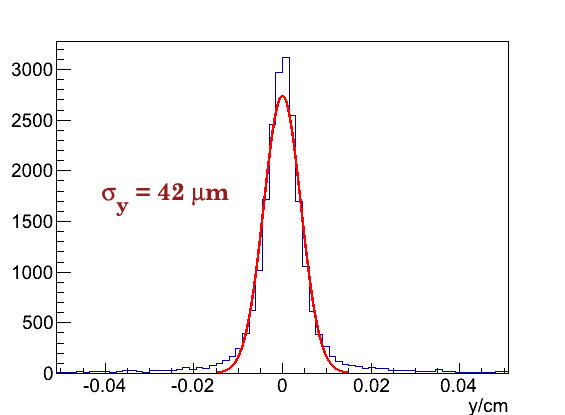}
\caption
[Distributions of the $J/\psi$ vertices reconstructed from their decay in the nominal interaction point]
{Distributions of the $J/\psi$ vertices reconstructed from their decay in the nominal interaction point.}
\label{JPsiVtx}
\end{center}
\end{figure*}

The first test has been performed reconstructing the following reaction:

\begin{centering} 
$\overline{p}p \rightarrow \psi(2S) \rightarrow J/\psi \pi^+ \pi^- \rightarrow \mu^+ \mu^- \pi^+ \pi^-$\\
\end{centering}
Running a full analysis and performing a vertex fit on the $J/\psi$ candidates it was possible to determine the vertex resolution. \Figref{JPsiVtx} shows the distributions of the x and y coordinates of the reconstructed vertices. The resolution in z is slightly worse ($\sigma_\text{z} = 64$~$\tcmu$m, while $\sigma_\text{x} = 43$~$\tcmu$m and $\sigma_\text{y} = 42$~$\tcmu$m) due to the forward boost of the decay.
%\footnote{The anti-proton beam collides on a fixed target, therefore the system has a forward boost in order to conserve the total momentum.}. 
The overall vertex reconstruction is excellent.

\begin{figure*}[p]
\begin{center}
\includegraphics[width=0.48\textwidth]{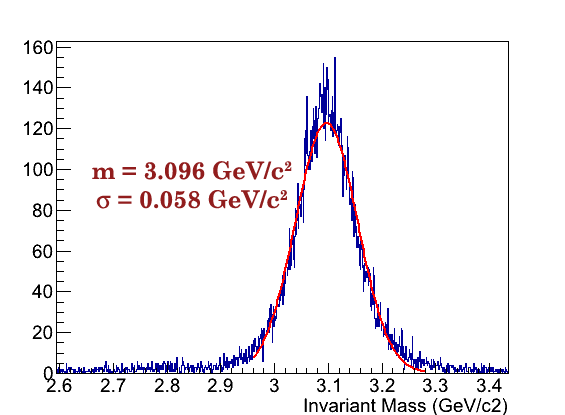}
\includegraphics[width=0.48\textwidth]{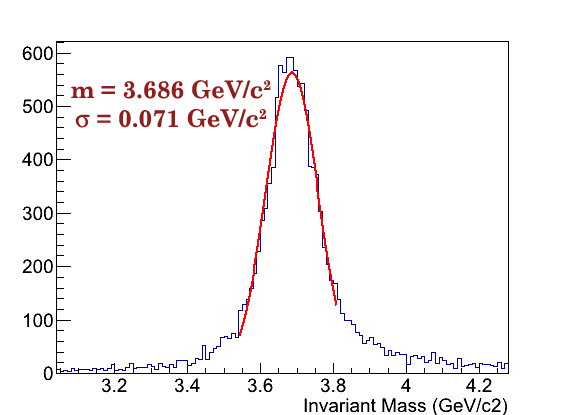}
\caption
[Mass distribution for $J/\psi$ and $\psi(2S)$ candidates]
{Mass distribution for $J/\psi$ and $\psi(2S)$ candidates.}
\label{PsiMassOne}
\end{center}
\end{figure*}

\Figref{PsiMassOne} shows the mass distribution obtained for the $J/\psi$ and the $\psi(2S)$ candidates, respectively. The masses obtained fitting these distributions are in full agreement with the expected values.

\begin{figure*}[p]
\begin{center}
\includegraphics[width=0.48\textwidth]{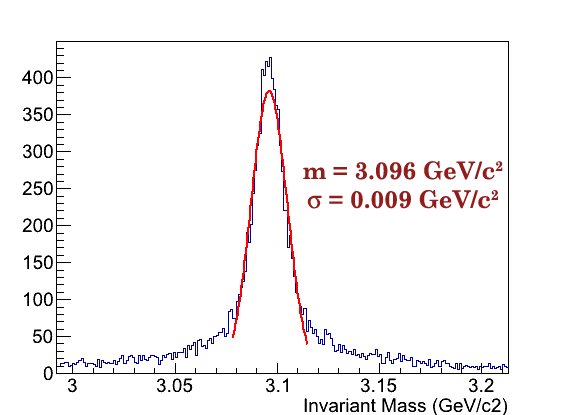}
\caption
[Distribution of the $J/\psi$ mass obtained with the missing mass method knowing the initial $\psi(2S)$ state]
{Distribution of the $J/\psi$ mass obtained with the missing mass method knowing the initial $\psi(2S)$ state.}
\label{MMass}
\end{center}
\end{figure*}

\Figref{MMass} shows the missing mass of the $\overline{p}p\rightarrow \pi^{+}\pi^{-}$ reaction at the $\psi(2S)$ peak, from which the $J/\psi$ signal emerges.
%is an example of the results obtained using the knowledge we have about the initial state; here the missing mass of the $J/\psi$ is plotted. These values were obtained subtracting from the initial $\Psi(2S)$ state the four-momenta of the two reconstructed pions. 
The $J/\psi$ signal obtained in this way is much narrower than that obtained by the invariant mass of the lepton pairs (\figref{PsiMassOne}).
%This gives a much narrower $J/\psi$ mass distribution than the one shown in \figref{PsiMassOne}.

%\begin{figure}[h]
%\begin{center}
%\includegraphics[width=0.48\textwidth]{simulations/pictures/DEF_Phys/MuMu/JPsiVertXOk}
%\includegraphics[width=0.48\textwidth]{simulations/pictures/DEF_Phys/MuMu/JPsiVertYOk}
%\caption
%[Distributions of the $J/\psi$ vertices reconstructed from their decay in the nominal interaction point]
%{Distributions of the $J/\psi$ vertices reconstructed from their decay in the nominal interaction point.}
%\label{JPsiVtx}
%\end{center}
%\end{figure}

%\clearpage

\subsubsection*{$J/\psi \rightarrow e^{+} e^{-}$}

\begin{figure*}[]
\begin{center}
\includegraphics[width=0.48\textwidth]{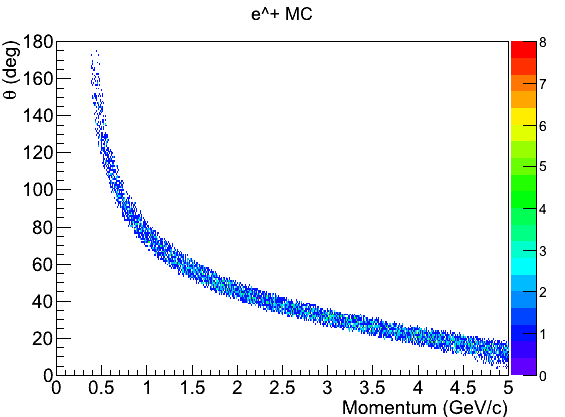}
\includegraphics[width=0.48\textwidth]{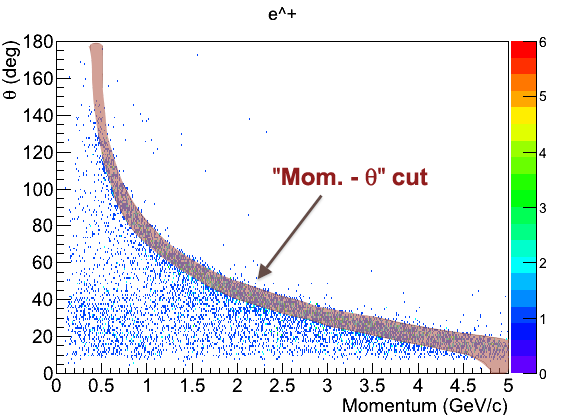}
\caption
[Electron-positron polar angle and momentum distributions from MC-truth and from the reconstruction]
{Electron-positron polar angle and momentum distributions from MC-truth (left) and from the reconstruction (right). In the second figure the reconstructed distribution is superposed to the kinematic cut applied during the analysis.}
\label{PsiPos}
\end{center}
\end{figure*}

\begin{figure*}[p]
\begin{center}
\includegraphics[trim=0 0 1.2cm 1.2cm,  clip, width=0.48\textwidth]{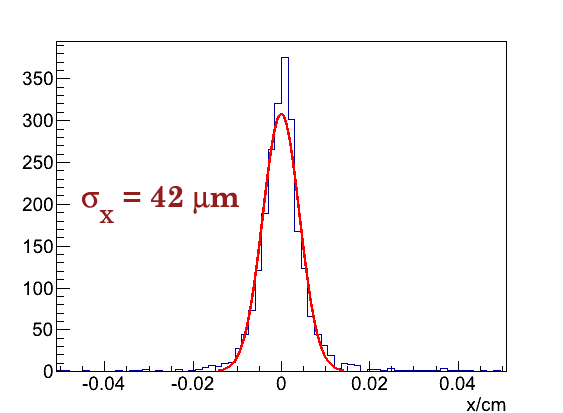}
\includegraphics[trim=0 0 1.2cm 1.2cm,  clip, width=0.48\textwidth]{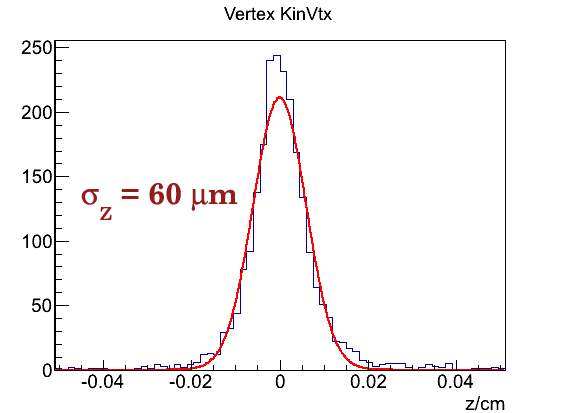}
\caption
[Reconstructed $J/\psi$ vertices]{Reconstructed $J/\psi$ vertices.}
\label{JPsiVtx2}
\end{center}
\end{figure*}

\begin{figure*}[p]
\begin{center}
\includegraphics[trim=0 0 1.2cm 1.2cm,  clip, width=0.48\textwidth]{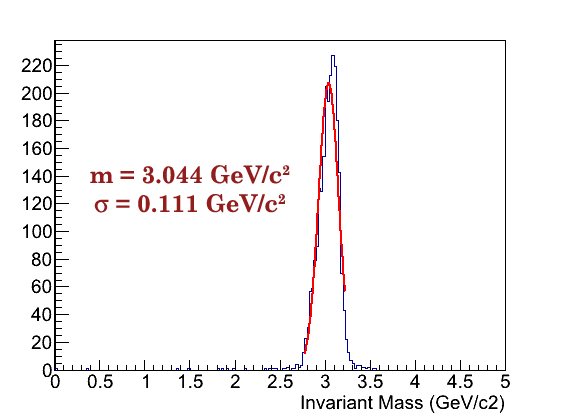}
\includegraphics[trim=0 0 1.2cm 1.2cm,  clip, width=0.48\textwidth]{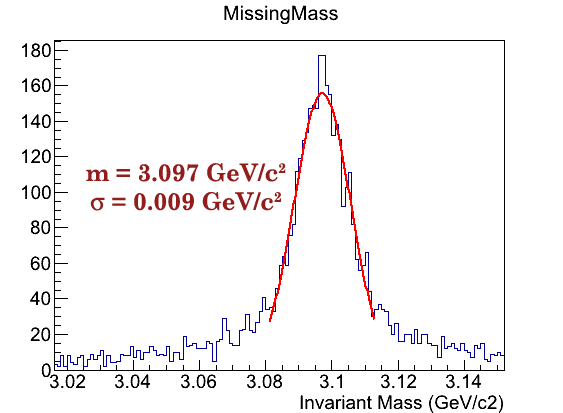}
\caption
[$J/\psi$ Reconstructed mass and missing mass distributions]{$J/\psi$ Reconstructed mass (left) and missing mass (right) distributions.}
\label{JPsiMasses}
\end{center}
\end{figure*}

The decay of the $J/\psi$ into $e^+$ $e^-$ has been studied in the same situation.
%starting from the same initial state. 
This decay is more effected by multiple scattering than the muonic one. Therefore some selections on the reconstructed candidates can improve the final results. 
%The kinematics of the decay have been studied. 
Only $\mathrm{\pi}$ and $\mathrm{e}$ candidates matching the allowed phase space regions\footnote{This cut was based on the MC-true phase space distributions for this decay} were taken into account for the rest of the analysis. 
\begin{figure*}[]
\begin{center}
\includegraphics[width=0.48\textwidth]{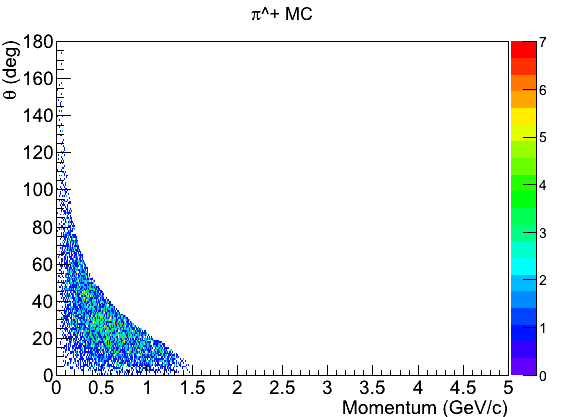}
\includegraphics[width=0.48\textwidth]{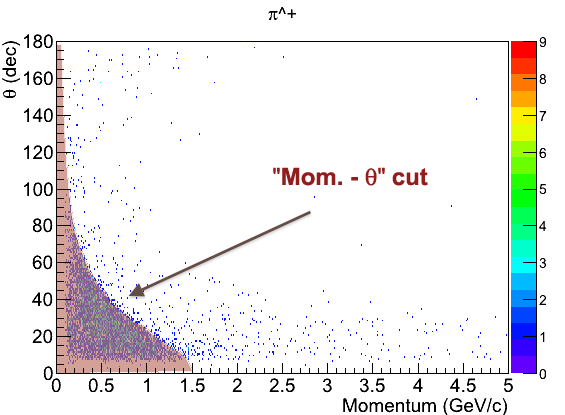}
\caption
[Pion polar angle and momentum distributions from MC-truth and from the reconstruction]
{Pion polar angle and momentum distributions from MC-truth (left) and from the reconstruction (right). In the second plot the kinematic cut is highlighted by a colored band.}
\label{PsiPions}
\end{center}
\end{figure*}
\Figref{PsiPions} and \figref{PsiPos} show the MC prediction for the polar angle and momentum distributions compared with what is obtained from the reconstruction. Two colored bands highlight the specific cuts applied on the kinematics of the track candidates.
Applying these selections, $J/\psi$ candidates have been reconstructed. The distributions of the reconstructed vertex positions are shown in \figref{JPsiVtx2}.
\Figref{JPsiMasses} shows the invariant mass of possible $J/\psi$ candidates and the missing mass built by the measured pions (which obviously does not depend on the the $J/\psi$ decay mode).
%the results obtained with the missing mass method. 

%Since the missing mass method does not depend on the $J/\psi$ decay, the results have to be similar to what was obtained with the previous channel. However, during this analysis the allowed kinematics selection was applied also to pion candidates. 

%Due to the kinematic selections imposed on the pions, the results here are slightly better (smaller width of the missing mass distribution) than obtained in the previous case. %with $J/\psi \rightarrow \mu^+ \mu^-$.
\begin{figure*}[p]
\begin{center}
\includegraphics[trim=0 0 1.2cm 1.2cm,  clip, width=0.48\textwidth]{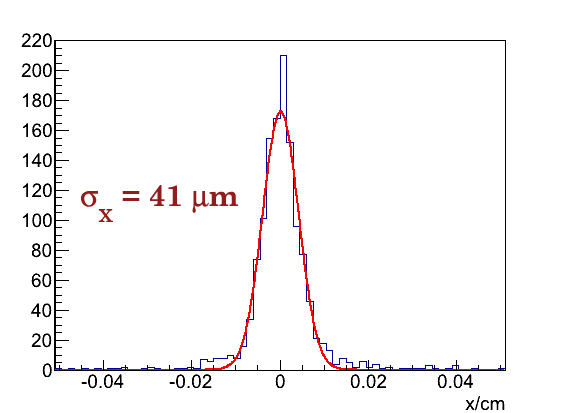}
\includegraphics[trim=0 0 1.2cm 1.2cm,  clip, width=0.48\textwidth]{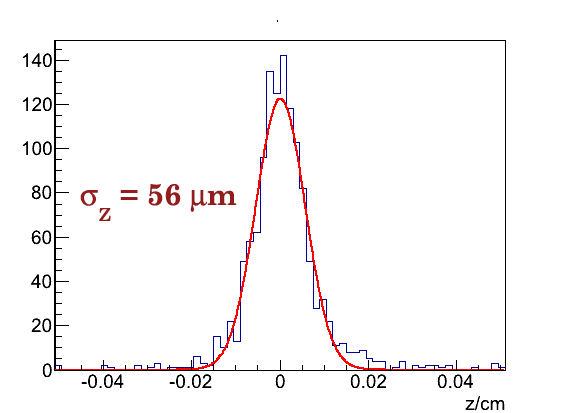}
\caption
[Coordinates of the reconstructed $\psi(2S)$ vertices]
{Coordinates of the reconstructed $\psi(2S)$ vertices.}
\label{PsiVtx}
\end{center}
\end{figure*}

\begin{figure*}[p]
\begin{center}
\includegraphics[trim=0 0 1.2cm 1.2cm,  clip, width=0.48\textwidth]{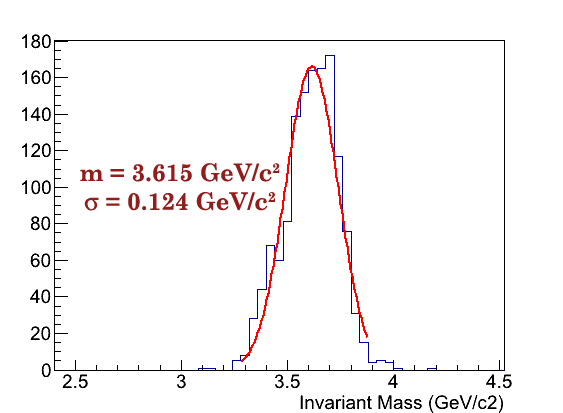}
\caption
[Mass distribution for the $\psi(2S)$ candidates]{Mass distribution for the $\psi(2S)$ candidates.}
\label{PsiMass}
\end{center}
\end{figure*}

\Figref{PsiVtx} shows the distribution of the reconstructed $\psi(2S)$ vertex positions. These vertices are calculated combining four particles. Therefore more constraints are applied than in the case of the $J/\psi$ vertex. This results in a better z resolution due to the forward boost of the system. These considerations are based on the fact that both the $\psi(2S)$ and the $J/\psi$ decay immediately in the interaction vertex.
The invariant mass of the $\psi(2S)$ candidates are shown in \figref{PsiMass} where the results of a Gaussian fit on the peak are reported. %These values are compatible with the Monte-Carlo input.
Since $\psi(2S)$ candidates are obtained combining $J/\psi$ ones with a $\pi^+$ and a $\pi^-$, which both have to satisfy track quality requirements, the final total number of reconstructed $\psi(2S)$ is smaller than the $J/\psi$ available statistics.

\clearpage

% ----------------------------------------

  \subsection{Benchmark Channel: \D mesons}
  \authalert{A: Ralf Kliemt}
  
In order to explore charm physics in \pnd the reconstruction of \D
mesons is a crucial point. The biggest challenge is to reduce the background from the signal channels, which have typically a low cross section. Because the decay lengths are $c\tau= 312$~$\tcmu$m for the charged and $c\tau= 123$~$\tcmu$m for the neutral ground state \D mesons, a selection by the decay vertex is envisaged.

 \subsubsection{$\ppbar\rightarrow D^+D^-$}

\begin{figure}[]
\begin{center}
%trim option's parameter order: left bottom right top
\includegraphics[width=0.75\columnwidth]
{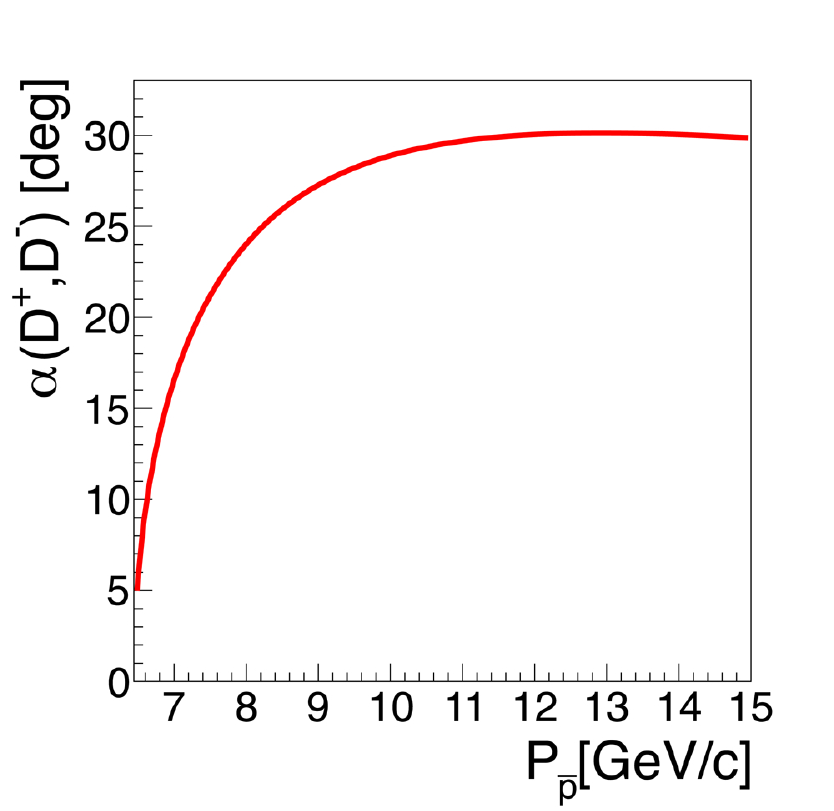}
\caption[\Dp/\Dm max. opening angle]{\Dp/\Dm maximum opening angle depending on the \pbar beam momentum.}
\label{channels:opening}
\end{center}
\end{figure}

The charged ground state $D$ mesons decay by 9.4\% branching fraction into a final state with three charged particles ($K\pi\pi$, see \cite{PDG:2010}). It is convenient to have only charged particles in the final state to be less dependent on other detector components in this study.
In the exit channel six charged tracks then have to be reconstructed: 
\begin{equation*}\pbarp \rightarrow \DpDm \rightarrow K^{-}\pi^{+}\pi^{+}K^{+}\pi^{-}\pi^{-}\end{equation*}
For this study 99200 signal events have been simulated at 9~\gevc incident antiproton momentum, enabling a large maximal opening angle between the two $D$ mesons (see \figref{channels:opening}).
Furthermore a set of 344525 non-resonant background events was simulated featuring the same final state: $K^{+}K^{-}2\pi^{+}2\pi^{-}$.
\begin{table*}[]
\begin{center}
\begin{tabular}{|c||c|c|c||c|c|c|}
\hline                                            & \Dp/\Dm & comb. & nonres.    & \Dz/\Dzbar & comb. & nonres. \\
\hline
\hline Simulated events            & 99200    & --         & 344525    & 98800        & --          & 198125 \\
\hline Simulated \D's                  & 198400 & --         & (689050)  & 197600     & --          & (396250) \\
\hline \D's After reconstruction & 120835 & 67039 & 688188    & 135024     & 27161 & 322771\\
\hline
\hline $\pm150$~\mevcc Mass window  & 61579    & 21963 & 67645      & 102763     & 2232   & 23301 \\
\hline  Vertex r and z cut               & 15004    & 689   & 1060         & 17255       & 194      & 543 \\
\hline  4C kinematic fit               & 1058      & 25        & 0             & 1592          & 18        & 2 \\
\hline
\hline  Suppression factor (fits \& cuts)        & 8.8$\cdot$10$^{-3}$ & 3.7$\cdot$10$^{-4}$ & $<$ 1.5$\cdot$10$^{-6}$ &  1.2$\cdot$10$^{-2}$  &  6.6$\cdot$10$^{-4}$  & 6.2$\cdot$10$^{-6}$ \\
\hline  Suppression factor (total)  & 5.3$\cdot$10$^{-3}$ & 1.3$\cdot$10$^{-4}$ & $<$ 1.5$\cdot$10$^{-6}$ &  8.6$\cdot$10$^{-3}$  &  9.1$\cdot$10$^{-5}$  & 5.1$\cdot$10$^{-6}$ \\
\hline
\end{tabular}
\caption[D meson counts]{D meson counts from different sources: Signal channel, combinatorial background (comb.) and non-resonant background (nonres.). All cuts and fits are applied sequentially.}
\label{channels:DCounts}
\end{center}
\end{table*}

\begin{figure}[]
\begin{center}
%trim option's parameter order: left bottom right top
%\includegraphics[trim = 0mm 111mm 0mm 0mm, clip,width=\textwidth]{simulations/pictures/DMesons/DmFitterVergleich.pdf}
\includegraphics[width= 0.8\columnwidth]
{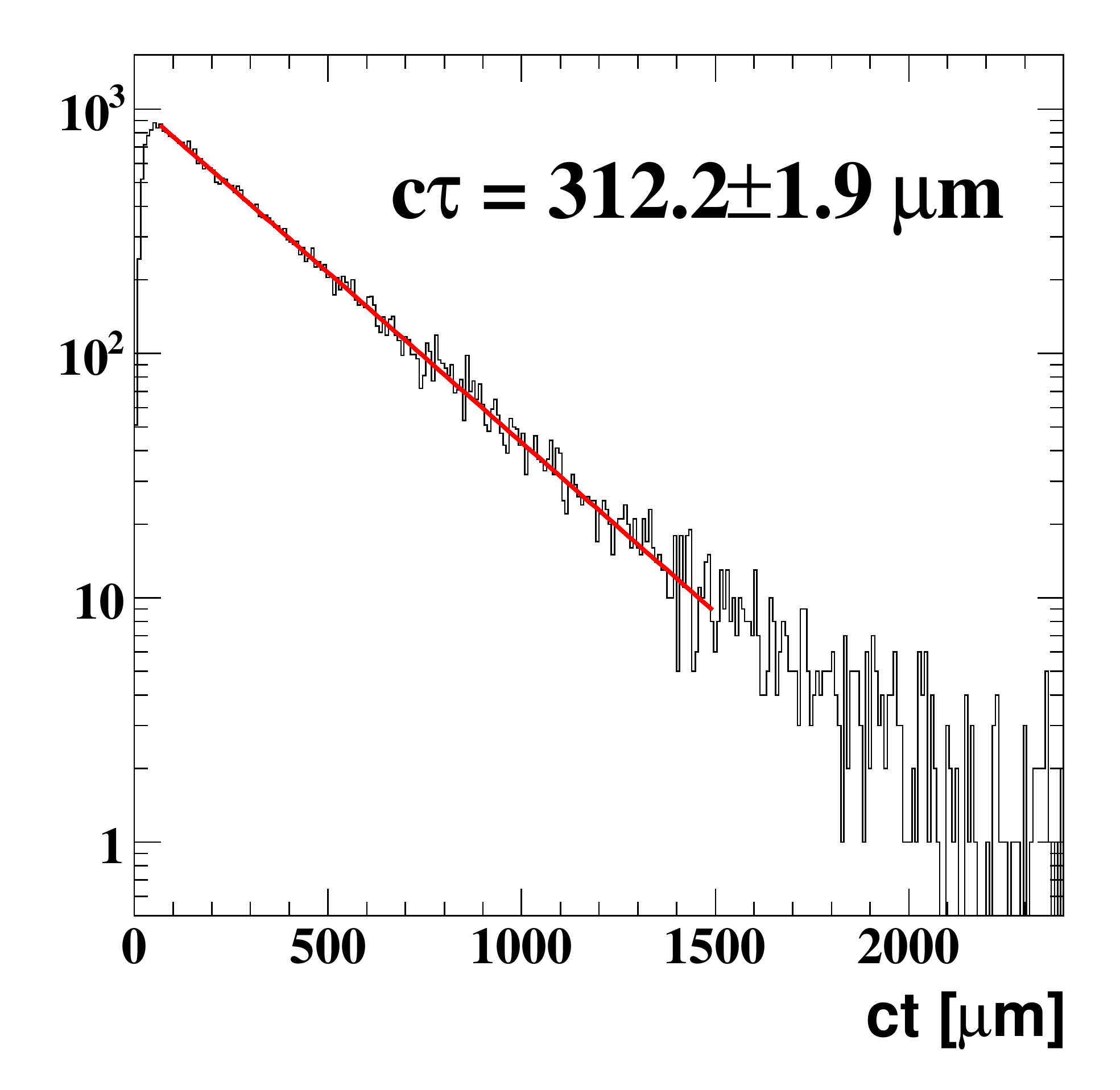}
\caption[$D^{+}$ and $D^{-}$ decay lengths]
{$D^{+}$ and $D^{-}$ decay lengths (PDG value: $c\tau = 311.8$~$\tcmu$m) \cite{PDG:2010}, obtained directly with the POCA finder without any correction.}
\label{channels:DpDmctau}
\end{center}
\end{figure}

\begin{figure}[]
\begin{center}
\includegraphics[width= 0.85\columnwidth]
{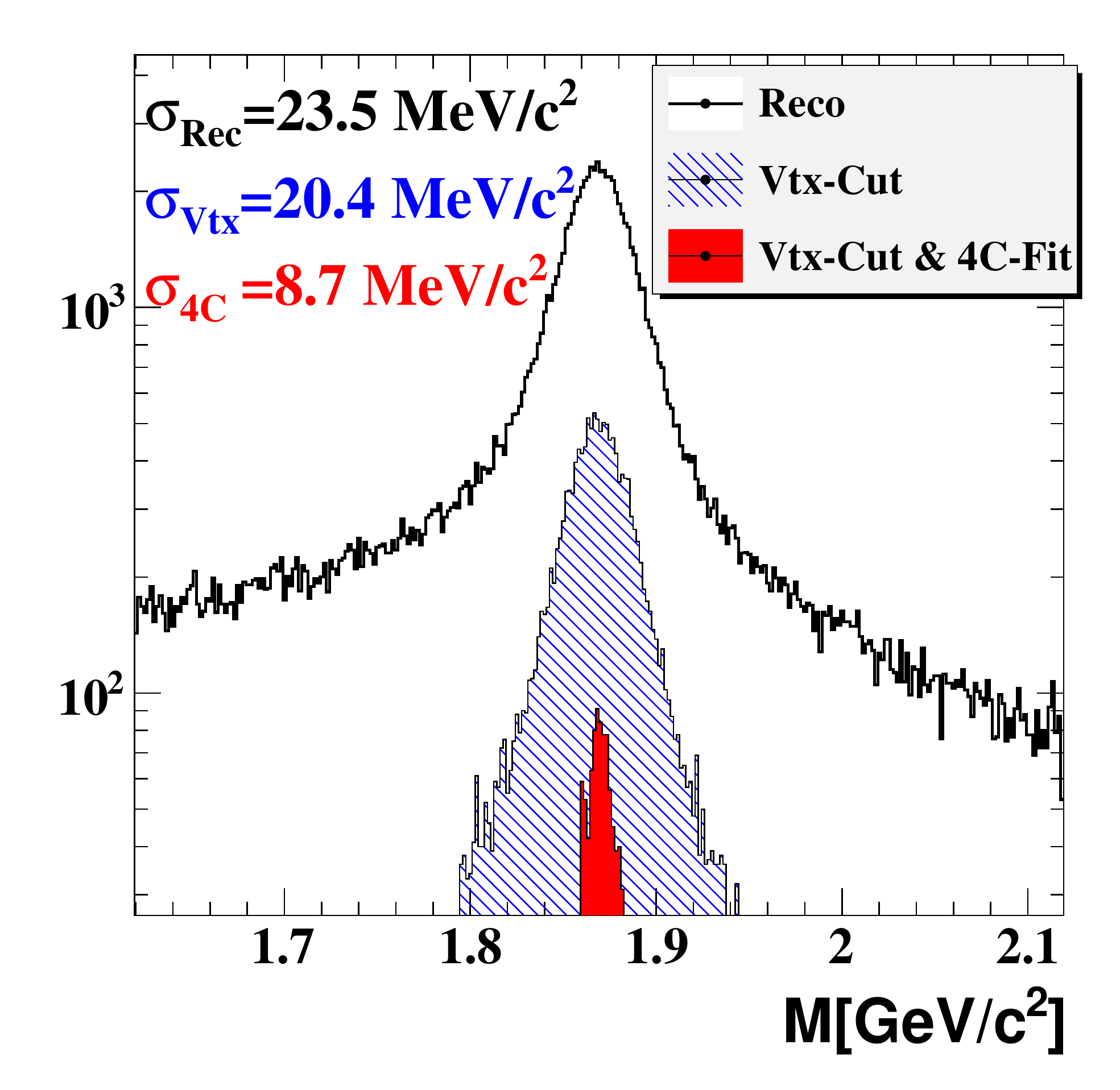}
\caption[\Dp and \Dm mass resolutions]{\Dp and \Dm invariant mass resolutions after vertex selection and after a fit constrained by the beam four-momentum. %The resolutions are $20.4\,\mevcc$ and $8.7\,\mevcc$ without and with the 4C fit, respectively. 
}
\label{channels:DpDmmass}
\end{center}
\end{figure}

After the reconstruction 60.9\% of the original \D mesons survived with 33.8\% of additional combinatorial background.
%Of the non-resonant background sample almost all of the possible \D candidates passed the detectors acceptance and reconstruction, because of the different phase space occupancy.
The exact numbers are compiled in \tabref{channels:DCounts}.
A rather wide mass window cut of $\pm$~150~\mevcc around the nominal charged \D mass (taken from \cite{PDG:2010}) was applied to reduce the non-resonant background by a large factor. 
Cutting on the vertex z coordinate (POCA) of the \D candidates (478~$\tcmu$m $< z $) and on the ``distance'' (see section \ref{sec::vtxvtx}) calculated by the POCA finder ($d < 300$~$\tcmu$m), much of the combinatorics as well as the non-resonant background is removed.
The vertex resolution values are given as the gaussian fit's width in \tabref{tab:dpdmv}. The three available vertex finder/fitter algorithms have been used. 
With a precise vertex position it is possible to measure the decay length of the \D's. We found $c\tau = $~($312.2\pm1.9$)~$\tcmu$m which is in agreement with the PDG value of $c\tau = 311.8$~$\tcmu$m (\cite{PDG:2010}, see \figref{channels:DpDmctau}). This includes also a correction for the tracking acceptance.
The POCA finder gives a good resolution, hence it was used to finally determine the mass resolution (shown in \figref{channels:DpDmmass}), which is 20.4~\mevcc by reconstruction only, and 8.7~\mevcc after the fits. 
The combinations of both \D mesons in each event are fitted with a four constraint kinematic fit. Here the knowledge of the initial states four-momentum, i.e.~the ideal beam, is introduced. After a cut on the fit's $\chi^{2}$ probability ($p(\chi^{2},\text{n.d.f.})>0.001$) most of the background sources are suppressed and the \D mass resolution is improved significantly (see \figref{channels:DpDmmass}). 

\begin{table}[]
\begin{center}
\begin{tabular}{|c|c|c|c|}
\hline  \Dp& Poca & PRG & KinVtx  \\
\hline $\sigma_\text{x}/\tcmu$m & 56.9 & 86.1 & 46.9\\
\hline $\sigma_\text{y}/\tcmu$m & 56.3 & 84.8 & 46.1\\
\hline $\sigma_\text{z}/\tcmu$m &  113 &  125 & 93.2\\
\hline
\end{tabular}
\end{center}
\begin{center}
\begin{tabular}{|c|c|c|c|}
\hline  \Dm& Poca & PRG & KinVtx  \\
\hline $\sigma_\text{x}/\tcmu$m & 57.4 & 85.3 & 46.3\\
\hline $\sigma_\text{y}/\tcmu$m & 56.0 & 84.4 & 45.7\\
\hline $\sigma_\text{z}/\tcmu$m & 110 & 123 & 94.1\\
\hline
\end{tabular}
\caption[$D^{+}$ and $D^{-}$ vertex resolutions]{$D^{+}$ and $D^{-}$ vertex resolution values obtained with the available vertex fitters.}
\label{tab:dpdmv}
\end{center}
\end{table}


An important number is the signal-to-background ratio. The number of simulated events has to be scaled by the channel cross section accordingly. The non-resonant background production cross section can be estimated to $\sigma_{\text{NR}}\approx $~23.4~$\tcmu\text{b}$ (value at 8.8~\gevc in \cite{Hill:1983se}) and for the signal to 0.04~$\tcmu\text{b}$ \cite{Kaidalov:1994mda}. 
Using the suppression factors from \tabref{channels:DCounts} and the final states branching fraction, one can estimate:
\begin{equation}
S/B 
= 
\frac{f_{\text{S}}\cdot\sigma_{\text{S}}\cdot BR}{f_{\text{BG}}\cdot\sigma_{\text{BG}}+f_{\text{C}}\cdot\sigma_{\text{S}}\cdot BR} 
\approx 
0.56
\end{equation}

 \subsubsection{$\pbarp\rightarrow \DzDzbar $}
 
\begin{figure}[]
\begin{center}
\includegraphics[width= 0.8\columnwidth]
{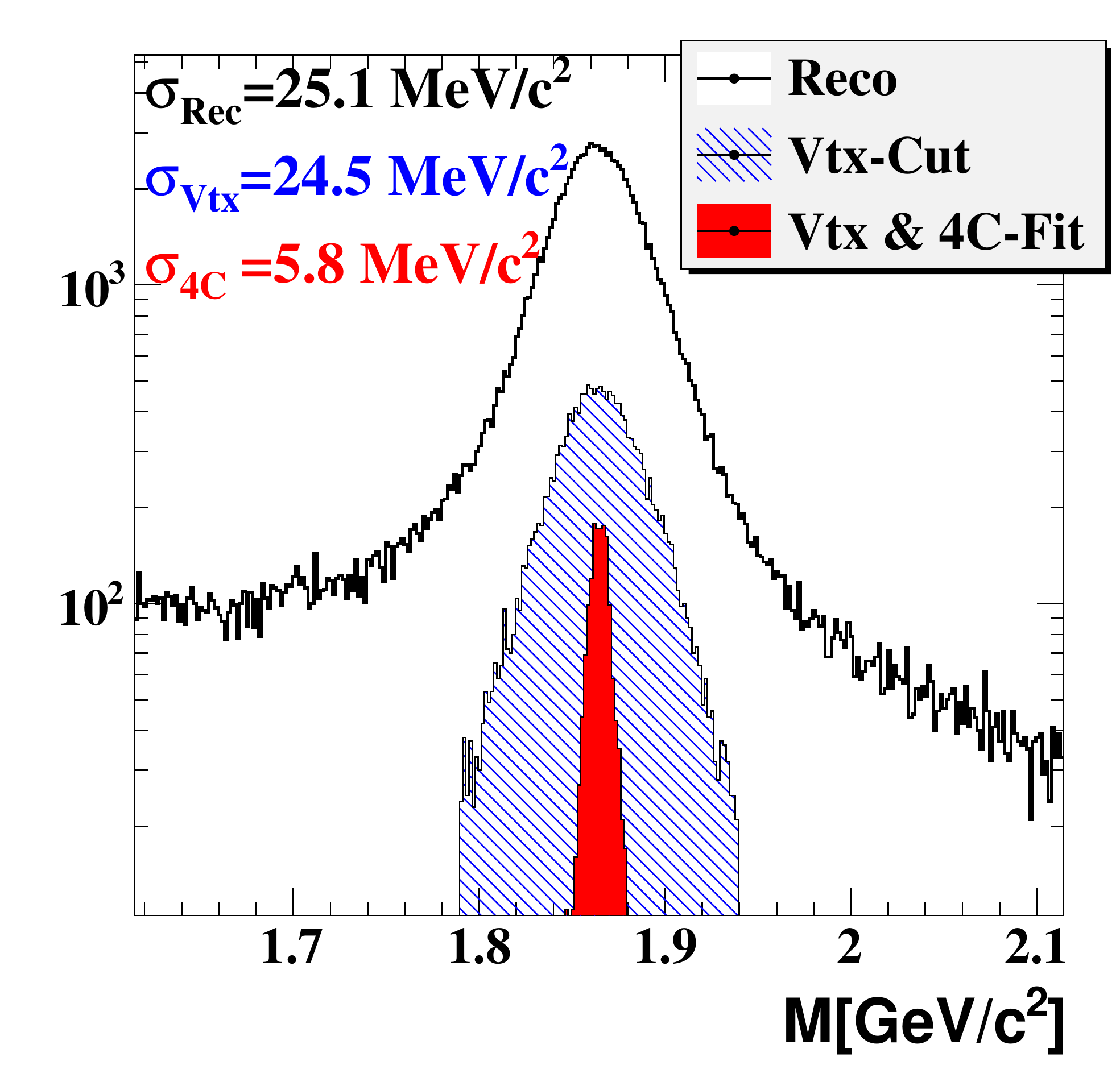}
\caption[\Dz and \Dzbar mass resolutions]{\Dz and \Dzbar Mass resolution values after Vertex selection and after a fit constrained by the beam four-momentum. 
%The resolutions are $24.5\,\mevcc$ and $5.8\,\mevcc$ without and with the 4C fit, respectively.
 }
\label{channels:DaDmass}
\end{center}
\end{figure}

%Repeating the charged case's analysis sequence, n
Neutral ground state \D mesons decay by 3.89\% into a two charged particle state (see \cite{PDG:2010}): 
\begin{equation*}\pbarp\rightarrow\DzDzbar\rightarrow K^{-}\pi^{+}K^{+}\pi^{-}\end{equation*}
 Having only two particles to form the \D a higher reconstruction efficiency is expected, due to the smaller number of particles which have to be inside the detectors acceptance. 
Out of the 98900 simulated signal events, 68.3\% where reconstructed while additional 13.7\% of combinatoric background is found. 
The numbers of surviving events after the cuts are compiled in \tabref{channels:DCounts}, while the vertex resolutions with the three fitters/finders are presented in \tabref{tab:dzdzbv}. Similar to the charged \D mesons one, the mass resolution achieved is 24.5~\mevcc before and 5.8~\mevcc after the fits (see \figref{channels:DaDmass}).
\begin{table}[]
\begin{center}
\begin{tabular}{|c|c|c|c|}
\hline  \Dz& Poca & PRG & KinVtx  \\
\hline $\sigma_\text{x}/\,\tcmu$m & 47.3 & 57.9 & 44.3\\
\hline $\sigma_\text{y}/\,\tcmu$m & 45.6 & 51.6 & 42.9\\
\hline $\sigma_\text{z}/\,\tcmu$m & 88.4 & 94.9 & 90.2\\
\hline
\end{tabular}
\end{center}
\begin{center}
\begin{tabular}{|c|c|c|c|}
\hline  \Dzbar & Poca & PRG & KinVtx  \\
\hline $\sigma_\text{x}/\,\tcmu$m & 47.5 & 58.3 & 44.6\\
\hline $\sigma_\text{y}/\,\tcmu$m & 46.3 & 51.9 & 43.5\\
\hline $\sigma_\text{z}/\,\tcmu$m & 88.4 & 94.1 & 89.3\\
\hline
\end{tabular}
\caption[\Dz and \Dzbar vertex resolutions]{\Dz and \Dzbar vertex resolution values obtained with the available vertex fitters.}
\label{tab:dzdzbv}
\end{center}
\end{table}
Extracting the decay length of the neutral \D's we find $c\tau = $~($119.7\pm0.8$)~$\tcmu$m which agrees with the PDG's input value $c\tau = 122.9$~$\tcmu$m \cite{PDG:2010}. 
 
 \begin{figure}[]
\begin{center}
%trim option's parameter order: left bottom right top
%\includegraphics[trim = 0mm 111mm 0mm 0mm, clip,width=\textwidth]{simulations/pictures/DMesons/DFitterVergleich.pdf}
\includegraphics[width= 0.8\columnwidth]
{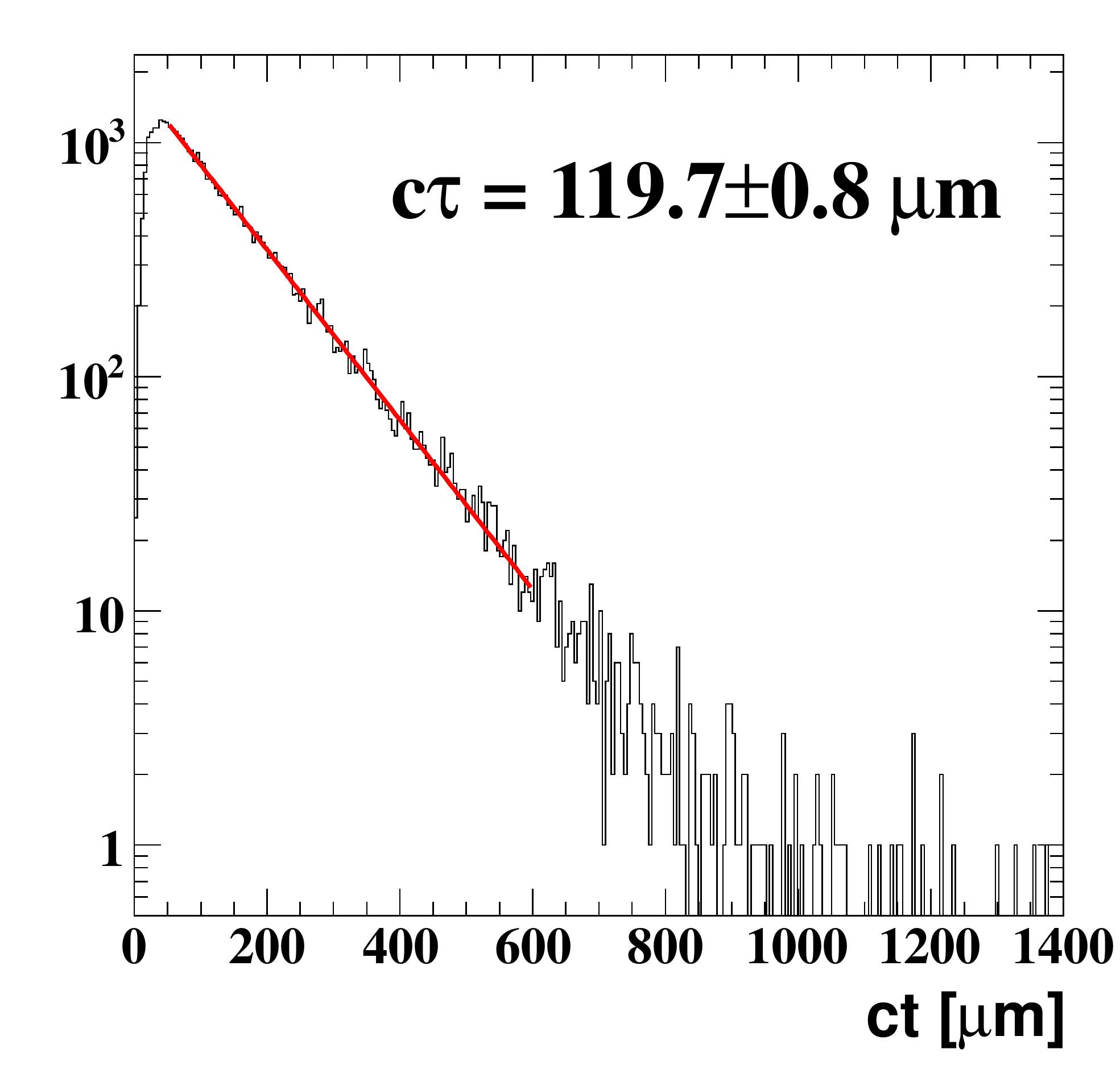}
\caption[\Dz and \Dzbar decay lengths]{Decay lengths of \Dz and \Dzbar (PDG value: $c\tau = $122.9~$\tcmu$m) \cite{PDG:2010}.}
\label{channels:DaDctau}
\end{center}
\end{figure}
 
\begin{figure}[]
\begin{center}
\includegraphics[width=\columnwidth]
{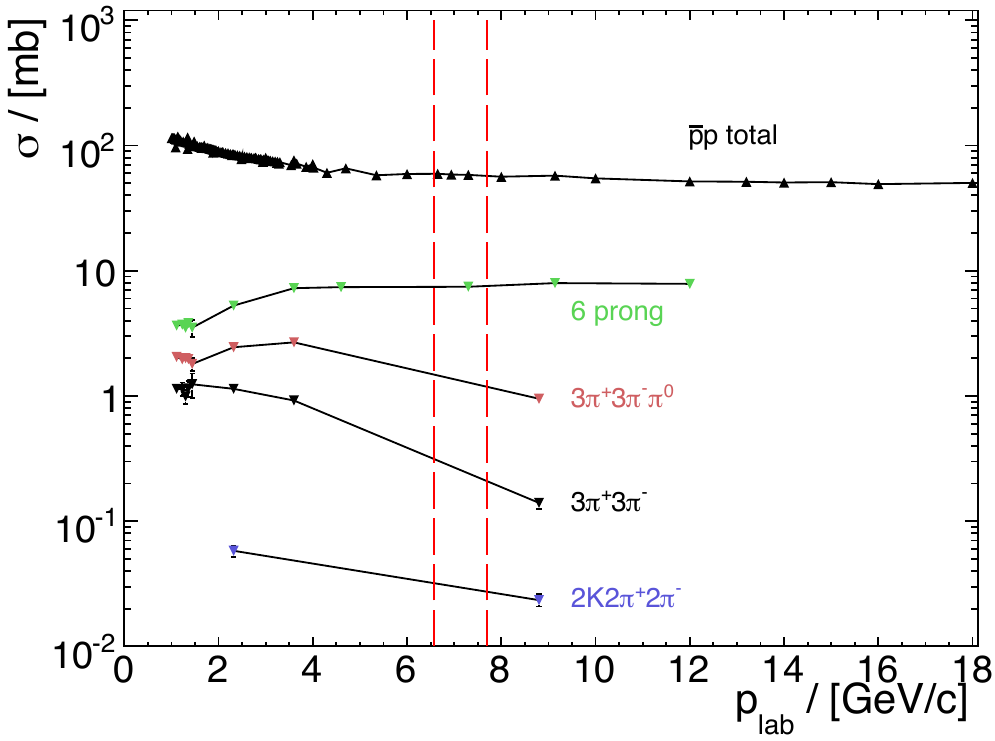}
\caption[6-prong background cross sections]{Overview about possible 6-prong background reactions and their total cross sections compared to the total \pbarp cross section \cite{thesis:rj09}.}
\label{channels:bgxsec}
\end{center}
\end{figure}

Calculating the signal-to-background ratio we approximate the cross section for the signals by using the charged \D mesons case, $\sigma_{\DzDzbar}\approx $~0.004~$\tcmu\text{b}$ and $\sigma_{\text{BG}}\approx$~8.5~$\tcmu\text{b}$ (value at 8.8~\gevc in \cite{Hill:1983se}, see also \figref{channels:bgxsec}) for the non-resonant background. The resulting signal-to-background ratio, using the numbers from \tabref{channels:DCounts}, is:
\begin{equation}
S/B 
%= 
%\frac{f_{S}\cdot\sigma_{SC}\cdot BR}{f_{BG}\cdot\sigma_{BG}} 
\approx 
%= 
0.03
\end{equation}

\subsection{Considerations About the Vertexing}

The physics channels studied in the previous sections proof the \Mvd vertexing capabilities. 
These results are not affected by the different x-y resolutions mentioned in section \ref{sec::vertexing}, since all the decays are boosted forward, where the vertexing performance of the \Mvd is homogeneous (see figure \figref{ThetaScan}).
The reconstruction of $J/\psi$ candidates showed the vertex resolutions achievable with two-body leptonic decays ($\sigma_\text{x}\sim\sigma_\text{y}\approx$ 43~$\tcmu$m, $\sigma_\text{z}\approx$ 60~$\tcmu$m).
The study of the full $\psi(2S)\rightarrow J / \psi \pi^+ \pi^-$ reaction demonstrates the improvement of the z vertex reconstruction with a bigger number of constraint (more tracks taken into account for the vertex determination). This effect is more evident in z the in the other coordinates due to the forward boost of the system.
The detailed study of the effect of several analysis techniques on the analysis of charged and neutral \D mesons highlights how crucial the \Mvd is when dealing with small cross sections and in the cases where the topology of the background are quite similar to the one of the signal under study. Vertex cuts result in being crucial for the background suppression (for example in the case of charged \D mesons nearly a factor one thousand), since they can provide an efficient particle flagging on the basis of the decay lengths. What was shown is demonstrating the key role of vertexing. Nonetheless more refined analysis strategies will be implemented in the whole scope of the full \panda experiment.

%% file: projectmanagement/projectmanagement.tex
\chapter{Project Management}

The participating institutions in the \Panda MVD project
are located at the University of Bonn, Germany (formerly
at TU Dresden), at the IKP 1 of the Forschungszentrum J\"ulich
and the University and INFN of Torino. As outlined in this 
Technical Design Report these institutions share the load
of constructing an operational inner tracker system for
the \panda detector as discussed in the following. In this
chapter, provisions taken or planned for quality control
of components upon delivery from the vendors chosen
and of subsystems will be discussed.

\section{Quality Control and Assembly}
Quality control of components upon delivery from various
vendors has already partly been discussed in the chapters
dealing with the selection of certain parts for the MVD
subsystems. Routines for test procedures such as radiation
hardness and electrical uniformity and other performance 
parameters have been developed and are described in the
previous chapters. They are briefly summarised in the following.

\subsection{Pixel}
The realisation of the hybrid pixel detector can be divided into a sequence of
elementary steps for different elements:
sensors and readout chips, modules, super modules completely equipped with 
cooling pipes and mechanics supports. Additional infrastructures are needed, such as 
electrical-optical converters, cables, power supplies and off-detector electronics.

Concerning sensors, at present epitaxial raw wafers can be expected to be produced
by ITME, which has provided the material for the sensor R\&D. 
The resistivity and the epitaxial layer thickness are directly evaluated by the 
manufacturer to confirm the specifications requested.

FBK was the manufacturer of the pixel sensor prototypes produced 
in the R\&D phase and performed the quality control of each patterned wafer.
In particular, I-V and C-V curves are deemed very sensitive and good indicators 
of device quality. Further control on test structures as
C-V on MOS structure, I-V trend on a gated diode, and 
C-V inter-pixel measurements are foreseen to investigate specific issues.
A visual inspection of metals and bonding pads 
of each sensor allows etching uniformity evaluation.

Three sensor production batches (about 15 wafers/batch) 
are sufficient to manufacture all pixel sensors needed for \panda.
An additional batch can be foreseen for spare sensors.
Wafer samples from production processes will be used to obtain diodes 
for radiation damage tests to monitor the radiation resistance uniformity, 
using neutrons from the nuclear reactor at the LENA laboratory in Pavia.

After testing, the sensor wafers are sent to manufacturers such 
as VTT or IZM in order to obtain the pixel assemblies with the 
readout chips. In particular, the Sn-Pb bump deposition is a 
delicate process but extensive screening was developed during 
the pixel detector realisation for the LHC experiments.
A failure during bump deposition will affect the whole batch under work 
and most of these failures can be detected with optical inspection,
while certain corrosion effects only show up after weeks. 
A peculiar process that has to be applied to the sensor wafer 
for the \panda MVD is the thinning in order to remove most of 
the Cz substrate. Failures during thinning can be destructive, even
breaking the wafer. An optical inspection allows to ascertain a 
positive result of the thinning process on the macroscopic level, while 
I-V measurements will be used to exclude micro breaks. 
Then, sensor dicing is the last step before the coupling to the readout 
chips.
Additional I-V tests assess the sensors to proceed to this working step.
Both VTT and IZM exhibit the capability to perform all the steps for 
target sensors, as screened during the R\&D phase.

The readout chip wafers will be produced with a dedicated 
engineering run. They will be tested using an
additional probe card on a semi-automatic probe station. 
This equipment is already present at INFN-Torino as well 
as in J\"ulich.
The test is delicate in particular due to the required
degree of cleanness to avoid contamination of the 
ASICs that pass the test and proceed to the 
bump deposition process.

The flip chip process joins pixel sensor and readout chips 
to obtain assemblies (modules). Afterwards, a test is mandatory to have 
a fast feedback into the bump bonding process, even if this test endangers 
the functionality of some of the finalised assemblies, as this process
has not yet been shown to produce fully reliable constant yields.
The test has to be planned at the vendor's lab facilities because this is 
the last step where the module can be reworked.

In parallel with the assembly of modules, production of all 
other components will be performed in order to speedily proceed to super modules.

Buses and cables are foreseen to be produced at CERN as a base solution.
Reliability tests, both with visual inspection and electrical measurements, including 
accelerated aging processes, are planned using suitable equipment at INFN-Torino.

Mechanical supports, carbon foam discs and barrel staves, will be assembled 
together with the cooling pipes at INFN-Torino using dedicated jigs. 
Each of these assemblies will be tested with a visual inspection 
and verified with a precision measurement machine. 

Dedicated jigs will be of fundamental importance for the assembly of the super modules, 
where the wire bonding process to connect readout chips to the bus and the glueing 
between the layers are critical steps at a stage where any mistake 
can potentially be dangerous to the whole system under work.
This mounting is possible at the INFN-Torino owing to the installations 
already used during the construction of the SDD of the ALICE experiment.
A final survey using a measurement machine is mandatory for mapping the system.

The powering of the detector is based on a system that relies on several steps 
from power supplies to DC-DC converters (under study at CERN) 
to the voltage regulators housed on the readout electronics.

Off-detector electronics and DAQ are under development and based 
also on electrical-optical converters under development at CERN 
(GBit project). J\"ulich is involved in the adoption of this design 
for \panda.

The MVD mechanics has been designed and evaluated using FEM analysis and
prototypes have been made for confirming the proposed concept.
Stress tests and behaviour of the mechanical setup as a function 
of humidity and temperature variations will be performed at INFN-Torino 
using suitable equipment such as a climatic chamber.

The cooling plant for the full MVD system is presently under study at INFN-Torino.

\subsection{Strips}
In the course of prototyping for the strip sensors, two
manufacturers have demonstrated their ability to deliver 
double-sided sensors of suitable quality. CiS (Erfurt/Germany)
was selected for the first prototyping run of full-size barrel 
sensors, whose performance during extensive testing has 
been 
described in the strip part of this TDR. FBK (Trento/Italy) also
delivered sensors that have been used in the course of these
tests. 

As detailed in the preceding text, parameters such as breakdown
voltage, leakage current etc. will be controlled past-production 
by the manufacturer in order to guarantee the quality of the 
sensors delivered according to the limits as discussed 
in section \ref{sec:dssd}. In the prototyping stage it was shown
that the measurement of global sensor I-V and C-V-curves is 
already a good indicator for the quality of a sensor. 
Hence, it is planned to record these characteristics 
together with visual inspection data right after the 
delivery of the sensors. Furthermore, dose-level diodes 
from each wafer will be irradiated in order to project 
the radiation fitness of the sensor from the respective wafer. 
This can be done e.g. at the fission reactor of the Reactor 
Instituut Delft, Technical University Delft/Netherlands.

Assembly of the hybrid modules will be accompanied by 
tests after each assembly stage, i.e. electrical test, 
mechanical/stress test, irradiation tests (on selected 
sub-samples) and thermal stress tests (``burn-in'').

\subsection{Integration}

The final integration of the components will be done at the \Fair site. The different pre-tested components will be shipped from the  mounting sites to Darmstadt. Here, a final electrical test will be performed to ensure that the components were not damaged during shipping. After that, the two half-shells of the MVD will be mounted starting from the outside to the inside with the strip barrel part, followed by the pixel barrel part and the forward disks.

After each mounting step the electrical and cooling connections will be tested to ensure that none of the connections broke during the handling.

Finally, a full system test will be performed before the detector can then be moved into \panda and commissioned using the \HESR beam interacting with a \panda target.

\section{Safety}

Both design and construction of the MVD including the infrastructure
for its operation will be done according to the safety requirements 
of \Fair and the European and German safety rules.
Specific aspects in the design are based on CERN guidelines, 
that are clearly appropriate for scientific installations.
Assembly and installation ask for detailed procedures 
to avoid interference with and take into account 
concomitant assembly and movement of other detectors.
Radiation damage aspects have to be checked to allow 
the protection of all people involved in the operation 
and maintenance phases.
A MVD risk analysis has to take into account mechanical and 
electrical parts, the cooling plant as well as potentially
hazardous materials, and laser diodes.

The mechanical design of the support structures of the 
silicon devices has been checked by FEM analysis, 
and the material has been selected in concordance 
with the request of limited material budget.
Cables, pipes and optical fibers are made of non-flammable 
halogen-free materials as well as are the other components 
of the MVD. 
In addition, the materials are chosen to be radiation 
tolerant at the radiation level expected in the \panda 
environment. Here, the specific CERN bibliography concerning 
the material classification in terms of radiation tolerance was 
used. All supplies have appropriate safety circuits and 
fuses against shorts and the power channels have to be 
equipped with over-current and over-voltage control circuits.
DC-DC converters have to be cooled to avoid overheating and the 
power supply cables will be dimensioned correctly to prevent 
overheating. Safety interlocks on the electrical part have to 
be planned to prevent accidents induced by cooling 
leakage from other detectors.
Laser diodes are planned in the electrical to optical 
converters, hence safe housings and proper instruction (including 
specific warning signs) have to be used.
The cooling system is a leakless circuit based on water 
as cooling fluid, working in a temperature 
range not hazardous to humans. 
A specific control system including interlocks for 
electronics manages the cooling circuit,
monitoring important parameters such as temperature, 
pressure, etc.
When a malfunctioning is detected the system 
acts to enable safety measures.
\vfill

\section{Timeline and Work Packages}
\begin{figure*}[]
				\centering
				  \includegraphics[width=0.95\textwidth]{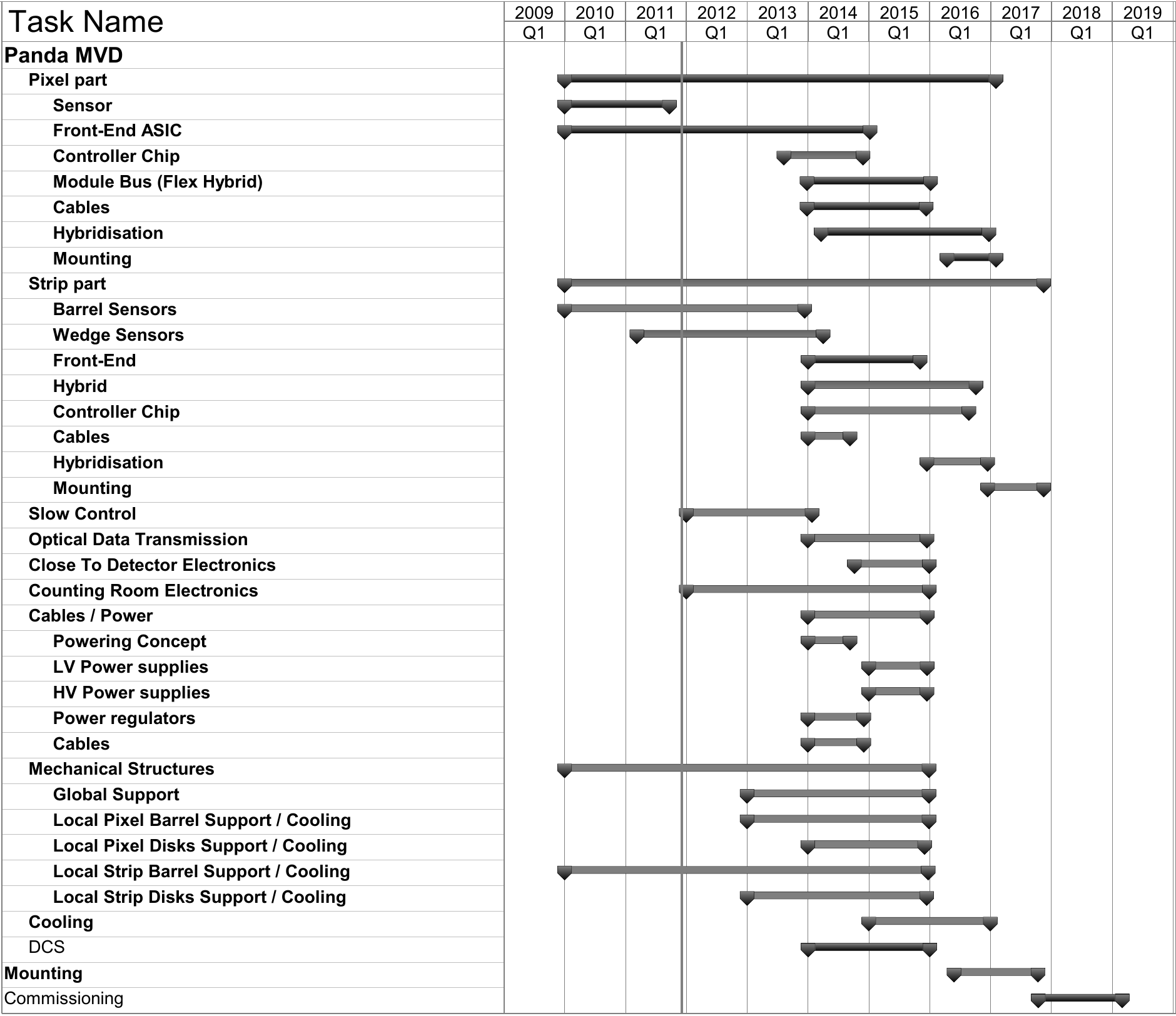}
				\caption[Generalised timeline of the \Panda MVD project]{Generalised timeline of the \Panda MVD project. Time spans needed for the realisation of different detector components are listed. The sequence of steps within a block of tasks is in general interdependent so that subsequent steps cannot be parallelised.}
				\label{fig:projectmanagement:timeline}
			\end{figure*}

The combined experience of the participating institutions and 
the available personpower is sufficient to realise the \Panda 
MVD within the allotted timeline that is summarised
schematically in \figref{fig:projectmanagement:timeline}. 
Note that the displayed time frame for the project starts in 2010.
Time spans needed for the realisation of different detector 
components are listed. The sequence of steps within a block 
of tasks is usually interdependent so that subsequent steps 
cannot be parallelised. While commissioning of the full system
is foreseen to proceed in the course of the first \Panda operation
as a complete system, smaller components will be available and
functional earlier for beam testing under realistic conditions
in order to verify the performance parameters.

Risk items have been identified and
fallback solutions are discussed in the chapters that deal
with technically challenging components whose performance
could not yet be fully predicted. This includes:
\begin{itemize}
\item {the sensors
of the hybrid pixel assembly, where the EPI sensor research
has been developed to a state where very thin sensors ($\approx 100\,\tcmu$m)
are within reach and can be expected to be produced by selected foundries
on a regular basis and with high yield. The process of bump bonding these
to the readout front-ends has successfully been performed for several
test assemblies but cannot yet be deemed fully operation for large-scale
production. A fallback solution readily at hand is the replacement of the
novel sensors by more conventional solutions for pixelised sensors that
have been used in the test phase of the hybrid pixel sensor assembly and
rest on well-tested production procedures at the cost of thicker Si layers.}
\item {the free-running front-end for the strip part of the \panda
MVD. Several solutions are under investigation and the need for a 
technical solution is common to all experiments planned at \Fair. 
Nevertheless, an ASIC fulfilling the specifications outlined in section \ref{stripASICreqs} of
this report has yet to be designed and produced in sufficiently large
quantities. A \panda -specific solution would require an enhanced effort
in ASIC design which is not part of the present project planning and 
could presumably only be achieved with additional personpower and cost.}
\end{itemize}
\vfill

\newpage
Future milestones include the 
realisation of full-size operational prototypes of the
building blocks of the \Panda MVD from the components 
whose functionality has been discussed in the chapters
of this TDR, and their assembly into operational 
small-scale systems. These will then be merged to
larger groups up to the final assembly of all parts for
the \panda experiment. 

\begin{figure*}[]
				\centering
				  \includegraphics[width=0.95\textwidth]{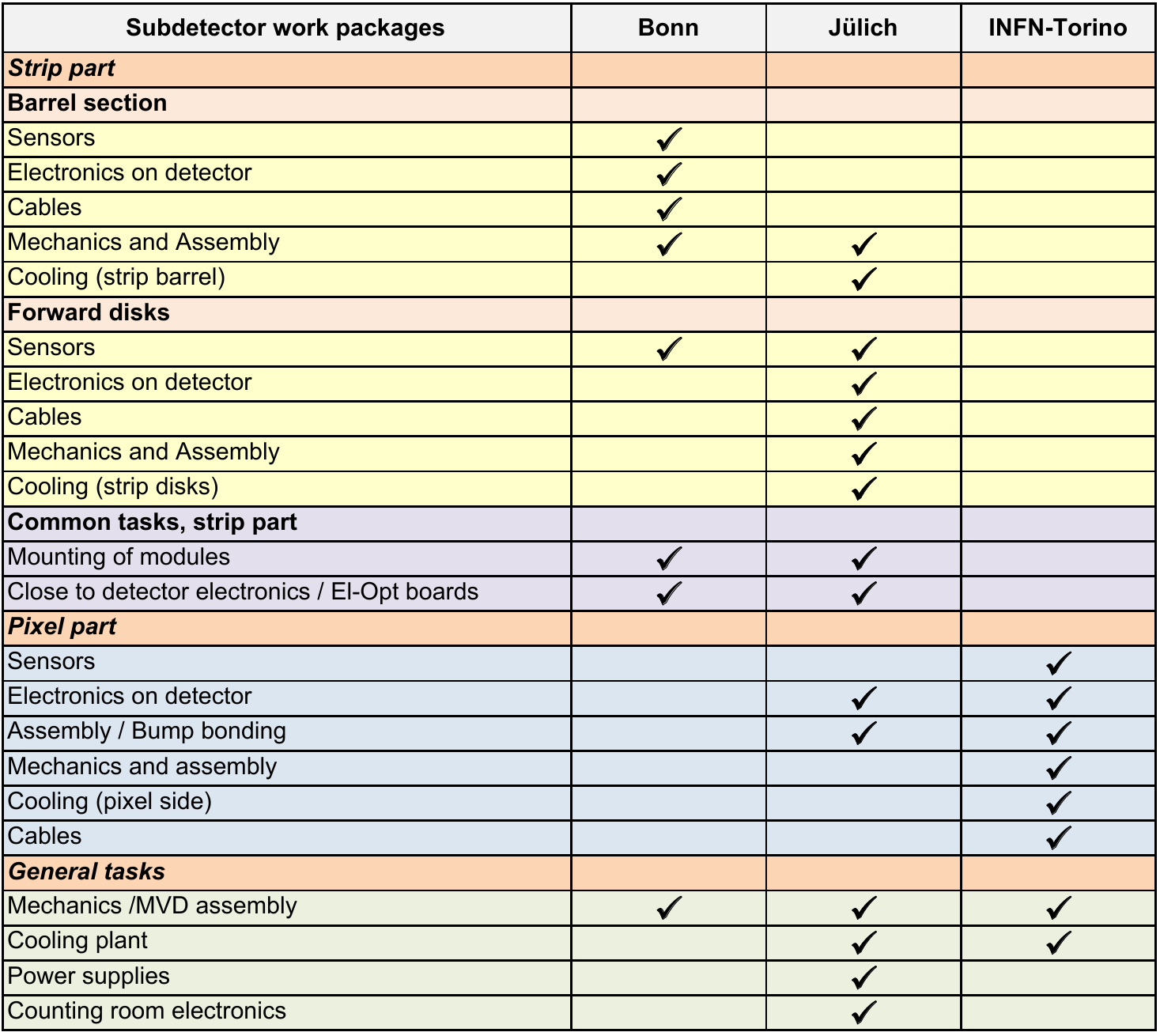}
				\caption[Work packages of the \Panda MVD project]{Work packages the \Panda MVD project and their distribution among the participating institutions.}
				\label{fig:projectmanagement:workpack}
			\end{figure*}

The work packages of the project have been 
defined and distributed among the participating institutions
as outlined in \figref{fig:projectmanagement:workpack}. 
Where responsibilities are shared,
which is also the case among the groups of Bonn and Iserlohn
in the readout of the strip sensors, one of the partners in
general provides the work package management. 

The MVD group 
will continue to work in close collaboration as demonstrated
by regular meetings using EVO and other electronic equipment
as well as frequent subgroup and specialised meetings at 
participating institutions. Reports to the \panda collaboration 
will be provided on a regular basis as done during the R\&D phase.

\section*{Acknowledgments}
We acknowledge financial support from 
the Bundesministerium f\"ur Bildung und Forschung (BmBF),
the Deutsche Forschungsgemeinschaft (DFG),
the Gesellschaft f\"ur Schwerionenforschung mbH GSI, Darmstadt, 
the Forschungszentrum J\"ulich GmbH,
the Helmholtz-Gemeinschaft Deutscher Forschungszentren (HGF), 
the Istituto Nazionale di Fisica Nucleare (INFN) and 
the Universit\`a degli Studi di Torino (Fisica).
The support of the National Funding agencies of all \panda groups and the European Union is gratefully acknowledged.
% the Swedish Research Council
% and the Polish Ministry of Science and Higher Education.

In addition, this work has been partially supported by the European Community Research
Infrastructure Integrating Activity:
HadronPhysics2 (grant agreement n. 227431) under the Sixth and Seventh Framework Program,
DIRAC secondary beams, the work-package WP26-ULISi and the FAIRnet I3.

Additionally, this work has been supported by the Helmholtz Association through funds provided to the Virtual Institute ``Spin and Strong QCD'' (VH-VI-231).

\hrulefill

\clearpage

%trim option's parameter order: left bottom right top
%\includegraphics[trim = 0mm 111mm 0mm 0mm, clip,width=\textwidth]{simulations/pictures/DMesons/DmFitterVergleich.pdf}

%% file: appendix/bonn-tracking-station.tex
%\begin{bibunit}[plain]

% ---
\chapter{The Bonn Tracking Station as a Validation for the Simulation Framework}
    
\authalert{Author: Simone Bianco}

\section{The BonnTracking Station}

\begin{figure}[!ht]
\begin{center}
\includegraphics[width=1.\columnwidth]{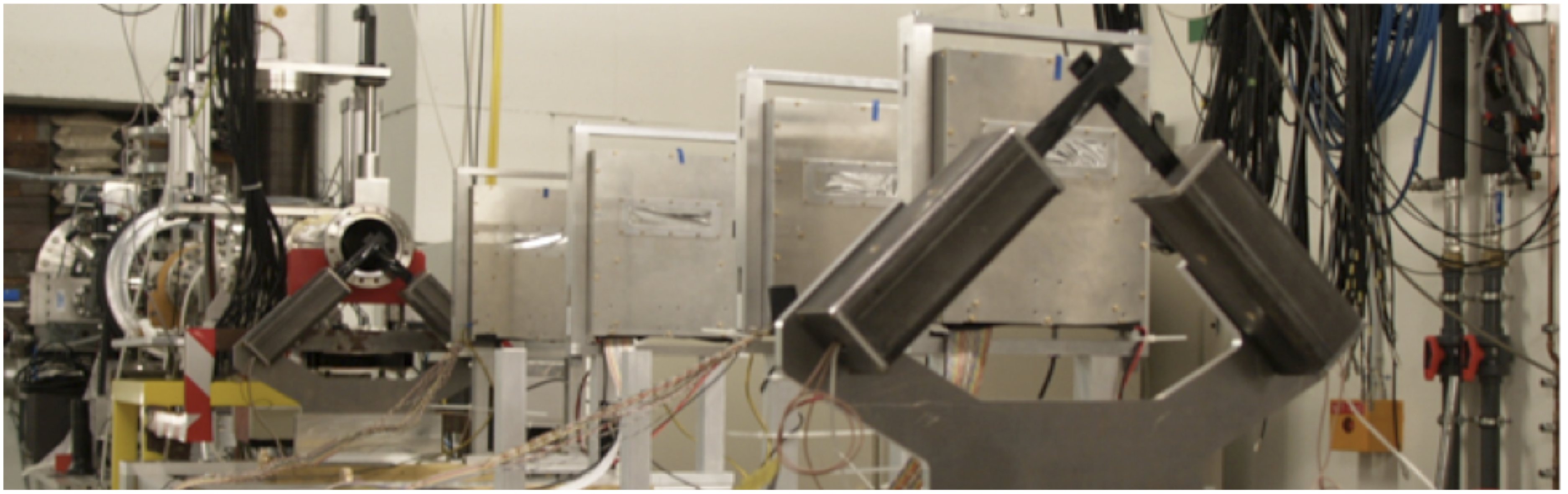}
\caption
[The Bonn tracking station set up at the COSY accelerator (J\"ulich)]
{The Bonn tracking station set up at the COSY accelerator (J\"ulich).}
\label{TSphoto}
\end{center}
\end{figure}

The implementation of silicon devices such as microstrips or pixels requires suitable tools to evaluate the main features either for sensors or for connected \frontend electronics. Among these studies those ones using a particle beam are very important. For that purpose a beam telescope based on silicon strips was designed and built by the Bonn MVD group.
It consists of four boxes positioned along the beam direction covering a maximum longitudinal range of nearly 200~cm. Each box is provided with silicon strips. Single sided and double sided sensor were used. Both kinds have an active area of $\mathrm{1.92~cm \times 1.92~cm}$, a thickness of 300~$\tcmu$m, a pitch of 50~$\tcmu$m and a 90$^{\circ}$ stereo angle. Four scintillators are disposed along the beam axis: two of them before and two after the silicon sensors. They generate signals which are used for triggering the readout. The tracking station allows to change the longitudinal position of each box and provides a specific holding frame permitting rotations of the device under test, which is located in the center of the telescope. The operative setup of the tracking station is shown in \figref{TSphoto}.

The tracking station has been tested in different beam conditions:
\begin{itemize}
\item	2.95~\gevc and 893~\mevc protons at COSY (J\"ulich);
\item	1 to 5~\gev electrons at DESY (Hamburg);
\item	Bremsstrahlung photons from an electron beam of 2~\gev at ELSA (Bonn).
\item	lepton pair production from Bremsstrahlung photons at ELSA (Bonn).
\end{itemize}

\subsection{Layout of the DAQ Chain}
\label{app:TS-DAQ}

\begin{figure}[!ht]
\begin{center}
\includegraphics[width=1.\columnwidth]{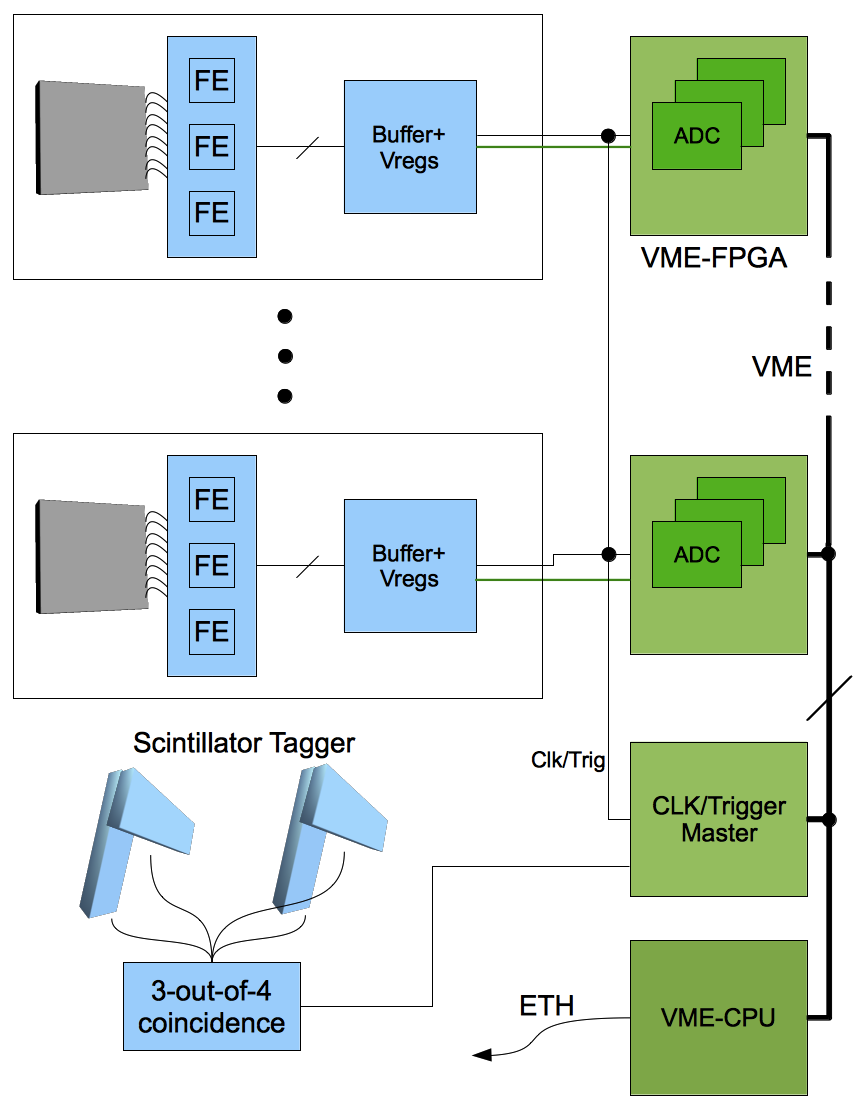}
\caption
[Layout of the DAQ chain]
{Layout of the DAQ chain.}
\label{daq}
\end{center}
\end{figure}

The data acquisition and reduction is done in several layers. First, all channels of the Si-Strip Sensors are time discretely sampled after amplification and pulse shaping through APV25 \frontend ASICs when an external trigger is asserted. The resulting analog samples are multiplexed in a single mixed (analog and digital) output per \frontend that is sampled by an ADC running synchronous to the \frontend clock. The ADCs are actually plug-in cards attached to universal FPGA-Modules with VME form factor (see \figref{daq})~\cite{Robert_twepp}. The FPGA configuration implements the following stages:
\begin{itemize}
\item frame decomposition
\item baseline compensation
\item pedestal tracking
\item hit-over-threshold assertion
\item cluster finding
\item FIFO buffering
\item noise statistics
\item slow-control interface controller
\end{itemize}
The resulting sparsified digitised cluster data is then read out via the VME-Controller through standard network link and stored to disk and processed for online monitoring.

\subsection{Structure of the Analysis Tools}

\begin{figure}[!ht]
\begin{center}
\includegraphics[width=1.\columnwidth]{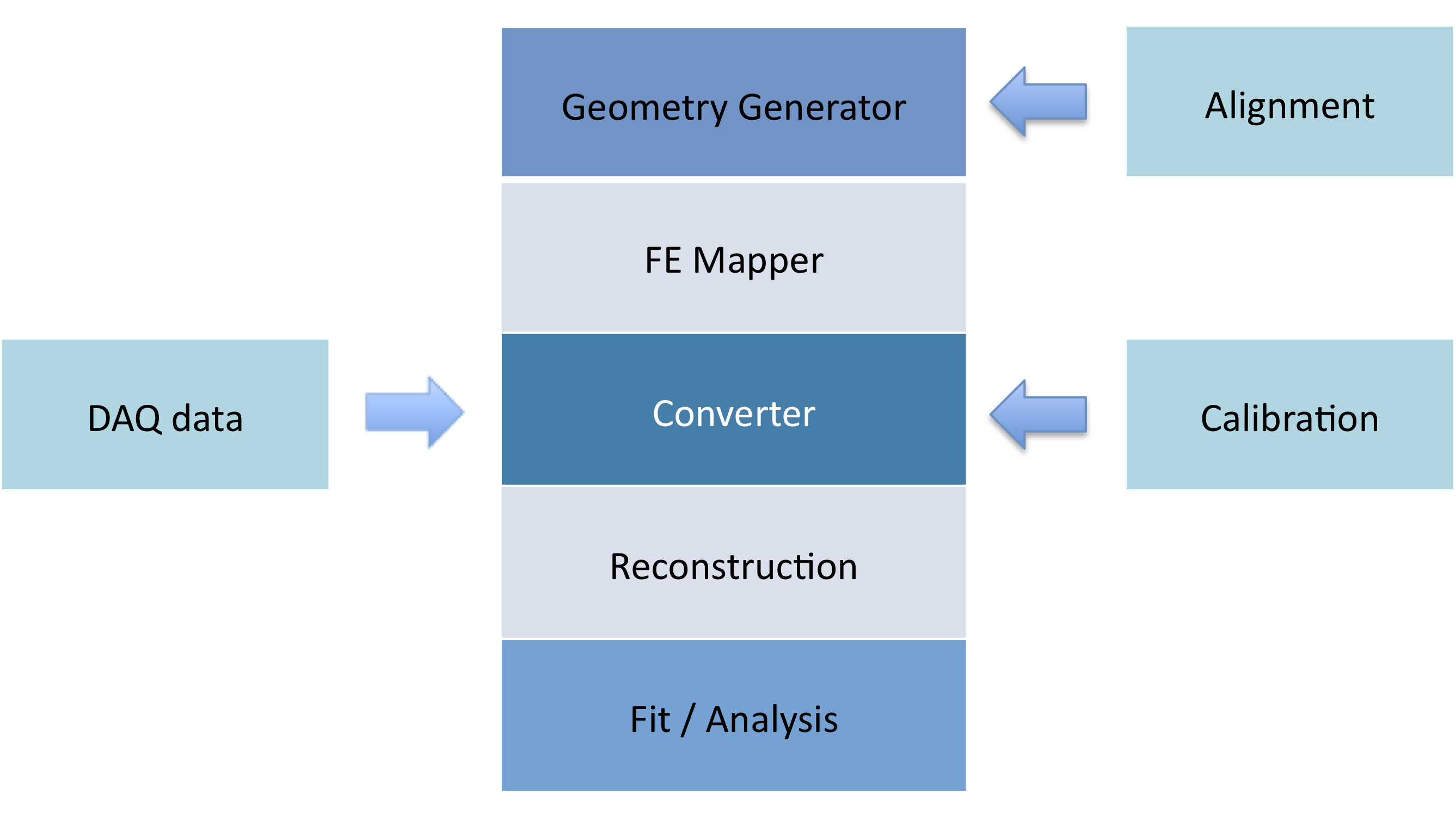}
\caption
[Structure of the analysis tools]
{Structure of the analysis tools.}
\label{anaTools}
\end{center}
\end{figure}

The PandaRoot framework has been used to analyse the results of the measurements performed with the tracking station (see~\cite{IEEE_Bianco}). A dedicated infra\-structure was required to import the DAQ raw data in the framework and there perform digitisation, hit reconstruction and analysis with the same tools usually applied to simulated data. The structure of these tools is shown in \figref{anaTools}. 

The first step toward the final analysis is the alignment. This was done in two steps: first an optical check by means of  a laser pointer was performed during the installation of the modules in the beam lines. Then an iterative off-line procedure based on the residuals method was applied to measured data in order to find the real positions of the sensors. This information was inserted in the geometry definition provided to the analysis framework. A map linking \frontend channel numbers to simulation-like strip identifiers was implemented.
The calibration of the sensors envisages the energy loss measurements available with this kind of detectors. The same charge was injected on each channel of the \frontend electronics to resolve differences in the response. 
%Then, with a minimum ionizing hypothesis on the particles used for these runs, an absolute ADC counts-to-energy-loss scale was fixed.

\begin{figure}[!ht]
\begin{center}
\includegraphics[width=1.\columnwidth]{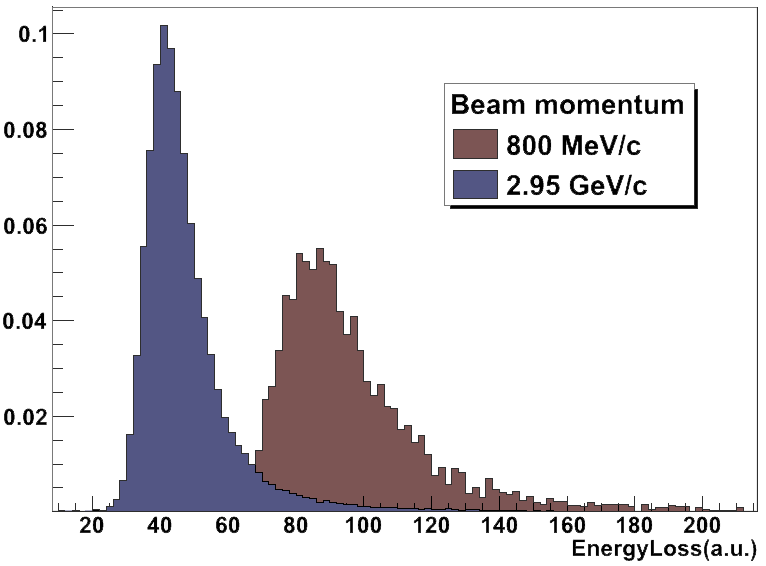}
\caption[Energy loss measured in one sensor with beams of protons of two different momenta]
{Energy loss measured in one sensor with beams of protons of two different momenta.}
\label{ElossComp}
\end{center}
\end{figure}

This set of tools allowed to perform clustering and hit reconstruction inside the PandaRoot simulation framework. This is a key feature to directly compare simulated and measured data, as shown in the next sections. An example of the functionality of this infrastructure is shown in \figref{ElossComp}, where energy losses obtained with protons of different momenta, are compared.

\subsection{Rotation of One Sensor}

\begin{figure}[!ht]
\begin{center}
\includegraphics[width=1.\columnwidth]{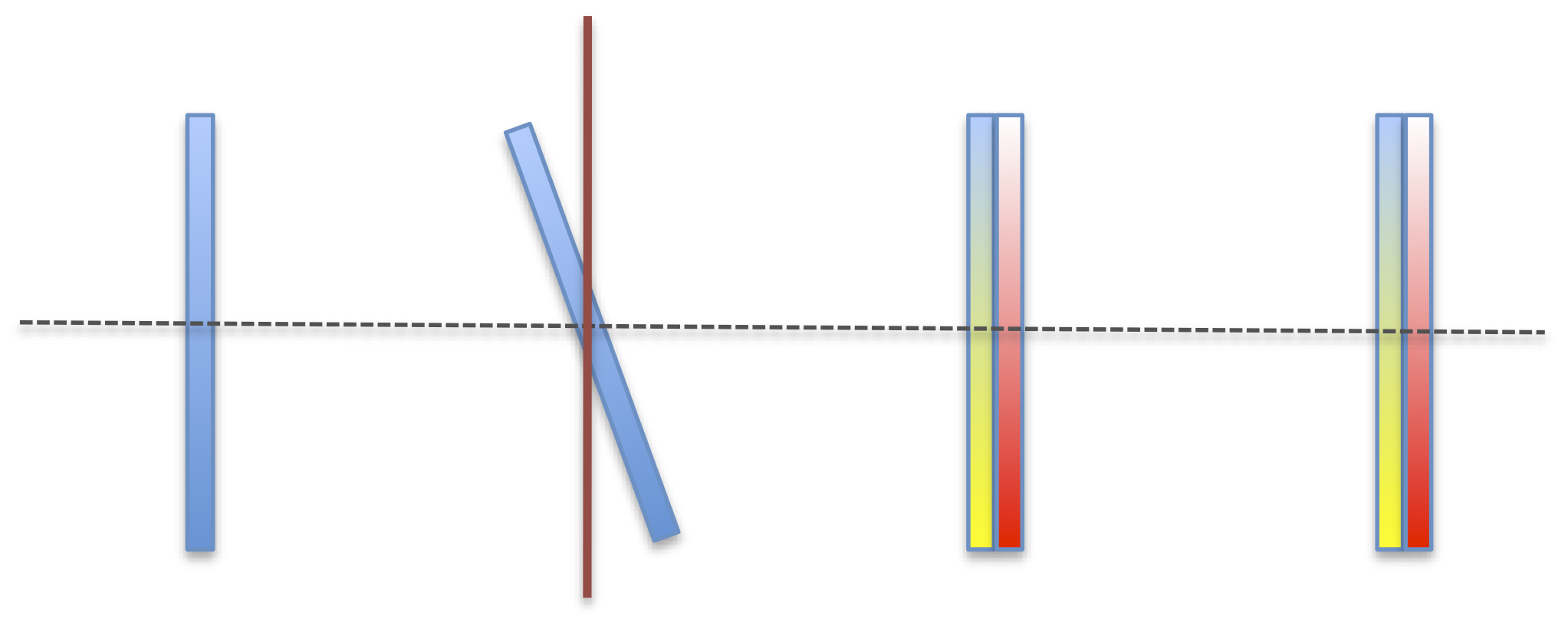}
\caption
[Top view of the setup used for the measurements performed rotating one double sided sensor]
{Top view of the setup used for the measurements performed rotating one double sided sensor.}
\label{Rotation}
\end{center}
\end{figure}

\begin{figure}[!ht]
\begin{center}
\includegraphics[width=1.\columnwidth]{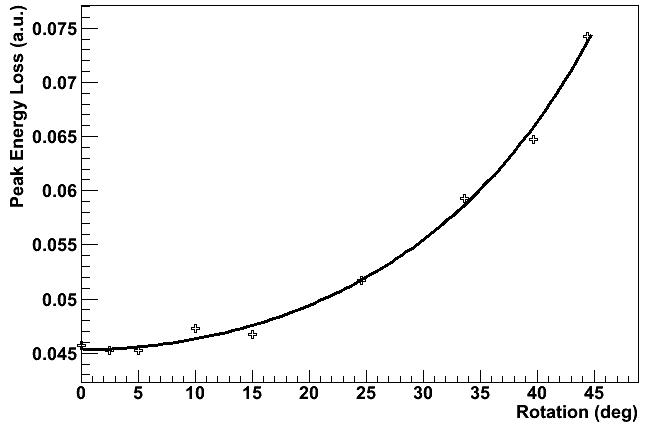}
\caption
[Peak energy loss of 4~\gev electrons measured in one sensor as a function of the rotation angle of the module with respect to the beam direction]
{Peak energy loss of 4~\gev electrons measured in one sensor as a function of the rotation angle of the module with respect to the beam direction.}
\label{RotMeas}
\end{center}
\end{figure}

\begin{figure}[!ht]
\begin{center}
\includegraphics[width=1.\columnwidth]{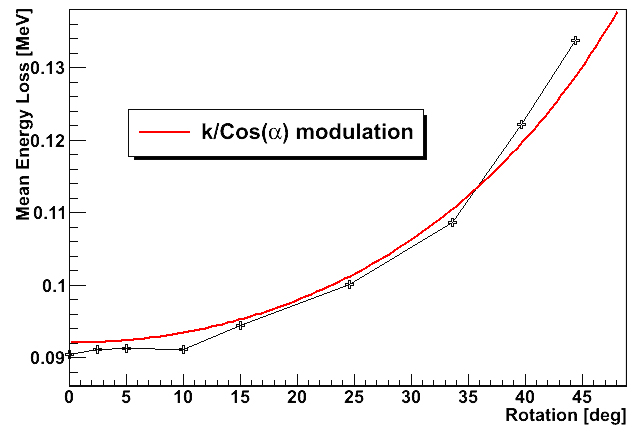}
\caption
[Simulation of the measurements shown in \figref{RotMeas}]
{Simulation of the measurements shown in \figref{RotMeas}.}
\label{RotSim}
\end{center}
\end{figure}

The tracking station is designed to permit the rotation of one sensor, changing in this way the incident angle of the beam to the detector plane (see \figref{Rotation}).
This feature is important to study the effect on cluster size and energy loss distributions. Increasing the incident angle a particle will cross the sensor along a longer path, thus having in average a bigger energy loss and hitting more strips.
These effects were studied during a beam test at DESY, where 4~\gev electrons were casted. There one of the double sided sensors was positioned on the rotating holder and several incident angles were set.
As foreseen, the peak position of the energy loss distributions moves to higher values as the incident angle gets bigger (see \figref{RotMeas}). 
A similar trend is predicted by simulations summarised in \figref{RotSim}.

\begin{figure}[!ht]
\begin{center}
\includegraphics[width=1.\columnwidth]{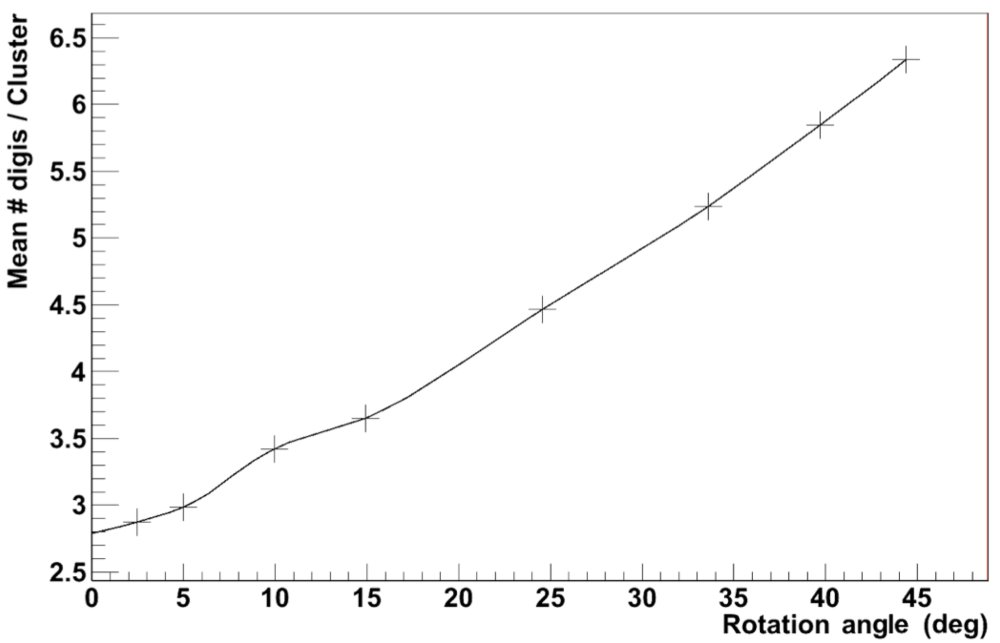}
\caption
[Measured cluster size (in terms of digis) as a function of the rotation angle]
{Measured cluster size (in terms of digis) as a function of the rotation angle.}
\label{CluMeas}
\end{center}
\end{figure}

The cluster size is another variable influenced by the incident angle of the beam. \Figref{CluMeas} shows the average size of the clusters (expressed in terms of number of strips on both sides belonging to the same cluster) as a function of the rotation angle. An enhancement is present for an angle of about 9$^{\circ}$. Such a particular value corresponds to cross the full width of a strip, while spanning the full thickness of the sensor\footnote{$\alpha=\arctan \left(\frac{pitch}{thickness}\right)=\arctan\left(\frac{50 \tcmu m}{300 \tcmu m}\right)=9.46^{\circ}$}. 
\begin{figure}[!ht]
\begin{center}
\includegraphics[width=1.\columnwidth]{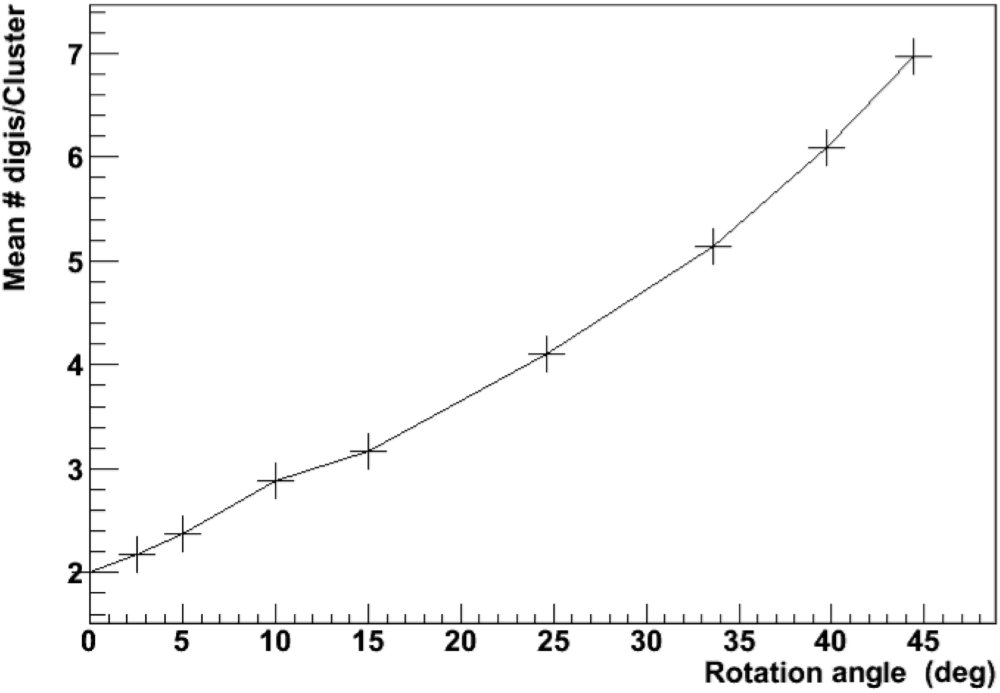}
\caption
[Simulation of the effect of rotation on the cluster size]
{Simulation of the effect of rotation on the cluster size.}
\label{CluSim}
\end{center}
\end{figure}
This feature is present in simulations as well, as shown in \figref{CluSim}. Simulations present some differences in the cluster sizes, although the trend is compatible with that one obtained with measured data. This is due to some thresholds set in the MC simulations, for example the minimum charge that can be collected by each strip. We could not set these thresholds according to the values observed with the real devices, since, at the moment, the absolute ADC energy loss calibration is not taken into account. Changes of these values can have a significant impact on the cluster size. A too high value for this threshold can result in a cluster with smaller multiplicity than that one obtained from measurements. On the other side too low thresholds in MC simulations can lead to bigger clusters than the measured ones, since strips where the collected charge is smaller than the hardware threshold are taken into account.

\subsection{Shift of One Sensor}

\begin{figure}[!htb]
\begin{center}
\includegraphics[width=1.\columnwidth]{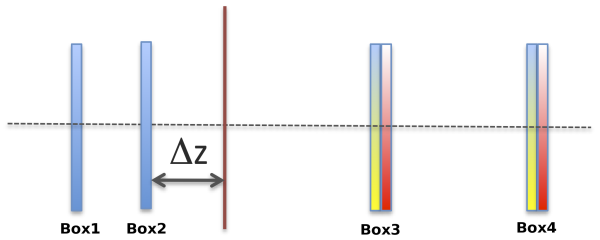}
\caption
[Use of the second box to scan different longitudinal positions]
{The figure shows how the second box was used to scan different longitudinal positions.}
\label{Shift}
\end{center}
\end{figure}

\begin{figure}[!ht]
\begin{center}
\includegraphics[width=1.\columnwidth]{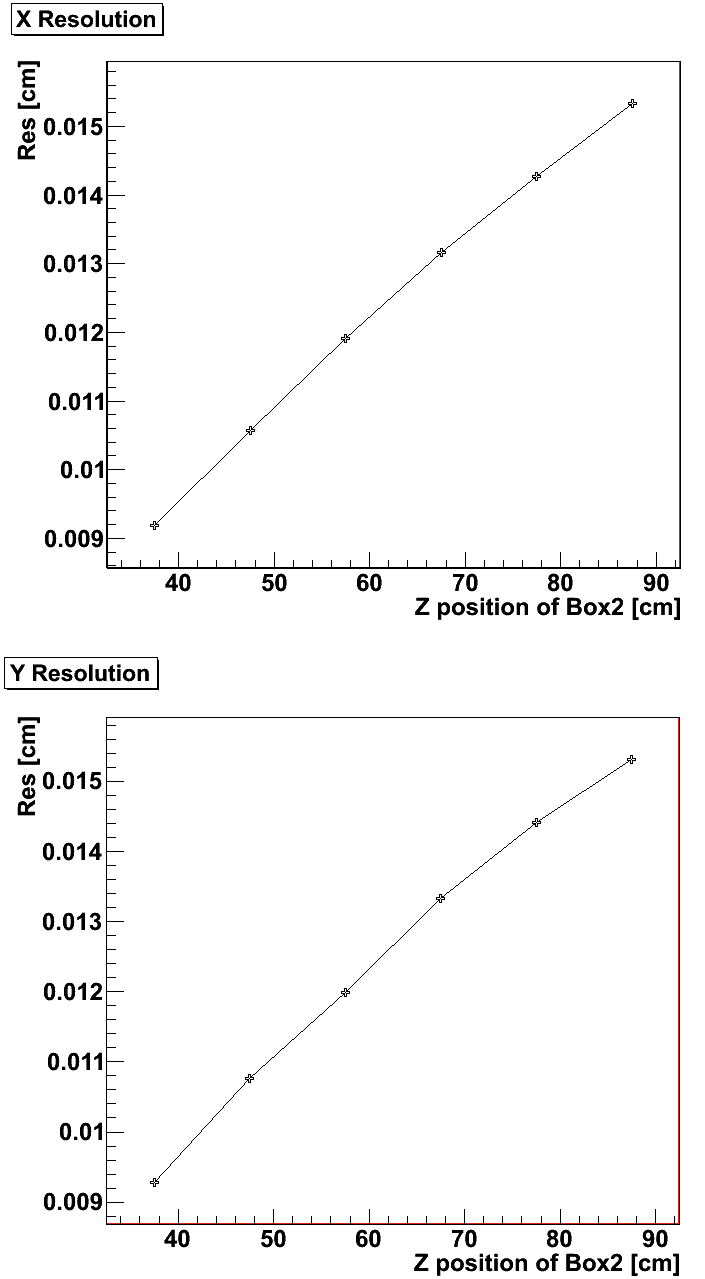}
\caption
[Results of the longitudinal scan performed at DESY with 3~\gev electrons]
{Results of the longitudinal scan performed at DESY with 3~\gev electrons. The second box was moved along the tracking station in a 50~cm wide range. The figures show the estimator previously defined as a function of the position of the second box.}
\label{ShiftMeas}
\end{center}
\end{figure}

The longitudinal position of the four boxes can be modified in a wide range. One of the sensors was moved along the beam axis during a beam time with 3~\gev electrons at DESY (see \figref{Shift}).
This scan was performed in order to understand the effect of the position of the second box on the tracking performance. Therefore, the unbiased resolution was extracted with different setups:
$$RES_{x}=\sqrt[4]{\sigma_{1}^x*\sigma_{2}^x*\sigma_{3}^x*\sigma_{4}^x}$$
where $\sigma_i^x$ is the standard deviation of the distribution of the x-residuals for the $i^{th}$ sensor.
Results are shown in \figref{ShiftMeas}.
Measurements show that the best unbiased resolution is obtained when the second box is as close as possible to the first one. This means that the effect of multiple scattering in the first box is not negligible at these energies and the best condition is the one where a second point is measured before the scattering cone diverges too much from the straight trajectory.

\subsection{Resolution Estimations}

The simulation framework allows to investigate the resolutions achievable with different setups. A comparison has been realised using three different setups and a beam of 5~\gev electrons. The simulations were performed placing and ideal silicon sensor with infinite precision at the center of the tracking station (see table \ref{tabTS}). This ideal sensor was set as a 300~$\tcmu$m thick silicon volume with square area. 
\begin{table}[!h]
\centering
\begin{tabular}{!{\vrule width 1.5pt}c|c|c|c|c|c!{\vrule width 1.5pt}}
 \noalign{\hrule height 1.5pt}
  & B1 & B2 & Device & B3 & B4 \\ \noalign{\hrule height 0.15pt}
 & z (cm)  & z (cm) & z (cm) & z (cm) & z (cm) \\ \hline \hline
A &16. & 86. & 110. & 145. & 185.5 \\ \hline 
B & 90. & 100. & 110. & 120. & 130.\\ \hline
C & 65. & 85. & 110. & 139. & 159.\\ \noalign{\hrule height 1.5pt}
\end{tabular}
\caption
[The position of the four boxes and of the ideal sensor]
{The position of the four boxes and of the ideal sensor are shown.}
\label {tabTS}
\end{table}
The position measured by this device was compared with the one extrapolated from the tracking station hits. This last evaluation was done calculating the intersection of the straight line fitted on the tracking station points with the plane of the ideal sensor.
Histograms were filled with the x and y distance between the measured and the extracted points and the distributions were fitted with Gaussians. The results are shown in \tabref{tabRES}.
\begin{table}[!h]
\centering
\begin{tabular}{!{\vrule width 1.5pt}c|c|c!{\vrule width 1.5pt}}
 \noalign{\hrule height 1.5pt}
Setup & $\sigma_x$ & $\sigma_y$\\ \noalign{\hrule height 0.15pt}
 & $\tcmu$m & $\tcmu$m\\ \hline \hline
A & 56 & 53 \\ \hline
B & 16 & 16 \\ \hline
C & 34 & 34 \\ \noalign{\hrule height 1.5pt}
\end{tabular}
\caption[Results of the simulations performed with different setups]
{Results of the simulations performed with different setups.}
\label {tabRES}
\end{table}
The best setup (B) is that one with all the sensors placed as close as possible to the device under study. This setup is not feasible with the actual tracking station due to some limitations of the holding structures. The setup C represents the positioning with the smallest achievable distances between the device and the tracking station boxes.

\subsection{Scattering Measurements}

\begin{figure}[!ht]
\begin{center}
\includegraphics[width=1.\columnwidth]{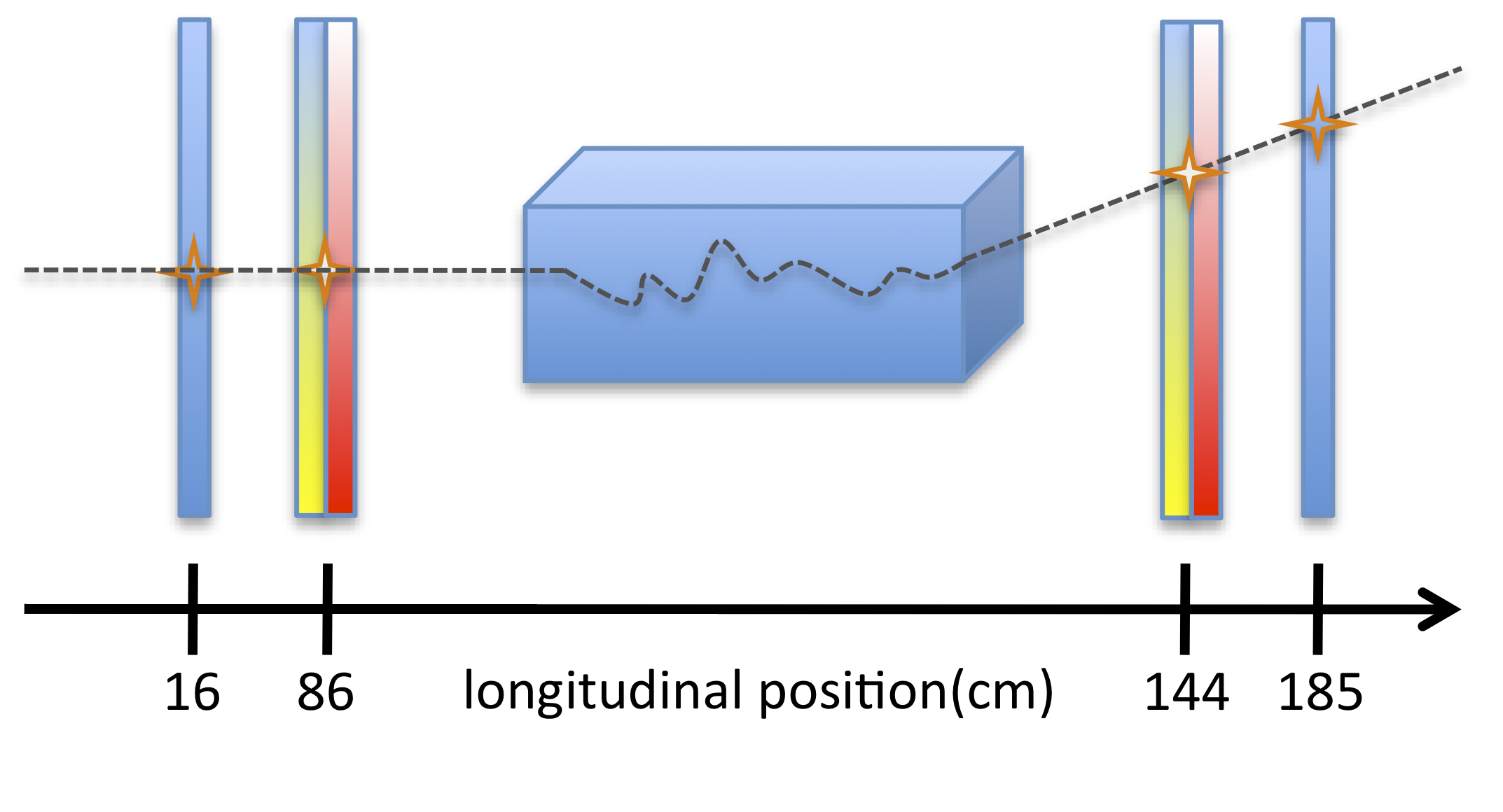}
\caption
[Setup used to measure the multiple in different carbon volumes]
{Setup used to measure the multiple in different carbon volumes. One double-sided strip sensor and two single-sided ones were used both before and behind the volumes.}
\label{ScattBox}
\end{center}
\end{figure}

The tracking station has been used to measure scattering in some volumes of different materials both with protons at COSY and with electrons at DESY. 
Measurements have been performed placing the scatterers on the central holder of the tracking station, with two hits measured upstream and two downstream of the volume (see \figref{ScattBox}). 
The samples used for these studies were carried out with materials foreseen in the implementation of the MVD support structures. Both carbon and carbon foams were characterised, as well as complex prototypes for the pixel staves:
\begin{itemize}
\item	empty volume: just air between the boxes;
\item	1~cm thick carbon volume with a density of $\sim$~1.79~g/cm$^3$;
\item	2~cm thick carbon volume with a density of $\sim$~1.70~g/cm$^3$;
\item	2.5~cm thick of a carbon foam with a density of $\sim$~0.52~g/cm$^3$;
\item a 4~mm thick pixel stave prototype realised with carbon foam and embedded cooling pipes (see \figref{fig:coolFig_07}). 
\end{itemize}

\begin{figure}[!ht]
\begin{center}
\includegraphics[width=1.\columnwidth]{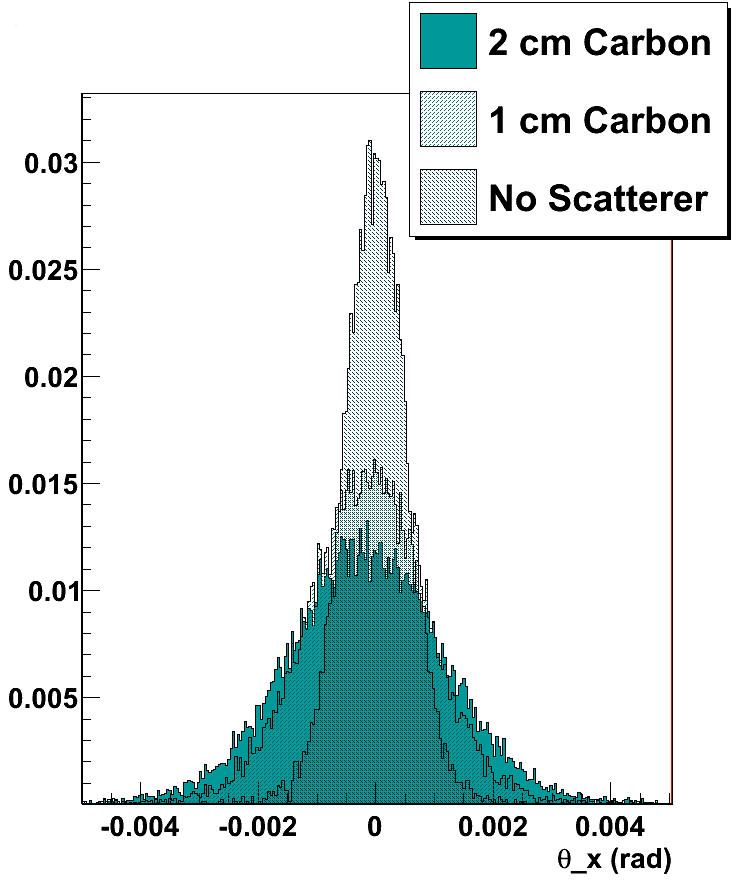}
\caption
[Distributions of the projected scattering angles obtained with 2.95~\gevc protons and the previously described scattering volumes]
{Distributions of the projected scattering angles obtained with 2.95~\gevc protons and the previously described scattering volumes.}
\label{ScattMeas}
\end{center}
\end{figure}

The analysis focused on the determination of the projected scattering distributions. For each setup the x and y projected scattering angle distributions were measured (same examples are shown in \figref{ScattMeas}). In their central part these distribution are well approximated by Gaussian centered around zero. In the following we will refer to $\sigma_x$ and $\sigma_y$  meaning the standard deviations obtained with a Gaussian fit, on the x and y projected distributions, respectively.

The results obtained allow to distinguish the effect of different scatterers, even with light ones. In order to validate the simulation framework as a tool for estimates of the effect of passive materials a full set of simulations has been performed. All the setups used in real measurements were reproduced allowing a direct comparison of simulations and real data. \Tabref{tabUno} summarises the results obtained from measurements and simulations using different scatterers and particle beams.
\begin{table}
\centering
\begin{tabular}{!{\vrule width 1.5pt}c|c|c|c|c!{\vrule width 1.5pt}}
 \noalign{\hrule height 1.5pt}
Vol. & Ptc & Mom. & Meas. $\sigma$ & Sim. $\sigma$ \\ \noalign{\hrule height 0.15pt}
  &  & \gevc & mrad & mrad \\ \hline \hline
1 cm C & p$^+$ & 2.95 & 1.02 & 1.01 \\ \hline
2 cm C & p$^+$ & 2.95 & 1.34 & 1.33\\ \hline
Air & e$^-$ & 1.0 & 1.24 & 1.40\\ \hline
Air & e$^-$ & 3.0 & 0.423 & 0.476\\ \hline
Air & e$^-$ & 5.4 & 0.243 & 0.284\\ \hline
Foam & e$^-$ & 1.0 & 2.18 & 2.54\\ \hline
Stave & e$^-$ & 1.0 & 1.76 & 1.87\\ \hline
Stave & e$^-$ & 3.0 & 0.601 & 0.611\\ \hline
Stave & e$^-$ & 4.0 & 0.471 & 0.483\\ \noalign{\hrule height 1.5pt}
\end{tabular}
\caption[Comparison between the sigma of Gaussian fits on the distributions of the projected scattering angles]
{Comparison between the sigma of Gaussian fits on the distributions of the projected scattering angles obtained with real measurements and Geant3 simulations using different scatterers.}
\label {tabUno}
\end{table}
Simulations are in full agreement with the measurements performed with protons. A discrepancy of the order of some 10\%, between measured and simulated values, has been found for electrons, due to a stronger effect foreseen in Geant3. Anyway, the scaling of scattering with momenta and different scatterers is really well described by simulations and we can rely on the predictions of the simulation framework to evaluate the effect of passive materials.

%\cite{IEEE_Bianco}

%\section*{Authors}
%\begin{tabbing}
%\hspace{3cm} \=S. Bianco\\
%\end{tabbing}

\putbib[lit_appendix]

%\clearpage

%\end{bibunit}

% - EOF

%% file: appendix/juelich-readout-system.tex
\chapter{J\"{u}lich Readout System}

\authalert{Author: Simone Esch}

\section{Overview}

For the development of the Micro Vertex Detector (MVD) the evaluation of prototypes and detector parts is very important.
Different prototypes of the pixel \frontend chip ToPix (\textbf{To}rino \textbf{Pix}el) need to be tested and characterised under similar conditions to improve the development.

To control these devices under test (DUT) and to save the taken data a suitable readout system is necessary.
To have similar conditions for different prototypes and development stages a modular concept of a readout system is required which can be adapted in a simple way to the specific interface of different types of electronics.
% At the same time high performance should be provided to handle the data volume of a hole detector systems. 
At the same time the system must provide a high performance to allow the evaluation of single \frontend chips as well as fully assembled modules. 
The possibility to implement routines for online data processing is also desirable. 

A digital readout system has been developed to meet all these requirements for the development of the MVD, the J\"{u}lich Readout System.

%For the development of a detector the evaluation of prototypes and detector parts is very important. 
%To control the devices unter test (DUT) and to save the taken data a suitable readout system is necessary.
%To have similar conditions for different prototypes and development stages a modular concept of a readout system is preferable which can be adapted in a simple way to the specific duties.
%At the same time high performance should be provided to handle the data volume of a hole detector systems. 

%A digital readout system has been developed to meet all these requirements for the development of the MVD, the J\"{u}lich Readout System.

\begin{figure}[th]
 \centering
 \includegraphics[width=0.7\columnwidth]{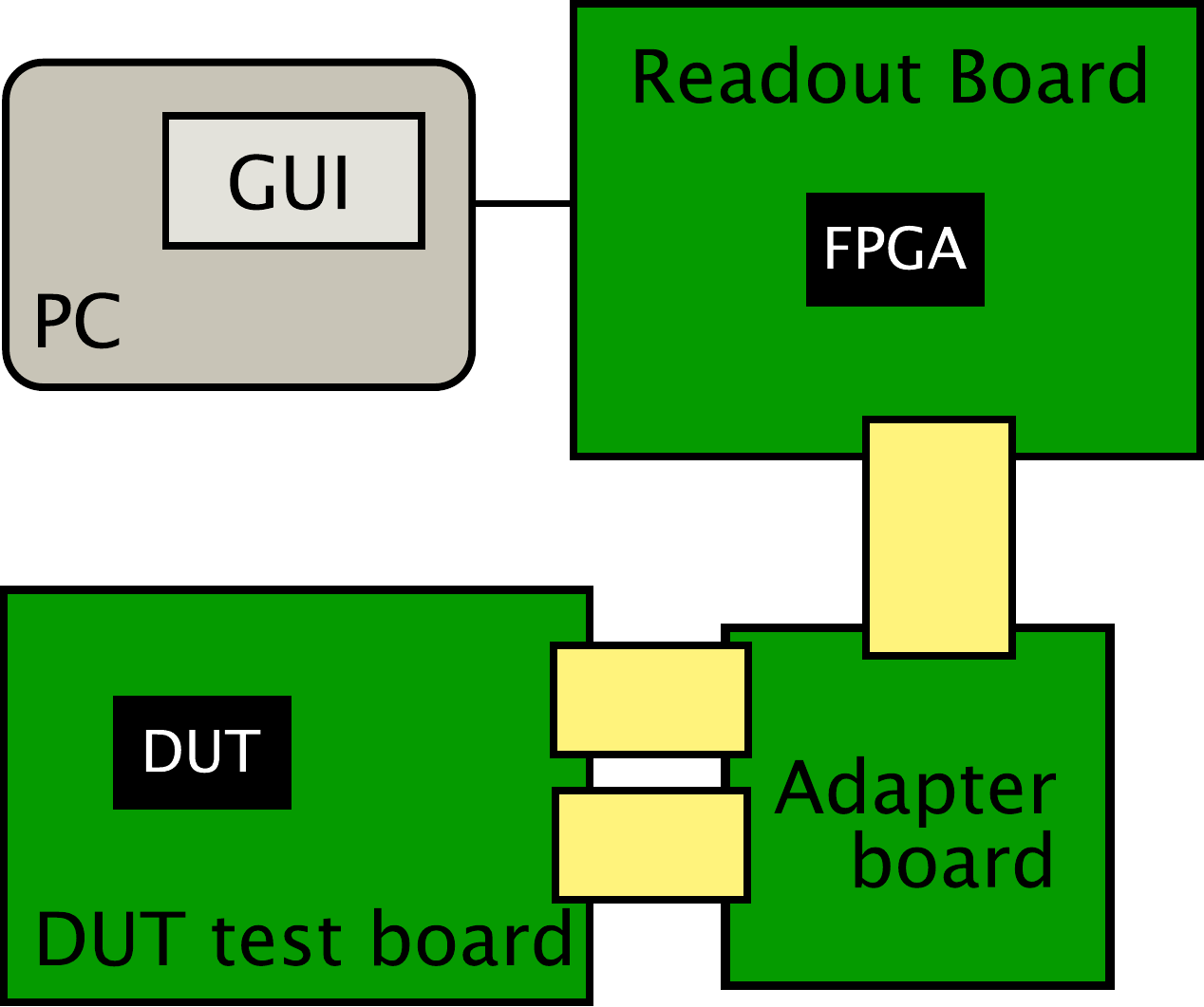}
 % mrf_setup_drawing.pdf: 365x305 pixel, 72dpi, 12.88x10.76 cm, bb=0 0 365 305
 \caption[Schematic drawing of the J\"{u}lich Readout System]{Schematic drawing of the J\"{u}lich Readout System.}
 \label{fig:mrf_overviwe_drawing}
\end{figure}
The J\"{u}lich Readout System is a compact and powerful table top setup which allows a quick testing of new detector components. 
The setup consists of  (see \figref{fig:mrf_overviwe_drawing}):

\begin{itemize}
\item Digital readout board
\item FPGA Firmware
\item MRF Software with GUI
\item Readout PC
\item Device under test
\item Adapter board PCB

\end{itemize}

%the  configured with the , a adapter board PCB to which the DUT will be connected and a readout PC with the MRF software framework installed .

The central component is the FPGA\footnote{\textbf{F}ield \textbf{P}rogrammable \textbf{G}ate \textbf{A}rray} based readout board.
It is connected to the readout PC via an optical connection which is used to receive commands from the user and to send data to the PC for further processing.
The corresponding software infrastructure is implemented within the MVD readout framework (MRF).
The prototype is controlled by the user with a graphical user interface (GUI).
On the other side the readout board is connected to the DUT on its test board.
An intermediate adapter board converts the DUT signal interface to the interface of the readout board.

\section{Basic Concepts}

%To achieve high flexibility to handle different detector components it is necessary to have a strong modular design. 

In the following we will outline the basic concept of the J\"{u}lich Readout System and introduce the individual components comprising the modular design. 

\subsubsection*{Modularity}

Complex procedures are broken down into simpler subtasks which are each handled by individual modules.
%Complex procedures could be broken down into subtasks which could be handled stand alone.
%Each of this subtasks will be treated by a special developed module.
%Modules of more general working subtasks could be easily reused 
To add or change functionality it is only necessary to add a new module or to change an existing one limiting required changes to individual modules which significantly improves stability and simplifies debugging.
%If a new procedure uses similar tasks existing modules could easily be reused.
The modules can also easily be reused when already implemented subtasks reappear in a different context.
This modular concept makes the readout system very flexible and easy to maintain. 

\subsubsection*{Communication Layers}

\begin{figure}[htb]
\centering
\includegraphics[width=\columnwidth]{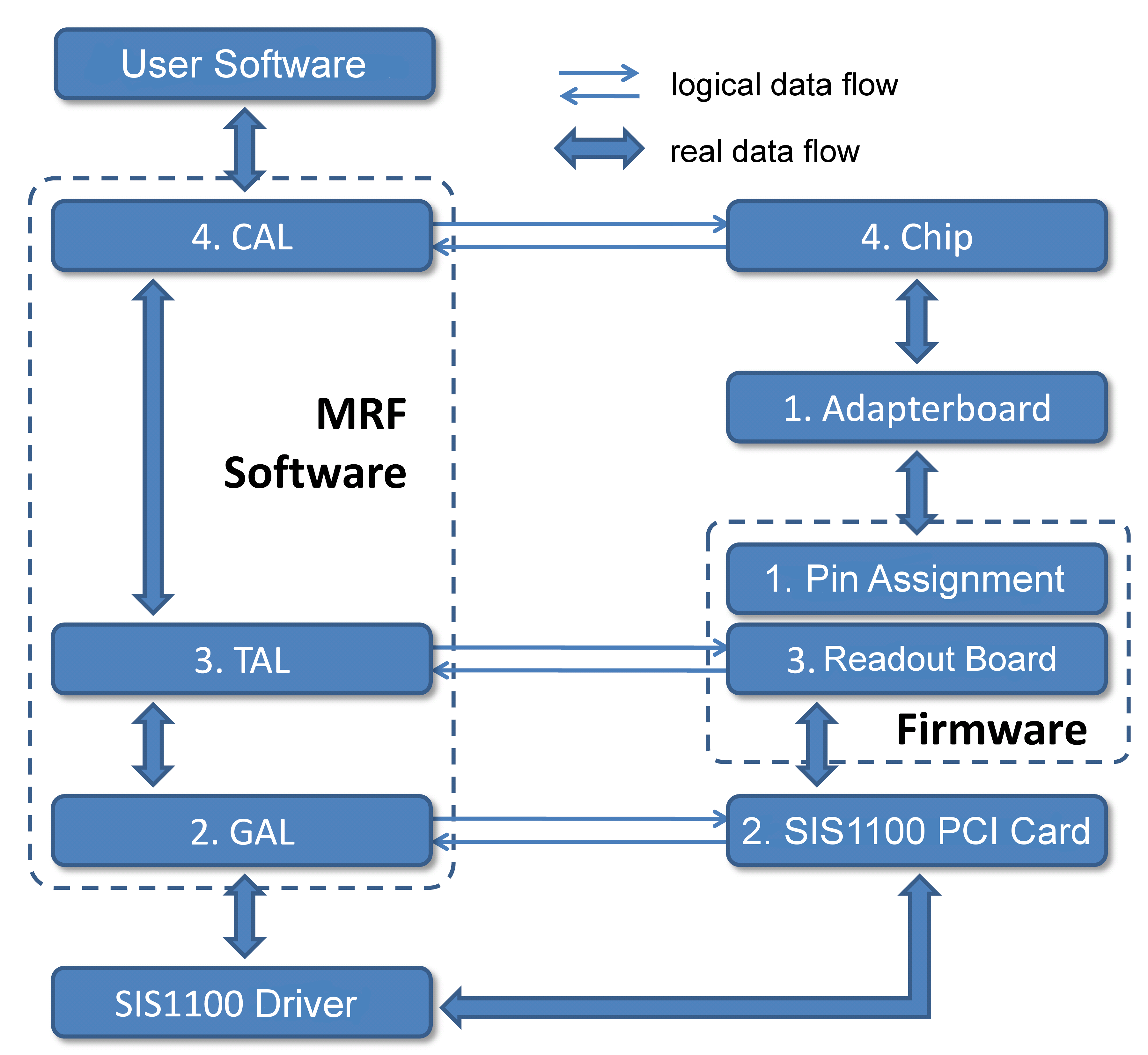}
 % .: 0x0 pixel, 0dpi, 0.00x0.00 cm, bb=
\caption[Schematic overview of the J\"{u}lich Readout System]{Schematic overview of the J\"{u}lich Readout System. 
Every hardware component of the system corresponds to one of the four defined communication layers and has a representation in software or firmware (after \cite{thesis_mertens}).}
\label{fig:appendix_mrf_software_schematic}
\end{figure}

In addition to the modular setup the MRF follows the principle of different abstraction layers based on the OSI model\footnote{\textbf{O}pen \textbf{S}ystems \textbf{I}nterconnection model} \cite{osi}.
Following this principle, the MRF defines four communication layers (see \figref{fig:appendix_mrf_software_schematic}).
Every hardware component of the readout system corresponds to a specific communication layer and has a representation in software or firmware.
%The software equivalent makes it possible to control and communicate with the hardware. 
Each layer uses the functionality of the layer directly below, without knowledge of the exact implementation, and provides a well-defined set of functions to the layer above.
Due to this approach layers can be replaced without changes to the other layers.
The MRF defines the following layers:  

\begin{itemize}
\item Layer 1 - Physical Layer: Establishing of signal connections;
\item Layer 2 - GAL (Generic Access Layer): Communication with readout board via SIS1100 optical connection;
\item Layer 3 - TAL (Transport Access Layer): Access to functionality of the readout board;
\item Layer 4 - CAL (Chip Access Layer): Communication with DUT.
\end{itemize}

In the following, we list the most important components of the readout system and indicate to which communication layer they belong.

\subsubsection*{Adapter Board, Layer 1}

The adapter board establishes the signal connection between the readout board and the DUT.
Due to the usually non-standard interface of the DUT testboard, a specific adapter board needs to be developed for each type of DUT.  

\subsubsection*{Digital Readout Board with Firmware, Layer 1 and 3}

The FPGA based digital readout board is the main hardware component of the system. 
The VHDL\footnote{\textbf{V}ery High Speed Integrated Circuit \textbf{H}ardware \textbf{D}escription \textbf{L}anguage} based firmware configures the FPGA with the implemented functionality (layer 3) and the pin assignment (layer 1) of the DUT interface.

\subsubsection*{Optical Connection to Readout Board, Layer 2}

The data transport from the digital readout board to the PC is done via a 1~Gbit/s optical connection (SIS1100, \cite{sis1100}). 
On the PC side a SIS1100~PCI card is installed.
%on the readout board side we have either a ontop pluggable PCB with SIS1100 functionality or a SIS1100 FPGA core which sends the protocoll over a SFP\footnote{\textbf{S}mall \textbf{F}orm-factor \textbf{P}luggable} gigabit transceiver.
The SIS1100 interface of the digital readout board is implemented as a SIS1100~CMC (Board~A) or as a combination of a SIS1100~FPGA core and an SFP\footnote{\textbf{S}mall \textbf{F}orm-factor \textbf{P}luggable} optical gigabit transceiver (Board~B).

\subsubsection*{MRF Readout Software, Layer 2, 3, 4 }

The MRF readout software (short MRF) is the main software component of the readout system. 
Here the communication layers 2, 3 and 4 are defined (see chapter~\ref{chap:software_framework}).
These layers establish connections to the DUT and the firmware of the digital readout board.

%The framework establishes the connection to the hardware and provides classes to handle the full functionality of the DUT and the readout board. 

\subsubsection*{MRF-Control: Graphic User Interface}

A graphical user interface (GUI) has been developed in C++ based on the Qt framework (see~\cite{qt}) for an easy control by the user. 
%Internally the MRF readout software takes data, which will be saved by the user interface. 
Full configuration and readout routines are available. 
Measured data can be easily saved to disk.
The settings of the DUT and the setup can be saved and reload in special configuration files. 

\subsection{Digital Readout Board A}

\begin{figure}[!th]
	\centering
	\includegraphics[width=\columnwidth]{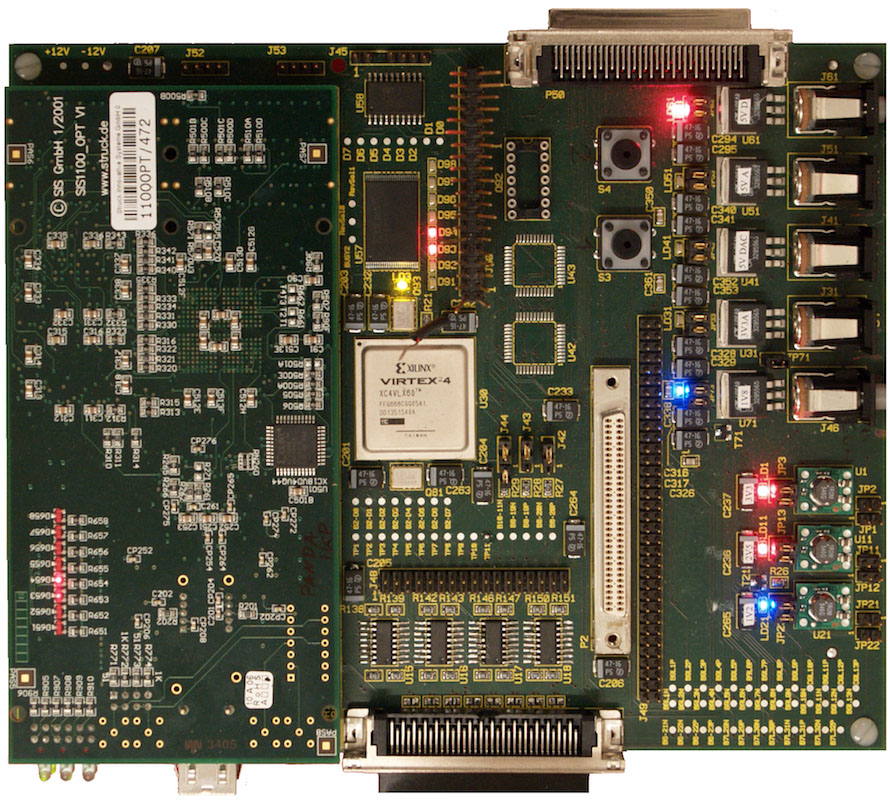}
	% .: 0x0 pixel, 0dpi, 0.00x0.00 cm, bb=
	\caption[The digital readout board A of the Readout System]{The digital readout board A of the Readout System with SIS1100~CMC (left), Xilinx Virtex 4 FPGA (middle), power connectors (right) and 68-pin connector to adapter board (middle right) (from \cite{thesis_mertens}). }
	\label{fig:appendix_mrf_readoutboard}
\end{figure}

The first digital readout board (see \figref{fig:appendix_mrf_readoutboard}) is a custom development and production of the Zentrallabor f\"{u}r Elektronik (ZEL, Central Laboratory for Electronics) of the Forschungszentrum J\"{u}lich \cite{development}. 
A Xilinx Virtex~4~FPGA, configured with the firmware, takes the control of the DUT communication and of the data buffering. 
The readout board is connected to the PC via a 1~GBit/s optical link. 
An optical fiber has the advantage of galvanic separation of PC and readout board.
The SIS1100 protocol is used to transfer the data from the buffer to the PC. 
To implement the SIS1100 protocol on the readout board A a Common Mezzanine Card (CMC) with SIS1100 functionality is connected to the readout board. 
The adapter board is connected by a 68-pin flat band cable.

The main features and external interfaces of the digital readout board are:
\begin{itemize}
\item Xilinx~Virtex~4 XC4VLX60 FPGA
\item 32~MB Xilinx Platform Flash for firmware storage
\item 64~free configurable FPGA input/output pads 
\item 16~LVDS inputs
\item 16~LVDS outputs
\item SIS1100 based optical connection to PC with 1~Gbit/s
\item 50~MHz single ended clock
\end{itemize}

\subsection{Digital Readout Board B}

\begin{figure}[b]
\centering
\includegraphics[width=\columnwidth]{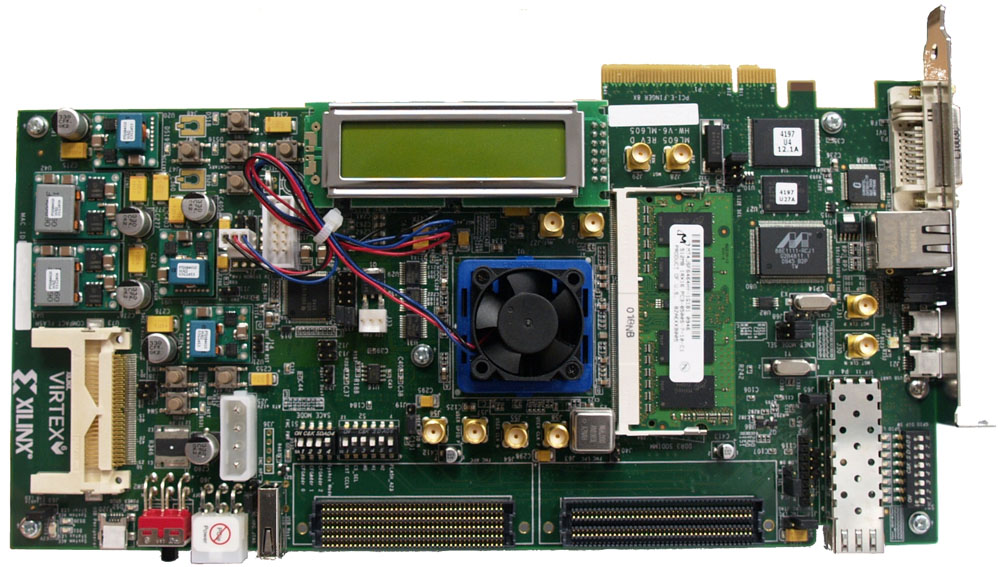}
 % .: 0x0 pixel, 0dpi, 0.00x0.00 cm, bb=
\caption[The digital readout board B ML605 from Xilinx]{The digital readout board~B ML605 from Xilinx with a Virtex~6 (middle), SFP port (up left), FMC connectors to adapter board (up middle), Compact flash slot (right) and DDR3 RAM (next to FPGA on the left).}
\label{fig:appendix_mrf_readoutboard_2}
\end{figure}
To meet the  requirements of an upcoming full size ToPix prototype and online analysis an upgrade of the digital readout board was developed.
The ML605 evaluation board is a commercial product by Xilinx and is comparatively cheap and easy to get.
It contains a DDR3 RAM slot for data storage and a Virtex~6 FPGA (see \figref{fig:appendix_mrf_readoutboard_2}). 
The connection to the PC is established via a SFP gigabit transceiver. 
The SIS1100 protocol is implemented as a IP core into the firmware. 

Specification of the ML605 are:
\begin{itemize}
 \item Xilinx Virtex~6 XC6VLX240T FPGA
\item DDR3 SODIMM memory (expandable)
\item SFP Module connector
\item FMC connectors for Standard Mezzanine Cards
\item LC-Display
\item Compact Flash Card for firmware storage
\item 66~MHz single ended and 200~MHz differential clock
\end{itemize}

\subsection{Firmware}\label{chap:firmware}

%To configure the FPGA and the functionality of the board a firmware is necessary. 
The FPGA on the digital readout board is configured with a firmware which implements the desired functionality and configures the external interfaces.

The firmware for both readout boards is written in VHDL, a hardware description language, and then synthesised with the Xilinx development tools, i.e. translated to a format which is understandable by the FPGA. 

\begin{figure}[!htb]
 \centering
 \includegraphics[width=\columnwidth]{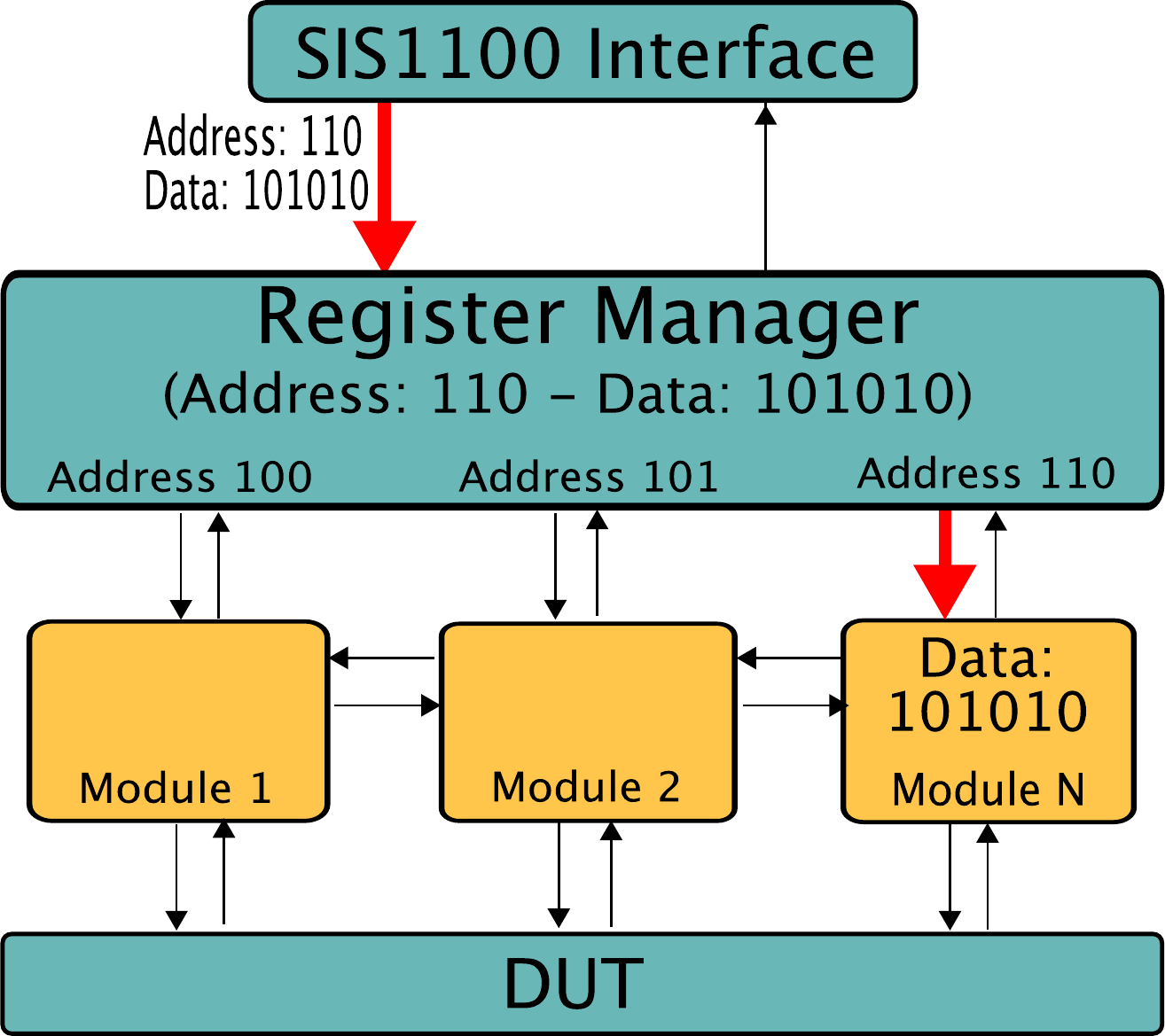}
 % register_manager.pdf: 378x521 pixel, 72dpi, 13.34x18.38 cm, bb=0 0 378 521
 \caption[Schematic firmware setup]{Schematic firmware setup. Data delivered via SIS1100 will be distributed to the modules indicated by the address. The red arrows indicate the path of the shown example data packet.}
 \label{fig:register_manager}
\end{figure}

The firmware is divided into modules which handle different subtask concerning the DUT, the board functionality or the communication.
All modules are connected to the register manager which is itself connected to the SIS1100 interface (see \figref{fig:register_manager}).
Data which arrive via the SIS1100 are formatted in address-data pairs.
The register manager will distribute the data depending on the address to the concerning modules.
If data should be read from the board, the register manager will take the data from the module indicated by its address. 

This modular design makes it easy to adapt the firmware to a different readout board or a different DUT respectively. 
\subsection{MVD Readout Framework}\label{chap:software_framework}

The MRF software framework implements the upper three levels of the communication model.
The MRF is written in C++ and has no external dependencies except the Standard Template Library (STL) to avoid conflicts with other libraries.
This makes the MRF software a decoupled package which can be used in any environment. 

For every communication layer a basic set of commands is implemented to provide the data transport between the layers.
In addition specific commands for the used hardware are implemented.

\subsubsection*{GAL: Generic Access Layer}

The generic access layer provides the basic functionality for communication with the digital readout board.
%It opens and closes connections to endpoints. 
The \textit{read} and \textit{write} commands provide a basic reading and writing to registers over an open connection. 

All more complex commands of the upper layers are implemented using these basic functions. 

%\textbf{GAL Layer for SIS1100}

Including the SIS1100 driver into the GAL more powerful functionality is implemented.
The SIS1100 driver provides DMA transfer\footnote{\textbf{D}irect \textbf{M}emory \textbf{A}ccess}. 
Data from a buffer on the readout board can be send directly to a data container structure in the MRF software.

%In addition the readout board is able to send interrupt signals to the PC where a specific action like ``readout data'' can be triggered.
In addition the readout board is able to send interrupt signals to the PC to indicate certain conditions such as data from the DUT being available. 
A transfer to the PC can then be triggered automatically.

\subsubsection*{TAL: Transport Access Layer}

The transport access layer provides access to the functionality of the firmware.
%For both versions of the digital readout board certain basic functionality is available e.g. the online configuration of the clock generator or controlling of status LEDs.
%For firmware functionality see Chapter \ref{chap:firmware}.
For both versions of the digital readout board certain basic functionality is available such as the online configuration of the clock generator and the controlling of status LEDs.

\subsubsection*{CAL: Chip Access Layer}

The Chip Access Layer implements the functionality of the DUT and needs to be adapted to the specific hardware e.g. configuration of the chip or the readout sequence.
Two versions have been made for the control of the Atlas FE-I3 and another one for the ToPix~v2.
%For both chips the full functionality could be implemented in the software package. 
The complete readout and all testing procedures are available for both chips.

\section{Conclusion}

The J\"{u}lich Readout System has been used to test the \frontend prototypes ToPix~v2 for the MVD and FE-I3 as a fallback solution.
%The FE-I3 and ToPix v2 could be brought to a running state with the readout system. 
The FE-I3 has been tested with the digital readout board A, the ToPix~v2 with both the readout boards A and B.
Due to the modular design of the digital readout system the adaption to the different chips and hardware was very fast.  
%The ToPix v2 has been readout with both hardware versions of the digital readout board.
Different measurements like stability tests with different clock frequencies, threshold scans and tuning of threshold dispersions have successfully been done \cite{pohl} \cite{thesis_mertens}.

%The MRF-A board has been used to readout the Atlas pixel \frontend chip FE-I3 and the ToPix2.

%For the ToPix2 automatic measurements of single pixel thresholds and of the threshold dispersion of all pixel have been done. 

%%%%%%%%%%%%%%%%%%%%%%%%%%%%%%%%%%%%%%%%%%%%%%%%%%%%%%%%%%%%%%%%%%%%%%%%%%%%%%%%%%%%%%%%%%%%%%%%%%%%%%%%%%%%%%%%%%%%%%%%%%%%%%%%%%%

%\section*{Authors}
%\begin{tabbing}
%\hspace{3cm} \=S. Esch\\
%\end{tabbing}

\putbib[lit_appendix]

\clearpage

%% file: appendix/sim-vertexing.tex
\chapter{Details on Vertexing\label{app::vertexing}}
    
\authalert{Author: Simone Bianco}

\begin{figure}[]
\centering
\begin{center}
\includegraphics[width=0.8\columnwidth]{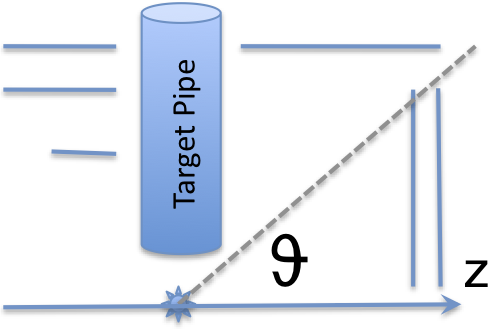}
\caption[Schematic side view of the barrel sensors covering the $\phi$ regions around the target pipe]{Schematic side view of the barrel sensors covering the $\phi$ regions around the target pipe.}
\label{sensors}
\end{center}
\end{figure}

In section~\ref{sec::vertexing} a characterisation of the vertexing performance was shown. Four pions were propagated though the MVD volume from different common start vertices and the achieved resolutions were mapped as a function of the position of the vertex. All the scans showed a different resolution for the x and y coordinates of the reconstructed vertices. When the polar angle ($\theta$) of the pions reaches values at around $50^{\circ}$, the vertex resolution for the x coordinate becomes worse than the y one. This is not expected because of the $\phi$ symmetry of the MVD design. In order to find out the reason of this effect several tests were performed.
First of all the vertex finders have been characterised in detail to find out possible critical behavior. \Tabref{tabRotPOCA} shows the effect of an artificial rotation of $90^{\circ}$ applied to tracks in the x-y plane, before applying the vertexing. This test was performed with four pions of 1~\gevc momentum distributed over the full angular ranges ($\phi \in [0^{\circ},360^{\circ}]$ and $\theta \in [10^{\circ},140^{\circ}]$). The rotation moves the worse reconstruction performance to the y coordinate, excluding a bias introduced by a different treatment of x and y by the vertexing tools.
\begin{table}[!h]
\centering
\begin{tabular}{!{\vrule width 1.5pt}c|c|c!{\vrule width 1.5pt}}
 \noalign{\hrule height 1.5pt} 
 $\sigma$ ($\tcmu$m) & No Rot & $90^{\circ}$ \\ \noalign{\hrule height 1.0pt}
 x & 88.3 & 69.4\\ \hline 
 y & 69.4 & 88.3\\ \noalign{\hrule height 1.5pt}
\end{tabular}
\caption[Results obtained with an artificial rotation of track candidates before the vertex finding (POCA)]
{Results obtained with an artificial rotation of track candidates before the vertex finding (POCA).}
\label{tabRotPOCA}
\end{table}
Several investigations lead to the conclusion that this effect is due to the constrains to the design of the inner barrel layers. Because of the presence of the target pipe and due to mechanical restrictions it was impossible to cover with pixel sensors the full $\phi$ acceptance in the area immediately downstream of the interaction point (a schematic view is shown in \figref{sensors}). 
The design of barrel 1 and 2 does not foresee ``forward" barrel sensors in the top and bottom regions. This translates into the fact that tracks flying into these holes in the acceptance create a smaller number of hits in the MVD and have bigger distances between their first hit and the interaction point (see \figref{pic-Coverage2DHisto-DifferentSetups} and \figref{pic-CoverageResult2D-Distance}). 
These particles will be reconstructed with worse performances, leading to worse vertex determination. Since these critical regions are in the top and bottom regions of the MVD, they effect much more the determination of the x coordinate of the vertex than y.
This appears clearly if one propagates four pions from the interaction point, setting their $\phi$ angles so that they are in an orthogonal disposition in the transverse plane. There one can rotate differently the set of tracks and check the effect on vertexing performances.  
\begin{figure}[]
\begin{center}
\includegraphics[width=0.8\columnwidth]{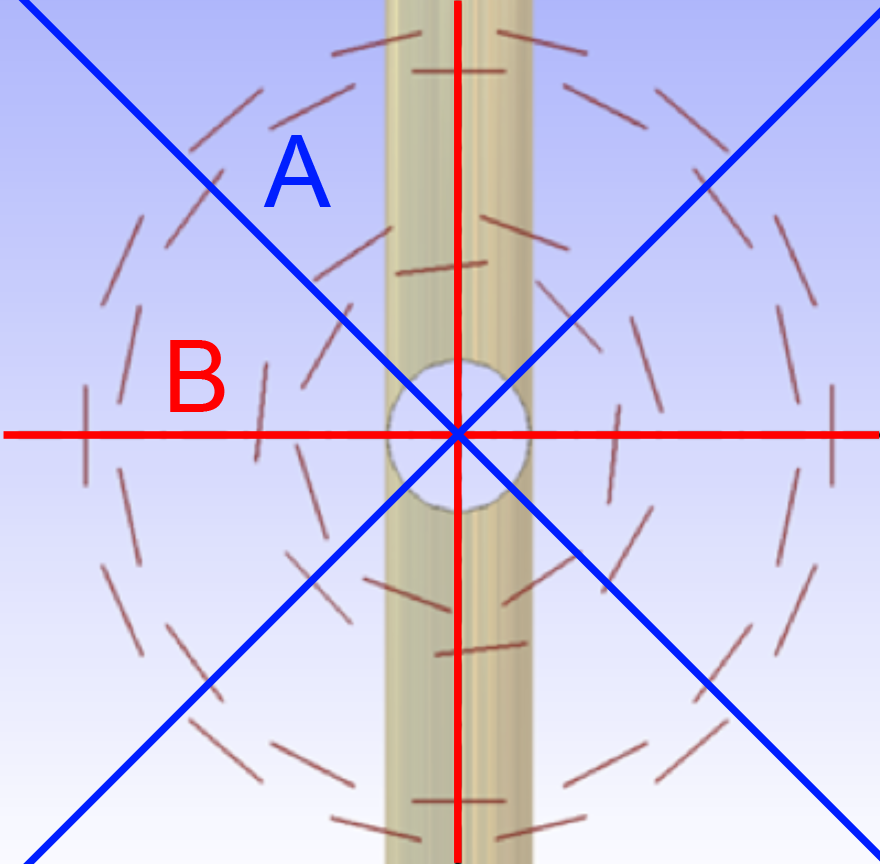}
\caption[Configurations used for the simulations]{Configurations used for the simulations.}
\label{crosses}
\end{center}
\end{figure}
\Figref{crosses} shows the two configurations chosen to investigate the effect of the gaps: in case ``A" the four pions do not enter the target pipe and they all create a hit in the first barrel layer, while in case ``B" two tracks are propagated with initial directions pointing inside the target pipe hole. In both cases $\theta$ was set to $75^{\circ}$ for all the tracks.
\begin{table}[!h]
\centering
\begin{tabular}{!{\vrule width 1.5pt}c|c|c!{\vrule width 1.5pt}}
 \noalign{\hrule height 1.5pt}
 Setup & $\sigma_{x}$($\tcmu$m) & $\sigma_{y}$($\tcmu$m) \\ \noalign{\hrule height 1.0pt}
 A cross, no pipes & 30 & 30\\ \hline 
 A cross, pipes in & 33 & 33\\ \hline 
 B cross, no pipes & 103 & 29\\ \hline 
 B cross, pipes in & 103 & 30\\ \hline 
 All $\phi$, pipes in & 67 & 47\\ \noalign{\hrule height 1.5pt}
\end{tabular}
\caption[Resolutions obtained with the different setups and kinematics]
{Resolutions obtained with the different setups and kinematics. $\theta$ was fixed to $75^{\circ}$.}
\label {tabPOCAxy}
\end{table}
\Tabref{tabPOCAxy} summarises the results obtained with the configurations A and B, both with and without the material of beam and target pipes defined in the geometry. It clearly appears that setup B produces a much worse vertexing performance for the x coordinate, while the configuration A shows the same behaviour for x and y. 
For these last studies a different vertexing tool has been used (the Kinematic Vertex Fit), to cross-check that the x-y different performances were not induced by the vertexing itself. This is the reason why results here are different from the ones obtained in the rest of this section (there the POCA method was used).
The different x-y resolutions appeared to be strictly related to the polar angle of the tracks. For particles flying in the forward region the performance is the same for x and y. At $\theta$ of about $50^{\circ}$ the difference starts to appear. This can be seen in \figref{ThetaScan} where the x and y resolutions obtained with different $\theta$ values (integrating over the full $\phi$) are summarised.

%\section*{Authors}
%\begin{tabbing}
%\hspace{3cm} \=S. Bianco\\
%\end{tabbing}

%% file: appendix/aliTech.tex
\chapter{Alignment}

\section{Introduction}
\authalert{Author: Francesca De Mori}

In track reconstruction and physics analysis the detector alignment is a 
mandatory task.
An accurate determination of a large number of parameters is required to 
allow precise track and vertex reconstruction. 

The momentum resolution of reconstructed tracks depends in part on the 
alignment of the detectors in space. It should not be degraded significantly
(commonly 20$\%$) with respect to the resolution expected in case of the 
ideal geometry without misalignment.

This implies that the detectors positions must be known to an accuracy 
better than 20$\%$ of the intrinsic resolution. This corresponds to 
an alignment precision of 10~$\tcmu$m or better in the bending plane. 

\begin{figure}[!ht]
\begin{center}
\includegraphics[width=0.45\textwidth]{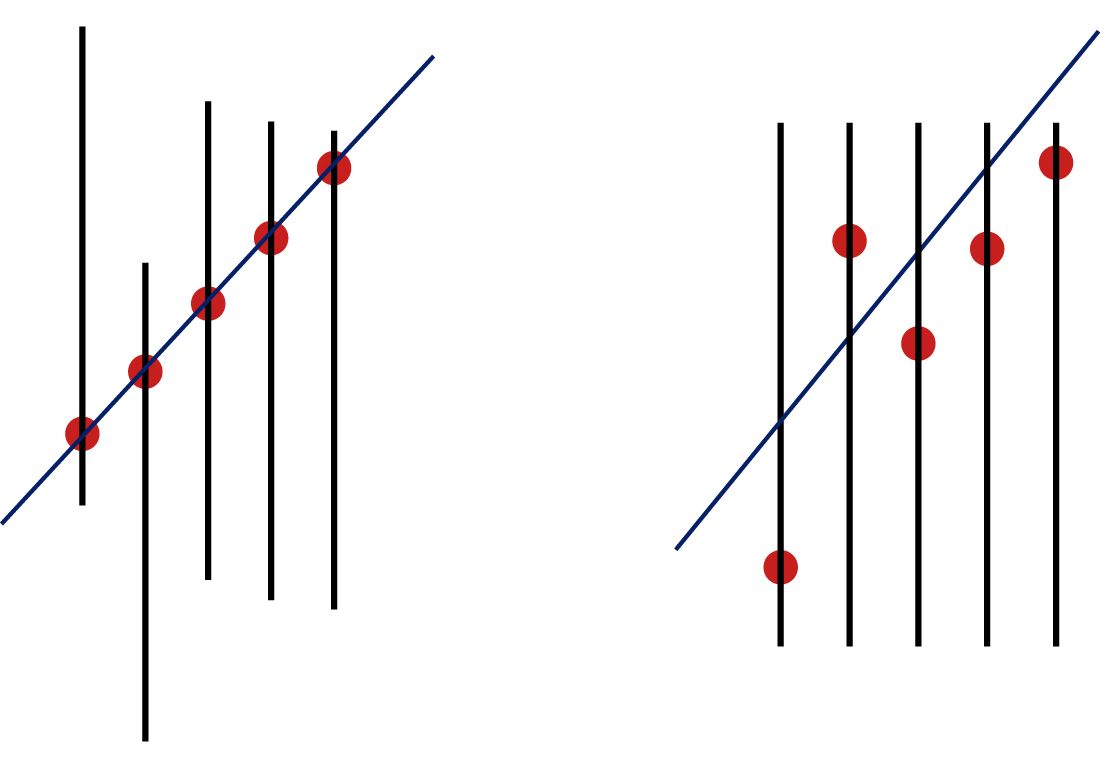}
\caption
[The basic alignment problem: the correct module geometry must be
deduced by studying the deviation of the hit positions with
respect to the tracks]
{The basic alignment problem: the correct module geometry (left) must be
deduced by studying the deviation of the hit positions (red points) with
respect to the tracks.}
\label{fig:aliGeneral}
\end{center}
\end{figure}

Optical surveys and full mapping of MVD modules and super-modules can be 
a very useful starting point and a constraint in the alignment procedure 
but do not provide the required accuracy and cannot correct the time 
dependent changes of the geometry after the tracker installation.
 Module position fluctuations (O(10~$\tcmu$m)) and  barrel/stave/disk 
deformation (O(100~$\tcmu$m)) can occurr.
Therefore alignment algorithms based on the track reconstruction should 
be used for the determination of the sensor positions and orientations.
The alignment procedure aims to determine the positions of the modules of 
each sub-detector. The basic problem is sketched in \figref{fig:aliGeneral}.
 
%Each silicon module as a rigid body has six degrees 
%of freedom (three of translation, three of rotation), neglecting
%module deformations such as bending and twisting.This
%means that the silicon modules (176 Pixel  and 140  Strip  modules) alone 
%have almost 2000 DoF, at least an order of magnitude less than LHC Inner 
%Detectors \cite{Atlas}\cite{Alice}\cite{Cms}.
% A track-based alignment algorithm  performs minimization of the 
%$\chi ^2$ of the hit-track residuals with respect to the parameters 
%describing the sensor position\cite{Blobel}. 

The barrels and the disks require different strategies and the last 
is the more difficult task.
The best candidates for barrel layers alignment are cosmic rays of 
high momentum(p $>$ 3~\gevc) with or without magnetic field. 
The availability of collision track samples will be investigated in 
particular to align the MVD disks since they are less often 
traversed by cosmic ray tracks.

\section{General 
Strategies and Techniques for Vertex/Tracking Detectors}
\authalert{Author: Alessandra Filippi}

Any tracking system can be considered as an assembly of separate modules,
whose position can be displaced during assembling, during detector 
maintenance and during data taking periods as well, due to possible varying
environmental conditions (temperature, humidity, elastic strains, gravitational
saggings, etc.) Therefore,
alignment procedures are needed to detect and correct possible distortions.
Any deformation of the mechanical structure of the modules or their
mislocation from the nominal position implies a depletion of the particle 
position determination. Therefore, mapping and recovering the position of the
any tracking detector (especially the vertex ones) 
is of central importance to preserve the intrinsic spatial resolution
of the detector during a long data taking.

The alignment of large detectors in particle physics usually requires the 
determination and tuning of a large number of alignment parameters, typically
not smaller than 1000. 
Alignment parameters for instance define the
space coordinates and orientation of detector components. Since they are common
to all data sets of real measurements produced by the tracking system, they are
called {\it global} parameters. Usually alignment parameters are corrections
to default values (coming from a topological survey following the detector
installation), so they are expected to be small and the value zero is their
initial approximation. Alignment parameters are usually locally defined, 
i.e.~in each module reference system.

\begin{figure*}[!ht]
\centering
\includegraphics[width=0.75\textwidth]{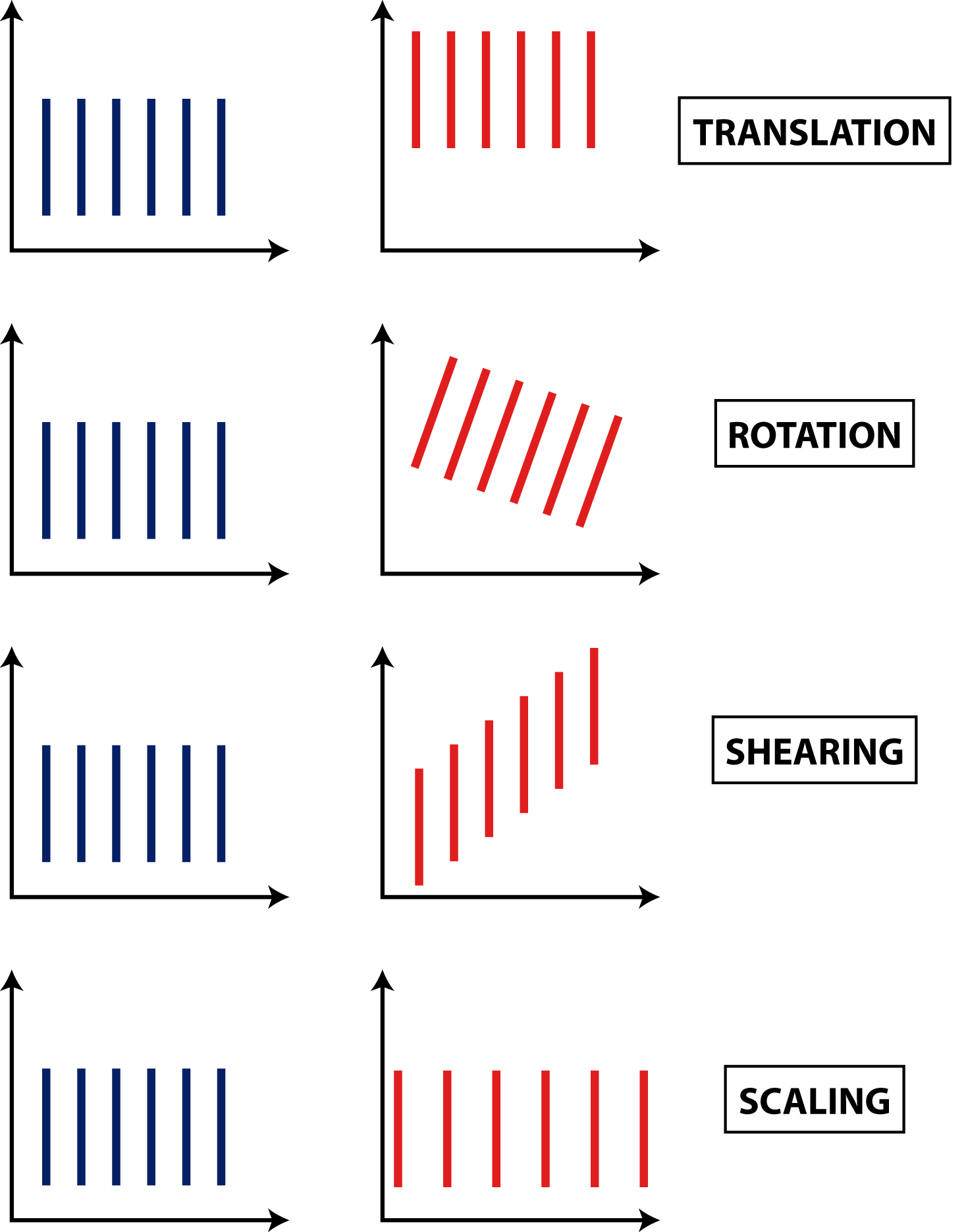}
\caption
[Schematic view of the four basic types of linear transformations]
{Schematic view of the four basic types of linear transformations.}
\label{fig:transform}
\end{figure*}

In a three-dimensional space, the most general set of
alignment parameters able to describe any displacement is constituted by
12 figures:
\begin{itemize}
\item three rescaling of coordinate axes,
\item three global traslations,
\item three rotations,
\item three shearings.
\end{itemize}
The four basic types of linear displacements are shown in 
\figref{fig:transform}.

Under certain conditions some of these parameters 
may be fixed and the number of
global parameters can be reduced. For instance, for a stack of rigid modules 
disposed vertically with respect to the beam axis ($z$, as can be  a
stack of forward moduli), the rescaling along 
$x$ and $y$, as well as a $xy$ shearing, would imply a deformation of the 
modules and of their pitch and so these degrees of freedom can be fixed. In
this particular geometrical layout, also, shearings and rotations basically 
coincide. 

Alignment procedures are usually data-driven, so that displacements are
determined comparing the measured values of hits by tracks from physics
interactions and from cosmics, and fitted track coordinates. The deviations 
between the fitted and the measured data, the {\it residuals}, are used to
determine the alignment parameters by means of proper least squares fits. 
Tracks are also defined through a series of parameters, which for instance
describe their slopes or curvatures, and are valid for each single track only.
They are called {\it local} parameters, to distinguish them from the global,
alignment, ones. Despite their locality, track parameters 
must be defined in the global
(i.e.~the whole detector) reference system. 
Together with alignment parameters,
they define the complete solution of the minimisation procedure. 

In general, the alignment strategy of a stack of tracking modules proceeds 
through a series of stages, which can be shortly summarised as follows:
\begin{enumerate}
\item{\bf Precision Mechanical Assembly:} The modules are accurately assembled and
mounted onto high precision base supports. The mounting tolerances define the
scope of the possible misalignment: for instance, if a module can be placed 
in a box with a 20~$\tcmu$m accuracy, one can expect misalignements to be 
distributed as a Gaussian centered on zero and 20~$\tcmu$m  wide.   
\item{\bf System Metrology:} At the time of construction, coordinated measuring
machines must be used to survey and fully map the surfaces of individual 
modules. Also mounted frames coordinates have to be fully surveyed, in standard
conditions of temperature, pressure and humidity. Results from survey have then
to be compared to nominal designs fixing macroscopic deviations. The outcomes 
from the mechanical surveys will provide the reference positions of all 
components.
\item{\bf Software Alignment:} If the quality of mechanical survey is 
good enough,
alignment may be performed offline, during data taking or even afterwards.
Otherwise, a fast alignment engine must be available to tune-up the starting
alignment configuration before data taking, and possibly run whenever critical 
conditions are met. In general, the software alignment must be flexible
enough to account for a wide range of possible occurrences arising during
the data taking. Statistical methods are used to 
recover the actual position of the modules, as will be described in the following.   
\item{\bf Alignment Monitoring:} A monitoring of the quality of detector alignments
during data taking can sometimes be advisable; a monitoring 
deamon checking periodically the quality
of track residuals during data taking could automatically trigger
the re-running of software alignment procedures to tune online the alignment
quality.
\end{enumerate}

In the following, the Software Alignment stages will be described in some
detail presenting the available tools commonly used in the alignment procedures
of some particle physics vertex detectors.

\subsection{Software Alignment of Tracking Modules and Modern Tools}

The analysis of the residual distributions and their minimisation
allow to position detector modules in their actual locations with an accuracy
which is only limited by the available statistics of tracks to be fitted,
and the detector intrinsic resolution.

Track residuals are functionally dependent on the both global and local 
parameters. The problem of alignment is in general formulated as a mathematical
problem of minimisation of the functional obtained by the squares of residuals
of both global and local parameters:
\begin{equation}
\chi^2 = \sum_{\mathrm{events}}\sum_{\mathrm{tracks}}\sum_{\mathrm{hits}} \Delta^2_i/\sigma^2_i
\end{equation}
where $\Delta_i = x_{\mathrm{fit}}-x_{\mathrm{meas}}$ is the hit residual, and $\sigma_i$ the 
accuracy of the hit coordinate measurement. The fitted track position $x_{\mathrm{fit}}$
has as well a functional dependence both of the geometrical misalignment of the module
and of the track parameters. A track model must be chosen to define the track
dependence on the free parameters. For instance, in the simplest linear model, 
the observables have expectation values
which are linearly dependent on the set of parameters.
In this case, the fitted track position may be given by the linear sum of
two contributions,
one pertaining to the $n$ track local parameters $a_i$ and one to the $m$ global
alignment ones, $\alpha_j$, together with their derivatives, $d_i$ and 
$\delta_j$ respectively:
\begin{equation}
  x_{\text{fit}} = \sum_{i=1}^n a_id_i + \sum_{j=1}^\nu \alpha_j\delta_j.
\end{equation}
Using a linear track model, the set $\mathbf{A} = \{a_i\}+\{\alpha_i\}$ 
of parameters minimising 
the $\chi^2$ quadratic functional
is obtained solving the set of normal linear equations of least squares, in
matrix form $\mathbf{C=Ab}$. $\mathbf{C}$ is a symmetric $N\times N$ matrix 
($N = n+m$), and $\mathbf{b}$ a $N$-vector, both of them being given by sums 
of independent contributions from each measurement.  
The solution of this system of linear equations poses 
the major problem of the alignment task, since the
number of parameters to be determined (i.e.~the dimension $N$
of the $\mathbf{C}$ matrix of the normal equations to be inverted, to find
the $\mathbf{A}$ covariance matrix) is at least of the 
order of $10^3$. This large parameter multiplicity requires a number
of tracks to be fitted by the alignment procedure of the order of $10^4-10^6$,
to get a reasonable accuracy. However, this is not a severe complication 
as $\mathbf{C}$
and $\mathbf{b}$ already contain the complete information from single 
measurements, which thus are not needed to be stored any further. 
A practical limit to
the number of parameters is given by the time necessary to perform the matrix
inversion, which is proportional to $N^3$, and to the storage/memory space 
($\propto N^2$, since double precision calculations are often mandatory). 
This means that a number of parameters larger than $10^4$ starts
being critical, if the matrix equation is to be solved by standard methods. 
With 10000 parameters, a typical computing time for matrix inversion
is about six hours, which becomes almost one year for $N=10^5$.

For the \panda MVD, the typical number of expected degrees of freedom (for
pixels stations and strips) is about 5000, about half as much as LHC 
experiments. Thus, at least execution times should not represent a problem. 

However, the normal equation matrix could 
easily become singular, in case of missing data (for instance, dead channels), 
or in case of a complete correlation between 
parameters (for instance, when tracks are
projected into themselves by geometrical transformations). In this case the
matrix inversion fails and the 
analytical solution of the problem is prevented. This requires alternative 
approximate or iterative
approaches, or special numerical algorithms to cope with the problem.

A first rough approximation is to perform a series of track 
fits which initially
neglect alignment parameters: in this case the fits depend only on the reduced
set of local parameters, and are easier to perform. However, this method is not
statistically correct since the track model is basically wrong and the obtained
results are biased, and possibly also unstable as affected by detector 
inefficiencies not taken into account. Through several iterations the method 
can effectively reduce the
residuals (though not necessarily converging), but the alignment 
parameters applied as corrections in iterative fits
will still be  biased. Moreover, the correlations between local and global 
fits, present in the functional, in this way are completely ignored.
 
A better solution to reduce the problem dimension is to use tracks extrapolated
from outer detectors, in order to fix at least the parameters of thet tracks. In this 
case one can align a given subdetector (for instance, stack of modules along
a given axis or in a detector sector) with respect to other parts of the system. 
The method has the
advantage of providing the alignment of the detector as a whole with high
precision. On the other hand, it works
effectively only when the detector dimensions are not too large, because with
large extrapolation lenghts the precision of the extrapolated track parameters
can soon become much worse than that for tracks reconstructed in the 
subdetectors only, due to multiple scattering effects and misalignements of the
reference detector out of control. This is actually the method adopted for
the first stage of alignment of the FINUDA vertex detector \cite{re:finuda},
composed by two layers of double-sided silicon microstrip 
modules (50~$\tcmu$m pitch) arranged in cylidrical geometry around the beam axis.
The applied procedure was based on a minimisation engine which aligned, by means of
straight cosmic rays, stacks of four modules at a time, two per each of the
two detector layers (starting from those crossed by a larger number of straight cosmic
rays). Then adjacent single modules of the inner/outer layers were aligned, one at a time, 
with respect to the already aligned stacks. The error in the alignment of a single module,
assuming multiple scattering effects are negligible, is given by the sum in quadrature of
the modulus intrinsic resolution ($\sigma_{\mathrm{intr}}$) and the alignment precision 
($\sigma_{\mathrm{align}}$). The minimum number of tracks needed to perform a reliable minimisation 
is determined by the condition $(\sigma_{\mathrm{align}}/\sigma_{\mathrm{intr}})^2 \ll 1$.
Once a satisfactory relative
alignment of the two layers was achieved, they were then 
aligned with respect to the outer 
tracker (a straw tube array), which could provide information on global
rotations and tilts of the vertex detector as a whole. With this method,
after having taken into account multiple scattering and sagging effects, a 
resolution of the vertex detector moduli of about 20~$\tcmu$m could be 
achieved for the $\phi$ coordinate, and better than 30~$\tcmu$m for $z$. 
%They result from an alignment precision of about 3~$\tcmu$m for the $\phi$
%coordinate, and of $\sim 5 \tcmu$m for $z$. 
For the sake of comparison, the
starting mechanical tolerance of the holder devices was $\sim 200$~$\tcmu$m.

The most efficient alignment method consists, of course, in 
making a simultaneous
fit of all global and local parameters, by means of dedicated algorithms
which could overtake the above mentioned matrix inversion problem.
To this purpose, the 
{\fontencoding{OT1}\fontfamily{ppl}\selectfont MILLEPEDE II} 
programme was developed \cite{re:blobel0, re:blobel1, re:blobel2},
and it is presently widely used for the alignment of large particle physics
detectors (among them, ALICE ITS \cite{re:aliceits}, LHCb VELO
\cite{re:lhcb}, HERA-B OTR \cite{re:herab, re:herab2} and H1 
SVD \cite{re:h1}). 
Also FINUDA used it for fine tuning purposes). 
The {\fontencoding{OT1}\fontfamily{ppl}\selectfont MILLEPEDE II} 
programme is based on partition techniques which split the
problem of inversion of a large matrix into a set of smaller inversions, 
exploiting the feature of the $\mathrm{C}$ matrix of being very sparse,
presenting many null submatrices whose computation can be skipped reducing
execution times. Depending on the fraction of vanishing submatrix elements,
alignment problems with a number of global parameters of the order of 10$^5$
can be attacked with standard computing power. Of course, the matrix size
ought to be conveniently reduced, whenever possible, before starting the
minimisation procedure. To this purpose, some preliminary iterative steps are
needed, in order to:
\begin{itemize}
\item linearise the equations (by means of Taylor expansions), if the 
dependence on the global parameters happens not to be linear, and
\item remove outliers (i.e.~data with large residuals) 
from the data by proper cuts (to get tighter in
several iterations), of properly down-weighting them.
\end{itemize}
Sometimes the minimisation is not sufficient to achieve an optimal alignment,
if some degrees of freedom are undefined, which call from additional 
conditions, expressed by linear (or linearised) constraints. The 
{\fontencoding{OT1}\fontfamily{ppl}\selectfont MILLEPEDE II} programme can
handle additional these linear constraints (and parameters) exploiting the Lagrange 
multiplier method. 

The {\fontencoding{OT1}\fontfamily{ppl}\selectfont MILLEPEDE II} 
package is originally built by two parts. The first
({\fontencoding{OT1}\fontfamily{ppl}\selectfont MILLE}) is called in 
user programs and basically prepares the data files 
(containing measurements and derivatives with respect to local and global 
parameters, in their proper reference systems). These files are
processed by the second, standalone, program 
({\fontencoding{OT1}\fontfamily{ppl}\selectfont PEDE}), which performs the fits
and determines the global parameters, steering the solutions. The solutions for
local parameters, which are track dependent, are not needed. The package, originally
written in FORTRAN-77,
is available for FORTRAN and C-based aligment procedures and can be easily
embedded in existing alignment frameworks (AliMillePede in AliRoot, LHCb 
VELO alignment software package \cite{re:velo1,re:velo2,re:velo3}).

The  {\fontencoding{OT1}\fontfamily{ppl}\selectfont MILLEPEDE II} 
package allows to
achieve very satisfactory alignment conditions. 
Using  this framework, the forward stations of the LHCb VELO detector, 
for instance, 
could be aligned with a precision
of 3~$\tcmu$m on $x$ and $y$ translational coordinates
(to be compared with an intrisic resolution larger than 5~$\tcmu$m),
and with an accuracy of 0.4~mrad on the rotational degrees of freedom 
(around the $z$ axis). The alignment of this
detector was performed using tracks from LHC pp collisions and tracks 
from beam halo particles crossing the
whole detector. Such an approach can also be envisaged for the 
\panda MVD forward disks. 
\vfill

\putbib[lit_appendix]
\clearpage

%% file: main/acronyms.tex
% acronyms.tex
%
\addcontentsline{toc}{chapter}{List of Acronyms}
\begin{acronym}[HERA-B OTR]
\acro{3D}{3-Dimensional}
\acro{ADC}{Analog to Digital Converter}
\acro{ALICE}{A Large Ion Collider Experiment}
\acro{ALICE ITS}{ALICE Inner Tracking System}
\acro{APV}{Analog Pipeline Voltage mode}
\acro{ASIC}{Application Specific Integrated Circuit}
\acro{ATCA}{Advanced Telecommunications Computing Architecture}
\acro{ATLAS}{A Toroidal LHC ApparatuS}
\acro{BERT}{Bit Error Rate Test}
\acro{BTeV}{B meson TeV}
\acro{CAL}{Chip Access Layer}
\acro{CCD}{Charge-Coupled Device}
\acro{CCU}{Chip Control Unit}
\acro{CERN}{Conseil Europ\'een pour la Recherche Nucl\'eaire}
\acro{CMC}{Common Mezzanine Card}
\acro{CMOS}{Complementary Metal Oxide Semiconductor}
\acro{CMP}{Chemical Mechanical Polishing}
\acro{CMS}{Compact Muon Solenoid}
\acro{CNT}{Counter}
\acro{COSY}{COoler SYnchrotron}
\acro{CR}{Configuration Register}
\acro{CRC}{Cyclic Redundant Check}
\acro{CRCU}{Column Readout Control Unit}
\acro{CSA}{Charge Sensitive Amplifier}
\acro{CT}{Central Tracker}
\acro{Cz}{Czochralski}
\acro{DAC}{Digital-Analog Converter}
\acro{DAFNE}{Double Annular ring For Nice Experiments}
\acro{DAQ}{Data Acquisition}
\acro{DC}{Direct Current}
\acro{DC-DC}{DC-to-DC Converter}
\acro{DCS}{Detector Control System}
\acro{DESY}{Deutsches ElektronenSYnchrotron}
\acro{DIRC}{Detector for Internally Reflected Cherenkov Light}
\acro{DMA}{Direct Memory Access}
\acro{DPM}{Dual Parton Model}
\acro{DVCS}{Deeply Virtual Compton Scattering}
\acro{ECC}{Error Correction Code}
\acro{EDM}{Electrical Discharge Machining}
\acro{EMC}{Electromagnetic Calorimeter}
\acro{ENC}{Equivalent Noise Charge}
\acro{EPICS}{Experimental Physics and Industrial Control System}
\acro{FAIR}{Facility for Antiproton and Ion Research}
\acro{FBK - ITC}{Fondazione Bruno Kessler - Istituto Trentino di Cultura}
\acro{FE}{Front-End}
\acro{FEE}{Front-End Electronics}
\acro{FEM}{Finite Element Method}
\acro{FET}{Field-Effect Transistor}
\acro{FIFO}{First In First Out}
\acro{FINUDA}{FIsica NUcleare a DAFNE}
\acro{FNAL}{Fermi National Accelerator Laboratory}
\acro{FPGA}{Field Programmable Gate Array}
\acro{FS}{Forward Spectrometer}
\acro{FTS}{Forward Tracking System}
\acro{FZ}{Floating Zone}
\acro{FZJ}{Forschungszentrum J\"ulich}
\acro{GAL}{Generic Access Layer}
\acro{GBLD}{GigaBit Laser Driver}
\acro{GBT}{GigaBit}
\acro{GBTX}{GigaBit Transceiver}
\acro{GBTIA}{GigaBit TransImpedance Amplifier}
\acro{GBT-SCA}{GigaBit Slow Control ASIC}
\acro{GEM}{Gas Electron Multiplier}
\acro{GR}{Guard Ring}
\acro{GSI}{Gesellschaft f\"ur Schwerionenforschnung}
\acro{GUI}{Graphical User Interface}
\acro{H1 - SVD}{H1 Silicon Vertex Detector}
\acro{HERA}{Hadron-Elektron-Ring-Anlage}
\acro{HERA-B OTR}{HERA-B Outer TRacker}
\acro{HESR}{High Energy Storage Ring}
\acro{HL}{High Luminosity (HESR operation mode)}
\acro{HR}{High Resolution (HESR operation mode)}
\acro{HR}{High Resistivity (wafers)}
\acro{HV}{High Voltage}
\acro{I$^2$C}{Inter-Integrated Circuit}
\acro{INFN}{Istituto Nazionale di Fisica Nucleare}
\acro{IP}{Interaction Point}
\acro{ITME}{Institute of Electronic Materials Technology}
\acro{LD}{Laser Driver}
\acro{LED}{Light Emission Diode}
\acro{LENA}{Laboratorio Energia Nucleare Applicata}
\acro{LHC}{Large Hadron Collider}
\acro{LHCb}{Large Hadron Collider beauty}
\acro{LHCb VELO}{LHCb VErtex LOcator}
\acro{LNL}{Laboratori Nazionali di Legnaro}
\acro{LR}{Low Resistivity (wafers)}
\acro{LVDS}{Low-Voltage Differential Signaling}
\acro{MC}{Monte-Carlo}
\acro{MCP PMT}{Multi-Channel Plate PMT}
\acro{MDC}{Module Data Concentrator}
\acro{MDT}{Mini Drift Tubes}
\acro{MicroTCA}{Micro Telecommunications Computing Architecture}
\acro{MIP}{Minimum Ionizing Particle}
\acro{MMB}{MVD Multiplexer Board}
\acro{MOS}{Metal Oxide Semiconductor}
\acro{MPV}{Most Probable Value}
\acro{MR}{Middle Resistivity (wafers)}
\acro{MRF}{MVD Readout Framework}
\acro{MVD}{Micro Vertex Detector}
\acro{NIEL}{Non-Ionizing Energy Loss}
\acro{OSI}{Open Systems Interconnection}
\acro{PANDA}{antiProton ANnihilation at DArmstadt}
\acro{PCB}{Printed Circuit Board}
\acro{PCI}{Peripheral Component Interconnect}
\acro{PCIe}{Peripheral Component Interconnect Express}
\acro{p.d.f.}{Probability Density Function}
\acro{PD}{Photo Diode}
\acro{PDG}{Particle Data Group}
\acro{PID}{Particle Identification}
\acro{PLL}{Phase Locked Loop}
\acro{PMT}{Photomultiplier}
\acro{POCA}{Point of Closest Approach}
\acro{PRBS}{Pseudo Random Bit Sequence}
\acro{PVD}{Physical Vapour Deposition}
\acro{QCD}{Quantum Chromo Dynamics}
\acro{RF}{Radio Frequency}
\acro{RICH}{Ring Imaging Cherenkov Counter}
\acro{RMS}{Root Mean Square}
\acro{SDS}{Silicon Detector Software}
\acro{SEU}{Single Event Upset}
\acro{SFP}{Small Form-factor Pluggable}
\acro{SIMS}{Secondary Ion Mass Spectrometry}
\acro{SLHC}{Super Large Hadron Collider}
\acro{SLVS}{Scalable Low Voltage Signaling}
\acro{SMD}{Surface Mount Device}
\acro{SODA}{Synchronisation Of Data Acquisition}
\acro{SP}{Separation Power}
\acro{SR}{Shift Register}
\acro{STI}{Shallow Trench Insulation}
\acro{STT}{Straw Tube Tracker}
\acro{TAL}{Transport Access Layer}
\acro{TIA}{TransImpedance Amplifier}
\acro{TMR}{Triple Modular Redundancy}
\acro{TOF}{Time-of-Flight}
\acro{ToPix}{Torino Pixel}
\acro{ToT}{Time-over-Threshold}
\acro{TRx}{Transceiver}
\acro{TS}{Target Spectrometer}
\acro{UI}{Unit Interval}
\acro{UMC}{United Microelectronics Corporation}
\acro{UrQMD}{Ultra-relativistic Quantum Molecular Dynamic}
\acro{VCSEL}{Vertical-Cavity Surface-Emitting Laser}
\acro{VHSIC}{Very-High-Speed Integrated Circuits}
\acro{VHDL}{VHSIC Hardware Description Language}
\acro{VL}{Versatile Link}
\acro{VMC}{Virtual Monte-Carlo}
\acro{VME}{Versa Module Eurocard}
\acro{WASA}{Wide Angle Shower Apparatus}
\acro{ZEL}{Zentrallabor f\"{u}r Elektronik at FZJ}
\end{acronym}
\vfill